\documentclass[aps,floats,twocolumn,prx,showpacs,10pt,longbibliography]{revtex4-1}
\usepackage{graphicx}
\usepackage{epstopdf}
\usepackage[colorlinks=true,urlcolor=blue,citecolor=blue,linkcolor=blue,breaklinks=true]{hyperref}
\usepackage{color}
\usepackage{colortbl}
\usepackage{setspace}
\usepackage[utf8]{inputenc}
\usepackage{amsmath}
\usepackage{placeins}
\usepackage{mathtools}

\newcommand{\Tr}{\operatorname{Tr}}

\newcommand{\omsf}{\omega_{\mathrm{sf}}}

\definecolor{highestT}{rgb}{1.0, 0.0, 0.0}
\definecolor{highT}{rgb}{1.0, 0.408, 0.0}
\definecolor{middleT}{rgb}{0.0, 1.0, 1.0}
\definecolor{lowT}{rgb}{0.0, 0.545, 1.0}
\definecolor{lowestT}{rgb}{0.158, 0.0, 1.0}

\begin{document}

\title{\texorpdfstring{Tracking the Footprints of Spin Fluctuations:\\A MultiMethod, MultiMessenger Study of the Two-Dimensional Hubbard Model}{Tracking the Footprints of Spin Fluctuations: A MultiMethod, MultiMessenger Study of the Two-Dimensional Hubbard Model}}

\author{Thomas Sch{\"a}fer$^{a,b,c}$}
\email{t.schaefer@fkf.mpg.de}
\author{Nils Wentzell$^d$}
\author{Fedor {\v S}imkovic IV$^{a,b}$}
\author{Yuan-Yao He$^{d,e}$}
\author{\\Cornelia Hille$^f$}
\author{Marcel Klett$^{f,c}$}
\author{Christian J. Eckhardt$^{g,h}$}
\author{Behnam Arzhang$^i$}
\author{Viktor Harkov$^{j,k}$}
\author{\\Fran{\c{c}}ois-Marie Le R{\'e}gent$^b$}
\author{Alfred Kirsch$^b$}
\author{Yan Wang$^l$}
\author{Aaram J. Kim$^m$}
\author{Evgeny Kozik$^m$}
\author{Evgeny A. Stepanov$^j$}
\author{\\Anna Kauch$^g$}
\author{Sabine Andergassen$^f$}
\author{Philipp Hansmann$^{n,o}$}
\author{Daniel Rohe$^{p}$}
\author{Yuri~M.~Vilk$^l$}
\author{James~P.~F.~LeBlanc$^i$}
\author{\\Shiwei Zhang$^{d,e}$}
\author{A.-M. S. Tremblay$^l$}
\author{Michel Ferrero$^{a,b}$}
\author{Olivier Parcollet$^{d,q}$}
\author{Antoine Georges$^{a,b,d,r}$\vspace{4mm}}

\affiliation{$^a$Coll{\`e}ge de France, 11 place Marcelin Berthelot, 75005 Paris, France}
\affiliation{$^b$CPHT, CNRS, {\'E}cole Polytechnique, Institut Polytechnique de Paris, Route de Saclay, 91128 Palaiseau, France}
\affiliation{$^c$Max-Planck-Institut f{\"u}r Festk{\"o}rperforschung, Heisenbergstra{\ss}e 1, 70569 Stuttgart, Germany}
\affiliation{$^d$Center for Computational Quantum Physics, Flatiron Institute, 162 Fifth avenue, New York, NY 10010, USA}
\affiliation{$^e$Department of Physics, College of William and Mary, Williamsburg, Virginia 23187, USA}
\affiliation{$^f$Institut für Theoretische Physik and Center for Quantum Science, Universität T{\"u}bingen, Auf der Morgenstelle 14, 72076 T{\"u}bingen, Germany}
\affiliation{$^g$Institute of Solid State Physics, TU Wien, A-1040 Vienna, Austria}
\affiliation{$^h$Institut f{\"u}r Theoretische Festk{\"o}rperphysik, RWTH Aachen University, 52074 Aachen, Germany}
\affiliation{$^i$Department of Physics and Physical Oceanography, Memorial University of Newfoundland, St John’s, Newfoundland and Labrador, Canada, A1B 3X7}
\affiliation{$^j$I. Institute of Theoretical Physics, Department of Physics, University of Hamburg, Jungiusstrasse 9, 20355 Hamburg, Germany}
\affiliation{$^k$European X-Ray Free-Electron Laser Facility, Holzkoppel 4, 22869 Schenefeld, Germany}
\affiliation{$^l$D{\'e}partment de Physique, Institut quantique and RQMP, Universit{\'e} de Sherbrooke, Sherbrooke, Qu{\'e}bec, Canada J1K 2R1}
\affiliation{$^m$Department of Physics, King's College London, Strand, London WC2R 2LS, United Kingdom}
\affiliation{$^n$Max-Planck-Institut f\"ur Chemische Physik fester Stoffe, N\"othnitzerstra{\ss}e 40, 01187 Dresden, Germany}
\affiliation{$^o$Department of Physics, University of Erlangen-Nuremberg, 91058 Erlangen, Germany}
\affiliation{$^p$Forschungszentrum Juelich GmbH, Juelich Supercomputing Centre (JSC), SimLab Quantum Materials, Juelich, Germany}
\affiliation{$^q$Universit\'e Paris-Saclay, CNRS, CEA, Institut de physique th\'eorique, 91191, Gif-sur-Yvette, France}
\affiliation{$^r$DQMP, Universit{\'e} de Gen{\`e}ve, 24 quai Ernest Ansermet, CH-1211 Gen{\`e}ve, Suisse}

\date{ \today }

\begin{abstract} 
The Hubbard model represents the fundamental model for interacting quantum systems and electronic correlations. 
Using the two-dimensional half-filled Hubbard model at weak coupling as testing grounds, we perform a comparative study of a comprehensive set of state of the art quantum many-body methods. 
Upon cooling into its insulating antiferromagnetic ground-state,
the model hosts a rich sequence of distinct physical regimes  
with crossovers between a high-temperature incoherent regime, an intermediate temperature metallic regime 
and a low-temperature insulating regime with a pseudogap created by antiferromagnetic fluctuations.
We assess the ability of each method to properly address these physical regimes and crossovers through the computation 
of several observables probing both quasiparticle properties and magnetic correlations, with 
two numerically exact methods (diagrammatic and determinantal quantum Monte Carlo) serving as a benchmark. 
By combining computational results and analytical insights, we elucidate the nature and role of spin fluctuations 
in each of these regimes. 
Based on this analysis, we explain how quasiparticles can coexist with increasingly 
long-range antiferromagnetic correlations, and why dynamical mean-field theory is found to provide a remarkably accurate approximation of local 
quantities in the metallic regime.
We also critically discuss whether imaginary time methods are able to capture the non-Fermi liquid singularities of this fully nested system.
\end{abstract}

\maketitle

\tableofcontents

\section{Introduction}

\subsection{Purpose of this article}
The Hubbard model \cite{Hubbard1963, Hubbard1964, Gutzwiller1963, Kanamori1963} has, for interacting quantum systems, a similar status as the Ising model for 
classical phase transitions and magnetism. It is the simplest possible model that can be considered, which nonetheless captures essential aspects of the physical 
phenomena of interest. 
In relation to materials with strong electronic correlations, the Hubbard model in its simplest form (especially, with a single 
band) is at best an approximation to reality.  
However, experimental progress in the field of cold atomic gases in optical lattices now yield rather accurate 
physical realizations of this simple model in the laboratory \cite{Jaksch2005, Bloch2005, Lewenstein2007}.

In contrast to the Ising model however, our current understanding of the Hubbard model is still lacunary. 
A thorough understanding can be reached in the limiting cases of one dimension \cite{Essler2005} and infinite dimensions (infinite 
lattice connectivity) \cite{Metzner1989, Georges1992a, Jarrell1992}, thanks to efficient analytical and computational methods available in these limits. In contrast, the two-dimensional case relevant to both cuprate superconductors \cite{Keimer2015, Timusk1999, Norman2005, Lee2006} and cold atomic gases \cite{Jaksch2005, Bloch2005, Lewenstein2007} still holds many open questions, both in relation to its phase diagram as a function of interaction, particle density and temperature, and regarding the nature of excited states as well as response functions and transport properties. 
There is a broad consensus in the community that progress on these outstanding issues is essential, even if addressed through the deceptively simple-looking Hubbard model.

In recent years, a number of computational methods have been developed to this end. In this context, it is of crucial importance to interrogate these methods regarding their respective 
ability to address regimes of physical interest. Furthermore, increasing emphasis is being put on establishing 
definite results with controlled computational methods, which can then serve as benchmarks \cite{LeBlanc2015, Motta2017, Williams2020} for approximate, often  
more flexible and computationally efficient methods.

Here we focus on a regime of the Hubbard model which is 
simple at first sight but, as we shall see, deceptively so: 
small interaction values (often referred to as ``weak coupling") and half-filling on the square lattice with nearest-neighbour hopping. The main purpose of this article is to assess 
the ability of state-of-the-art computational methods to address 
the finite-temperature physics of the model in this regime. We provide an extensive comparison 
between basically {\it all} methods currently available for this purpose, 
with two distinct Monte Carlo methods serving as reference benchmarks.
Despite the apparent simplicity of this regime, we shall see that all methods 
face rather severe limitations, especially regarding the lowest temperature that can 
currently be reached. 
Our study also interrogates the model through a set of different physical observables, spanning 
thermodynamic properties, single-particle correlation functions (the Green function and associated 
self-energy) as well as two-particle correlations (the spin correlation function and correlation length). 
Because of this wide spectrum of both methods and observables, we have borrowed terminology from the 
astrophysics community in designating our work as a `multi-method, multi messenger study'~\cite{Neronov2019}.

Despite the deceptive simplicity of the two-dimensional Hubbard model in this parameter regime, the physics is quite rich and non-trivial. As is well established, the ground-state is 
an antiferromagnetic insulator which can be qualitatively understood using Slater's classic description \cite{Slater1951}. 
However, finite-temperature properties display a rich sequence of 
interesting crossovers between physically distinct regimes as the system is cooled down towards its 
antiferromagnetic insulating ground-state. 
Two key features make these finite-temperature properties and crossovers non-trivial: 
(i) the fact that, despite perfect nesting, fluctuations destroy antiferromagnetic long-range order at any non-zero temperature 
- while the correlation length being exponentially large (Mermin-Wagner theorem \cite{Mermin1966, Hohenberg1967}) 
and 
(ii) the van Hove singularity present at the `antinodal' points of the 
Fermi surface, which further suppresses coherence of single-particle excitations near these points. Furthermore, the perfect nesting of the Fermi surface in combination with two-dimensionality leads to a departure from Fermi-liquid behavior in the metallic regime.

Another important goal of this article is therefore to discuss and characterize these different physical 
regimes and crossovers in details, with a particular focus on assessing the ability of the different 
methods to capture their physics properly. 
By combining computational results and analytical insights, we elucidate in particular the role and nature 
of spin fluctuations in the different regimes. As we shall see, this reveals some unexpected features of 
the metallic regime which are likely to have broader implications for materials with electronic correlations. 
Furthermore, we discuss whether the imaginary time computational methods considered in this article are 
able to probe the subtle non-Fermi liquid singularities of the metallic state caused by the perfect nesting of the 
Fermi surface.
This rich physics thus makes this regime of the Hubbard model a perfect opportunity for systematic benchmarks, as well as useful testing grounds for future work on more complex regimes of parameters as well as real materials.

\subsection{Overview of the methods assessed}
\noindent
We categorize the algorithms considered in this article into the following groups: 

\begin{itemize}
\item {\bf Benchmark methods}.
We consider two very different Monte Carlo methods.
The first one is determinantal quantum Monte Carlo (DQMC, \cite{Blankenbecler1981}), the other one is the diagrammatic Monte Carlo method (further referred to as DiagMC) \cite{Prokofev1998} (in its recent connected determinant implementation [CDet] for connected one-particle reducible quantities \cite{Rossi2017} and $\Sigma$DDMC for one-particle irreducible quantities \cite{Simkovic2019, Moutenet2018}, respectively). The reason for their application is twofold: 
(i) The methods are controlled and, hence, in the regime where they can be applied and converged, these methods are numerically exact. Therefore they are  serving as a benchmark for the other approximate methods considered in this article. (ii) The regimes where the methods actually break down 
will be assessed - a crucial information in relation to their application to more challenging regimes. For both benchmark methods, we show error bars in the figures (which may, however, be smaller than the respective symbol sizes). All data points in this manuscript obtained by these methods are numerically exact.

\item {\bf Mean field methods}. In Sec.~\ref{sec:mf}, we discuss dynamical mean-field theory (DMFT, \cite{Georges1996}) 
as a reference point beyond which spatial fluctuations must be included to properly address the two-dimensional model. 
In that section, we also briefly discuss simple static mean-field theory (MFT). 
As we shall see, DMFT provides a good starting point for our study and, remarkably, yields an accurate approximation of local 
observables through most of the metallic regime.

\item {\bf Cluster extensions of DMFT}. The dynamical cluster approximation (DCA), and cellular DMFT (CDMFT) 
provide one possible route to systematically include spatial correlations within the DMFT framework, beyond the 
single-site approximation \cite{Maier2005,Kotliar2006,Tremblay2006,Lichtenstein2000}. Note that cluster-based methods (like CDMFT, DCA and also cluster-TRILEX, see next paragraph) are controlled methods with the control parameter being the size of the cluster. 
However, in some regimes shown in this paper (low temperatures) these algorithms 
could not be converged as a function of cluster size (for reasons explained in App.~\ref{app:dca} and \ref{app:cdmft}).

\item {\bf Vertex based extensions of DMFT}. Another route for including spatial correlations beyond single-site DMFT is relying on higher order Green functions (vertex functions). In the main text we  present results from the dynamical vertex approximation (D$\Gamma$A, ladder version), the triply irreducible local expansion (TRILEX) in various flavors, the dual fermion (DF, ladder version) and the dual boson (DB, single-shot) approach \cite{RMPVertex}.

\item {\bf Other approaches}: in this category we show results from the two-particle self-consistent approach (TPSC, TPSC+, \cite{Vilk1997, Tremblay2011, Wang2019}), the functional renormalization group (fRG~\cite{Metzner2012}, here considered up to one loop with Katanin substitution) and the parquet approximation (PA \cite{DeDominicis1964b, Bickers2004}).
\end{itemize}
This list covers the vast majority of currently available computational methods able to address finite temperature properties. 
One notable exception is the minimally entangled typical thermal state method (METTS) and related approaches, which combine together 
tensor network representations and stochastic sampling~\cite{White2009,Stoudenmire2010,Bruognolo2015,wietek2020}. A systematic exploration of this method as applied to the Hubbard model is currently being actively 
pursued by several groups and comparisons with the present methods will have to be performed in future work \cite{Wietek2021}.

The basic principles of each of these methods, useful references for further reading and results from slightly differing implementations of the respective methods and algorithms are summarized in App.~\ref{app:methods}. 
Throughout the paper we consider the interaction value of $U=2t$. Let us stress from the outset that, despite this rather moderate interaction value, each of these methods encounters limitations in their regime of applicability. 
These limitations stem either from (i) the approximation performed or (ii) algorithmic 
obstacles.

We find that the lowest reachable temperature for the DiagMC algorithm is  $T^{\text{DiagMC}}_{\text{min}} \approx 0.06t$. In this case, reaching lower temperatures 
is hindered by the difficulty in summing the perturbative series. 
Interestingly, we find that the limitation of the DQMC algorithm is similar, 
$T^{\text{DQMC}}_{\text{min}} \approx 0.06t$. 
In that case, the limitations originate from the exponentially 
growing correlation length which would require the simulation of prohibitively 
large systems at lower $T$. 
 DMFT, in contrast, can be converged to very low temperatures and also at $T=0$. Self-consistent methods (e.g. TRILEX) suffer from convergence problems at low-$T$,  
 whereas calculations involving a `single-shot' correction beyond DMFT without self-consistency 
 such as D$\Gamma$A, DB or DF can be performed as long as the correlation length can be accurately resolved (D$\Gamma$A, DB) 
 or, for DF, as long as the starting point - paramagnetically restricted DMFT - 
 remains reasonably accurate.  
  The finite momentum grid also limits the application of fRG and PA, and, to a lesser degree, TPSC and TPSC+. 
  An intrinsic limitation of TPSC occurs in the renormalized classical regime 
(see App.~\ref{app:tpsc}) leading to a rather severe overestimate of the onset temperature of the pseudogap. 
  TPSC+ has been proposed to remedy this: in the present paper, the first application of TPSC+ is actually presented, 
  but its applicability has yet to be explored more widely. 
  An obvious limitation of quantum cluster theories are the cluster sizes which they can reach, 
  which have to be compared to the correlation length - a very demanding criterion 
  in the present case, as will be shown later.

\subsection{Definition of the model, the role of the van Hove singularity and nesting}
\label{sec:model}

We consider the single-band Hubbard model defined by the following Hamiltonian:
\begin{equation}
 H\!=\!-t\sum\limits_{\left<ij\right>,\sigma} c^{\dagger}_{i\sigma} c_{j\sigma} + U\sum\limits_{i}n_{i\uparrow}n_{i\downarrow},
 \label{eqn:Hubbard}
\end{equation}
where $t$ is the (nearest-neighbor) hopping amplitude, $\left<ij\right>$ denotes summation over nearest-neighbor lattice sites, $\sigma \in \{\uparrow, \downarrow\}$ 
the electron's spin, 
$U$ the strength of the (purely local) Coulomb interaction and $n_{i\sigma}\!=\!c^{\dagger}_{i\sigma}c_{i\sigma}$ the spin resolved number operator. Throughout the paper all energies are given in units of $t\!=\!1$. Furthermore we set $\hbar=\!1\!$ and $k_\text{B}\!=\!1$.
We consider the case of $U\!=\!2$ (usually regarded as ``weak coupling'') at half-filling $n\!=\!\left<n_\uparrow + n_\downarrow\right>\!=\!1$, corresponding to a chemical potential of $\mu\!=\!U/2\!=\!1$ and the simple square lattice, resulting in the following dispersion relation for the electrons (lattice constant $a=1$):
 \begin{equation}
  \label{eqn:dispersion} \varepsilon_{\mathbf{k}} = -2\left[\cos\left(k_{x}\right) + \cos\left(k_{y}\right)\right].
 \end{equation}
 
 \begin{figure}[t!]
\centering
      \includegraphics[width=0.45\textwidth,angle=0]{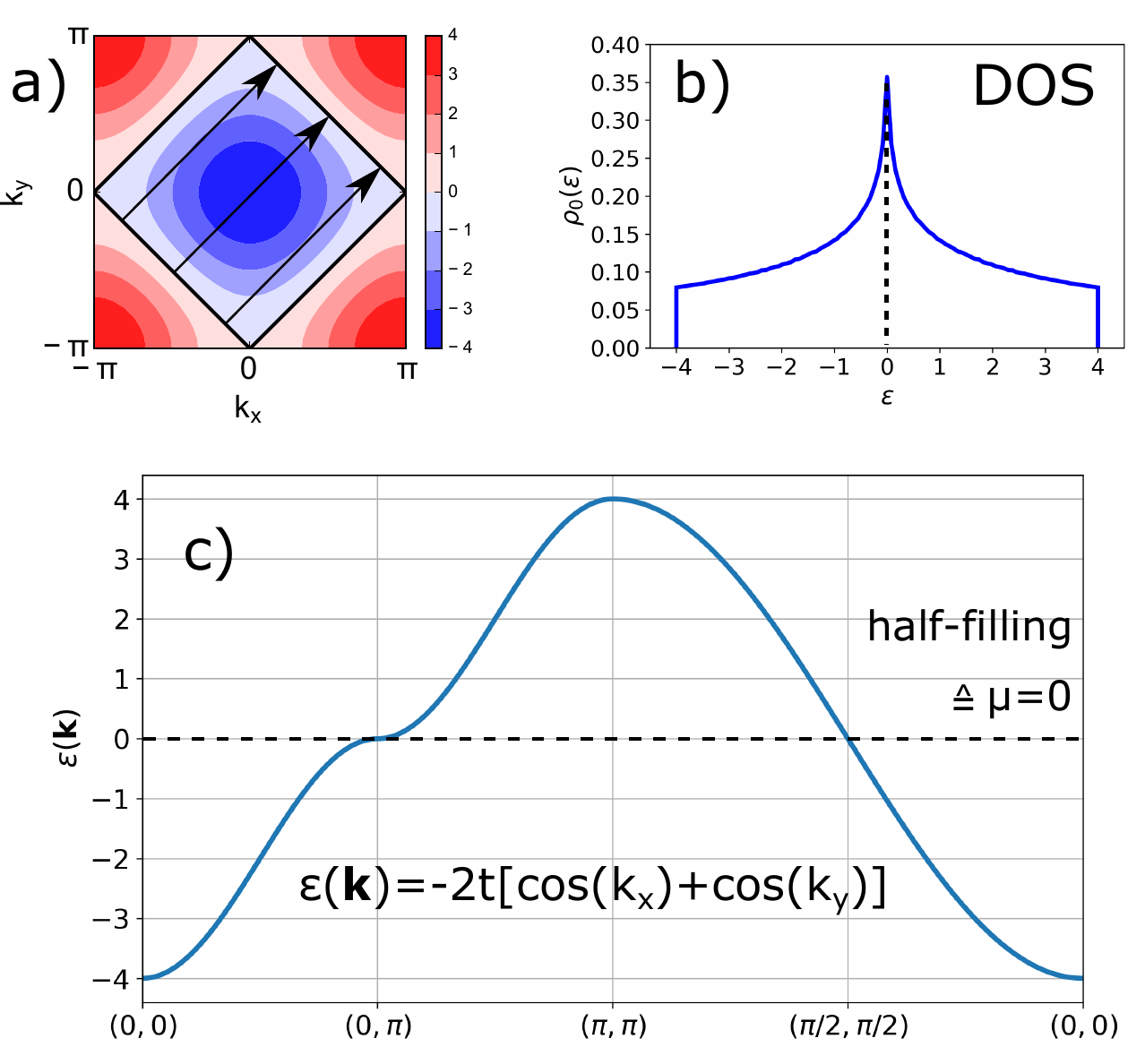}    
      \caption{\label{fig:fs_dos}(Color online.) a) Momentum distribution of the (non-interacting) dispersion relation of the simple square lattice, Eq.~(\ref{eqn:dispersion}), for $t\!=\!1$ throughout the first Brillouin zone. The Fermi surface of the half-filled system ($\mu\!=\!0$) is  diamond-shaped (bold black), the black arrows indicate the nesting vectors, mutually connecting Fermi surface points. b) The corresponding (particle-hole symmetric) density of states (DOS) as a function of energy $\rho_{0}(\varepsilon)$. $\varepsilon\!=\!0$ corresponds to half-filling. c) The value of the dispersion relation along a high-symmetry path  exhibits a plateau around $(\pi,0)$, leading to a vanishing Fermi velocity $v_{\text{F}}$.}
\end{figure}

\begin{figure*}[t!]
        \centering
               \includegraphics[width=0.98\textwidth]{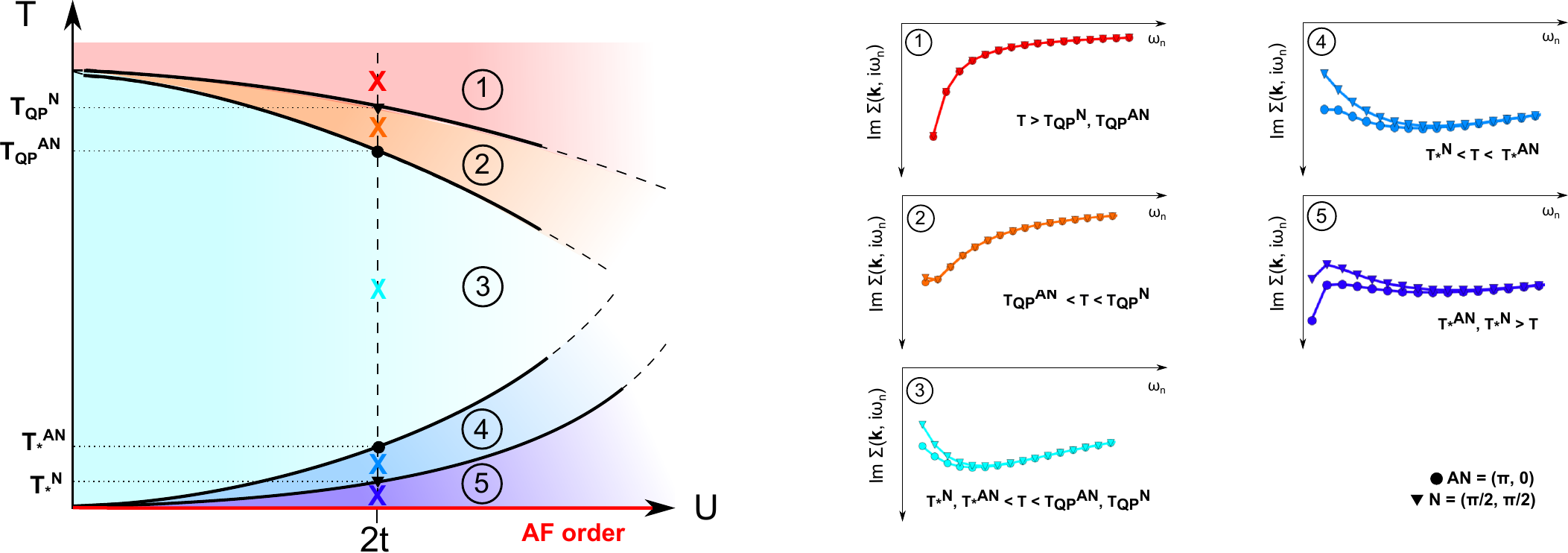}
        \caption{\label{fig:phase_diagram}(Color online.) Left: Schematic phase diagram of the two-dimensional Hubbard model on a simple square lattice in the weak-coupling regime around $U=2t$. 
        $T^{\text{N}}_{\text{QP}}$ and $T^{\text{AN}}_{\text{QP}}$ denote the onset of coherence at the nodal $\mathbf{k}=(\pi/2,\pi/2)$ (triangle) and 
        antinodal $\mathbf{k}=(\pi,0)$ (dot) Brillouin zone point respectively. The onset of the (pseudo-)gap at antinode and node is denoted by $T_{*}^{\text{AN}}$ 
        and $T_{*}^{\text{N}}$ respectively. Right: Qualitative sketches of imaginary parts of the self-energy on the Matsubara axis (extracted from D$\Gamma$A calculations) for temperatures corresponding to the colors given in the phase diagram.
        }
\end{figure*}

The particular form of the dispersion and the case of half-filling leads to a very peculiar diamond-shaped Fermi surface, already resulting in an 
interesting behavior without interactions present: (i) it exhibits a (``perfect'') nesting by the momentum 
vector $\mathbf{Q}=(\pi,\pi)$, which connects every Fermi surface point to another respective one (see Fig.~\ref{fig:fs_dos}), leading 
to an enhanced susceptibility at $\mathbf{q}=\mathbf{Q}$ and (ii) the momenta around the (stationary) antinodal Fermi surface point $\mathbf{k}_{\text{AN}}\!=\!(\pi,0)$ 
imply a logarithmic divergence in the density of states $\rho_{0}(\varepsilon)$ at the Fermi level (van Hove singularity, Fig.~\ref{fig:fs_dos}), leading to a larger scattering phase space than at the nodal point $\mathbf{k}_{\text{N}}\!=\!(\pi/2,\pi/2)$. 
Furthermore, because we consider only nearest-neighbor hopping, the diamond-shaped Fermi surface displays perfect nesting 
by the whole family of wave vectors of the form $(q_x,\pm q_x)$, with consequences for the nature of the metallic regime.

\subsection{Organization of this article}
\label{sec:outline}
This article is organized as follows: In Sec.~\ref{sec:qual}, we describe 
the different physical regimes encountered in this model as a function of temperature, 
using results from our two benchmarks methods (DiagMC and DQMC).
In Sec.~\ref{sec:mf}, we discuss dynamical mean-field theory, which serves as 
a starting point of several approximate methods considered in this article. 
In Sec.~\ref{sec:results_single_fluct}, we discuss the calculation of single-particle 
properties using all the different methods introduced above. 
In Sec.~\ref{sec:docc}, we discuss the $T$-dependence of the double occupancy and 
its physical significance. 
In Sec.~\ref{sec:two_particle}, we discuss two-particle response functions and the 
$T$-dependence of the magnetic correlation length. 
In Sec.~\ref{sec:spin_fluct}, we discuss the implications of our computational results for 
the physics of spin fluctuations in this model. Finally, a discussion and conclusions 
are provided in Sec.~\ref{sec:conclusion}. 
A number of appendices present more technical points as well as details of the 
different methods. 
As Supplemental Material \cite{Suppl} we provide all the numerical data used in the figures of the main text.
\setlength{\tabcolsep}{10pt}
\begin{table*}
\caption{\label{tab:scales} Temperature scales discriminating the regions \textcircled{1}-\textcircled{5} of Fig.~\ref{fig:phase_diagram}
calculated by various many-body techniques.}
\begin{tabular}{|lr!{\color{highestT}\vrule  width 2pt}r!{\color{highT}\vrule  width 2pt}r!{\color{middleT}\vrule  width 2pt}r!{\color{lowT}\vrule  width 2pt}r!{\color{lowestT}\vrule  width 2pt}r!{\color{black}\vrule  width 2pt}r}
\hline
\hline
method && $T_{\text{QP}}^{\text{N}}$ & $T_{\text{QP}}^{\text{AN}}$ & $T_{\text{*}}^{\text{AN}}$ & $T_{\text{*}}^{\text{N}}$ & $T_{\text{N{\'e}el}}$
\\\hline
DiagMC && 0.42	& 0.35 &	0.065 &	0.0625 &	0 \\
DQMC && 0.42    & 0.35 &	0.065&	0.0625	&	0\\\hline				
MFT & & $\infty$ & $\infty$ & 0.2  & 0.2  & 0.2\\
DMFT &&	0.45	& 0.45 & 0.08 & 0.08 & 0.08\\
DCA, N$_{\text{c}}$=128 (PM enforced) &&	0.42 	& 0.35 &	?	&?	&-	\\
CDMFT, N$_{\text{c}}$=64 (PM enforced) &&	0.45 	& 0.42 &	?	&?	&see Sec.~\ref{sec:two_particle} \\
CDMFT+CFE, N$_{\text{c}}$=64 (PM enforced) &&	0.42 	& 0.35 &	?	&?	&see Sec.~\ref{sec:two_particle}
\\\hline
D$\Gamma$A (ladder)            &&        0.42	& 0.35 &  0.065 &	0.059 &0 \cite{Katanin2009, Rohringer2016, RohringerPC2020}\\
DF (ladder)&&0.44&0.37&0.062&0.06&0 (?)\\
DB (single-shot)&&0.42&0.35&$<$0.07&$<$0.07&0 (?)\\
TRILEX &&	0.44 &	0.35	&$<$0.055	&	$<$0.055	&	0 (?)\\
TRILEX, $\Lambda^2$ &&	        0.44 &	0.35	&$<$0.055	&	$<$0.055	&	0 (?) \\
TRILEX, N$_{\text{c}}$=2 &&	0.44 &	0.35	&$<$0.055	&	$<$0.055	&	0 (?)\\
TRILEX, N$_{\text{c}}$=4 &&	0.44 &	0.35	&$<$0.055	&	$<$0.055	&	0 (?)\\\hline
TPSC && 0.42 & 0.29 &0.13& 0.1&0\\
TPSC+ &&0.44 & 0.37 &0.07&$<$0.07&	0 \\\hline
fRG (one-loop Katanin)&&	0.42 &	0.35	& 0.08 &- \cite{Baier2004, Kugler2017, Kugler2018,Hille2020, Hille2020b} &	$>$0 \cite{Bickers1992,Baier2004, Eckhardt18,Eckhardt2020}\\
PA &&0.44&0.37&$<$0.05&$<$0.05&	0 \cite{Bickers1992,Eckhardt18,Eckhardt2020}\\\hline
\end{tabular}
\end{table*}
\begin{figure*}[htb!]
\centering
				\includegraphics[width=0.32\textwidth,angle=0]{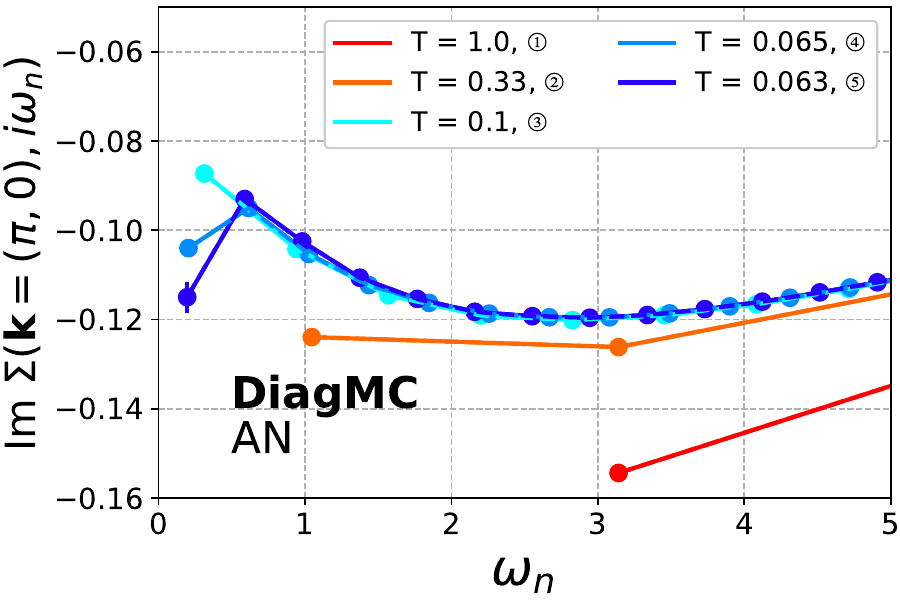}
                \includegraphics[width=0.32\textwidth,angle=0]{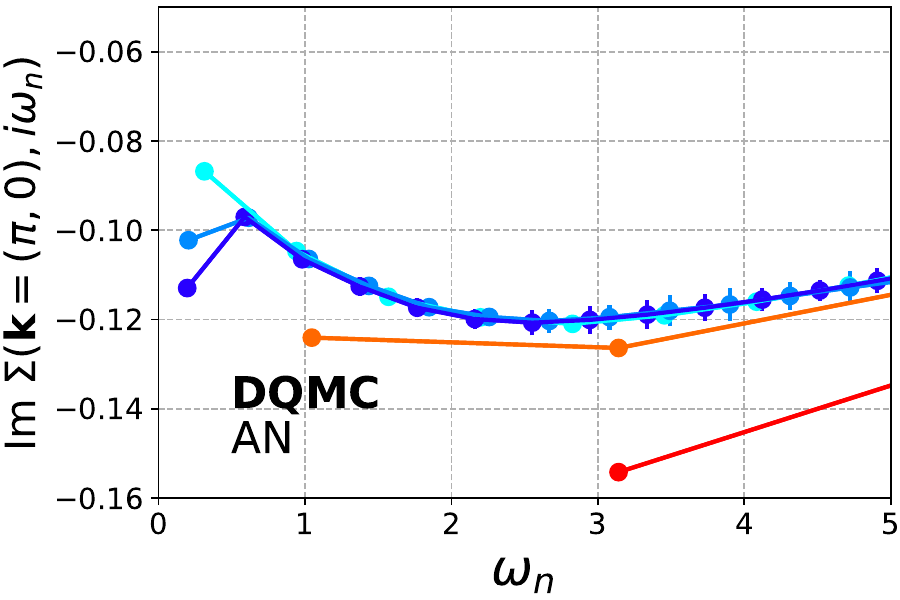} \\         
                \includegraphics[width=0.32\textwidth,angle=0]{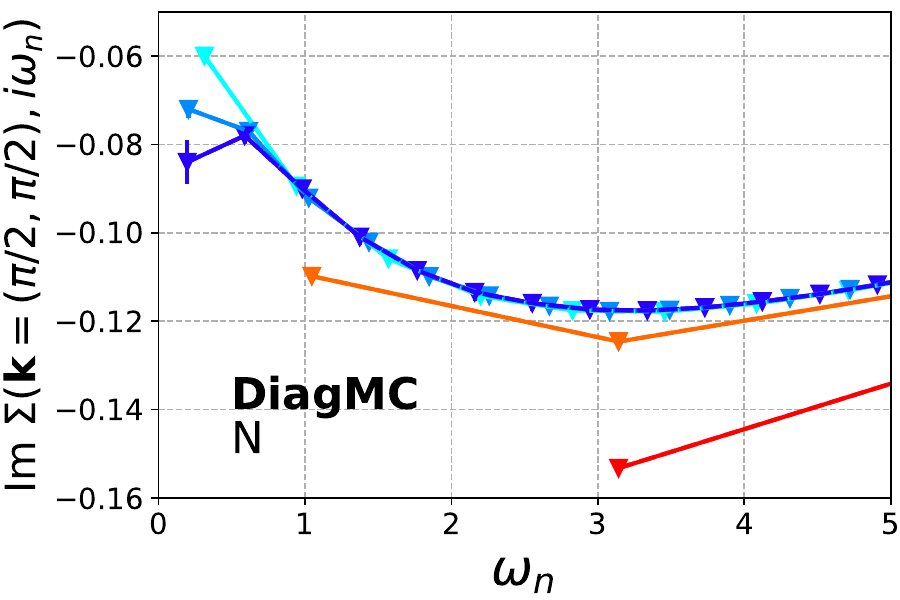}
                \includegraphics[width=0.32\textwidth,angle=0]{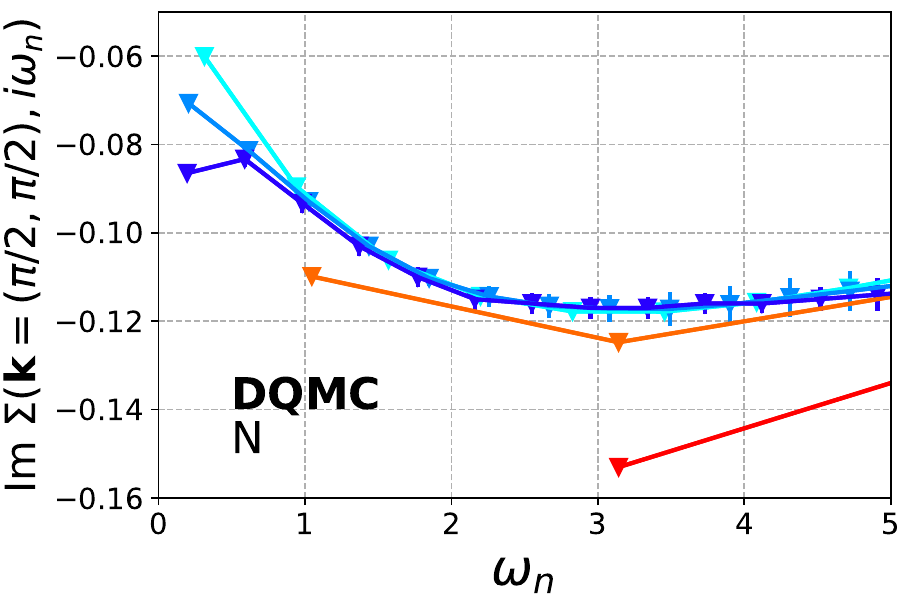}
      \caption{\label{fig:sigma}(Color online.) 
      Imaginary part of the self-energy for antinode (upper row) and node (lower row) 
      as a function of Matsubara frequencies for the two numerically exact techniques DiagMC (left panels) and DQMC (right panels). Please note that the error bars may be of the order of the marker size.}
      \end{figure*}
\section{Qualitative description: physical regimes and crossovers}
\label{sec:qual}
Before presenting detailed results from a variety of many-body approaches in Sec.~\ref{sec:mf}-\ref{sec:two_particle}, in this Section we sketch the overall physical picture which emerges from this study in Fig.~\ref{fig:phase_diagram} (see also \cite{Schaefer2015b, Simkovic2020, Kim2020}). The left panel  indicates in a schematic manner 
the key crossover scales that delimit different physical regions as a function of temperature $T$, for a given 
value of $U$. Our quantitative study focuses on $U=2$ but the qualitative statements made here are expected to 
apply throughout the weak to intermediate coupling regime (see e.g. \cite{Tanaka2019} for a study of the evolution to higher couplings).
As the system is cooled down from high temperature, we observe several regimes with qualitatively different physical properties (a quantitative criterion for the onset of these scales will be given at the end of the section). 

At high temperature, thermal fluctuations prevent the formation of long-lived quasiparticles: 
this regime can be thought as an `incoherent soup' of fermions above their degeneracy temperature and is depicted as the red shaded 
area \textcircled{1} in Fig.~\ref{fig:phase_diagram}. 
Cooling the system progressively extinguishes these thermal fluctuations, leading to increased coherence 
in the single-particle spectrum and the appearance of long-lived quasiparticles.
Here and below, we use the term `quasiparticle' in a general and somewhat loose sense of a dispersing single-particle 
excitation with a `long enough' lifetime. 
For the specific model at hand, because of perfect nesting, the `quasiparticles' do not obey Landau Fermi liquid 
theory however: this is discussed in more details in Sec.~\ref{sec:nesting}.
At the node, this quasiparticle coherence scale $T^N_{QP}$ corresponds to the temperature at 
which the thermal de Broglie wavelength $v^*_{\text{F}}/(\pi T)$ along the nodal direction becomes larger than the lattice spacing, 
with $v_F^*$ being the effective Fermi velocity renormalized by interactions. The metallic regime is depicted as region \textcircled{3} (light blue) in Fig.~\ref{fig:phase_diagram}.

The crossover scale associated with the passage from region \textcircled{1} to region \textcircled{3} 
is not the same all along the Fermi surface however. Because of the van Hove singularity stemming from the antinodal points of the Fermi surface such as $(\pi,0)$ (see Sec.~\ref{sec:model}), 
the coherence temperature $T^{\text{N}}_{\text{QP}}$ at the nodal point 
$\mathbf{k}_{\text{N}}\!=\!(\pi/2,\pi/2)$ is higher than the coherence temperature at the antinodal point $T^{\text{AN}}_{\text{QP}}\!<\!T^{\text{N}}_{\text{QP}}$. 
This defines an extended crossover region \textcircled{2} in which the system 
is coherent near the nodes but still incoherent near the antinodes (orange shaded area in Fig.~\ref{fig:phase_diagram}).
\begin{figure*}[ht!]
\centering
                \includegraphics[width=0.31\textwidth,angle=0]{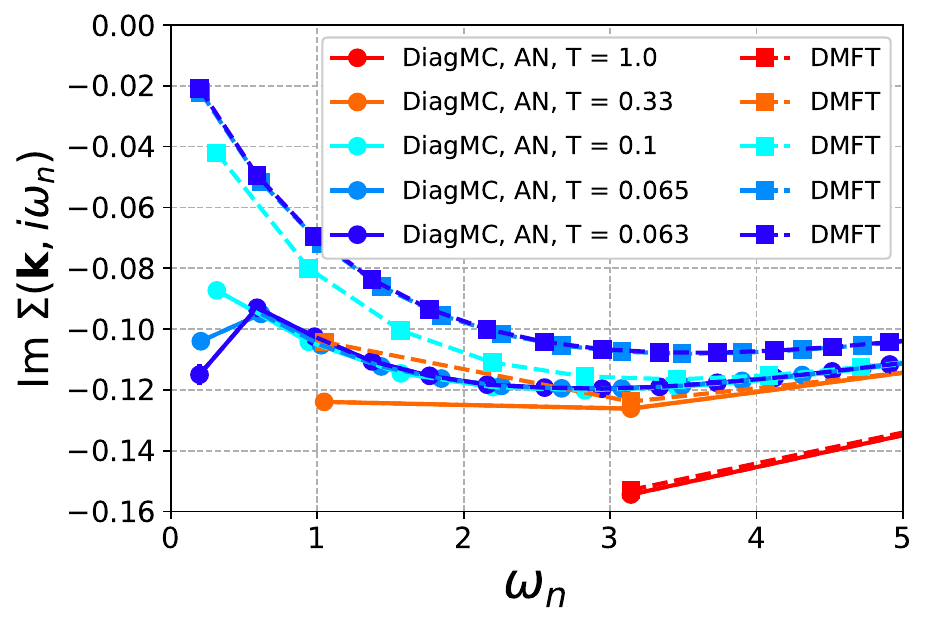}
                \hspace{2mm}
                \includegraphics[width=0.31\textwidth,angle=0]{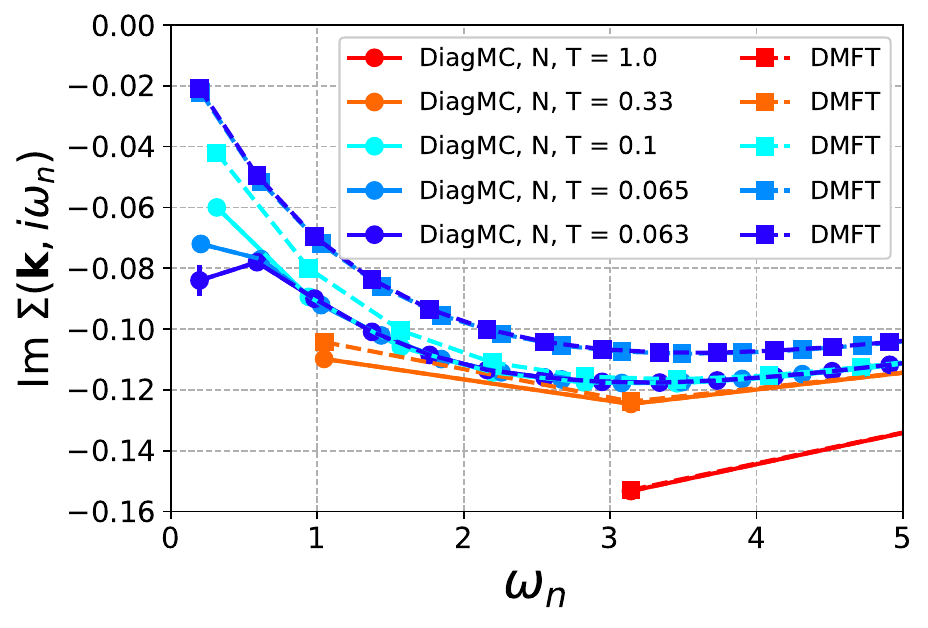}
                \hspace{2mm}
                \includegraphics[width=0.31\textwidth,angle=0]{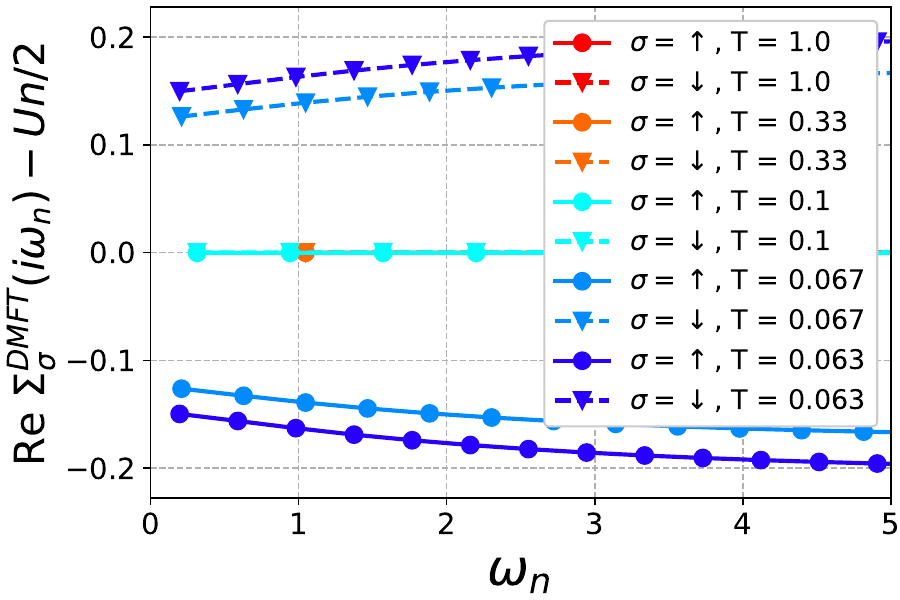}
      \caption{\label{fig:sigma_dmft_CDet}(Color online.) Left and central panel: Imaginary part of the self-energy as a function of frequency at the antinode (left panel) and node (central panel) in DiagMC (circles, solid lines) 
      and the DMFT impurity self-energy (squares, dashed lines) for various temperatures. \label{fig:re_sigma_dmft}Right panel: Spin resolved real part of the DMFT self-energy as a function of Matsubara frequency for several temperatures (the Hartree term of $Un/2=1$ has been subtracted). Please note that $T_N^{\text{DMFT}} \approx 0.08$ and for $T<T_N^{\text{DMFT}}: \Sigma_\uparrow \neq \Sigma_\downarrow$.
      }
\end{figure*}
Although further lowering the temperature in the metallic regime \textcircled{3} initially results 
in freezing out thermal fluctuations and hence in
an increase of the quasiparticle lifetime, this does not persist down to the lowest temperature.
Indeed, antiferromagnetic correlations develop as $T$ is lowered, with an exponentially growing 
correlation length, eventually diverging at $T=0$ when the ground-state with antiferromagnetic long-range order is reached.

In this low-$T$ regime, long-wavelength antiferromagnetic fluctuations (Slater paramagnons) lead 
to an enhancement of the quasiparticle scattering rate upon cooling and to the formation of a pseudogap in the 
single-particle spectrum, which evolves into a sharp gap in the Slater-like insulator at $T=0$ \cite{Slater1951}.
Once again, the crossover temperature $T_{*}$ corresponding to the suppression of coherence and the opening of the pseudogap 
is not uniform along the Fermi surface: it is larger at the antinodes where the destruction of coherence occurs first upon 
cooling, and smaller at the nodes: $T^{\text{N}}_{\text{*}}\!<\!T^{\text{AN}}_{\text{*}}$. 
Hence, in the dark blue shaded area \textcircled{4} where $T^{\text{N}}_{\text{*}}\!<T\!<\!T^{\text{AN}}_{\text{*}}$, 
one has a partially (pseudo-)gapped Fermi surface. Eventually, all states of the Fermi surface 
are suppressed by antiferromagnetic fluctuations for $T\!<\!T^{\text{N}}_{\text{*}}$ resulting in a fully open pseudogap 
everywhere on the Fermi surface (purple shaded area \textcircled{5}). 
Let us stress again that long-range antiferromagnetic order and a true gap only sets in at $T=T_{\text{N\'eel}}\!=\!0$ as a 
consequence of the Mermin-Wagner theorem \cite{Mermin1966, Hohenberg1967}. 

Since all the temperature scales described above correspond to crossovers, an appropriate criterion must be defined 
to identify and quantify them. 
These scales mostly refer to the presence or absence of characteristic spectral features in the single-particle 
properties, and hence an obvious observable would be 
the momentum- and energy- resolved spectral function: 
$A(\mathbf{k}, \omega)=-\frac{1}{\pi}\text{Im }G(\textbf{k}, \omega + i0^+)$ 
and corresponding self-energy as a function of real frequency $\omega$ (see also App.~\ref{app:fermi_liquid} and \cite{Rohringer2016}). 
However, as all the methods considered in the following are formulated on the Matsubara (imaginary) frequency axis, 
a much more practical criterion can be obtained via the imaginary frequency dependence of the (imaginary part of the) momentum-resolved self-energy $\text{Im }\Sigma(\textbf{k},i\omega_n)$, 
consistently with previous work \cite{Schaefer2015b, Schaefer2015c, Simkovic2020}. 
Representative results for this quantity in the five different regimes discussed above are displayed on 
the right panel of  Fig.~\ref{fig:phase_diagram}.
At high temperature, the thermal fluctuations lead to a divergent behavior of $\text{Im }\Sigma(\textbf{k},i\omega_n)$ at low frequencies. 
$T_{\text{QP}}$ can thus be defined as the temperature where this divergent behavior is eased, i.e. when 
the slope between the first and second Matsubara frequency changes sign and becomes negative~\cite{Schaefer2015b, Schaefer2015c, Simkovic2020}. 
In the metallic regime (region \textcircled{3}), the behavior of $\text{Im }\Sigma(\textbf{k},i\omega_n)$ over the lowest 
Matsubara frequencies can be 
approximated by a Taylor series $\text{Im }\Sigma(\textbf{k},i\omega_n)=-\gamma_{\textbf{k}}/Z_{\textbf{k}}+ \omega_n (1-1/Z_{\textbf{k}})+\cdots$. 
In a standard Fermi liquid, $Z_{\textbf{k}}$ and $\gamma_{\textbf{k}}$ are the quasiparticle spectral weight ($Z_{\mathbf{k}}\!<\!1$)
and inverse quasiparticle lifetime, respectively (see App.~\ref{app:fermi_liquid} for details and a critical discussion). 
Likewise, the onset of insulating/pseudogapped behaviour at $T_{\text{*}}^{\text{N},\text{AN}}$ is signalled by an additional 
change of slope, which becomes positive again at lower temperatures.
The actual quantitative criterion for extracting the data of Tab.~\ref{tab:scales} using the quasiparticle weight $Z_{\mathbf{k}}$ is also discussed in App.~\ref{app:fermi_liquid}.

\begin{figure}[t!]
\centering
                \includegraphics[width=0.45\textwidth,angle=0]{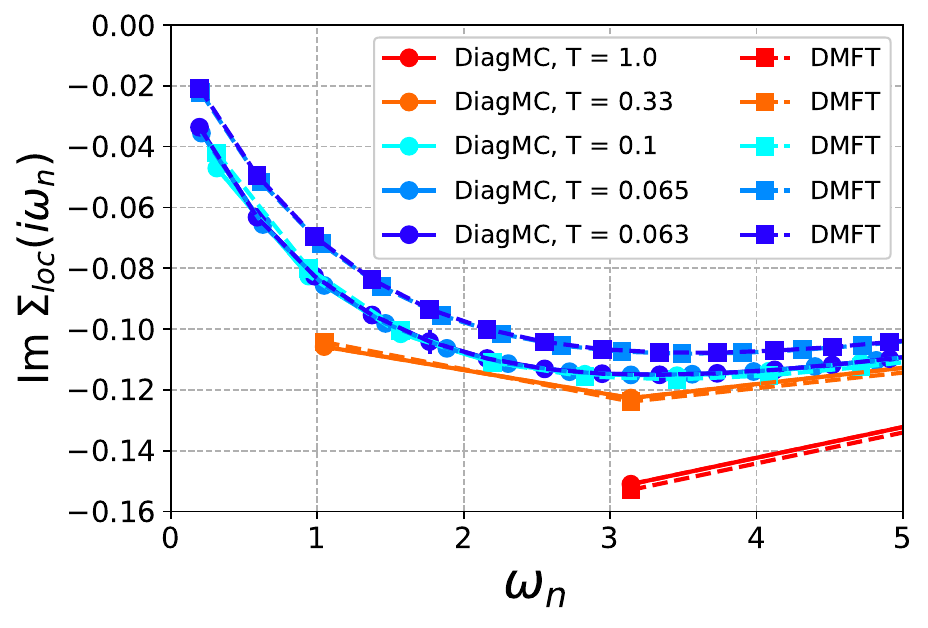}
      \caption{\label{fig:sigma_loc}(Color online.) Imaginary part of the local self-energy calculated by DiagMC compared to DMFT. Note that the two lowest temperatures displayed, at which a discrepancy appears, 
      correspond to the DMFT ordered phase. 
      }
\end{figure}

To illustrate the onset of these different regimes at the one particle level, we display in Fig.~\ref{fig:sigma} results for the imaginary part of the self-energy on the Matsubara axis calculated with the two numerically exact methods used in this paper, namely diagrammatic Monte Carlo (DiagMC, left panels, see App.~\ref{app:diagmc}) and determinantal quantum Monte Carlo (DQMC, right panels, for technical details, including finite size scaling and Trotter extrapolation see App.~\ref{sec:dqmc}, for preceding works see also \cite{Rost2012, Schaefer2015b, Rost2015}).
Both exact methods shown in Fig.~\ref{fig:sigma} exhibit $T_{\text{QP}}$ and $T_{\text{*}}$ with a momentum differentiation between antinode (upper panels) and node (lower panels) with $T_{\text{QP}}^{\text{N}} \approx 0.42$ and $T_{\text{QP}}^{\text{AN}} \approx 0.35$ for the onset of the quasiparticle coherence and $T_{*}^{\text{AN}} \approx 0.065$ and $T_{\text{*}}^{\text{N}} \approx 0.0625$ for the onset of the insulating behavior (more intermediate temperatures have been calculated for the extraction, see Fig.~\ref{fig:fl} and App.~\ref{app:fermi_liquid}).
DiagMC and DQMC are in agreement within error bars and (for the one-particle quantities) are able to be converged until $T \approx 0.063$. 
With the aim of giving an overview and for further reference, we summarize in Table~\ref{tab:scales} 
the results of each of the different methods investigated below for the crossover temperatures delimiting the 
five distinct regimes discussed above.
\section{The dynamical mean-field reference point and the role of fluctuations}
\label{sec:mf}

Several methods considered in the following use dynamical mean-field theory (DMFT) as a starting point. 
Within DMFT, local fluctuations are taken into account (i.e. quantum and thermal fluctuations between the 
four possible local states on each site), but spatial fluctuations are not. 
In the two-dimensional model considered here, these fluctuations are strong and DMFT should 
be viewed as a zeroth-order approximation, which needs to be extended in order to take these fluctuations into account. 
DMFT is exact in the formal limit of infinite dimensions or infinite lattice connectivity, in which 
spatial fluctuations do become negligible \cite{Metzner1989, Georges1992a}.
\subsection{Self-energies and quasiparticle coherence}
\label{sec:DMFT_self}
Because non-local fluctuations are neglected, the self-energy is approximated within DMFT by a function which is 
local in real space, i.e. independent of momentum. 
In Fig.~\ref{fig:sigma_dmft_CDet}, we display the DMFT self-energy for several temperatures, and compare it to the  
benchmark (DiagMC) result for both the antinodal and nodal points (left and central panel, respectively).
It is seen that the DMFT approximation is accurate at very high temperatures, where indeed correlations are very local 
(as shown below, the magnetic correlation length is only a couple of lattice spacing down to $T\simeq 0.2$). 
Deviations between the DMFT self-energy and the DiagMC benchmark are already apparent at $T\simeq 0.33$: at this 
temperature, these deviations are small at the nodal point but already significant at  
the antinodal point due to the proximity of the van Hove singularity and the resulting momentum-dependence 
of the self-energy.

It is interesting to note, however, that the {\it local} component of the self-energy 
(i.e on-site or, equivalently, momentum integrated) 
is in good agreement with DMFT down to a much lower temperature. 
As shown in Fig.~\ref{fig:sigma_loc}, the local 
self-energy obtained from DiagMC is still on top of the DMFT result for temperatures as low as 
$T=0.1$, close to the DMFT ordering temperature. 
This may come as a surprise in 
view of the fact that the antiferromagnetic correlation length (Sec.~\ref{sec:two_particle}) is sizeable 
at this temperature, of order five lattice spacings. 
In Sec.~\ref{sec:spin_fluct},  we provide an explanation to this 
observation and discuss further its physical significance.

\begin{figure}[t!]
        \centering
                \includegraphics[width=0.45\textwidth]{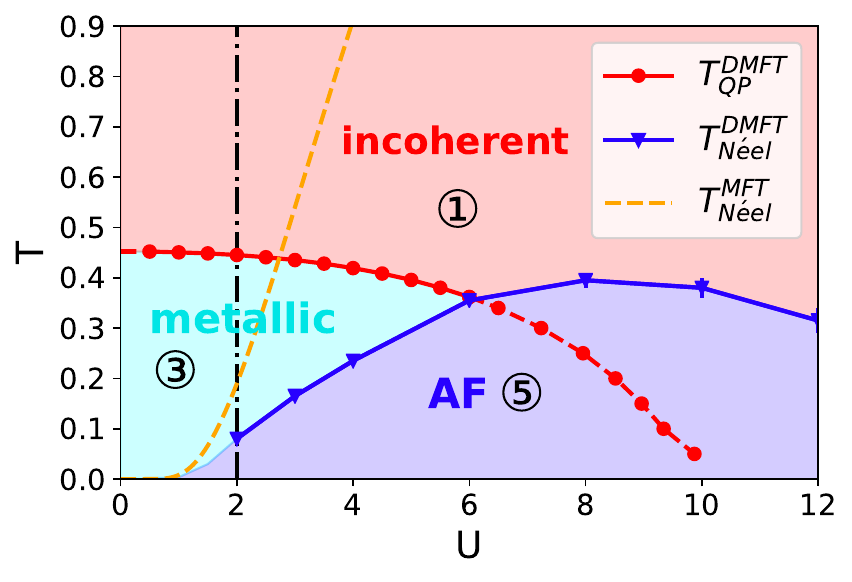}
                \hspace{4mm}
                \includegraphics[width=0.45\textwidth,angle=0]{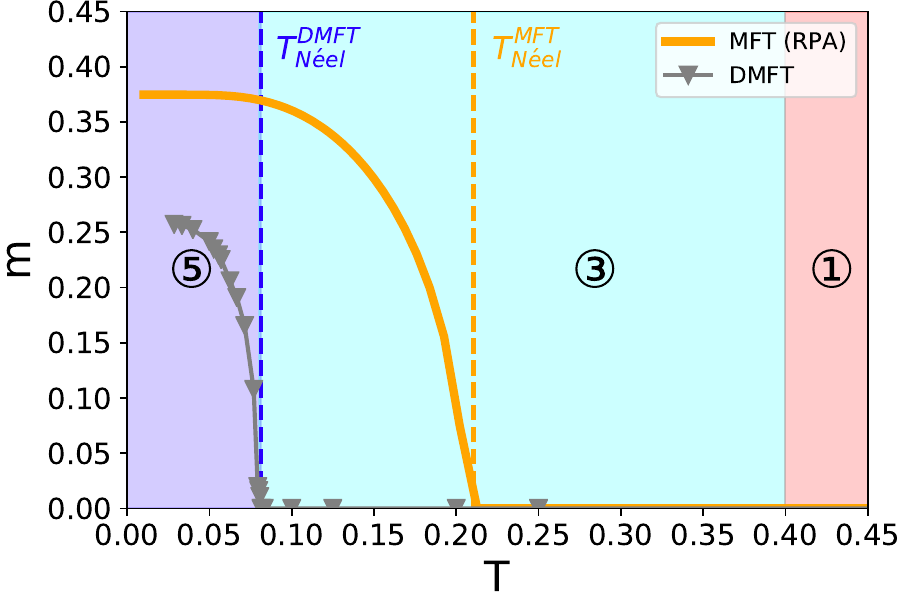}
        \caption{\label{fig:phase_diagram_dmft}
        (Color online.) Upper panel: Phase diagram of the half-filled Hubbard model on a square lattice as a function of $U$ within DMFT. The dotted dashed line marks $U\!=\!2$.
        \label{fig:magn}
        Lower panel: Magnetization at $U\!=\!2$ as a function of $T$ calculated in (symmetry-broken) DMFT (dark-gray curve) and MFT (orange curve).
        }
\end{figure}
From the results in Fig.~\ref{fig:sigma_dmft_CDet} (and more temperature points, not shown) one observes that, within DMFT, the onset of quasiparticle coherence 
associated with the crossover from the high-$T$ incoherent regime into the lower-$T$ metallic regime 
(regime \textcircled{1} to regime \textcircled{2} in Fig.~\ref{fig:phase_diagram}) occurs at $T_{\text{QP}}^{\text{DMFT}}\simeq 0.45$. 
This is in reasonable agreement with the QP coherence scale at the node from our benchmark calculations 
$T_{\text{QP}}^{\text{N}}\simeq 0.42$, but somewhat higher than the antinodal value 
$T_{\text{QP}}^{\text{AN}}\simeq 0.35$, again due to the lack of momentum dependence of the self-energy, which 
is essential to account for nodal-antinodal differentiation. 
We have calculated the quasiparticle coherence temperature within DMFT as a function of coupling $U$, and display the 
result in Fig.~\ref{fig:phase_diagram_dmft} (red line in upper panel). 
As expected, this scale decreases as the coupling is increased. 

\subsection{Crossover scales: the DMFT viewpoint}
\label{sec:DMFT_crossovers}
When used as an approximation for the two-dimensional half-filled Hubbard model of interest here, 
DMFT yields a symmetry breaking phase transition into an insulating phase with antiferromagnetic (AF) 
long-range order at finite temperatures for any value of $U$. 
This is expected because DMFT does not take into account spatial fluctuations which destroy finite-T AF ordering 
in two dimensions (and, hence, does not satisfy the Mermin-Wagner theorem \cite{Mermin1966, Hohenberg1967}). 
The N\'eel temperature $T_{\text{N{\'e}el}}$ obtained within DMFT is displayed in Fig.~\ref{fig:phase_diagram_dmft} 
as a function of $U$ (blue curve in upper panel), as 
well as the corresponding staggered magnetization $m$ as a function of temperature (lower panel). 
As expected, the N\'eel temperature displays a maximum, which signals the crossover between the 
weak-coupling regime in which the N\'eel temperature and magnetization are exponentially small, and 
the strong coupling regime. For very large $U$, the charge gap is of order $U$ and the magnetization saturates. 
In that regime, for $T\ll U$, spin degrees of freedom are described  by an effective Heisenberg model with 
superexchange $J=4t^2/U$, and the DMFT N\'eel temperature is proportional to $J$.
For reference, we also display in Fig.~\ref{fig:phase_diagram_dmft} the result of the standard static mean-field theory 
(i.e. the Hartree mean-field for the transition into the spin-density wave phase, denoted below by `MFT'). 
MFT considerably overestimates the N\'eel temperature as well as $m$ for most values of $U$. This is particularly pronounced at large $U$ where static MFT does 
not correctly separate spin and charge degrees of freedom so that the ordering temperature incorrectly coincides 
with the charge gap $Um$ ($\sim U$ at large $U$)  (for a comparison of MFT and DMFT see \cite{Sangiovanni2006}). 

The key observation is that, taken together, the DMFT quasiparticle coherence scale (red line in Fig.~\ref{fig:phase_diagram_dmft}) 
and the DMFT N\'eel temperature (blue line) can be taken as semi-quantitative mean-field estimates for the crossover 
lines separating the different regimes discussed in Fig.~\ref{fig:phase_diagram} above. 
The DMFT quasiparticle coherence scale is an estimate for the crossover into the metallic regime \textcircled{3} in Fig.~\ref{fig:phase_diagram} 
(with $T_{\text{QP}}^{\text{DMFT}}\simeq 0.45$ comparable within error bars to the nodal value from our benchmark, while 
the antinodal one is about $12\%$ smaller). 
In turn, since in this weak-coupling regime the insulating gap is associated with magnetic quasi-long range order,  
the DMFT N\'eel temperature is an estimate for the onset of the insulating behavior - regime \textcircled{5} in Fig.~\ref{fig:phase_diagram}.
Note that $T_{\text{N{\'e}el}}^{\text{DMFT}}\simeq 0.08$, compared to $T_{*}^{\text{AN}}\simeq 0.065$ while 
$T_{*}^{\text{N}}\simeq 0.0625$ is about $4\%$ smaller.
\begin{figure}[t!]
\includegraphics[width=0.45\textwidth,angle=0]{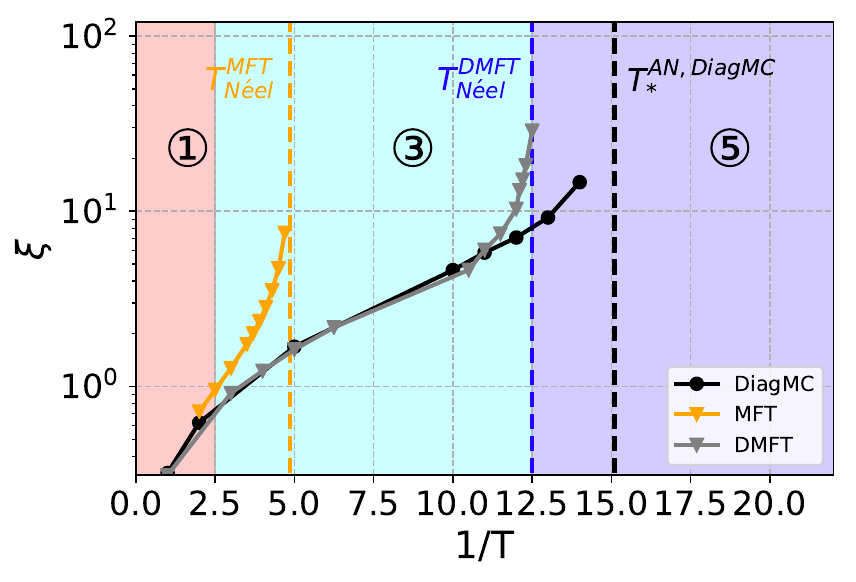}
      \caption{\label{fig:xi_MFT}(Color online.) Magnetic correlation length as a function of (inverse) temperature for DiagMC, MFT, and DMFT on a logarithmic scale.
      }
\end{figure}

The magnetic correlation length can be calculated within both MFT and DMFT. 
In both cases this is achieved by the appropriate Bethe-Salpeter equations 
for the correlation function: with a static vertex equal to the bare $U$ within MFT 
(the calculation then reduces to the random phase approximation, RPA), 
and with a fully frequency-dependent but spatially local vertex within DMFT, see \citep{Georges1996}.
It is often overlooked that mean-field methods (MFT or DMFT) provide us with a determination of 
correlation functions and indeed it naively appears as somewhat paradoxical that spatially dependent correlation functions can be obtained from a mean-field 
ansatz which is local at the one-particle level. 
This is actually a classic question in statistical mechanics, whose resolution lies in a careful interpretation of linear-response theory \cite{Parisi1998}.
The mean-field correlation lengths are displayed in Fig.~\ref{fig:xi_MFT} in comparison 
to the DiagMC benchmark. It is apparent that the DMFT correlation length is in excellent 
agreement with the benchmark down to $T\simeq 0.1$. The figure also clearly illustrates the connection between long-range magnetic correlations and insulating behavior.
\begin{figure}[t!]
\centering
                \includegraphics[width=0.45\textwidth,angle=0]{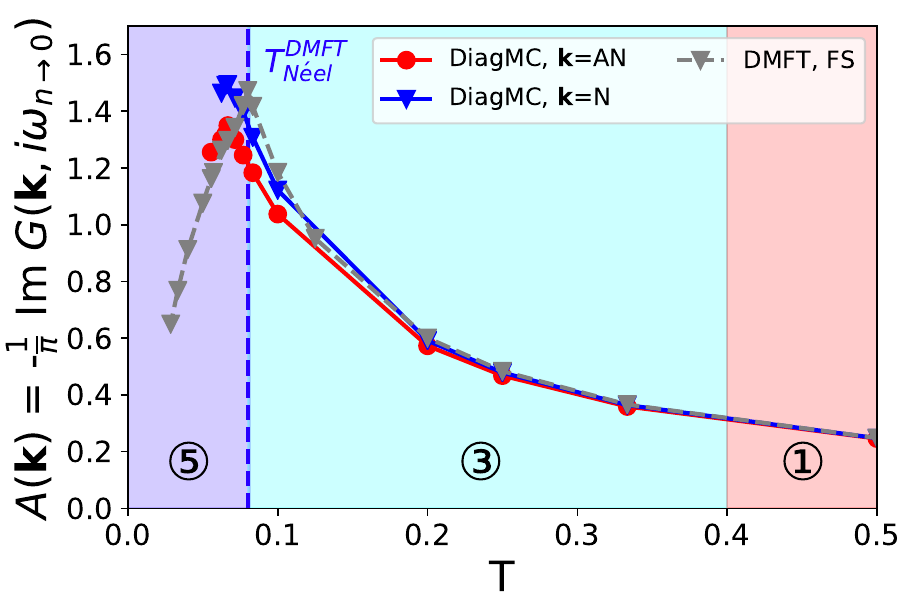}
      \caption{\label{fig:spectrum}(Color online.) Spectral function extrapolated to zero frequency at the antinode and node, obtained from the imaginary part of the Green function in DiagMC (solid lines, red circles antinode, blue triangles node) 
      and DMFT (on the Fermi surface, dashed line with gray triangles).}
\end{figure}
\begin{figure*}[ht!]
\centering
				\includegraphics[width=0.24\textwidth,angle=0]{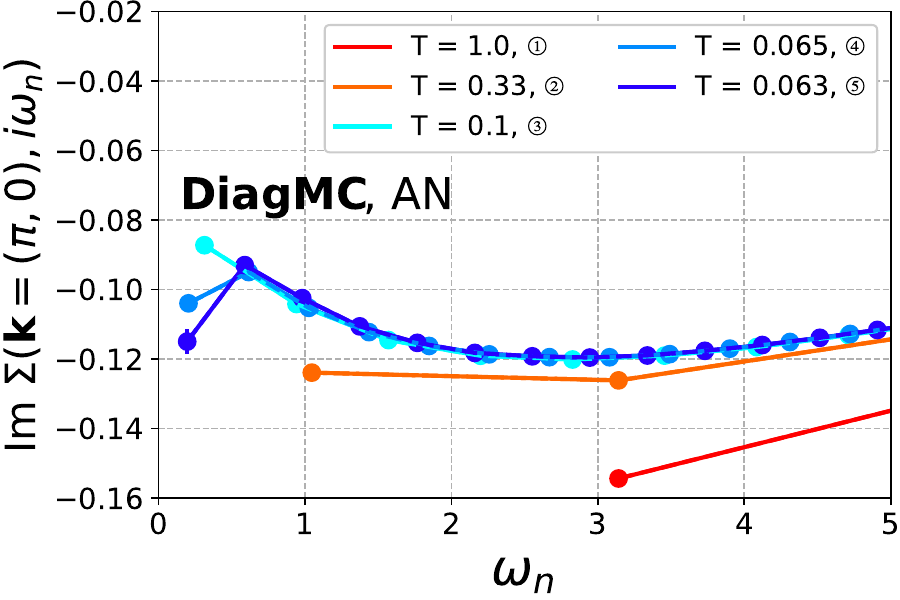}
                \includegraphics[width=0.24\textwidth,angle=0]{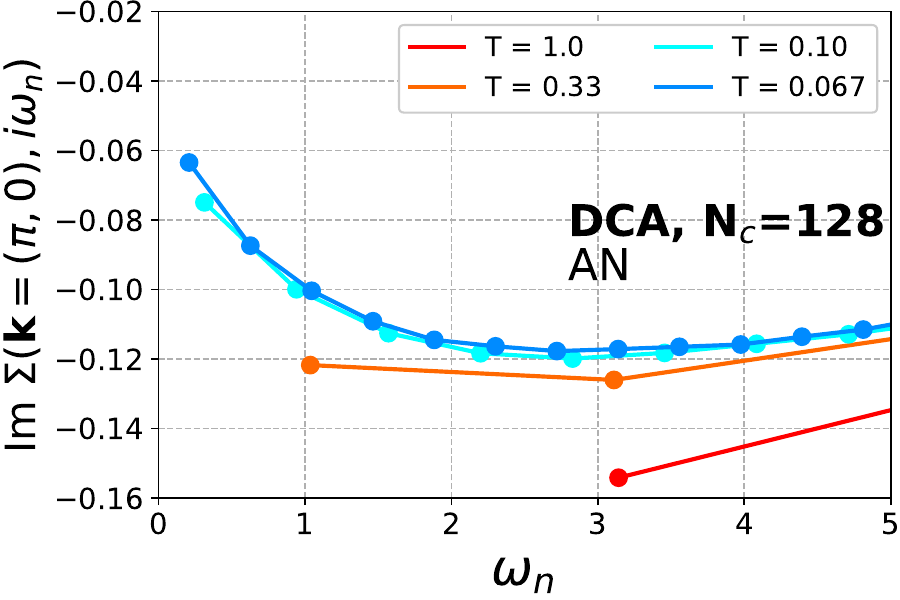}    
                \includegraphics[width=0.24\textwidth,angle=0]{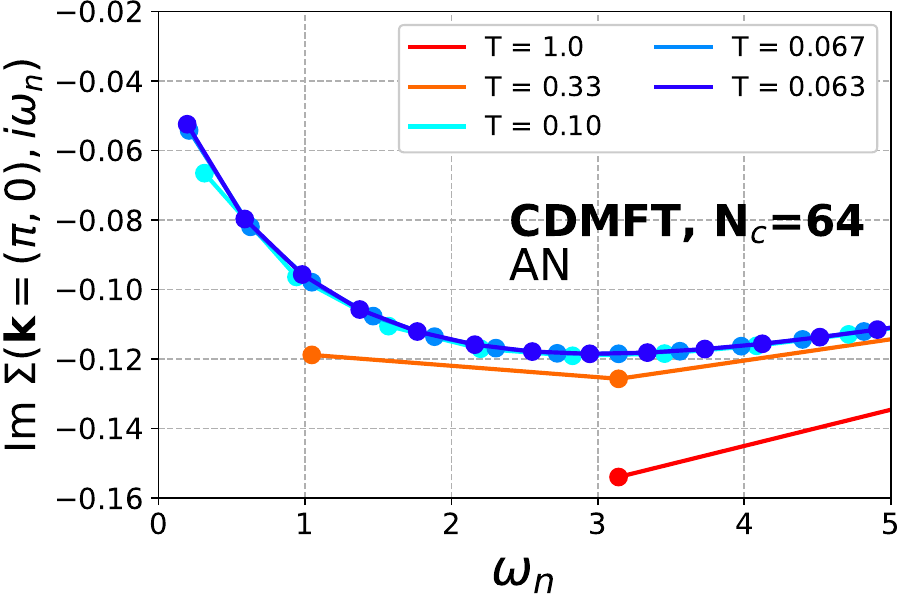}
                \includegraphics[width=0.24\textwidth,angle=0]{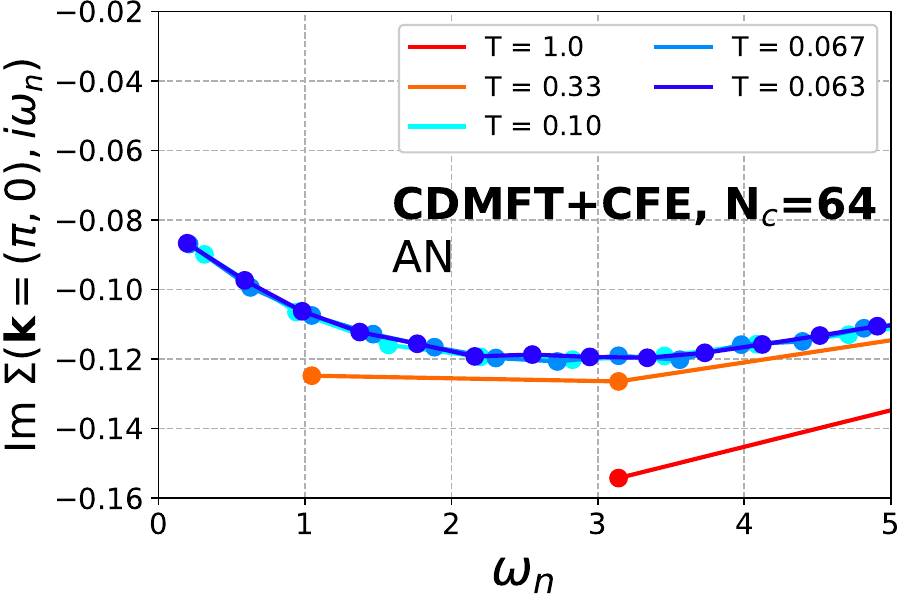}
                \\
                \includegraphics[width=0.24\textwidth,angle=0]{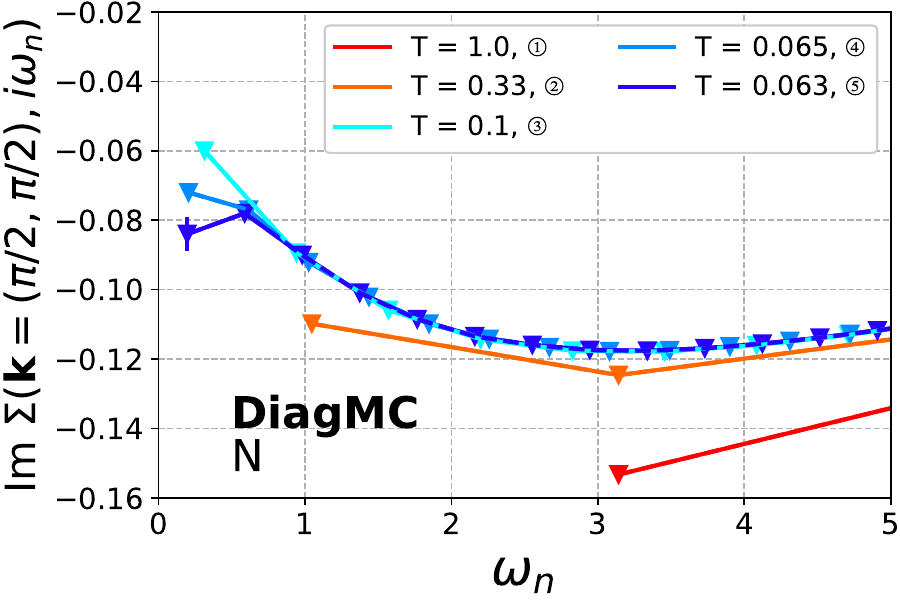}
                \includegraphics[width=0.24\textwidth,angle=0]{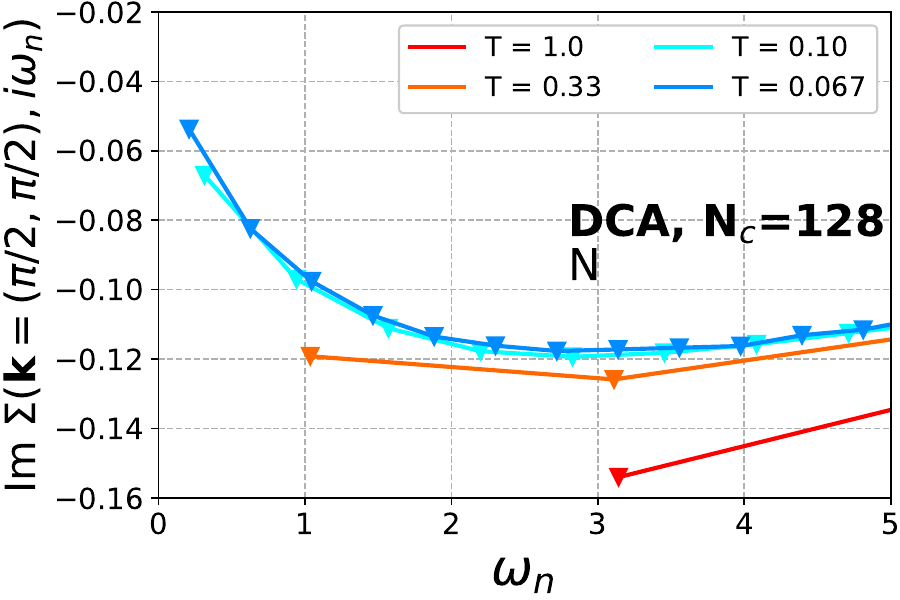}
                \includegraphics[width=0.24\textwidth,angle=0]{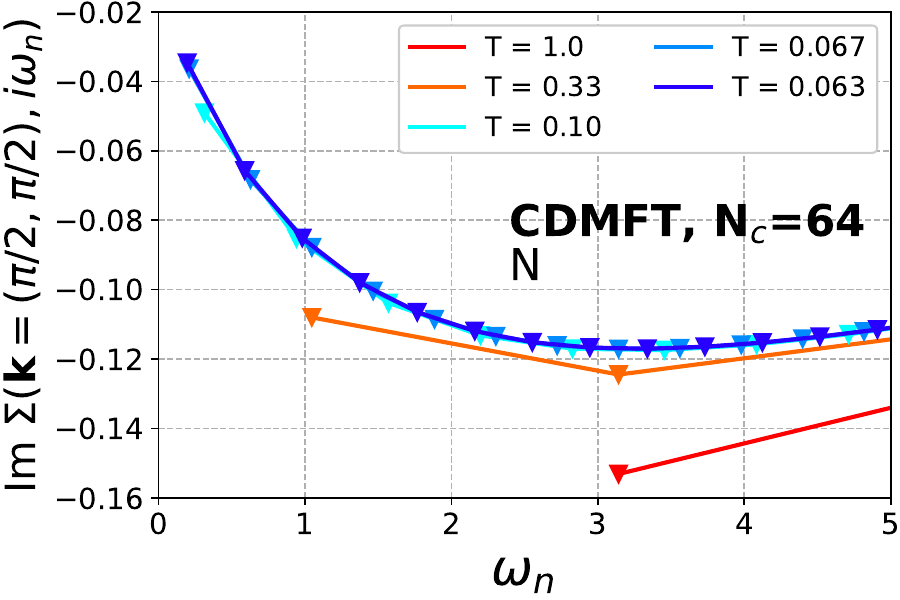}
                \includegraphics[width=0.24\textwidth,angle=0]{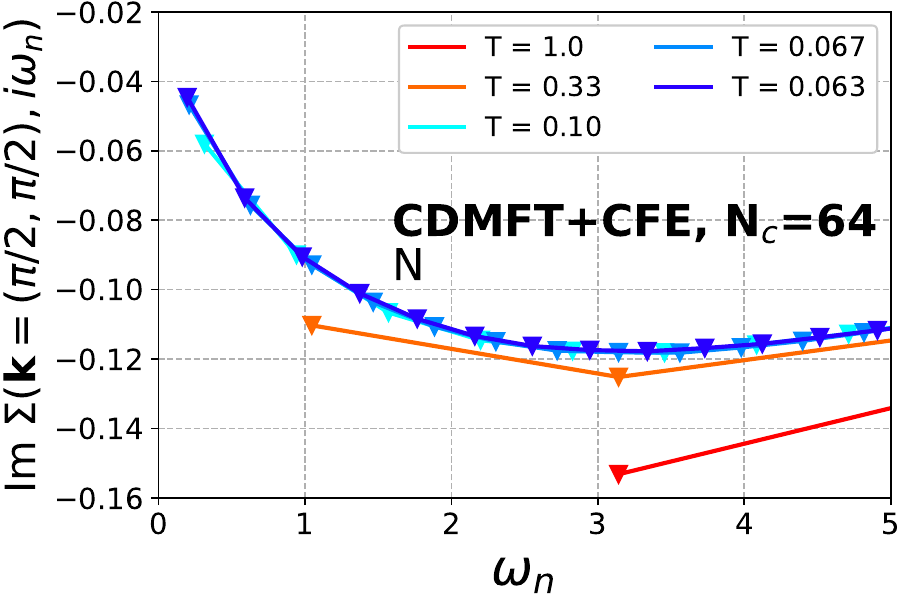}
      \caption{\label{fig:sigma_dca}(Color online.) Imaginary part of the self-energy for antinode (upper row) and node (lower row) 
      as a function of Matsubara frequencies for DiagMC (first panel on the left-hand side), DCA (N$_c$=128, second panel), CDMFT (N$_c$=64, third panel) and center-focused extrapolated CDMFT (CDMFT+CFE, N$_c$=64, fourth panel).
      }
\end{figure*}
\subsection{The insulating regime}
\label{sec:DMFT_insulating}
We finally comment on the comparison between the DMFT self-energies and the true solution in the low-$T$ insulating regime. 
Within DMFT, the insulating gap in this weak-coupling regime corresponds to a Slater mechanism associated with 
AF long-range order \cite{Slater1951}. It is associated with spin polarization and a non-zero value of the real part of the self-energy 
$\mathrm{Re}\Sigma_\uparrow(0)=-\mathrm{Re}\Sigma_\downarrow(0)\neq 0$ such 
that the quasiparticle equation $\omega+\mu-\mathrm{Re }\Sigma(\omega)=0$ has solutions only for $\omega$ 
outside the gap. The low-frequency behavior of the DMFT self-energy is non-singular, and in particular 
it has (using also particle-hole symmetry) a spin-independent linear term $\mathrm{Im }\Sigma(i\omega_n) \propto (1-1/Z)\omega_n+\cdots$, 
corresponding to $\mathrm{Re }\Sigma_\sigma(\omega)=\mathrm{Re}\Sigma_\sigma(0)+(1-1/Z) \omega+\cdots$ on 
the real frequency axis, which is similar in structure in the metallic phase and in the AF-ordered insulating phase. 
This is clearly apparent in Fig.~\ref{fig:sigma_dmft_CDet}. 
In contrast, because a spin polarization is absent at any non-zero $T$ in the true solution, the self-energy must be 
much more singular at low frequencies to open the insulating gap, as becomes clear from 
Fig.~\ref{fig:sigma} and Fig.~\ref{fig:sigma_dmft_CDet}. The precise nature of this 
singularity is discussed in Sec.~\ref{sec:spin_fluct}. 
At $T=0$, however, the model does order and the general structure of the self-energy
is expected to be similar to the Slater (and DMFT) one discussed above. 

We finally display in Fig.~\ref{fig:spectrum} the value of the spectral function extrapolated to zero frequency, as obtained from DMFT as well as from the DiagMC benchmark at the node and antinode. 
Interestingly, there is rather good agreement between these different methods for this quantity, despite the  different low-frequency behavior of the self-energy. 
In particular, we note that the crossover into the insulating regime is very sharp when seen from this 
physical observable, which is rather well approximated by the DMFT solution that has long range order 
in the insulating regime and hence displays a singularity at the N\'eel temperature. 
Detailed analysis close to the crossover temperature would of course reveal differences.

Summarizing, we see that DMFT provides a reasonable approximate description of the key crossovers 
encountered as a function of temperature. As a mean-field theory, it of course mimicks the 
crossover into an insulating phase with a large AF correlation length as a phase transition into a 
phase with long-range AF order. Including fluctuations beyond DMFT is therefore especially crucial 
in two dimensions, in which long-range order exists only at $T=0$. This is the purpose of the cluster and 
diagrammatic (vertex-based) extensions of DMFT discussed in the following sections.
\section{Including fluctuations beyond mean-field: single-particle properties} 
\label{sec:results_single_fluct}
In this section, we show how the inclusion of (non-Gaussian bosonic) spatial fluctuations influences the one-particle properties (self-energies) beyond the mean field picture. This can be achieved by several methods: cluster extensions (dynamical cluster approximation DCA, cellular DMFT CDMFT, Sec.~\ref{sec:dca}), and diagrammatic extensions of DMFT (Sec.~\ref{sec:diag}) as well as with other approaches (Sec.~\ref{sec:uncontrolled}). In this section, the results for the self-energy are presented and compared to the exact (benchmark) methods.

\subsection{Cluster extensions of DMFT: DCA and CDMFT}
\label{sec:dca}

Quantum cluster methods~\cite{Maier2005} are quite natural extensions of DMFT in which the non-local 
components of the self-energy are computed by considering a cluster of $N_c$ sites self-consistently embedded 
in the lattice. This provides a controlled sequence of approximations which converge to the exact result 
when $N_c\rightarrow\infty$. Depending on the regime of parameters, convergence in this asymptotic limit may 
or may not be attained in practice however. 

There are several flavors available within the broad family of quantum cluster theories, depending on how exactly the interacting cluster is described and how the embedding is performed. First we consider the dynamical cluster approximation (DCA, \cite{Hettler1998, Hettler2000, Aryanpour2002}), where the Brillouin zone is paved with patches, 
and the self-energy is approximated as a piece-wise constant function over these patches whose components are 
calculated from a cluster of $N_c$ sites with periodic boundary conditions, hence defining the discrete momenta associated with each patch. For methodological details see \cite{Maier2005} and App.~\ref{app:dca}.

Fig.~\ref{fig:sigma_dca} (left) shows the self-energy on the Matsubara axis of a DCA calculation with N$_c=128$ momentum patches for several temperatures and the benchmark DiagMC data as reference. For these calculations symmetry breaking to an AF long-range ordered phase was not allowed. 
We see that the improvement brought over single-site DMFT by these calculations is that the momentum-dependence of the quasiparticle coherence scale along the Fermi surface and nodal-antinodal differentiation is quantitatively reproduced (regimes \textcircled{1}-\textcircled{3}). 

However, the (non-magnetic) DCA with N$_c=128$ is not able to open the pseudogap at low temperatures 
(shown in  Fig.~\ref{fig:sigma_dca} until $T=0.067$, i.e. $\beta=1/T=16$). 
The reason for this failure is easily understood by noting that the correlation length, previously displayed in  
Fig.~\ref{fig:xi_MFT}, is $\xi \approx 15$ lattice sites at the lowest T available ($\beta=14$) and even larger at lower T.
A periodic cluster of finite size $N_c$ cannot resolve a correlation length larger than 
$\xi^{\text{max}}_{\text{DCA}}\sim \sqrt{N_c}/2$ which is $\simeq 5.7a$ for $N_c=128$. 
From this argument, one would expect that a cluster of order $1000$ sites would be required to capture 
the opening of the insulating pseudogap with DCA in the case at hand. 
As discussed in Appendix~\ref{app:dca}, this is beyond practical reach of the algorithm that we use here in the context of DCA.

A different strategy in the family of cluster extensions of DMFT is to consider a real-space embedding 
and hence a cluster with open boundary conditions, as in the cellular DMFT (CDMFT) approach~\cite{Lichtenstein2000, Kotliar2001}. Fig.~\ref{fig:sigma_dca} also displays data from CDMFT with $N_c=64=8\times 8$. As apparent 
from these data, the real-space embedding method again captures the momentum differentiation of the coherence scale, but fails in opening the pseudogap for the same reasons as DCA ($\xi^{\text{max}}_{\text{CDMFT}} \approx 8 / 2 = 4$). 
From a quantitative perspective, a recently introduced extrapolation scheme (center focused extrapolation, CDMFT+CFE \cite{Klett2020}, App.~\ref{app:cdmft}) improves the comparison with the benchmark, however for the $U=2$ case also fails to open a  gap. Hence, the size of the clusters has to be increased much beyond the values of N$_c$ considered here in order to reproduce the pseudogap at this small interactions strength (see also \cite{Simkovic2020}). 

Summarizing, we conclude that extensions of DMFT based on cluster embedding methods succeed in reproducing the 
momentum dependence of the quasiparticle coherence scale (nodal-antinodal differentiation) in the metallic regime, but that much 
larger cluster sizes would be necessary in order to capture the opening of the insulating pseudogap, due to the 
very large correlation length in this weak coupling regime. Cluster embedding approaches perform much better in regimes with a smaller correlation length, such as larger values of $U$ (i.e. in the strong coupling regime)  \cite{Rohringer2011} or (disregarding the sign-problem) when doping away from half-filling, as documented by the success of these methods in capturing the strong coupling pseudogap of the doped Hubbard model \cite{Gull2008, Gull2009, Gull2010, Gull2013, Parcollet2004, Macridin2006, Tremblay2006, Senechal2004, Haule2007b, Park2008, Ferrero2009, Ferrero2009b, Werner2009, Sordi2012, Sordi2012b, Sordi2013, Gunnarsson2015, Wu2018, Scheurer2018, Reymbaut2019}.
\begin{figure*}[ht!]
\centering		
				\includegraphics[width=0.30\textwidth,angle=0]{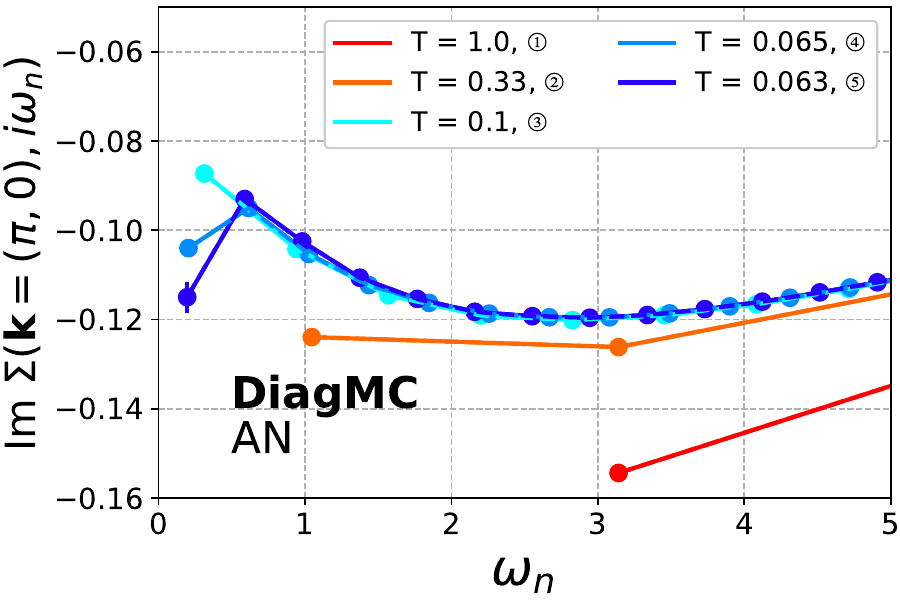}
                \includegraphics[width=0.30\textwidth,angle=0]{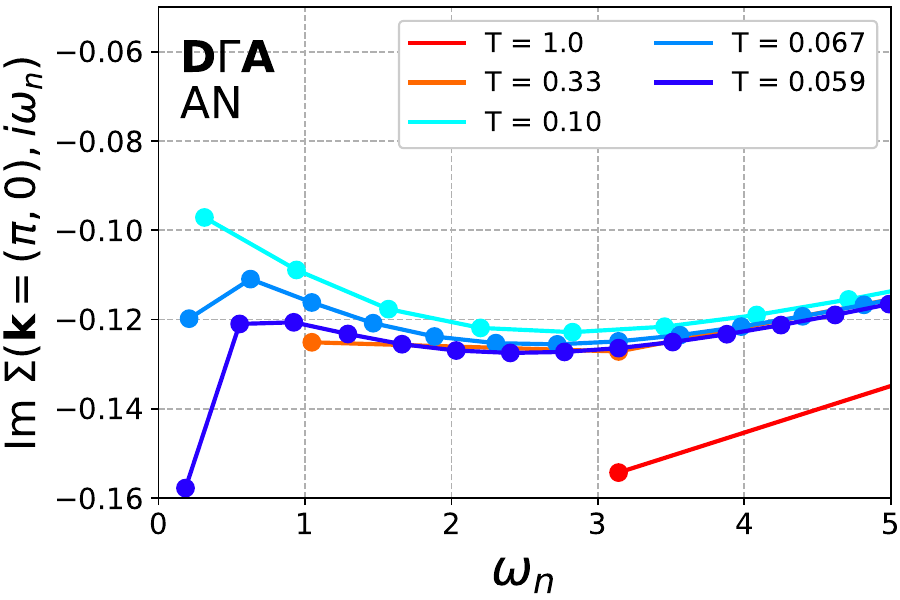}
                \includegraphics[width=0.30\textwidth,angle=0]{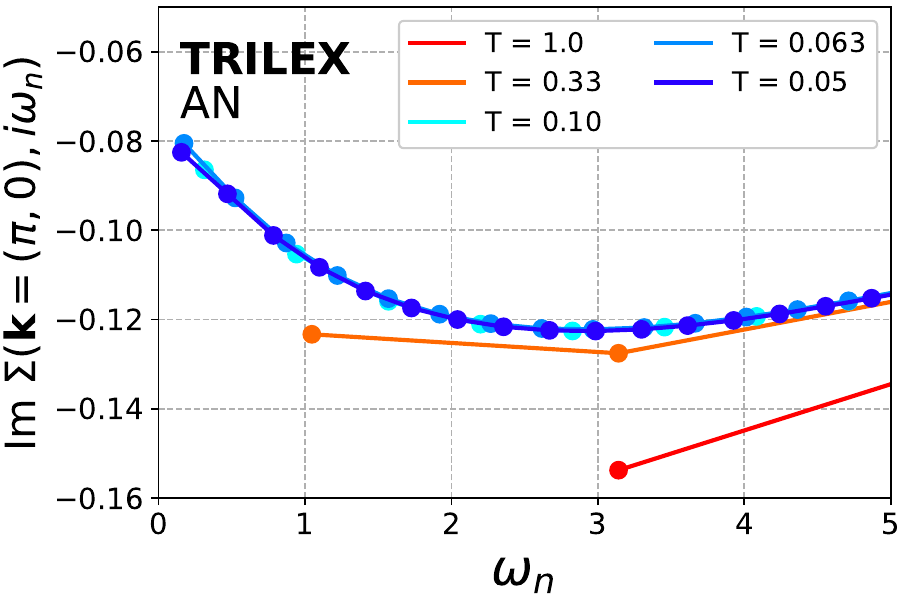}
                \includegraphics[width=0.30\textwidth,angle=0]{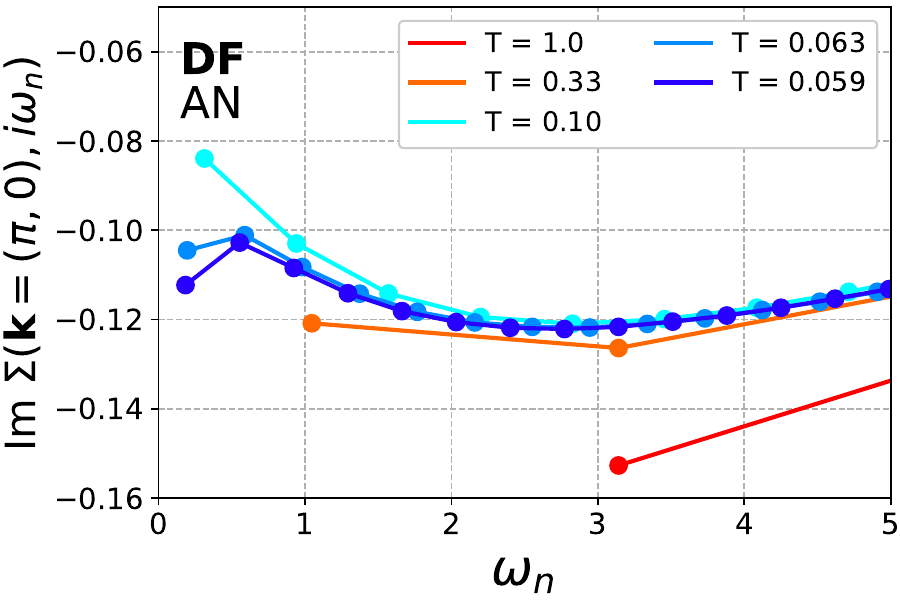}
                \includegraphics[width=0.30\textwidth,angle=0]{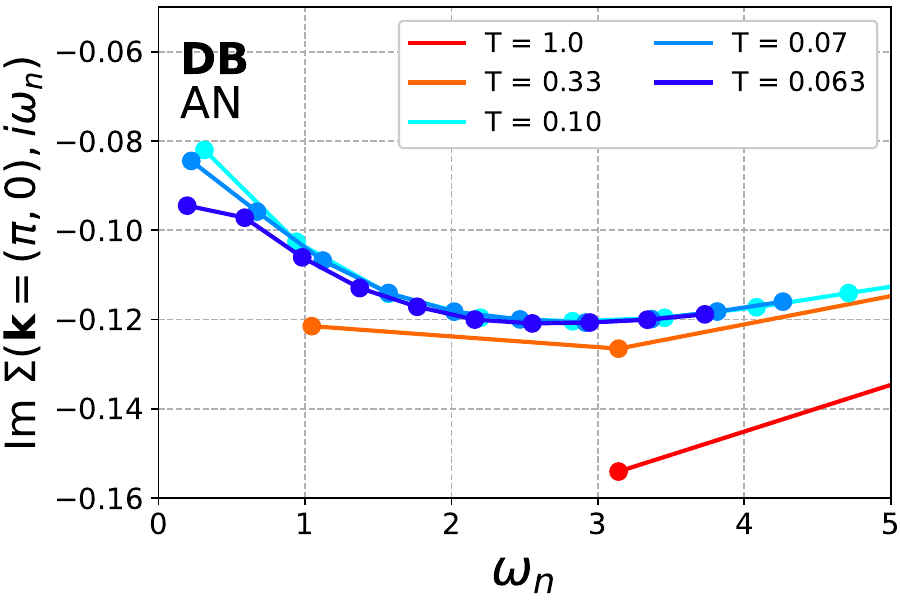}
                \includegraphics[width=0.30\textwidth,angle=0]{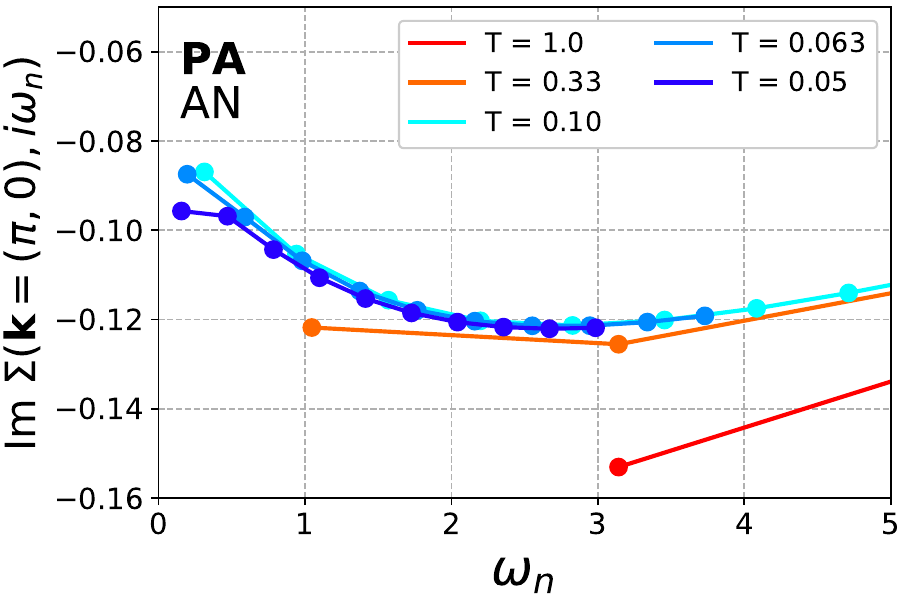}\\
                \includegraphics[width=0.30\textwidth,angle=0]{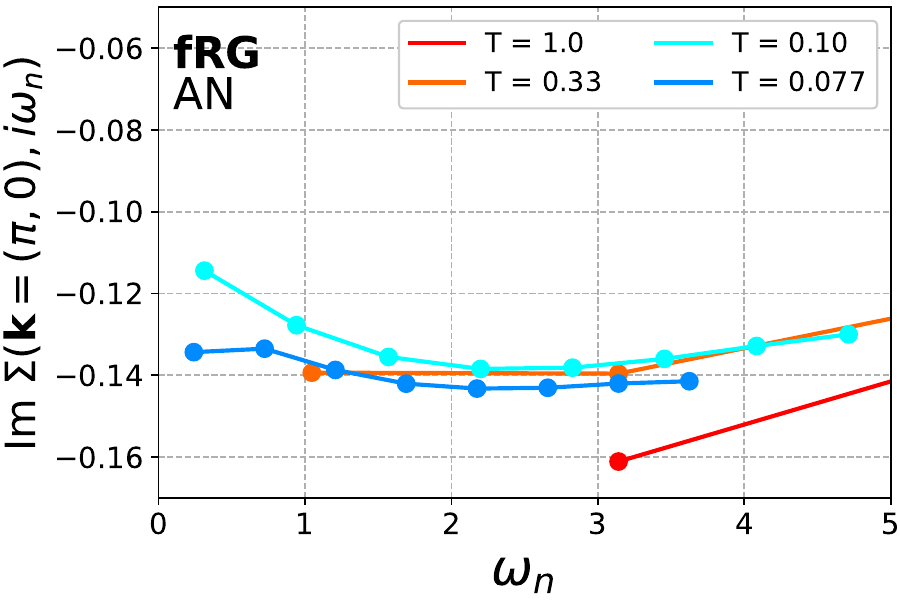}
                \includegraphics[width=0.30\textwidth,angle=0]{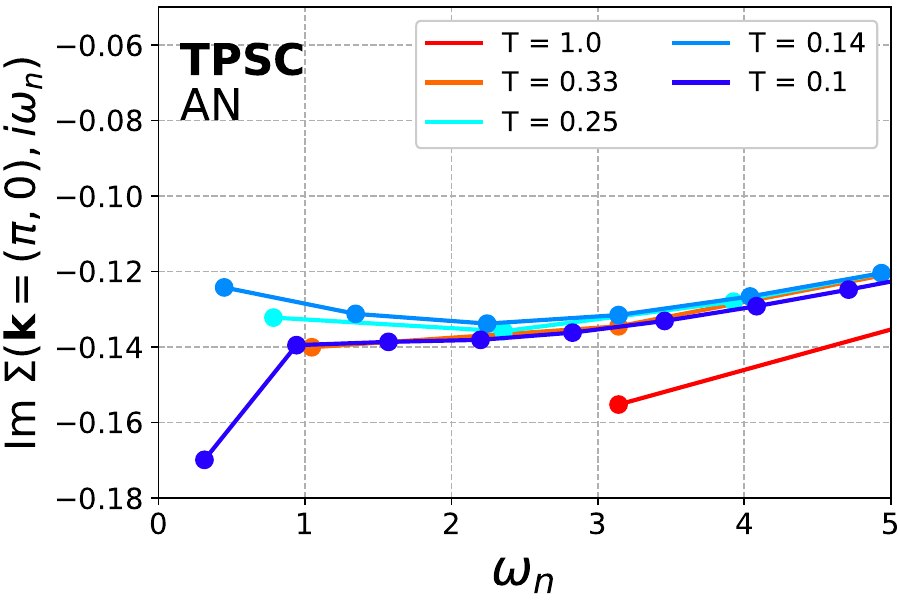}
                \includegraphics[width=0.30\textwidth,angle=0]{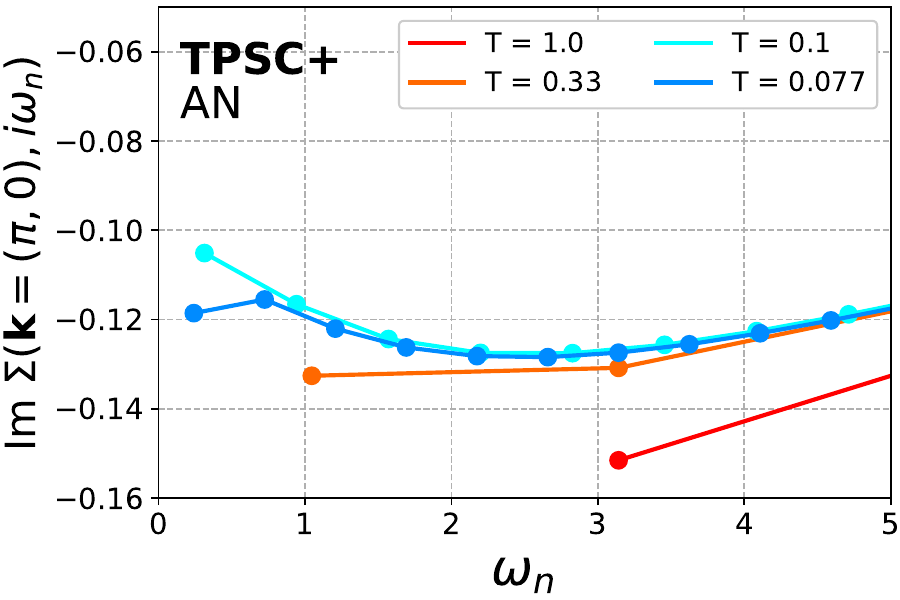}
      \caption{\label{fig:sigma_fluct_an}(Color online.) Imaginary parts of the self energies at the antinode as a function of Matsubara frequencies calculated by  various many body methods. Please note that the lowest temperatures shown sometimes differ for the respective methods in order to show as many of the respective temperature regimes as possible. Also note that the vertical axis is different for the last row of three figures.}
\end{figure*}
\begin{figure*}[ht!]
\centering		
				\includegraphics[width=0.30\textwidth,angle=0]{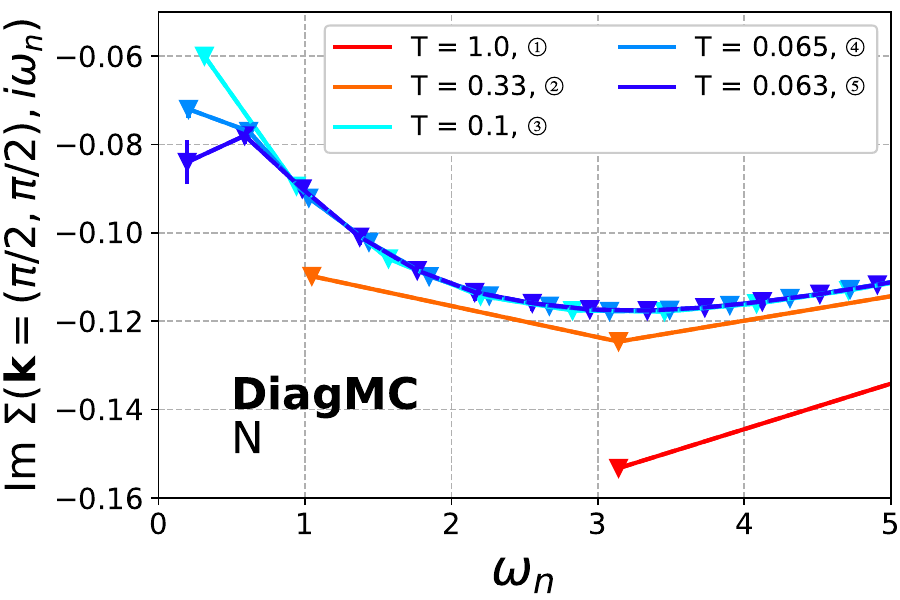}
                \includegraphics[width=0.30\textwidth,angle=0]{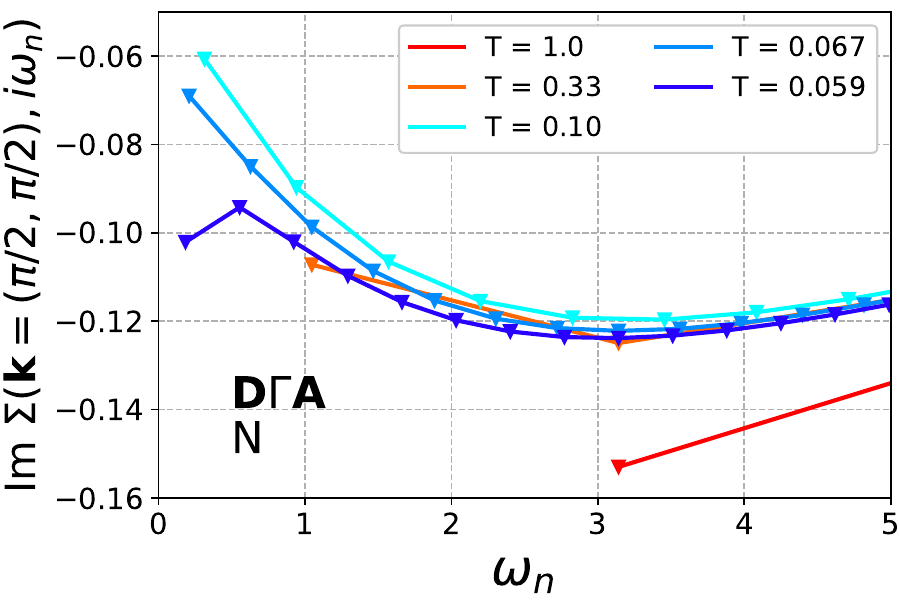}
                \includegraphics[width=0.30\textwidth,angle=0]{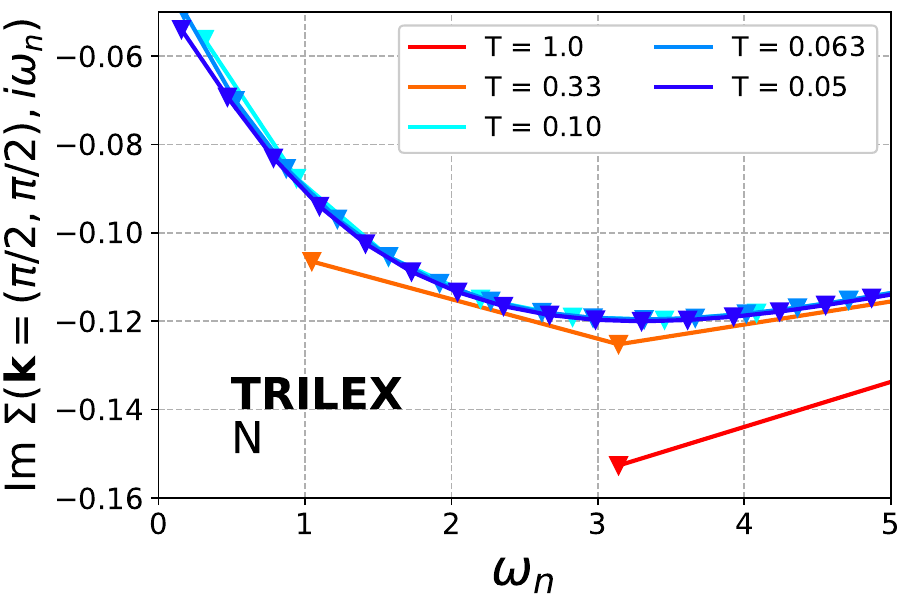}
                
                \includegraphics[width=0.30\textwidth,angle=0]{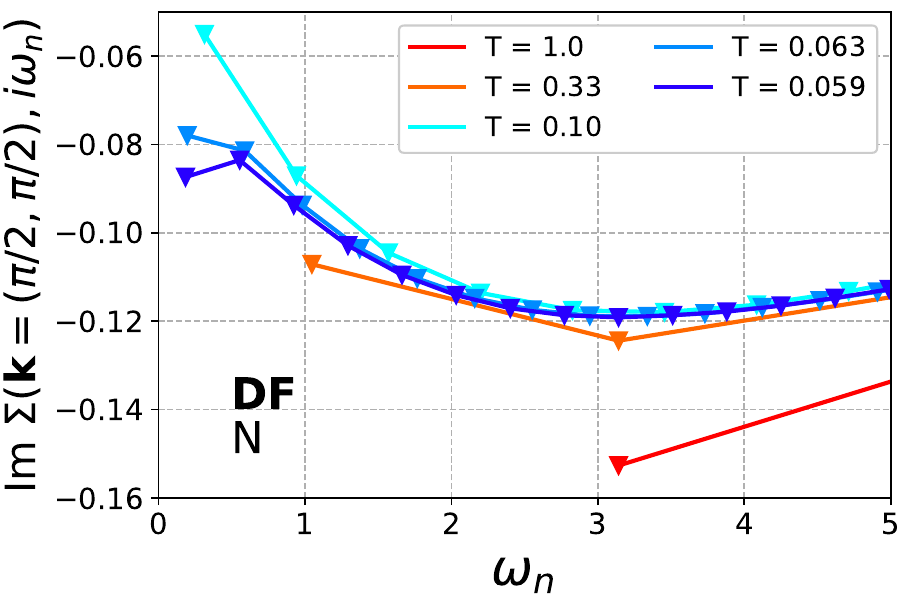}
                \includegraphics[width=0.30\textwidth,angle=0]{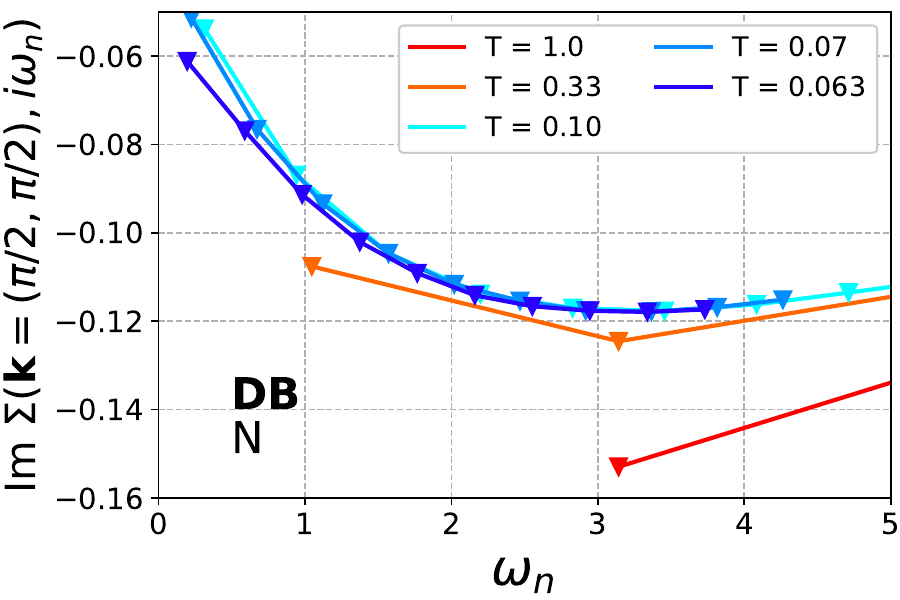}
                \includegraphics[width=0.30\textwidth,angle=0]{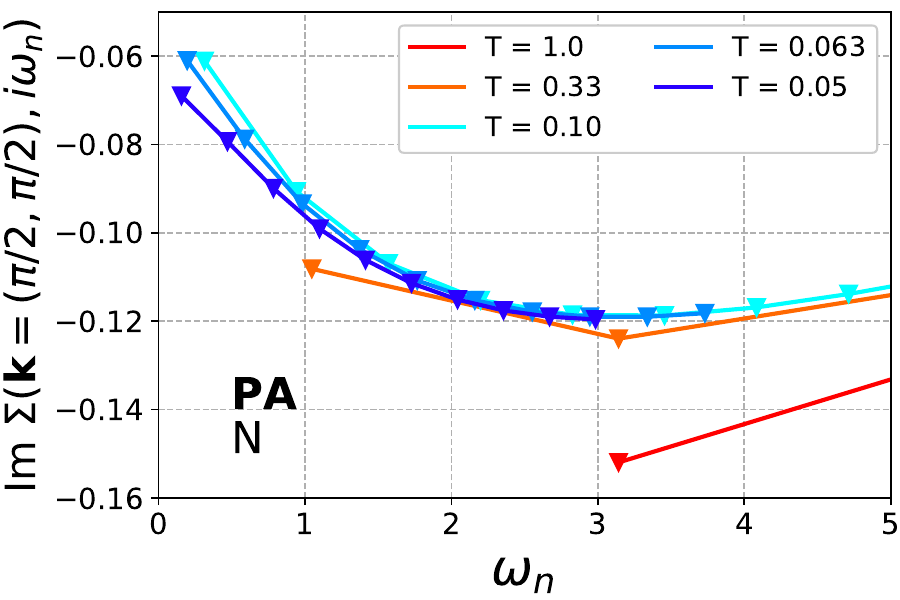}\\
                \includegraphics[width=0.30\textwidth,angle=0]{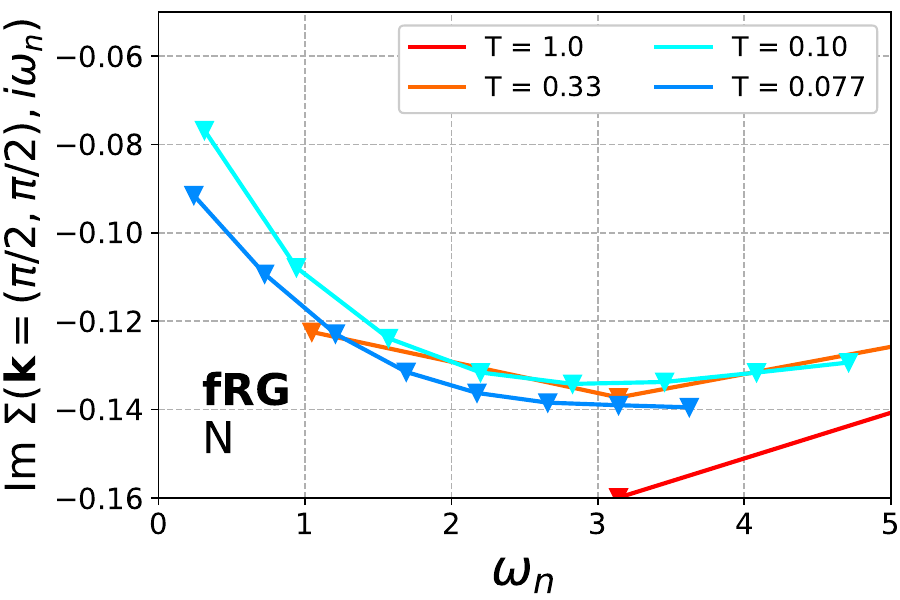}
                \includegraphics[width=0.30\textwidth,angle=0]{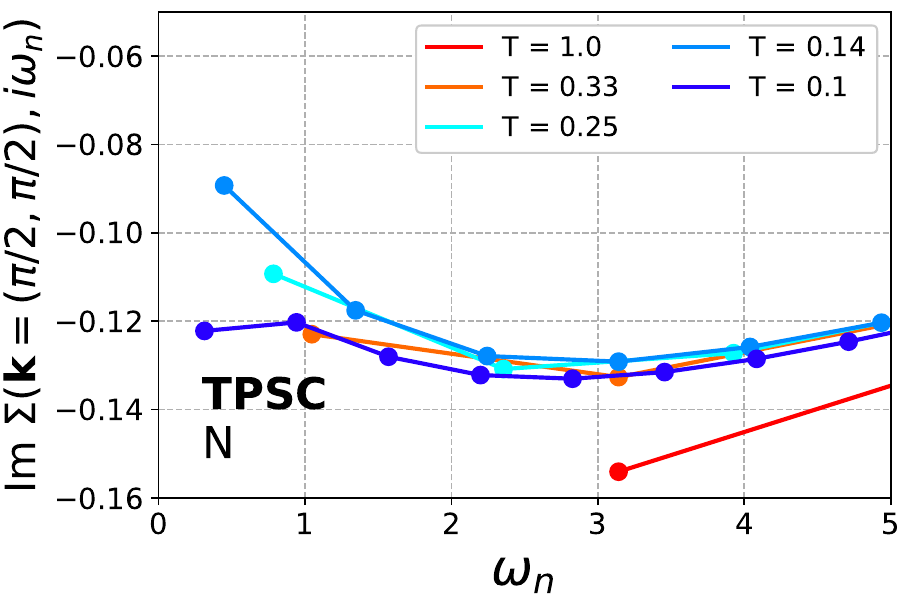}
                \includegraphics[width=0.30\textwidth,angle=0]{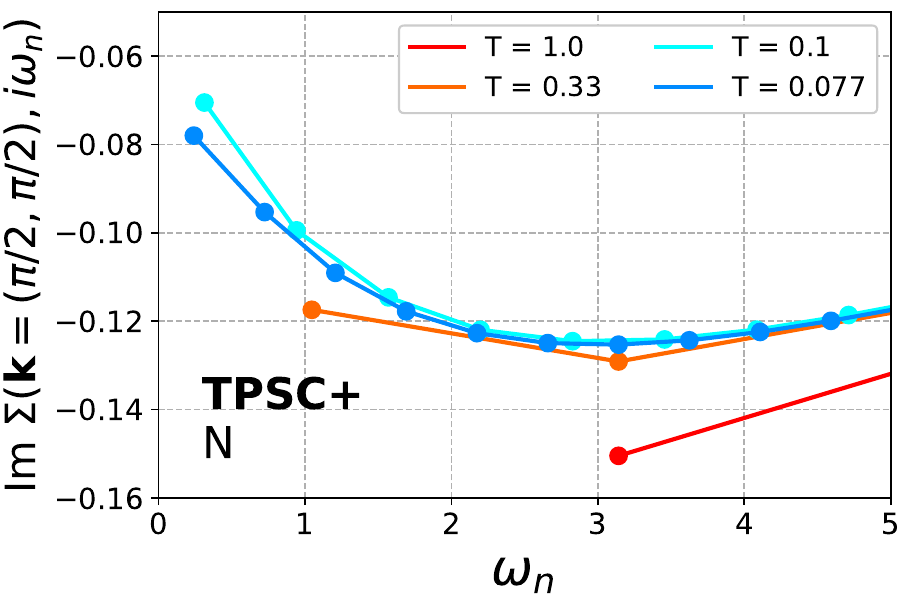}
      \caption{\label{fig:sigma_fluct_n}(Color online.) Imaginary parts of the self energies at the node as a function of Matsubara frequencies calculated by various many body methods. Please note that the lowest temperatures shown sometimes differ for the respective methods in order to show as many of the respective temperature regimes as possible.}
\end{figure*}
Let us also note here that, analogously to what was done with DMFT in Fig.~\ref{fig:spectrum}, 
allowing for AF order within these cluster methods - and hence mimicking the large correlation length regime as an ordered state - 
would most likely improve the agreement, see also \cite{Fratino2017}. 
We do not attempt this here - see however Sec.~\ref{sec:two_particle} for additional information about ordering within CDMFT.

\subsection{\texorpdfstring{D$\Gamma$A, TRILEX, and dual fermions/bosons}{DGA, TRILEX, and dual fermions/bosons}}
\label{sec:diag}

In view of the limitations of cluster embedding theories for this $U\!=\!2$ weak coupling regime, we  
now turn to an alternative way of treating long-ranged correlations more efficiently here, 
namely diagrammatic extensions of DMFT.

We display in Fig.~\ref{fig:sigma_fluct_an} the results of several such diagrammatic extensions for 
the frequency-dependence of the self-energy at the antinode at different temperatures (apart from the diagrammatic extensions of DMFT also other approximation methods are considered there, see next subsection). 
Fig.~\ref{fig:sigma_fluct_n} displays the results for the nodal point. Please note that the lowest temperature displayed is not always the same for the different methods, and it is useful to refer to Table~\ref{tab:scales} 
as a reminder of the important crossover scales.
For the sake of comparison, the first panel shows again the data from the DiagMC benchmark.

We observe that all of the diagrammatic extensions of DMFT presented here [ladder-D$\Gamma$A with a Moriya $\lambda$-correction in the spin channel (App.~\ref{app:dga}), TRILEX $\Lambda^2$ (App.~\ref{app:trilex}), ladder DF (App.~\ref{app:df}) and single-shot DB (please note that, in the absence of the nonlocal interaction, the fully self-consistent DB theory would coincide with the DF approach when the bosonic hybridization function is discarded, see App.~\ref{app:db})] are able to correctly reproduce the crossover from 
the incoherent to the metallic regime. Indeed, all methods display incoherent behavior (region \textcircled{1}) at high temperatures, before the onset of quasiparticles becomes visible first for the nodal point (region \textcircled{2}) and then, at lower temperatures, for the antinode (region \textcircled{3}). The temperatures of this onset $T_{\text{QP}}$, if at all, only slightly deviate from each other and the benchmarks within the numerical accuracy. 

Larger deviations, both on a qualitative and quantitative level become visible, however, when lowering the temperature into the insulating pseudogap regime associated with growing magnetic correlations. 
Let us remind the reader that this crossover is  
signalled by a second change of slope in the self-energies - first at the antinode (region \textcircled{4}) and then at the node (region \textcircled{5}) - corresponding to a scattering rate that grows upon cooling. 

Whereas D$\Gamma$A and DF correctly reproduce these crossovers into the pseudogap regime, 
TRILEX does not exhibit these changes of slope, down to the lowest temperatures where we could converge the method. 
The DF method also succeeds rather quantitatively, both at the node and antinode, while the DB method appears to perform better at the antinode than at the node (but does not open the gap at the accessible temperatures).
From a more quantitative point of view, DF and DB slightly underestimate the scattering rate at the node with respect to DiagMC whereas D$\Gamma$A seems to slightly overestimate the scattering rate at the antinode and simultaneously exhibits a slightly lower $T_{*}^{\text{N}}$ than the benchmark.

Summarizing, we conclude that among the diagrammatic extensions of DMFT presented here, the D$\Gamma A$ and the DF method appear to be best at capturing the different crossover regimes for the self-energy.
In terms of the practical ability of performing calculations in this parameter regime, 
we must point out that all methods suffer from convergence problems when going down to lower and lower temperatures. 
The reason for these problems varies from method to method. 
For the benchmark methods: in DiagMC the series cannot be summed at low-$T$ and the DQMC suffers from the exponentially growing correlation length for $T<T_{\text{min}} \approx 0.063$.  
In the case of the D$\Gamma$A ($T_{\text{min}} \approx 0.05$), lower temperatures can be reached if one is able to converge in the internal momentum grids. The same is true for TRILEX ($T_{\text{min}} \approx 0.05$), DF ($T_{\text{min}} \approx 0.05$) and DB ($T_{\text{min}} \approx 0.063$). Please also note that within DiagMC the lowest reachable temperature is different for node and antinode ($1/T_{\text{min}}^{\text{AN}}\!=\!18$ vs. $1/T_{\text{min}}^{\text{N}}\!=\!16$).

\subsection{Other approaches: TPSC, TPSC+, fRG, PA}
\label{sec:uncontrolled}

Figs.~\ref{fig:sigma_fluct_an} and \ref{fig:sigma_fluct_n} also show results for three other approaches: 
TPSC/TPSC+, fRG and the parquet approximation (PA). 
Like the diagrammatic extensions of DMFT, all of them are able to reproduce the two distinct quasiparticle coherence scales $T_{\text{QP}}^{\text{N,AN}}$ at the node and antinode. 
However, there are significant deviations from the benchmark regarding the onset of the insulating 
pseudogap behavior. 

TPSC is one of the first methods in which a detailed understanding of the mechanism responsible for 
the weak-coupling pseudogap was achieved early on (see Refs.~\cite{Vilk1995,Vilk1996,Vilk1997} and 
Sec.~\ref{sec:spin_fluct}). As seen from Figs.~\ref{fig:sigma_fluct_an} and \ref{fig:sigma_fluct_n}, 
the change of slope in the self energies associated with the pseudogap opening is indeed qualitatively 
captured by TPSC, but the onset temperatures $T_{*}^{\text{AN,N}}$ are severely overestimated. As discussed 
in Sec.~\ref{sec:two_particle}, this is due to an overestimation of spin fluctuations in this method. 
A recent improvement of the method, TPSC+~\cite{Wang2019}, leads to a definite improvement 
in this respect as shown 
on the figures. TPSC+ partially feeds back the self-energy into the fluctuation propagators, mimicking frequency-dependent vertex corrections. 
\begin{figure*}[ht!]
      \includegraphics[width=0.32\textwidth,angle=0]{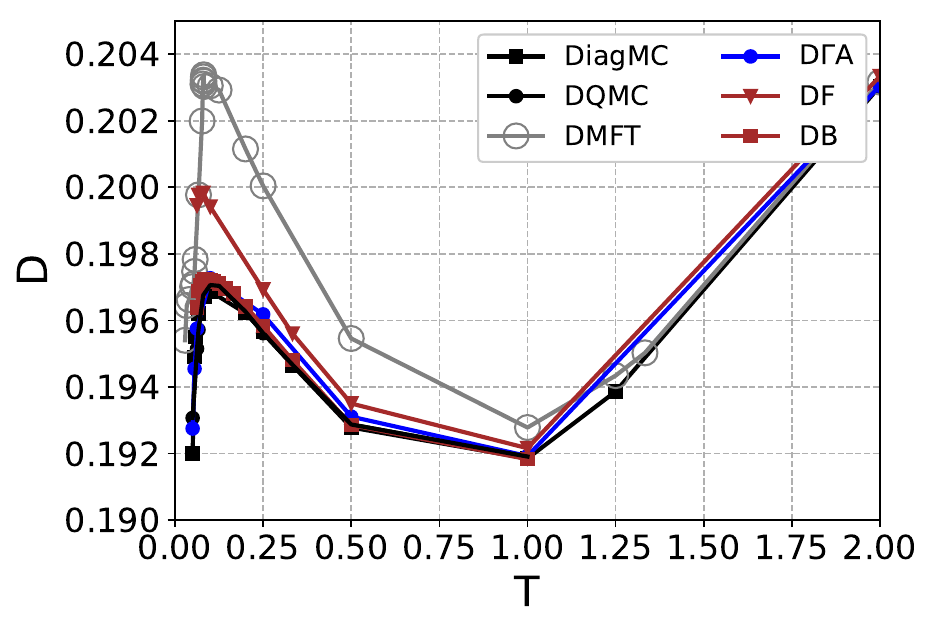}
      \includegraphics[width=0.32\textwidth,angle=0]{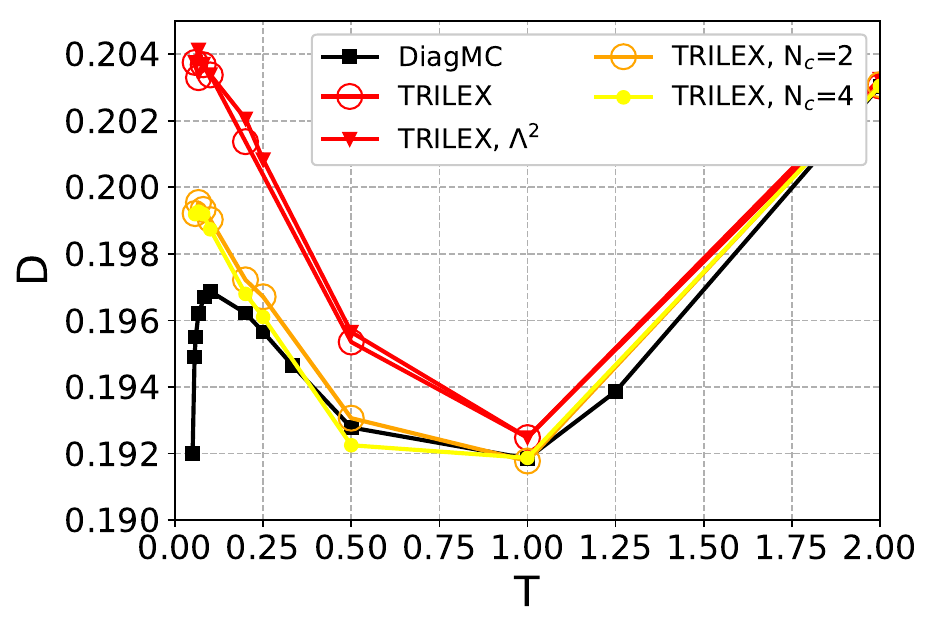}
      \includegraphics[width=0.32\textwidth,angle=0]{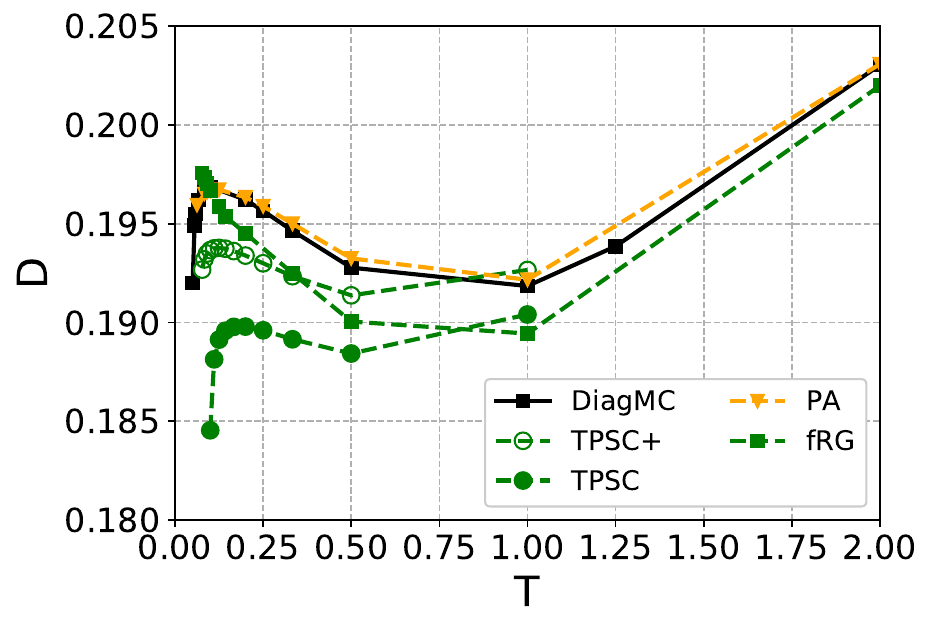}
      \caption{\label{fig:docc}(Color online.) 
      Double occupancy $D$ as a function of temperature from various methods.}
\end{figure*}
The PA appears to eventually capture insulating behavior at the antinode, although at lower temperatures $T<0.05$ in comparison to DiagMC, but doesn't open a pseudogap at the node at this temperature.

The fRG calculations are possible only down to a ``pseudocritical" temperature scale $T\simeq 0.07$
at which the running coupling constants diverge (see also the discussion in Ref.~\cite{Hille2020b}). 
Down to this temperature, however, fRG is in qualitative agreement with the benchmark and shows a non-metallic behavior at the antinode (regime \textcircled{4}).

\section{Double occupancy and Pomeranchuk effect}
\label{sec:docc}

In view of its physical significance discussed below, we present in this section the temperature-dependence 
of the double occupancy: 
\begin{equation}
 D = \left<n_\uparrow n_\downarrow\right>.
 \label{eqn:docc}
\end{equation}
It is displayed as a function of temperature in Fig.~\ref{fig:docc}, as obtained from different methods. 
We see (left panel, DQMC and DiagMC benchmarks) that three regimes are found: at high $T$ 
(down to about $T\simeq 1$) $D(T)$ decreases 
upon cooling and then reaches a minimum, at intermediate temperatures $D(T)$ actually increases upon cooling and, finally, $D(T)$ sharply drops 
when entering the gapped regime. 
The high $T$ regime is expected and easy to understand: as $T$ is raised, an increasing number of high-energy doubly occupied 
configurations are thermally populated. This is apparent from the simple expression of $D(T)$ in the atomic (zero hopping) limit: 
\begin{equation}
 D_{\text{at}}=\frac{1}{2+2\exp{(U/(2T))}},
 \label{eqn:docc_at}
\end{equation}
which approaches $D_{0}=0.25$ for $U/T\rightarrow{0}$, i.e. at very high temperatures or, alternatively, in the non-interacting limit. 
The intermediate regime in which $D(T)$ increases upon cooling is more interesting - note that this regime includes in particular large parts of 
the metallic region \textcircled{3}. As observed early on in Ref.~\citep{Georges1993},  this apparently counter-intuitive 
non-monotonic behaviour of $D(T)$ can be understood qualitatively from entropy considerations. Observing that the 
entropy $S$ is obtained from the free energy $F$ as $S=-\frac{\partial F}{\partial T}$ and that $D=\frac{\partial F}{\partial U}$, one 
obtains the thermodynamic Maxwell relation: 
\begin{equation}
\left.\frac{\partial D}{\partial T}\right|_U\,=\,-\left.\frac{\partial S}{\partial U}\right|_T.
\label{eq:maxwell}
\end{equation}
Increasing $U$ at fixed temperature in the metallic regime leads to an increase in entropy. Indeed, in this regime, the entropy is linear in $T$ with a slope related to the effective mass which 
grows as $U$ is increased. Hence, the right-hand side of Eq.~(\ref{eq:maxwell}) is negative and thus $D(T)$ 
must increase upon cooling/decrease upon heating in this regime. 
This is the same phenomenon as the famous `Pomeranchuk effect' in liquid $^3$He (Clausius-Clapeyron equation): upon heating the system, a tendency to increased localization (smaller $D$) is found, because localization leads to a higher (spin) entropy and is thus favorable thermodynamically (until thermal population of doubly 
occupied sites kicks in at higher $T$). Note that the sign of $\partial D/\partial T$ also directly determines the shape of the isentropy lines in 
the $(U,T)$ plane, which are defined by $S(T_i(U),U)=\text{const.}$ and hence obey~\cite{Werner2005,Dare2007}:
\begin{equation}
    c(T_i)\frac{\partial T_i}{\partial U}=T_i\frac{\partial D}{\partial T}\bigg\rvert_{T=T_i},
\end{equation}
with $c=T\frac{\partial S}{\partial T}$ the specific heat per lattice site. Hence cooling can in principle 
be achieved by increasing the interaction strength adiabatically in the regime where $\partial D/\partial T<0$, as 
initially suggested in \cite{Werner2005}, further discussed in \cite{Dare2007}, and experimentally realized
in cold atomic systems with extended $SU(6)$-symmetry~\cite{Taie2012}. 

Finally, the low $T$ behaviour in which $D$ sharply decreases again upon cooling corresponds to temperatures around regime \textcircled{4} in which the system behaves as an antiferromagnetc insulator. 
We note that the total energy of the Hubbard model is given by: 
\begin{eqnarray}
    \left<H\right>&=& -t\sum\limits_{\left<ij\right>,\sigma} \left<c^{\dagger}_{i\sigma} c_{j\sigma}\right> + U\sum\limits_{i}\left<n_{i\uparrow}n_{i\downarrow}\right>\\
    &=&E_{\text{kin}}+E_{\text{pot}} \nonumber,
\end{eqnarray}
so that $D=E_{\text{pot}}/U$. The drop in $D$ can be understood from the fact that at small $U$ the crossover into the (Slater) antiferromagnetic correlation regime  
corresponds to a gain in potential energy~\cite{Fratino2017,Schaefer2015b, Rohringer2016}. It was shown in Ref.~\cite{Fratino2017} that at strong coupling, in contrast, $D$ increases when entering the (Heisenberg) antiferromagnetic correlations regime, corresponding to a gain in kinetic energy. 
 
We see (left panel of Fig.~\ref{fig:docc}) that DMFT reproduces all three regimes qualitatively, but overestimates the amplitude of the `Pomeranchuk effect' 
by approximately a factor of two, as discussed in Ref.~\cite{Dare2007}. The reason is that the spin entropy in the localized state at high-$T$ is overestimated in DMFT, due to its neglect of spatial correlations. 
Obviously, in DMFT, the sharp drop at low $T$ corresponds to the phase transition into an ordered phase - 
a mean-field description of the actual crossover (see above). 
This can also be seen as an increase of the magnetic local moment, since via $\langle S_z^2\rangle=1-2D$ the double occupancy is related to the increase of AF correlations.

All the methods displayed in Fig.~\ref{fig:docc} qualitatively reproduce the non-monotonous behaviour of $D(T)$ but the sharp drop at low $T$ is, as expected, only present in the methods which can describe the low $T$ antiferromagnetic insulator regime. 
It is, for this reason, absent in the TRILEX based methods for the range of temperatures studied 
(middle panel). Single-site TRILEX also overestimates $D$ significantly, but cluster extensions 
of TRILEX are closer to the benchmark at intermediate $T$.
DF similarly overestimates the double occupancy. 
For details, including differences between different variants of these `dual' methods, see the discussion about self-consistency in App.~\ref{app:db}. fRG qualitatively captures the Pomeranchuk effect. In the present implementation, however, it does not display a drop at low temperatures.

In contrast, D$\Gamma$A and DB (single-shot) follow the benchmark quite accurately. 
Also, the parquet approximation (PA) appears to capture all three described regimes, while TPSC and TPSC+ qualitatively predict the correct physical picture albeit less accurate quantitatively than PA in comparison to the benchmarks.

\begin{figure}[t!]
        \centering
                \includegraphics[width=0.39\textwidth]{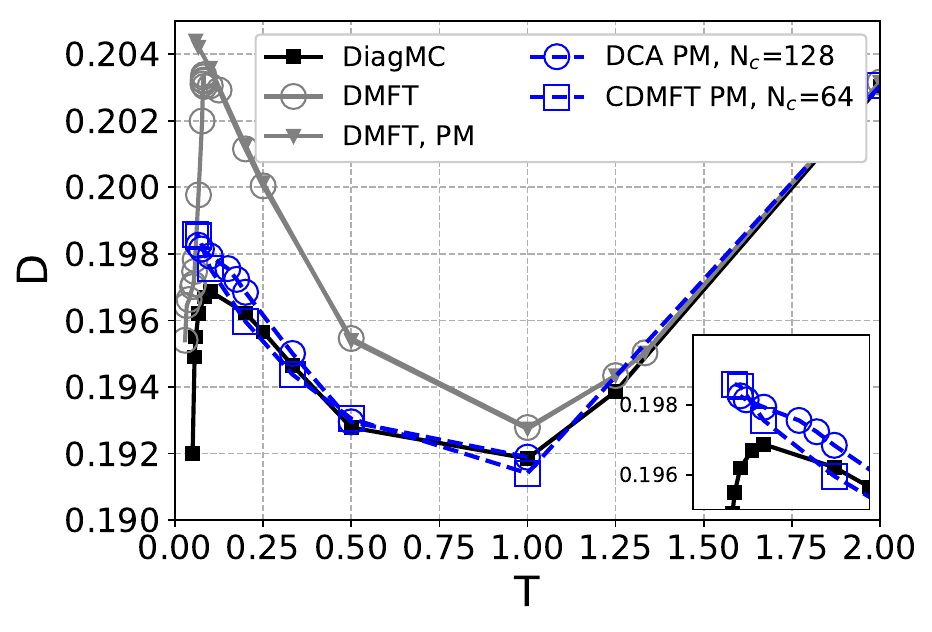}
        \caption{\label{fig:Docc_cluster}(Color online.) 
        Results for the double occupancy from DMFT, DCA and CDMFT compared to DiagMC. For DMFT, we display both the (metastable) solution obtained 
        by constraining the system not to order (PM) and the lower free-energy solution with AF magnetic ordering. 
        The zoom in the inset confirms the absence of a downturn in the cluster techniques when constrained to the PM solution, due to the insufficient cluster size.}
\end{figure}

The above results suggest that long-range magnetic correlations are the cause of the downturn of the double occupancy at low temperatures. 
To further establish this point, we display in Fig.~\ref{fig:Docc_cluster} the double occupancy vs. temperature calculated with DMFT by either constraining the 
solution to remain paramagnetic or allowing for the solution with long-range AF order (which is the lower-energy stable solution at low-$T$ within DMFT). 
One sees that only the latter displays the downturn, emphasizing that the AF ordered solution, despite violating Mermin-Wagner theorem, is a better approximation 
to thermodynamic quantities 
in a regime where the correlation length is large. We also display in Fig.~\ref{fig:Docc_cluster} the results obtained with DCA and CDMFT, 
restricted to the PM solution. 
In that case, as expected, the downturn is not reproduced, since the cluster sizes investigated here are too small as compared to the correlation length 
(see also the discussion in Sec.~\ref{sec:dca}).

\begin{figure}[b!]
        \centering
                \includegraphics[width=0.35\textwidth]{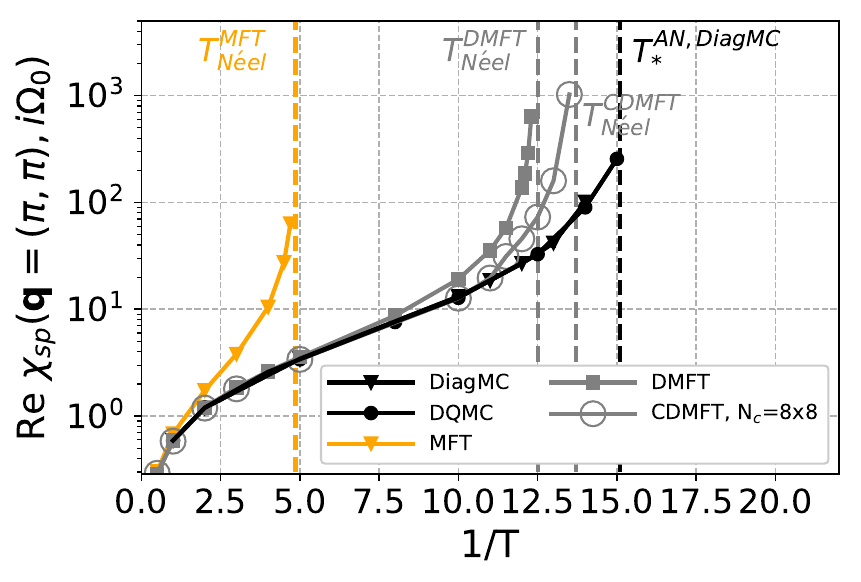}
        \caption{\label{fig:chi_cdmft}(Color online.) Antiferromagnetic static susceptibility $\chi_{\text{sp}}(\mathbf{q}\!=\!(\pi,\pi), i\Omega_{n}\!=\!0)$ as a function of (inverse) temperature 
      for various methods on a logarithmic scale.}
\end{figure}

\begin{figure*}[ht!]

                \includegraphics[width=0.32\textwidth,angle=0]{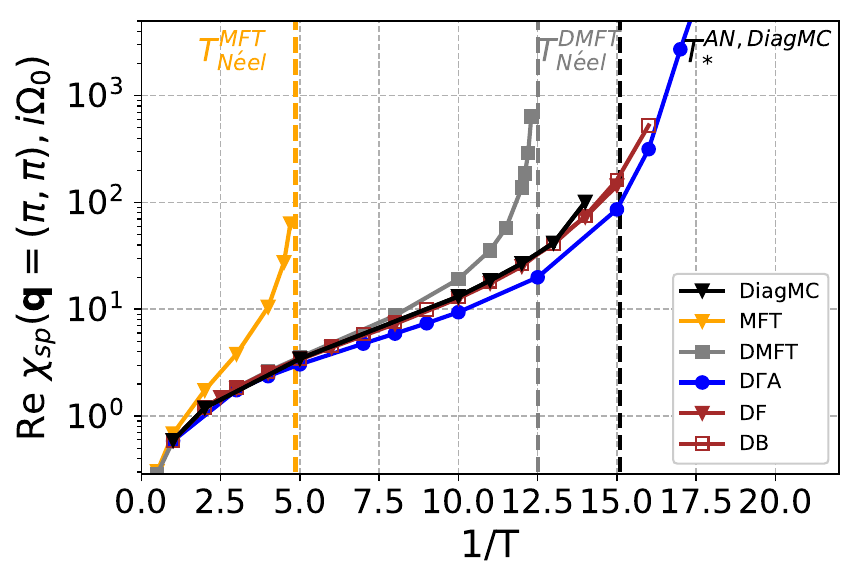}
                \includegraphics[width=0.32\textwidth,angle=0]{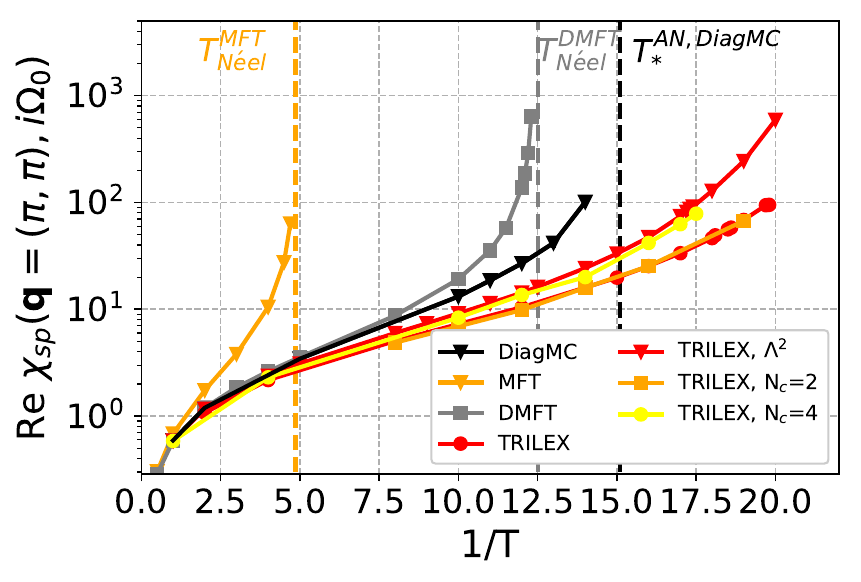}
                \includegraphics[width=0.32\textwidth,angle=0]{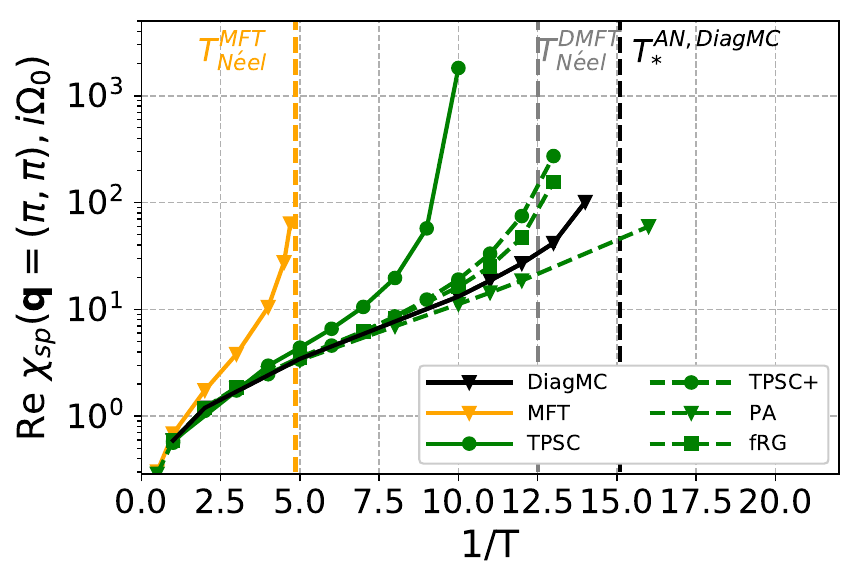}
      \caption{\label{fig:chi_sp_pipi}(Color online.) Antiferromagnetic static susceptibility $\chi_{\text{sp}}(\mathbf{q}\!=\!(\pi,\pi), i\Omega_{n}\!=\!0)$ as a function of (inverse) temperature 
      for various 
      methods on a logarithmic scale (continued).}
\end{figure*}

We end this section with a technical remark on the actual calculation of $D$ in the investigated methods. In DQMC and DiagMC $D$ can be directly calculated as an equal-time correlation function. For the other methods the calculation of the double occupancy is possible either via the (lattice) Green's function and self-energy, utilizing the equation of motion technique based on the Galitskii-Migdal formula \cite{Galitskii1958}:
\begin{equation}
 D = \frac{T}{U}\sum\limits_{\mathbf{k},\omega_{n}}\Sigma\left(\mathbf{k},i\omega_{n}\right)G\left(\mathbf{k},i\omega_{n}\right),
 \label{eqn:docc_2}
\end{equation}
(used in the D$\Gamma$A, DF, PA and TPSC/TPSC+, for details on the latter see App.~\ref{app:tpsc}) or, when allowed by the algorithm used for the impurity solver, from a direct computation on the impurity (for DMFT, its cluster extensions and TRILEX, see also \cite{vanLoon2016}). The DB result was obtained via the local part of the lattice susceptibility as discussed in Ref.~\cite{vanLoon2016}. For a comparison to self-consistent DB we refer to App.~\ref{app:db}. For details on the calculation of $D$ in fRG we refer to App.~\ref{app:frg}. We note in passing that, when applying Eq.~(\ref{eqn:docc_2}), treating the high frequency tails is very important to obtain accurate results.

\section{Including fluctuations beyond mean-field: magnetic correlations}
\label{sec:two_particle}

In this section, we turn to two-particle correlation functions and compare the different methods for 
two important observables probing the magnetic correlations: the static antiferromagnetic spin susceptibility $\chi_{\text{sp}}(\mathbf{q}\!=\!(\pi,\pi), i\Omega_{n}=0)$ and the magnetic correlation length $\xi$.

\subsection{Antiferromagnetic static spin susceptibility}
The spin susceptibility (or spin correlation function) is given by: 
\begin{equation}
 \chi_{\text{sp}}(\mathbf{q}, i\Omega_{n})\!=\!\int\limits_{0}^{\beta}{d\tau}\sum\limits_{\mathbf{r}} e^{i\tau\Omega_{n}}{e^{-i\mathbf{q}\mathbf{r}}\left< S_{z}(\mathbf{r},\tau)S_{z}(0,0) \right>},
 \label{eqn:chi_sp}
\end{equation}
where we define here
\begin{equation}
 S_{z}(\mathbf{r},\tau)= n_{\uparrow}(\mathbf{r}, \tau) - n_{\downarrow}(\mathbf{r}, \tau),
 \label{eqn:sz}
\end{equation}
and $\beta=1/T$. Note that we omit for simplicity an additional prefactor of $1/2$ for the spin operator and that $\left<n_\uparrow\right>\!=\!\left<n_\downarrow\right>\!=\!0.5$ at half-filling in the paramagnetic 
phase ($T\neq 0$). 

The temperature dependence of the static spin susceptibility 
$\chi_{\text{sp}}(\mathbf{q}, i\Omega_{n}\!=\!0)$ is of particular interest for the case of the half-filled Hubbard model on the square lattice, as the perfect nesting (see Fig.~\ref{fig:fs_dos}) leads to a strong enhancement at $\mathbf{Q}\!=\!(\pi,\pi)$. 
The temperature dependence of $\chi_{\text{sp}}(\mathbf{q}\!=\!\mathbf{Q}, i\Omega_{n}\!=\!0)$ reflects the increasing 
dominance of antiferromagnetic spin fluctuations upon cooling. 
One starts at high $T$ with almost independent fluctuating moments and a Curie law 
$\chi_{\text{sp}}(\mathbf{q}\!=\!\mathbf{Q}, i\Omega_{n}\!=\!0) \propto T^{-1}$ 
(bosonic mean-field behavior). Approaching the $T=0$ ground-state with antiferromagnetic long-range 
order, the range of spin correlations grows and nonlocal spin fluctuations in the paramagnetic phase 
(antiferromagnetic paramagnons) develop. 
At low-$T$, a regime with an exponentially growing correlation length is found (see below).
\\\\
Figs.~\ref{fig:chi_cdmft} and \ref{fig:chi_sp_pipi} displays $\chi_{\text{sp}}(\mathbf{q}\!=\!\mathbf{Q}, i\Omega_{n}=0)$ for various methods as a function of (inverse) temperatures on a logarithmic scale. We start our analysis with Fig.~~\ref{fig:chi_cdmft}. As already described in the with the mean-field picture in Sec.~\ref{sec:mf} both MFT ($T_{\text{N{\'e}el}}^{\text{MFT}}\! \approx 0.21\!$, orange triangles) and DMFT ($T^{\text{DMFT}}_{\text{N{\'e}el}}\! \approx 0.08\!$, gray squares), due to their mean-field nature, 
incorrectly predict finite N{\'e}el temperatures: the crossover is mimicked as a true (continuous) phase transition (left panel). The thermodynamic transition manifests itself as a divergence of the susceptibility at the corresponding wave vector.

Of course, the Mermin-Wagner theorem prohibits finite temperature ordering in 2D, which is reflected in the data of both benchmark methods DiagMC (black triangles) and DQMC (black circles), whose susceptibilities do not diverge at finite $T$. 
Rather, after following the mean-field curves at high temperatures, the benchmark data enter an intermediate regime 
in which $\chi(\mathbf{Q})$ appears to increase approximately exponentially. This regime  largely coincides with 
the metallic regime \textcircled{3}, eventually crossing over into a second exponential regime at low-$T$ (see also Sec.~\ref{sec:two_slopes}), which sets in at a temperature close to the DMFT N\'eel ordering. 
\begin{figure*}[ht!]
                \includegraphics[width=0.32\textwidth,angle=0]{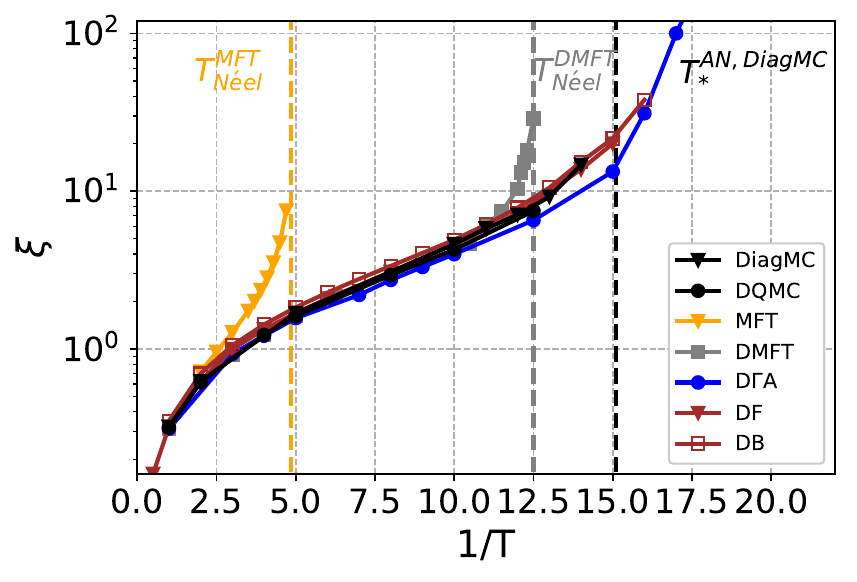}
                \includegraphics[width=0.32\textwidth,angle=0]{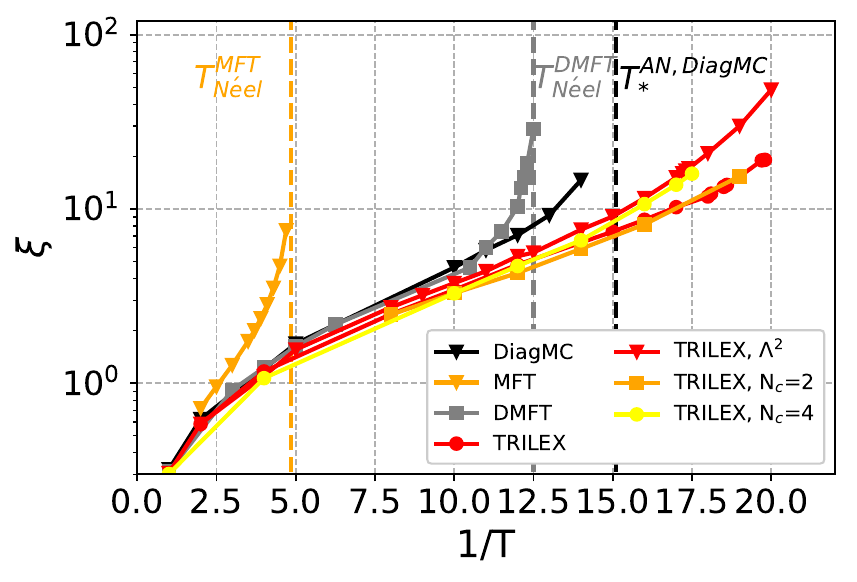}
                \includegraphics[width=0.32\textwidth,angle=0]{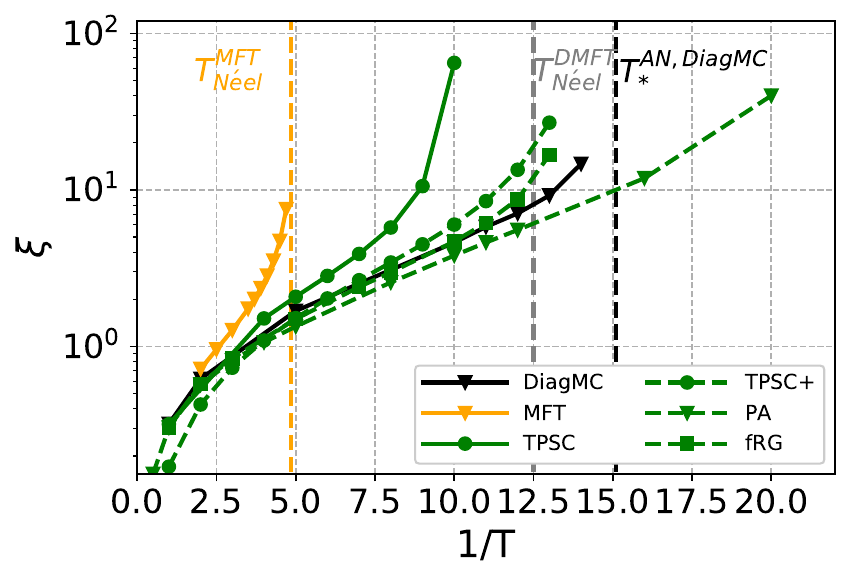}
      \caption{\label{fig:xi}(Color online.) Magnetic correlation lengths $\xi$ extracted from the magnetic susceptibility as a function of (inverse) temperature for various 
      methods on a logarithmic scale.}
\end{figure*}
This low $T$ exponential regime is to be expected, since there, the charge degrees of freedom 
are frozen out by the gap and the system enters the insulating regime \textcircled{5}, as indicated by the black dashed line $T_{*}^{\text{AN,DiagMC}}$. 
In this regime, the effective spin dynamics is expected to be 
described by a non-linear sigma model and this exponential growth is typical of the lower 
critical dimension $d=2$~\cite{Chakravarty1988,Chakravarty1989,Vilk1997,Borejsza2003, Borejsza2004}. 
The first exponential observed in the metallic regime is more surprising and will be 
discussed in more details below. 
Let us stress that, due to the reasons already mentioned for single-particle quantities, both benchmark methods are limited in terms of the temperatures they can reach ($T_{\text{min}} \approx 0.07$). Please note that the lowest reachable temperature differs from the one for the self-energy.

The inclusion of short-range correlations with CDMFT leads to (i) a quantitative agreement with the benchmark until $T \approx 0.1$ and 
(ii) only a slight drop of the N{\'e}el temperature in comparison to DMFT, to $T_{\text{N{\'e}el}} \approx 0.073$. In principle, as the cluster extensions of DMFT are controlled methods, the Mermin-Wagner theorem is restored in the infinite cluster size limit. However, this restoration has been shown to be logarithmic in the strong-coupling regime \cite{Maier2005a}, which is in agreement with this very small change in $T_{\text{N{\'e}el}}$ (see also \cite{Klett2020}).

Turning to the diagrammatic extensions of DMFT, in the left panel of Fig.~\ref{fig:chi_sp_pipi} one can see that D$\Gamma$A (left panel), which respects the Mermin-Wagner theorem \cite{Toschi2007, Katanin2009, Rohringer2016, RohringerPC2020}, captures well the different regimes of the benchmark 
(Curie law at high temperatures and the two exponential regimes). The small quantitative underestimation observed here may potentially be cured by an improved version of the Moriya $\lambda$-correction \cite{Rohringer2016} or a more thorough treatment of the asymptotics of the vertex function as a function of frequency~\cite{Wentzell2020,Kaufmann2017,Katanin2020}. 
DF and single-shot DB agree well with the benchmark where the algorithm can be reliably converged. 

In the case of TRILEX (central panel) we present results for different variants of the method, all of which seem to capture a low temperature exponential scaling, however, with different degrees of accuracy: whereas single-site TRILEX (red circles, solid line) and cluster TRILEX with two cluster sites (TRILEX N$_{c}=2$, orange squares) largely underestimate the values of the susceptibility, the result is significantly improved by increasing the cluster size to four (TRILEX N$_{c}=4$, yellow circles). Remarkably, the TRILEX variant with the electron-boson vertex inserted on both sides (TRILEX $\Lambda^2$, red triangles, see App.~\ref{app:trilex}) seems to be on top of the cluster N$_{c}=4$ results. Investigating how these values eventually converge to the exact results with $N_c \rightarrow \infty$ is left to future studies.

We finally turn to the other methods (right panel): 
Both PA and fRG capture quantitatively the high-$T$ Curie regime. 
The PA appears to systematically underestimate $\chi_{\text{sp}}$ from $1/T=10$ on, whereas fRG is overestimating it.
Let us just comment here that although the presented fRG scheme does not fulfill the Mermin-Wagner theorem, its recent multiloop extension \cite{Kugler2017, Kugler2018,Hille2020} and the PA does \cite{Bickers1992, Eckhardt2020} (see also \cite{Baier2004}).
Both TPSC and TPSC+ are in agreement with the benchmarks at high temperature, but, as already mentioned in the previous section about the self-energy, the spin fluctuation intensity is significantly overestimated in TPSC with respect to the benchmark (right panel). This overestimation is improved again in the TPSC+.

Our results also demonstrate that, remarkably, neither the fact that the Mermin-Wagner theorem is respected by a theory nor that a theory 
uses self-consistent interacting Green functions,   
guarantees a quantitatively better result, as can be inferred by a comparison of the TPSC or PA result with the benchmark, respectively.

\subsection{Magnetic correlation length}
\label{subsec:corrlength}

The spatial range of antiferromagnetic spin correlations can be quantified by the magnetic correlation length 
$\xi$, which in practice can be extracted by a fitting procedure using the Ornstein-Zernike (OZ) form of the 
bosonic propagator \cite{Zernike1916}:
\begin{equation}
 \chi_{\text{sp}}(\mathbf{q}, i\Omega_{n}=0) = \frac{A}{(\mathbf{q}-\mathbf{Q)^2 + \xi^{-2}}}.
 \label{eqn:ornstein}
\end{equation}
The results presented in this section were obtained using a slightly modified form of the OZ expression, 
appropriate for a model on a lattice, as described in App.~\ref{app:ornstein}. 
Note that we neglect (small) deviations from the OZ form beyond mean-field.
Within the OZ form, the AF static susceptibility discussed previously obeys: 
$\chi(\mathbf{Q},i\Omega_n=0) \sim A \xi^2$.

Fig.~\ref{fig:xi} shows the correlation length $\xi$ obtained via such a fitting procedure applied to the susceptibility data from the different methods. One can immediately see that the curves are, to a large extent, qualitatively similar to the susceptibility curves of Fig.~\ref{fig:chi_sp_pipi}. 
Interestingly, the intermediate-$T$ exponential behaviour in the metallic regime is clearly visible in the 
DiagMC benchmark data. The lowest-$T$ exponential regime in the insulator is hard to reach with DiagMC, but 
is obtained in D$\Gamma$A (and, less clearly, in PA and TRILEX) which can be used down to lower temperature than the benchmark methods. We note, however, that sizeable quantitative differences do exist between the different methods at low-$T$, 
and hence we conclude that the precise determination of the correlation length in the low-$T$ regime, where it becomes 
exponentially large, is a challenge for all state of the art computational methods currently available.

\subsection{The three regimes of magnetic correlations}
\label{sec:two_slopes}
Summarizing, we observe three successive regimes for magnetic correlations as the temperature is lowered. 
At high-$T$ ($\gtrsim 1/5$), a Curie mean-field behaviour is found with a small correlation length ($\lesssim 2$), 
and all methods (except static MFT) are basically in quantitative agreement in this regime. 
At intermediate temperatures $1/5\gtrsim T\gtrsim1/12.5 \simeq T_{\text{N\'eel}}^{\text{DMFT}}$, 
we find a correlation length that appears to increase exponentially to a good approximation, with 
values $\xi(T=1/5)\simeq 2$ reaching $\xi(T=1/13)\simeq 10$. 
This is somewhat surprising since this regime is metallic with quasiparticles that become more 
coherent as temperature is lowered, at least in the nodal region. 
We discuss further the physical significance of this finding in Sec.~\ref{sec:spin_fluct} by comparing to spin-fluctuation theories. 
All methods reproduce this intermediate exponential regime qualitatively. 
On the quantitative level, excellent agreement is found between the benchmark DiagMC and especially 
DF and DB throughout this regime. 
We also note very good agreement for the determination of the correlation length with DMFT and TPSC+ down to 
$T\simeq 1/10$. Most other methods (D$\Gamma$A, the various variants of TRILEX and the PA) also provide 
a satisfactory determination of the susceptibility and correlation length in this intermediate regime. 

\begin{figure*}[ht!] \includegraphics[width=0.32\textwidth,angle=0]{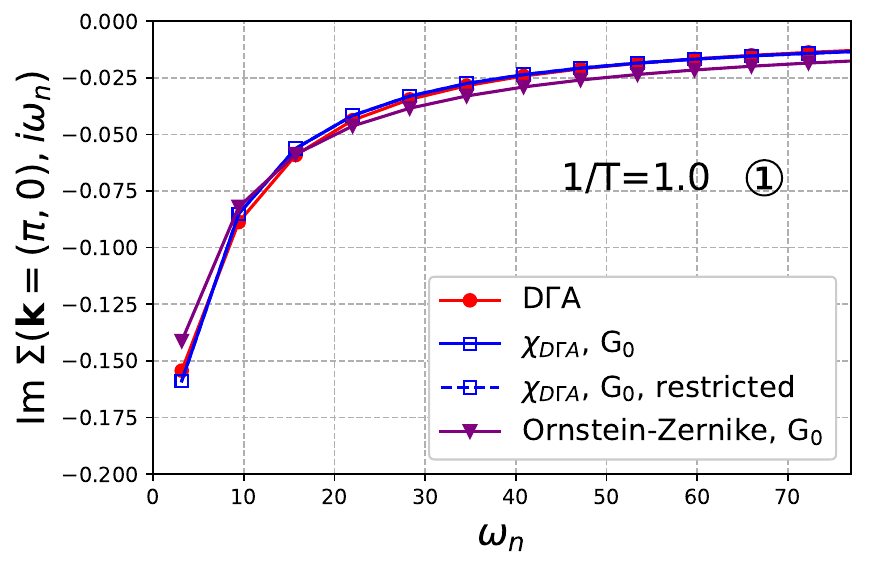} \includegraphics[width=0.32\textwidth,angle=0]{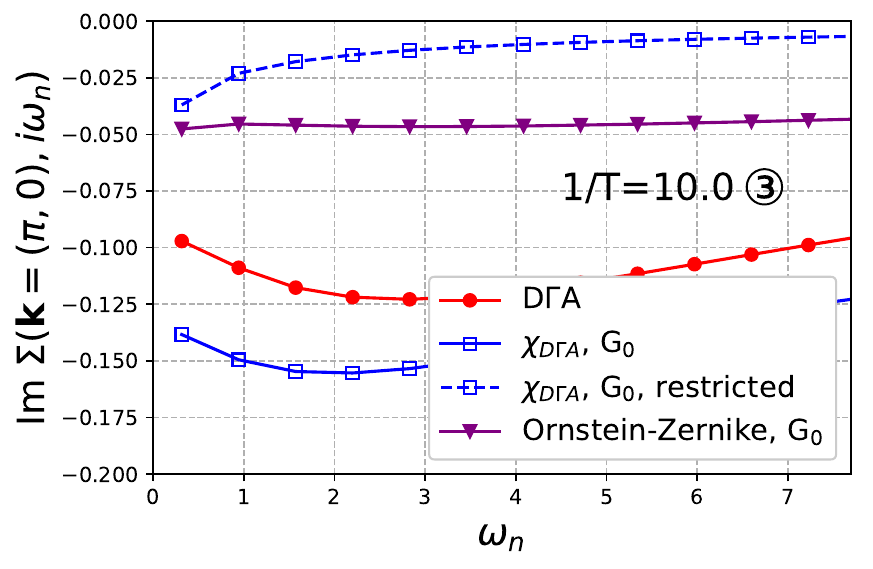} \includegraphics[width=0.305\textwidth,angle=0]{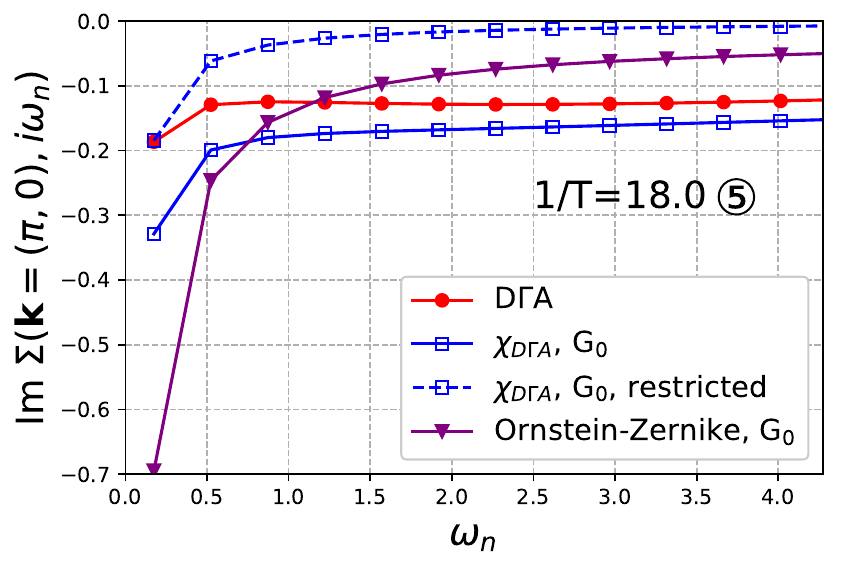}
      \caption{\label{fig:sf}(Color online.) 
Spin fluctuation theory put to the test:  comparison between the D$\Gamma$A self-energy and the result 
of weak-coupling spin fluctuation theory for three temperatures corresponding to 
regimes \textcircled{1} (incoherent), \textcircled{3} (metallic) and \textcircled{5} (pseudogap).
Both the full D$\Gamma$A susceptibility and its Ornstein-Zernike approximation are being considered, as 
well as the effect of restricting the fluctuations to the long-wavelength antiferromagnetic collective mode 
(see text). 
}
\end{figure*}

The low-temperature insulating regime $T\lesssim 1/12.5 \simeq T_{\text{N\'eel}}^{\text{DMFT}}$ can hardly be 
probed with current state of the art benchmark methods. The D$\Gamma$A results down to $T=1/20$ are consistent with the 
exponential growth of the correlation length \cite{Schaefer2015b, Rohringer2016}, expected from the low-energy description of the spin degrees of freedom by 
a non-linear sigma model once a charge (pseudo)gap opens up. 
We note significant discrepancies between the different approximate methods available in this regime however.
The growth rate of this exponential regime appears to 
be different (and faster) than the one in the intermediate-$T$ metallic regime, a finding which will have to be confirmed 
in future work by exact computational methods when they become capable of reaching lower temperatures.

\section{Insights into the nature and role of spin fluctuations}
\label{sec:spin_fluct}

In this section, we provide understanding into the physical nature of the different regimes 
highlighted above, especially the metallic regime and the (pseudo)gapped insulating regime. 
We focus in particular on the nature and role of spin fluctuations. 
We explore whether spin fluctuation theory of the weak-coupling type 
is able to describe qualitatively 
the various regimes: this analysis also provides analytical insights into the physics of these different regimes. 
We focus in particular on the following key questions:
\begin{itemize}
    \item[(i)] What is the physical mechanism for the opening of the pseudogap at low temperature? 
    \item[(ii)] What are the implications of the growing antiferromagnetic correlation length upon cooling for the coherence of quasiparticles in the metallic regime?
    \item[(iii)] To what extent can this metallic regime be characterized as a Fermi liquid?
\end{itemize}

\subsection{Weak-coupling spin fluctuation theory}
\label{subsec:sf_comparison}

In this subsection, we interrogate weak-coupling spin-fluctuation theory and ask whether it provides a satisfactory description of the behaviour of the self-energy, at least on 
a qualitative level, in the different regimes of temperature. 
This will also provide guidance throughout the rest of this section in identifying which fluctuations 
contribute most to the self-energy. 

In the simplest version of spin fluctuation theory the self-energy can be expressed 
as (omitting the Hartree term): 
\begin{eqnarray}
\label{eqn:spin_fluct}
&\Sigma&_{\mathrm{SF}}(\mathbf{k},i\omega_{n})\,=\,\\
&g^2T&\,\int \frac{d^2q}{(2\pi)^2}\sum\limits_{\Omega_{n}}G(\mathbf{k}+\mathbf{q},i\omega_{n}+i\Omega_{n})\chi(\mathbf{q},i\Omega_{n}).\nonumber
\end{eqnarray}
In this expression, $\chi(\mathbf{q},i\Omega_n)$ is the momentum- and frequency-dependent spin 
susceptibility and $G(\mathbf{k},i\omega_n)$ is the (lattice) Green function. 
In the following we always choose $G=G_0$ to be the non-interacting Green function, see the end of Sec.~\ref{subsec:insights_pseudogap} for comments about the drawbacks of self-consistent schemes. 

The coupling constant $g$ in Eq.~(\ref{eqn:spin_fluct}) characterizes the coupling between electrons and 
the spin collective modes. It is renormalized as compared to its bare value as $U$ increases.
In the following, we are more interested in asking whether weak coupling spin 
fluctuation theory {\it qualitatively} captures the different regimes than in quantitative statements. However, 
when a comparison is attempted, a value of $g$ has to be chosen. 
For example, in the version of TPSC considered in the present article, $g$ 
is given by (focusing on the spin channel only): 
$g^2_{\mathrm{TPSC}}\,=\,\frac{3}{8}\,U\,U_{\mathrm{sp}}$, with 
$U_{\mathrm{sp}}\,=\,U \langle n_\uparrow n_\downarrow \rangle/
\langle n_\uparrow \rangle \langle n_\downarrow\rangle$.
The factor $3/8$ ensures rotational invariance by properly accounting for the relative 
contributions of longitudinal and transverse spin fluctuations~\cite{Moukouri2000}, while the 
expression of $U_{\mathrm{sp}}$ is key to the TPSC scheme, see also App.~\ref{app:tpsc}. 
In the following we shall, for simplicity, use $g^2=3U^2/8$ when performing quantitative comparisons. 

Fig.~\ref{fig:sf} presents a comparison between the D$\Gamma$A results for the self-energy 
at the antinode $\mathbf{k}=(\pi,0)$ and the spin-fluctuation expression Eq.~(\ref{eqn:spin_fluct}) 
in which different choices are made for the spin susceptibility $\chi(\mathbf{q},i\Omega_n)$. We perform the comparison using the susceptibility data from D$\Gamma$A due to its good agreement with the benchmark, the availability of the $\mathbf{q}$-resolved susceptibility and its ability to enter the highly insulating regime \textcircled{5}.  
This comparison is performed for three different temperatures, corresponding to the incoherent regime 
\textcircled{1}, the metallic regime \textcircled{3} and the pseudogapped insulating regime \textcircled{5}.

We begin the discussion with the result obtained by using the full (solid blue curve, square markers) 
D$\Gamma$A susceptibility. 
We see that, at a qualitative level, the spin-fluctuation approximation succeeds in reproducing the 
characteristic low-frequency dependence of the self-energy associated with all three regimes. 
There is furthermore good quantitative agreement in the incoherent regime \textcircled{1}. 
With the chosen value of $g$, the self-energy is overestimated in both the metallic and 
pseudogap regimes. Using $g=g_{\mathrm{TPSC}}$ together with the value of 
the double occupancy calculated in Fig.~\ref{fig:docc} largely remedies this overestimation for most of the metallic regime.

In order to better understand which spin fluctuations dominate the different regimes, we 
have also applied Eq.~(\ref{eqn:spin_fluct}) by restricting the integration over the momentum transfer 
$\mathbf{q}$ to the vicinity of the antiferromagnetic wave vector $\mathbf{Q}=(\pi,\pi)$ according 
to $|q_i-Q_i|<2\xi^{-1}, i\in \left\{x,y\right\}$ (dashed blue curves in Fig.~\ref{fig:sf}). 
We see that in the pseudogap regime \textcircled{5} the qualitative frequency dependence of the antinodal 
self-energy is not modified by this restriction (although obviously, as a result of the restriction in the integration, 
the self-energy is underestimated by an amount which is approximately frequency independent). 
This indicates that, in this regime, the self-energy is indeed dominated by the antiferromagnetic collective 
modes associated with the vicinity of $\mathbf{Q}=(\pi,\pi)$. 
In contrast, a completely different situation is found in the metallic regime \textcircled{3}: restricting the 
momentum sum to the vicinity of $\mathbf{Q}$ yields a spin-fluctuation self-energy which has a qualitatively different 
frequency dependence than the actual one, inconsistent with the properties of a metal (and actually closer to 
the shape associated with a pseudogap). It also underestimates the overall order of magnitude of the 
self-energy by more than a factor of four. 
We thus conclude that the single-particle properties in the metallic regime are not 
controlled only by the spin fluctuations associated with the 
antiferromagnetic wave vector: taking into account spin fluctuations at other wave vectors is crucial. 
We will come back in detail to this observation in Sec.~\ref{subsec:insights_metal}. 
We finally note that the incoherent regime is not affected by the momentum restriction, simply because the 
correlation length is so small there that our criterion on the transfer momentum actually does not restrict 
the integration domain significantly.

Finally, we consider whether the spin susceptibility in Eq.~(\ref{eqn:spin_fluct}) can be approximated, 
as often done in spin fluctuation theories, by an Ornstein-Zernike form emphasizing long-wavelength 
antiferromagnetic fluctuations: 
\begin{equation}\label{eqn:chiOZ}
	\chi(\mathbf{q}, i\Omega_{n}) = \frac{A}{(\mathbf{q}-\mathbf{Q})^2 + \xi^{-2} + 
	\frac{\left|\Omega_n\right|}{\gamma}}.
\end{equation}
As detailed in App.~\ref{app:ornstein}, we use a lattice generalization of this expression 
and perform a fit of the momentum and frequency dependence of the D$\Gamma$A susceptibility 
in order to determine the amplitude $A$, correlation length $\xi$ (Sec.~\ref{sec:two_particle}) and 
Landau damping coefficient $\gamma$ as a function of temperature.
The antinodal self-energy obtained by inserting  expression (\ref{eqn:chiOZ}) into the spin-fluctuation 
expression (\ref{eqn:spin_fluct}) is displayed in Fig.~\ref{fig:sf} (purple data points). 
This approximation is quantitatively accurate in the high-temperature incoherent regime.  
It captures qualitatively the characteristic frequency dependence of a pseudogap in the 
low-temperature regime, as also detailed analytically in the following section, but considerably 
overestimates its magnitude. 
In contrast, in the metallic regime, it fails quite severely both qualitatively and quantitatively, 
indicating that the Ornstein-Zernike approximation fails to capture important spectral weight in 
$\chi(\mathbf{q},i\Omega_n)$ at $\mathbf{q}$ far from $(\pi,\pi)$ which is crucial in the metallic 
regime as emphasized above and further detailed in Sec.~\ref{subsec:insights_metal}. 

\subsection{Analytical insights into the pseudogap insulating regime} 
\label{subsec:insights_pseudogap}

We provide here some analytical insights into the low-$T$ pseudogap regime, based on the weak 
coupling spin fluctuation theory and the Ornstein-Zernike approximation for the spin susceptibility in Eq.~(\ref{eqn:chiOZ}).
As seen above, these approximations are qualitatively reasonable (although not quantitatively accurate) in this regime. 
For pioneering works on the formation of a pseudogap due to antiferromagnetic correlation, see Refs.~\cite{Vilk1997} 
and \cite{Borejsza2004}. 
In the latter, the authors focused on orientational fluctuations of the antiferromagnetic order described 
by a non-linear sigma model and, expressing the physical electron as a product of a Schwinger boson and an auxiliary fermion, were able to describe the crossover from a weak-coupling pseudogap in the Slater regime to a strong-coupling 
insulating gap in the Mott-Hubbard regime.

\begin{figure}[t!]
                \includegraphics[width=0.40\textwidth,angle=0]{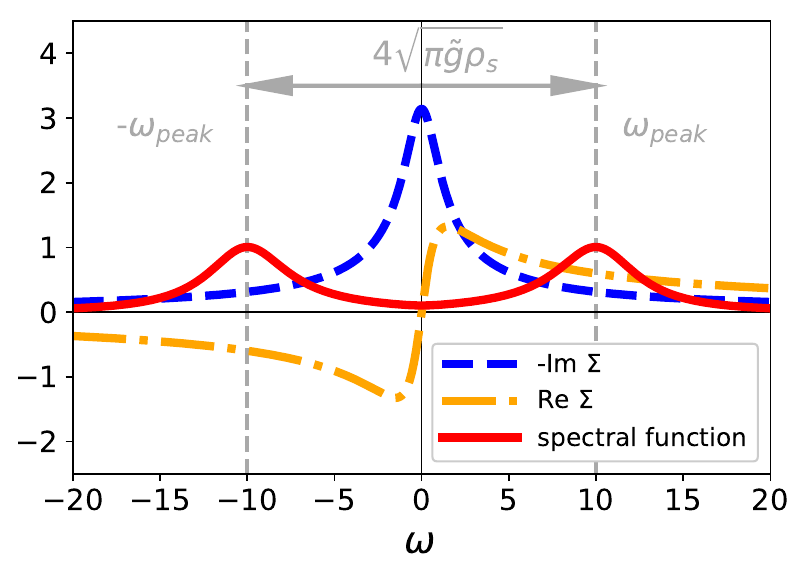}
      \caption{\label{fig:sf_spectrum}(Color online.) Pseudogap regime: Real (orange) and imaginary (blue) part of the self-energy on the real axis phenomenologically calculated in weak-coupling spin fluctuation theory 
      and its corresponding spectral function (red).
      }
\end{figure}
In this `renormalized classical' regime, only the lowest Matsubara frequency needs 
to be retained in Eq.~(\ref{eqn:spin_fluct}) leading to~\cite{Vilk1996, Vilk1997,Wu2018}:
\begin{equation}
\label{eqn:sf_matsubara}
\text{Im} \Sigma(\mathbf{k}_\text{F},i\omega_n)\,=\,
-\tilde{g}\frac{T}{\sqrt{\omega_n^2-\omega_c^2}}
\ln\,\frac{\omega_n+\sqrt{\omega_n^2-\omega_c^2}}
{\omega_n-\sqrt{\omega_n^2-\omega_c^2}}.
\end{equation}
In this expression, the prefactor $A$ of the spin susceptibility has been combined with $g^2$ and 
numerical prefactors to yield a coupling $\tilde{g}\propto Ag^2$ with the dimension of an energy.
In Eq.~(\ref{eqn:sf_matsubara}), 
only the dominant term has been retained, $\mathbf{k}_\text{F}$ is a point on the Fermi surface away from the van Hove singularity 
(i.e. from the antinode) so that the Fermi velocity $v_F$ is non-vanishing, and $\omega_c$ designates the important low-energy scale:
\begin{equation}
\omega_c\,\equiv\,\frac{v_F}{\xi}\,=\,\frac{v_F}{\xi_0}e^{-2\pi\rho_s/T},
\end{equation}
which explicitly depends on the Fermi velocity (i.e. $\mathbf{k}_\text{F}$). In the last expression, we have postulated an exponential growth of the correlation length, 
characteristic of the low-$T$ gapped regime, with $\rho_s$ as the spin stiffness. 
Expression (\ref{eqn:sf_matsubara}) has the following behaviour at low and high (imaginary) 
frequency:
\begin{eqnarray}
\mathrm{Im}\Sigma(\mathbf{k}_\text{F},i\omega_n)
&\sim& -2\tilde{g}T\frac{1}{\omega_n}\ln\frac{\omega_n}{\omega_c}\,\,\,\,\,\,(\omega_n\gg\omega_c)
\label{eqn:sf_high}
\\
&\sim& -\pi\tilde{g}\frac{T}{\omega_c}\,\,\,\,\,\, (\omega_n\ll\omega_c).
\label{eqn:sf_low}
\end{eqnarray}
These expressions shed light on the dependence of the self-energy on Matsubara frequency 
reported above in the low-$T$ regime. The lowest Matsubara frequency in this 
regime is always larger than the tiny low-energy scale $\omega_c$: in fact the condition 
$\pi T \gg \omega_c$ defines the range of low temperatures in which the insulating/pseudogap
behaviour is observed~\cite{Vilk1996, Vilk1997}. 
In this regime, the Matsubara frequency  
self-energy seemingly displays a downwards divergent behaviour as in Eq.~(\ref{eqn:sf_high}). 
However, if a calculation for low-$\omega \lesssim\omega_c$ 
was possible, it would actually display a saturation to a finite value as in Eq.~(\ref{eqn:sf_low}). 

For completeness, we discuss the implications of this expression for the low-frequency spectral 
function and self-energy on the real frequency axis. Analytically continuing the above expression 
yields (for $\mathbf{k}\neq\mathbf{k}_{\text{AN}}$): 
\begin{eqnarray}
\label{eqn:im_spin_fluct}
\text{Re }\Sigma_{\text{ret}}(\mathbf{k}_\text{F},\omega)&=&
\tilde{g}\frac{T}{\sqrt{\omega^2+\omega_c^2}}
\text{ln}\left|
\frac{\omega+\sqrt{\omega^2+\omega_c^2}}
{\omega-\sqrt{\omega^2+\omega_c^2}}\right| \\
\text{Im }\Sigma_{\text{ret}}(\mathbf{k}_\text{F},\omega)&=&
-\pi\tilde{g}\frac{T}{\sqrt{\omega^2+\omega_c^2}},
\label{eqn:im_spin_fluct_2}
\end{eqnarray}
with the following asymptotic forms:
\begin{eqnarray}
|\omega|&\gg&\omega_c\,:\,\,\, 
\text{Re}\Sigma \simeq 2\frac{\tilde{g}T}{\omega} \ln \frac{\omega}{\omega_c}\,\,\,,\,\,\,
-\text{Im}\Sigma \simeq \pi\frac{\tilde{g}T}{|\omega|},\label{eq:exp2}\\
|\omega|&\ll&\omega_c\,:\,\,\, 
\text{Re}\Sigma \simeq 2\frac{\tilde{g}T}{\omega_c^2}\, \omega\,\,\,,\,\,\,
-\text{Im}\Sigma \simeq \pi\frac{\tilde{g}T}{\omega_c}\label{eq:exp1}
\end{eqnarray}
with a slope obviously having a sign opposite to that of a Fermi liquid (`$Z>1$'). 
The corresponding self-energy and spectral function on the Fermi surface  
$A(\mathbf{k},\omega)=-\mathrm{Im}[\omega+\mu-\varepsilon_{\mathbf{k}}-\Sigma]^{-1}/\pi$ 
are plotted in Fig.~\ref{fig:sf_spectrum}. 
The two prominent peaks at the edge of the pseudogap in the spectral function can be understood by 
using the asymptotic form of $\mathrm{Re}\Sigma$ for 
$\omega>\omega_c$ in the quasiparticle equation $\omega-\text{Re}\Sigma(\omega)=0$.  
We see that the poles of the 
Green function are located at a frequency $\omega$ such that $\omega^2=2\tilde{g}T\,\ln \omega/\omega_c$. The dominant term on the r.h.s. of this equation is from  
$-2\tilde{g}T\ln\omega_c$, and we finally obtain the locations 
of the main peaks at:
\begin{equation}
\omega_{\text{peak}}\,=\,\pm 2\sqrt{\pi\tilde{g}\rho_s} + \cdots,
\label{eqn:peak}
\end{equation} 
where the corrections are linear in temperature. 
Hence, the location of these peaks become temperature-independent in the low-$T$ limit. 
As $T\rightarrow 0$ these two peaks evolve continuously into the gap-edge peaks 
defining the single-particle excitations across the insulating gap due 
to long-range antiferromagnetic order. 
Note that, within mean-field theory the $T=0$ gap is given by $Um_s$ with $m_s$ as the staggered 
magnetization, and that close to a quantum critical point the scaling 
$m_s\propto\sqrt{\rho_s}$ indeed holds, establishing consistency with Eq.~(\ref{eqn:peak}) 
(in the present case of $d=2$ additional logarithmic corrections intervene). 
Note that the above derivation of the pseudogap behaviour down to $T\!=\!0$ relies on the exponential growth of the correlation length 
and hence would not apply if the ground-state did not have long-range order. For an alternative approach based on thermal fluctuations starting from the $T\!=\!0$ ordered state, see Ref.~\cite{chubukov2010}. 

The width of the AF insulating peaks \cite{Slater1951, Borejsza2003, Borejsza2004} are of order $(\tilde{g}T^2/\rho_s)^{1/2}$ (for a DMFT analysis see also \cite{Sangiovanni2006}). It is furthermore interesting to understand the pseudogap regime in terms of lengths instead of the energy $\omega_c$ \cite{Vilk1997}. The one-particle thermal de Broglie wavelength $v_F/(\pi T)$ determines the maximum distance over which an electron wave packet remains coherent despite thermal agitation. Eq.~(\ref{eq:exp1}) shows that the imaginary part of the self-energy at zero frequency becomes very large when this length is smaller than the correlation length for spin fluctuations, leading to the decrease of spectral weight at zero frequency. For electrons in the pseudogap regime, the spins look ordered over the length where temperature does not destroy its coherence. This argument holds only in two dimensions. In higher dimensions, the phase space associated with the integration over wave vectors in Eq.~(\ref{eqn:spin_fluct}) diminishes the importance of long wavelength fluctuations \cite{Vilk1997} for the self-energy. As is well known from DMFT, in infinite dimensions, the self-energy is not influenced by long-wavelength spin fluctuations \cite{Metzner1989, Georges1992a}.  

A similar analysis can be performed for Fermi surface points in the vicinity of a van Hove singularity, i.e. when $v_{F}\!=\!0$. The expansion of the dispersion relation then is stopped only after the second order term, giving for $\mathbf{k}_\text{AN}=(\pi,0)$ after integration over $q_x$
\begin{eqnarray}
    -\frac{1}{\pi}&\text{ Im }& \Sigma_{\text{ret}}(\mathbf{k}_\text{AN}, \omega)=\\
    &=&\frac{\tilde{g}^2}{(2\pi)^2} T \int dq_y \frac{1}{\sqrt{\omega + q_y^2}}\frac{1}{\omega + \xi^{-2} + 2q_y^2},
\end{eqnarray} which yields logarithmic divergences. Even if these divergences could be cut off, for $\omega \rightarrow 0$
\begin{equation}
    -\frac{1}{\pi}\text{ Im } \Sigma_{\text{ret}}(\mathbf{k}_\text{AN}, \omega \rightarrow 0) \!\propto\! 
    \frac{1}{2}\tilde{g}^2 T \xi^2,
\label{eqn:sf_nested}
\end{equation}
so that the imaginary part of $\Sigma$ is now proportional to $\xi^2$ instead of $\xi$ (see also \cite{Vilk1997, Vilk1997b, Rohringer2016}, \cite{Lemay2000, Zlatic1997} for second order perturbation theory and \cite{Katanin2005b} in the context of ferromagnetic fluctuations), i.e. way larger than for momenta away from van Hove points.

To conclude this subsection on the pseudogap insulating regime, we recall some findings and comment on the ability of the description of this regime in self-consistent theories. First we note, that  if one uses $G$ (with a finite self-energy) instead of $G_0$ as fermionic propagator in Eq.~(\ref{eqn:spin_fluct}), one obtains Im $\Sigma(\mathbf{k}, 0) \propto T^{-2}$, hence missing the exponential growth of Eqs.~(\ref{eqn:sf_low},\ref{eqn:sf_nested}) \cite{Rohringer2016}, see also \cite{Borejsza2003}.

On more general grounds, we note that the self-consistency at the level of vertex corrections should be consistent with the one of the Green function. This remark serves as a warning about imposing self-consistency solely over the Green function. When using approximate schemes which truncate perturbation theory or select a specific class of diagrams without taking vertex corrections into account in a consistent manner, self-consistency may lead to an incorrect physical description - especially in relation to the opening of insulating gaps. For an insightful discussion of these limitations of self-consistent perturbation theory, 
see Sec.~6.1 of Ref.~\cite{Vilk1997}. Observations about the inadequacy of self-consistency in truncated perturbation theory approximations have previously been made also in other contexts, such as the GW approximation and DMFT where it is well-known that self-consistent perturbation theory truncated to second-order fails in yielding Hubbard bands and the opening of a Mott gap~\cite{Georges1992a, MuellerHartmann1989, Ku2002, Delaney2004, Ku2004}. An additional, however, completely different problem can arise when self-consistent theories (e.g. bold diagrammatic Monte Carlo \cite{Prokofev2007, Deng2015}) are applied: due to the multivaluedness of the Luttinger-Ward functional \cite{Kozik2015, Stan2015, Vucicevic2018, Rossi2015, Kim2020b}, intimately related \cite{Gunnarsson2017} to divergences in vertex functions \cite{Schaefer2013, Janis2014, Schaefer2016c, Chalupa2017, Thunstrom2018, Springer2019, Reitner2020, Chalupa2020}, the bold approaches may converge to an unphysical branch.

\subsection{Insights into the metallic regime}
\label{subsec:insights_metal}

\begin{figure}[t!]
\centering
                \includegraphics[width=0.47\textwidth,angle=0]{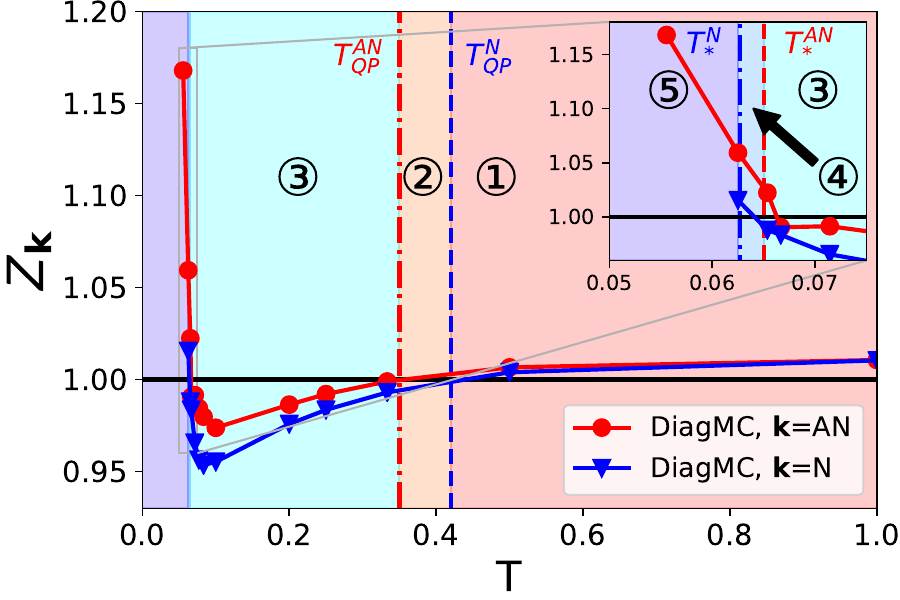}
                \includegraphics[width=0.45\textwidth,angle=0]{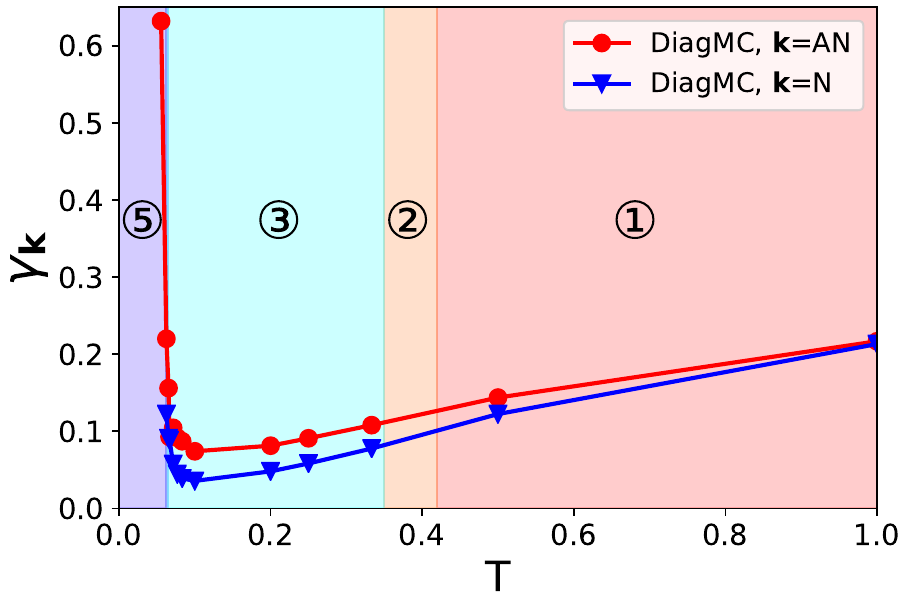}
      \caption{\label{fig:fl}(Color online.) Quasiparticle parameters $Z_{\mathbf{k}}$ (upper panel, the gray box denotes the boundaries of the inset) and $\gamma_{\mathbf{k}}$ (lower panel) at the antinode (red circles) and node (blue triangles) for the DiagMC benchmark as a function of temperature.}
\end{figure}

\subsubsection{Temperature dependence and characteristic scales}

Finally, in this section, we provide insights into the physics of the metallic regime. 
We display in Fig.~\ref{fig:fl} the quantities
\begin{eqnarray}
 Z_{\mathbf{k}}&=&\left[1-\frac{\partial\text{Im }\Sigma(\mathbf{k},i\omega)}{\partial\omega}
 \bigg\vert_{\omega\rightarrow{0}}\right]^{-1} \label{eqn:z_main} \\
 \gamma_{\mathbf{k}}&=&\tau_{\mathbf{k}}^{-1}=-Z_{\mathbf{k}} \cdot \text{Im }\Sigma(\mathbf{k}, i\omega)\bigg\vert_{\omega\rightarrow{0}} \label{eqn:gamma_main}
\end{eqnarray}
as a function of temperature, at the nodal and antinodal points on the Fermi surface, 
obtained by a fit to the DiagMC self-energy on Matsubara frequencies. 
For the details about how these parameters are extracted, see App.~\ref{app:fermi_liquid}. 
For a Fermi liquid, $Z_{\mathbf{k}}$ would correspond to the quasiparticle spectral weight and 
$\gamma_{\mathbf{k}}$ to the inverse of the quasiparticle lifetime. We are dealing however with a 
perfectly nested system in which Fermi liquid behaviour does not strictly apply, and this interpretation 
has to be taken with care, as further discussed in Sec.~\ref{sec:nesting}. 
Indeed, we observe that $\gamma_{\mathbf{k}}$ at the nodal point follows 
an approximately $T$-linear temperature dependence for $T\gtrsim 0.1$. 
The antinodal value of $\gamma_{\mathbf{k}}$ is significantly larger, highlighting momentum 
differentiation. 

The metallic regime \textcircled{3} was conventionally defined above as the regime where the self-energy 
has a negative slope at low frequency, thus allowing to define a $Z_{\mathbf{k}}$ smaller than unity, as indicated 
on the upper panel of Fig.~\ref{fig:fl}. 
From the $T$-dependence of $\gamma_{\mathbf{k}}$, we see that region \textcircled{3} defined in this manner
actually consists of two sub-regimes (see also {\cite{Rohe2020, Afchain2005}}): 
For $0.1 \lesssim T \lesssim T_{\text{QP}}^{\text{AN}}$ (regime \textcircled{3}a covering most of region 
\textcircled{3}), the inverse quasiparticle lifetime $\gamma_{\mathbf{k}}$ decreases upon reducing $T$, 
indicating an increase of quasiparticle coherence upon cooling (as characteristic of a metal). 
When cooling below $T\simeq 1/10$ a minimum of $\gamma_{\mathbf{k}}$ (and $Z_{\mathbf{k}}$) is found,
below which the lifetime decreases upon cooling as a precursor of the pseudogap regime (regime \textcircled{3}b). 

\begin{figure}[t!]
                \includegraphics[width=0.50\textwidth]{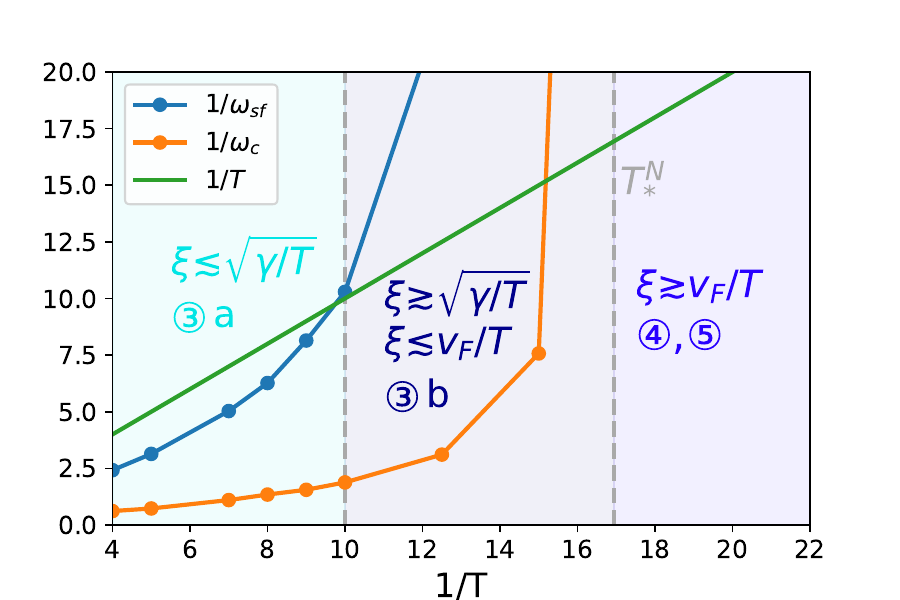}
      \caption{\label{fig:sf_scales}(Color online.) 
      Comparison of the two energy scales $\omega_c=v_F/\xi$ and $\omsf=\gamma\xi^{-2}$ to inverse 
      temperature (or equivalently of the correlation length to the two length scales $\sqrt{\gamma/T}$ and 
      $v_F/T$). This delimitates three regimes: the metallic regime \textcircled{3}a, a narrow 
      precursor regime of the pseudogap \textcircled{3}b
      and a non-metallic pseudogap regime \textcircled{4},\textcircled{5}. 
      The scales are determined by a fit to the frequency and momentum dependence of the D$\Gamma$A 
      susceptibility (see text).
      }
\end{figure}
\begin{figure*}[t!] \includegraphics[width=0.32\textwidth,angle=0]{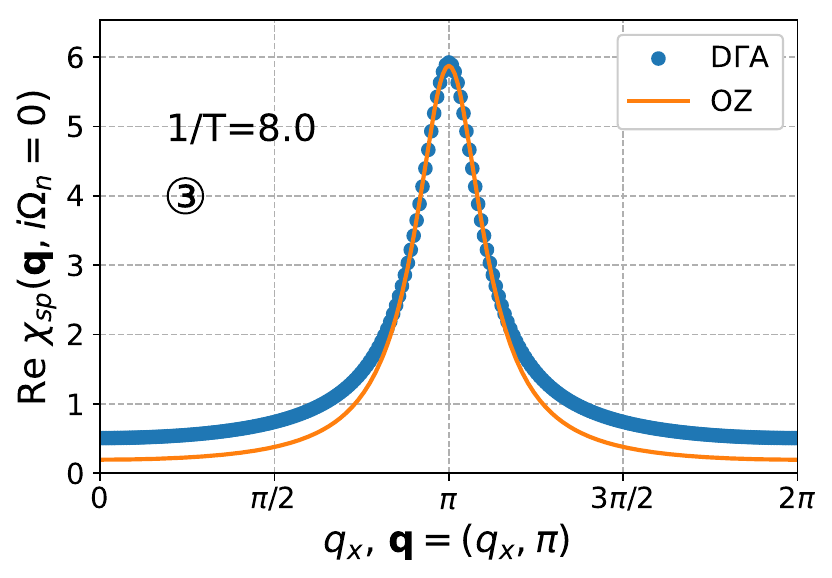}
             \includegraphics[width=0.32\textwidth,angle=0]{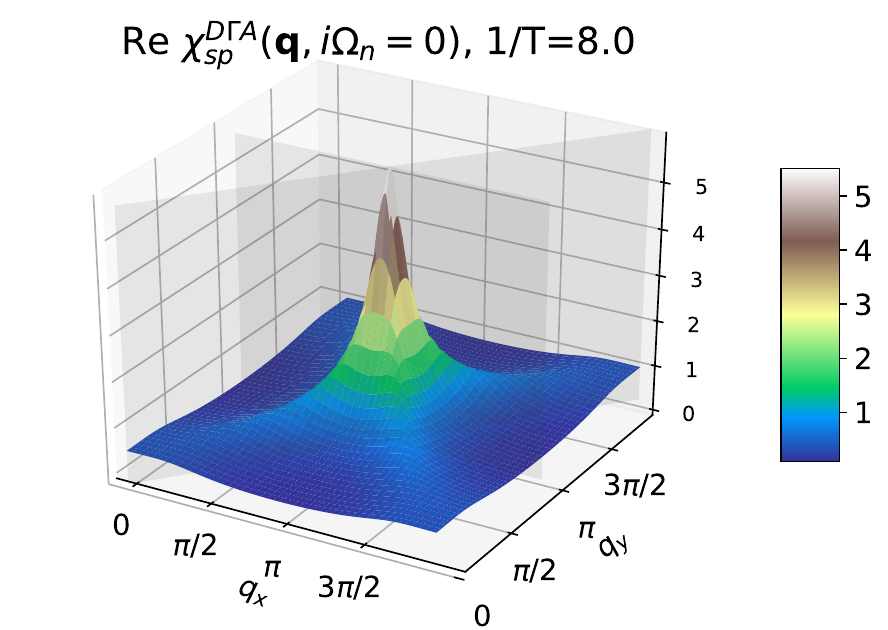}
                \includegraphics[width=0.34\textwidth,angle=0]{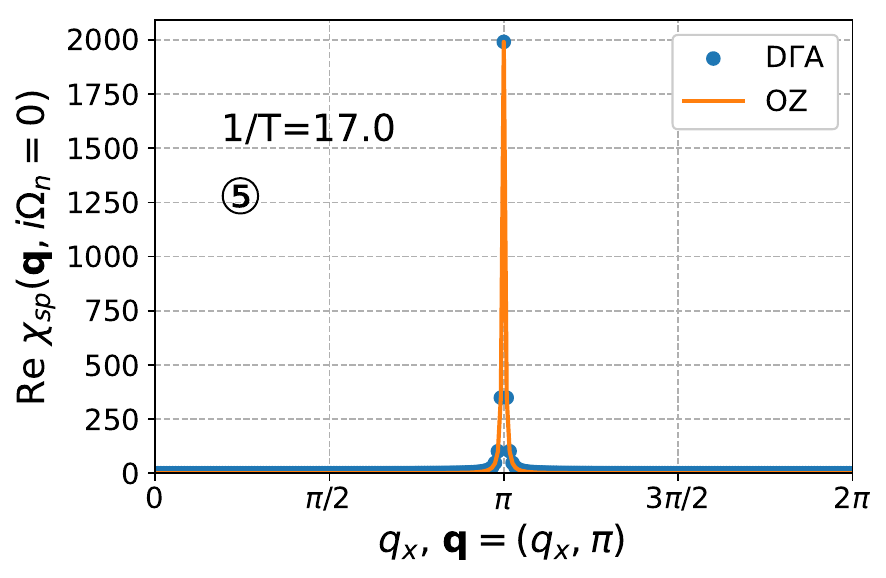}
             \includegraphics[width=0.32\textwidth,angle=0]{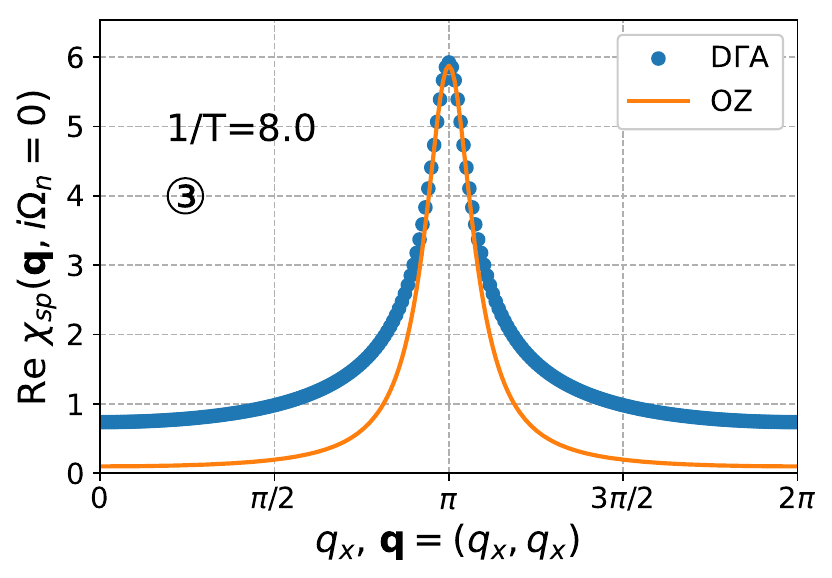}
             \includegraphics[width=0.32\textwidth,angle=0]{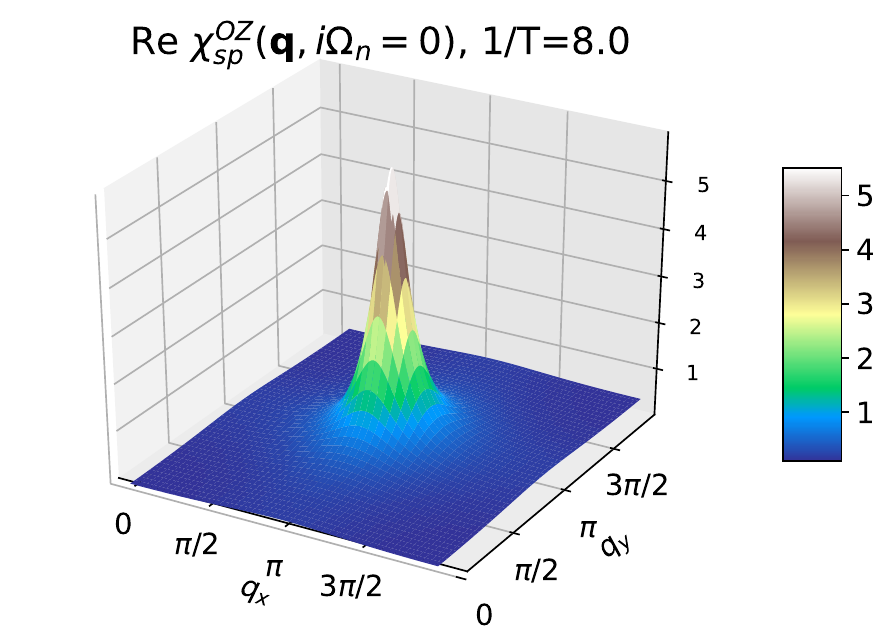}
                \includegraphics[width=0.34\textwidth,angle=0]{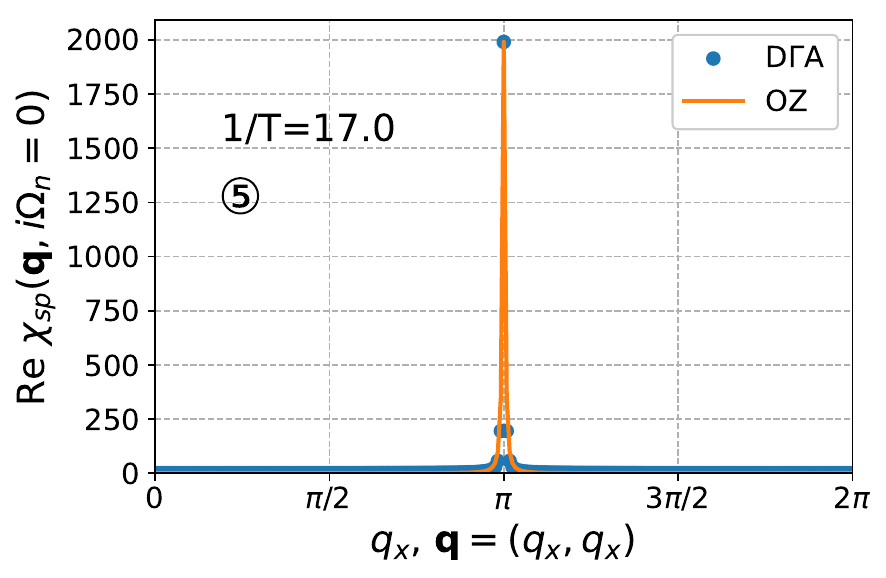}
      \caption{\label{fig:chi_compOZ}(Color online.)
Comparison between the static antiferromagnetic susceptibilities obtained from D$\Gamma$A and fits to an Ornstein-Zernike form (see Appendix~\ref{app:ornstein} and Figs.~\ref{fig:ornstein}, \ref{fig:ornstein_dynamical} 
for the corresponding fit parameters). 
In the temperature regime around $1/T=8$, the D$\Gamma$A susceptibility has, in addition to the 
antiferromagnetic peak at $\mathbf{Q}=(\pi,\pi)$, important contributions associated with the nesting 
vectors of the form $(q_x,\pm q_x)$ (`ridges' in the upper middle panel). 
Although describing the vicinity of $\mathbf{Q}=(\pi,\pi)$ accurately,
the Ornstein-Zernike form does not account for these contributions (lower middle panel). At lower temperatures, these contributions become negligible. The gray planes in the middle upper panel indicate the cuts shown, i.e. $\mathbf{q}=(q_x,\pi)$ and $\mathbf{q}=(q_x,q_x)$.}
\end{figure*}

\begin{figure}[b!]
 \includegraphics[width=0.458\textwidth,angle=0]{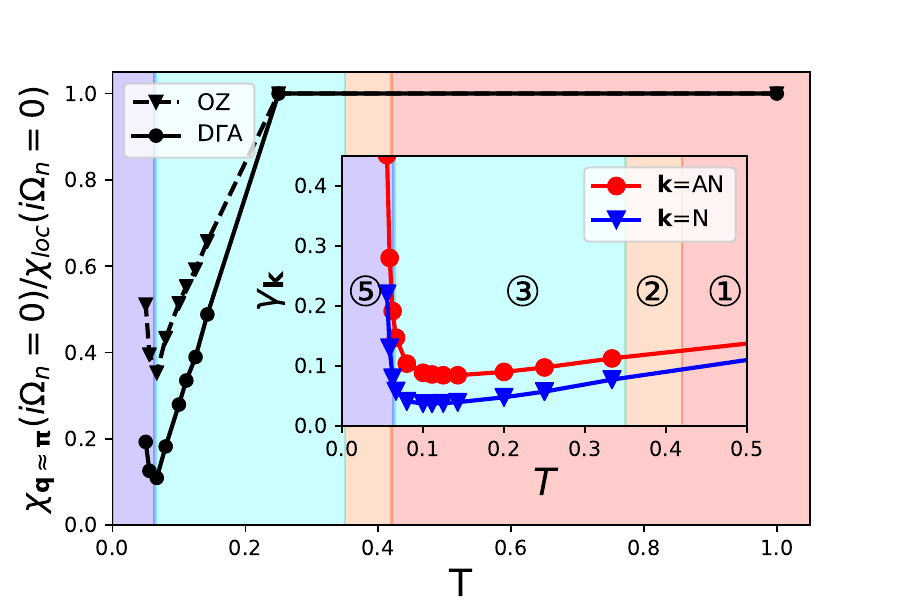}               
      \caption{\label{fig:ratio_susceptibilities}(Color online.) 
Main panel: Relative contributions to the AF susceptibility from wave vectors in the vicinity of $\mathbf{Q}=(\pi,\pi)$ for D$\Gamma$A (solid line, black circles) and the Ornstein-Zernike fit (dashed line, black triangles) as a function of $T$. Inset: For completeness $\gamma_{\mathbf{k}}$ from D$\Gamma$A is shown for the different regimes as a function of temperature for the antinode (red circles) and node (blue triangles).}
\end{figure}

In order to rationalize these findings, we turn again to spin fluctuation theory, Eq.~(\ref{eqn:spin_fluct}).
In contrast to the pseudogap regime, it is crucial here to perform the full convolution over all Matsubara frequencies. 
Using the spectral representation of $\chi(\mathbf{q},i\Omega_n)$, the single-particle scattering rate 
at zero frequency is obtained from Eq.~(\ref{eqn:spin_fluct}) as: 
\begin{eqnarray}\label{eq:rate_FL}
&-&\mathrm{Im}\Sigma(\mathbf{k},i0^+)\,\sim\,
\\
&g^2& \int d^2q\,\int d\omega\, \frac{1}{\sinh (\beta\omega)}\,\mathrm{Im}\chi(\mathbf{q},\omega)\, 
\delta\left(\omega-\varepsilon_{\mathbf{k+q}}\right). \nonumber
\end{eqnarray}
The key point here is to compare the width in frequency of the two factors entering this expression: 
$1/\sinh\beta\omega$ and $\mathrm{Im}\chi(\mathbf{q},\omega)$. 
The width of the former is of order temperature $T$, while the width of the latter is set by Landau 
damping, as clear from the Ornstein-Zernike form Eq.~(\ref{eqn:chiOZ}) applicable close to the antiferromagnetic 
wavevector which reads for real frequencies: 
\begin{equation}\label{eq:OZ_real_freq}
	\mathrm{Im}\chi(\mathbf{q}, \omega) = A\,\frac{\omega/\gamma}
	{\left[(\mathbf{q}-\mathbf{Q})^2 + \xi^{-2}\right]^2 + (\omega/\gamma)^2}.
\end{equation}
This expression peaks at the characteristic spin fluctuation frequency:
\begin{equation}
    \omsf\,=\,\gamma\xi^{-2}.
\end{equation}
Hence, for $T\lesssim\omsf$, only the low-frequency behaviour of $\mathrm{Im}\chi(\mathbf{q},\omega)$ 
matters in Eq.~(\ref{eq:rate_FL}). 
This analysis shows that, besides the scale $\omega_c=v_F/\xi$ which controls the pseudogap regime, 
the spin-fluctuation scale $\omsf=\gamma \xi^{-2}$ is important for the physics of the metallic regime (see \cite{Abanov2003} and references therein). 
In Fig.~\ref{fig:sf_scales}, we display these two scales as a function 
of temperature, with $\xi(T)$ and $\gamma(T)$ determined from a fit to the D$\Gamma$A susceptibility as 
described above and in App.~\ref{app:ornstein}. 
We compare these two scales to temperature itself, and observe that this indeed defines three distinct regimes, which correspond to 
a very good approximation to the regimes observed in our numerical results, namely 
(note that $\omsf<\omega_c$):
\begin{itemize}
    \item $T\lesssim \omsf$ ($\xi\lesssim \sqrt{\gamma/T}$): metallic regime  \textcircled{3}a.
    \item $T\gtrsim \omega_c$ ($\xi \gtrsim v_F/T$, Vilk criterion): insulating pseudogap regime \textcircled{4}, \textcircled{5}.
    \item $\omsf\lesssim T\lesssim \omega_c$ ($v_F/T\gtrsim\xi\gtrsim \sqrt{\gamma/T}$). 
    In this narrow intermediate regime ($0.06 \lesssim T \lesssim 0.1$, \textcircled{3}b) the pseudogap 
    is not yet fully  opened but the 
    scattering rate no longer behaves as in a metal: it displays a minimum and increases at lower $T$ as the pseudogap regime is entered, as displayed in Fig.~\ref{fig:fl}. 
    We note in passing that the upper boundary of this regime coincides with the temperature at which the double occupancy displays a local maximum (see Fig.~\ref{fig:docc}), corresponding to the onset temperature above which local magnetic moments are formed (see also \cite{Kim2020, Chalupa2020}). It also has corresponding signatures 
in other thermodynamic observables \cite{Kim2020}.
\end{itemize}

\subsubsection{Which fluctuations dominate the metallic regime ?}
\label{sec:metallic}

Expression (\ref{eq:rate_FL}) also allows to understand an important observation made in 
Sec.~\ref{subsec:sf_comparison}, namely that in the metallic regime the spin fluctuations  contributing to the self-energy 
are not dominated only by the vicinity of the antiferromagnetic wave vector. 

In order to further document and validate this point, we display in Fig.~\ref{fig:chi_compOZ} a comparison 
between the momentum dependence of the spin susceptibility as obtained in D$\Gamma$A and that of 
its Ornstein-Zernike fit (the latter privileges the vicinity of the antiferromagnetic wave vector). 
We see (left panel) that at $T=1/8$ which lies in the metallic regime, the OZ fit captures very poorly the 
momentum dependence of the susceptibility far from $\mathbf{Q}$. In particular, the OZ fit misses the 
sizeable weight that is visible in the D$\Gamma$A susceptibility 
along the diagonals $(q_x,\pm q_x)$. 
That weight comes from scattering associated with nesting vectors parallel to the antiferromagnetic zone boundaries, as is apparent already in second-order perturbation theory \cite{Lemay2000}. In contrast, the OZ form is much more accurate at lower temperatures when $\xi$ has grown to larger values 
and the system is in the pseudogap regime (right panel): indeed, the physics of the pseudogap 
is dominated by antiferromagnetic fluctuations (Sec.~\ref{subsec:insights_pseudogap}).

This is further documented in Fig.~\ref{fig:ratio_susceptibilities} which displays the ratio between the 
susceptibility integrated over a region of the order of $2\xi^{-1}$ around $\mathbf{Q}$ to its integral 
over the whole zone. It is seen that this ratio decreases as $T$ is lowered throughout the metallic regime, while 
it increases again in the pseudogap regime: hence a large fraction of the  spectral weight is missed by 
focusing on the vicinity of $\mathbf{Q}$ in the metallic regime. Fig.~\ref{fig:ratio_susceptibilities} also 
shows that the OZ fit overestimates the relative spectral weight associated with the vicinity of the antiferromagnetic wave vector.

Using the expressions  (\ref{eq:rate_FL}) and (\ref{eq:OZ_real_freq}), it can be shown that the contribution of the 
$(\pi,\pi)$ spin-fluctuations to 
the zero-frequency scattering rate on the FS (away from the van Hove point) behaves in the metallic regime 
as $\sim\,g^2\,(A/\gamma)\,T^2\xi^3/v_F$. As shown in Appendix~\ref{app:ornstein}, the ratio $A/\gamma$ decreases 
quickly upon cooling in this regime, which insures that this contribution to the scattering rate also decreases 
despite the fast increase of the correlation length. 
A similar remark applies to the local (momentum-integrated) scattering rate, for which this contribution can be estimated as $\sim\,g^2\rho_0\,(A/\gamma)\,T^2\xi^2$ (up to logarithmic terms). 
However, importantly, the local self-energy involves contributions from fluctuations 
at all wave-vectors, as already clear at the level of the spin-fluctuation approximation by 
integrating Eq.~(\ref{eq:rate_FL}) over $\mathbf{k}$. This explains why DMFT provides, in the metallic regime, an excellent approximation to the local component of the self-energy. In contrast, the DMFT approximation 
does not properly capture the contributions to the self-energy of fluctuations which are strongly peaked at a specific wave vector. For a plot of the (negative) imaginary part of the local  self-energy  extrapolated  to  zero  Matsubara  frequency  a  function  of  temperature we refer to Fig.~\ref{fig:im_sigma_zero} in App.~\ref{app:ornstein}.

Summarizing, we have shown that a comparison of the two key energy scales 
$\omega_c=v_F/\xi$ and $\omsf=\gamma\xi^{-2}$ to temperature allows for a determination of the 
different physical regimes as a function of temperature, in excellent quantitative agreement with our 
numerical results. 
The momentum-differentiated metallic regime evolves at low-$T$ 
into a precursor regime of the pseudogap where the quasiparticle lifetime reaches a maximum. 
We have also shown that the metallic regime is not dominated by the spin fluctuations associated with only the 
antiferromagnetic wave vector, and explained why quasiparticles with a lifetime increasing upon cooling can 
coexist with antiferromagnetic fluctuations characterized by a correlation length which strongly increases  
(approximately exponentially). 

\subsubsection{Perfect nesting and non-Fermi liquid behaviour}
\label{sec:nesting}

\begin{figure}[t!]
\centering
               \includegraphics[width=0.459\textwidth,angle=0]{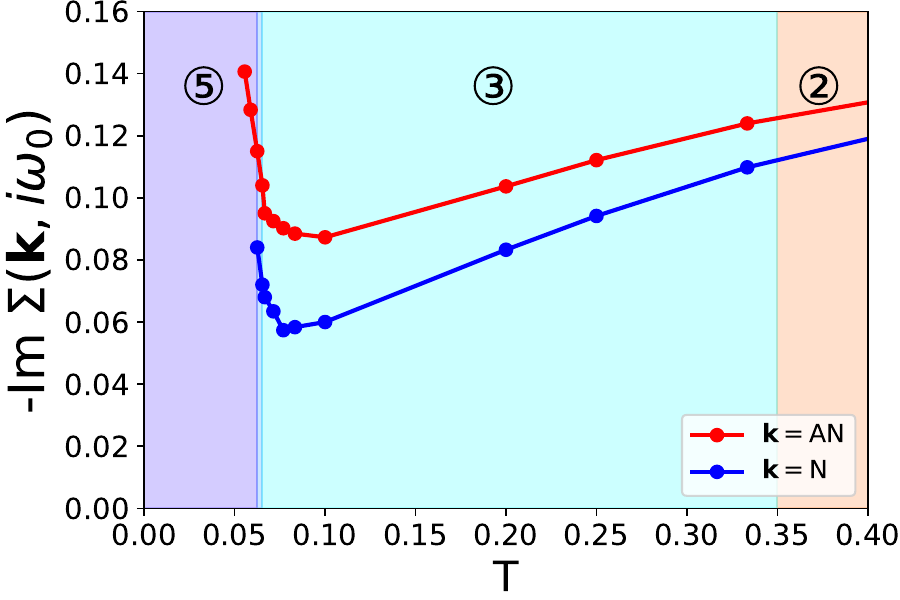}
      \caption{\label{fig:im_sigma_n0}(Color online.) 
      Diagnosing non-Fermi liquid behavior: the imaginary part of the self-energy at the first Matsubara frequency 
      calculated by DiagMC for the nodal and antinodal point displays deviation from linearity as a function of $T$ (see text).}
\end{figure}

In the half-filled model with $t'=0$ considered in this paper, we do not expect Fermi liquid behaviour to hold in 
the metallic phase because of perfect nesting. Indeed, the whole one-dimensional set of wave vectors 
with $q_x=\pm q_y$ maps a point on the diamond-shaped Fermi surface onto another one.  In this section, 
we briefly discuss whether our computations can detect hints of non-Fermi liquid behaviour, despite 
providing data restricted to imaginary (Matsubara) frequencies.

In a Fermi liquid, the self-energy takes the following form, when considered on the Fermi surface and at low frequency and temperature:
\begin{equation}
\label{eq:FLsigma}
  -\mathrm{Im }\Sigma(\mathbf{k}_{\text{F}},\omega)\,\propto\,\omega^2+(\pi T)^2 + \cdots
\end{equation}
On the Matsubara axis, the r.h.s. of this expression reads $-\omega_n^2+(\pi T)^2$. This contribution vanishes when considering the 
first Matsubara frequency $\omega_0=\pi T$. Hence, the temperature dependence of $\mathrm{Im}\Sigma(\mathbf{k}_{\text{F}},i\omega_0)$ at the 
first Matsubara frequency  is controlled by the dominant term $(1-1/Z)\omega_0=(1-1/Z)\pi T+\cdots$ with no quadratic 
correction of order $T^2$: linear dependence of this quantity on temperature is a hallmark of Fermi liquid behaviour 
(`first Matsubara frequency rule', see Ref.~\cite{Chubukov_2012}). 
Fig.~\ref{fig:im_sigma_n0}, displays the $T$-dependence of this quantity as obtained from the benchmark DiagMC at both the nodal 
and antinodal points on the Fermi surface. We see visible deviation from linearity throughout the metallic regime, indeed hinting at 
a departure from Fermi liquid behavior.  
Further evidence for non-Fermi liquid behaviour is provided by  the dependence on temperature of the 
parameters $Z_{\mathbf{k}}$ and $\gamma_{\mathbf{k}}$ obtained by a fit of the self-energy on Matsubara frequencies 
and displayed above on Fig.~\ref{fig:fl}. Indeed, 
(i) $Z_{\mathbf{k}}$  does not stabilize to a T-independent value before the pseudogap kicks in, but 
rather slowly decreases throughout the metallic regime and 
(ii) the temperature dependence of $\gamma_{\mathbf{k}}$ is better described as $T$-linear rather that 
quadratic like in a Fermi liquid. 
Further details about the proper interpretation of the Matsubara 
fitted parameters $Z_{\mathbf{k}}$ and $\gamma_{\mathbf{k}}$ when Fermi liquid behaviour does not apply 
are given in Appendices~\ref{app:fermi_liquid} and \ref{app:2pt}.  

The breakdown of Fermi liquid theory in a perfectly nested system has been explored early on in 
Ref.~\cite{Virosztek1990} and subsequent works. To understand its origin, it is useful to consider 
the spin-fluctuation expression (\ref{eq:rate_FL}) and note that, 
when $\mathrm{Im}\chi$ is free from singularities and behaves linearly as a function of $\omega$ at low frequency, 
this equation  yields the characteristic Fermi liquid $T^2$ behavior for $T\lesssim \omsf$ and $\omega=0$:
\begin{equation}
\label{eq:rate_FLT2}
  -\mathrm{Im }\Sigma(\mathbf{k},i0^+)\,\sim\,  g^2\,T^2\, \int\,d^2q\, 
  \frac{1}{\Gamma_\mathbf{q}}\,  \delta\left(\varepsilon_{\mathbf{k+q}}\right)\,,
\end{equation}
in which $1/\Gamma_\mathbf{q}\equiv\mathrm{Im}\chi(\mathbf{q}, \omega)/\omega|_{\omega=0}$. 
In turn, non-Fermi liquid behaviour in a perfectly nested system can be traced to singularities in $\mathrm{Im}\chi(\mathbf{q},\omega)$. 
These singularities are already manifest at the level of second order perturbation theory, as studied in details 
in the PhD thesis of Lemay~\cite{Lemay2000} and also recently in Ref.~\cite{Rohe2020}. 
In Appendix~\ref{app:2pt}, we discuss second order perturbation theory in more details, together 
with a comparison to our computational results. 
Theoretical approaches of the non-Fermi liquid singularities for nested systems beyond lowest order perturbation theory 
have been attempted using parquet~\cite{Zheleznyak1997} and fRG~\cite{Vistulo1997,Zanchi2000} methods, but we are not 
aware of a full solution of the case considered in the present paper. 

From a computational standpoint, a precise characterisation of non-Fermi liquid behaviour in the metallic phase 
would require algorithms able to determine the behaviour of the self-energy and 
response functions directly on the real frequency axis, in both the $\omega >T$ and $\omega <T$ regimes 
(the latter being inaccessible to methods which provide data at Matsubara frequencies only). 
This goes beyond the scope of the present work and is a major challenge motivating the development of 
new computational methods (see, e.g., \cite{Bertrand2019, Vucicevic2020, Taheridehkordi2020, Taheridehkordi2020b}).

\subsection{A simple approximation}
\label{subsec:simple}

\begin{figure*}[t!]     
\includegraphics[width=0.32\textwidth,angle=0]{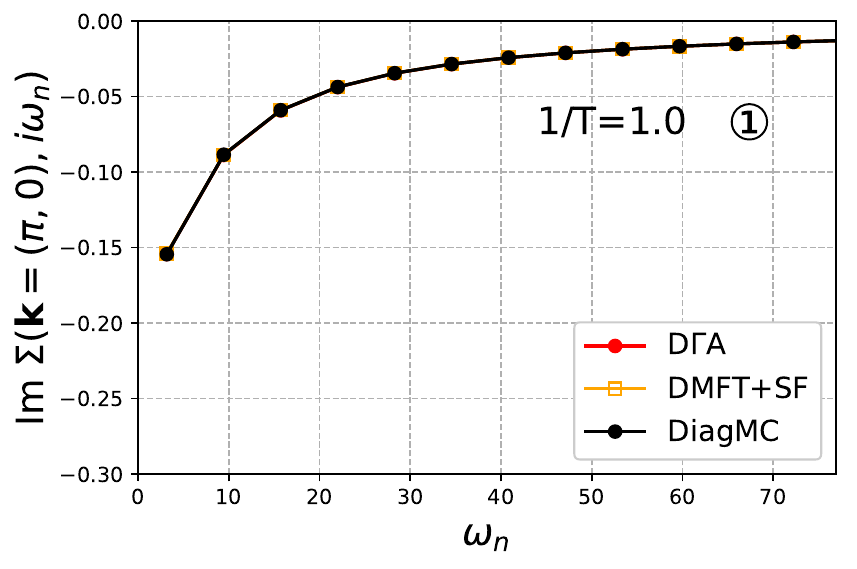}
\includegraphics[width=0.32\textwidth,angle=0]{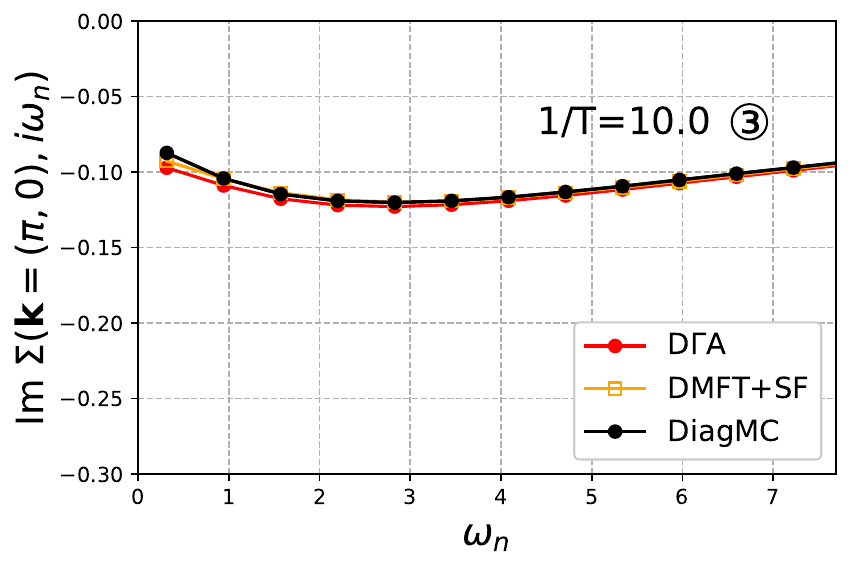}
\includegraphics[width=0.32\textwidth,angle=0]{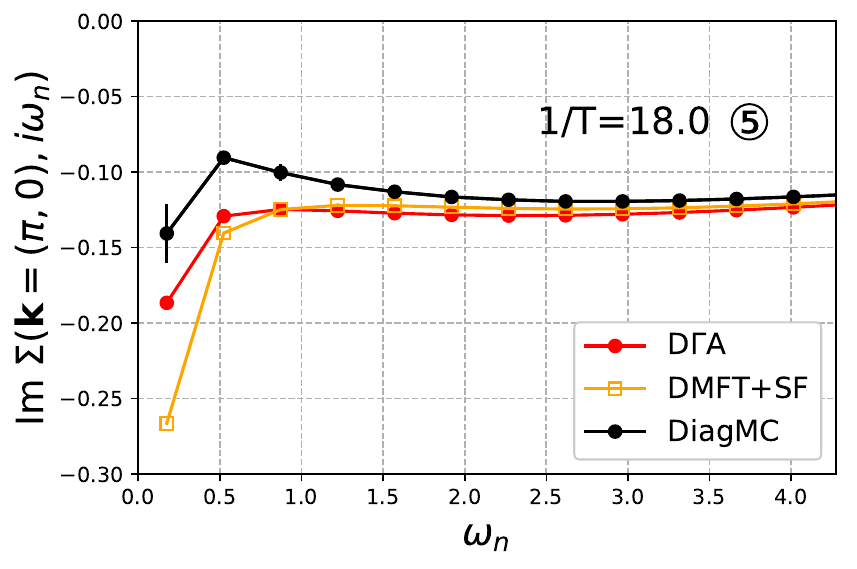}
\includegraphics[width=0.32\textwidth,angle=0]{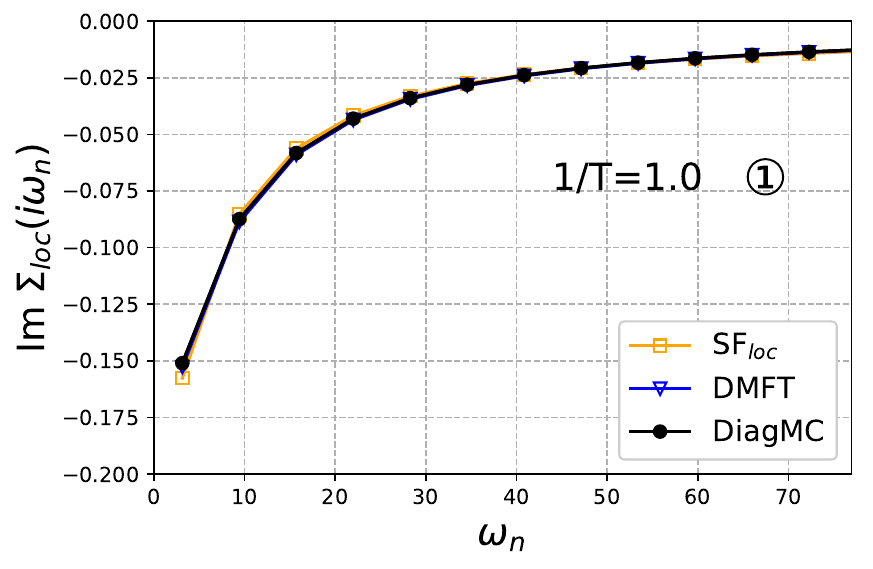}
\includegraphics[width=0.32\textwidth,angle=0]{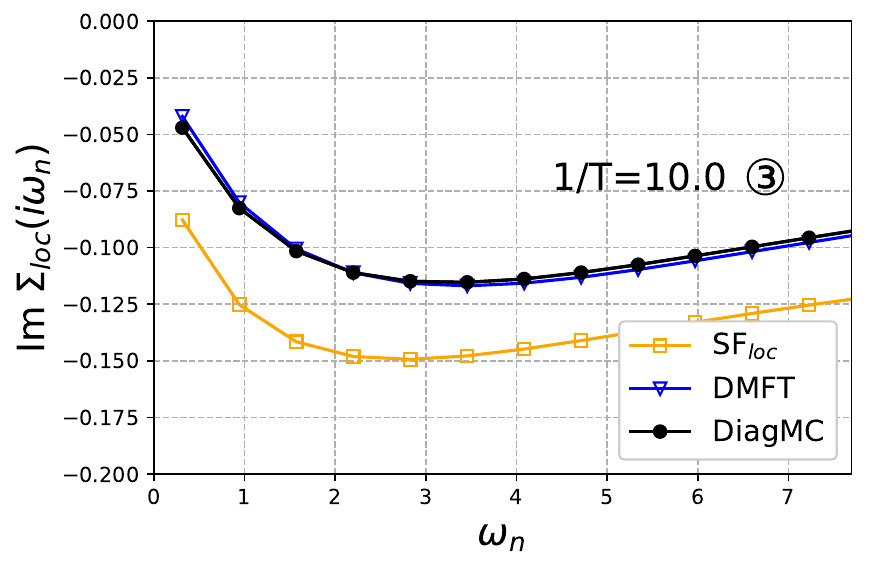}
\includegraphics[width=0.32\textwidth,angle=0]{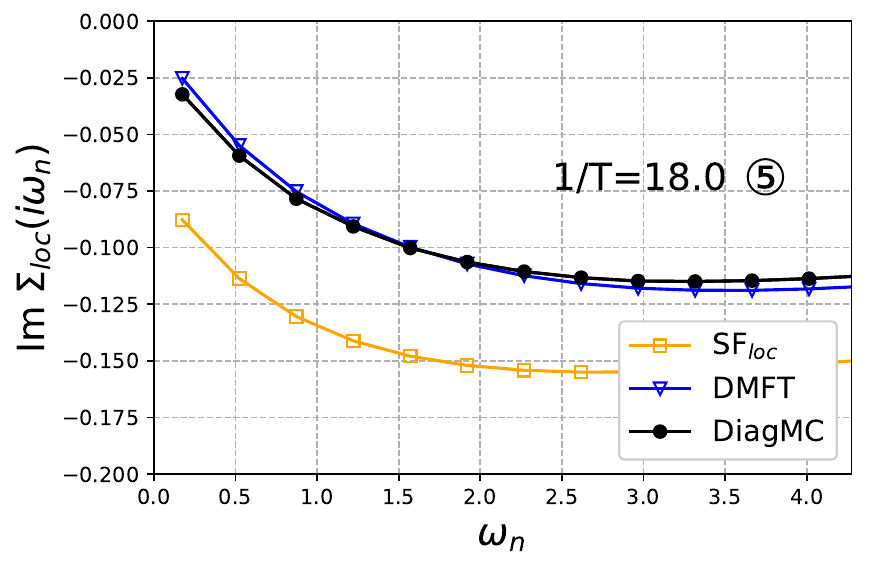}
      \caption{\label{fig:dmft+sf}(Color online.) 
Upper row: Weak coupling spin fluctuation theory combined with DMFT according to 
Eq.~(\ref{eq:simpleapprox}) - a simplified version of D$\Gamma$A ignoring vertex corrections (see text) and comparison to the DiagMC benchmark (black circles). 
Lower row: Comparison between the local self energy as obtained from DMFT and 
the corresponding weak-coupling spin fluctuation approximation to the DiagMC benchmark (black circles).}
\end{figure*}

In the previous sections, we have seen that fluctuations from all momentum transfers $\mathbf{q}$  
contribute to the local (momentum-integrated) part of the self-energy in the metallic regime. 
Long-wavelength antiferromagnetic fluctuations associated with $\mathbf{q}\simeq\mathbf{Q}$ become 
increasingly important as temperature is lowered and are responsible for the formation of the pseudogap. 
This suggests to try a simple approximation in which one relies on DMFT for the local part of the self-energy, 
while a weak-coupling approximation Eq.~(\ref{eqn:spin_fluct}) is used to account for the non-local part, namely: 
\begin{equation}\label{eq:simpleapprox}
\Sigma(\mathbf{k},i\omega_n)\,\simeq\, \Sigma_{\mathrm{DMFT}}(i\omega_n)\,+\,
\left[\Sigma_{\mathrm{SF}}(\mathbf{k},i\omega_n) - \Sigma^{\mathrm{loc}}_{\mathrm{SF}}(i\omega_n)\right].
\end{equation}
In this expression, the last term removes the local contribution from the spin-fluctuation formula, so that the 
local, $\mathbf{k}$-averaged part of the self-energy is entirely accounted for by DMFT. Please note that, in this subsection, we take the self-energy from paramagnetically restricted DMFT. To evaluate Eq.~(\ref{eq:simpleapprox}), we use the bare value $g^2=3U^2/8$ of the coupling to spin fluctuations, and 
the D$\Gamma$A spin susceptibility. We recall that the latter is simply given by: 
$\chi^{-1}(\mathbf{q},\omega)=\chi_{\mathrm{DMFT}}^{-1}(\mathbf{q},\omega)+\lambda$ with 
$\lambda$ the Moriya correction which ensures consistency with the Mermin-Wagner theorem and is determined 
such that $T\sum_{\mathbf{q},n}\chi(\mathbf{q},i\Omega_n)=\chi_{\mathrm{loc}}$, with 
$\chi_{\mathrm{loc}}$ being the local susceptibility as obtained in DMFT from the effective impurity model (for details see App.~\ref{app:dga}).
Expression (\ref{eq:simpleapprox}) is actually a simplified version of the D$\Gamma$A, in which vertex functions 
are set to unity (see also App.~\ref{app:dga} to recall of the D$\Gamma$A equation of motion. For an analysis in the context of superconductivity in cuprates see \cite{Kitatani2018}).

The result of this simple `DMFT+SF' approximation of Eq.~(\ref{eq:simpleapprox}) is compared in Fig.~\ref{fig:dmft+sf} (top row) 
to the DiagMC benchmark and to D$\Gamma$A for three temperatures. 
We see that it performs excellently in the incoherent and in the metallic regime. In the pseudogap regime it is qualitatively 
reasonable, but overestimates the magnitude of the self-energy at low-frequency (in  accordance with the observations 
made above.) This indicates that, for this rather weak value of the coupling $U=2$, the vertex correction terms included in 
D$\Gamma$A are not essential in the metallic regime, but become increasingly important as temperature is lowered into the 
pseudogap regime. 

The lower panels in Fig.~\ref{fig:dmft+sf} display the local part of the self-energy as obtained from DMFT together with its approximation 
from weak-coupling spin fluctuation theory [the local term removed in Eq.~(\ref{eq:simpleapprox})], in comparison to DiagMC. 
This clearly demonstrates that a weak-coupling spin fluctuation approximation does not provide a good estimate of the local 
component, even at this weak value of $U$. In contrast, the non-perturbative DMFT provides an excellent description 
of the $\mathbf{k}$-averaged (local) self-energy, while non-local contributions can be reasonably 
accounted for by weak coupling spin fluctuation theory at this value of $U$.

\section{Conclusion and outlook}
\label{sec:conclusion}

In this article, we have purposefully focused on an apparently `simple' regime of the two dimensional Hubbard model 
(simple square lattice, weak coupling, half filling) with three goals in mind: 
(i) assessing the ability of state-of-the-art computational methods to address the physics of this model at finite temperature, 
through the computation of a range of one-particle and two-particle observables 
(ii) providing an extensive assessment of basically {\it all} many-body methods currently available for this purpose, by comparing them to 
two very different Monte Carlo methods (DQMC and DiagMC) serving as benchmarks and, importantly 
(iii) investigating the rich physics associated with the different regimes and crossovers found in this model, as it 
evolves upon cooling 
from a high-temperature incoherent regime into a momentum-differentiated metal and eventually into an insulating regime 
with a pseudogap, and elucidating the nature and role of spin fluctuations in these different regimes. 

It is satisfying to observe that the two benchmark methods considered in this article are successful at computing 
the properties of this model throughout the metallic regime and are in excellent agreement 
with each other. However, we also found that they
face rather severe limitations when attempting to reach low temperatures, to the extent that only the onset of the pseudogap regime 
can be reached. Indeed, the pseudogap is associated with an exponentially growing antiferromagnetic correlation length 
which requires prohibitively large systems to be simulated with DQMC, while DiagMC, which in contrast 
works directly in the thermodynamic limit, is faced with convergence issues when summing the perturbative series at low-$T$. We are hopeful that this can be overcome by using recent improvements of diagrammatic Monte Carlo based on the CDet algorithm \cite{Rossi2017}, together with improved schemes for summing the perturbative series.

Many of the methods considered in this article use the dynamical mean field theory (DMFT) as a starting point and treat spatial fluctuations beyond this 
starting point to various degrees of approximation. Obviously, the chosen regime of parameters 
puts DMFT very much out of its 'comfort zone' since the physics at low-$T$ is dominated by long-range 
spatial fluctuations  which in two dimensions also prevent long-range order at any non-zero temperature. 
Nonetheless we have emphasized and documented that, when properly interpreted, DMFT provides 
a useful initial description of the different regimes. 
Indeed, in the low-$T$ pseudogap regime when the correlation length is large, describing the system as long-range ordered is a reasonable mean-field starting point.  
Furthermore, remarkably, we found that DMFT provides a highly accurate determination of 
the local (momentum-averaged) self-energy, not only at high-$T$ as expected, but also through most of the metallic regime.  
This is a striking observation given that the correlation length reaches values as large as ten lattice spacing in this 
regime, and we further comment below on the physical reason explaining this finding.

Because of the very large correlation length at low-$T$, 
cluster extensions of DMFT are not the best route to follow in this parameter regime of the Hubbard model. 
Although they succeed in describing the momentum-differentiated metallic regime in satisfactory agreement with 
the benchmarks, huge cluster sizes would be required to properly capture the pseudogap regime. 
Cluster extensions of DMFT are much better equipped for addressing strong coupling regimes with shorter 
correlation lengths. 

In contrast, we found that extensions of DMFT (ladder-D$\Gamma$A, TRILEX, ladder-DF and single-shot DB), that make use of response and vertex functions in order to take into account the contributions of fluctuations to the self-energy, perform much better here. 
We assessed their respective degree of accuracy and found that the D$\Gamma$A and DF/DB methods 
are particularly successful at describing all physical regimes of interest in satisfactory agreement 
with the benchmarks, 
including the onset of the pseudogap regime. They also allow to enter deeper in the pseudogap regime: 
it is an open question for future work to assess their degree of validity in this low-$T$ regime when accurate and controlled benchmarks become available.

We have also considered many-body methods that do not make use of DMFT as a starting point, such as TPSC/TPSC+,  
fRG and the PA. TPSC played an important role early on in elucidating the physical origin of the 
weak-coupling pseudogap in relation to long-wavelength antiferromagnetic correlations. 
We found that it captures the different regimes qualitatively, but strongly overestimates the 
onset of the pseudogap temperature and the correlation length itself. 
TPSC+, a recent extension of the method, leads to significant improvements. 
In contrast, the PA underestimates the pseudogap onset temperature, especially at the nodal point,  whereas it captures the behavior of double occupancy very accurately. The conventional one-loop fRG shown here is in qualitative agreement with the benchmark until its running coupling constants diverge. We remark, however, that the present implementation can be systematically improved by the recently introduced multiloop extension of fRG that converges to the parquet approximation \cite{Kugler2018, Tagliavini2019, Kugler2018a}, as well as expanding around the DMFT solution (DMF$^2$RG) \cite{Taranto2014, Vilardi2019}. Their combination, a multi-loop expansion of the fRG around DMFT, provides a very promising direction for future studies.


On the physics side, perhaps the most intriguing finding of our study is that, in the metallic regime, 
single-particle excitations with a lifetime that increases upon cooling 
appear to happily coexist with an 
antiferromagnetic correlation length which increases steeply, reaching about ten lattice spacings 
at the onset of the pseudogap. We have provided an explanation to this finding and also shown 
that spin fluctuations associated with the vicinity of the antiferromagnetic wave-vector are not the only important contributions to the self-energy in this regime. 
Fluctuations from all nesting vectors also contribute to the self-energy on the Fermi surface, 
and fluctuations from all momentum transfers contribute especially to the local, momentum-integrated, self-energy.
This explains, importantly, why the latter is so accurately captured by single-site DMFT, even in the presence of strong antiferromagnetic 
fluctuations which are responsible for the momentum differentiation on the Fermi surface. 

At the lowest temperatures, antiferromagnetic fluctuations eventually dominate, destroying the coherent metal  
which gives way to the pseudogap regime (with a narrow precursor regime at which the quasiparticle lifetime reaches 
a maximum). 
This provides a rationale for the success of methods such as D$\Gamma$A and the DF, which 
incorporate fluctuations of all wave vectors beyond DMFT. In the parameter regime of interest, we have 
proposed and tested a simplified version of such an approach. 

\begin{figure*}[t!]     
\includegraphics[width=0.24\textwidth,angle=0]{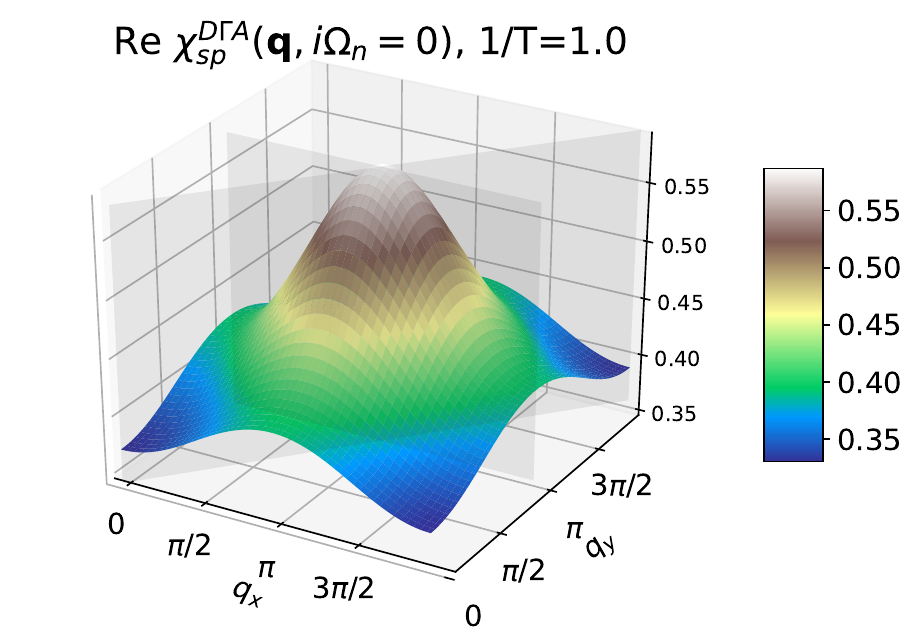}
\includegraphics[width=0.24\textwidth,angle=0]{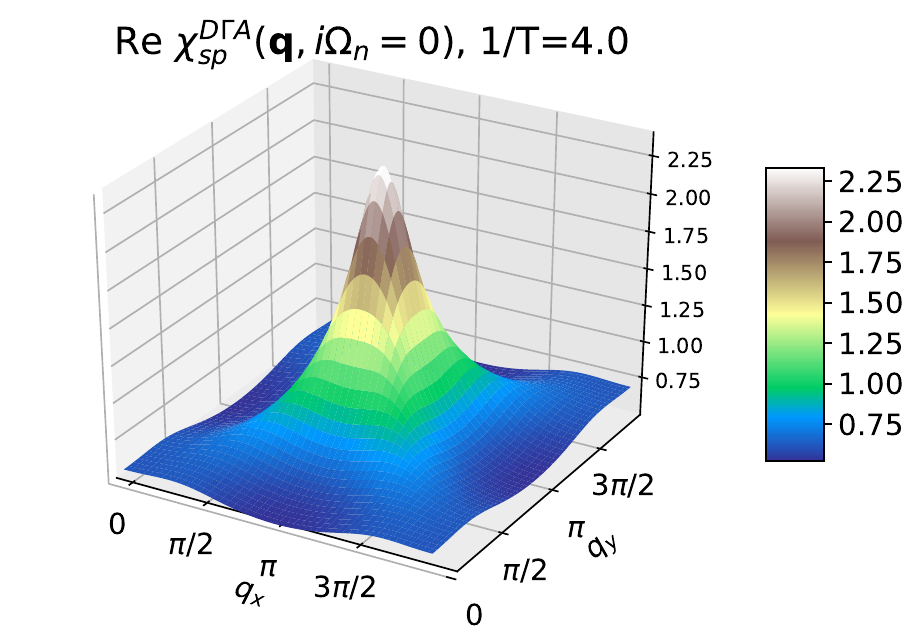}
\includegraphics[width=0.24\textwidth,angle=0]{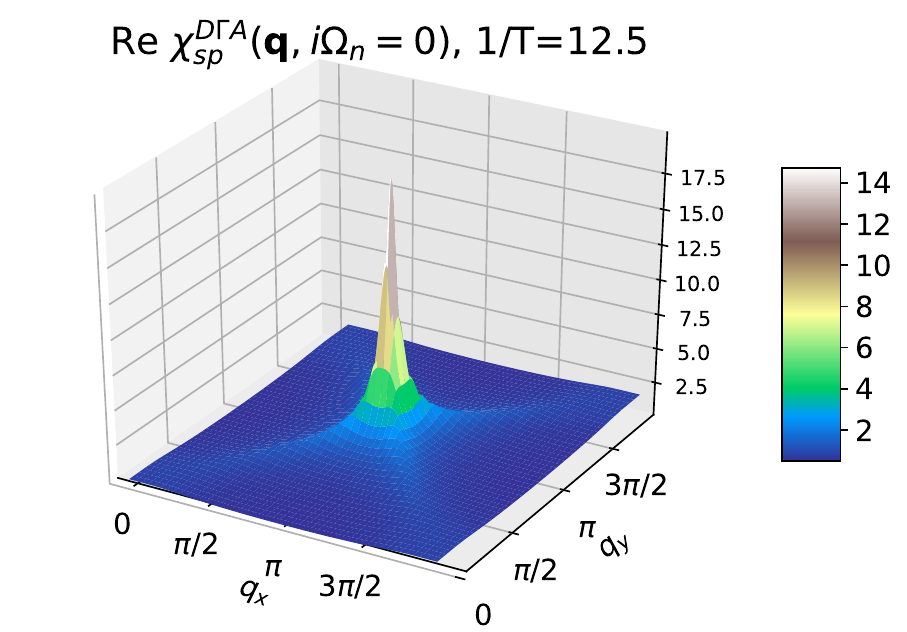}
\includegraphics[width=0.24\textwidth,angle=0]{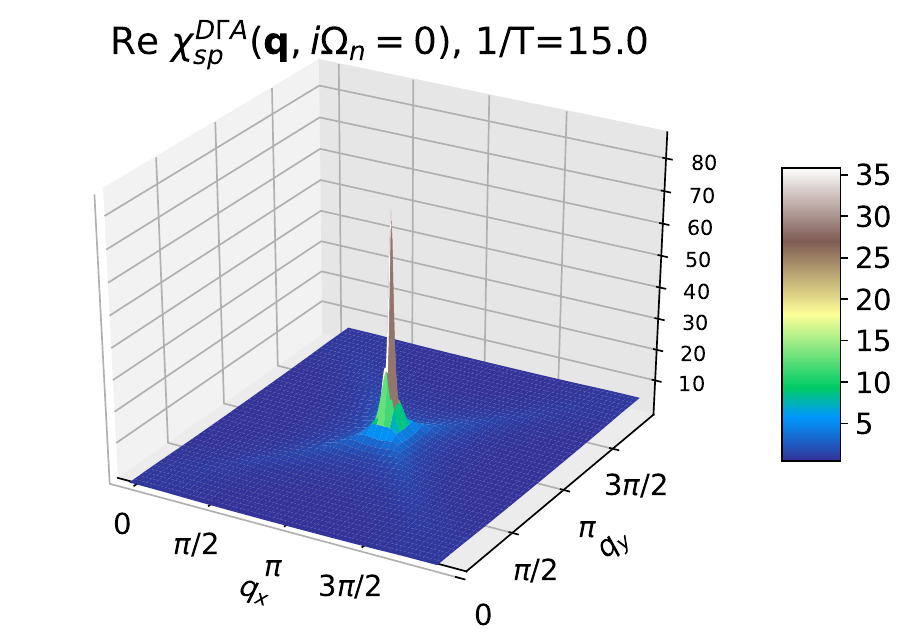} \includegraphics[width=0.24\textwidth,angle=0]{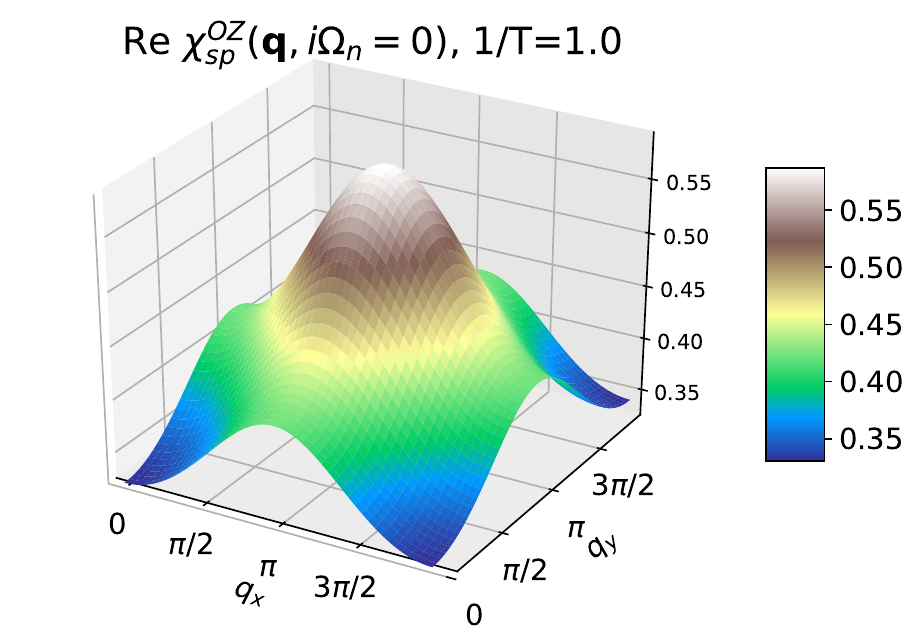}
\includegraphics[width=0.24\textwidth,angle=0]{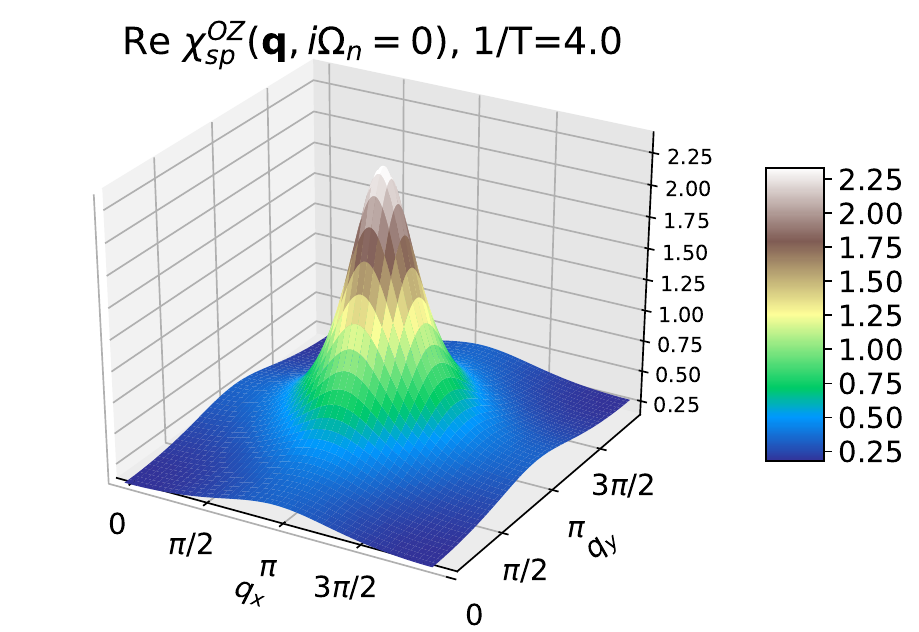}
\includegraphics[width=0.24\textwidth,angle=0]{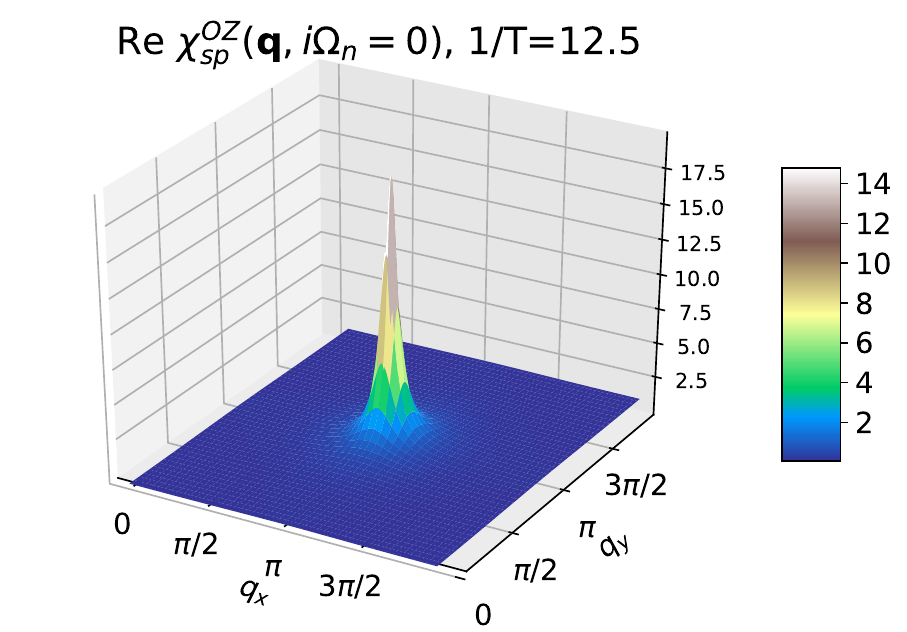}
\includegraphics[width=0.24\textwidth,angle=0]{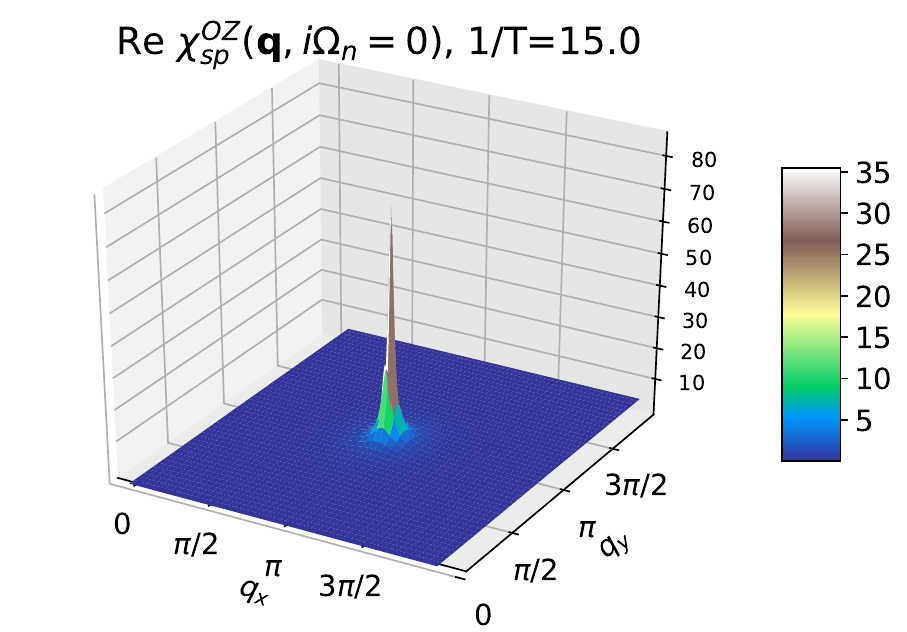}
\includegraphics[width=0.24\textwidth,angle=0]{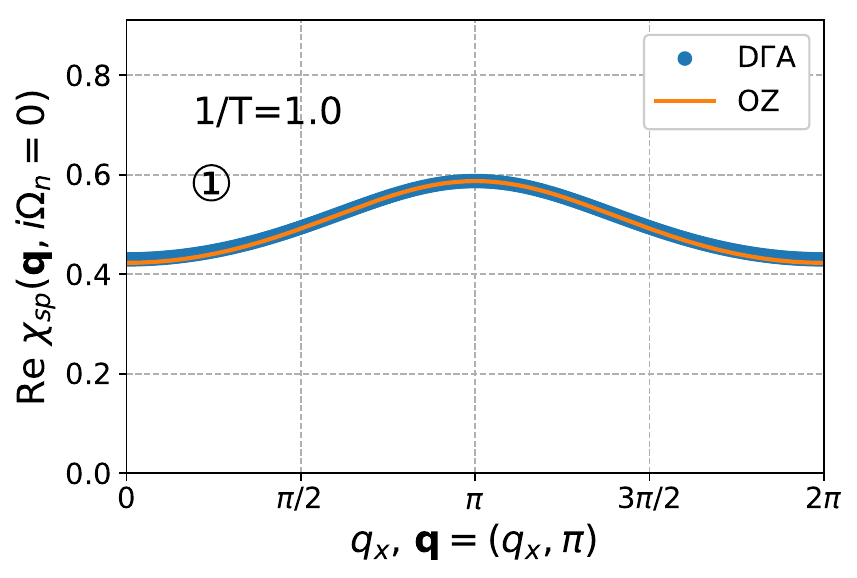}
\includegraphics[width=0.24\textwidth,angle=0]{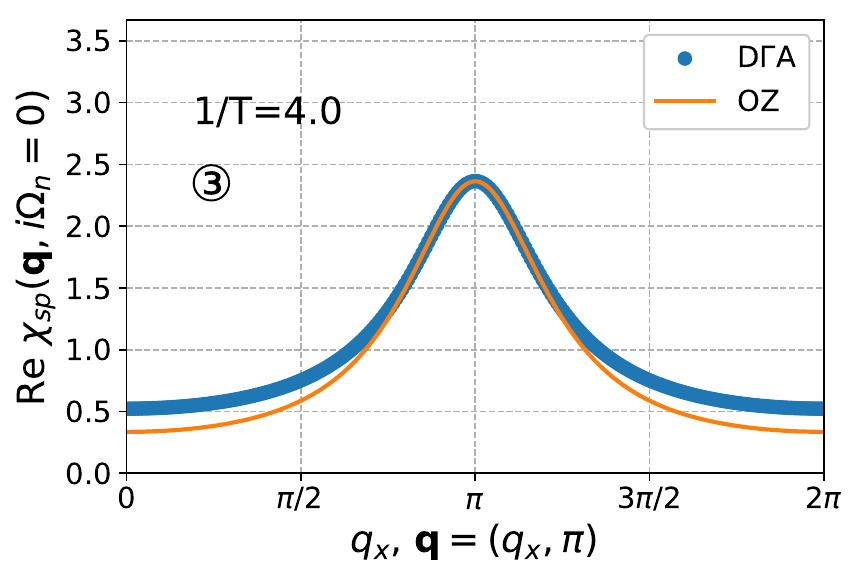}
\includegraphics[width=0.24\textwidth,angle=0]{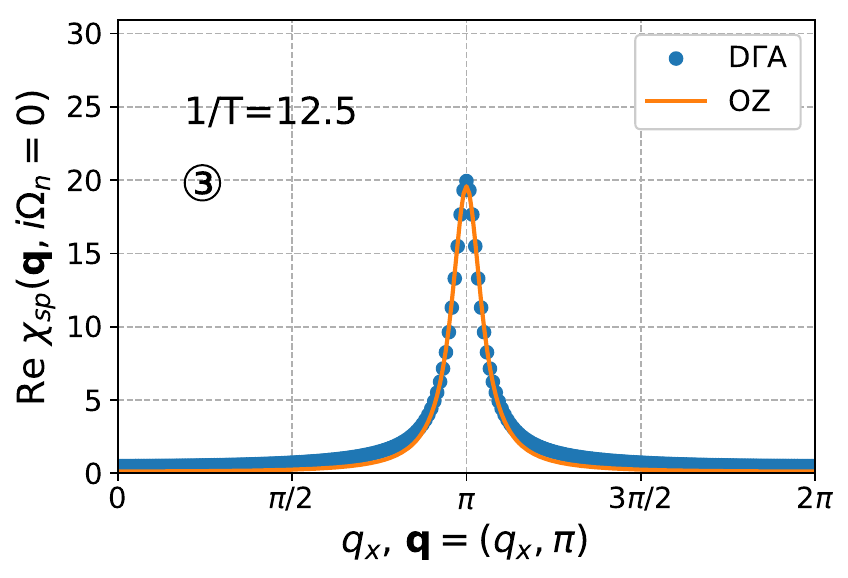}
\includegraphics[width=0.24\textwidth,angle=0]{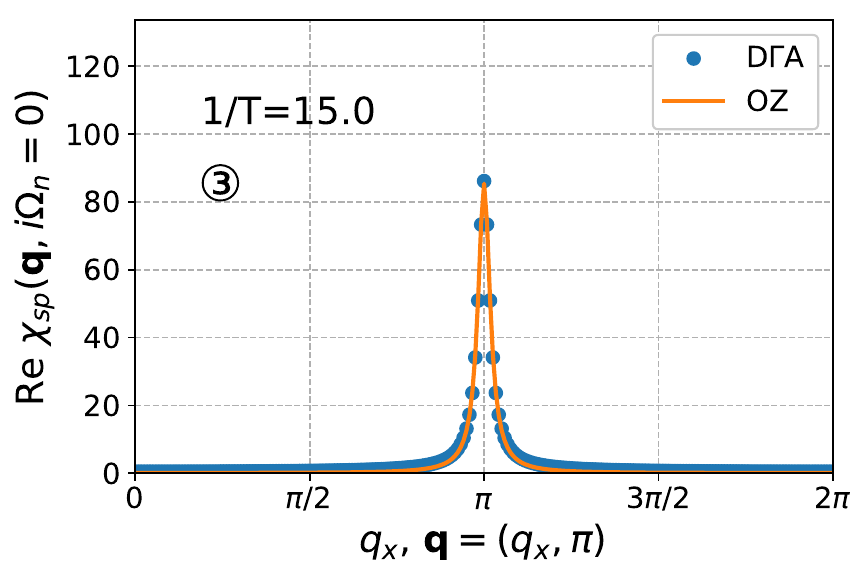}
\includegraphics[width=0.24\textwidth,angle=0]{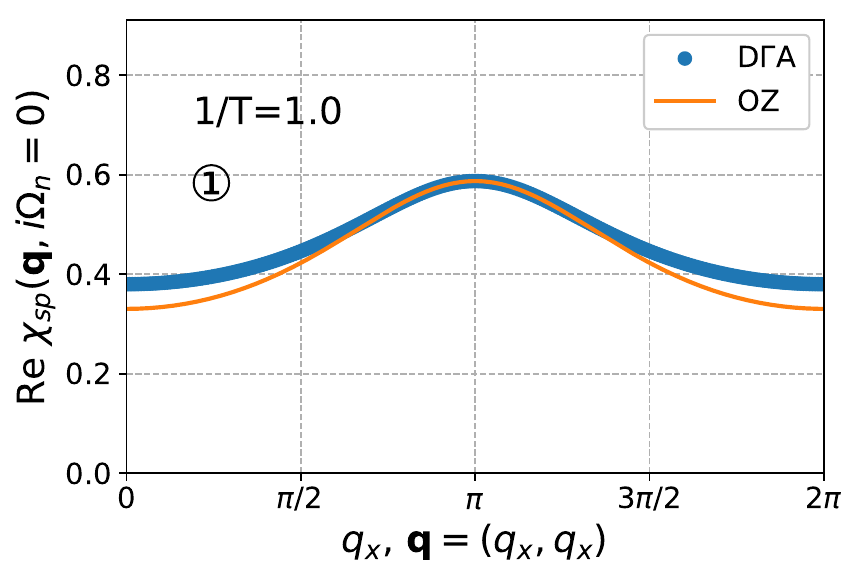}
\includegraphics[width=0.24\textwidth,angle=0]{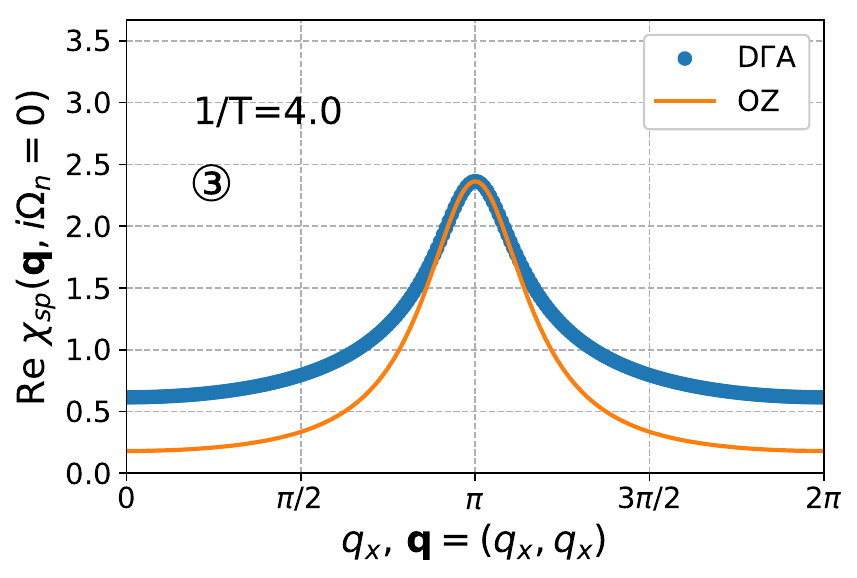}
\includegraphics[width=0.24\textwidth,angle=0]{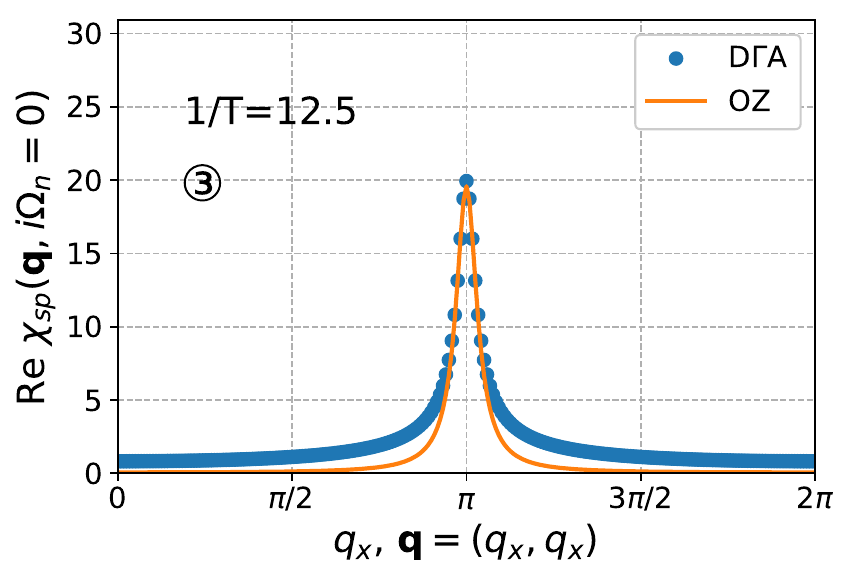}
\includegraphics[width=0.24\textwidth,angle=0]{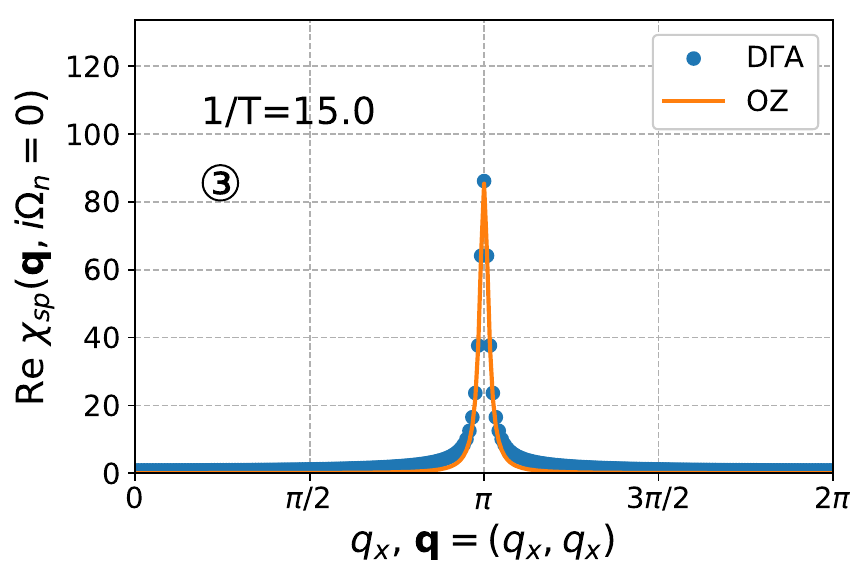}
      \caption{\label{fig:ornstein}(Color online.) 
The static (zero-frequency) antiferromagnetic susceptibilities obtained from D$\Gamma$A (uppermost row), the fitted data from an Ornstein-Zernike form (second row) and a comparison of both at the momentum cuts $\mathbf{q}=(q_x,\pi)$ [third row] and  $\mathbf{q}=(q_x,q_x)$ [lowest row] for various temperatures.}
\end{figure*}

We have documented in details the rich sequence of crossovers occurring in this model upon reducing temperature, 
from an incoherent regime at high $T$ all the way down to the pseudogap regime at low-$T$, through the intermediate 
temperature momentum-differentiated metal, and assessed the ability of the different computational methods 
to capture these crossovers. By analyzing the contribution of spin fluctuations, we showed that these different regimes
are delimited by the comparison of temperature itself to the two characteristic energy scales $v_F/\xi$ and 
$\gamma\xi^{-2}$, or equivalently of the correlation length $\xi$ to the thermal de Broglie wavelength $~\sim v_F/T$ and 
to the length scale associated with Landau damping $\sim \sqrt{\gamma/T}$ (with $\gamma$ being the Landau damping constant). 

Finally, we have also emphasized (Sec.~\ref{sec:nesting} and Appendices~\ref{app:fermi_liquid} and \ref{app:2pt}) the limitations 
of imaginary time/Matsubara computational methods in probing the delicate low energy non-Fermi liquid singularities of this perfectly nested model 
in the metallic regime.

Looking forward, the numerical, theoretical and physical insights obtained in this article 
should find direct applications in computational studies~\cite{Kitatani2020, Galler2016, Kitatani2018} 
of materials for which the interplay between local electronic correlations and long-wavelength spin 
fluctuations play a crucial role, such as Sr$_2$RuO$_4$ \cite{Strand2019, Tamai2019, Acharya2019}
or the iron-based superconductors \cite{Zantout2019}. 
On the model level, our findings will be immediately useful for describing how physical fluctuations 
(density, magnetic, pairing) on the two-particle level influence one-particle spectral functions and 
self-energies, and how quasiparticles are affected and sometimes destroyed altogether by these fluctuations~\cite{Gunnarsson2016,  Gunnarsson2017, Wu2016, Rohringer2018b, Arzhang2020, Gunnarsson2018, Schaefer2020}.

\acknowledgments 
We would like to thank Andrey Chubukov, Chlo{\'e} Gauvin-Ndiaye, Emanuel Gull, Karsten Held, Carsten Honerkamp, Andrey A. Katanin, Alexander I. Lichtenstein, Walter Metzner, Andrew J. Millis, Alice Moutenet, Georg Rohringer, Subir Sachdev, Michael Thoennessen, and Alessandro Toschi for useful and valuable discussions, as well as T. S. Barry for support and encouragement. 
T. S. and A. G. are indebted to Katharina K{\"o}lbl and Sandrine Kott for their infinite patience during the writing phase of this manuscript.

\begin{figure*}[t!]     
\includegraphics[width=0.32\textwidth,angle=0]{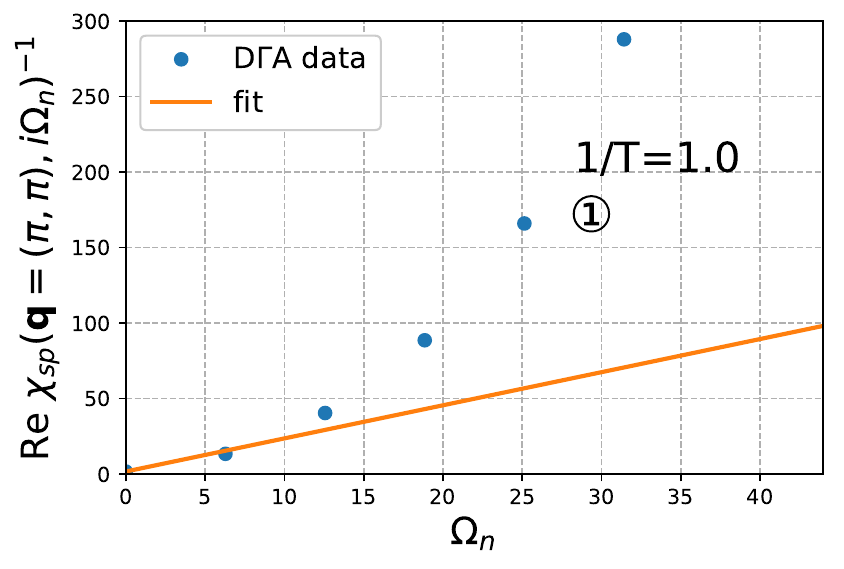}
\includegraphics[width=0.3155\textwidth,angle=0]{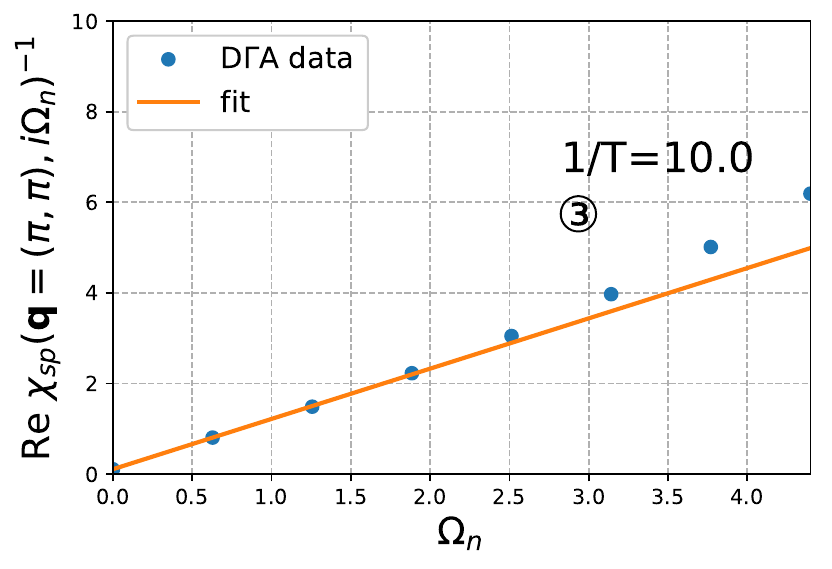}
\includegraphics[width=0.31\textwidth,angle=0]{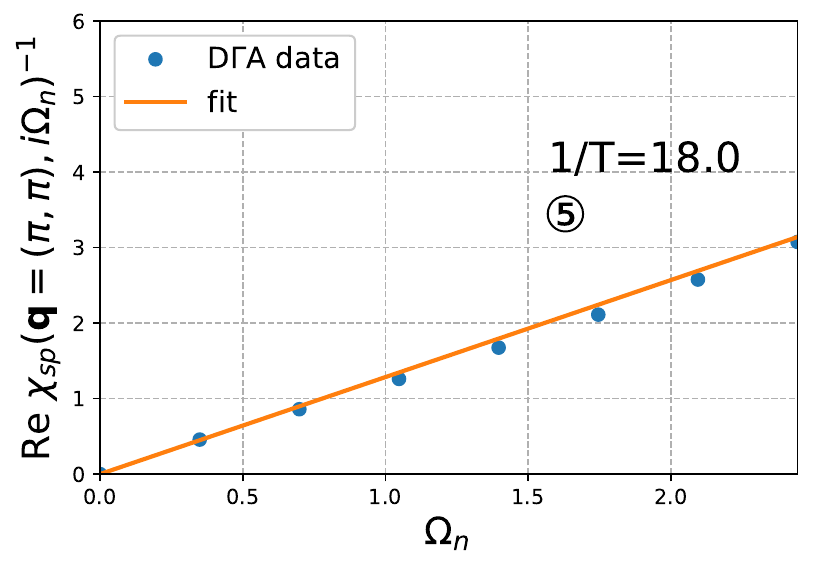}
\includegraphics[width=0.32\textwidth,angle=0]{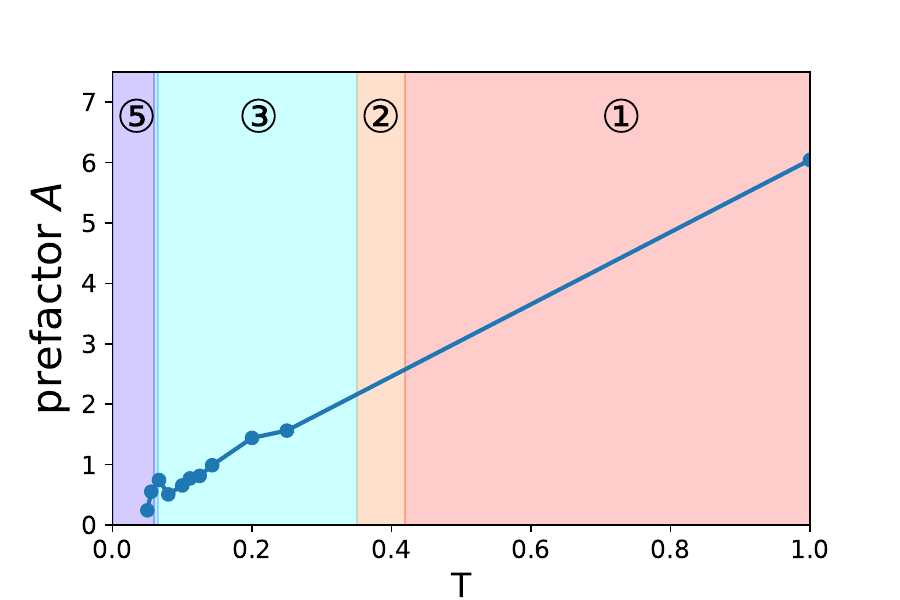}
\includegraphics[width=0.32\textwidth,angle=0]{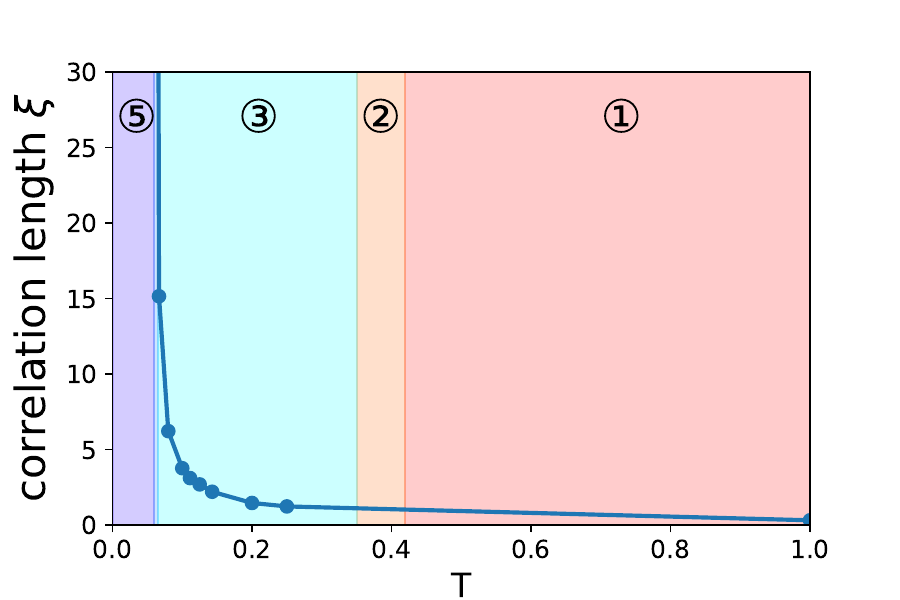}
\includegraphics[width=0.31\textwidth,angle=0]{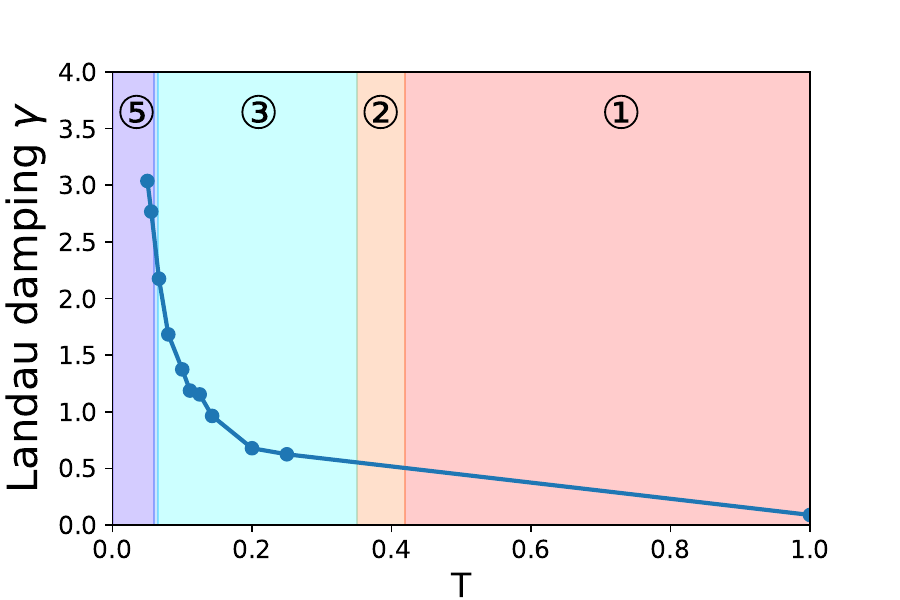}
      \caption{\label{fig:ornstein_dynamical}(Color online.) 
Upper row: Fits of the frequency-dependent part of the (inverse) antiferromagnetic susceptibility from D$\Gamma$A by an Ornstein-Zernike form at three different temperatures. Lower row: The prefactor $A$ and the magnetic correlation length $\xi$ have been fixed to the ones of the static propagator and their temperature dependencies together with the one of the Landau damping $\gamma$ are shown.}
\end{figure*}

The present work was supported by: 
the Austrian Science Fund (FWF) through the Erwin-Schr\"odinger Fellowship J 4266 - ``{\sl Superconductivity in the 
vicinity of Mott insulators}'' (SuMo, T.S.), and projects P 30997 (C.E.) and P 30819 (A. Kauch); the European Research Council for the European Union Seventh Framework Program (FP7/2007-2013) with ERC Grant No. 319286 (QMAC, T.S. and A.G.); 
the Simons Foundation through the Simons Collaboration on the Many Electron Problem (M.F., F.{\v S}., E.K.); 
the Natural Sciences and Engineering Research Council (Canada) under grants RGPIN-2019-05312 (A.-M.S.T.) and  RGPIN-2017-04253 (J.P.F.L.); the Canada First Research Excellence Fund, the Research Chair on the Theory of Quantum Materials, Compute Canada and Calcul Québec (Y.W., A.-M.S.T.).
A. Kim and E.K. acknowledge funding from EPSRC through grant EP/P003052/1. V.H. and E.A.S. acknowledge the support by the North-German Supercomputing Alliance (HLRN) under the Project No. hhp00042. C.H. and S.A. acknowledge financial support from the Deutsche Forschungsgemeinschaft (DFG) through Projects No. AN 815/4-1 and No. AN 815/6-1. 
T.S., F.{\v S}. and M.F. acknowledge the hospitality of the Center for Computational Quantum Physics at the Flatiron Institute. The Flatiron Institute is a division of the Simons Foundation.
The authors acknowledge the computer support teams at CPHT {\'E}cole Polytechnique, CEA Saclay (IPHT-LSCE) and at the Flatiron Institute as well as the computer service facility of the MPI-FKF for their help. This work was granted access to the HPC resources of TGCC and IDRIS under the allocations A0070510609 and A0090510609 attributed by GENCI (Grand Equipement National de Calcul Intensif). Parts of the simulations were performed on computers provided by Calcul Qu\'ebec, and Compute Canada. J.P.F.L. acknowledges computational support from ACENET and Compute Canada. The PA calculations have been done on Vienna Scientific Cluster (VSC) and JURECA at Forschungszentrum J{\"u}lich \cite{Jureca2018}. D.~R. gratefully acknowledges the computing time granted through JARA on the supercomputer JURECA \cite{Jureca2018} at Forschungszentrum J{\"u}lich.

\appendix
\section*{Appendices}

\section{Ornstein-Zernike form of the magnetic susceptibility}
\label{app:ornstein}
Close to a magnetic phase transition, the magnetic susceptibility (bosonic propagator) assumes the following  Ornstein-Zernike form \cite{Zernike1916}:
\begin{equation}\label{eqn:ornstein_app}
	\chi(\mathbf{q}, i\Omega_{n}) = \frac{A}{(\mathbf{q}-\mathbf{Q})^2 + \xi^{-2} + 
	\frac{\left|\Omega_n\right|}{\gamma}},
\end{equation}
with $A$ being a prefactor, $\gamma$ the Landau damping constant (the dynamical critical exponent has been set to $z=2$ here), and $\xi$ the magnetic correlation length. This form can be used to evaluate the spin-fluctuation diagram given in Eq.~(\ref{eqn:spin_fluct}). In the renormalized classical regime \textcircled{5}, the sum may be restricted to the lowest Matsubara frequency without great loss of precision, which is, however, not possible in the metallic regime \textcircled{3}.
The estimation of the correlation length from the static magnetic susceptibility in this manuscript has been performed by an empirically more robust formula (especially in the limit of small $\xi$) of the Ornstein-Zernike form, which incorporates the lattice (tight-binding) structure of the problem \cite{Rohringer2011}, neglecting a possible anomalous exponent $\eta$ (cf. $\eta\!\approx\! 0.037$ for the three-dimensional Heisenberg model \cite{Huang1987}):
\begin{equation}
 \chi_{\text{sp}}(\mathbf{q}, i\Omega_{n}=0) = \frac{A}{4\text{sin}^{2}\left(\frac{q_x-Q_x}{2}\right) +  4\text{sin}^{2}\left(\frac{q_y-Q_y}{2}\right) + \xi^{-2}},
 \label{eqn:ornstein_sin}
\end{equation}
and reduces to the original form in the long wavelength  limit (small arguments of the sine). For the fits shown we fixed $q_y\!=\!\pi$.
 Fig.~\ref{fig:ornstein} shows data for the static antiferromagnetic susceptibility in D$\Gamma$A (upper row) and the respective Ornstein-Zernike fits for various temperatures. Please note again, as discussed in the main text, that the Ornstein-Zernike fit is most accurate only around the vicinity of $\mathbf{Q}=(\pi,\pi)$ and may not capture the physics at other $\mathbf{q}$ points, important e.g. in the metallic regime (see Sec.~\ref{subsec:insights_metal}). These fits of the static antiferromagnetic susceptibility are used to determine the prefactor $A$ and the magnetic correlation length $\xi$. For extracting the Landau damping $\gamma$, $A$ and $\xi$ have been fixed and $\gamma$ has been fitted from the frequency dependence of the propagator. The upper panels of Fig.~\ref{fig:ornstein_dynamical} show these fits of D$\Gamma$A data for three representative temperatures. The lower panels show the obtained parameters $A$, $\xi$ and $\gamma$ as a function of temperature.
 
\begin{figure}[b!]
              \includegraphics[width=0.45\textwidth,angle=0]{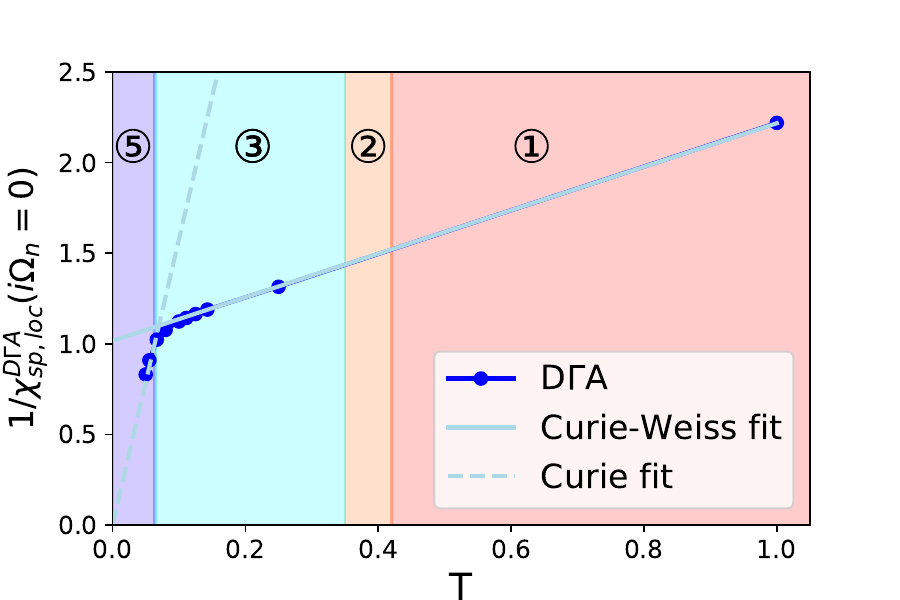}
      \caption{\label{fig:chi_loc}(Color online.) The inverse local spin susceptibility calculated in D$\Gamma$A as a function of temperature, which exhibits Curie-Weiss (Curie) behavior in the metallic (insulating) regime.}
\end{figure}

\begin{figure}[t!]
              \includegraphics[width=0.45\textwidth,angle=0]{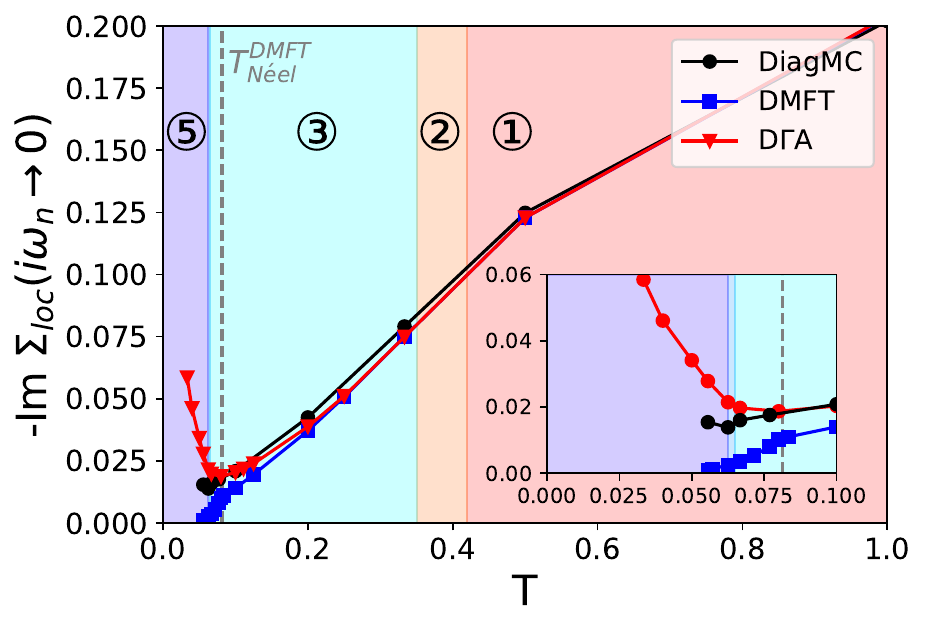}
      \caption{\label{fig:im_sigma_zero}(Color online.) The (negative) imaginary part of the local self-energy extrapolated to zero Matsubara frequency as a function of temperature calculated in DiagMC (black circles), DMFT (blue squares) and D$\Gamma$A (red triangles).}
\end{figure}

 We see that the prefactor $A(T)$ rapidly decreases with temperature in the metallic regime. 
 This can be rationalized by observing that the local susceptibility 
 $\chi_{\mathbf{loc}}=\int d^2q/(2\pi)^2\,\chi(\mathbf{q},\Omega_n=0)$ obeys a Curie-Weiss law in this regime, 
 implying that $A\ln\xi \sim C/(T+\Theta)$ and hence that $A$ decreases linearly with $T$. 
 In contrast, in the low-$T$ insulating regime, the sum-rule 
 $T\sum_n\int d^2q/(2\pi)^2 \chi(\mathbf{q},\Omega_n=0)=n-2\langle n_\uparrow n_\downarrow\rangle$ implies 
 $TA\ln\xi \sim \mathrm{const.}$ and hence $A$ reaches a constant value at low $T$. For the temperature dependence of the local spin susceptibility in D$\Gamma$A see Fig.~\ref{fig:chi_loc}. The (negative) imaginary part of the local  self-energy  extrapolated  to  zero  Matsubara frequency as a function  of  temperature, relevant for the discussion in Sec.~\ref{sec:metallic}, is shown in  Fig.~\ref{fig:im_sigma_zero}.

\section{Low-frequency expansion of the self-energy}
\label{app:fermi_liquid}
In a Fermi liquid, the self-energy on the Fermi surface can be expanded at low (imaginary) frequency 
as a Taylor series:  
\begin{eqnarray}
 \Sigma(\mathbf{k}_{\text{F}},i\omega\rightarrow 0) 
 &\simeq& \text{Re } \Sigma(\mathbf{k}_{\text{F}},0) + i\text{Im } \Sigma(\mathbf{k}_{\text{F}},0)+\nonumber\\
  &&+ i\omega\,\frac{\partial\text{Im }\Sigma(\mathbf{k}_{\text{F}},i\omega)}{\partial\omega} +\cdots
 \label{eqn:FL}
\end{eqnarray}
Inserting this expression into Dyson's equation $G^{-1}=G_0^{-1}-\Sigma$, we see that the spectral function
displays a quasiparticle peak with a spectral weight $Z_{\mathbf{k}}$ and width $\gamma_{\mathbf{k}}$
(inverse of the quasiparticle lifetime $\tau_{\mathbf{k}}$) given by: 
\begin{eqnarray}
 Z_{\mathbf{k}}&=&\left[1-\frac{\partial\text{Im }\Sigma(\mathbf{k},i\omega)}{\partial\omega}
 \bigg\vert_{\omega\rightarrow{0}}\right]^{-1} \label{eqn:z} \\
 \gamma_{\mathbf{k}}&=&\tau_{\mathbf{k}}^{-1}=-Z_{\mathbf{k}} \cdot \text{Im }\Sigma(\mathbf{k}, i\omega)\bigg\vert_{\omega\rightarrow{0}} \label{eqn:gamma}
\end{eqnarray}
Performing also a Taylor expansion for momenta $\mathbf{k}$ close to  Fermi surface, the renormalization of the 
quasiparticle effective mass is obtained as: 
\begin{equation}
\left.\left(\frac{m^*}{m^{\phantom{*}}}\right)^{-1}\right|_{\mathbf{k}_F} 
= Z_{\mathbf{k}_F} \left[ 1 + \frac{\hat{\mathbf{e}}_{\mathbf{k}_F}\cdot\nabla_k 
\text{Re }\Sigma(\mathbf{k},i\omega\rightarrow 0)}
{\hat{\mathbf{e}}_{\mathbf{k}_F}\cdot\nabla_k \varepsilon_{\mathbf{k}}} \right]_{\mathbf{k}=\mathbf{k}_F}.
\end{equation}
In practice, $Z_{\mathbf{k}}$ and $\gamma_{\mathbf{k}}$, as displayed e.g. in our discussion 
of the metallic regime in Fig.~\ref{fig:fl}, are obtained in this article by a fit of the imaginary part of the self-energy 
$\text{Im }\Sigma(\mathbf{k},i\omega_n)$ 
calculated on the discrete set of Matsubara frequencies, by a fourth-order polynomial. 
This is common practice in analyzing computational results available in imaginary time at 
finite temperature. 
The sign of the slope of the self-energy at low Matsubara frequencies was also used to distinguish between 
`metallic' behaviour ($Z_{\mathbf{k}}<1$) and the pseudogap and incoherent regimes in which 
the positive slope corresponds formally to $Z_{\mathbf{k}}>1$. 
Care should be applied, however, in interpreting the values of $Z_{\mathbf{k}}$ 
and $\gamma_{\mathbf{k}}$ obtained from such Matsubara fits in terms of quasiparticles properties, when the system does not 
obey Fermi liquid behaviour at low frequencies or low temperatures 
(as is the case here, because of perfect nesting and of the gradual opening of the pseudogap). 
We illustrate and clarify this point by considering, for simplicity, a linear fit of the imaginary part of the 
self-energy over the first two Matsubara frequencies $\omega_0=\pi T$, $\omega_1=3\pi T$, namely: 
\begin{equation}
\text{Im}\Sigma(i\omega_n)\,=\,
\text{Im}\Sigma(i0^+)|_{\text{linfit}}\,
+\,\omega_n (1-\frac{1}{Z_{\text{linfit}}}) +\cdots
\end{equation}
with:
\begin{eqnarray}\label{eqn:linfit1}
\text{Im}\Sigma(i0^+)|_{\text{linfit}}\,&=&\,
\frac{\omega_1\text{Im}\Sigma(i\omega_0)-\omega_0\text{Im}\Sigma(i\omega_1)}{\omega_1-\omega_0},\nonumber \\
1-\frac{1}{Z_{\text{linfit}}}\,&=&\,\frac{\text{Im}\Sigma(i\omega_1)-\text{Im}\Sigma(i\omega_0)}{\omega_1-\omega_0}.
\end{eqnarray}
For simplicity, we have dropped the momentum dependence in these expressions: 
it is understood that the analysis is performed at a given value of $\mathbf{k}$.  
We now use the spectral representation of the self-energy 
$\Sigma(i\omega_n)\,=\,\int d\omega \frac{\sigma(\omega)}{i\omega_n-\omega}$ 
with $\sigma(\omega)\equiv -\frac{1}{\pi}\text{Im}\Sigma(\omega+i0^+)$, yielding:
\begin{equation}
-\text{Im}\Sigma(i\omega_n)\,=\,\omega_n\int d\omega \frac{\sigma(\omega)}{\omega_n^2+\omega^2}\,\,\,\,\,\,.
\end{equation}
Substituting this into Eqs.~(\ref{eqn:linfit1}), after simple algebra, we finally obtain:
\begin{eqnarray}\label{eqn:linfit2}\nonumber
-\text{Im}\Sigma(i0^+)|_{\text{linfit}}\,&=&\,
12\pi^3T^3\,\int d\omega\,
\frac{\sigma(\omega)}{(\pi^2T^2+\omega^2)(9\pi^2T^2+\omega^2)}
\\ \nonumber
1-\frac{1}{Z_{\text{linfit}}}\,&=&\,
\int d\omega\,\sigma(\omega)
\frac{3\pi^2T^2-\omega^2}{(\pi^2T^2+\omega^2)(9\pi^2T^2+\omega^2)}
\end{eqnarray}
We now analyze how these two quantities behave in the low-temperature limit. 
To this aim, we consider cases in which 
$\omega/T$ scaling applies for small $\omega$ and $T$ (but an arbitrary ratio $\omega/T$), and that 
$\sigma(\omega,T)=T^\alpha \phi(\omega/T)$ with $\phi$ a scaling function such that 
$\phi(x\ll 1)\sim\text{const.}$ and $\phi(x\gg 1)\sim |x|^\alpha$. 
The case $\alpha=2$ corresponds to a Fermi liquid for which the scaling function is known to be 
$\phi(x)=A(\pi^2+x^2)$. 
A value of $\alpha <2$ corresponds to non-Fermi liquid behaviour with 
$\omega/T$ scaling, as found for example close to quantum critical points controlled by strong coupling 
fixed points~\cite{Sachdev1999}.
For an analysis of non-Fermi liquid behaviour and $\omega/T$ scaling 
within second order perturbation theory, see Appendix~\ref{app:2pt}. 

To obtain the low-$T$ behaviour of $-\text{Im}\Sigma(i0^+)|_{\text{linfit}}$, we observe that the 
$\omega/T$ scaling form can be directly substituted in the above expression while preserving the 
convergence of the integral. This leads to, for $T\rightarrow 0$:
\begin{equation}
-\text{Im}\Sigma(i0^+)|_{\text{linfit}}\,\sim\,
12\pi^3T^\alpha\,\int dx\,
\frac{\phi(x)}{(\pi^2+x^2)(9\pi^2+x^2)}.
\end{equation}
This expression demonstrates that a fit to the self-energy on Matsubara frequencies is able to 
correctly capture the non-Fermi liquid temperature dependence of the scattering rate. Note that 
for this to hold, it is crucial that $\omega/T$ scaling indeed applies. This derivation has been performed for 
simplicity for a linear fit, but extends to a higher order polynomial fit. 

The low-$T$ analysis of $Z_{\text{linfit}}$ proceeds along slightly different lines. Indeed, direct substitution 
of the $\omega/T$ scaling form into the equation above would lead to an ultraviolet divergent 
integral when $\alpha>1$. In that case, the limit $T\rightarrow 0$ can directly be taken to yield:
\begin{equation}
\alpha > 1\,\,\,:\,\,\,Z_{\text{linfit}}|_{T\rightarrow 0}\,=\,
\left[1+\int d\omega \frac{\sigma(\omega,T=0)}{\omega^2}\right]^{-1}.
\end{equation}
This is indeed the exact $T=0$ expression of the quasiparticle weight resulting from the spectral representation. 
Note that, for $\sigma(\omega)\sim \omega^\alpha$ with $\alpha>1$ at low frequency, the above integral converges in the infrared, 
and that in that case there are coherent quasiparticles, with a finite spectral weight and a 
scattering rate $\sim T^\alpha$ smaller than their energy. 
In the opposite case $\alpha <1$, the $\omega/T$ scaling function must be used which yields 
the low-$T$ behaviour:
\begin{equation}\nonumber
Z_{\text{linfit}}
\sim\left[1+\frac{1}{T^{1-\alpha}}\times\int dx \phi(x) \frac{x^2-3\pi^2}{(\pi^2+x^2)(9\pi^2+x^2)}\right]^{-1}.
\end{equation}
At low-$T$, $Z_{\text{linfit}}$ vanishes at $T^{1-\alpha}$ indicating the breakdown of the quasiparticle 
concept. 
For $\alpha=1$ a very slow logarithmic vanishing of $Z$ is expected. 

In conclusion, this analysis demonstrates that when $\omega/T$ scaling applies together with a 
simple power-law behaviour, a fit to the 
self-energy over Matsubara frequencies is able to pickup the correct $T$-dependence of the 
scattering rate and quasiparticle weight in both the Fermi liquid and non-Fermi liquid case, in spite 
of the fact that the frequency dependence of the self-energy for $\omega < T$ is inaccessible from 
Matsubara frequencies. 
$\omega/T$ scaling typically applies in the quantum critical regime associated with 
quantum critical points (QCP) controlled by a strong coupling fixed point~\cite{Sachdev1999}. 
In this work, the QCP is the one associated with the disappearance of antiferromagnetism at $U=0$, 
and hence it is not clear whether $\omega/T$ scaling applies in the metallic regime. Logarithmic 
violations may be expected, for example.
\section{Self-energy from second order perturbation theory}
\label{app:2pt}
For further reference, and to facilitate the discussion of the consequences of perfect nesting in Sec.~\ref{sec:nesting}, here we show the results of second order perturbation theory (2PT) for the self-energy.
\subsection{Real frequencies}
We first consider the self-energy in 2PT on the {\it real frequency} axis (see also \cite{Lemay2000, Rohe2020}). Its functional form can be gained by 
careful analytic calculations performed in Sec.~4.4 of the PhD thesis of F. Lemay \cite{Lemay2000} that we summarize in this subsection.
We start by considering the expression on the Matsubara axis
\begin{equation}
\text{Im } \Sigma(\mathbf{k},i\omega_n)=U^2T \sum\limits_{\mathbf{q},i\Omega_n}G_0(\mathbf{k}+\mathbf{q},i\omega_n+i\Omega_n)\chi_0(\mathbf{q},i\Omega_n),
\label{eqn:2pt_iwn}
\end{equation}
with the (non-interacting) bubble
\begin{equation}
    \chi_0(\mathbf{q},i\Omega_n)=-T\sum\limits_{\mathbf{k}',i\omega_{n'}}G_0(\mathbf{k}'+\mathbf{q},i\omega_{n'}+i\Omega_n)G_0(\mathbf{k}',i\omega_{n'}),
    \label{eqn:lindhard}
\end{equation}
and $G_0$ being the non-interacting Green function.

\subsubsection{Analytical considerations}
Analytic continuation of Eq.~(\ref{eqn:2pt_iwn}) gives
\begin{eqnarray}
\text{Re }\Sigma(\mathbf{k},\omega)&=&\frac{U^2}{8\pi^3}P\int d^2q\int d\omega'\text{Im }\chi_0(\mathbf{q},\omega')\\
&\times& \frac{n_{\text{B}}(\omega')+n_{\text{F}}(\xi^0_{
\mathbf{k}+\mathbf{q}})}{\omega+\omega'-\xi^0_{
\mathbf{k}+\mathbf{q}}},\nonumber\\
\text{Im }\Sigma(\mathbf{k},\omega)&=&-\frac{U^2}{8\pi^2}\int d^2q\int d\omega'\text{Im }\chi_0(\mathbf{q},\omega')\label{eqn:2PT_real_im}\\
&\times& \left[n_{\text{B}}(\omega')+n_{\text{F}}(\xi^0_{
\mathbf{k}+\mathbf{q}})\right]\delta(\omega+\omega'-\xi^0_{
\mathbf{k}+\mathbf{q}}),\nonumber
\end{eqnarray}
where $n_{\text{F,B}}$ denote Fermi- and Bose distribution functions, respectively, $P$ is the principal value and $\xi_{\mathbf{k}}^{0}=\varepsilon_\mathbf{k}-\mu$. These expression show that at frequencies lower than the temperature  $|\omega|\!\ll\!\pi T$ (and, therefore, smaller than the first Matsubara frequency), the self-energy takes a clear non-Fermi-liquid form for Fermi-surface vectors $\mathbf{k}_{\text{F}}$ (see also Sec.~\ref{subsec:insights_metal}):
\begin{equation}
\text{Re }\Sigma(\mathbf{k}_{\text{F}}, \omega)  \overset{|\omega| \ll \pi T}{\sim} \begin{cases} 
      -\omega\ln{|\omega|},& \mathbf{k}_{\text{F}} \neq \mathbf{k}_{\text{N}}\\
      \text{sgn}(\omega)\sqrt{|\omega|}\,& \mathbf{k}_{\text{F}} = \mathbf{k}_{\text{N}}
   \end{cases}
\label{eqn:nesting_re},
\end{equation}
and
\begin{equation}
    \text{Im }\Sigma(\mathbf{k}_{\text{F}}, \omega) - \text{Im }\Sigma(\mathbf{k}_{\text{F}}, 0) \overset{|\omega| \ll \pi T}{\sim} \begin{cases} 
      |\omega|,& \mathbf{k}_{\text{F}} \neq \mathbf{k}_{\text{N}}\\
      \sqrt{|\omega|}\,& \mathbf{k}_{\text{F}} = \mathbf{k}_{\text{N}}
   \end{cases},
\label{eqn:nesting_im}
\end{equation}
with $\mathbf{k}_{\text{N}}\!=\!(\pi/2,\pi/2)$.

The singular behavior of the self-energy at low frequency $\left|\omega\right|\ll\pi T$
comes from the singular behavior of the imaginary part of the susceptibility%
\begin{equation}
\text{Im }\chi_{0}\left(  \mathbf{q},\omega\right)  \sim-\frac{\omega}{T}%
\ln\left\vert \frac{\left\vert \omega\right\vert -\omega_{s}\left(
\mathbf{q}\right)  }{\Delta^{2}}\right\vert\label{eqn:2PT_chi0}
\end{equation}
where $\Delta\sim\pi T$ and
\begin{equation}
\omega_{s}\left(  \mathbf{q}\right)  =4\left\vert \sin\left(  \left\vert
q_{x}\right\vert /2\right)  -\sin\left(  \left\vert q_{y}\right\vert
/2\right)  \right\vert .
\end{equation}
The origin of these singularities, that appear at small frequency $\omega$ when
$\mathbf{q}$\textbf{ }is near the diagonal, can be understood roughly from the
fact that for any external momentum $\mathbf{q}$ located near the diagonal of
the Brillouin zone, there are electrons near the half-filled diamond Fermi
surface that can be scattered to electrons also near the Fermi surface.

More precisely, suppose that an electron below the Fermi surface at $\mathbf{k-q/}%
2$ is excited above the Fermi surface at $\mathbf{k+q/}2$ such that
$\omega=\xi_{\mathbf{k+q/}2}^{0}-\xi_{\mathbf{k-q/}2}^{0}$. All internal
momenta $\mathbf{k}$ in the bubble must be considered, while $\omega$ and
$\mathbf{q}$ are fixed. But the velocity of the particle-hole excitation
$\nabla_{\mathbf{k}}\left(  \xi_{\mathbf{k+q/}2}^{0}-\xi_{\mathbf{k-q/}2}%
^{0}\right)  $ vanishes at $\mathbf{k}=\left(  \pm\pi/2,\pm\pi/2\right)  $ for
any value of $\mathbf{q.}$\textbf{ }That leads to van Hove-like singularities
in the particle-hole excitations. The integral over $\mathbf{k}$ gives
logarithmic contributions to $\text{Im }\chi_{0}\left(  \mathbf{q}%
,\omega\right)  $ that are finite, except for $\omega=\omega_{s}\left(
\mathbf{q}\right)  $ where they diverge when $\omega_{s}\left(  \mathbf{q}%
\right)  \ $is different from zero. Note that the singular frequency
$\omega_{s}\left(  \mathbf{q}\right)  $ vanishes when $\mathbf{q}$ is on the
diagonal.

These singularities influence the low frequency behavior of the imaginary part
of the self-energy $\text{Im }\Sigma(  \mathbf{k}_{F},\omega)
$ as follows. We are considering now the imaginary part in Eq.~(\ref{eqn:2PT_real_im}) and integrate over the $\delta$-distribution to arrive at
\begin{eqnarray}
\text{Im }\Sigma(\mathbf{k},\omega)&=&-\frac{U^2}{8\pi^2}\int d^2q\int d\omega'\text{Im }\chi_0(\mathbf{q},\xi^0_{
\mathbf{k}+\mathbf{q}/2}-\omega)\nonumber\\
&\times& \left[n_{\text{B}}(\xi^0_{
\mathbf{k}+\mathbf{q}/2}-\omega)+n_{\text{F}}(\xi^0_{
\mathbf{k}+\mathbf{q}/2})\right],
\end{eqnarray}
For the sake of detecting non-Fermi liquid behavior, we are interested in the regime where the inequality
$\left|\omega\right|\ll\pi T$ is satisfied.  In that regime, it is the classical limit of
the Bose function that dominates and singularities come only from integrals of
the logarithms in Eq.~(\ref{eqn:2PT_chi0}), whose arguments then are $\vert \vert
\xi_{\mathbf{k+q/2}}^{0}-\omega\vert -\omega_{s}(  \mathbf{q}%
)\vert .$

\begin{figure}[b!]
\centering
\includegraphics[width=0.4\textwidth,angle=0]{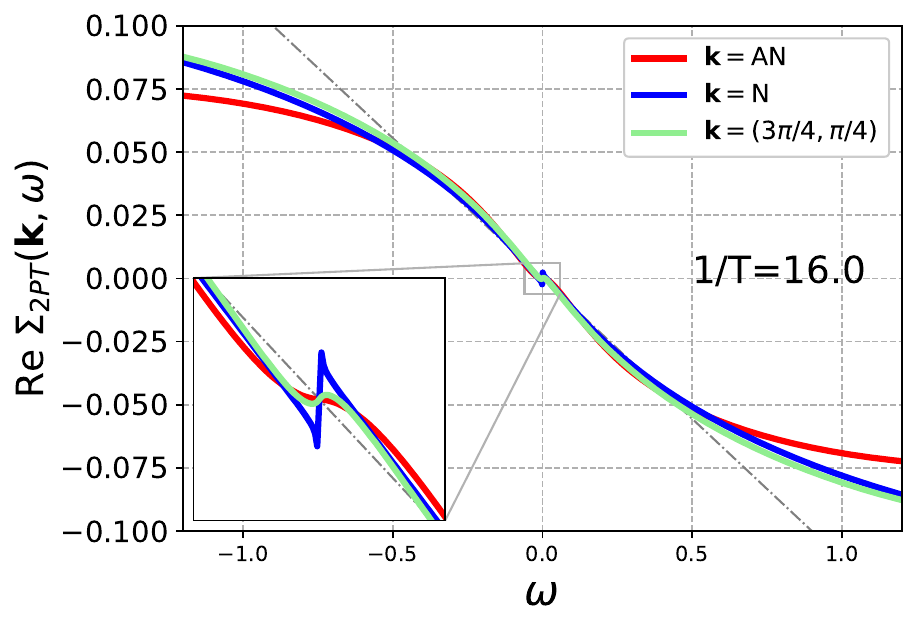}
\includegraphics[width=0.4\textwidth,angle=0]{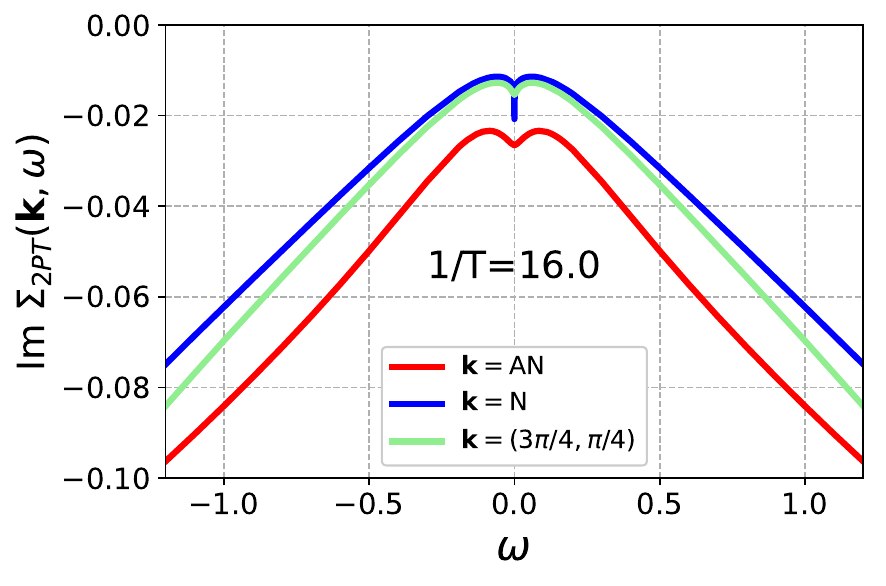}
               \includegraphics[width=0.42\textwidth,angle=0]{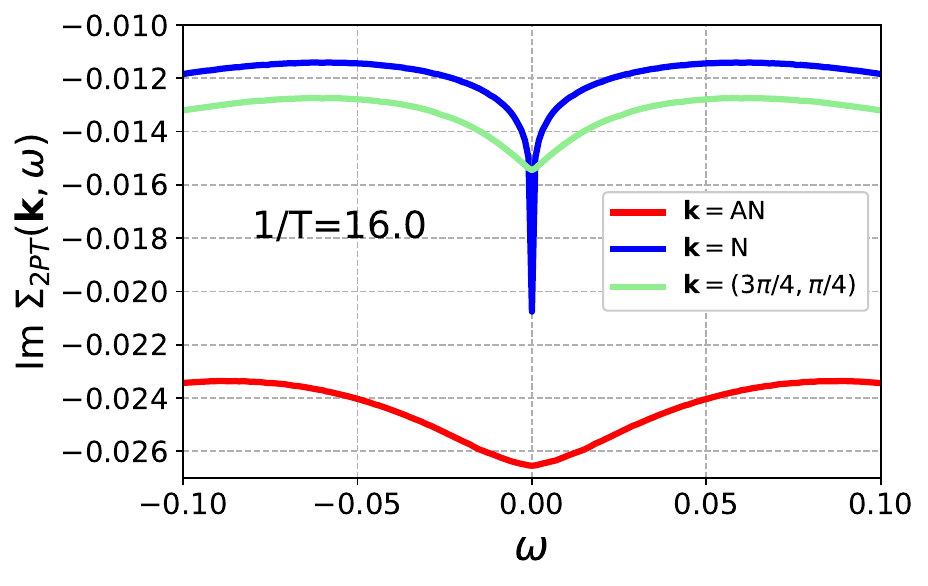}

      \caption{\label{fig:2pt_real}(Color online.) Upper panels: real and imaginary parts of the self-energy from 2PT at $1/T=16$ for antinode (red), node (blue) and the generic point $\mathbf{k}\!=\!(3\pi/4,\pi/4)$ [light-green] on the real frequency axis. The dotted-dashed line shows the  respective result from a Fermi-liquid fit from the Matsubara axis.}
\end{figure}

We take the integral over $q_{x}$ to run over the Brillouin zone, then the
integral over $q_{y}$ is restricted by the Bose function to a range near the
diagonals where the inequality $\left\vert \xi_{\mathbf{k+q/}2}^{0}\right\vert
\ll\pi T$ is satisfied. This determines the limits of integration over $q_{y}$
to be of order $\delta q_{y}\sim$ $\pi T/v_{F}$ away from the diagonal. The
argument of the logarithm $\left\vert \left\vert \xi_{\mathbf{k+q/}2}%
^{0}-\omega\right\vert -\omega_{s}\left(  \mathbf{q}\right)  \right\vert $,
when expanded near the diagonal, can be approximated by
\begin{equation}
-\omega+(  2\sin(k_{y}-q_{x})\pm2\cos(  q_{x}/2)  )
\delta q_{y}\equiv-\omega+\alpha(  k_{y},q_{x})  \delta q_{y}.
\end{equation}
The integral over $q_{y}$ can be done, leading to other logarithms that depend
on $\alpha(  k_{y},q_{x})  $. The singular part of the remaining
$q_{x}$ integral comes from regions where $\alpha(  k_{y},q_{x})  $
is small. The final result is obtained by expanding $\alpha(  k_{y}%
,q_{x})  $ around values of $q_{x}$ where $\alpha(  k_{y}%
,q_{x})  $ vanishes. In the Taylor expansion of $\alpha(
k_{y},q_{x})  $ there are terms linear in $\delta q_{x}$ that give the
final $-\vert \omega\vert $ result for $\text{Im }\Sigma(\mathbf{k}=(  k_{x},k_{y}),\omega)  $ everywhere, except when the linear term
in the expansion of $\alpha(  k_{y},q_{x})  $ vanishes, which
occurs for $\mathbf{k}_{F}=$ $(  \pm\pi/2,\pm\pi/2)  .$ There, the
leading term scales as $\delta q_{x}^{2},$ which modifies the integral over
$q_{x}$ and leads to $\text{Im }\Sigma(  k_{x},k_{y,}\omega)
$ proportional to $-\sqrt{\vert \omega\vert }.$      
The Fermi points $\mathbf{k}_{F}=$ $(  \pm\pi/2,\pm\pi/2)  $ seem
special because it is there that the energy  $\xi_{\mathbf{k+q/}2}^{0}%
-\xi_{\mathbf{k-q/}2}^{0}$ associated with particle-hole excitations can be
small for the largest range of values of $\mathbf{q}$ near the diagonal.

\subsubsection{Numerical evaluation}

In Fig.~\ref{fig:2pt_real} we show the numerical evaluation of these expressions for the real (first panel) and imaginary part (second and third panel) for $U=2t$ and $1/T=16$, i.e. deep inside the metallic regime for the antinode (red), node (blue) and generic point $\mathbf{k}\!=\!(3\pi/4,\pi/4)$ (light-green). In both real and imaginary parts one can clearly see non-Fermi liquid behavior at low frequencies: at low frequencies, the slope of the real part is reversed for the node and the generic $\mathbf{k}$-point and non-linear also for the antinode. For comparison we also show (with the dotted-dashed line) the respective low frequency result from Fermi-liquid fit from the Matsubara axis (see later in this Appendix), which are oblivious of the features at $|\omega|\!\ll\!\pi T$.

\begin{figure}[t!]
\centering
               \includegraphics[width=0.459\textwidth,angle=0]{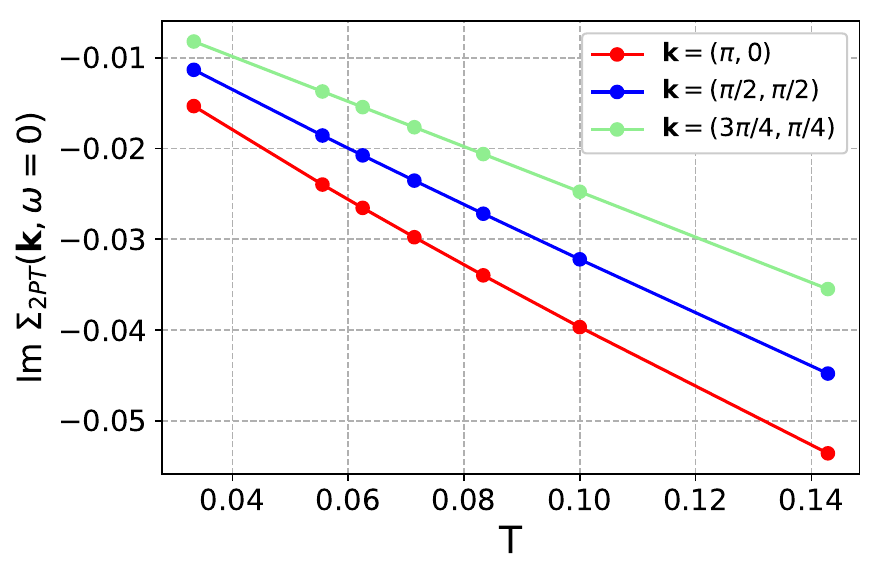}
               \includegraphics[width=0.459\textwidth,angle=0]{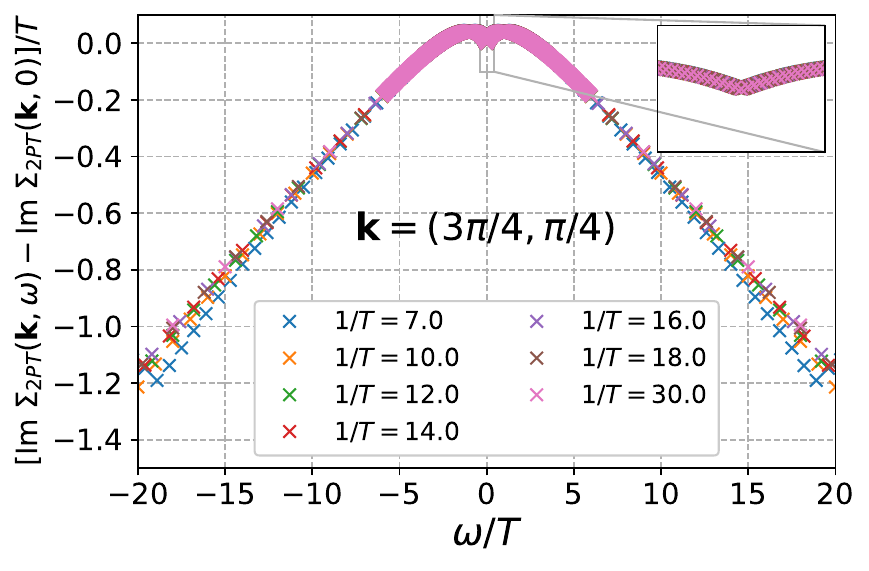}
      \caption{\label{fig:2pt_real_omega_0}(Color online.) Upper panel: The imaginary part of the self-energy on the real axis at zero frequency and various Fermi surface points calculated in second order perturbation theory as a function of temperature. Lower panel: data collapse from $\omega/T$-scaling at the generic momentum point for several temperatures in the metallic regime. The inset shows a zoom to low frequencies.}
\end{figure}

\begin{figure*}[t!]     
\includegraphics[width=0.24\textwidth,angle=0]{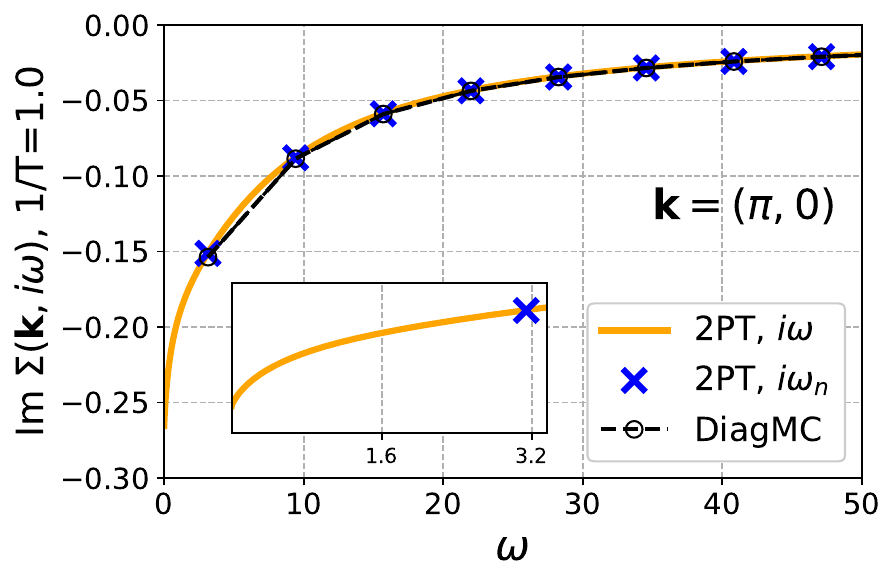}
\includegraphics[width=0.24\textwidth,angle=0]{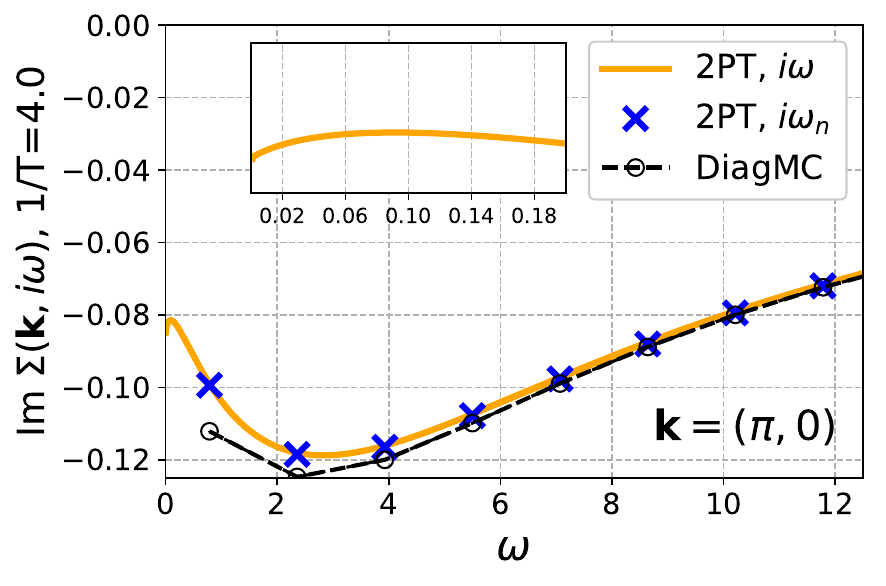}
\includegraphics[width=0.24\textwidth,angle=0]{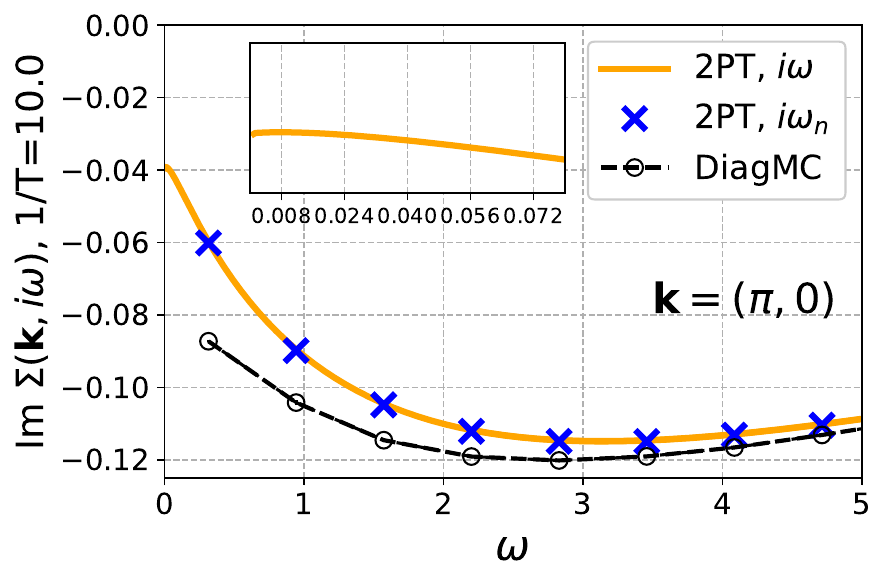}
\includegraphics[width=0.24\textwidth,angle=0]{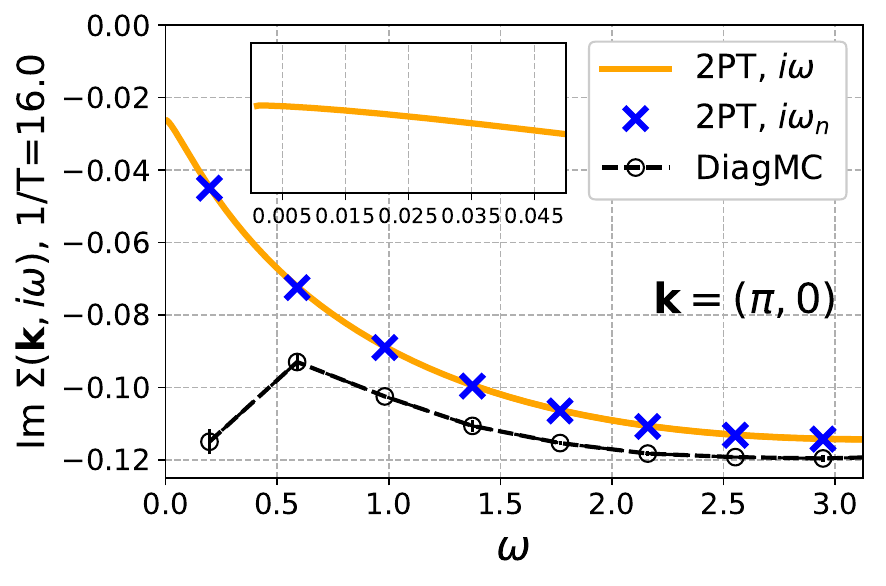} \includegraphics[width=0.24\textwidth,angle=0]{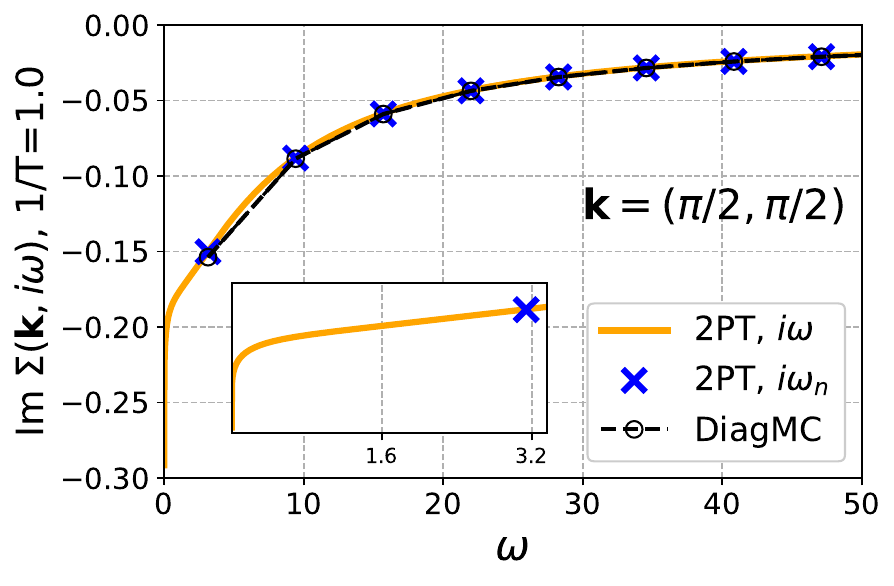}
\includegraphics[width=0.24\textwidth,angle=0]{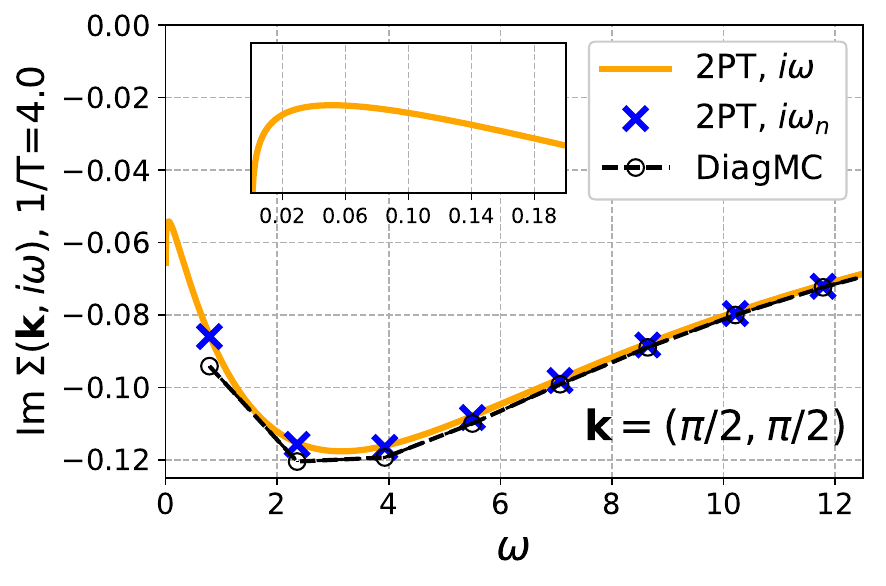}
\includegraphics[width=0.24\textwidth,angle=0]{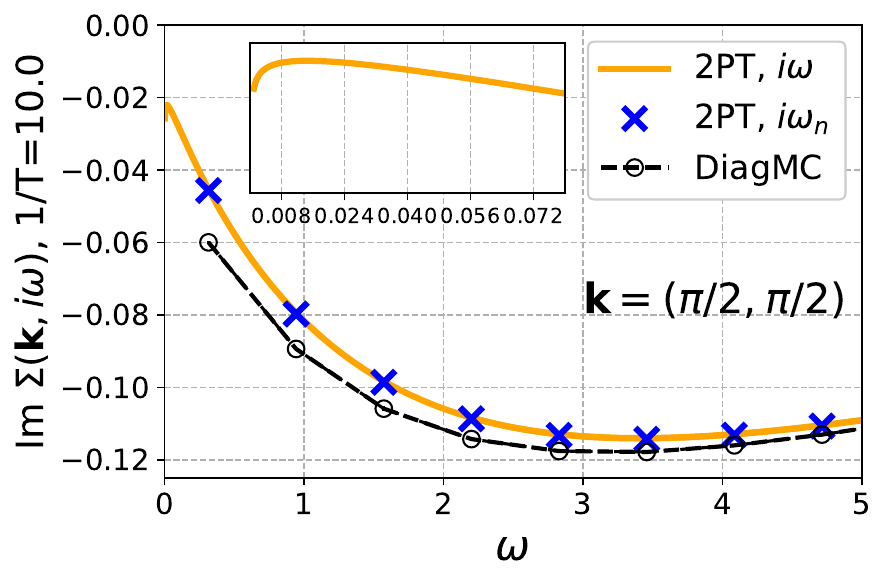}
\includegraphics[width=0.24\textwidth,angle=0]{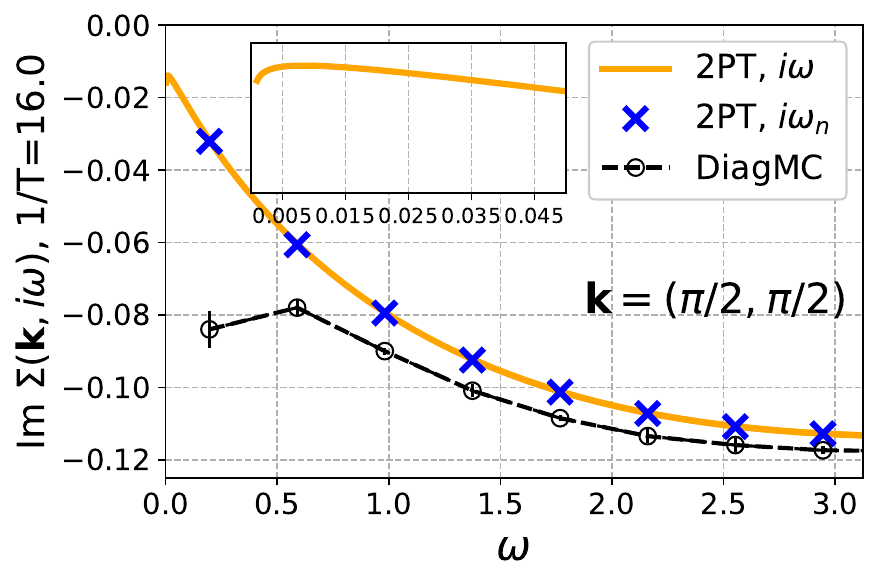}
\includegraphics[width=0.24\textwidth,angle=0]{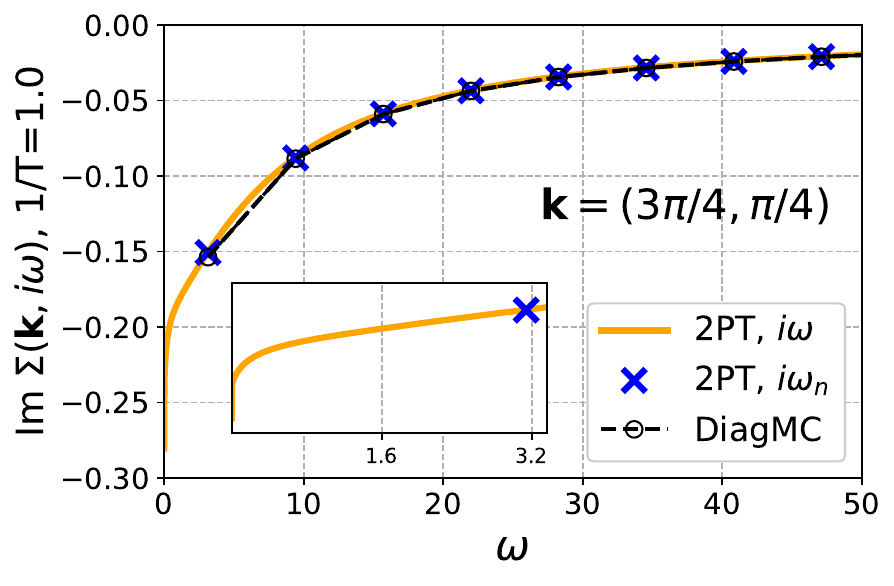}
\includegraphics[width=0.24\textwidth,angle=0]{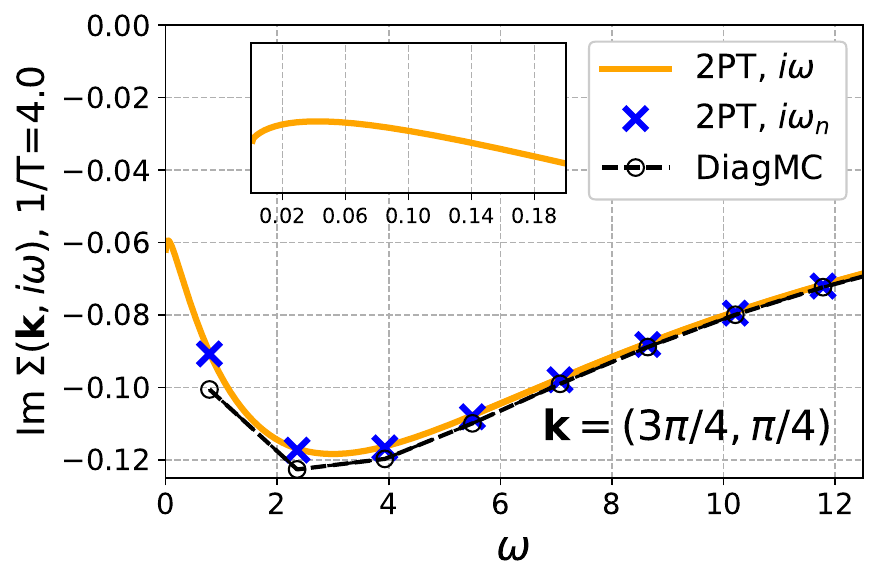}
\includegraphics[width=0.24\textwidth,angle=0]{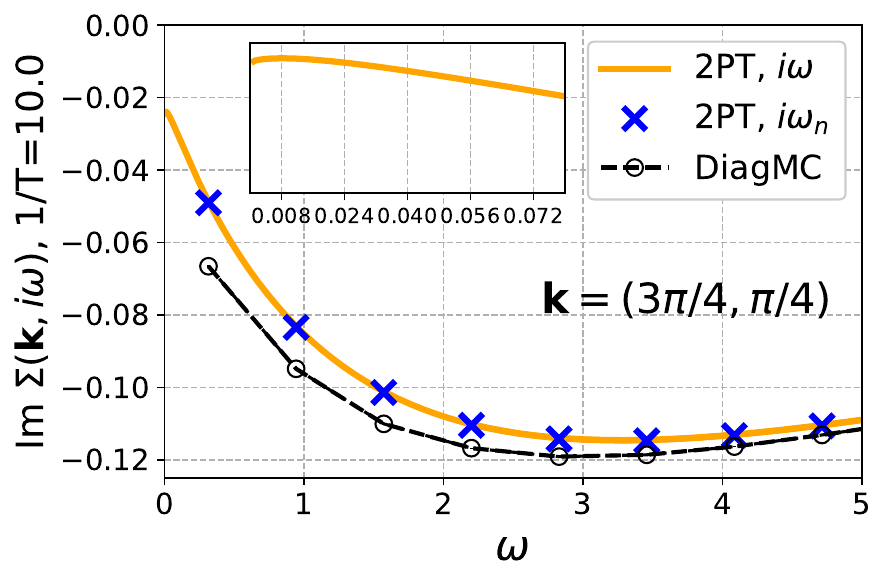}
\includegraphics[width=0.24\textwidth,angle=0]{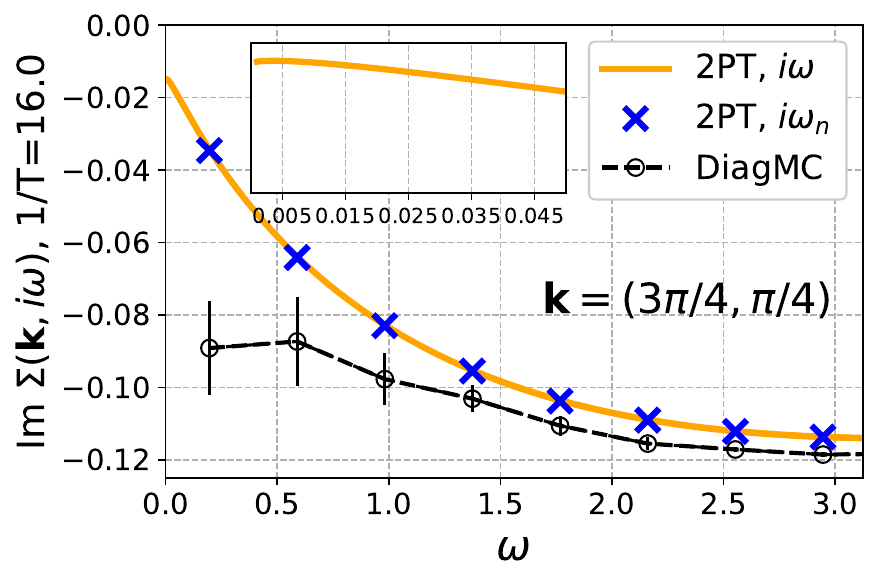}
\includegraphics[width=0.24\textwidth,angle=0]{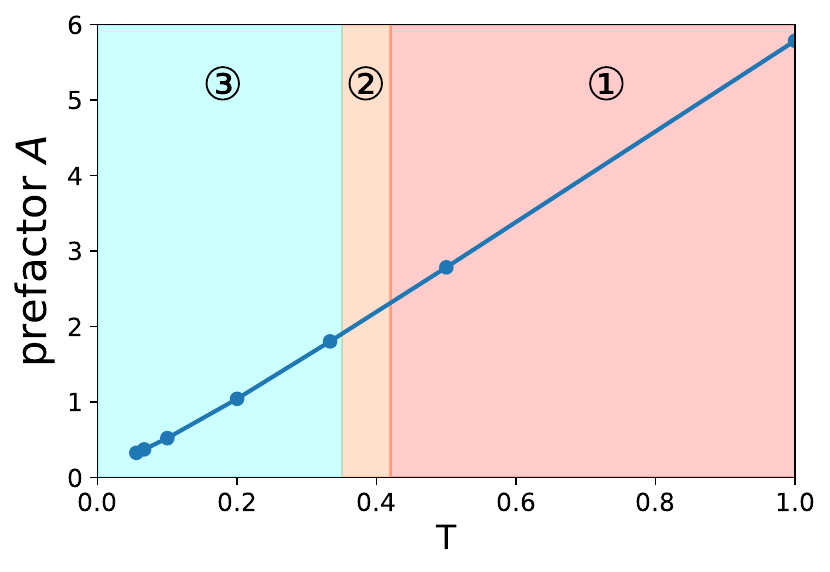}
\includegraphics[width=0.24\textwidth,angle=0]{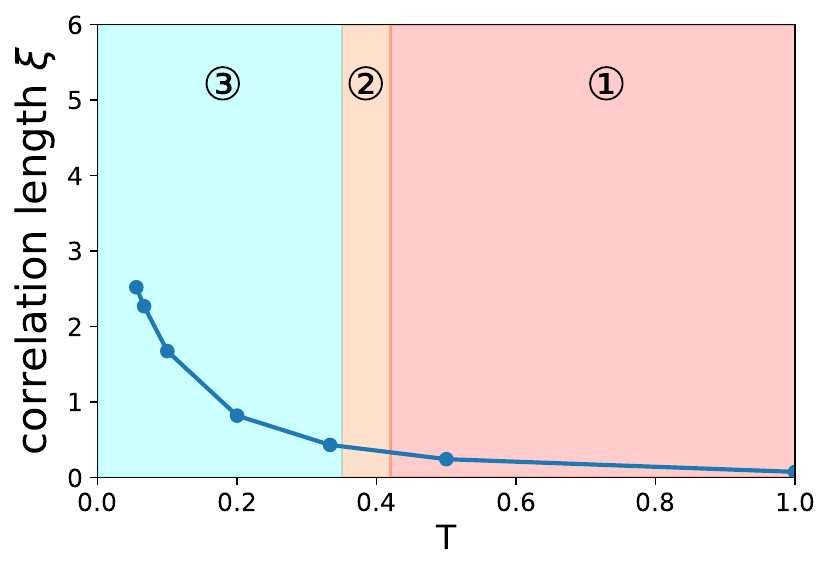}
\includegraphics[width=0.24\textwidth,angle=0]{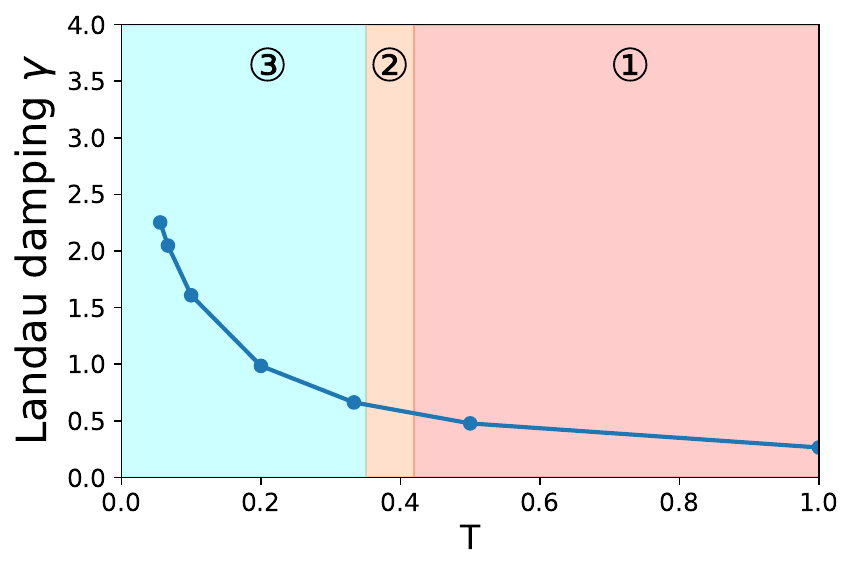}
\includegraphics[width=0.24\textwidth,angle=0]{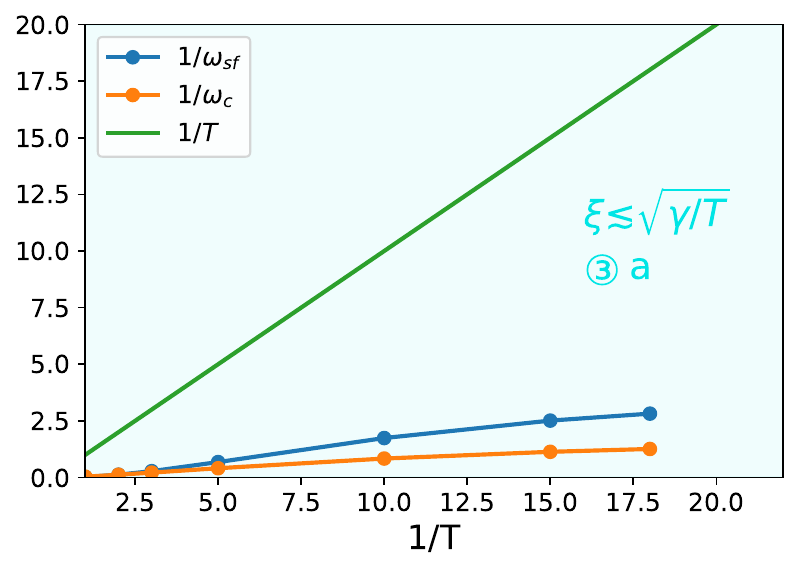}
      \caption{\label{fig:2pt}(Color online.) 
Comparison of the imaginary part of the self-energy as a function of imaginary frequency for the antinode and node (first and second row, respectively) obtained by second order perturbation theory 2PT (orange solid lines, blue crosses mark the Matsubara frequencies) to the DiagMC benchmark (black open circles, dashed lines) for various temperatures ($1/T=1,4,10,16$ from left to right). The insets show a zoom of the lowest frequency data (the interval range of the y-axis is constant for all $\mathbf{k}$-points shown in order to ensure comparability). Third row: results for $\mathbf{k}\!=\!(3\pi/4,\pi/4)$. Lowest row: Ornstein-Zernike parameters (for the bubble) and energy scales for 2PT (obtained from a Matsubara fit) as functions of (inverse) temperature.}
\end{figure*}

Focusing on the imaginary part, one can observe that, despite their different magnitudes, the functional forms of the antinode and the generic point are very similar. The node, however, exhibits a completely different functional behavior, eventually resulting in $|\text{Im }\Sigma(\mathbf{k}=\text{N}, 0)|>|\text{Im }\Sigma(\mathbf{k}=(3\pi/4,\pi/4), 0)|$ for very small frequencies, and indeed compatible with a square root behavior.
Additionally to its frequency behavior we extracted the zero-frequency dependence on temperature, shown in the upper panel of Fig.~\ref{fig:2pt_real_omega_0}. Although $\log$-corrections are expected \cite{Lemay2000}, the dependence appears rather linear. Eventually, with our precise numerical data, we can numerically demonstrate that $\text{Im} \Sigma(\mathbf{k},\omega)$ indeed exhibits $\omega/T$ scaling as discussed in the main text, by plotting $\left(\text{Im} \Sigma(\mathbf{k},\omega)- \text{Im }\Sigma(\mathbf{k}, 0)\right)/T$ over $\omega/T$ (lower panel of Fig.~\ref{fig:2pt_real_omega_0}), which results in an excellent data collapse over a broad temperature (and frequency) range inside the metallic regime. However, as already stated in App.~\ref{app:fermi_liquid}, $\omega/T$ scaling typically applies in the quantum critical regime associated with 
quantum critical points (QCP) controlled by a strong coupling fixed point~\cite{Sachdev1999}. The QCP present is associated with the disappearance of antiferromagnetism at $U=0$, and hence it is not clear whether $\omega/T$ scaling generally applies (e.g. logarithmic corrections may be expected).

\subsection{Imaginary frequencies and comparison}
In order to compare our findings on the real axis with the DiagMC benchmark, we directly evaluate Eq.~(\ref{eqn:2pt_iwn}) on the {\it Matsubara} axis. We used meshes of $N_{i\omega}\!=\!200$  Matsubara frequencies and a linear momentum mesh of $N_{q}\!=\!N_{k}\!=\!200$ for the summations performed. Fig.~\ref{fig:2pt} shows the 2PT data (blue crosses) in comparison to the DiagMC benchmark (black circles and dashed line) for the antinode (uppermost row) and node (second row) as well as for the generic point $\mathbf{k}\!=\!(3\pi/4,\pi/4)$ [third row] for various temperatures. The benchmark result is described quantitatively by 2PT in the incoherent regime and fairly well inside the high temperature metallic regime. In particular, 2PT exhibits a clear momentum differentiation between node and antinode. However it fails in the description of the pseudogap, i.e. the regimes \textcircled{4} and \textcircled{5} are absent. This is particularly apparent comparing the emerging energy and length scales obtained from Ornstein-Zernike fits (last row, cf. discussion in Sec.~\ref{subsec:insights_metal} and Fig.~\ref{fig:sf_scales}).

However, as already discussed before, this may be quite hard to judge by looking at data from the Matsubara axis only. Therefore, we additionally computed the self-energy in 2PT on a {\it continuous} imaginary frequency grid (solid orange lines in Fig.~\ref{fig:2pt}, see insets for a zoom \footnote{For obtaining the data on the continuous imaginary frequency axis, in a first step, $\text{Im }\Sigma(\mathbf{k},\omega)$ has been calculated on the real axis. After retrieving $\text{Re }\Sigma(\mathbf{k},\omega)$ via the Kramers-Kronig relations, the Cauchy formula has been used for the forward continuation onto the imaginary axis. The calculation on the real axis profits from analytical treatments of the delta function which reduces the quadrature dimensions.}). One indeed observes that, for frequencies way smaller than the ones accessible by a Matsubara grid $|\omega|\!\ll\!\pi T$, the slope of the imaginary part of the self-energy changes, at least in this fully nested case, to being {\it positive}, yielding clear non-Fermi liquid behavior. This behavior is particularly emphasized at the nodal point (second row).

\begin{figure}[t!]
\centering
                \includegraphics[width=0.45\textwidth,angle=0]{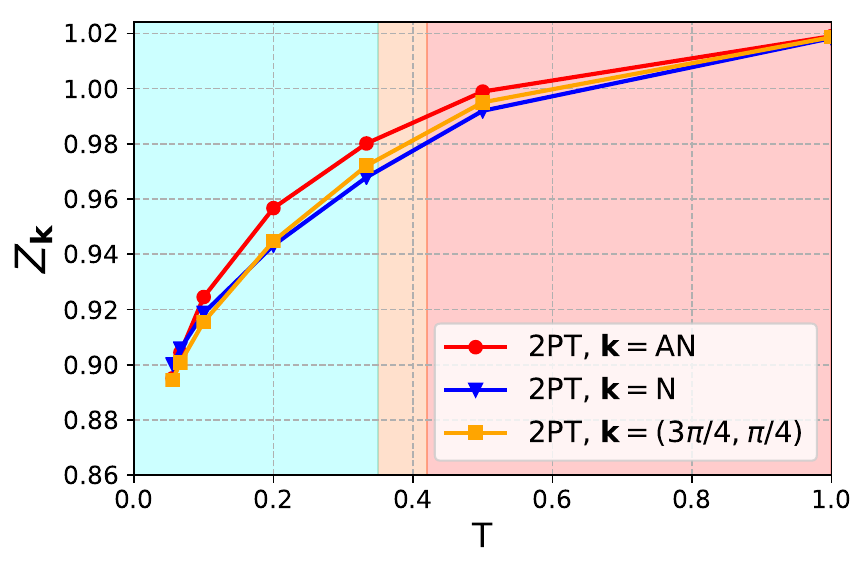}
                \includegraphics[width=0.45\textwidth,angle=0]{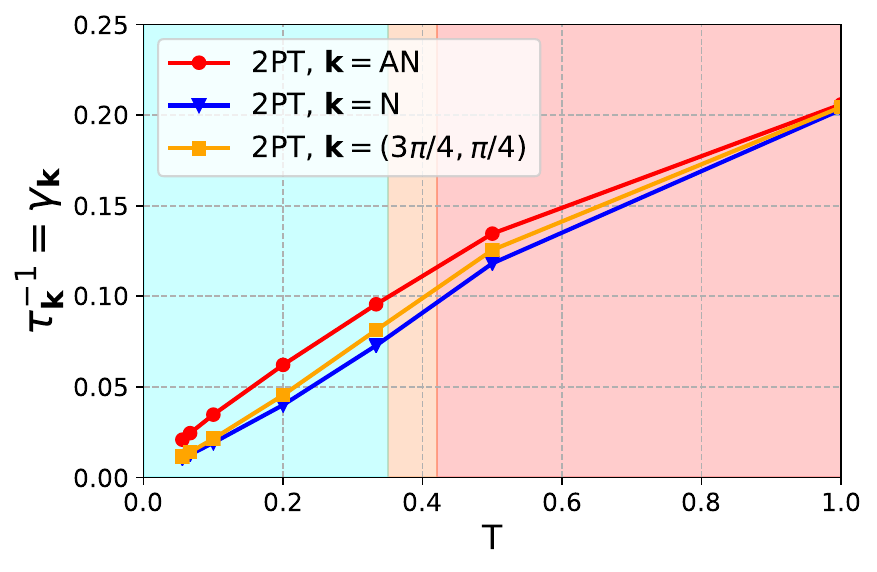}
                \includegraphics[width=0.45\textwidth,angle=0]{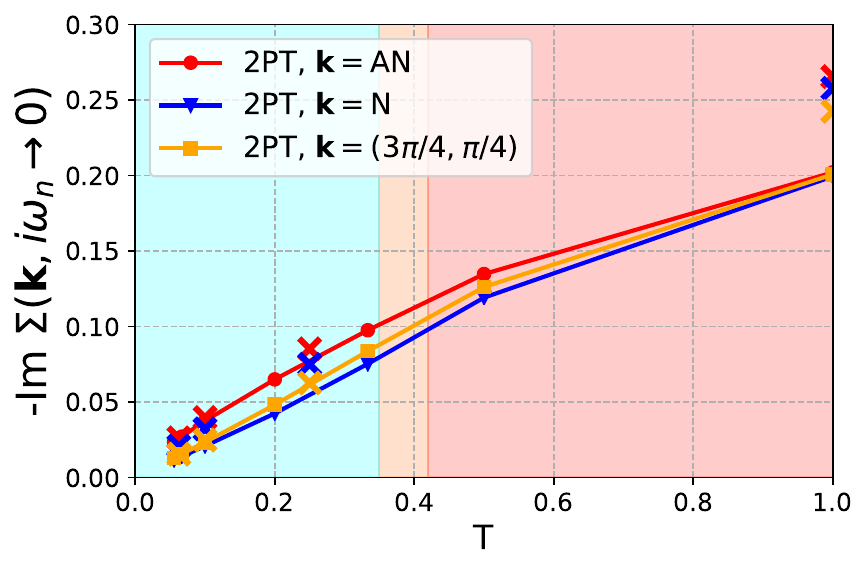}
      \caption{\label{fig:2pt_qp}(Color online.) Quasiparticle parameters $Z_{\mathbf{k}}$ (upper panel) and $\gamma_{\mathbf{k}}$ (middle panel) and the imaginary part of the self-energy extrapolated to zero frequency (lower panel) at the antinode (red circles), node (blue triangles) and $\mathbf{k}\!=\!(3\pi/4,\pi/4)$ [orange squares] in second order perturbation theory as a function of temperature. The values for the crosses in the lowest panel have been obtained directly from the real frequency axis.}
\end{figure}

Using a Fermi-liquid ansatz (see App.~\ref{app:fermi_liquid}), one can extract a momentum-dependent quasiparticle weight $Z_{\mathbf{k}}$ and inverse quasiparticle lifetime $\gamma_{\mathbf{k}}$ from the Matsubara axis (Fig.~\ref{fig:2pt_qp}). As already discussed in the main text, these parameters are only meaningful (in the strict sense of describing Landau quasiparticles) if the low-frequency behavior of the self-energy is the "correct" one, i.e. compatible with the one expected by a Fermi liquid (e.g. in exhibiting a negative slope and a long quasiparticle lifetime).
In order to estimate the quality of this plain Matsubara fit, we show the "offset" $-\text{Im }\Sigma(\mathbf{k},\omega\rightarrow 0)$ from both the real frequency data (crosses in the lowest panel of Fig.~\ref{fig:2pt_qp}) and the Matsubara fits. We notice that both methods of extraction are in a very good agreement inside the metallic regime, except at the nodal point, where a severe underestimation by the Matsubara data of $\sim 50\%$ is found. 

\section{Brief description and references of the presented methods}
\label{app:methods}
As one of the goals of this manuscript is a synopsis of a variety of different modern many-body techniques, in this Appendix, we will give a short overview 
on these. Of course, the Appendix cannot serve as an exhaustive review, but will contain the basic idea of each method as well as the necessary references for further 
information and, if necessary, computational details for obtaining the results presented in this paper.

\subsection{DiagMC}
\label{app:diagmc}
Diagrammatic Monte Carlo (DiagMC), first developed by B.~Svistunov and N.~Prokofiev \cite{Prokofev1998}, is based on the idea of stochastically sampling Feynman diagrams contributing to a perturbative expansion directly in the thermodynamic limit. In a broad sense, the Diagrammatic Monte Carlo approach consists of the choice of a diagrammatic expansion for a given observable, a numerical algorithm computing the corresponding expansion coefficients and, if necessary, the resummation techniques applied to the resulting perturbative series.

The physically most natural \cite{Abrikosov1975}, skeleton, expansions are potentially dangerous when applied to the two-dimensional fermionic Hubbard model as they can converge to wrong results in the vicinity of half-filling or whenever the local magnetic moment develops \cite{Kozik2015, Rossi2015, Gunnarsson2017, Kim2020b, Chalupa2020}. In contrast, bare expansions have a well-defined analytic structure and generally have a non-zero convergence radius in the two-dimensional fermionic Hubbard model. Such expansions can be constructed around an arbitrary starting point \cite{Wu2016, Rossi2016, chen2019combined, Rossi2020, Simkovic2020b}. The default choice \cite{VanHoucke2010,Kozik2010} is to expand around the non-interacting Hamiltonian with the chemical potential shifted by a constant ($\alpha=U/2$ at half-filling) \cite{Rubtsov2005}, which effectively eliminates all diagrams with Hartree-type insertions, leading to a much improved variance, which is what we adopt here.

The choice of algorithm used to compute the expansion coefficients determines how large expansion orders can be obtained with reasonable error bars. In this publication we employ the currently most efficient class of algorithms, based on the Connected Determinants Diagrammatic Monte Carlo (CDet) introduced for connected quantities by R.~Rossi \cite{Rossi2017} and generalized to the evaluation of one-particle irreducible self-energy diagrams in Refs.~\cite{Simkovic2020, Moutenet2018, Rossi2018}. The particular self-energy implementation used in this publication is $\Sigma$DDMC from Ref.~\cite{Simkovic2019} as it is very well-suited for the computation of specific $\mathrm{k}$-points directly in momentum-space. In its original formulation, CDet only works with bare diagrammatic expansions (consisting exclusively of interaction vertices and bare propagators $G_0(\mathrm{r},\tau)$), although a generalization to semi-bold \cite{Rossi2016} and bold \cite{VanHoucke2012,Deng2015} schemes has recently been developed \cite{Rossi2020,Simkovic2020b}. Contrary to previously developed Diagrammatic Monte Carlo algorithms \cite{Prokofev2008, Prokofev2007, VanHoucke2010, Kozik2010, VanHoucke2012, Deng2015, Simkovic2019b} which evaluate a single Feynman diagram at each Monte Carlo step, CDet computes a factorial number of relevant diagrams permuted over all internal vertices at exponential cost, scaling with the expansion order $m$ as $o(3^m)$ for connected quantities \cite{Rossi2017} and $o(m^2 3^m)$ for one-particle irreducible quantities \cite{Simkovic2020, Moutenet2018, Rossi2018} (or, alternatively, scaling as $o(m^2 2^m)$ and $o(m^4 2^m)$, respectively \cite{Koivisto2007}). This is achieved by grouping all possible diagram topologies of a particular diagram order into determinants and subsequently subtracting all disconnected, and in the case of the self-energy also all one-particle-reducible, diagrams from the sum by using a recursive formula. As a consequence of the exponential scaling, the total error of a convergent series scales polynomially with computational time \cite{Rossi2017b}. The statistical variance of the algorithm only allows for the computation of a finite number of expansion coefficients, typically $m\sim 10\!-\!12$ (compared to $m\sim 6\!-\!7$ for previous Diagrammatic Monte Carlo algorithms).

In challenging cases, such as low temperatures and/or high values of physical coupling, resummation techniques are needed to evaluate the series close to, or beyond the,  radius of convergence. This constitutes the only source of systematic error in this approach. A protocol for the resummation of extrapolation, which gives control over this systematic error, has been developed in Ref.~\cite{Simkovic2019}.

At half-filling, every other order is equal to zero for the double occupancy (even) as well as for the self-energy evaluated at every $\mathrm{k}$-point on the non-interacting Fermi-surface (odd). A consequence of particle-hole symmetry at half-filling is that a single calculation of the expansion coefficients at a given temperature yields results for all values of $U$, provided the series is resummable (away from half-filling this is still the case, however the evaluated density $n$ also changes as a function of interaction strength $U$). For all presented observables, the limiting factor to the applicability of this approach at high enough interactions and low enough temperatures is the presence of singularities in the vicinity of the positive real axis in the complex plane of interaction strength $U$, which render the resummation of the series prohibitively hard. Despite the fact that, in this publication, series have only been evaluated within their radius of convergence, resummation by means of Pad\'e and Dlog-Pad\'e approximants  \cite{Simkovic2019, Brezinski1996, Gonnet2013} has been performed in order to accelerate the convergence of series whilst controlling the systematic error of the extrapolations.
In the half-filled two-dimensional Hubbard model, this approach has previously allowed for numerically exact results that shed light on the metallic to quasi-antiferromagnet crossover as described by the self-energy \cite{Simkovic2020} as well as various other thermodynamic observables \cite{Kim2020, Lenihan2021}.

\subsection{DQMC}
\label{sec:dqmc}

\begin{figure}[t!]
                \includegraphics[width=0.45\textwidth,angle=0]{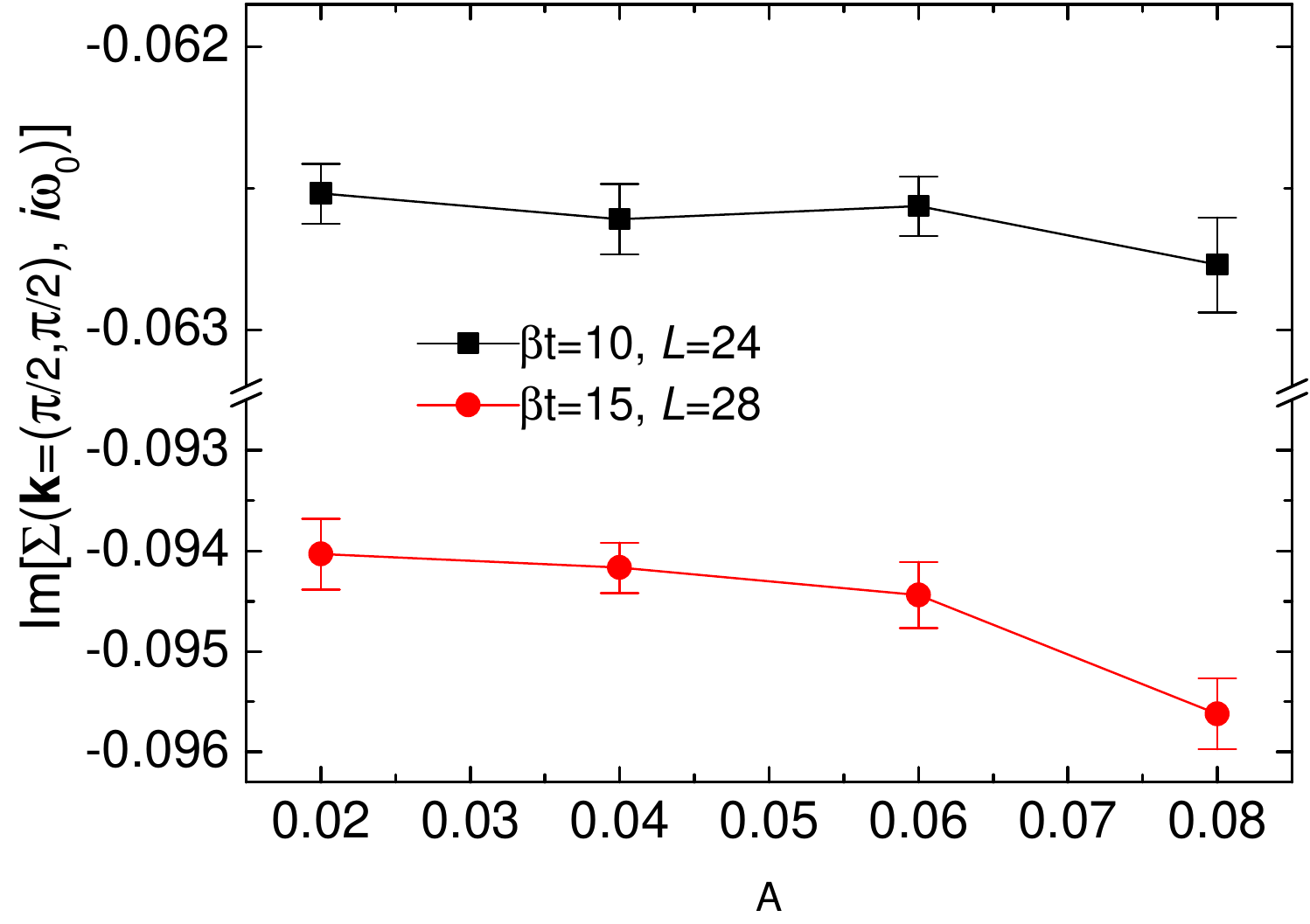}
                \includegraphics[width=0.45\textwidth,angle=0]{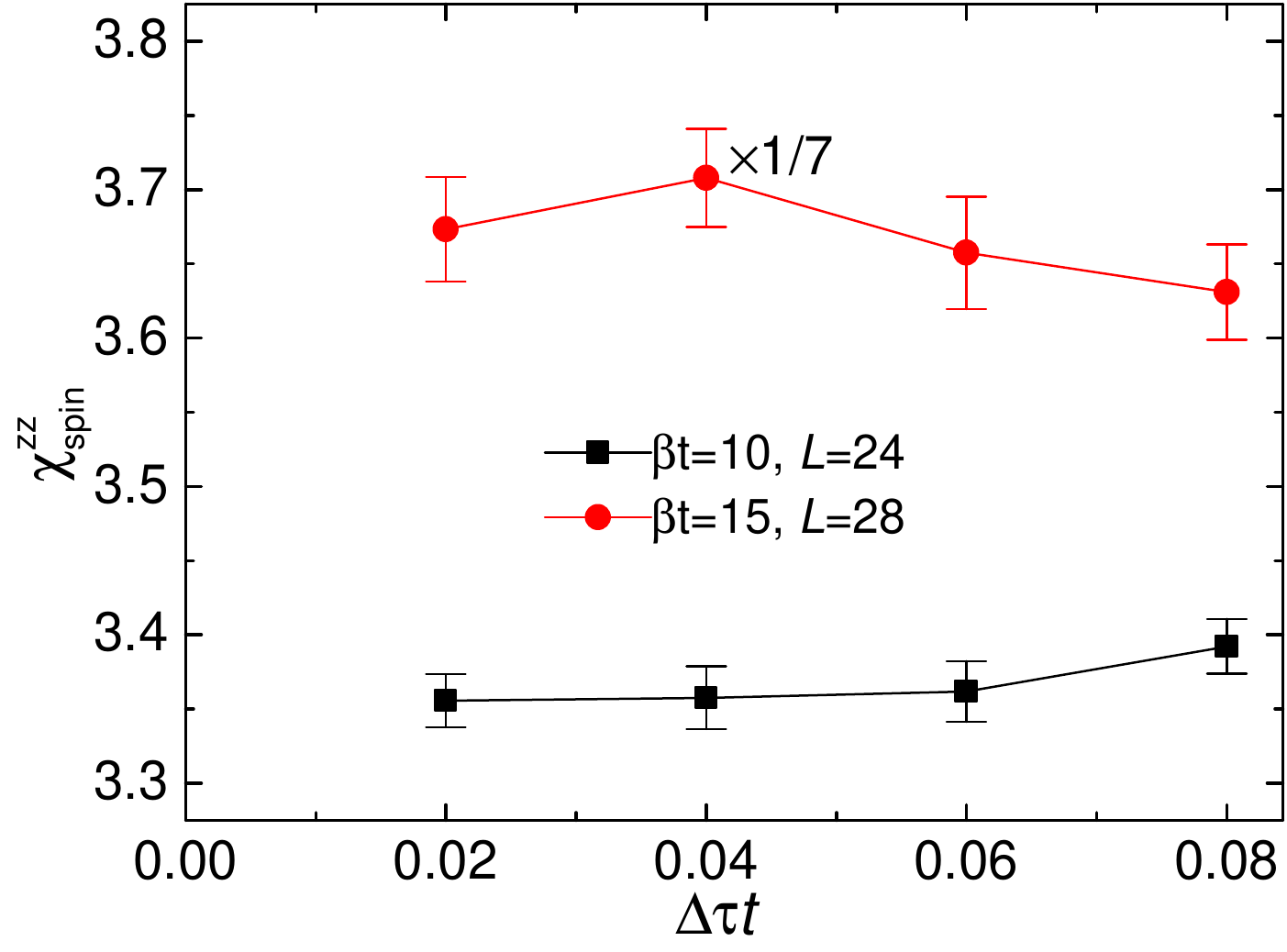}
      \caption{\label{fig:trotter}(Color online.) DQMC: Trotter extrapolation for the lowest Matsubara frequency of the imaginary part of the self-energy at the node (upper panel) and the magnetic susceptibility at the antiferromagnetic wave vector (lower panel).}
\end{figure}

The determinantal Quantum Monte Carlo (DQMC) algorithm, formulated by Blankenbecler, Scalapino, and Sugar~\cite{Blankenbecler1981}, is a controlled method and has been widely applied to finite-temperature simulations of correlated fermion systems. Its basic methodology is to transform two-body interactions into free fermions coupled with auxiliary fields and then sample the fields to compute fermionic observables. To achieve this, a Trotter decomposition (such as $e^{-\Delta\tau\hat{H}}=e^{-\Delta\tau\hat{H}_0/2}e^{-\Delta\tau\hat{H}_I}e^{-\Delta\tau\hat{H}_0/2}+\mathcal{O}[(\Delta\tau)^3]$ is used in this work, where $\hat{H}_0$ and $\hat{H}_1$ are the tight-binding and interacting parts of the model Hamiltonian), and a Hubbard-Stratonovich (HS) transformation (the discrete spin decomposition~\cite{Hirsch1983} for on-site Hubbard interaction is used in this work) are applied within the discretization for the inverse temperature as $\beta=M\Delta\tau$. When convergence $\Delta\tau \rightarrow 0$ can be established, the method is numerically exact.

The systematic error from finite $\Delta\tau$ is controllable and can be eliminated by extrapolating simulations with several different $\Delta\tau$ values.
In Fig.~\ref{fig:trotter}, we present representative results of systematic Trotter errors in DQMC simulations. Two systems, $L=24$ with $\beta t=10$ and $L=28$ with $\beta t=15$, are studied. We concentrate on the self-energy at $\mathbf{k}=(\pi/2,\pi/2)$ (upper panel) and the spin susceptibility (lower panel). The results for both parameters explicitly show the convergences to the $\Delta\tau=0$ limit for $\Delta\tau t\le 0.06$ within our statistical uncertainty. This means that our choice of $\Delta\tau t\le 0.02$ is reliable. 

Further details about DQMC algorithm can be found in several review papers~\cite{AssaadEvertz2008,Chang2015}. We have also implemented our most recent improvements~\cite{Yuan19a,Yuan19b} of this method in this work. The minus sign problem is absent for the Hubbard model at half filling studied in this work and DQMC method can reach large system sizes. For the calculation of dynamical quantities, the imaginary-time single-particle Green function and correlation functions are measured in DQMC simulation, and then the imaginary-frequency observables are obtained through the Fourier transformation. The Dyson equation was then applied to compute the self-energy. For the $U/t=2$ case, convergence to thermodynamic limit of all the physical observables is observed for up to $L=32$ for $\beta t\le10$. For lower temperatures, we perform finite-size scalings using second-order polynomials with constraints of monotonic behaviors to reach the thermodynamic limit. We typically use between $10^4$ to $3\times 10^5$ measurement samples for systems with linear system size from $L=20$ to $L=48$. In DQMC the number of Matsubara frequencies is determined by the number of imaginary time slices $M$, which in turn depends on the Trotter step, chosen here to be $\Delta\tau=0.02$ as mentioned earlier.

\subsection{MFT}
\label{app:mft}
The calculation of the static mean-field theory (MFT) susceptibility is analytically equivalent to the (non-selfconsistent) random phase approximation (RPA). The idea of the RPA, originally introduced in \cite{Bohm1951}, is building ladders for the susceptibilities with the bare interaction $U$ in the physically relevant channel, utilizing the non-interacting susceptibility Eq.~(\ref{eqn:lindhard}).
The fermionic propagator, thus, is the non-interacting Green function $G_{0}(\mathbf{k},i\omega_{n})$. The interacting susceptibility is  obtained via:
\begin{equation}
	\chi^\text{RPA}_{\text{sp}}(\mathbf{q},i\Omega_{n})=\frac{2\chi_{0}(\mathbf{q},i\Omega_{n})}{1-U\chi_{0}(\mathbf{q},i\Omega_{n})}.
\end{equation}
We used momentum grid with a maximum linear mesh sizes of $N_{q}=128$ and $N_{k}=128$ and the number of fermionic Matsubara frequencies being $N_{i\omega}=1000$.

\subsection{DMFT}
\label{app:dmft}
The dynamical mean field theory (DMFT) has become one of the standard techniques for tackling strongly correlated systems over the past decades. Its basic idea consists in mapping the full lattice Hamiltonian Eq.~(\ref{eqn:Hubbard}) onto a self-consistently determined single site Anderson impurity model. This procedure is exact in infinite dimensions, but represents an approximation in finite dimensions due to its neglection of  spatial correlations. Hence, the self-energy in DMFT is purely local. For further reading about the details of the algorithm we recommend the seminal papers \cite{Georges1992a,Metzner1989} and the review \cite{Georges1996}. The self-energy and magnetization data presented in this paper have been produced using a state of the art continuous time quantum Monte Carlo impurity solver in its interaction expansion (CT-INT \cite{Rubtsov2004,Rubtsov2005,Gull2011a}), which is, like the DMFT self-consistency scheme, entirely implemented in the TRIQS framework \cite{TRIQS}. For the calculation of the (magnetic) susceptibility and correlation length we used data from both (i) an exact diagonalization (with four bath sites) impurity solver and the implementation of the Bethe-Salpeter equations presented in \cite{Rohringer2018} and (ii) the CT-INT solver and tprf framework of TRIQS \cite{tprf}, carefully crosschecking that they obtain the same results. As the magnetic correlations (and correlation lengths) grow exponentially reaching low temperatures, we used a very dense momentum grid for these data with a maximum linear mesh sizes of $N_{q}=200$ and $N_{k}=200$ and the number of fermionic Matsubara frequencies being $N_{i\omega}=160$. In the case of the CT-INT impurity solver, we used $N_{\text{cycles}}=1.2 \times 10^7$ Monte Carlo steps.

\subsection*{Cluster extensions of DMFT: DCA, CDMFT}
\subsection{DCA}
\label{app:dca}
The dynamical cluster approximation (DCA) \cite{Hettler1998, Hettler2000, Maier2005} is an embedding technique wherein the electron self-energy is obtained from the solution of an impurity model with $N_c$ interacting sites coupled to an infinite bath. As such, the DCA is just one particular generalization to $N_c>1$  of the `single site' dynamical mean field method that becomes exact as the number of impurity sites $N_c\rightarrow \infty$ \cite{Georges1992a, Georges1996, Maier2005}.
The DCA formulation partitions the Brillouin zone into $N_c$ equal area tiles and approximates the self-energy in each tile $a$, $\Sigma_a$ as a piecewise constant function of momentum as
\begin{equation}
\Sigma(\mathbf{k},\omega)=\sum_{a=1...N_c}\phi_a(\mathbf{k})\Sigma_a(\omega),
\label{coarsegraining}
\end{equation}
where $\phi_a(\mathbf{k})=1$ if $\mathbf{k}\in a$ and $\phi_a(\mathbf{k})=0$ for $\mathbf{k}\notin a$. Similar to DMFT, DCA invokes a self consistency loop where an initial guess for the impurity model parameters produces a set of $\Sigma_a$ which are then used to update the bath. 
In this work we obtain results for large values of $N_c$ in the paramagnetic phase but note that at weak-coupling and low temperatures studied in this work the DCA expansion in cluster size $N_c$ does not appear to be in a scaling regime and therefore one needs much larger clusters in order to estimate the infinite system size limit \cite{Simkovic2020, LeBlanc2013}.

To solve the $N_c$ site impurity problem we use the continuous time auxiliary field method \cite{Gull2008a,Gull2011a} with submatrix updates \cite{Gull2011}. Similar to other methods that employ Hubbard-Stratonovich decoupling, the primary computational hurdle is finding determinants of an $m \times m$ matrix where $m$ is referred to as the expansion order. Unlike Hirsch-Fye solvers that employ a fixed $\Delta\tau$ the expansion order and time steps in CT-AUX are not fixed, an important improvement that removes the necessity for a $\Delta\tau \rightarrow 0$ extrapolation. In the case of DCA, applying CT-AUX to clusters of size $N_c$ results in an expansion order $m\propto \beta{UN_c}$ \cite{Gull2008}.  In this work we have presented data primarily for fixed cluster size $N_c=128$ on a bipartite cluster. The DCA self-consistency is presumed to be paramagnetic and is iterated $\approx 15 \to 20 $times until the deviation between iterations is much less than the statistical error obtained from $\approx 2\times 10^6$ samples of each frequency up to $n=1024$. Our DCA and CT-AUX codes are based on the ALPS libraries \cite{ALPSCore,ALPSCore_updated}.

\begin{figure}[b!]
                \includegraphics[width=0.45\textwidth,angle=0]{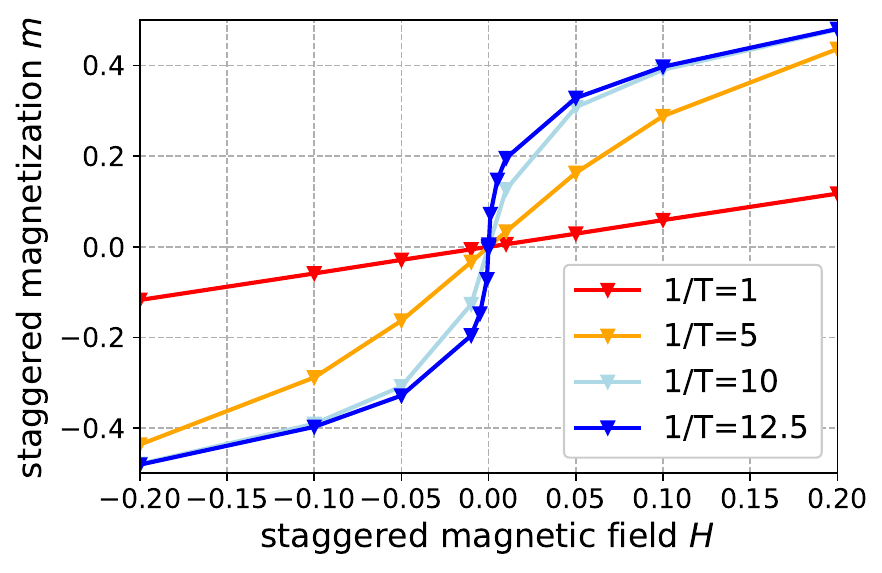}
      \caption{\label{fig:m_H}(Color online.) Antiferromagnetic staggered magnetization $m$ vs. staggered applied field $H$ in CDMFT on a $8 \times 8$ lattice for various temperatures.}
\end{figure}

\subsection{CDMFT}
\label{app:cdmft}
\begin{figure*}[ht!]

                \includegraphics[width=0.32\textwidth,angle=0]{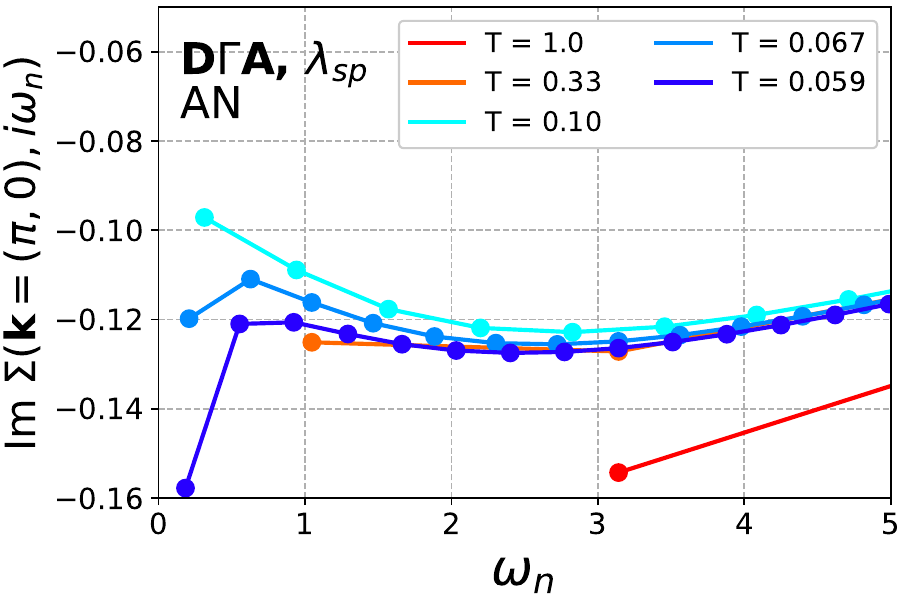}
                \includegraphics[width=0.32\textwidth,angle=0]{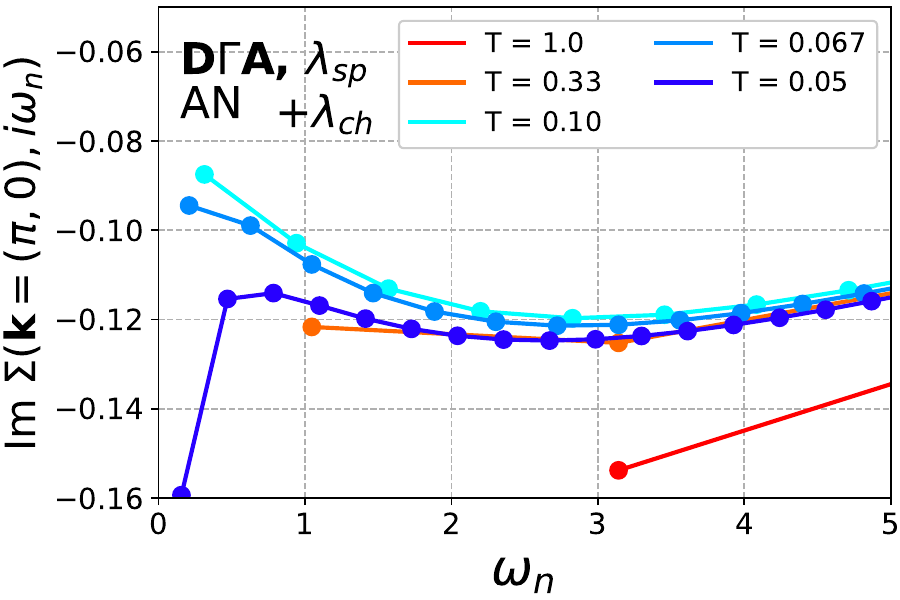}
                \includegraphics[width=0.32\textwidth,angle=0]{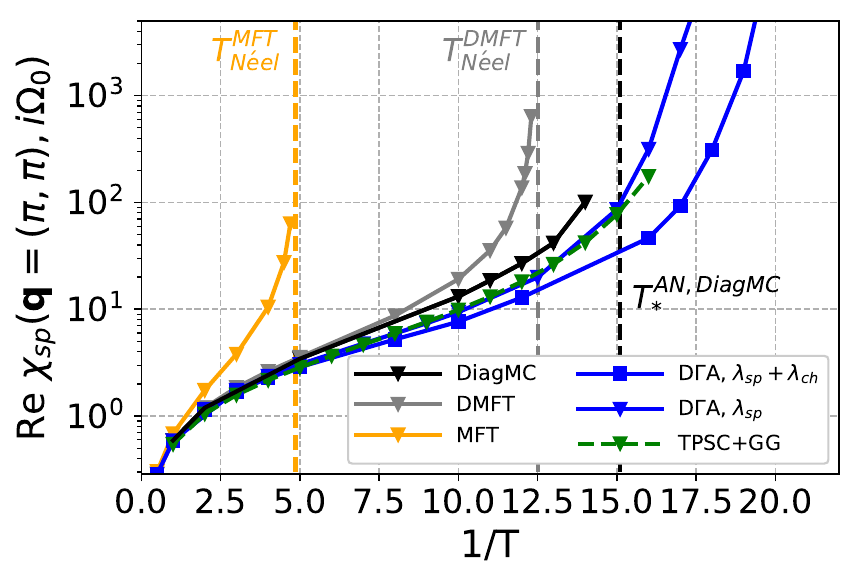}
                \includegraphics[width=0.32\textwidth,angle=0]{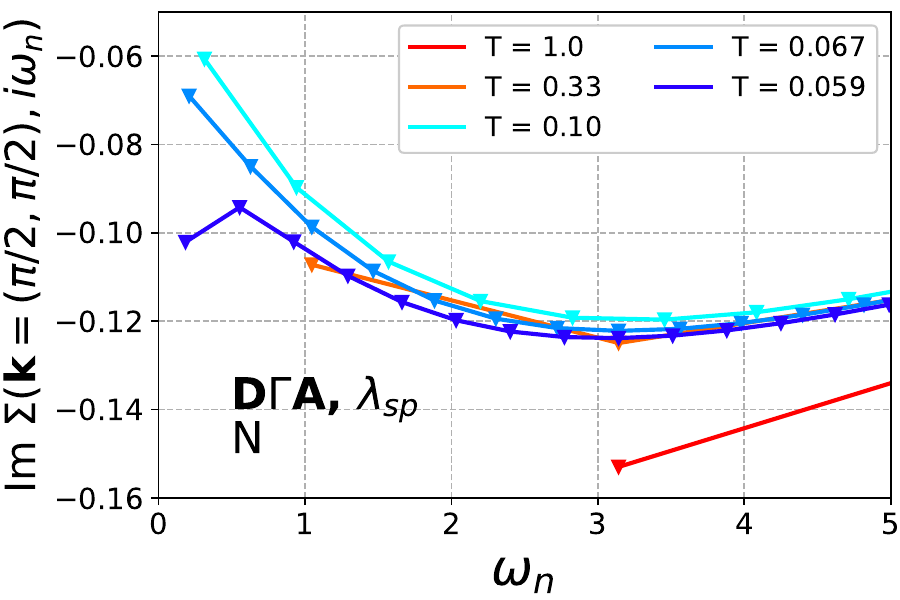}
                \includegraphics[width=0.32\textwidth,angle=0]{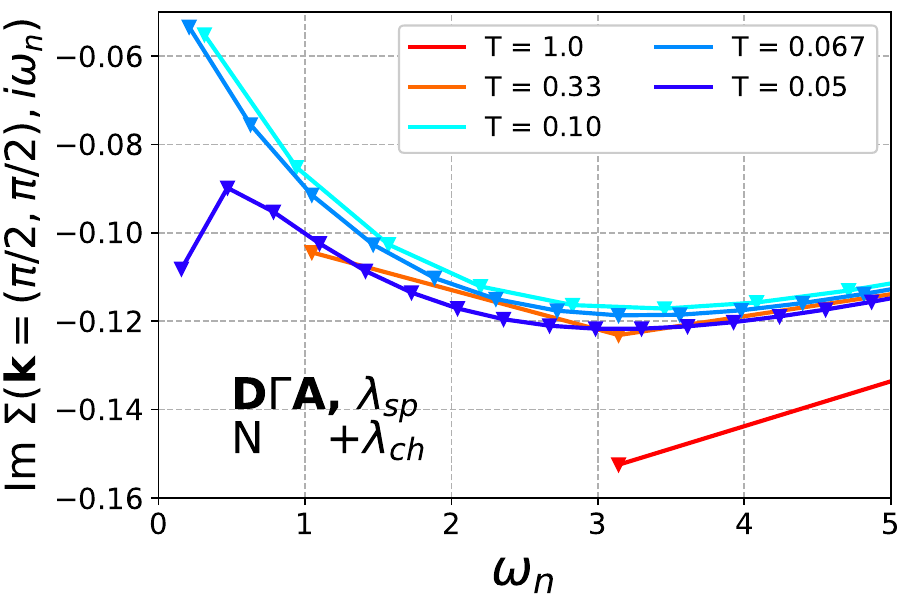}
                \includegraphics[width=0.32\textwidth,angle=0]{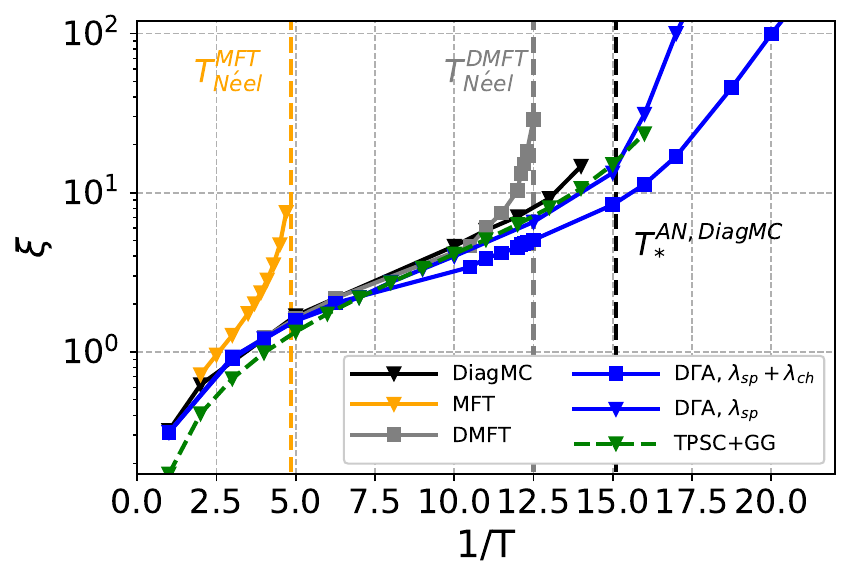}
      \caption{\label{fig:dga}(Color online.) Results from alternative Moriya correction schemes in ladder-D$\Gamma$A for various quantities. Please note that the lowest temperatures shown for the self-energy do not agree in order to show the full gap in both schemes.}
\end{figure*}
Cellular dynamical mean-field theory (CDMFT) is the real space cluster
extension of
DMFT \cite{Maier2005,Parcollet2004,Sakai2012,Park2008}. While
proposed around the same time, it has up to today been less spread
than the complementary dynamical cluster approximation which is
formulated in momentum space. The impurity of the auxiliary Anderson
model in CDMFT consists usually, contrary to single site DMFT, of
superstructures obtained by upfolding the lattices unit-cell. On the
square lattice, e.g., these may consist of $N\times N$ quadratic
patches. Hence, non-local correlations are included on length scales
given by the cluster size as intersite single-particle self
energies. Even within the aforementioned patch geometries, which typically
retain point group symmetry, translational symmetry is broken (for
$N>2$ even within the cluster). Re-periodization schemes suffer from
ambiguity and may even lead to convergence problems when attempted inside the
self-consistency loop.  The lattice quantities  are  therefore  approximated from  the  converged CDMFT solution by restoring the translational invariance for the Green function, the self-energy, or its cumulants \cite{Lichtenstein2000, Kotliar2001, Stanescu2006, Sakai2012, Sakai2009, Staar2013a, Sakai2016, Verret2019}. This study uses a cumulant scheme presented in App.~B of \cite{Klett2020}.

A recent study \cite{Klett2020} has shown that a
so-called center-focused extrapolation of self energies (and even
susceptibilities) to infinite cluster sizes (with a linear regression
in $1/N$) yields the best results in comparison to numerically exact
diagrammatic quantum Monte Carlo calculations. The CDMFT+CFE
self-energies presented in this paper have been obtained by
considering quadratic patches of up to $N=8$ for the extrapolation
scheme. The auxiliary impurity models were solved using a
state-of-the-art continuous time quantum Monte Carlo impurity solver
in its interaction expansion (CT-INT \cite{Rubtsov2004,Rubtsov2005,Gull2011a}) implemented in the TRIQS
framework \cite{TRIQS}. We used 20 CDMFT self-consistency loops (until convergence). Afterwards we performed 30 simulations with different random seeds starting from this previously converged solution. We take the self-energy as the mean of these 30 simulations. The solver parameters were a cycle length of 300, 10000 warmup cycles and $4 \times  10^6$ cycles.

For the calculation of the magnetic susceptibility we applied an antiferromagnetic staggered field $H$ and measured the staggered magnetization $m$. We then linearized the $m-H$ curve at small fields:

\begin{equation}
    \text{Re }\chi(\mathbf{q}=(\pi,\pi),i\Omega_n=0) = \left.\frac{\partial m}{\partial H}\right\vert_{H=0}.
\end{equation}

Fig.~\ref{fig:m_H} shows a sample of these curves for various temperatures. The transition to an antiferromagnetic ordered state is signalled as a vertical tangent at $H=0$, i.e. as divergence of the susceptibility, at $1/T_{\text{N{\'e}el}} \approx 13.7$. This represents a very small correction w.r.t. DMFT, where $1/T_{\text{N{\'e}el}} \approx 12.5$.

\subsection*{Diagrammatic extensions of DMFT}

A different route for the inclusion of spatial correlations on top of DMFT is taken by its diagrammatic extensions. Their principle is to extract a higher order correlation function from an auxiliary impurity model and consistently calculate the desired observables from it. For a very recent overview of diagrammatic extension of DMFT see \cite{RMPVertex}. More detailed information for each method is given below.

\subsection{\texorpdfstring{D$\Gamma$A}{DGA}}
\label{app:dga}
In the dynamical vertex approximation (D$\Gamma$A) \cite{Toschi2007, Katanin2009} the two-particle analog of the self-energy, the fully irreducible two-fermion scattering vertex $\Lambda$, in a DMFT spirit, is assumed to be purely local. In order to obtain the corresponding susceptibilities and self-energies, without the knowledge of leading instabilities, the parquet equations have to be solved self-consistently (see \cite{Valli2010, Valli2015} for nanoscopics, \cite{Kauch2020} for $\pi$-tons and \cite{Eckhardt2020, Kauch2019} for the two-dimensional Hubbard model). However, if (like in the present case of the Hubbard model), the leading instability is known, the scheme can be significantly facilitated by considering only the Bethe-Salpeter in the associated scattering channel and the Dyson-Schwinger equation, in order to obtain susceptibilities and self-energies (for successful applications see \cite{Rohringer2011, Valli2015, Schaefer2015a, Galler2016, Schaefer2015b, Schaefer2015c, Pudleiner2016, Schaefer2017, Schaefer2019, DelRe2019, Kitatani2020, Klebel2021}). In the vicinity of a phase transition, in order to restore the proper sum rules for the susceptibilities (and, thus, the asymptotics of the self-energy and therefore the two-particle self-consistency), within the ladder-version of the D$\Gamma$A, a Moriyaesque so-called $\lambda$ correction is used \cite{Toschi2007, Katanin2009, SchaeferThesis, Rohringer2016}. For the results presented in the main text of the current manuscript, this correction is only done for the spin susceptibility ($\lambda_{\text{sp}}$, see below). For the calculation of the (magnetic) susceptibility and correlation length we used data from an exact diagonalization (with four bath sites) impurity solver and the implementation of the Bethe-Salpeter equations and Dyson-Schwinger equation presented in \cite{Rohringer2018} (again, like in the DMFT case) were crosschecked with CT-INT. We used momentum grid with a maximum linear mesh sizes of $N_{q}=200$ and $N_{k}=200$ and the total number of fermionic as well as bosonic Matsubara frequencies being $N_{i\omega}=N_{i\Omega}=160$.
In the perspective of a future improved version of a Moriyaesque $\lambda$-correction \cite{Katanin2009, Rohringer2016}, it is instructive to also compare the results obtained with the correction in both particle-hole channels (i.e. charge and spin, $\lambda_{\text{sp}}+\lambda_{\text{ch}}$, see Eq.~(6) in \cite{Rohringer2016}) 
\begin{equation}
    \frac{1}{\beta}\sum\limits_{\Omega,\mathbf{q}}\chi_{r,\mathbf{q}}^{\Omega, \lambda_{r}}=\frac{1}{\beta}\sum\limits_{\Omega}\chi_{r}^{\Omega}
\end{equation}
(with $r=(\text{ch},\text{sp})$) to the ones obtained only with a correction in the spin channel ($\lambda_{\text{sp}}$, Eq~(5) in \cite{Rohringer2016}):
\begin{equation}
    \frac{1}{\beta}\sum\limits_{\Omega,\mathbf{q}}\frac{1}{2}\left[\chi_{\text{sp},\mathbf{q}}^{\Omega,\lambda_{\text{sp}}}+\chi_{\text{ch},\mathbf{q}}^{\Omega}\right]=\frac{1}{\beta}\sum\limits_{\Omega}\chi^{\Omega,\text{AIM}}_{\uparrow\uparrow}=\frac{n}{2}\left(1-\frac{n}{2}\right).
\end{equation}
Results for both schemes are presented in  Fig.~\ref{fig:dga}. In comparison with DiagMC in Fig.~\ref{fig:sigma_fluct_an}, the self-energy obtained with $\lambda_{\text{sp}}+\lambda_{\text{ch}}$ would lead to a slightly better agreement of the imaginary part at low frequencies at the antinode in comparison to the benchmark (left and central panel), but a slightly worse agreement for the susceptibility, the correlation length and $T_{\text{*}}$ (right panel).
It should be noted that the spin susceptibility and correlation length of D$\Gamma$A with $\lambda_{\text{sp}}$ agrees almost perfectly with a version of TPSC+ coined TPSC+GG down to low $T$. A more detailed comparison of different TPSC variants is done in App.~\ref{app:tpsc}.
Eventually, in order to be able to compare the simple D$\Gamma$Aesque approximation discussed in Sec.~\ref{subsec:simple}, we recall the Dyson-Schwinger equation of motion for the ladder-version of the D$\Gamma$A (see \cite{Toschi2007, Katanin2009, Rohringer2013a}, here omitting the Hartree term):
\begin{eqnarray}
    \Sigma(\mathbf{k},i\omega_n)&=&\frac{UT^2}{2}\sum\limits_{\mathbf{q},i\Omega_n}[\gamma^{\omega\Omega}_{\text{ch},\mathbf{q}}-3\gamma^{\omega\Omega}_{\text{sp},\mathbf{q}} + 2\\
    &+&U\gamma^{\omega\Omega}_{\text{ch},\mathbf{q}}\chi^{\Omega}_{\text{ch},\mathbf{q}}+3U\gamma^{\omega\Omega}_{\text{sp},\mathbf{q}}\chi^{\Omega}_{\text{sp},\mathbf{q}}\nonumber\\
    &-&\sum\limits_{\mathbf{k}',i\omega'_{n}}\left(F_{\text{ch}}^{\omega\omega'\Omega}-F_{\text{sp}}^{\omega\omega'\Omega}\right)\nonumber\\
    &\cdot&G(\mathbf{k}',i\omega'_n)G(\mathbf{k}'+\mathbf{q},i\omega'_n+i\Omega_n)]\nonumber\\
    &\cdot&G(\mathbf{k}+\mathbf{q},i\omega_n+i\Omega_n)\nonumber,
\end{eqnarray}
where $\chi$ refers to the physical, Moriya-corrected susceptibility, $F$ denotes the full two-particle vertex from the self-consistently determined Anderson impurity model in DMFT and  $G\!\coloneqq\!G_{\text{DMFT}}$ and the last line of the equation accounts for double counting corrections of the local part of the self-energy. $\gamma$ denotes the electron-boson coupling vertex (already integrated over the internal momentum $\mathbf{k}$). For the simple approximation in Sec.~\ref{subsec:simple}, the electron-boson coupling vertex is set to unity for all momenta and frequencies. A precise definition of the components of the equation of motion can be found at Eq.~(4.19) of \cite{Rohringer2013a}.

\subsection{TRILEX}
\label{app:trilex}
The triply irreducible local expansion (TRILEX) method is a relatively recent approach to strong correlations, which utilizes the decoupling of the fermion-fermion interaction of the Hubbard Hamiltonian into auxiliary bosons \cite{Ayral2015, Ayral2016a}. For these bosons, the fermion-boson vertex of a self-consistently determined impurity model is extracted and the lattice polarization and self-energy are calculated via Hedin's equations. The beauty of this approach consists in the facts that (i) the fermion-boson vertex is relatively easy to calculate (compared to the four-point vertex) and (ii) the convergence of the method to the exact solution can be achieved via a cluster extension of TRILEX \cite{Ayral2017} (we used DCA clusters with two and four sites in this paper) and (iii) the extension for treating superconductivity is straight forward \cite{Vucicevic2017}. However, the interaction is not unique and comes at the price of introducing an additional (Fierz) parameter $\alpha$, which in this paper was set in such a way that we decouple only the spin channel (i.e. $\alpha=1/3$ in Heisenberg spin decoupling \cite{Ayral2016a}). This choice can be motivated by a so-called fluctuation diagnostics analysis of the self-energy in the pseudogap regime of the Hubbard model \cite{Gunnarsson2015}. A different route to enhance the results of TRILEX can be made by inserting the fermion-boson vertex on both sites of the Hedin equations (an approach which we called TRILEX $\Lambda^2$, see also \cite{Stepanov2016a, Stepanov2019} for a similar idea in the dual theories, see also \cite{Krien2019} for an efficient evaluation of the polarization bubble in DMFT and \cite{Krien2017} for the question of conservation in two-particle self-consistent theories). The self-energies in the main text using this method show results which turn out to be similar to $N_c=4$ cluster TRILEX. We used momentum grids with a maximum linear mesh sizes of $N_{q}=128$ and $N_{k}=128$ and the number of fermionic and bosonic Matsubara frequencies were chosen as $N_{i\omega}=200$ and $N_{i\Omega}=20$, respectively. For the impurity solver (CT-INT \cite{Rubtsov2004,Rubtsov2005,Gull2011a}) we used $N_{\text{cycles}}=6.9 \times 10^7$ Monte Carlo steps.

\begin{figure*}[ht!]

                \includegraphics[width=0.23\textwidth,angle=0]{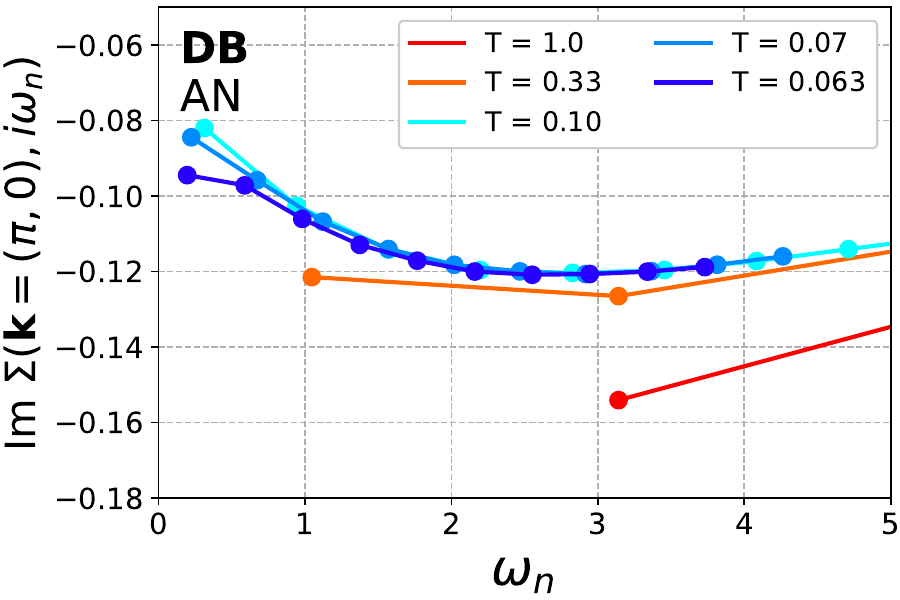}
                \includegraphics[width=0.23\textwidth,angle=0]{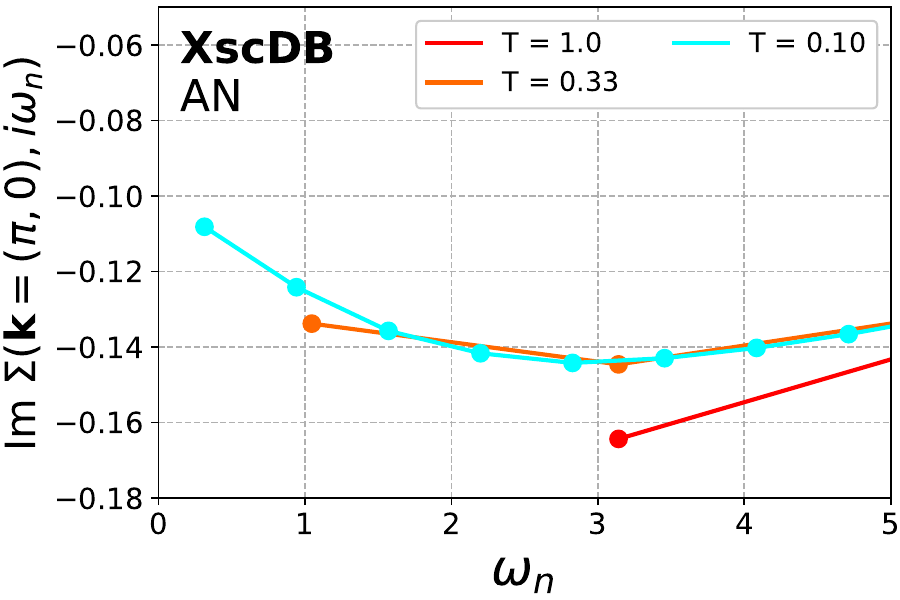}
                \includegraphics[width=0.23\textwidth,angle=0]{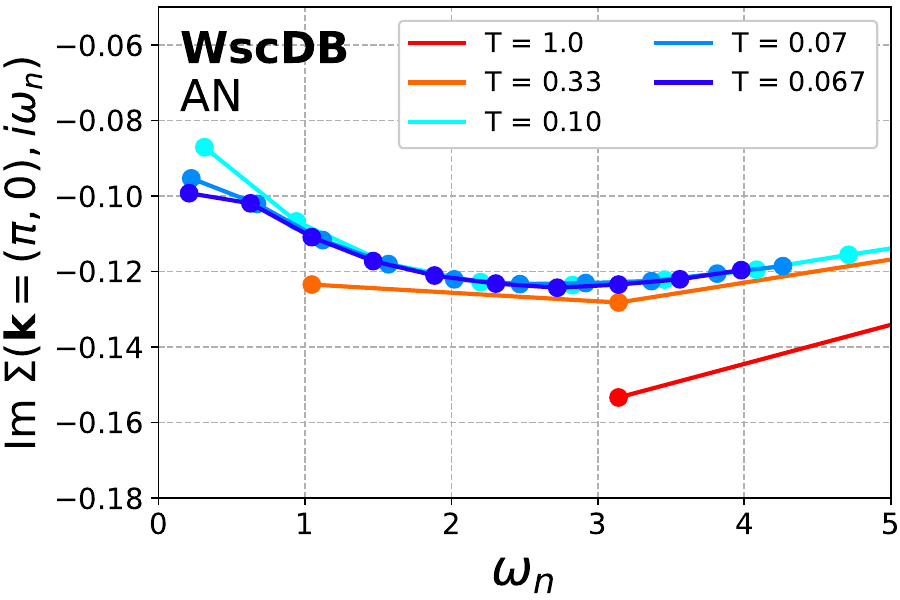}
                \includegraphics[width=0.23\textwidth,angle=0]{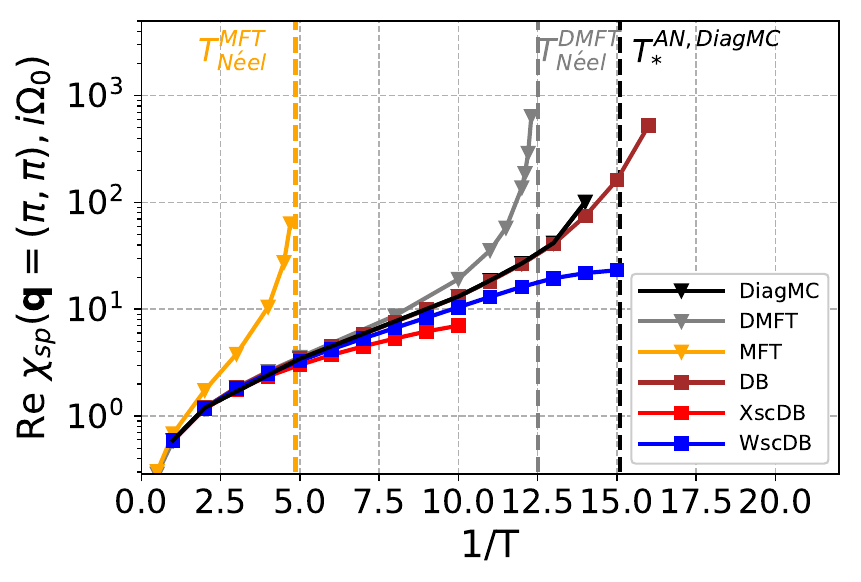}
                \includegraphics[width=0.23\textwidth,angle=0]{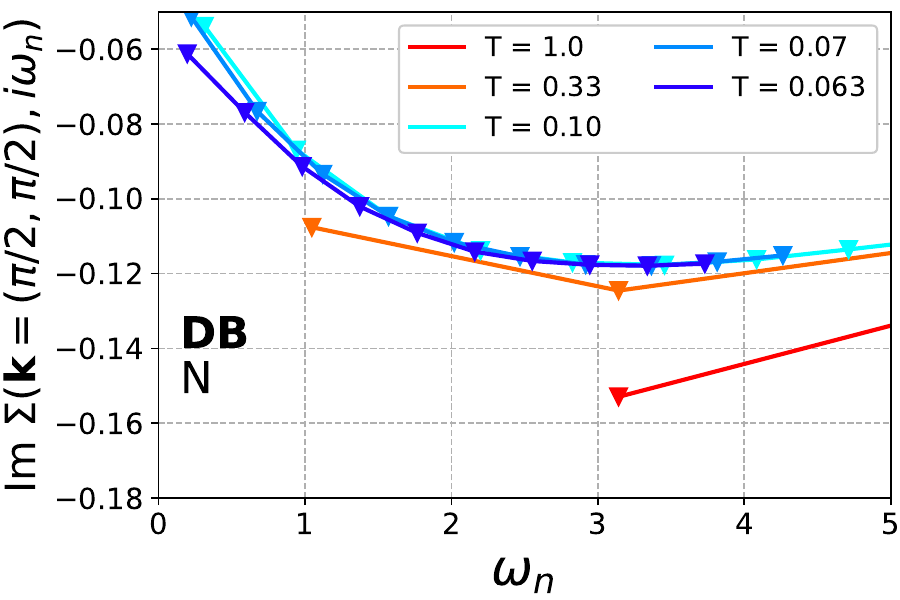}
                \includegraphics[width=0.23\textwidth,angle=0]{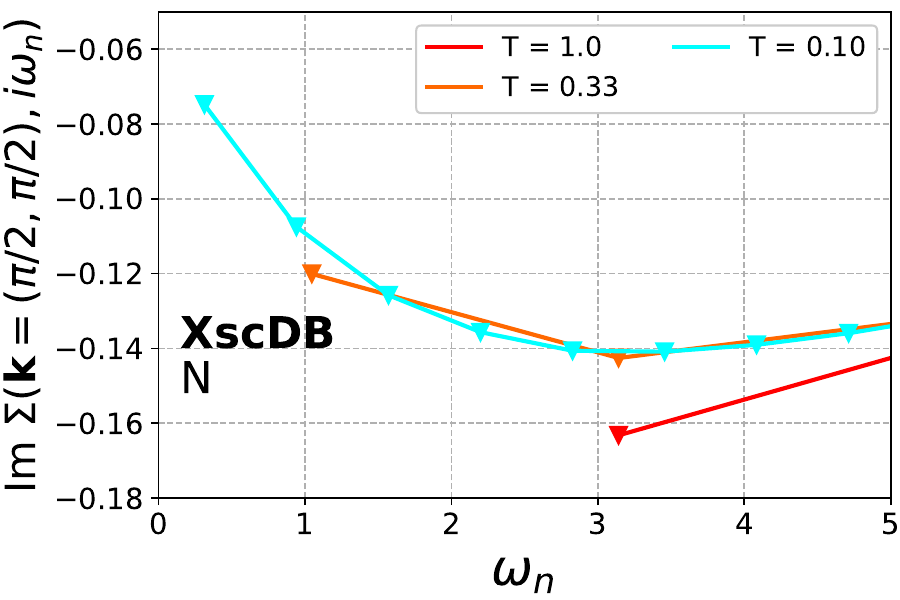}
                \includegraphics[width=0.23\textwidth,angle=0]{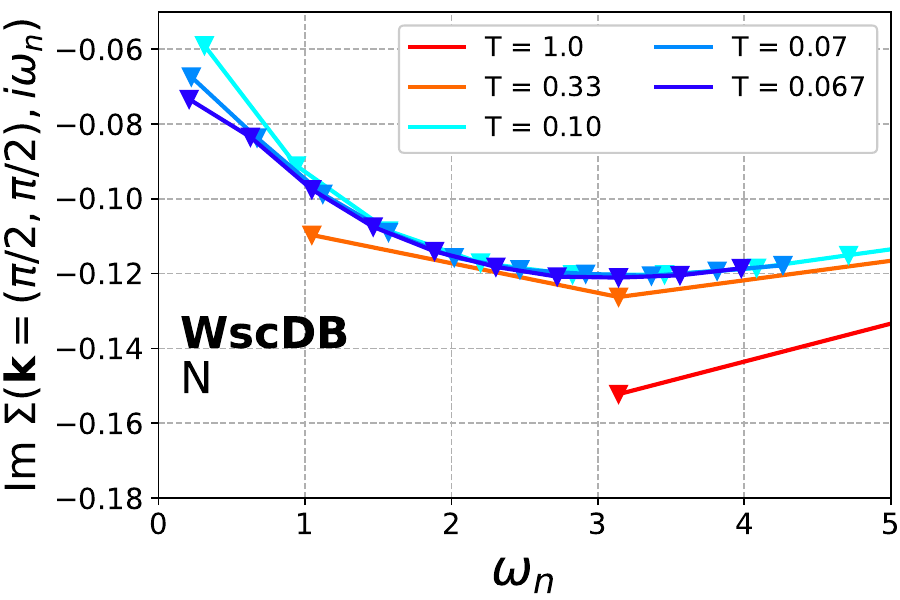}
                \includegraphics[width=0.23\textwidth,angle=0]{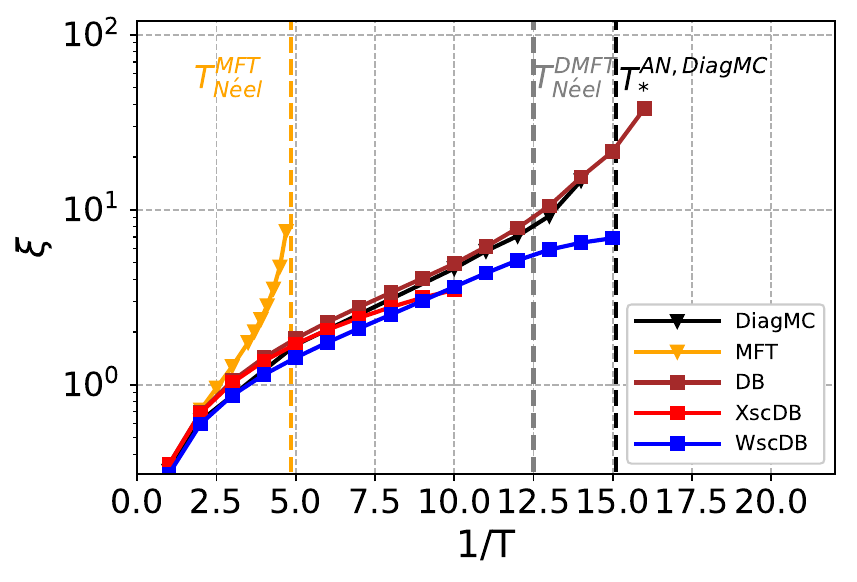}
      \caption{\label{fig:DB}(Color online.) Results from alternative DB schemes single-shot (DB), self-consistent (XscDB) and W-self-consistent (WscDB) for various quantities.}
\end{figure*}

\subsection{DF}
\label{app:df}
The dual fermion approach \cite{Rubtsov2008, Rubtsov2009,Hafermann2009} is a diagrammatic extension of the single site dynamical mean field theory (DMFT), motivated by the idea that non-local corrections to DMFT can be captured by a perturbative expansion around a solution of the dynamical mean field impurity problem. 

In the dual-fermion formalism \cite{Rubtsov2008,Antipov15} one replaces the lattice problem with a lattice of coupled Anderson impurity problems resulting in an action of the form
\begin{equation}
    S[f,f^*]= \sum_{\omega k \sigma}\text{g}_\omega^{-2} \left((\Delta_\omega - \epsilon_k)^{-1}
    +\text{g}_\omega\right) f^*_{\omega k \sigma} f_{\omega
    k \sigma} + \sum_i V_i,
\end{equation}
with $V_i\equiv V[f^*_i,f_i]$, and where $\text{g}_w$ is the momentum independent Green function of the Anderson impurity problem, and $\Delta_\omega$ is the hybridization function between the impurity and the bath \cite{Georges1996,Rubtsov2008}. The dual-fermion action now depends on $f$ and $f^*$ which are dual operators obtained via a Hubbard-Stratonovich transformation and in this dual space the interaction terms have become local, collected into the function $V_i$ \cite{Sarker1988, Pairault2000, Bourbonnais1985}.
Correctly representing $V_i$ remains problematic due to the complexity of higher-leg vertex functions \cite{Rohringer2012, Ribic2017b} and so we truncate the vertex at the level of 4-leg operators although higher order contributions may be important \cite{Iskakov2016, Gukelberger2017, RMPVertex}.
Once the dual self-energy, $\tilde{\Sigma}(k,\omega)$ is obtained, the lattice self-energy is given by
\begin{equation}
\Sigma(k,\omega)=\frac{\tilde{\Sigma}(k,\omega)}{1+\text{g}_{\omega}\tilde{\Sigma}(k,\omega) }+\Sigma^{\text{DMFT}}(\omega).
\label{eqn:dualselfenergy}
\end{equation}

We present DF results from the open-source \texttt{opendf} code \cite{Antipov15}, starting from a single-site ($N_c\!=\!1$)  dynamical mean field solution obtained with a continuous-time auxiliary-field (CT-AUX, \cite{Gull2008,Gull2011a,Gull2011}) impurity solver. The method treats all local correlations in a non-perturbative manner via DMFT and then perturbatively includes non-local correlations via a restricted set of self-consistent ladder diagrams for the non-local (`dual') self 
energy  in the charge and spin channels - this is known as the self-consistent ladder dual fermion method \cite{Hafermann2009}.
From this method one can also extract two particle spin susceptibilities which we present.  However, there is no self-consistency on the two-particle level and as such the status of two particle susceptibilities from DF is known to be approximate and expected to maintain only qualitative correctness \cite{Leblanc2019}. In all cases the input DMFT solutions (Green functions and Vertices) are obtained using a continuous time auxiliary field method (CT-AUX) \cite{Gull2011a}.

DMFT results for the full vertex $F^{\nu,\nu^\prime,\omega}$ are obtained with a frequency truncation in $\nu,\nu^\prime,\omega$.  We present results with truncations $|\nu_c|=|\nu_c^\prime|=96$ in fermionic frequencies and $|\omega_c|=64$ in bosonic frequencies.  The DF result is not strongly dependent upon this truncation.  More important is the known sensitivity of the result to the momentum resolution \cite{vanLoon2017}.  As such we employ a $64\times 64$ k-space grid in the DF dual self-consistency.  Further, the result for the lattice self energy is extremely sensitive to the value of the dual self energy due to the inversion in equation (C7). We iterate the DF self-consistency until convergence on the scale of $1\times 10^{-8}$. We also want to refer to a recent study of the doped Hubbard model in a dual parquet scheme \cite{Astretsov2020, Krien2020} and a dual parquet solution of the Falicov-Kimball model \cite{Astleithner2020}.

\begin{figure}[ht!]
        \centering
                \includegraphics[width=0.45\textwidth]{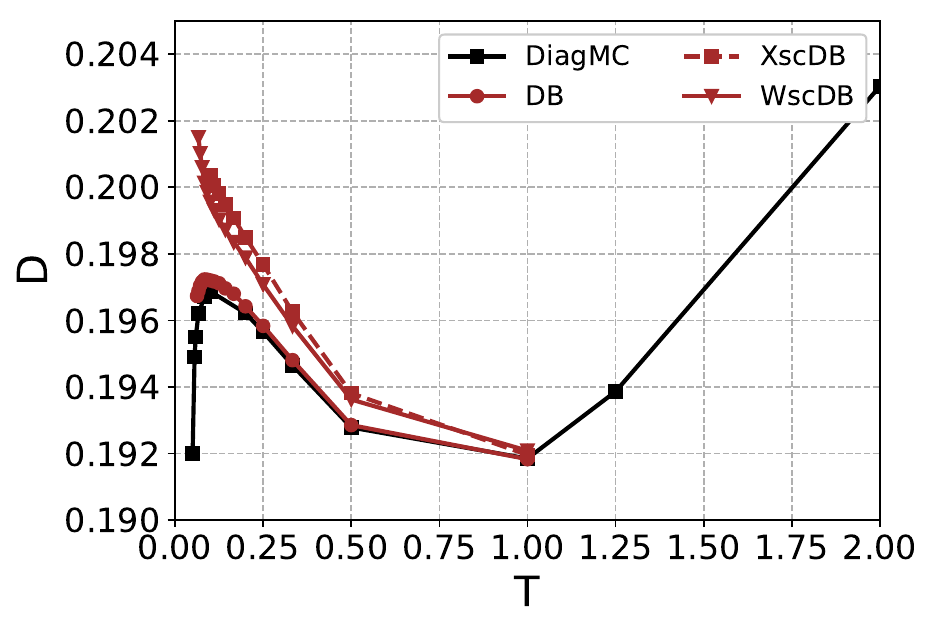}
        \caption{\label{fig:DB_docc}(Color online.) Results from alternative DB schemes single-shot (DB), self-consistent (XscDB) and W-self-consistent (WscDB) for the double occupancy compared to DiagMC.}
\end{figure}

\subsection{DB}
\label{app:db}
\noindent
The Dual Boson (DB) approach \cite{Rubtsov2012, vanLoon2014a, Stepanov2016} is an extension of the DF approach that additionally to the local Coulomb interaction accounts for the effect of nonlocal interactions in different bosonic channels ($\varsigma$). Within the DB approach, local electronic correlations are considered exactly in the framework of the (extended) dynamical mean-field theory (EDMFT) \cite{Sun2002, Sun2004}. Nonlocal collective fluctuations are treated diagrammatically beyond the EDMFT level. For this aim, dual boson fields $\varphi^{\varsigma}$ are introduced in addition to dual fermion variables $f^{(*)}$ that are already present in the DF approach. The DB theory is derived analytically using a path integral formalism, so many existing EDMFT-based approaches can be obtained as a certain approximation of the DB method \cite{Stepanov2019, Stepanov2016}. Also, the DB theory fulfills the Mermin-Wagner theorem, which allows to avoid unphysical phase transitions in two dimensions \cite{Peters2019}.

The action of the DB theory is following
\begin{equation}
{\cal \tilde{S}}
= -\sum_{{\bf k},\omega,\sigma} f^{*}_{{\bf k}\omega\sigma}\tilde{\cal G}^{-1}_{{\bf k}\omega\sigma}f^{\phantom{*}}_{{\bf k}\omega\sigma} - \frac12\sum_{{\bf q},\Omega,\varsigma}  \varphi^{\varsigma}_{{\bf q}\Omega} \tilde{\cal W}^{\varsigma~-1}_{{\bf q}\Omega} \varphi^{\varsigma}_{-{\bf q},-\Omega} + \tilde{\cal F}.
\end{equation}

Here, the bare fermion $\tilde{\mathcal{G}}_{\mathbf{k},\nu,\sigma}$ and boson $\tilde{\mathcal{W}}^{\varsigma}_{\mathbf{q},\omega}$ propagators are given by nonlocal parts of EDMFT Green function and renormalized interaction \cite{Stepanov2016}, respectively.

The interaction part $\tilde{\cal F}[f^{*}, f, \varphi]$ of the dual action contains all possible exact local fermion-fermion and fermion-boson vertex functions of the impurity problem. Here, as well as in most of DB approximations, we restrict ourselves to the lowest-order (two-particle) interaction terms that are given by the 4-leg fermion-fermion and 3-leg fermion-boson vertex functions. This truncation of the interaction allows to describe collective charge \cite{vanLoon2014, Stepanov2019a} and spin \cite{Stepanov2018, Stepanov2018b} degrees of freedom in a conserving way using the ladder DB approximation \cite{Rubtsov2012, Stepanov2016, vanLoon2014a, Hafermann2014a}.

In the main part of the text only single-shot ladder Dual Boson results are discussed. These calculations are performed on the basis of the converged DMFT solution of the problem, where the bosonic hybridization function is equal to zero. Importantly, in the latter case the DB theory fully coincides with the DF approach if only the local Coulomb interaction is considered. The corresponding local impurity problem is solved using the open source CT-HYB solver \cite{Hafermann2013, Hafermann2014} based on the ALPS libraries \cite{ALPS2}. This solution requires $N_{\text{cycles}} = 8.1 \times 10^{7}$ Monte Carlo steps. After that, we calculate the dual self-energy and polarization operator diagrammatically, and perform only the inner self-consistency loop in order to obtain the dressed Green’s function and renormalized interaction using the Dyson equation. For this, we use momentum grid with a maximum linear mesh size of $N_{k} = 128$, and the number of fermionic and bosonic Matsubara frequencies $N_{i\omega} = 256$ and $N_{i\Omega} = 64$, respectively. The expression for the lattice self-energy of the DB approach coincides with the one of the DF theory in Eq.~(\ref{eqn:dualselfenergy}).
The lattice polarization function can be found using a similar expression \cite{Stepanov2016a}:
\begin{equation}
    \Pi({\bf q},i\Omega_n) = \frac{\tilde\Pi({\bf q}, i\Omega_n)}{1+{\cal W}(i\Omega_n)\tilde\Pi({\bf q}, i\Omega_n)} + \Pi^{\rm EDMFT}(i\Omega_n).
\end{equation}

The fully self-consistent DB calculations can be performed as follows. To obtain the fermionic hybridization of the effective impurity problem, we use the outer self-consistency condition that equates the local part of the lattice Green function and local impurity Green function $\sum_{\bf k} G_{{\bf k}\omega\sigma} = g_{\omega\sigma}$. Regarding the bosonic hybridization function, there is no clear receipt how this quantity has to be determined. Here, we investigate two different self-consistency schemes that fix the bosonic hybridization. For the X-self-consistent (Xsc) result the local part of the lattice susceptibility is equated to the corresponding local susceptibility of the impurity problem $\sum_{\bf q} X^{\varsigma}_{{\bf}\Omega} = \chi^{\varsigma}_{\Omega}$. The other self-consistency can be imposed on a renormalized (screened) interaction (Wsc) $\sum_{\bf q} W^{\varsigma}_{{\bf}\Omega} = w^{\varsigma}_{\Omega}$. 
The renormalized interaction $W$ of the lattice problem can be defined as
\begin{align}
W^{\varsigma \,-1}({\bf q},i\Omega_{n}) = U^{\varsigma\,-1} - \Pi^{\varsigma}({\bf q},i\Omega_{n}),\label{eqn:db_w}
\end{align} 
where $U^{\rm ch/sp} = \pm U/2$~\cite{Stepanov2019}. The EDMFT renormalized interaction can be obtained neglecting the dual contribution to polarization operator in Eq.~(\ref{eqn:db_w}), so that $\Pi^{\varsigma}({\bf q},i\Omega_{n})=\Pi^{\varsigma\,{\rm EDMFT}}(i\Omega_{n})$. The renormalized interaction of the impurity problem can be found as 
\begin{align}
w^{\varsigma \,-1}(i\Omega_{n}) = \left(U^{\varsigma} + Y^{\varsigma}(i\Omega_{n})\right)^{-1} - \Pi^{\varsigma\,{\rm EDMFT}}(i\Omega_{n}),
\end{align}
where $Y^{\varsigma}(i\Omega_{n})$ is the bosonic hybridization function.
Corresponding results are shown in Figs.~\ref{fig:DB} and \ref{fig:DB_docc}. We note as well that the comparisons between self-consistent DB and self-consistent DF schemes are in good agreement, but differ from the exact result. As we point out in the main text, the single-shot DB approach correctly reproduces exact DiagMC results at almost all temperatures. Surprisingly, we observe that the Xsc DB calculations strongly deviate from the exact result presented in both Figures. At the same time, we find that the Wsc DB result for the self-energy agrees with DiagMC calculations even better than the single-shot DB one. However, two particle quantities, such as the lattice susceptibility and double occupancy, get worse when the self-consistency on the renormalized interaction is utilized. This can be explained by the fact that considered bosonic self-consistencies cannot fix all desired single- and two-particle quantities at the same time. Therefore, the question of a good self-consistency for the bosonic hybridization function remains open.

\subsection*{Other approximations}

\subsection{TPSC and TPSC+}
\label{app:tpsc}
\subsubsection{TPSC}
The two-particle self-consistent (TPSC)~\cite{Vilk1994,Vilk1995,Vilk1996} approach is a
nonperturbative approach that is only valid from weak to intermediate interaction strength;
hence, it does not describe the Mott transition. Nevertheless, there are a large number of physical phenomena
that it allows to study with an accuracy comparable to other numerically exact methods. In particular, in two
dimensions, it describes the entry into the renormalized classical regime of antiferromagnetic fluctuations.
There are no adjustable parameters as opposed to, for example, self-consistent renormalized spin-fluctuation theory. It satisfies
conservation laws for total spin and total charge, the Mermin-Wagner theorem, the Pauli principle in the form
$\langle n_{\sigma}^{2}\rangle = \langle n_{\sigma}\rangle$, as well as the local spin and local charge sum
rules. The two local sum rules in addition to the ansatz that the renormalized irreducible spin interaction
vertex is given by $U_\text{sp}\langle n_{\uparrow}\rangle \langle n_{\downarrow}\rangle = U\langle
n_{\uparrow} n_{\downarrow}\rangle$ suffice to obtain the irreducible vertices
\begin{align}
  U_\text{sp} \equiv
    \frac{\delta \Sigma_{\uparrow}^{(1)}}{\delta G_{\downarrow}^{(1)}} -
    \frac{\delta \Sigma_{\uparrow}^{(1)}}{\delta G_{\uparrow}^{(1)}};\
  U_\text{ch} \equiv
    \frac{\delta \Sigma_{\uparrow}^{(1)}}{\delta G_{\downarrow}^{(1)}} +
    \frac{\delta \Sigma_{\uparrow}^{(1)}}{\delta G_{\uparrow}^{(1)}},
  \label{dSigma/dG}
\end{align}
and then compute the spin and charge susceptibilities from
\begin{subequations} \label{chi_sp_ch}
\begin{align}
  \chi_\text{sp}(q) &= \frac{\chi^{(1)}(q)}{1 - \frac{1}{2}U_\text{sp}\chi^{(1)}(q)}; \\
  \chi_\text{ch}(q) &= \frac{\chi^{(1)}(q)}{1 + \frac{1}{2}U_\text{ch}\chi^{(1)}(q)}.
\end{align}
\end{subequations}
The local spin and charge sum rules are given by $\Tr\chi_\text{sp}(q) = n - 2\langle
n_{\uparrow}n_{\downarrow}\rangle$ and $\Tr\chi_\text{ch}(q) = n + 2\langle n_{\uparrow}n_{\downarrow}\rangle
- n^2$, respectively, and the trace is over the spin and momentum space. The above ansatz for $U_\text{sp}$ is inspired from Ref.~\onlinecite{Singwi1981,Ichimaru1982} and was found independently in Ref.~\onlinecite{Hedeyati1989}.

\begin{figure*}[ht!]

                \includegraphics[width=0.23\textwidth,angle=0]{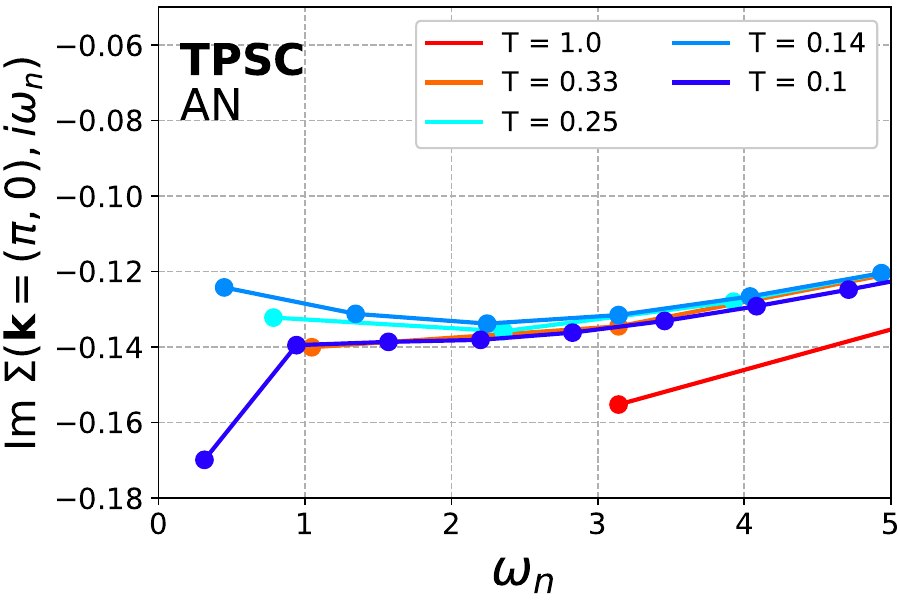}
                \includegraphics[width=0.23\textwidth,angle=0]{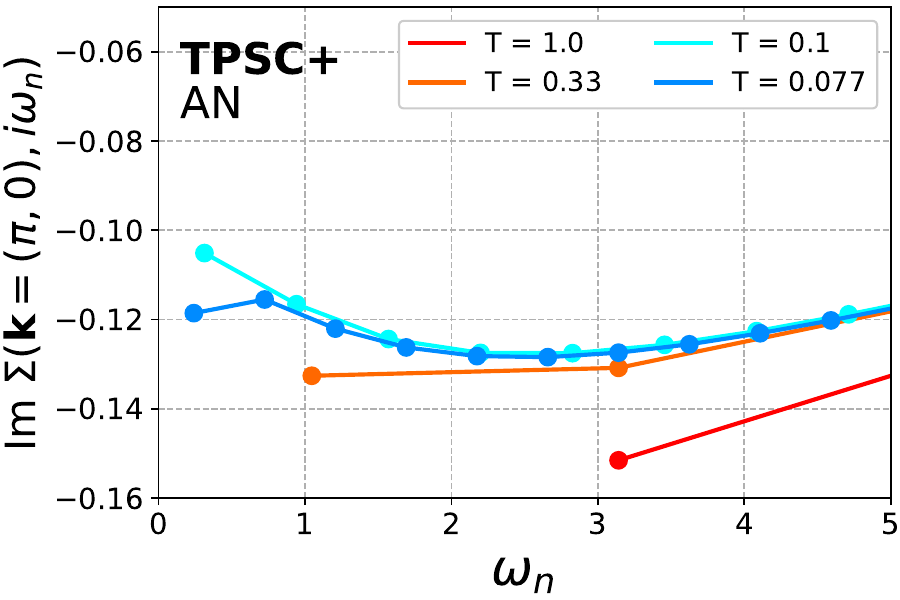}
                \includegraphics[width=0.23\textwidth,angle=0]{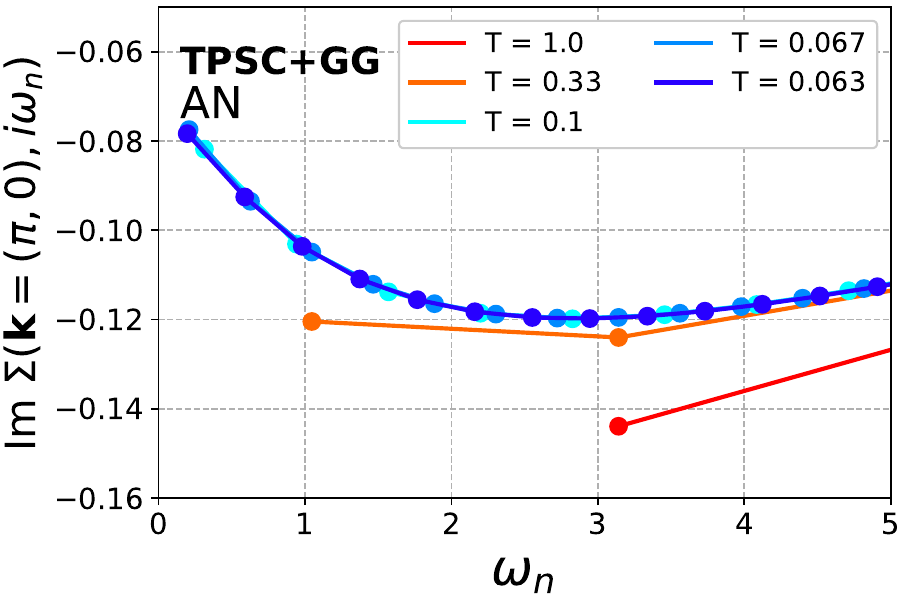}
                \includegraphics[width=0.23\textwidth,angle=0]{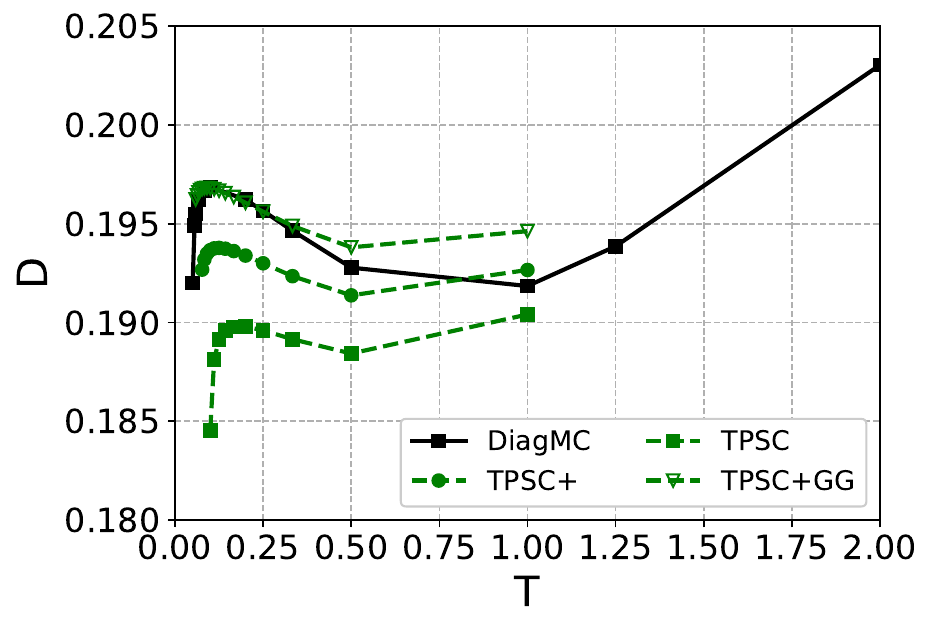}
                \includegraphics[width=0.23\textwidth,angle=0]{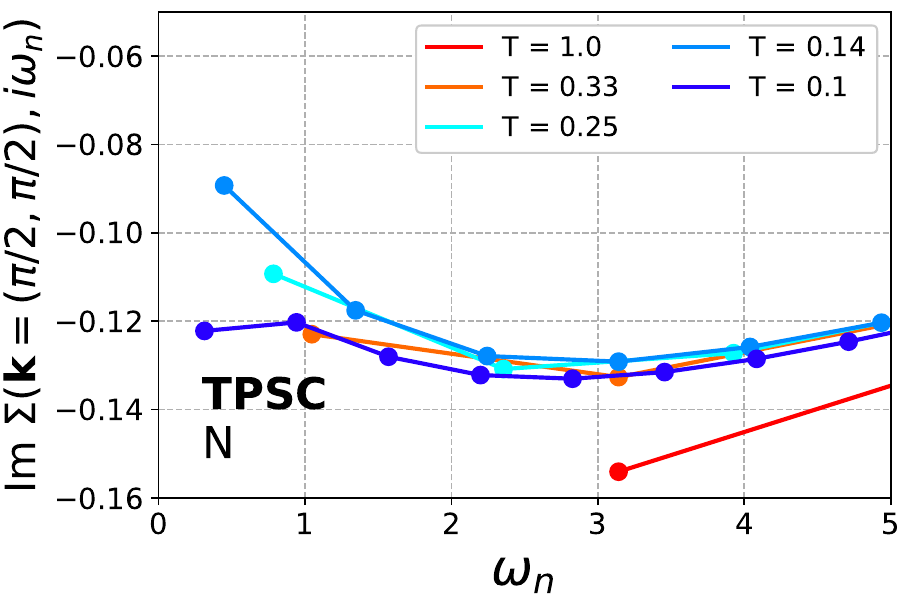}
                \includegraphics[width=0.23\textwidth,angle=0]{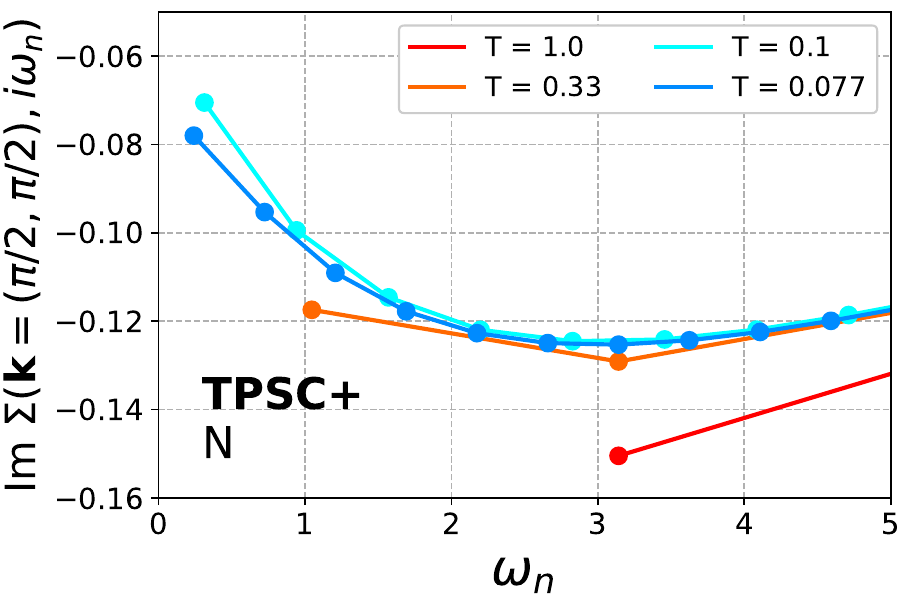}
                \includegraphics[width=0.23\textwidth,angle=0]{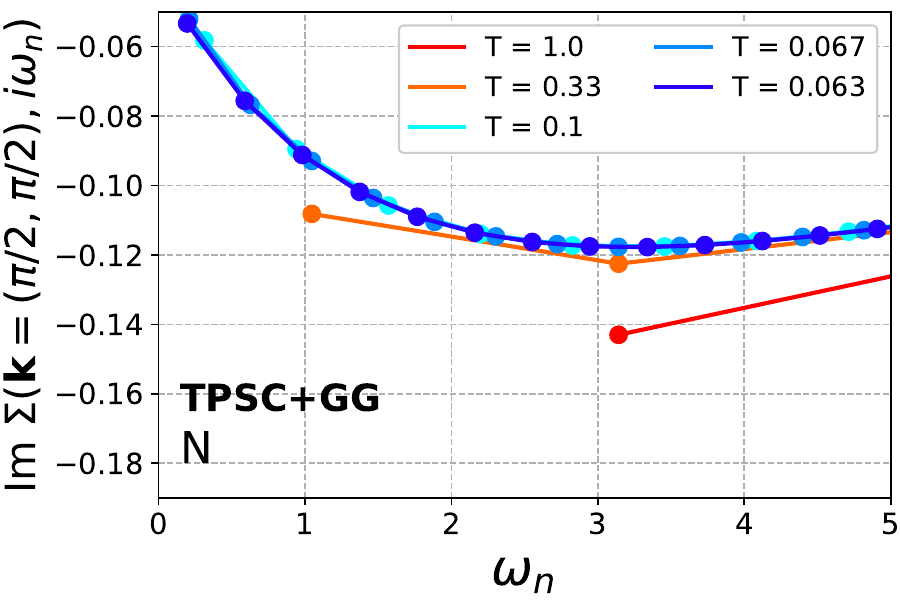}
                \hspace{4.09cm}
      \caption{\label{fig:tpsc}(Color online.) Results from TPSC, TPSC+ and TPSC+GG.}
\end{figure*}

At the first-level approximation (hence the superscript ``$(1)$'') of TPSC, the irreducible
particle-hole bubble diagram is obtained from
\begin{align}
  \chi^{(1)}(q) &= -\Tr \left[ G^{(1)} G^{(1)} \right] = -{2T}\sum_{k} G_0(k) G_0(k+q),
\end{align}
where the Green function $G_\sigma^{(1)}(k)$ includes a constant self-energy $\Sigma_\sigma^{(1)}$ and
renormalized chemical potential that lead to a non-interacting form $G_0(k)$. The two irreducible vertices
suffice to find an improved self-energy~\cite{Moukouri2000} that does not assume Migdal's theorem and that
takes into account rotational invariance and crossing symmetry
\begin{align} \label{self-2}
  \Sigma_{\sigma}^{(2)}(k) ={}
    & U n_{-\sigma} +{} \notag \\
    & \frac{UT}{8} \sum_{q}
      \left[ 3U_\text{sp}\chi_\text{sp}(q) + U_\text{ch}\chi_\text{ch}(q) \right]
      G_{\sigma}^{(1)}(k-q).
\end{align}

The consistency condition between one-particle and two-particle properties (Galitskii-Migdal formula referred in the main text)
\begin{align}
  \frac{1}{2}\Tr \left[ \Sigma ^{(2)} G^{(1)} \right] &= U\langle n_{\uparrow} n_{\downarrow}\rangle
  \label{Consistency 1-2}
\end{align}
serves as a guide for the domain of validity of TPSC. (Double occupancy obtained from sum rules
on spin and charge equals that obtained from the self-energy and the Green function. When $\frac{1}{2}\Tr \left[ \Sigma ^{(2)} G^{(2)} \right]$ starts to deviate from the above, the method starts to fail.)
See~\cite{Vilk1997,Allen2003} for detailed comparisons with QMC calculations and other approaches and
discussions of the physics. Ref.~\cite{Tremblay2006} reviews the work related to the pseudogap and
superconductivity up to 2005 including detailed comparisons with quantum cluster approaches in the regime of
validity that overlaps with TPSC (intermediate coupling). A pedagogical review of results and derivation
appears in Ref.~\cite{Tremblay2011}.

TPSC numerical results in the present paper have been obtained on a $256\times 256$ lattice and $2\times
1024$ Matsubara frequencies with tails treated analytically. Due to the self-energy iterations, we use a $128 \times 128$ lattice and $2 \times 512$ Matsubara frequencies in TPSC+ calculations, so that the sufficient precision can be obtained using moderate computational resources, as in TPSC.

\subsubsection{TPSC+}
One of the main limitations of TPSC is that even for weak to intermediate interaction strength,
it is not valid deep in the renormalized classical regime~\cite{Vilk1994}. For example, for the half-filled
Hubbard model with only nearest-neighbor hopping, one finds that $U_\text{sp}$ tends to zero as temperature
goes to zero, contrary to the fact that $U_\text{sp}$ and the site double occupancy $\langle n_{\uparrow}
n_{\downarrow}\rangle = (U_\text{sp}/U) \langle n_{\uparrow}\rangle \langle n_{\downarrow}\rangle$ must
saturate to a finite value as $T\to 0$ due to virtual states.

To remedy this problem, an improved TPSC has recently been proposed~\cite{Wang2019}. We refer to this extension as TPSC+.  It is based on an
extension of Kadanoff and Martin's ideas~\cite{Kadanoff1962} who treated the normal state of
superconductors in such a way that it connects smoothly to the superconducting state described by the BCS
equation. The key idea is that the pair susceptibility takes an asymmetric form $\chi(q)=-\Tr[G(k)G_0(q-k)]$.
It is called the pairing approximation or $GG_0$ theory in this context and it has been extensively used by Levin's group to
study pairing pseudogap and related phenomena~\cite{Chen_Stajic_Tan_Levin_2005,Varlamov_Levin_2018}.

We apply that idea to the repulsive Hubbard model and use the asymmetric form in the particle-hole bubble
\begin{align}
  \widetilde{\chi}^{(2)} &= -\frac{1}{2} \Tr \left[\widetilde{G}^{(2)}(k+q)G^{(1)}(k)+\widetilde{G}^{(2)}(k-q){G}^{(1)}(k) \right],
\end{align}
where the self-energy $\widetilde{\Sigma}^{(2)}$ entering $\widetilde{G}^{(2)}$ has the same form as Eq.~(\ref{self-2}): Contrary to the original Kadanoff-Martin approach, the
irreducible vertices are computed from the same sum rules and ansatz but the susceptibilities in Eq.~(\ref{chi_sp_ch}) are
obtained from $\widetilde{\chi}^{(2)}$ instead of from $\chi^{(1)}$. The tilde symbol ($\widetilde{\
}$) indicates that $\widetilde{\Sigma}^{(2)}$, $\widetilde{G}^{(2)}$, and $\widetilde{\chi}^{(2)}$
self-consistently depend on each other. Note that the trace also includes the spin.

The advantages of this approach are as follows. (a) The generalized Stoner criterion for the phase transition temperature $T_\text{N}$ becomes identical to the mean-field
gap equation in the antiferromagnetic state with the interaction vertex reduced from the bare $U$ to
$U_\text{sp}$~\cite{Wang2019}. 

(b) The Mermin-Wagner theorem and the Pauli principle are satisfied.
Analytical arguments that demonstrate these results proceed in a manner analogous to those in TPSC.
(c) One- and two-particle properties are consistent in the sense
that $\frac{1}{2}\Tr \left[ \widetilde{\Sigma}^{(2)} \widetilde{G}^{(2)} \right] = U\langle n_{\uparrow}
n_{\downarrow} \rangle$ is satisfied exactly with double-occupancy equal to that obtained from the local spin
and charge sum rules. (d) Analytical arguments analogous to those in TPSC show that there is a pseudogap in
two dimensions that is a precursor to the antiferromagnetically ordered state at zero temperature. Because the susceptibility dressed by self-energy remains finite at zero temperature, renormalized $U_\text{sp}$ remains finite while pushing $T_\text{N}$ to zero. Furthermore, at zero temperature, the size of the pseudogap $\Delta_\text{pseudogap}(0)$
becomes equal to the finite magnetic gap $\Delta(0)$.

On the
down side, the susceptibilities at zero wave vector do not vanish at finite frequency, as conservation laws
say they should. However, the values of the zero-wave vector susceptibilities at finite frequency are
generally small. Finally, the $f$-sum rule is also slightly violated.

\begin{figure*}[ht!]

                \includegraphics[width=0.32\textwidth,angle=0]{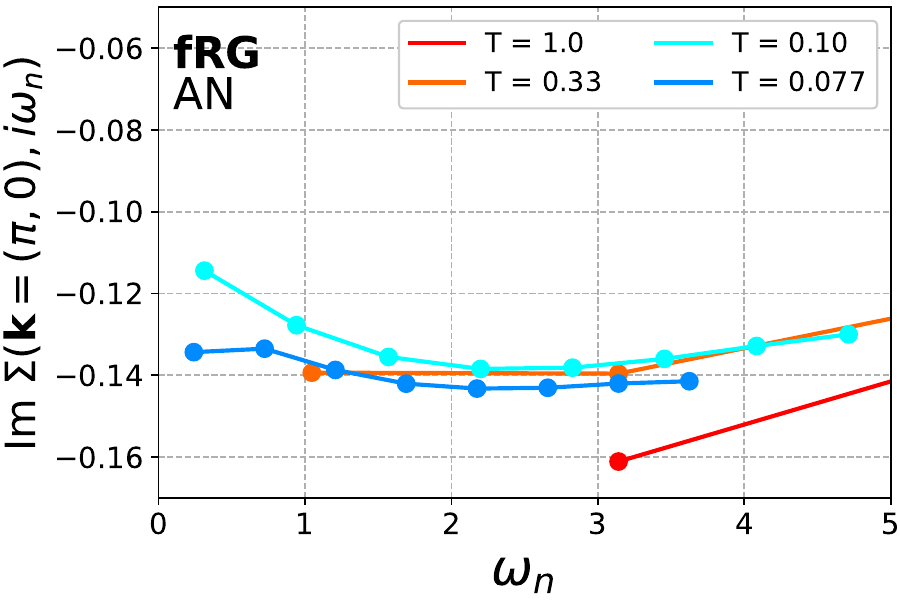}
                \includegraphics[width=0.32\textwidth,angle=0]{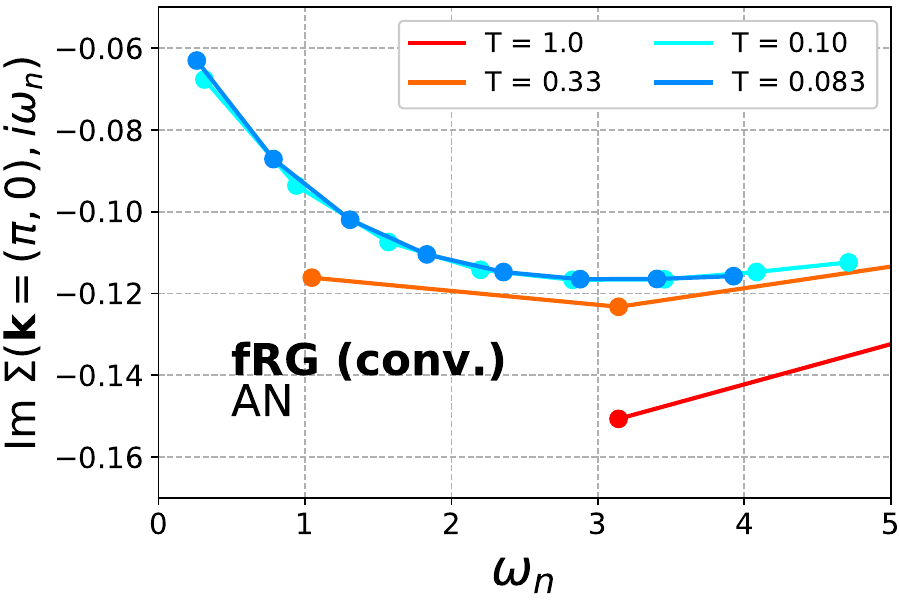}
                \includegraphics[width=0.32\textwidth,angle=0]{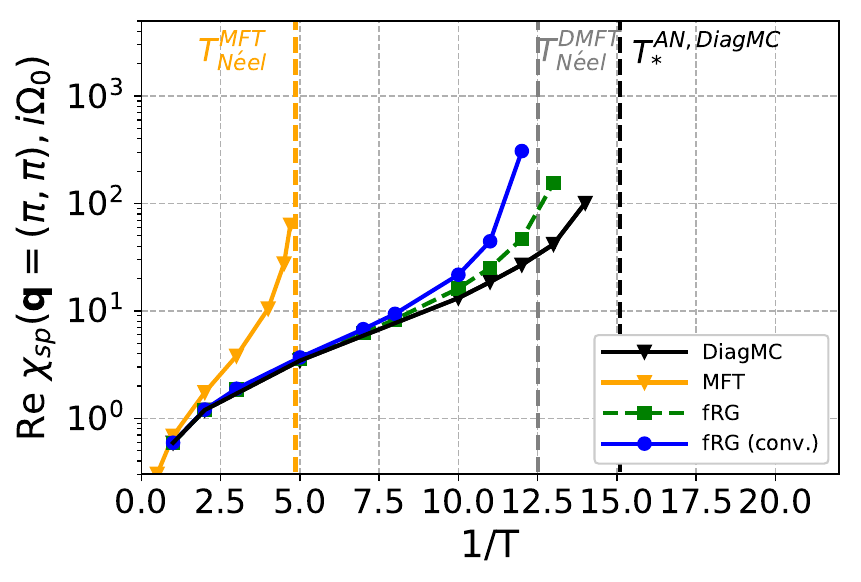}
                \includegraphics[width=0.32\textwidth,angle=0]{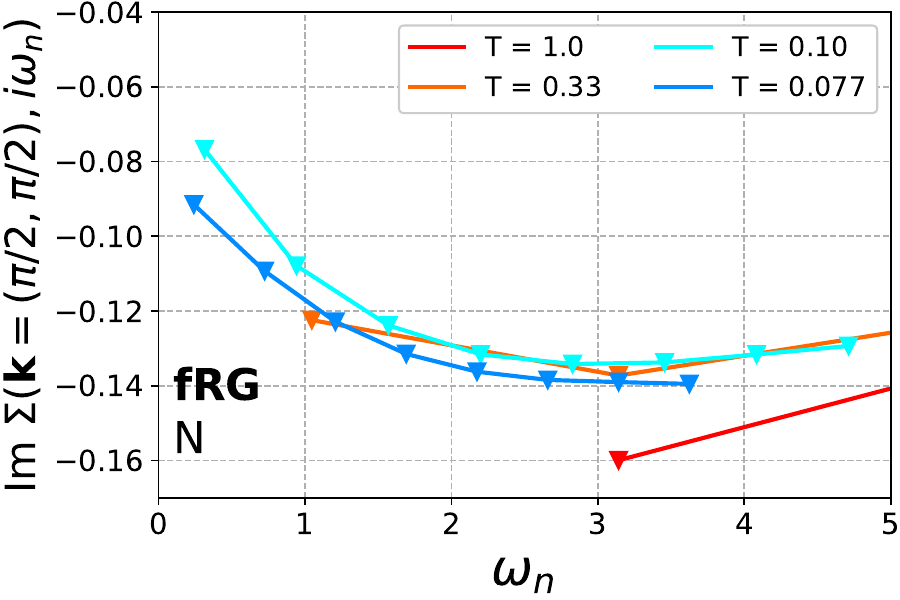}
                \includegraphics[width=0.32\textwidth,angle=0]{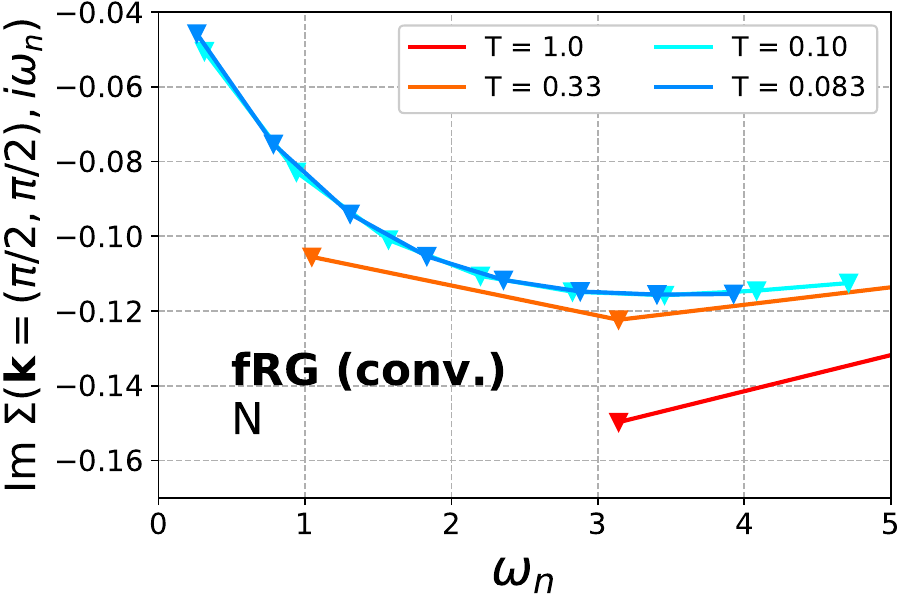}
                \includegraphics[width=0.32\textwidth,angle=0]{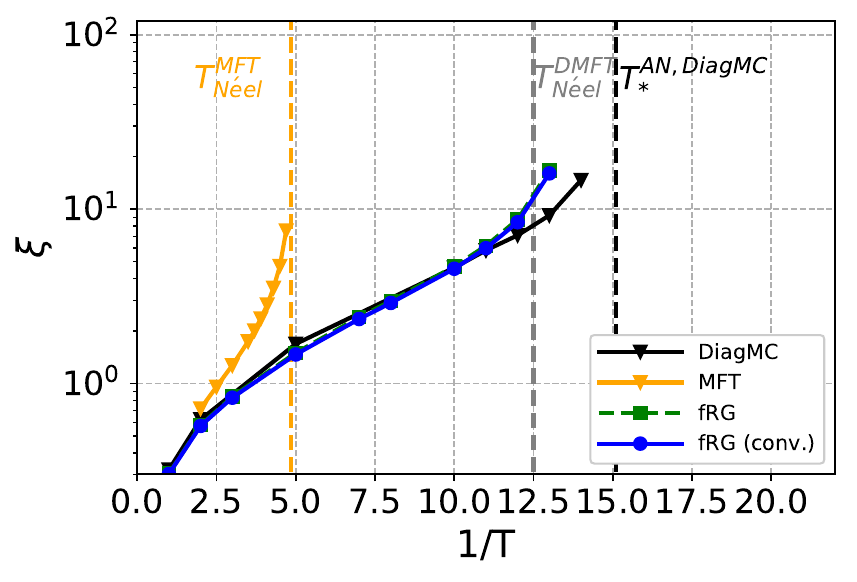}
      \caption{\label{fig:frg}(Color online.) Comparison of the imaginary parts of the self energies at the antinode (upper panel) and node (lower panel) as a function of Matsubara frequencies calculated by using the scale derivative of the Schwinger-Dyson equation (fRG, left panels, presented in the main text) and by the conventional self-energy flow [fRG (conv.), middle panels]. In contrast to the results shown in Figs.~\ref{fig:sigma_fluct_an} and \ref{fig:sigma_fluct_n} the reduced renormalization of the self-energy in fRG (conv.) leads to an enhanced spin susceptibility $\text{Re } \chi_{\text{sp}}(\mathbf{q}=(\pi,\pi), i\Omega_0)$ (right upper panel) and correlation length $\xi$ (lower right panel). The self-energy implementation in the conventional flow does not capture the pseudogap opening.}
\end{figure*}

Heuristically, another explanation of the approach comes from the expected cancellations between vertices and self-energy. Consider the Bethe-Salpeter equation for the
four-point function (susceptibilities)
\begin{align}
  \frac{\delta G}{\delta \phi} &= -G \frac{\delta G^{-1}}{\delta\phi} G
  \label{RPA_phi}
\end{align}
where we have used the identity $GG^{-1} = \delta(1-2)$.
Suppose a quasiparticle picture applies, namely $G = ZG_{0}$ where $Z$ is the quasiparticle weight. 

In some approximation, the $Z^{-1}$ from the vertices $\frac{\delta G^{-1}}{\delta\phi}$ cancels $Z$ from one of the Green functions. In some sense, $U_{sp}$ and $U_{ch}$ account for the average momentum-space screening of the interaction vertices while $Z^{-1}$ accounts for their frequency dependence.

Numerical results for TPSC+ in the present paper have been obtained on a $128\times 128$ lattice and
$2\times 512$ Matsubara frequencies. Fast Fourier transforms with a uniform grid for imaginary time $\tau$
are used in the self-consistent calculations of $\widetilde{\Sigma}_{\text{2PT}}$ and $\widetilde{\chi}_{\text{2PT}}$.

Eventually, we would like to comment on result from a TPSC+ variety, which utilizes the full Green function $G$ instead of $G_0$ and is therefore coined TPSC+GG. In Fig.~\ref{fig:dga} we already saw that there is a very good agreement of the TPSC+GG with D$\Gamma$A for the spin susceptibility and correlation length. Fig.~\ref{fig:tpsc} now also shows that at low $T$ the double occupancy agrees well with the benchmark, although it does not open the pseudogap.

\subsection{fRG}
\label{app:frg}

The characteristic scale-dependent behavior of numerous strongly correlated electron systems can be treated in a flexible and unbiased way by the functional renormalization group (fRG), see Refs.~\cite{Metzner2012,Salmhofer1999,Berges2002,Kopietz2010} for a review, (see also \cite{Deisz1996, Katanin2004b, Rohe2005, Kampf2006} for the context of the pseudogap). Its starting point is an exact functional flow equation, which yields the gradual evolution from a microscopic model action to the final effective action as a function of a continuously decreasing energy scale. Expanding in powers of the fields one obtains an exact hierarchy of flow equations for the $n$-particle irreducible vertex functions, which in practical implementations is truncated at the two-particle level.
\begin{table}
\begin{tabular}{|c|cccc|}
\hline
$1/T$ & $n$ & $k_x$ & $k_x^{\textrm{refine}}$ & $p_x$ \\
\hline
1     &  4 &  8      & 15 &  40 \\
2     &  4 &  8      & 15 &  40 \\
3     &  4 &  8      & 15 &  40 \\
5     &  6 & 12      & 15 &  60 \\
7     &  8 & 16      & 15 &  80 \\
8     &  8 & 16      & 15 &  80 \\
10    &  8 & 16      & 15 &  80 \\
11    &  8 & 16      & 15 &  80 \\
12    &  8 & 16      & 15 &  80 \\
13	  &  8 & 20      & 15 & 100 \\
\hline
\end{tabular}
\caption{Technical parameters for the fRG calculations for all temperatures, performed with a smooth frequency cutoff. $n$ is the number of positive fermionic frequencies that determined the parametrization of the two-particle vertex (see Ref.~\protect\cite{Tagliavini2019} for the definitions). The rest function contains $(4n+1)\times(2n)\times(2n)$, the $K_2$-function and the fermion-boson vertex $(4n+1)\times(2n)$, the $K_1$-function $(128n+1)$, and the self-energy $(8n)$ frequencies. The fermionic momentum dependence of the vertices and response functions is accounted for by a form factor expansion, where we considered only the local $s$-wave  contribution since at half filling the physics is dominated by antiferromagnetic fluctuations. 
The remaining momentum dependence of the vertices, the response functions and the self-energy are calculated on $(k_x \times k_x)$ equally spaced momentum patches, while the Green's functions and their summation in the particle-hole and particle-particle excitation are calculated on a $(p_x \times p_x)$ grid. The grid for the vertices and response functions is refined around $\mathbf{q}=(\pi,\pi)$ adding $(k_x^{\textrm{refine}}\times k_x^{\textrm{refine}})$ patches from the $(p_x \times p_x)$-grid. This accounts for $216$ additional patches with respect to the ones already included in the $(k_x \times k_x)$ grid. 
}
\label{tab:fRG_parameters}
\end{table}
Neglecting the renormalization of three- and higher order particle vertices yields approximate 1-loop flow equations for the self-energy and two-particle vertex~\footnote{Strongly correlated parameter regimes are beyond the 1-loop flow but might become accessible by exploiting the DMFT as a starting point for the fRG flow \cite{Taranto2014,Vilardi2019}.}. 

The underlying approximations are devised for the weak to moderate coupling regime. A substantial improvement with respect to previous fRG-based computation schemes relies on an efficient parametrization of the two-particle vertex, where we combine the so-called ”truncated unity” fRG \cite{Husemann2009,Wang2012,Lichtenstein2017} using the channel decomposition in conjunction with a form factor expansion for the momentum dependence with the full frequency treatment \cite{Vilardi2017} that includes the high-frequency asymptotics \cite{Rohringer2012,Wentzell2020}. We here use the Katanin replacement \cite{Katanin2004} in the flow equation for the two-particle vertex as a first step towards the multiloop extension of the fRG which would allow to sum up all the diagrams of the parquet approximation with their exact weight \cite{Kugler2018a,Kugler2018,Tagliavini2019}. In order to account for the form-factor truncation, the self-energy flow is determined by the scale derivative of the Schwinger-Dyson equation \cite{Hille2020,Hille2020b} replacing the conventional one-loop flow equation (in Fig.~\ref{fig:frg}). We refer to Refs. \cite{Hille2020,Hille2020b,Tagliavini2019} for the details of the algorithmic implementation and in particular also for the post-processed computation of the susceptibilities from the flowing vertex and self-energy, the technical parameters are reported in Tab.~\ref{tab:fRG_parameters}. The double occupancy was calculated from two-particle quantities according to
\begin{align} 
D=\frac{1}{2\beta}\sum_{{\bf q}, \Omega_n}  \left[ \chi_{\text{ch}}({\bf q}, i\Omega_n) -\chi_{\text{sp}}({\bf q}, i\Omega_n) \right] +\frac{n^2}{4} \,.
\end{align}
We note that taking into account also multiloop corrections, these algorithmic advancements have been shown to bring the fRG for interacting fermions on 2D lattices to a quantitatively reliable level, in particular recovering the PA \cite{Hille2020, Tagliavini2019}.
\subsection{PA}
\label{app:pa}
The parquet approach is a diagrammatic scheme first introduced by DeDominicis and Martin in 1964~\cite{DeDominicis1964b}. It relies on the classification of diagrams contributing to the full 1-particle irreducible 2-particle vertex in terms of their 2-particle reducibility. These diagrams can be either fully 2-particle irreducible or 2-particle reducible in one of three channels~\cite{DeDominicis1964, Bickers2004, Vasiliev74}. The method is unbiased with respect to the channels and exact if using the exact fully 2-particle irreducible vertex as an input.

The parquet approximation (PA) applied here, consists in setting this fully 2-particle irreducible vertex to the bare Hubbard $U$ i.e. its lowest order contribution. This is correct only up to fourth order in perturbation theory. The PA has previously been used to study several condensed matter problems including the Hubbard model \cite{Yang09, Tam2013, Kauch2019, Kauch2020, Li19, Pudleiner19a, Pudleiner19b}.
The here presented results have been obtained with the TUPS~\cite{Eckhardt2020} implementation which relies on the method of Truncated Unities~\cite{Lichtenstein2017, Eckhardt18} and vertex asymptotics~\cite{Wentzell2020,Li16} in order to represent vertex functions as well as a previously developed parallelization scheme in order to perform computations efficiently \cite{Tam2013, Li19}.

All results are converged in the
number of discrete lattice momenta (Nq) and positive
fermionic Matsubara frequencies (Nf+) used.  Concretely we use a linear
mesh size of $N_q = 48$ and $N_{f+} = 40$ at $\beta t= 16$ and
fewer frequencies and momenta for higher temperatures. The results for
$\beta t=20$ have been extrapolated to infinite momentum grid
size. The number of basis functions ($N_{FF}$) inherent to the
Truncated Unity method \cite{Eckhardt18} was set to $N_{FF} = 9$ for all calculations. As has been shown in \cite{Eckhardt18} it is sufficient to reproduce the full basis results for $\beta t \leq 5$ but constitutes an additional approximation for lower temperatures. This approximation is quite likely responsible for the PA underestimation of AFM susceptibility (and, as a consequence, the pseudogap temperature) in comparison with the benchmark. The AFM susceptibility is also underestimated in the full parquet D$\Gamma$A scheme, when the number of basis functions $N_{FF}$ is limited to  $N_{FF} \leq 9$, as has been shown in \cite{Kaufmann2020}. A very promising new computational scheme has recently been proposed \cite{Krien2020b} that may allow for significant reduction of frequency box sizes and thus make computations with more basis functions ($N_{FF} > 9$) feasible.
The double occupancy was calculated using Eq.~(\ref{eqn:docc_2}). Please note, that in the PA the sum rule relating the double occupancy obtained from one-particle and two-particle quantities
\begin{eqnarray}
D &=&  \frac{1}{2\beta} \sum_{{\bf q},\Omega_n}\left[ \chi_{\text{ch}}({\bf q},i\Omega_n) - \chi_{\text{sp}} ({\bf q},i\Omega_n)\right] + \frac{n^2}{4}\nonumber\\
&=& \frac{1}{\beta} \sum_{{\bf k},\omega_n} \Sigma({\bf k},i\omega_n) G({\bf k},i\omega_n)
\end{eqnarray}
is fulfilled by construction. The PA also fulfills \cite{Bickers1992, Eckhardt2020} the  Mermin-Wagner theorem \cite{Mermin1966, Hohenberg1967}.

\bibliography{main}

\begin{thebibliography}{288}%
\makeatletter
\providecommand \@ifxundefined [1]{%
 \@ifx{#1\undefined}
}%
\providecommand \@ifnum [1]{%
 \ifnum #1\expandafter \@firstoftwo
 \else \expandafter \@secondoftwo
 \fi
}%
\providecommand \@ifx [1]{%
 \ifx #1\expandafter \@firstoftwo
 \else \expandafter \@secondoftwo
 \fi
}%
\providecommand \natexlab [1]{#1}%
\providecommand \enquote  [1]{``#1''}%
\providecommand \bibnamefont  [1]{#1}%
\providecommand \bibfnamefont [1]{#1}%
\providecommand \citenamefont [1]{#1}%
\providecommand \href@noop [0]{\@secondoftwo}%
\providecommand \href [0]{\begingroup \@sanitize@url \@href}%
\providecommand \@href[1]{\@@startlink{#1}\@@href}%
\providecommand \@@href[1]{\endgroup#1\@@endlink}%
\providecommand \@sanitize@url [0]{\catcode `\\12\catcode `\$12\catcode
  `\&12\catcode `\#12\catcode `\^12\catcode `\_12\catcode `\%12\relax}%
\providecommand \@@startlink[1]{}%
\providecommand \@@endlink[0]{}%
\providecommand \url  [0]{\begingroup\@sanitize@url \@url }%
\providecommand \@url [1]{\endgroup\@href {#1}{\urlprefix }}%
\providecommand \urlprefix  [0]{URL }%
\providecommand \Eprint [0]{\href }%
\providecommand \doibase [0]{http://dx.doi.org/}%
\providecommand \selectlanguage [0]{\@gobble}%
\providecommand \bibinfo  [0]{\@secondoftwo}%
\providecommand \bibfield  [0]{\@secondoftwo}%
\providecommand \translation [1]{[#1]}%
\providecommand \BibitemOpen [0]{}%
\providecommand \bibitemStop [0]{}%
\providecommand \bibitemNoStop [0]{.\EOS\space}%
\providecommand \EOS [0]{\spacefactor3000\relax}%
\providecommand \BibitemShut  [1]{\csname bibitem#1\endcsname}%
\let\auto@bib@innerbib\@empty
\bibitem [{\citenamefont {Hubbard}(1963)}]{Hubbard1963}%
  \BibitemOpen
  \bibfield  {author} {\bibinfo {author} {\bibfnamefont {J.}~\bibnamefont
  {Hubbard}},\ }\bibfield  {title} {\enquote {\bibinfo {title} {{Electron
  Correlations in Narrow Energy Bands}},}\ }\href {\doibase
  10.1098/rspa.1963.0204} {\bibfield  {journal} {\bibinfo  {journal}
  {Proceedings of the Royal Society of London. Series A, Mathematical and
  Physical Sciences}\ }\textbf {\bibinfo {volume} {276}},\ \bibinfo {pages}
  {238--257} (\bibinfo {year} {1963})}\BibitemShut {NoStop}%
\bibitem [{\citenamefont {Hubbard}(1964)}]{Hubbard1964}%
  \BibitemOpen
  \bibfield  {author} {\bibinfo {author} {\bibfnamefont {J.}~\bibnamefont
  {Hubbard}},\ }\bibfield  {title} {\enquote {\bibinfo {title} {{Electron
  Correlations in Narrow Energy Bands. III. An Improved Solution}},}\ }\href
  {\doibase 10.1098/rspa.1964.0190} {\bibfield  {journal} {\bibinfo  {journal}
  {Proc R. Soc. London}\ }\textbf {\bibinfo {volume} {281}},\ \bibinfo {pages}
  {401--419} (\bibinfo {year} {1964})}\BibitemShut {NoStop}%
\bibitem [{\citenamefont {Gutzwiller}(1963)}]{Gutzwiller1963}%
  \BibitemOpen
  \bibfield  {author} {\bibinfo {author} {\bibfnamefont {Martin~C.}\
  \bibnamefont {Gutzwiller}},\ }\bibfield  {title} {\enquote {\bibinfo {title}
  {{Effect of Correlation on the Ferromagnetism of Transition Metals}},}\
  }\href {\doibase 10.1103/PhysRevLett.10.159} {\bibfield  {journal} {\bibinfo
  {journal} {Phys. Rev. Lett.}\ }\textbf {\bibinfo {volume} {10}},\ \bibinfo
  {pages} {159--162} (\bibinfo {year} {1963})}\BibitemShut {NoStop}%
\bibitem [{\citenamefont {Kanamori}(1963)}]{Kanamori1963}%
  \BibitemOpen
  \bibfield  {author} {\bibinfo {author} {\bibfnamefont {Junjiro}\ \bibnamefont
  {Kanamori}},\ }\bibfield  {title} {\enquote {\bibinfo {title} {{Electron
  Correlation and Ferromagnetism of Transition Metals}},}\ }\href {\doibase
  10.1143/PTP.30.275} {\bibfield  {journal} {\bibinfo  {journal} {Progress of
  Theoretical Physics}\ }\textbf {\bibinfo {volume} {30}},\ \bibinfo {pages}
  {275--289} (\bibinfo {year} {1963})},\ \Eprint
  {http://arxiv.org/abs/https://academic.oup.com/ptp/article-pdf/30/3/275/5278869/30-3-275.pdf}
  {https://academic.oup.com/ptp/article-pdf/30/3/275/5278869/30-3-275.pdf}
  \BibitemShut {NoStop}%
\bibitem [{\citenamefont {Jaksch}\ and\ \citenamefont
  {Zoller}(2005)}]{Jaksch2005}%
  \BibitemOpen
  \bibfield  {author} {\bibinfo {author} {\bibfnamefont {D.}~\bibnamefont
  {Jaksch}}\ and\ \bibinfo {author} {\bibfnamefont {P.}~\bibnamefont
  {Zoller}},\ }\bibfield  {title} {\enquote {\bibinfo {title} {The cold atom
  {Hubbard} toolbox},}\ }\href {\doibase
  https://doi.org/10.1016/j.aop.2004.09.010} {\bibfield  {journal} {\bibinfo
  {journal} {Annals of Physics}\ }\textbf {\bibinfo {volume} {315}},\ \bibinfo
  {pages} {52 -- 79} (\bibinfo {year} {2005})},\ \bibinfo {note} {special
  Issue}\BibitemShut {NoStop}%
\bibitem [{\citenamefont {{Bloch}}(2005)}]{Bloch2005}%
  \BibitemOpen
  \bibfield  {author} {\bibinfo {author} {\bibfnamefont {Immanuel}\
  \bibnamefont {{Bloch}}},\ }\bibfield  {title} {\enquote {\bibinfo {title}
  {{Ultracold quantum gases in optical lattices}},}\ }\href {\doibase
  10.1038/nphys138} {\bibfield  {journal} {\bibinfo  {journal} {Nature
  Physics}\ }\textbf {\bibinfo {volume} {1}},\ \bibinfo {pages} {23--30}
  (\bibinfo {year} {2005})}\BibitemShut {NoStop}%
\bibitem [{\citenamefont {Lewenstein}\ \emph {et~al.}(2007)\citenamefont
  {Lewenstein}, \citenamefont {Sanpera}, \citenamefont {Ahufinger},
  \citenamefont {Damski}, \citenamefont {Sen(De)},\ and\ \citenamefont
  {Sen}}]{Lewenstein2007}%
  \BibitemOpen
  \bibfield  {author} {\bibinfo {author} {\bibfnamefont {Maciej}\ \bibnamefont
  {Lewenstein}}, \bibinfo {author} {\bibfnamefont {Anna}\ \bibnamefont
  {Sanpera}}, \bibinfo {author} {\bibfnamefont {Veronica}\ \bibnamefont
  {Ahufinger}}, \bibinfo {author} {\bibfnamefont {Bogdan}\ \bibnamefont
  {Damski}}, \bibinfo {author} {\bibfnamefont {Aditi}\ \bibnamefont {Sen(De)}},
  \ and\ \bibinfo {author} {\bibfnamefont {Ujjwal}\ \bibnamefont {Sen}},\
  }\bibfield  {title} {\enquote {\bibinfo {title} {Ultracold atomic gases in
  optical lattices: mimicking condensed matter physics and beyond},}\ }\href
  {\doibase 10.1080/00018730701223200} {\bibfield  {journal} {\bibinfo
  {journal} {Advances in Physics}\ }\textbf {\bibinfo {volume} {56}},\ \bibinfo
  {pages} {243–379} (\bibinfo {year} {2007})}\BibitemShut {NoStop}%
\bibitem [{\citenamefont {Essler}\ \emph {et~al.}(2005)\citenamefont {Essler},
  \citenamefont {Frahm}, \citenamefont {G{\"o}hmann}, \citenamefont
  {Kl{\"u}mper},\ and\ \citenamefont {Korepin}}]{Essler2005}%
  \BibitemOpen
  \bibfield  {author} {\bibinfo {author} {\bibfnamefont {F.~H.~L.}\
  \bibnamefont {Essler}}, \bibinfo {author} {\bibfnamefont {H.}~\bibnamefont
  {Frahm}}, \bibinfo {author} {\bibfnamefont {F.}~\bibnamefont {G{\"o}hmann}},
  \bibinfo {author} {\bibfnamefont {A.}~\bibnamefont {Kl{\"u}mper}}, \ and\
  \bibinfo {author} {\bibfnamefont {V.}~\bibnamefont {Korepin}},\ }\href@noop
  {} {\emph {\bibinfo {title} {The One-Dimensional Hubbard model}}}\ (\bibinfo
  {publisher} {Cambridge University Press},\ \bibinfo {year}
  {2005})\BibitemShut {NoStop}%
\bibitem [{\citenamefont {Metzner}\ and\ \citenamefont
  {Vollhardt}(1989)}]{Metzner1989}%
  \BibitemOpen
  \bibfield  {author} {\bibinfo {author} {\bibfnamefont {Walter}\ \bibnamefont
  {Metzner}}\ and\ \bibinfo {author} {\bibfnamefont {Dieter}\ \bibnamefont
  {Vollhardt}},\ }\bibfield  {title} {\enquote {\bibinfo {title} {{Correlated
  Lattice Fermions in $d=\infty$ Dimensions}},}\ }\href {\doibase
  10.1103/PhysRevLett.62.324} {\bibfield  {journal} {\bibinfo  {journal} {Phys.
  Rev. Lett.}\ }\textbf {\bibinfo {volume} {62}},\ \bibinfo {pages} {324--327}
  (\bibinfo {year} {1989})}\BibitemShut {NoStop}%
\bibitem [{\citenamefont {Georges}\ and\ \citenamefont
  {Kotliar}(1992)}]{Georges1992a}%
  \BibitemOpen
  \bibfield  {author} {\bibinfo {author} {\bibfnamefont {Antoine}\ \bibnamefont
  {Georges}}\ and\ \bibinfo {author} {\bibfnamefont {Gabriel}\ \bibnamefont
  {Kotliar}},\ }\bibfield  {title} {\enquote {\bibinfo {title} {Hubbard model
  in infinite dimensions},}\ }\href {\doibase 10.1103/PhysRevB.45.6479}
  {\bibfield  {journal} {\bibinfo  {journal} {Phys. Rev. B}\ }\textbf {\bibinfo
  {volume} {45}},\ \bibinfo {pages} {6479--6483} (\bibinfo {year}
  {1992})}\BibitemShut {NoStop}%
\bibitem [{\citenamefont {Jarrell}(1992)}]{Jarrell1992}%
  \BibitemOpen
  \bibfield  {author} {\bibinfo {author} {\bibfnamefont {M.}~\bibnamefont
  {Jarrell}},\ }\bibfield  {title} {\enquote {\bibinfo {title} {{Hubbard model
  in infinite dimensions: A quantum Monte Carlo study}},}\ }\href {\doibase
  10.1103/PhysRevLett.69.168} {\bibfield  {journal} {\bibinfo  {journal} {Phys.
  Rev. Lett.}\ }\textbf {\bibinfo {volume} {69}},\ \bibinfo {pages} {168--171}
  (\bibinfo {year} {1992})}\BibitemShut {NoStop}%
\bibitem [{\citenamefont {{Keimer}}\ \emph {et~al.}(2015)\citenamefont
  {{Keimer}}, \citenamefont {{Kivelson}}, \citenamefont {{Norman}},
  \citenamefont {{Uchida}},\ and\ \citenamefont {{Zaanen}}}]{Keimer2015}%
  \BibitemOpen
  \bibfield  {author} {\bibinfo {author} {\bibfnamefont {B.}~\bibnamefont
  {{Keimer}}}, \bibinfo {author} {\bibfnamefont {S.~A.}\ \bibnamefont
  {{Kivelson}}}, \bibinfo {author} {\bibfnamefont {M.~R.}\ \bibnamefont
  {{Norman}}}, \bibinfo {author} {\bibfnamefont {S.}~\bibnamefont {{Uchida}}},
  \ and\ \bibinfo {author} {\bibfnamefont {J.}~\bibnamefont {{Zaanen}}},\
  }\bibfield  {title} {\enquote {\bibinfo {title} {{{From quantum matter to
  high-temperature superconductivity in copper oxides}}},}\ }\href {\doibase
  10.1038/nature14165} {\bibfield  {journal} {\bibinfo  {journal} {\nat}\
  }\textbf {\bibinfo {volume} {518}},\ \bibinfo {pages} {179--186} (\bibinfo
  {year} {2015})}\BibitemShut {NoStop}%
\bibitem [{\citenamefont {Timusk}\ and\ \citenamefont
  {Statt}(1999)}]{Timusk1999}%
  \BibitemOpen
  \bibfield  {author} {\bibinfo {author} {\bibfnamefont {Tom}\ \bibnamefont
  {Timusk}}\ and\ \bibinfo {author} {\bibfnamefont {Bryan}\ \bibnamefont
  {Statt}},\ }\bibfield  {title} {\enquote {\bibinfo {title} {The pseudogap in
  high-temperature superconductors: an experimental survey},}\ }\href {\doibase
  10.1088/0034-4885/62/1/002} {\bibfield  {journal} {\bibinfo  {journal}
  {Reports on Progress in Physics}\ }\textbf {\bibinfo {volume} {62}},\
  \bibinfo {pages} {61--122} (\bibinfo {year} {1999})}\BibitemShut {NoStop}%
\bibitem [{\citenamefont {{Norman}}\ \emph {et~al.}(2005)\citenamefont
  {{Norman}}, \citenamefont {{Pines}},\ and\ \citenamefont
  {{Kallin}}}]{Norman2005}%
  \BibitemOpen
  \bibfield  {author} {\bibinfo {author} {\bibfnamefont {M.~R.}\ \bibnamefont
  {{Norman}}}, \bibinfo {author} {\bibfnamefont {D.}~\bibnamefont {{Pines}}}, \
  and\ \bibinfo {author} {\bibfnamefont {C.}~\bibnamefont {{Kallin}}},\
  }\bibfield  {title} {\enquote {\bibinfo {title} {{The pseudogap: friend or
  foe of high Tc?}}}\ }\href {\doibase 10.1080/00018730500459906} {\bibfield
  {journal} {\bibinfo  {journal} {Advances in Physics}\ }\textbf {\bibinfo
  {volume} {54}},\ \bibinfo {pages} {715--733} (\bibinfo {year} {2005})},\
  \Eprint {http://arxiv.org/abs/cond-mat/0507031} {arXiv:cond-mat/0507031
  [cond-mat.supr-con]} \BibitemShut {NoStop}%
\bibitem [{\citenamefont {Lee}\ \emph {et~al.}(2006)\citenamefont {Lee},
  \citenamefont {Nagaosa},\ and\ \citenamefont {Wen}}]{Lee2006}%
  \BibitemOpen
  \bibfield  {author} {\bibinfo {author} {\bibfnamefont {Patrick~A.}\
  \bibnamefont {Lee}}, \bibinfo {author} {\bibfnamefont {Naoto}\ \bibnamefont
  {Nagaosa}}, \ and\ \bibinfo {author} {\bibfnamefont {Xiao-Gang}\ \bibnamefont
  {Wen}},\ }\bibfield  {title} {\enquote {\bibinfo {title} {{Doping a {Mott}
  insulator: Physics of high-temperature superconductivity}},}\ }\href
  {\doibase 10.1103/RevModPhys.78.17} {\bibfield  {journal} {\bibinfo
  {journal} {Rev. Mod. Phys.}\ }\textbf {\bibinfo {volume} {78}},\ \bibinfo
  {pages} {17--85} (\bibinfo {year} {2006})}\BibitemShut {NoStop}%
\bibitem [{\citenamefont {LeBlanc}\ \emph {et~al.}(2015)\citenamefont
  {LeBlanc}, \citenamefont {Antipov}, \citenamefont {Becca}, \citenamefont
  {Bulik}, \citenamefont {Chan}, \citenamefont {Chung}, \citenamefont {Deng},
  \citenamefont {Ferrero}, \citenamefont {Henderson}, \citenamefont
  {Jim\'enez-Hoyos}, \citenamefont {Kozik}, \citenamefont {Liu}, \citenamefont
  {Millis}, \citenamefont {Prokof'ev}, \citenamefont {Qin}, \citenamefont
  {Scuseria}, \citenamefont {Shi}, \citenamefont {Svistunov}, \citenamefont
  {Tocchio}, \citenamefont {Tupitsyn}, \citenamefont {White}, \citenamefont
  {Zhang}, \citenamefont {Zheng}, \citenamefont {Zhu},\ and\ \citenamefont
  {Gull}}]{LeBlanc2015}%
  \BibitemOpen
  \bibfield  {author} {\bibinfo {author} {\bibfnamefont {J.~P.~F.}\
  \bibnamefont {LeBlanc}}, \bibinfo {author} {\bibfnamefont {Andrey~E.}\
  \bibnamefont {Antipov}}, \bibinfo {author} {\bibfnamefont {Federico}\
  \bibnamefont {Becca}}, \bibinfo {author} {\bibfnamefont {Ireneusz~W.}\
  \bibnamefont {Bulik}}, \bibinfo {author} {\bibfnamefont {Garnet Kin-Lic}\
  \bibnamefont {Chan}}, \bibinfo {author} {\bibfnamefont {Chia-Min}\
  \bibnamefont {Chung}}, \bibinfo {author} {\bibfnamefont {Youjin}\
  \bibnamefont {Deng}}, \bibinfo {author} {\bibfnamefont {Michel}\ \bibnamefont
  {Ferrero}}, \bibinfo {author} {\bibfnamefont {Thomas~M.}\ \bibnamefont
  {Henderson}}, \bibinfo {author} {\bibfnamefont {Carlos~A.}\ \bibnamefont
  {Jim\'enez-Hoyos}}, \bibinfo {author} {\bibfnamefont {E.}~\bibnamefont
  {Kozik}}, \bibinfo {author} {\bibfnamefont {Xuan-Wen}\ \bibnamefont {Liu}},
  \bibinfo {author} {\bibfnamefont {Andrew~J.}\ \bibnamefont {Millis}},
  \bibinfo {author} {\bibfnamefont {N.~V.}\ \bibnamefont {Prokof'ev}}, \bibinfo
  {author} {\bibfnamefont {Mingpu}\ \bibnamefont {Qin}}, \bibinfo {author}
  {\bibfnamefont {Gustavo~E.}\ \bibnamefont {Scuseria}}, \bibinfo {author}
  {\bibfnamefont {Hao}\ \bibnamefont {Shi}}, \bibinfo {author} {\bibfnamefont
  {B.~V.}\ \bibnamefont {Svistunov}}, \bibinfo {author} {\bibfnamefont
  {Luca~F.}\ \bibnamefont {Tocchio}}, \bibinfo {author} {\bibfnamefont {I.~S.}\
  \bibnamefont {Tupitsyn}}, \bibinfo {author} {\bibfnamefont {Steven~R.}\
  \bibnamefont {White}}, \bibinfo {author} {\bibfnamefont {Shiwei}\
  \bibnamefont {Zhang}}, \bibinfo {author} {\bibfnamefont {Bo-Xiao}\
  \bibnamefont {Zheng}}, \bibinfo {author} {\bibfnamefont {Zhenyue}\
  \bibnamefont {Zhu}}, \ and\ \bibinfo {author} {\bibfnamefont {Emanuel}\
  \bibnamefont {Gull}} (\bibinfo {collaboration} {Simons Collaboration on the
  Many-Electron Problem}),\ }\bibfield  {title} {\enquote {\bibinfo {title}
  {{Solutions of the Two-Dimensional Hubbard Model: Benchmarks and Results from
  a Wide Range of Numerical Algorithms}},}\ }\href {\doibase
  10.1103/PhysRevX.5.041041} {\bibfield  {journal} {\bibinfo  {journal} {Phys.
  Rev. X}\ }\textbf {\bibinfo {volume} {5}},\ \bibinfo {pages} {041041}
  (\bibinfo {year} {2015})}\BibitemShut {NoStop}%
\bibitem [{\citenamefont {Motta}\ \emph {et~al.}(2017)\citenamefont {Motta},
  \citenamefont {Ceperley}, \citenamefont {Chan}, \citenamefont {Gomez},
  \citenamefont {Gull}, \citenamefont {Guo}, \citenamefont {Jim\'enez-Hoyos},
  \citenamefont {Lan}, \citenamefont {Li}, \citenamefont {Ma}, \citenamefont
  {Millis}, \citenamefont {Prokof'ev}, \citenamefont {Ray}, \citenamefont
  {Scuseria}, \citenamefont {Sorella}, \citenamefont {Stoudenmire},
  \citenamefont {Sun}, \citenamefont {Tupitsyn}, \citenamefont {White},
  \citenamefont {Zgid},\ and\ \citenamefont {Zhang}}]{Motta2017}%
  \BibitemOpen
  \bibfield  {author} {\bibinfo {author} {\bibfnamefont {Mario}\ \bibnamefont
  {Motta}}, \bibinfo {author} {\bibfnamefont {David~M.}\ \bibnamefont
  {Ceperley}}, \bibinfo {author} {\bibfnamefont {Garnet Kin-Lic}\ \bibnamefont
  {Chan}}, \bibinfo {author} {\bibfnamefont {John~A.}\ \bibnamefont {Gomez}},
  \bibinfo {author} {\bibfnamefont {Emanuel}\ \bibnamefont {Gull}}, \bibinfo
  {author} {\bibfnamefont {Sheng}\ \bibnamefont {Guo}}, \bibinfo {author}
  {\bibfnamefont {Carlos~A.}\ \bibnamefont {Jim\'enez-Hoyos}}, \bibinfo
  {author} {\bibfnamefont {Tran~Nguyen}\ \bibnamefont {Lan}}, \bibinfo {author}
  {\bibfnamefont {Jia}\ \bibnamefont {Li}}, \bibinfo {author} {\bibfnamefont
  {Fengjie}\ \bibnamefont {Ma}}, \bibinfo {author} {\bibfnamefont {Andrew~J.}\
  \bibnamefont {Millis}}, \bibinfo {author} {\bibfnamefont {Nikolay~V.}\
  \bibnamefont {Prokof'ev}}, \bibinfo {author} {\bibfnamefont {Ushnish}\
  \bibnamefont {Ray}}, \bibinfo {author} {\bibfnamefont {Gustavo~E.}\
  \bibnamefont {Scuseria}}, \bibinfo {author} {\bibfnamefont {Sandro}\
  \bibnamefont {Sorella}}, \bibinfo {author} {\bibfnamefont {Edwin~M.}\
  \bibnamefont {Stoudenmire}}, \bibinfo {author} {\bibfnamefont {Qiming}\
  \bibnamefont {Sun}}, \bibinfo {author} {\bibfnamefont {Igor~S.}\ \bibnamefont
  {Tupitsyn}}, \bibinfo {author} {\bibfnamefont {Steven~R.}\ \bibnamefont
  {White}}, \bibinfo {author} {\bibfnamefont {Dominika}\ \bibnamefont {Zgid}},
  \ and\ \bibinfo {author} {\bibfnamefont {Shiwei}\ \bibnamefont {Zhang}}
  (\bibinfo {collaboration} {Simons Collaboration on the Many-Electron
  Problem}),\ }\bibfield  {title} {\enquote {\bibinfo {title} {{Towards the
  Solution of the Many-Electron Problem in Real Materials: Equation of State of
  the Hydrogen Chain with State-of-the-Art Many-Body Methods}},}\ }\href
  {\doibase 10.1103/PhysRevX.7.031059} {\bibfield  {journal} {\bibinfo
  {journal} {Phys. Rev. X}\ }\textbf {\bibinfo {volume} {7}},\ \bibinfo {pages}
  {031059} (\bibinfo {year} {2017})}\BibitemShut {NoStop}%
\bibitem [{\citenamefont {Williams}\ \emph {et~al.}(2020)\citenamefont
  {Williams}, \citenamefont {Yao}, \citenamefont {Li}, \citenamefont {Chen},
  \citenamefont {Shi}, \citenamefont {Motta}, \citenamefont {Niu},
  \citenamefont {Ray}, \citenamefont {Guo}, \citenamefont {Anderson},
  \citenamefont {Li}, \citenamefont {Tran}, \citenamefont {Yeh}, \citenamefont
  {Mussard}, \citenamefont {Sharma}, \citenamefont {Bruneval}, \citenamefont
  {van Schilfgaarde}, \citenamefont {Booth}, \citenamefont {Chan},
  \citenamefont {Zhang}, \citenamefont {Gull}, \citenamefont {Zgid},
  \citenamefont {Millis}, \citenamefont {Umrigar},\ and\ \citenamefont
  {Wagner}}]{Williams2020}%
  \BibitemOpen
  \bibfield  {author} {\bibinfo {author} {\bibfnamefont {Kiel~T.}\ \bibnamefont
  {Williams}}, \bibinfo {author} {\bibfnamefont {Yuan}\ \bibnamefont {Yao}},
  \bibinfo {author} {\bibfnamefont {Jia}\ \bibnamefont {Li}}, \bibinfo {author}
  {\bibfnamefont {Li}~\bibnamefont {Chen}}, \bibinfo {author} {\bibfnamefont
  {Hao}\ \bibnamefont {Shi}}, \bibinfo {author} {\bibfnamefont {Mario}\
  \bibnamefont {Motta}}, \bibinfo {author} {\bibfnamefont {Chunyao}\
  \bibnamefont {Niu}}, \bibinfo {author} {\bibfnamefont {Ushnish}\ \bibnamefont
  {Ray}}, \bibinfo {author} {\bibfnamefont {Sheng}\ \bibnamefont {Guo}},
  \bibinfo {author} {\bibfnamefont {Robert~J.}\ \bibnamefont {Anderson}},
  \bibinfo {author} {\bibfnamefont {Junhao}\ \bibnamefont {Li}}, \bibinfo
  {author} {\bibfnamefont {Lan~Nguyen}\ \bibnamefont {Tran}}, \bibinfo {author}
  {\bibfnamefont {Chia-Nan}\ \bibnamefont {Yeh}}, \bibinfo {author}
  {\bibfnamefont {Bastien}\ \bibnamefont {Mussard}}, \bibinfo {author}
  {\bibfnamefont {Sandeep}\ \bibnamefont {Sharma}}, \bibinfo {author}
  {\bibfnamefont {Fabien}\ \bibnamefont {Bruneval}}, \bibinfo {author}
  {\bibfnamefont {Mark}\ \bibnamefont {van Schilfgaarde}}, \bibinfo {author}
  {\bibfnamefont {George~H.}\ \bibnamefont {Booth}}, \bibinfo {author}
  {\bibfnamefont {Garnet Kin-Lic}\ \bibnamefont {Chan}}, \bibinfo {author}
  {\bibfnamefont {Shiwei}\ \bibnamefont {Zhang}}, \bibinfo {author}
  {\bibfnamefont {Emanuel}\ \bibnamefont {Gull}}, \bibinfo {author}
  {\bibfnamefont {Dominika}\ \bibnamefont {Zgid}}, \bibinfo {author}
  {\bibfnamefont {Andrew}\ \bibnamefont {Millis}}, \bibinfo {author}
  {\bibfnamefont {Cyrus~J.}\ \bibnamefont {Umrigar}}, \ and\ \bibinfo {author}
  {\bibfnamefont {Lucas~K.}\ \bibnamefont {Wagner}} (\bibinfo {collaboration}
  {Simons Collaboration on the Many-Electron Problem}),\ }\bibfield  {title}
  {\enquote {\bibinfo {title} {{Direct Comparison of Many-Body Methods for
  Realistic Electronic Hamiltonians}},}\ }\href {\doibase
  10.1103/PhysRevX.10.011041} {\bibfield  {journal} {\bibinfo  {journal} {Phys.
  Rev. X}\ }\textbf {\bibinfo {volume} {10}},\ \bibinfo {pages} {011041}
  (\bibinfo {year} {2020})}\BibitemShut {NoStop}%
\bibitem [{\citenamefont {Neronov}(2019)}]{Neronov2019}%
  \BibitemOpen
  \bibfield  {author} {\bibinfo {author} {\bibfnamefont {Andrii}\ \bibnamefont
  {Neronov}},\ }\bibfield  {title} {\enquote {\bibinfo {title} {Introduction to
  multi-messenger astronomy},}\ }\href {\doibase
  10.1088/1742-6596/1263/1/012001} {\bibfield  {journal} {\bibinfo  {journal}
  {Journal of Physics: Conference Series}\ }\textbf {\bibinfo {volume}
  {1263}},\ \bibinfo {pages} {012001} (\bibinfo {year} {2019})}\BibitemShut
  {NoStop}%
\bibitem [{\citenamefont {Slater}(1951)}]{Slater1951}%
  \BibitemOpen
  \bibfield  {author} {\bibinfo {author} {\bibfnamefont {J.~C.}\ \bibnamefont
  {Slater}},\ }\bibfield  {title} {\enquote {\bibinfo {title} {{Magnetic
  Effects and the Hartree-Fock Equation}},}\ }\href {\doibase
  10.1103/PhysRev.82.538} {\bibfield  {journal} {\bibinfo  {journal} {Phys.
  Rev.}\ }\textbf {\bibinfo {volume} {82}},\ \bibinfo {pages} {538--541}
  (\bibinfo {year} {1951})}\BibitemShut {NoStop}%
\bibitem [{\citenamefont {Mermin}\ and\ \citenamefont
  {Wagner}(1966)}]{Mermin1966}%
  \BibitemOpen
  \bibfield  {author} {\bibinfo {author} {\bibfnamefont {N.~D.}\ \bibnamefont
  {Mermin}}\ and\ \bibinfo {author} {\bibfnamefont {H.}~\bibnamefont
  {Wagner}},\ }\bibfield  {title} {\enquote {\bibinfo {title} {{Absence of
  Ferromagnetism or Antiferromagnetism in One- or Two-Dimensional Isotropic
  Heisenberg Models}},}\ }\href {\doibase 10.1103/PhysRevLett.17.1307}
  {\bibfield  {journal} {\bibinfo  {journal} {Phys. Rev. Lett.}\ }\textbf
  {\bibinfo {volume} {17}},\ \bibinfo {pages} {1307--1307} (\bibinfo {year}
  {1966})}\BibitemShut {NoStop}%
\bibitem [{\citenamefont {Hohenberg}(1967)}]{Hohenberg1967}%
  \BibitemOpen
  \bibfield  {author} {\bibinfo {author} {\bibfnamefont {P.~C.}\ \bibnamefont
  {Hohenberg}},\ }\bibfield  {title} {\enquote {\bibinfo {title} {{Existence of
  Long-Range Order in One and Two Dimensions}},}\ }\href {\doibase
  10.1103/PhysRev.158.383} {\bibfield  {journal} {\bibinfo  {journal} {Phys.
  Rev.}\ }\textbf {\bibinfo {volume} {158}},\ \bibinfo {pages} {383--386}
  (\bibinfo {year} {1967})}\BibitemShut {NoStop}%
\bibitem [{\citenamefont {Blankenbecler}\ \emph {et~al.}(1981)\citenamefont
  {Blankenbecler}, \citenamefont {Scalapino},\ and\ \citenamefont
  {Sugar}}]{Blankenbecler1981}%
  \BibitemOpen
  \bibfield  {author} {\bibinfo {author} {\bibfnamefont {R.}~\bibnamefont
  {Blankenbecler}}, \bibinfo {author} {\bibfnamefont {D.~J.}\ \bibnamefont
  {Scalapino}}, \ and\ \bibinfo {author} {\bibfnamefont {R.~L.}\ \bibnamefont
  {Sugar}},\ }\bibfield  {title} {\enquote {\bibinfo {title} {{Monte Carlo
  calculations of coupled boson-fermion systems. I}},}\ }\href {\doibase
  10.1103/PhysRevD.24.2278} {\bibfield  {journal} {\bibinfo  {journal} {Phys.
  Rev. D}\ }\textbf {\bibinfo {volume} {24}},\ \bibinfo {pages} {2278--2286}
  (\bibinfo {year} {1981})}\BibitemShut {NoStop}%
\bibitem [{\citenamefont {Prokof'ev}\ and\ \citenamefont
  {Svistunov}(1998)}]{Prokofev1998}%
  \BibitemOpen
  \bibfield  {author} {\bibinfo {author} {\bibfnamefont {Nikolai~V.}\
  \bibnamefont {Prokof'ev}}\ and\ \bibinfo {author} {\bibfnamefont {Boris~V.}\
  \bibnamefont {Svistunov}},\ }\bibfield  {title} {\enquote {\bibinfo {title}
  {{Polaron Problem by Diagrammatic Quantum Monte Carlo}},}\ }\href {\doibase
  10.1103/PhysRevLett.81.2514} {\bibfield  {journal} {\bibinfo  {journal}
  {Phys. Rev. Lett.}\ }\textbf {\bibinfo {volume} {81}},\ \bibinfo {pages}
  {2514--2517} (\bibinfo {year} {1998})}\BibitemShut {NoStop}%
\bibitem [{\citenamefont {Rossi}(2017)}]{Rossi2017}%
  \BibitemOpen
  \bibfield  {author} {\bibinfo {author} {\bibfnamefont {Riccardo}\
  \bibnamefont {Rossi}},\ }\bibfield  {title} {\enquote {\bibinfo {title}
  {{Determinant Diagrammatic Monte Carlo Algorithm in the Thermodynamic
  Limit}},}\ }\href {\doibase 10.1103/PhysRevLett.119.045701} {\bibfield
  {journal} {\bibinfo  {journal} {Phys. Rev. Lett.}\ }\textbf {\bibinfo
  {volume} {119}},\ \bibinfo {pages} {045701} (\bibinfo {year}
  {2017})}\BibitemShut {NoStop}%
\bibitem [{\citenamefont {\ifmmode~\check{S}\else \v{S}\fi{}imkovic}\ and\
  \citenamefont {Kozik}(2019)}]{Simkovic2019}%
  \BibitemOpen
  \bibfield  {author} {\bibinfo {author} {\bibfnamefont {Fedor}\ \bibnamefont
  {\ifmmode~\check{S}\else \v{S}\fi{}imkovic}}\ and\ \bibinfo {author}
  {\bibfnamefont {Evgeny}\ \bibnamefont {Kozik}},\ }\bibfield  {title}
  {\enquote {\bibinfo {title} {{Determinant Monte Carlo for irreducible Feynman
  diagrams in the strongly correlated regime}},}\ }\href {\doibase
  10.1103/PhysRevB.100.121102} {\bibfield  {journal} {\bibinfo  {journal}
  {Phys. Rev. B}\ }\textbf {\bibinfo {volume} {100}},\ \bibinfo {pages}
  {121102(R)} (\bibinfo {year} {2019})}\BibitemShut {NoStop}%
\bibitem [{\citenamefont {Moutenet}\ \emph {et~al.}(2018)\citenamefont
  {Moutenet}, \citenamefont {Wu},\ and\ \citenamefont
  {Ferrero}}]{Moutenet2018}%
  \BibitemOpen
  \bibfield  {author} {\bibinfo {author} {\bibfnamefont {Alice}\ \bibnamefont
  {Moutenet}}, \bibinfo {author} {\bibfnamefont {Wei}\ \bibnamefont {Wu}}, \
  and\ \bibinfo {author} {\bibfnamefont {Michel}\ \bibnamefont {Ferrero}},\
  }\bibfield  {title} {\enquote {\bibinfo {title} {{Determinant Monte Carlo
  algorithms for dynamical quantities in fermionic systems}},}\ }\href
  {\doibase 10.1103/PhysRevB.97.085117} {\bibfield  {journal} {\bibinfo
  {journal} {Phys. Rev. B}\ }\textbf {\bibinfo {volume} {97}},\ \bibinfo
  {pages} {085117} (\bibinfo {year} {2018})}\BibitemShut {NoStop}%
\bibitem [{\citenamefont {Georges}\ \emph {et~al.}(1996)\citenamefont
  {Georges}, \citenamefont {Kotliar}, \citenamefont {Krauth},\ and\
  \citenamefont {Rozenberg}}]{Georges1996}%
  \BibitemOpen
  \bibfield  {author} {\bibinfo {author} {\bibfnamefont {Antoine}\ \bibnamefont
  {Georges}}, \bibinfo {author} {\bibfnamefont {Gabriel}\ \bibnamefont
  {Kotliar}}, \bibinfo {author} {\bibfnamefont {Werner}\ \bibnamefont
  {Krauth}}, \ and\ \bibinfo {author} {\bibfnamefont {Marcelo~J.}\ \bibnamefont
  {Rozenberg}},\ }\bibfield  {title} {\enquote {\bibinfo {title} {{Dynamical
  mean-field theory of strongly correlated fermion systems and the limit of
  infinite dimensions}},}\ }\href {\doibase 10.1103/RevModPhys.68.13}
  {\bibfield  {journal} {\bibinfo  {journal} {Rev. Mod. Phys.}\ }\textbf
  {\bibinfo {volume} {68}},\ \bibinfo {pages} {13} (\bibinfo {year}
  {1996})}\BibitemShut {NoStop}%
\bibitem [{\citenamefont {Maier}\ \emph
  {et~al.}(2005{\natexlab{a}})\citenamefont {Maier}, \citenamefont {Jarrell},
  \citenamefont {Pruschke},\ and\ \citenamefont {Hettler}}]{Maier2005}%
  \BibitemOpen
  \bibfield  {author} {\bibinfo {author} {\bibfnamefont {T.~A.}\ \bibnamefont
  {Maier}}, \bibinfo {author} {\bibfnamefont {M.}~\bibnamefont {Jarrell}},
  \bibinfo {author} {\bibfnamefont {T.}~\bibnamefont {Pruschke}}, \ and\
  \bibinfo {author} {\bibfnamefont {M.}~\bibnamefont {Hettler}},\ }\bibfield
  {title} {\enquote {\bibinfo {title} {{Quantum Cluster Theories}},}\ }\href
  {\doibase 10.1103/RevModPhys.77.1027} {\bibfield  {journal} {\bibinfo
  {journal} {Rev. Mod. Phys.}\ }\textbf {\bibinfo {volume} {77}},\ \bibinfo
  {pages} {1027} (\bibinfo {year} {2005}{\natexlab{a}})}\BibitemShut {NoStop}%
\bibitem [{\citenamefont {Kotliar}\ \emph {et~al.}(2006)\citenamefont
  {Kotliar}, \citenamefont {Savrasov}, \citenamefont {Haule}, \citenamefont
  {Oudovenko}, \citenamefont {Parcollet},\ and\ \citenamefont
  {Marianetti}}]{Kotliar2006}%
  \BibitemOpen
  \bibfield  {author} {\bibinfo {author} {\bibfnamefont {G.}~\bibnamefont
  {Kotliar}}, \bibinfo {author} {\bibfnamefont {S.~Y.}\ \bibnamefont
  {Savrasov}}, \bibinfo {author} {\bibfnamefont {K.}~\bibnamefont {Haule}},
  \bibinfo {author} {\bibfnamefont {V.~S.}\ \bibnamefont {Oudovenko}}, \bibinfo
  {author} {\bibfnamefont {O.}~\bibnamefont {Parcollet}}, \ and\ \bibinfo
  {author} {\bibfnamefont {C.~A.}\ \bibnamefont {Marianetti}},\ }\bibfield
  {title} {\enquote {\bibinfo {title} {Electronic structure calculations with
  dynamical mean-field theory},}\ }\href {\doibase 10.1103/RevModPhys.78.865}
  {\bibfield  {journal} {\bibinfo  {journal} {Rev. Mod. Phys.}\ }\textbf
  {\bibinfo {volume} {78}},\ \bibinfo {pages} {865} (\bibinfo {year}
  {2006})}\BibitemShut {NoStop}%
\bibitem [{\citenamefont {Tremblay}\ \emph {et~al.}(2006)\citenamefont
  {Tremblay}, \citenamefont {Kyung},\ and\ \citenamefont
  {S\'en\'echal}}]{Tremblay2006}%
  \BibitemOpen
  \bibfield  {author} {\bibinfo {author} {\bibfnamefont {A.~M.~S.}\
  \bibnamefont {Tremblay}}, \bibinfo {author} {\bibfnamefont {B.}~\bibnamefont
  {Kyung}}, \ and\ \bibinfo {author} {\bibfnamefont {D.}~\bibnamefont
  {S\'en\'echal}},\ }\bibfield  {title} {\enquote {\bibinfo {title} {Pseudogap
  and high-temperature superconductivity from weak to strong coupling. towards
  a quantitative theory},}\ }\href {http://dx.doi.org/10.1063/1.2199446}
  {\bibfield  {journal} {\bibinfo  {journal} {Low Temp. Phys.}\ }\textbf
  {\bibinfo {volume} {32}},\ \bibinfo {pages} {424--451} (\bibinfo {year}
  {2006})}\BibitemShut {NoStop}%
\bibitem [{\citenamefont {Lichtenstein}\ and\ \citenamefont
  {Katsnelson}(2000)}]{Lichtenstein2000}%
  \BibitemOpen
  \bibfield  {author} {\bibinfo {author} {\bibfnamefont {A.~I.}\ \bibnamefont
  {Lichtenstein}}\ and\ \bibinfo {author} {\bibfnamefont {M.~I.}\ \bibnamefont
  {Katsnelson}},\ }\bibfield  {title} {\enquote {\bibinfo {title}
  {{Antiferromagnetism and \textit{d}-wave superconductivity in cuprates: A
  cluster dynamical mean-field theory}},}\ }\href {\doibase
  10.1103/PhysRevB.62.R9283} {\bibfield  {journal} {\bibinfo  {journal} {Phys.
  Rev. B}\ }\textbf {\bibinfo {volume} {62}},\ \bibinfo {pages} {R9283--R9286}
  (\bibinfo {year} {2000})}\BibitemShut {NoStop}%
\bibitem [{\citenamefont {Rohringer}\ \emph
  {et~al.}(2018{\natexlab{a}})\citenamefont {Rohringer}, \citenamefont
  {Hafermann}, \citenamefont {Toschi}, \citenamefont {Katanin}, \citenamefont
  {Antipov}, \citenamefont {Katsnelson}, \citenamefont {Lichtenstein},
  \citenamefont {Rubtsov},\ and\ \citenamefont {Held}}]{RMPVertex}%
  \BibitemOpen
  \bibfield  {author} {\bibinfo {author} {\bibfnamefont {G.}~\bibnamefont
  {Rohringer}}, \bibinfo {author} {\bibfnamefont {H.}~\bibnamefont
  {Hafermann}}, \bibinfo {author} {\bibfnamefont {A.}~\bibnamefont {Toschi}},
  \bibinfo {author} {\bibfnamefont {A.~A.}\ \bibnamefont {Katanin}}, \bibinfo
  {author} {\bibfnamefont {A.~E.}\ \bibnamefont {Antipov}}, \bibinfo {author}
  {\bibfnamefont {M.~I.}\ \bibnamefont {Katsnelson}}, \bibinfo {author}
  {\bibfnamefont {A.~I.}\ \bibnamefont {Lichtenstein}}, \bibinfo {author}
  {\bibfnamefont {A.~N.}\ \bibnamefont {Rubtsov}}, \ and\ \bibinfo {author}
  {\bibfnamefont {K.}~\bibnamefont {Held}},\ }\bibfield  {title} {\enquote
  {\bibinfo {title} {Diagrammatic routes to nonlocal correlations beyond
  dynamical mean field theory},}\ }\href {\doibase
  10.1103/RevModPhys.90.025003} {\bibfield  {journal} {\bibinfo  {journal}
  {Rev. Mod. Phys.}\ }\textbf {\bibinfo {volume} {90}},\ \bibinfo {pages}
  {025003} (\bibinfo {year} {2018}{\natexlab{a}})}\BibitemShut {NoStop}%
\bibitem [{\citenamefont {Vilk}\ and\ \citenamefont
  {Tremblay}(1997)}]{Vilk1997}%
  \BibitemOpen
  \bibfield  {author} {\bibinfo {author} {\bibfnamefont {Y.~M.}\ \bibnamefont
  {Vilk}}\ and\ \bibinfo {author} {\bibfnamefont {A.-M.~S.}\ \bibnamefont
  {Tremblay}},\ }\bibfield  {title} {\enquote {\bibinfo {title}
  {{Non-Perturbative Many-Body Approach to the Hubbard Model and
  Single-Particle Pseudogap}},}\ }\href {\doibase 10.1051/jp1:1997135}
  {\bibfield  {journal} {\bibinfo  {journal} {J. Phys. I France}\ }\textbf
  {\bibinfo {volume} {7}},\ \bibinfo {pages} {1309--1368} (\bibinfo {year}
  {1997})}\BibitemShut {NoStop}%
\bibitem [{\citenamefont {Tremblay}(2011)}]{Tremblay2011}%
  \BibitemOpen
  \bibfield  {author} {\bibinfo {author} {\bibfnamefont {A.~M.~S.}\
  \bibnamefont {Tremblay}},\ }\bibfield  {title} {\enquote {\bibinfo {title}
  {{Two-particle-Self-Consistent Approach for the Hubbard model}},}\ }in\
  \href@noop {} {\emph {\bibinfo {booktitle} {Strongly Correlated Systems:
  Theoretical Methods}}},\ \bibinfo {editor} {edited by\ \bibinfo {editor}
  {\bibfnamefont {F.}~\bibnamefont {Mancini}}\ and\ \bibinfo {editor}
  {\bibfnamefont {A.}~\bibnamefont {Avella}}}\ (\bibinfo  {publisher} {Springer
  series},\ \bibinfo {year} {2011})\ Chap.~\bibinfo {chapter} {13}, pp.\
  \bibinfo {pages} {409--455}\BibitemShut {NoStop}%
\bibitem [{\citenamefont {Wang}\ \emph {et~al.}()\citenamefont {Wang},
  \citenamefont {Vilk},\ and\ \citenamefont {Tremblay}}]{Wang2019}%
  \BibitemOpen
  \bibfield  {author} {\bibinfo {author} {\bibfnamefont {Y.}~\bibnamefont
  {Wang}}, \bibinfo {author} {\bibfnamefont {Y.M.}\ \bibnamefont {Vilk}}, \
  and\ \bibinfo {author} {\bibfnamefont {A.-M.~S.}\ \bibnamefont {Tremblay}},\
  }\bibfield  {title} {\enquote {\bibinfo {title} {{An improved two-particle
  self-consistent approach}},}\ }\href@noop {} {\bibinfo  {journal} {To be
  published}\ }\BibitemShut {NoStop}%
\bibitem [{\citenamefont {Metzner}\ \emph {et~al.}(2012)\citenamefont
  {Metzner}, \citenamefont {Salmhofer}, \citenamefont {Honerkamp},
  \citenamefont {Meden},\ and\ \citenamefont {Sch\"onhammer}}]{Metzner2012}%
  \BibitemOpen
\bibfield  {journal} {  }\bibfield  {author} {\bibinfo {author} {\bibfnamefont
  {Walter}\ \bibnamefont {Metzner}}, \bibinfo {author} {\bibfnamefont
  {Manfred}\ \bibnamefont {Salmhofer}}, \bibinfo {author} {\bibfnamefont
  {Carsten}\ \bibnamefont {Honerkamp}}, \bibinfo {author} {\bibfnamefont
  {Volker}\ \bibnamefont {Meden}}, \ and\ \bibinfo {author} {\bibfnamefont
  {Kurt}\ \bibnamefont {Sch\"onhammer}},\ }\bibfield  {title} {\enquote
  {\bibinfo {title} {Functional renormalization group approach to correlated
  fermion systems},}\ }\href {\doibase 10.1103/RevModPhys.84.299} {\bibfield
  {journal} {\bibinfo  {journal} {Rev. Mod. Phys.}\ }\textbf {\bibinfo {volume}
  {84}},\ \bibinfo {pages} {299--352} (\bibinfo {year} {2012})}\BibitemShut
  {NoStop}%
\bibitem [{\citenamefont {{De Dominicis}}\ and\ \citenamefont
  {Martin}(1964{\natexlab{a}})}]{DeDominicis1964b}%
  \BibitemOpen
  \bibfield  {author} {\bibinfo {author} {\bibfnamefont {Cyrano}\ \bibnamefont
  {{De Dominicis}}}\ and\ \bibinfo {author} {\bibfnamefont {Paul~C.}\
  \bibnamefont {Martin}},\ }\bibfield  {title} {\enquote {\bibinfo {title}
  {{Stationary entropy principle and renormalization in normal and superfluid
  systems. II. diagrammatic formulation}},}\ }\href {\doibase
  10.1063/1.1704064} {\bibfield  {journal} {\bibinfo  {journal} {J. Math.
  Phys.}\ }\textbf {\bibinfo {volume} {5}},\ \bibinfo {pages} {31} (\bibinfo
  {year} {1964}{\natexlab{a}})}\BibitemShut {NoStop}%
\bibitem [{\citenamefont {Bickers}(2004)}]{Bickers2004}%
  \BibitemOpen
  \bibfield  {author} {\bibinfo {author} {\bibfnamefont {N.~E.}\ \bibnamefont
  {Bickers}},\ }\enquote {\bibinfo {title} {Theoretical methods for strongly
  correlated electrons},}\ \ (\bibinfo  {publisher} {Springer-Verlag New York
  Berlin Heidelbert},\ \bibinfo {year} {2004})\ Chap.~\bibinfo {chapter} {6},
  pp.\ \bibinfo {pages} {237--296}\BibitemShut {NoStop}%
\bibitem [{\citenamefont {White}(2009)}]{White2009}%
  \BibitemOpen
  \bibfield  {author} {\bibinfo {author} {\bibfnamefont {Steven~R.}\
  \bibnamefont {White}},\ }\bibfield  {title} {\enquote {\bibinfo {title}
  {{Minimally Entangled Typical Quantum States at Finite Temperature}},}\
  }\href {\doibase 10.1103/PhysRevLett.102.190601} {\bibfield  {journal}
  {\bibinfo  {journal} {Phys. Rev. Lett.}\ }\textbf {\bibinfo {volume} {102}},\
  \bibinfo {pages} {190601} (\bibinfo {year} {2009})}\BibitemShut {NoStop}%
\bibitem [{\citenamefont {Stoudenmire}\ and\ \citenamefont
  {White}(2010)}]{Stoudenmire2010}%
  \BibitemOpen
  \bibfield  {author} {\bibinfo {author} {\bibfnamefont {E~M}\ \bibnamefont
  {Stoudenmire}}\ and\ \bibinfo {author} {\bibfnamefont {Steven~R}\
  \bibnamefont {White}},\ }\bibfield  {title} {\enquote {\bibinfo {title}
  {Minimally entangled typical thermal state algorithms},}\ }\href {\doibase
  10.1088/1367-2630/12/5/055026} {\bibfield  {journal} {\bibinfo  {journal}
  {New Journal of Physics}\ }\textbf {\bibinfo {volume} {12}},\ \bibinfo
  {pages} {055026} (\bibinfo {year} {2010})}\BibitemShut {NoStop}%
\bibitem [{\citenamefont {Bruognolo}\ \emph {et~al.}(2015)\citenamefont
  {Bruognolo}, \citenamefont {von Delft},\ and\ \citenamefont
  {Weichselbaum}}]{Bruognolo2015}%
  \BibitemOpen
  \bibfield  {author} {\bibinfo {author} {\bibfnamefont {Benedikt}\
  \bibnamefont {Bruognolo}}, \bibinfo {author} {\bibfnamefont {Jan}\
  \bibnamefont {von Delft}}, \ and\ \bibinfo {author} {\bibfnamefont {Andreas}\
  \bibnamefont {Weichselbaum}},\ }\bibfield  {title} {\enquote {\bibinfo
  {title} {Symmetric minimally entangled typical thermal states},}\ }\href
  {\doibase 10.1103/PhysRevB.92.115105} {\bibfield  {journal} {\bibinfo
  {journal} {Phys. Rev. B}\ }\textbf {\bibinfo {volume} {92}},\ \bibinfo
  {pages} {115105} (\bibinfo {year} {2015})}\BibitemShut {NoStop}%
\bibitem [{\citenamefont {Wietek}\ \emph {et~al.}(2020)\citenamefont {Wietek},
  \citenamefont {He}, \citenamefont {White}, \citenamefont {Georges},\ and\
  \citenamefont {Stoudenmire}}]{wietek2020}%
  \BibitemOpen
  \bibfield  {author} {\bibinfo {author} {\bibfnamefont {Alexander}\
  \bibnamefont {Wietek}}, \bibinfo {author} {\bibfnamefont {Yuan-Yao}\
  \bibnamefont {He}}, \bibinfo {author} {\bibfnamefont {Steven~R.}\
  \bibnamefont {White}}, \bibinfo {author} {\bibfnamefont {Antoine}\
  \bibnamefont {Georges}}, \ and\ \bibinfo {author} {\bibfnamefont {E.~Miles}\
  \bibnamefont {Stoudenmire}},\ }\href@noop {} {\enquote {\bibinfo {title}
  {{Stripes, Antiferromagnetism, and the Pseudogap in the Doped Hubbard Model
  at Finite Temperature}},}\ } (\bibinfo {year} {2020}),\ \Eprint
  {http://arxiv.org/abs/2009.10736} {arXiv:2009.10736 [cond-mat.str-el]}
  \BibitemShut {NoStop}%
\bibitem [{\citenamefont {Wietek}\ \emph {et~al.}(2021)\citenamefont {Wietek},
  \citenamefont {Rossi}, \citenamefont {IV}, \citenamefont {Klett},
  \citenamefont {Hansmann}, \citenamefont {Ferrero}, \citenamefont
  {Stoudenmire}, \citenamefont {Sch{\"a}fer},\ and\ \citenamefont
  {Georges}}]{Wietek2021}%
  \BibitemOpen
  \bibfield  {author} {\bibinfo {author} {\bibfnamefont {Alexander}\
  \bibnamefont {Wietek}}, \bibinfo {author} {\bibfnamefont {Riccardo}\
  \bibnamefont {Rossi}}, \bibinfo {author} {\bibfnamefont {Fedor~{\v
  S}imkovic}\ \bibnamefont {IV}}, \bibinfo {author} {\bibfnamefont {Marcel}\
  \bibnamefont {Klett}}, \bibinfo {author} {\bibfnamefont {Philipp}\
  \bibnamefont {Hansmann}}, \bibinfo {author} {\bibfnamefont {Michel}\
  \bibnamefont {Ferrero}}, \bibinfo {author} {\bibfnamefont {E.~Miles}\
  \bibnamefont {Stoudenmire}}, \bibinfo {author} {\bibfnamefont {Thomas}\
  \bibnamefont {Sch{\"a}fer}}, \ and\ \bibinfo {author} {\bibfnamefont
  {Antoine}\ \bibnamefont {Georges}},\ }\href@noop {} {\enquote {\bibinfo
  {title} {{Mott insulating states with competing orders in the triangular
  lattice Hubbard model}},}\ } (\bibinfo {year} {2021}),\ \Eprint
  {http://arxiv.org/abs/2102.12904} {arXiv:2102.12904 [cond-mat.str-el]}
  \BibitemShut {NoStop}%
\bibitem [{Sup()}]{Suppl}%
  \BibitemOpen
  \href@noop {} {}\bibinfo {note} {See Supplemental Material available under
  this
  \href{https://journals.aps.org/prx/supplemental/10.1103/PhysRevX.11.011058}{link}.}\BibitemShut
  {Stop}%
\bibitem [{\citenamefont {Katanin}\ \emph {et~al.}(2009)\citenamefont
  {Katanin}, \citenamefont {Toschi},\ and\ \citenamefont {Held}}]{Katanin2009}%
  \BibitemOpen
  \bibfield  {author} {\bibinfo {author} {\bibfnamefont {A.~A.}\ \bibnamefont
  {Katanin}}, \bibinfo {author} {\bibfnamefont {A.}~\bibnamefont {Toschi}}, \
  and\ \bibinfo {author} {\bibfnamefont {K.}~\bibnamefont {Held}},\ }\bibfield
  {title} {\enquote {\bibinfo {title} {{Comparing pertinent effects of
  antiferromagnetic fluctuations in the two- and three-dimensional Hubbard
  model}},}\ }\href {\doibase 10.1103/PhysRevB.80.075104} {\bibfield  {journal}
  {\bibinfo  {journal} {Phys. Rev. B}\ }\textbf {\bibinfo {volume} {80}},\
  \bibinfo {pages} {075104} (\bibinfo {year} {2009})}\BibitemShut {NoStop}%
\bibitem [{\citenamefont {Rohringer}\ and\ \citenamefont
  {Toschi}(2016)}]{Rohringer2016}%
  \BibitemOpen
  \bibfield  {author} {\bibinfo {author} {\bibfnamefont {G.}~\bibnamefont
  {Rohringer}}\ and\ \bibinfo {author} {\bibfnamefont {A.}~\bibnamefont
  {Toschi}},\ }\bibfield  {title} {\enquote {\bibinfo {title} {{Impact of
  non-local correlations over different energy scales: A Dynamical Vertex
  Approximation study}},}\ }\href {\doibase 10.1103/PhysRevB.94.125144}
  {\bibfield  {journal} {\bibinfo  {journal} {Phys. Rev. B}\ }\textbf {\bibinfo
  {volume} {94}},\ \bibinfo {pages} {125144} (\bibinfo {year}
  {2016})}\BibitemShut {NoStop}%
\bibitem [{\citenamefont {Rohringer}\ and\ \citenamefont
  {Toschi}(2020)}]{RohringerPC2020}%
  \BibitemOpen
  \bibfield  {author} {\bibinfo {author} {\bibfnamefont {G.}~\bibnamefont
  {Rohringer}}\ and\ \bibinfo {author} {\bibfnamefont {A.}~\bibnamefont
  {Toschi}},\ }\href@noop {} {}\bibinfo {howpublished} {private communication}
  (\bibinfo {year} {2020})\BibitemShut {NoStop}%
\bibitem [{\citenamefont {Baier}\ \emph {et~al.}(2004)\citenamefont {Baier},
  \citenamefont {Bick},\ and\ \citenamefont {Wetterich}}]{Baier2004}%
  \BibitemOpen
  \bibfield  {author} {\bibinfo {author} {\bibfnamefont {Tobias}\ \bibnamefont
  {Baier}}, \bibinfo {author} {\bibfnamefont {Eike}\ \bibnamefont {Bick}}, \
  and\ \bibinfo {author} {\bibfnamefont {Christof}\ \bibnamefont {Wetterich}},\
  }\bibfield  {title} {\enquote {\bibinfo {title} {{Temperature dependence of
  antiferromagnetic order in the Hubbard model}},}\ }\href {\doibase
  10.1103/PhysRevB.70.125111} {\bibfield  {journal} {\bibinfo  {journal} {Phys.
  Rev. B}\ }\textbf {\bibinfo {volume} {70}},\ \bibinfo {pages} {125111}
  (\bibinfo {year} {2004})}\BibitemShut {NoStop}%
\bibitem [{\citenamefont {Kugler}\ and\ \citenamefont {von
  Delft}(2018{\natexlab{a}})}]{Kugler2017}%
  \BibitemOpen
  \bibfield  {author} {\bibinfo {author} {\bibfnamefont {Fabian~B.}\
  \bibnamefont {Kugler}}\ and\ \bibinfo {author} {\bibfnamefont {Jan}\
  \bibnamefont {von Delft}},\ }\bibfield  {title} {\enquote {\bibinfo {title}
  {{Multiloop Functional Renormalization Group That Sums Up All Parquet
  Diagrams}},}\ }\href {\doibase 10.1103/PhysRevLett.120.057403} {\bibfield
  {journal} {\bibinfo  {journal} {Phys. Rev. Lett.}\ }\textbf {\bibinfo
  {volume} {120}},\ \bibinfo {pages} {057403} (\bibinfo {year}
  {2018}{\natexlab{a}})}\BibitemShut {NoStop}%
\bibitem [{\citenamefont {Kugler}\ and\ \citenamefont {von
  Delft}(2018{\natexlab{b}})}]{Kugler2018}%
  \BibitemOpen
  \bibfield  {author} {\bibinfo {author} {\bibfnamefont {Fabian~B.}\
  \bibnamefont {Kugler}}\ and\ \bibinfo {author} {\bibfnamefont {Jan}\
  \bibnamefont {von Delft}},\ }\bibfield  {title} {\enquote {\bibinfo {title}
  {{Multiloop functional renormalization group for general models}},}\ }\href
  {\doibase 10.1103/PhysRevB.97.035162} {\bibfield  {journal} {\bibinfo
  {journal} {Phys. Rev. B}\ }\textbf {\bibinfo {volume} {97}},\ \bibinfo
  {pages} {35162} (\bibinfo {year} {2018}{\natexlab{b}})},\ \Eprint
  {http://arxiv.org/abs/1703.06505} {arXiv:1703.06505} \BibitemShut {NoStop}%
\bibitem [{\citenamefont {Hille}\ \emph
  {et~al.}(2020{\natexlab{a}})\citenamefont {Hille}, \citenamefont {Kugler},
  \citenamefont {Eckhardt}, \citenamefont {He}, \citenamefont {Kauch},
  \citenamefont {Honerkamp}, \citenamefont {Toschi},\ and\ \citenamefont
  {Andergassen}}]{Hille2020}%
  \BibitemOpen
  \bibfield  {author} {\bibinfo {author} {\bibfnamefont {Cornelia}\
  \bibnamefont {Hille}}, \bibinfo {author} {\bibfnamefont {Fabian~B.}\
  \bibnamefont {Kugler}}, \bibinfo {author} {\bibfnamefont {Christian~J.}\
  \bibnamefont {Eckhardt}}, \bibinfo {author} {\bibfnamefont {Yuan-Yao}\
  \bibnamefont {He}}, \bibinfo {author} {\bibfnamefont {Anna}\ \bibnamefont
  {Kauch}}, \bibinfo {author} {\bibfnamefont {Carsten}\ \bibnamefont
  {Honerkamp}}, \bibinfo {author} {\bibfnamefont {Alessandro}\ \bibnamefont
  {Toschi}}, \ and\ \bibinfo {author} {\bibfnamefont {Sabine}\ \bibnamefont
  {Andergassen}},\ }\bibfield  {title} {\enquote {\bibinfo {title}
  {{Quantitative functional renormalization group description of the
  two-dimensional Hubbard model}},}\ }\href {\doibase
  10.1103/PhysRevResearch.2.033372} {\bibfield  {journal} {\bibinfo  {journal}
  {Phys. Rev. Research}\ }\textbf {\bibinfo {volume} {2}},\ \bibinfo {pages}
  {033372} (\bibinfo {year} {2020}{\natexlab{a}})}\BibitemShut {NoStop}%
\bibitem [{\citenamefont {Hille}\ \emph
  {et~al.}(2020{\natexlab{b}})\citenamefont {Hille}, \citenamefont {Rohe},
  \citenamefont {Honerkamp},\ and\ \citenamefont {Andergassen}}]{Hille2020b}%
  \BibitemOpen
  \bibfield  {author} {\bibinfo {author} {\bibfnamefont {Cornelia}\
  \bibnamefont {Hille}}, \bibinfo {author} {\bibfnamefont {Daniel}\
  \bibnamefont {Rohe}}, \bibinfo {author} {\bibfnamefont {Carsten}\
  \bibnamefont {Honerkamp}}, \ and\ \bibinfo {author} {\bibfnamefont {Sabine}\
  \bibnamefont {Andergassen}},\ }\bibfield  {title} {\enquote {\bibinfo {title}
  {{Pseudogap opening in the two-dimensional Hubbard model: A functional
  renormalization group analysis}},}\ }\href {\doibase
  10.1103/PhysRevResearch.2.033068} {\bibfield  {journal} {\bibinfo  {journal}
  {Phys. Rev. Research}\ }\textbf {\bibinfo {volume} {2}},\ \bibinfo {pages}
  {033068} (\bibinfo {year} {2020}{\natexlab{b}})}\BibitemShut {NoStop}%
\bibitem [{\citenamefont {Bickers}\ and\ \citenamefont
  {Scalapino}(1992)}]{Bickers1992}%
  \BibitemOpen
  \bibfield  {author} {\bibinfo {author} {\bibfnamefont {N.~E.}\ \bibnamefont
  {Bickers}}\ and\ \bibinfo {author} {\bibfnamefont {D.~J.}\ \bibnamefont
  {Scalapino}},\ }\bibfield  {title} {\enquote {\bibinfo {title} {Critical
  behavior of electronic parquet solutions},}\ }\href {\doibase
  10.1103/PhysRevB.46.8050} {\bibfield  {journal} {\bibinfo  {journal} {Phys.
  Rev. B}\ }\textbf {\bibinfo {volume} {46}},\ \bibinfo {pages} {8050--8056}
  (\bibinfo {year} {1992})}\BibitemShut {NoStop}%
\bibitem [{\citenamefont {Eckhardt}\ \emph {et~al.}(2018)\citenamefont
  {Eckhardt}, \citenamefont {Schober}, \citenamefont {Ehrlich},\ and\
  \citenamefont {Honerkamp}}]{Eckhardt18}%
  \BibitemOpen
  \bibfield  {author} {\bibinfo {author} {\bibfnamefont {C.~J.}\ \bibnamefont
  {Eckhardt}}, \bibinfo {author} {\bibfnamefont {G.~A.~H.}\ \bibnamefont
  {Schober}}, \bibinfo {author} {\bibfnamefont {J.}~\bibnamefont {Ehrlich}}, \
  and\ \bibinfo {author} {\bibfnamefont {C.}~\bibnamefont {Honerkamp}},\
  }\bibfield  {title} {\enquote {\bibinfo {title} {{Truncated-unity parquet
  equations: Application to the repulsive Hubbard model}},}\ }\href {\doibase
  10.1103/PhysRevB.98.075143} {\bibfield  {journal} {\bibinfo  {journal} {Phys.
  Rev. B}\ }\textbf {\bibinfo {volume} {98}},\ \bibinfo {pages} {075143}
  (\bibinfo {year} {2018})}\BibitemShut {NoStop}%
\bibitem [{\citenamefont {Eckhardt}\ \emph {et~al.}(2020)\citenamefont
  {Eckhardt}, \citenamefont {Honerkamp}, \citenamefont {Held},\ and\
  \citenamefont {Kauch}}]{Eckhardt2020}%
  \BibitemOpen
  \bibfield  {author} {\bibinfo {author} {\bibfnamefont {Christian~J.}\
  \bibnamefont {Eckhardt}}, \bibinfo {author} {\bibfnamefont {Carsten}\
  \bibnamefont {Honerkamp}}, \bibinfo {author} {\bibfnamefont {Karsten}\
  \bibnamefont {Held}}, \ and\ \bibinfo {author} {\bibfnamefont {Anna}\
  \bibnamefont {Kauch}},\ }\bibfield  {title} {\enquote {\bibinfo {title}
  {{Truncated unity parquet solver}},}\ }\href {\doibase
  10.1103/PhysRevB.101.155104} {\bibfield  {journal} {\bibinfo  {journal}
  {Phys. Rev. B}\ }\textbf {\bibinfo {volume} {101}},\ \bibinfo {pages}
  {155104} (\bibinfo {year} {2020})}\BibitemShut {NoStop}%
\bibitem [{\citenamefont {Sch\"afer}\ \emph
  {et~al.}(2015{\natexlab{a}})\citenamefont {Sch\"afer}, \citenamefont {Geles},
  \citenamefont {Rost}, \citenamefont {Rohringer}, \citenamefont {Arrigoni},
  \citenamefont {Held}, \citenamefont {Bl\"umer}, \citenamefont {Aichhorn},\
  and\ \citenamefont {Toschi}}]{Schaefer2015b}%
  \BibitemOpen
  \bibfield  {author} {\bibinfo {author} {\bibfnamefont {T.}~\bibnamefont
  {Sch\"afer}}, \bibinfo {author} {\bibfnamefont {F.}~\bibnamefont {Geles}},
  \bibinfo {author} {\bibfnamefont {D.}~\bibnamefont {Rost}}, \bibinfo {author}
  {\bibfnamefont {G.}~\bibnamefont {Rohringer}}, \bibinfo {author}
  {\bibfnamefont {E.}~\bibnamefont {Arrigoni}}, \bibinfo {author}
  {\bibfnamefont {K.}~\bibnamefont {Held}}, \bibinfo {author} {\bibfnamefont
  {N.}~\bibnamefont {Bl\"umer}}, \bibinfo {author} {\bibfnamefont
  {M.}~\bibnamefont {Aichhorn}}, \ and\ \bibinfo {author} {\bibfnamefont
  {A.}~\bibnamefont {Toschi}},\ }\bibfield  {title} {\enquote {\bibinfo {title}
  {{Fate of the false Mott-Hubbard transition in two dimensions}},}\ }\href
  {\doibase 10.1103/PhysRevB.91.125109} {\bibfield  {journal} {\bibinfo
  {journal} {Phys. Rev. B}\ }\textbf {\bibinfo {volume} {91}},\ \bibinfo
  {pages} {125109} (\bibinfo {year} {2015}{\natexlab{a}})}\BibitemShut
  {NoStop}%
\bibitem [{\citenamefont {\ifmmode~\check{S}\else \v{S}\fi{}imkovic}\ \emph
  {et~al.}(2020)\citenamefont {\ifmmode~\check{S}\else \v{S}\fi{}imkovic},
  \citenamefont {LeBlanc}, \citenamefont {Kim}, \citenamefont {Deng},
  \citenamefont {Prokof'ev}, \citenamefont {Svistunov},\ and\ \citenamefont
  {Kozik}}]{Simkovic2020}%
  \BibitemOpen
  \bibfield  {author} {\bibinfo {author} {\bibfnamefont {Fedor}\ \bibnamefont
  {\ifmmode~\check{S}\else \v{S}\fi{}imkovic}}, \bibinfo {author}
  {\bibfnamefont {J.~P.~F.}\ \bibnamefont {LeBlanc}}, \bibinfo {author}
  {\bibfnamefont {Aaram~J.}\ \bibnamefont {Kim}}, \bibinfo {author}
  {\bibfnamefont {Youjin}\ \bibnamefont {Deng}}, \bibinfo {author}
  {\bibfnamefont {N.~V.}\ \bibnamefont {Prokof'ev}}, \bibinfo {author}
  {\bibfnamefont {B.~V.}\ \bibnamefont {Svistunov}}, \ and\ \bibinfo {author}
  {\bibfnamefont {Evgeny}\ \bibnamefont {Kozik}},\ }\bibfield  {title}
  {\enquote {\bibinfo {title} {{Extended Crossover from a Fermi Liquid to a
  Quasiantiferromagnet in the Half-Filled 2D Hubbard Model}},}\ }\href
  {\doibase 10.1103/PhysRevLett.124.017003} {\bibfield  {journal} {\bibinfo
  {journal} {Phys. Rev. Lett.}\ }\textbf {\bibinfo {volume} {124}},\ \bibinfo
  {pages} {017003} (\bibinfo {year} {2020})}\BibitemShut {NoStop}%
\bibitem [{\citenamefont {Kim}\ \emph {et~al.}(2020)\citenamefont {Kim},
  \citenamefont {Simkovic},\ and\ \citenamefont {Kozik}}]{Kim2020}%
  \BibitemOpen
  \bibfield  {author} {\bibinfo {author} {\bibfnamefont {Aaram~J.}\
  \bibnamefont {Kim}}, \bibinfo {author} {\bibfnamefont {Fedor}\ \bibnamefont
  {Simkovic}}, \ and\ \bibinfo {author} {\bibfnamefont {Evgeny}\ \bibnamefont
  {Kozik}},\ }\bibfield  {title} {\enquote {\bibinfo {title} {{Spin and Charge
  Correlations across the Metal-to-Insulator Crossover in the Half-Filled 2D
  Hubbard Model}},}\ }\href {\doibase 10.1103/PhysRevLett.124.117602}
  {\bibfield  {journal} {\bibinfo  {journal} {Phys. Rev. Lett.}\ }\textbf
  {\bibinfo {volume} {124}},\ \bibinfo {pages} {117602} (\bibinfo {year}
  {2020})}\BibitemShut {NoStop}%
\bibitem [{\citenamefont {Tanaka}(2019)}]{Tanaka2019}%
  \BibitemOpen
  \bibfield  {author} {\bibinfo {author} {\bibfnamefont {Arata}\ \bibnamefont
  {Tanaka}},\ }\bibfield  {title} {\enquote {\bibinfo {title} {{Metal-insulator
  transition in the two-dimensional Hubbard model: Dual fermion approach with
  Lanczos exact diagonalization}},}\ }\href {\doibase
  10.1103/PhysRevB.99.205133} {\bibfield  {journal} {\bibinfo  {journal} {Phys.
  Rev. B}\ }\textbf {\bibinfo {volume} {99}},\ \bibinfo {pages} {205133}
  (\bibinfo {year} {2019})}\BibitemShut {NoStop}%
\bibitem [{\citenamefont {Sch\"afer}\ \emph
  {et~al.}(2016{\natexlab{a}})\citenamefont {Sch\"afer}, \citenamefont
  {Toschi},\ and\ \citenamefont {Held}}]{Schaefer2015c}%
  \BibitemOpen
  \bibfield  {author} {\bibinfo {author} {\bibfnamefont {T.}~\bibnamefont
  {Sch\"afer}}, \bibinfo {author} {\bibfnamefont {A.}~\bibnamefont {Toschi}}, \
  and\ \bibinfo {author} {\bibfnamefont {K.}~\bibnamefont {Held}},\ }\bibfield
  {title} {\enquote {\bibinfo {title} {{Dynamical vertex approximation for the
  two-dimensional Hubbard model}},}\ }\href {\doibase
  10.1016/j.jmmm.2015.07.103} {\bibfield  {journal} {\bibinfo  {journal}
  {Journal of Magnetism and Magnetic Materials}\ }\textbf {\bibinfo {volume}
  {400}},\ \bibinfo {pages} {107--111} (\bibinfo {year}
  {2016}{\natexlab{a}})}\BibitemShut {NoStop}%
\bibitem [{\citenamefont {Rost}\ \emph {et~al.}(2012)\citenamefont {Rost},
  \citenamefont {Gorelik}, \citenamefont {Assaad},\ and\ \citenamefont
  {Bl\"umer}}]{Rost2012}%
  \BibitemOpen
  \bibfield  {author} {\bibinfo {author} {\bibfnamefont {D.}~\bibnamefont
  {Rost}}, \bibinfo {author} {\bibfnamefont {E.~V.}\ \bibnamefont {Gorelik}},
  \bibinfo {author} {\bibfnamefont {F.}~\bibnamefont {Assaad}}, \ and\ \bibinfo
  {author} {\bibfnamefont {N.}~\bibnamefont {Bl\"umer}},\ }\bibfield  {title}
  {\enquote {\bibinfo {title} {{Momentum-dependent pseudogaps in the
  half-filled two-dimensional Hubbard model}},}\ }\href {\doibase
  10.1103/PhysRevB.86.155109} {\bibfield  {journal} {\bibinfo  {journal} {Phys.
  Rev. B}\ }\textbf {\bibinfo {volume} {86}},\ \bibinfo {pages} {155109}
  (\bibinfo {year} {2012})}\BibitemShut {NoStop}%
\bibitem [{\citenamefont {Rost}\ and\ \citenamefont
  {Blümer}(2015)}]{Rost2015}%
  \BibitemOpen
  \bibfield  {author} {\bibinfo {author} {\bibfnamefont {D}~\bibnamefont
  {Rost}}\ and\ \bibinfo {author} {\bibfnamefont {N}~\bibnamefont {Blümer}},\
  }\bibfield  {title} {\enquote {\bibinfo {title} {{Deciding the fate of the
  false Mott transition in two dimensions by exact quantum Monte Carlo
  methods}},}\ }\href {\doibase 10.1088/1742-6596/640/1/012047} {\bibfield
  {journal} {\bibinfo  {journal} {Journal of Physics: Conference Series}\
  }\textbf {\bibinfo {volume} {640}},\ \bibinfo {pages} {012047} (\bibinfo
  {year} {2015})}\BibitemShut {NoStop}%
\bibitem [{\citenamefont {Sangiovanni}\ \emph {et~al.}(2006)\citenamefont
  {Sangiovanni}, \citenamefont {Toschi}, \citenamefont {Koch}, \citenamefont
  {Held}, \citenamefont {Capone}, \citenamefont {Castellani}, \citenamefont
  {Gunnarsson}, \citenamefont {Mo}, \citenamefont {Allen}, \citenamefont {Kim},
  \citenamefont {Sekiyama}, \citenamefont {Yamasaki}, \citenamefont {Suga},\
  and\ \citenamefont {Metcalf}}]{Sangiovanni2006}%
  \BibitemOpen
  \bibfield  {author} {\bibinfo {author} {\bibfnamefont {G.}~\bibnamefont
  {Sangiovanni}}, \bibinfo {author} {\bibfnamefont {A.}~\bibnamefont {Toschi}},
  \bibinfo {author} {\bibfnamefont {E.}~\bibnamefont {Koch}}, \bibinfo {author}
  {\bibfnamefont {K.}~\bibnamefont {Held}}, \bibinfo {author} {\bibfnamefont
  {M.}~\bibnamefont {Capone}}, \bibinfo {author} {\bibfnamefont
  {C.}~\bibnamefont {Castellani}}, \bibinfo {author} {\bibfnamefont
  {O.}~\bibnamefont {Gunnarsson}}, \bibinfo {author} {\bibfnamefont {S.-K.}\
  \bibnamefont {Mo}}, \bibinfo {author} {\bibfnamefont {J.~W.}\ \bibnamefont
  {Allen}}, \bibinfo {author} {\bibfnamefont {H.-D.}\ \bibnamefont {Kim}},
  \bibinfo {author} {\bibfnamefont {A.}~\bibnamefont {Sekiyama}}, \bibinfo
  {author} {\bibfnamefont {A.}~\bibnamefont {Yamasaki}}, \bibinfo {author}
  {\bibfnamefont {S.}~\bibnamefont {Suga}}, \ and\ \bibinfo {author}
  {\bibfnamefont {P.}~\bibnamefont {Metcalf}},\ }\bibfield  {title} {\enquote
  {\bibinfo {title} {{Static versus dynamical mean-field theory of Mott
  antiferromagnets}},}\ }\href {\doibase 10.1103/PhysRevB.73.205121} {\bibfield
   {journal} {\bibinfo  {journal} {Phys. Rev. B}\ }\textbf {\bibinfo {volume}
  {73}},\ \bibinfo {eid} {205121} (\bibinfo {year} {2006})}\BibitemShut
  {NoStop}%
\bibitem [{\citenamefont {Parisi}(1998)}]{Parisi1998}%
  \BibitemOpen
  \bibfield  {author} {\bibinfo {author} {\bibfnamefont {Giorgio}\ \bibnamefont
  {Parisi}},\ }\href@noop {} {\emph {\bibinfo {title} {Statistical Field
  Theory}}}\ (\bibinfo  {publisher} {CRC Press (Advanced Book Classics},\
  \bibinfo {year} {1998})\BibitemShut {NoStop}%
\bibitem [{\citenamefont {Hettler}\ \emph {et~al.}(1998)\citenamefont
  {Hettler}, \citenamefont {Tahvildar-Zadeh}, \citenamefont {Jarrell},
  \citenamefont {Pruschke},\ and\ \citenamefont {Krishnamurthy}}]{Hettler1998}%
  \BibitemOpen
  \bibfield  {author} {\bibinfo {author} {\bibfnamefont {M.~H.}\ \bibnamefont
  {Hettler}}, \bibinfo {author} {\bibfnamefont {A.~N.}\ \bibnamefont
  {Tahvildar-Zadeh}}, \bibinfo {author} {\bibfnamefont {M.}~\bibnamefont
  {Jarrell}}, \bibinfo {author} {\bibfnamefont {T.}~\bibnamefont {Pruschke}}, \
  and\ \bibinfo {author} {\bibfnamefont {H.~R.}\ \bibnamefont
  {Krishnamurthy}},\ }\bibfield  {title} {\enquote {\bibinfo {title} {Nonlocal
  dynamical correlations of strongly interacting electron systems},}\ }\href
  {\doibase 10.1103/PhysRevB.58.R7475} {\bibfield  {journal} {\bibinfo
  {journal} {Phys. Rev. B}\ }\textbf {\bibinfo {volume} {58}},\ \bibinfo
  {pages} {R7475--R7479} (\bibinfo {year} {1998})}\BibitemShut {NoStop}%
\bibitem [{\citenamefont {Hettler}\ \emph {et~al.}(2000)\citenamefont
  {Hettler}, \citenamefont {Mukherjee}, \citenamefont {Jarrell},\ and\
  \citenamefont {Krishnamurthy}}]{Hettler2000}%
  \BibitemOpen
  \bibfield  {author} {\bibinfo {author} {\bibfnamefont {M.~H.}\ \bibnamefont
  {Hettler}}, \bibinfo {author} {\bibfnamefont {M.}~\bibnamefont {Mukherjee}},
  \bibinfo {author} {\bibfnamefont {M.}~\bibnamefont {Jarrell}}, \ and\
  \bibinfo {author} {\bibfnamefont {H.~R.}\ \bibnamefont {Krishnamurthy}},\
  }\bibfield  {title} {\enquote {\bibinfo {title} {Dynamical cluster
  approximation: Nonlocal dynamics of correlated electron systems},}\ }\href
  {\doibase 10.1103/PhysRevB.61.12739} {\bibfield  {journal} {\bibinfo
  {journal} {Phys. Rev. B}\ }\textbf {\bibinfo {volume} {61}},\ \bibinfo
  {pages} {12739--12756} (\bibinfo {year} {2000})}\BibitemShut {NoStop}%
\bibitem [{\citenamefont {Aryanpour}\ \emph {et~al.}(2002)\citenamefont
  {Aryanpour}, \citenamefont {Hettler},\ and\ \citenamefont
  {Jarrell}}]{Aryanpour2002}%
  \BibitemOpen
  \bibfield  {author} {\bibinfo {author} {\bibfnamefont {K.}~\bibnamefont
  {Aryanpour}}, \bibinfo {author} {\bibfnamefont {M.~H.}\ \bibnamefont
  {Hettler}}, \ and\ \bibinfo {author} {\bibfnamefont {M.}~\bibnamefont
  {Jarrell}},\ }\bibfield  {title} {\enquote {\bibinfo {title} {Analysis of the
  dynamical cluster approximation for the {Hubbard} model},}\ }\href {\doibase
  10.1103/PhysRevB.65.153102} {\bibfield  {journal} {\bibinfo  {journal} {Phys.
  Rev. B}\ }\textbf {\bibinfo {volume} {65}},\ \bibinfo {pages} {153102}
  (\bibinfo {year} {2002})}\BibitemShut {NoStop}%
\bibitem [{\citenamefont {Kotliar}\ \emph {et~al.}(2001)\citenamefont
  {Kotliar}, \citenamefont {Savrasov}, \citenamefont {P\'alsson},\ and\
  \citenamefont {Biroli}}]{Kotliar2001}%
  \BibitemOpen
  \bibfield  {author} {\bibinfo {author} {\bibfnamefont {Gabriel}\ \bibnamefont
  {Kotliar}}, \bibinfo {author} {\bibfnamefont {Sergej~Y.}\ \bibnamefont
  {Savrasov}}, \bibinfo {author} {\bibfnamefont {Gunnar}\ \bibnamefont
  {P\'alsson}}, \ and\ \bibinfo {author} {\bibfnamefont {Giulio}\ \bibnamefont
  {Biroli}},\ }\bibfield  {title} {\enquote {\bibinfo {title} {{Cellular
  Dynamical Mean Field Approach to Strongly Correlated Systems}},}\ }\href
  {\doibase 10.1103/PhysRevLett.87.186401} {\bibfield  {journal} {\bibinfo
  {journal} {Phys. Rev. Lett.}\ }\textbf {\bibinfo {volume} {87}},\ \bibinfo
  {pages} {186401} (\bibinfo {year} {2001})}\BibitemShut {NoStop}%
\bibitem [{\citenamefont {Klett}\ \emph {et~al.}(2020)\citenamefont {Klett},
  \citenamefont {Wentzell}, \citenamefont {Sch\"afer}, \citenamefont
  {Simkovic}, \citenamefont {Parcollet}, \citenamefont {Andergassen},\ and\
  \citenamefont {Hansmann}}]{Klett2020}%
  \BibitemOpen
  \bibfield  {author} {\bibinfo {author} {\bibfnamefont {Marcel}\ \bibnamefont
  {Klett}}, \bibinfo {author} {\bibfnamefont {Nils}\ \bibnamefont {Wentzell}},
  \bibinfo {author} {\bibfnamefont {Thomas}\ \bibnamefont {Sch\"afer}},
  \bibinfo {author} {\bibfnamefont {Fedor}\ \bibnamefont {Simkovic}}, \bibinfo
  {author} {\bibfnamefont {Olivier}\ \bibnamefont {Parcollet}}, \bibinfo
  {author} {\bibfnamefont {Sabine}\ \bibnamefont {Andergassen}}, \ and\
  \bibinfo {author} {\bibfnamefont {Philipp}\ \bibnamefont {Hansmann}},\
  }\bibfield  {title} {\enquote {\bibinfo {title} {{Real-space cluster
  dynamical mean-field theory: Center-focused extrapolation on the one- and two
  particle-levels}},}\ }\href {\doibase 10.1103/PhysRevResearch.2.033476}
  {\bibfield  {journal} {\bibinfo  {journal} {Phys. Rev. Research}\ }\textbf
  {\bibinfo {volume} {2}},\ \bibinfo {pages} {033476} (\bibinfo {year}
  {2020})}\BibitemShut {NoStop}%
\bibitem [{\citenamefont {Rohringer}\ \emph {et~al.}(2011)\citenamefont
  {Rohringer}, \citenamefont {Toschi}, \citenamefont {Katanin},\ and\
  \citenamefont {Held}}]{Rohringer2011}%
  \BibitemOpen
  \bibfield  {author} {\bibinfo {author} {\bibfnamefont {G.}~\bibnamefont
  {Rohringer}}, \bibinfo {author} {\bibfnamefont {A.}~\bibnamefont {Toschi}},
  \bibinfo {author} {\bibfnamefont {A.}~\bibnamefont {Katanin}}, \ and\
  \bibinfo {author} {\bibfnamefont {K.}~\bibnamefont {Held}},\ }\bibfield
  {title} {\enquote {\bibinfo {title} {{Critical Properties of the Half-Filled
  Hubbard Model in Three Dimensions}},}\ }\href {\doibase
  10.1103/PhysRevLett.107.256402} {\bibfield  {journal} {\bibinfo  {journal}
  {Phys. Rev. Lett.}\ }\textbf {\bibinfo {volume} {107}},\ \bibinfo {pages}
  {256402} (\bibinfo {year} {2011})}\BibitemShut {NoStop}%
\bibitem [{\citenamefont {Gull}\ \emph
  {et~al.}(2008{\natexlab{a}})\citenamefont {Gull}, \citenamefont {Werner},
  \citenamefont {Wang}, \citenamefont {Troyer},\ and\ \citenamefont
  {Millis}}]{Gull2008}%
  \BibitemOpen
  \bibfield  {author} {\bibinfo {author} {\bibfnamefont {E.}~\bibnamefont
  {Gull}}, \bibinfo {author} {\bibfnamefont {P.}~\bibnamefont {Werner}},
  \bibinfo {author} {\bibfnamefont {X.}~\bibnamefont {Wang}}, \bibinfo {author}
  {\bibfnamefont {M.}~\bibnamefont {Troyer}}, \ and\ \bibinfo {author}
  {\bibfnamefont {A.~J.}\ \bibnamefont {Millis}},\ }\bibfield  {title}
  {\enquote {\bibinfo {title} {{Local order and the gapped phase of the Hubbard
  model: A plaquette dynamical mean-field investigation}},}\ }\href
  {http://stacks.iop.org/0295-5075/84/i=3/a=37009} {\bibfield  {journal}
  {\bibinfo  {journal} {EPL (Europhysics Letters)}\ }\textbf {\bibinfo {volume}
  {84}},\ \bibinfo {pages} {37009} (\bibinfo {year}
  {2008}{\natexlab{a}})}\BibitemShut {NoStop}%
\bibitem [{\citenamefont {Gull}\ \emph {et~al.}(2009)\citenamefont {Gull},
  \citenamefont {Parcollet}, \citenamefont {Werner},\ and\ \citenamefont
  {Millis}}]{Gull2009}%
  \BibitemOpen
  \bibfield  {author} {\bibinfo {author} {\bibfnamefont {Emanuel}\ \bibnamefont
  {Gull}}, \bibinfo {author} {\bibfnamefont {Olivier}\ \bibnamefont
  {Parcollet}}, \bibinfo {author} {\bibfnamefont {Philipp}\ \bibnamefont
  {Werner}}, \ and\ \bibinfo {author} {\bibfnamefont {Andrew~J.}\ \bibnamefont
  {Millis}},\ }\bibfield  {title} {\enquote {\bibinfo {title}
  {{Momentum-sector-selective metal-insulator transition in the eight-site
  dynamical mean-field approximation to the Hubbard model in two
  dimensions}},}\ }\href {\doibase 10.1103/PhysRevB.80.245102} {\bibfield
  {journal} {\bibinfo  {journal} {Phys. Rev. B}\ }\textbf {\bibinfo {volume}
  {80}},\ \bibinfo {pages} {245102} (\bibinfo {year} {2009})}\BibitemShut
  {NoStop}%
\bibitem [{\citenamefont {Gull}\ \emph {et~al.}(2010)\citenamefont {Gull},
  \citenamefont {Ferrero}, \citenamefont {Parcollet}, \citenamefont {Georges},\
  and\ \citenamefont {Millis}}]{Gull2010}%
  \BibitemOpen
  \bibfield  {author} {\bibinfo {author} {\bibfnamefont {E.}~\bibnamefont
  {Gull}}, \bibinfo {author} {\bibfnamefont {M.}~\bibnamefont {Ferrero}},
  \bibinfo {author} {\bibfnamefont {O.}~\bibnamefont {Parcollet}}, \bibinfo
  {author} {\bibfnamefont {A.}~\bibnamefont {Georges}}, \ and\ \bibinfo
  {author} {\bibfnamefont {A.~J.}\ \bibnamefont {Millis}},\ }\bibfield  {title}
  {\enquote {\bibinfo {title} {{Momentum-space anisotropy and pseudogaps: A
  comparative cluster dynamical mean-field analysis of the doping-driven
  metal-insulator transition in the two-dimensional Hubbard model}},}\ }\href
  {\doibase 10.1103/PhysRevB.82.155101} {\bibfield  {journal} {\bibinfo
  {journal} {Phys. Rev. B}\ }\textbf {\bibinfo {volume} {82}},\ \bibinfo
  {pages} {155101} (\bibinfo {year} {2010})}\BibitemShut {NoStop}%
\bibitem [{\citenamefont {Gull}\ \emph {et~al.}(2013)\citenamefont {Gull},
  \citenamefont {Parcollet},\ and\ \citenamefont {Millis}}]{Gull2013}%
  \BibitemOpen
  \bibfield  {author} {\bibinfo {author} {\bibfnamefont {Emanuel}\ \bibnamefont
  {Gull}}, \bibinfo {author} {\bibfnamefont {Olivier}\ \bibnamefont
  {Parcollet}}, \ and\ \bibinfo {author} {\bibfnamefont {Andrew~J.}\
  \bibnamefont {Millis}},\ }\bibfield  {title} {\enquote {\bibinfo {title}
  {{Superconductivity and the Pseudogap in the Two-Dimensional Hubbard
  Model}},}\ }\href {\doibase 10.1103/PhysRevLett.110.216405} {\bibfield
  {journal} {\bibinfo  {journal} {Phys. Rev. Lett.}\ }\textbf {\bibinfo
  {volume} {110}},\ \bibinfo {pages} {216405} (\bibinfo {year}
  {2013})}\BibitemShut {NoStop}%
\bibitem [{\citenamefont {Parcollet}\ \emph {et~al.}(2004)\citenamefont
  {Parcollet}, \citenamefont {Biroli},\ and\ \citenamefont
  {Kotliar}}]{Parcollet2004}%
  \BibitemOpen
  \bibfield  {author} {\bibinfo {author} {\bibfnamefont {O.}~\bibnamefont
  {Parcollet}}, \bibinfo {author} {\bibfnamefont {G.}~\bibnamefont {Biroli}}, \
  and\ \bibinfo {author} {\bibfnamefont {G.}~\bibnamefont {Kotliar}},\
  }\bibfield  {title} {\enquote {\bibinfo {title} {{Cluster Dynamical Mean
  Field Analysis of the Mott Transition}},}\ }\href {\doibase
  10.1103/PhysRevLett.92.226402} {\bibfield  {journal} {\bibinfo  {journal}
  {Phys. Rev. Lett.}\ }\textbf {\bibinfo {volume} {92}},\ \bibinfo {pages}
  {226402} (\bibinfo {year} {2004})}\BibitemShut {NoStop}%
\bibitem [{\citenamefont {Macridin}\ \emph {et~al.}(2006)\citenamefont
  {Macridin}, \citenamefont {Jarrell}, \citenamefont {Maier}, \citenamefont
  {Kent},\ and\ \citenamefont {D'Azevedo}}]{Macridin2006}%
  \BibitemOpen
  \bibfield  {author} {\bibinfo {author} {\bibfnamefont {Alexandru}\
  \bibnamefont {Macridin}}, \bibinfo {author} {\bibfnamefont {M.}~\bibnamefont
  {Jarrell}}, \bibinfo {author} {\bibfnamefont {Thomas}\ \bibnamefont {Maier}},
  \bibinfo {author} {\bibfnamefont {P.~R.~C.}\ \bibnamefont {Kent}}, \ and\
  \bibinfo {author} {\bibfnamefont {Eduardo}\ \bibnamefont {D'Azevedo}},\
  }\bibfield  {title} {\enquote {\bibinfo {title} {{Pseudogap and
  Antiferromagnetic Correlations in the Hubbard Model}},}\ }\href {\doibase
  10.1103/PhysRevLett.97.036401} {\bibfield  {journal} {\bibinfo  {journal}
  {Phys. Rev. Lett.}\ }\textbf {\bibinfo {volume} {97}},\ \bibinfo {pages}
  {036401} (\bibinfo {year} {2006})}\BibitemShut {NoStop}%
\bibitem [{\citenamefont {S\'en\'echal}\ and\ \citenamefont
  {Tremblay}(2004)}]{Senechal2004}%
  \BibitemOpen
  \bibfield  {author} {\bibinfo {author} {\bibfnamefont {David}\ \bibnamefont
  {S\'en\'echal}}\ and\ \bibinfo {author} {\bibfnamefont {A.-M.~S.}\
  \bibnamefont {Tremblay}},\ }\bibfield  {title} {\enquote {\bibinfo {title}
  {{Hot Spots and Pseudogaps for Hole- and Electron-Doped High-Temperature
  Superconductors}},}\ }\href {\doibase 10.1103/PhysRevLett.92.126401}
  {\bibfield  {journal} {\bibinfo  {journal} {Phys. Rev. Lett.}\ }\textbf
  {\bibinfo {volume} {92}},\ \bibinfo {pages} {126401} (\bibinfo {year}
  {2004})}\BibitemShut {NoStop}%
\bibitem [{\citenamefont {Haule}\ and\ \citenamefont
  {Kotliar}(2007)}]{Haule2007b}%
  \BibitemOpen
  \bibfield  {author} {\bibinfo {author} {\bibfnamefont {Kristjan}\
  \bibnamefont {Haule}}\ and\ \bibinfo {author} {\bibfnamefont {Gabriel}\
  \bibnamefont {Kotliar}},\ }\bibfield  {title} {\enquote {\bibinfo {title}
  {{Strongly correlated superconductivity: A plaquette dynamical mean-field
  theory study}},}\ }\href {\doibase 10.1103/PhysRevB.76.104509} {\bibfield
  {journal} {\bibinfo  {journal} {Phys. Rev. B}\ }\textbf {\bibinfo {volume}
  {76}},\ \bibinfo {pages} {104509} (\bibinfo {year} {2007})}\BibitemShut
  {NoStop}%
\bibitem [{\citenamefont {Park}\ \emph {et~al.}(2008)\citenamefont {Park},
  \citenamefont {Haule},\ and\ \citenamefont {Kotliar}}]{Park2008}%
  \BibitemOpen
  \bibfield  {author} {\bibinfo {author} {\bibfnamefont {H.}~\bibnamefont
  {Park}}, \bibinfo {author} {\bibfnamefont {K.}~\bibnamefont {Haule}}, \ and\
  \bibinfo {author} {\bibfnamefont {G.}~\bibnamefont {Kotliar}},\ }\bibfield
  {title} {\enquote {\bibinfo {title} {{Cluster Dynamical Mean Field Theory of
  the Mott Transition}},}\ }\href {\doibase 10.1103/PhysRevLett.101.186403}
  {\bibfield  {journal} {\bibinfo  {journal} {Phys. Rev. Lett.}\ }\textbf
  {\bibinfo {volume} {101}},\ \bibinfo {pages} {186403} (\bibinfo {year}
  {2008})}\BibitemShut {NoStop}%
\bibitem [{\citenamefont {Ferrero}\ \emph
  {et~al.}(2009{\natexlab{a}})\citenamefont {Ferrero}, \citenamefont
  {Cornaglia}, \citenamefont {De~Leo}, \citenamefont {Parcollet}, \citenamefont
  {Kotliar},\ and\ \citenamefont {Georges}}]{Ferrero2009}%
  \BibitemOpen
  \bibfield  {author} {\bibinfo {author} {\bibfnamefont {Michel}\ \bibnamefont
  {Ferrero}}, \bibinfo {author} {\bibfnamefont {Pablo~S.}\ \bibnamefont
  {Cornaglia}}, \bibinfo {author} {\bibfnamefont {Lorenzo}\ \bibnamefont
  {De~Leo}}, \bibinfo {author} {\bibfnamefont {Olivier}\ \bibnamefont
  {Parcollet}}, \bibinfo {author} {\bibfnamefont {Gabriel}\ \bibnamefont
  {Kotliar}}, \ and\ \bibinfo {author} {\bibfnamefont {Antoine}\ \bibnamefont
  {Georges}},\ }\bibfield  {title} {\enquote {\bibinfo {title} {{Pseudogap
  opening and formation of Fermi arcs as an orbital-selective Mott transition
  in momentum space}},}\ }\href {\doibase 10.1103/PhysRevB.80.064501}
  {\bibfield  {journal} {\bibinfo  {journal} {Phys. Rev. B}\ }\textbf {\bibinfo
  {volume} {80}},\ \bibinfo {pages} {064501} (\bibinfo {year}
  {2009}{\natexlab{a}})}\BibitemShut {NoStop}%
\bibitem [{\citenamefont {Ferrero}\ \emph
  {et~al.}(2009{\natexlab{b}})\citenamefont {Ferrero}, \citenamefont
  {Cornaglia}, \citenamefont {Leo}, \citenamefont {Parcollet}, \citenamefont
  {Kotliar},\ and\ \citenamefont {Georges}}]{Ferrero2009b}%
  \BibitemOpen
  \bibfield  {author} {\bibinfo {author} {\bibfnamefont {M.}~\bibnamefont
  {Ferrero}}, \bibinfo {author} {\bibfnamefont {P.~S.}\ \bibnamefont
  {Cornaglia}}, \bibinfo {author} {\bibfnamefont {L.~De}\ \bibnamefont {Leo}},
  \bibinfo {author} {\bibfnamefont {O.}~\bibnamefont {Parcollet}}, \bibinfo
  {author} {\bibfnamefont {G.}~\bibnamefont {Kotliar}}, \ and\ \bibinfo
  {author} {\bibfnamefont {A.}~\bibnamefont {Georges}},\ }\bibfield  {title}
  {\enquote {\bibinfo {title} {{Valence bond dynamical mean-field theory of
  doped Mott insulators with nodal/antinodal differentiation}},}\ }\href
  {\doibase 10.1209/0295-5075/85/57009} {\bibfield  {journal} {\bibinfo
  {journal} {{EPL} (Europhysics Letters)}\ }\textbf {\bibinfo {volume} {85}},\
  \bibinfo {pages} {57009} (\bibinfo {year} {2009}{\natexlab{b}})}\BibitemShut
  {NoStop}%
\bibitem [{\citenamefont {Werner}\ \emph {et~al.}(2009)\citenamefont {Werner},
  \citenamefont {Gull}, \citenamefont {Parcollet},\ and\ \citenamefont
  {Millis}}]{Werner2009}%
  \BibitemOpen
  \bibfield  {author} {\bibinfo {author} {\bibfnamefont {Philipp}\ \bibnamefont
  {Werner}}, \bibinfo {author} {\bibfnamefont {Emanuel}\ \bibnamefont {Gull}},
  \bibinfo {author} {\bibfnamefont {Olivier}\ \bibnamefont {Parcollet}}, \ and\
  \bibinfo {author} {\bibfnamefont {Andrew~J.}\ \bibnamefont {Millis}},\
  }\bibfield  {title} {\enquote {\bibinfo {title} {{Momentum-selective
  metal-insulator transition in the two-dimensional Hubbard model: An 8-site
  dynamical cluster approximation study}},}\ }\href {\doibase
  10.1103/PhysRevB.80.045120} {\bibfield  {journal} {\bibinfo  {journal} {Phys.
  Rev. B}\ }\textbf {\bibinfo {volume} {80}},\ \bibinfo {pages} {045120}
  (\bibinfo {year} {2009})}\BibitemShut {NoStop}%
\bibitem [{\citenamefont {Sordi}\ \emph
  {et~al.}(2012{\natexlab{a}})\citenamefont {Sordi}, \citenamefont {S\'emon},
  \citenamefont {Haule},\ and\ \citenamefont {Tremblay}}]{Sordi2012}%
  \BibitemOpen
  \bibfield  {author} {\bibinfo {author} {\bibfnamefont {G.}~\bibnamefont
  {Sordi}}, \bibinfo {author} {\bibfnamefont {P.}~\bibnamefont {S\'emon}},
  \bibinfo {author} {\bibfnamefont {K.}~\bibnamefont {Haule}}, \ and\ \bibinfo
  {author} {\bibfnamefont {A.-M.~S.}\ \bibnamefont {Tremblay}},\ }\bibfield
  {title} {\enquote {\bibinfo {title} {{Strong Coupling Superconductivity,
  Pseudogap, and Mott Transition}},}\ }\href {\doibase
  10.1103/PhysRevLett.108.216401} {\bibfield  {journal} {\bibinfo  {journal}
  {Phys. Rev. Lett.}\ }\textbf {\bibinfo {volume} {108}},\ \bibinfo {pages}
  {216401} (\bibinfo {year} {2012}{\natexlab{a}})}\BibitemShut {NoStop}%
\bibitem [{\citenamefont {Sordi}\ \emph
  {et~al.}(2012{\natexlab{b}})\citenamefont {Sordi}, \citenamefont {S{\'e}mon},
  \citenamefont {Haule},\ and\ \citenamefont {Tremblay}}]{Sordi2012b}%
  \BibitemOpen
  \bibfield  {author} {\bibinfo {author} {\bibfnamefont {G}~\bibnamefont
  {Sordi}}, \bibinfo {author} {\bibfnamefont {P}~\bibnamefont {S{\'e}mon}},
  \bibinfo {author} {\bibfnamefont {Kristjan}\ \bibnamefont {Haule}}, \ and\
  \bibinfo {author} {\bibfnamefont {A-MS}\ \bibnamefont {Tremblay}},\
  }\bibfield  {title} {\enquote {\bibinfo {title} {{Pseudogap temperature as a
  Widom line in doped Mott insulators}},}\ }\href
  {https://www.nature.com/articles/srep00547} {\bibfield  {journal} {\bibinfo
  {journal} {Scientific reports}\ }\textbf {\bibinfo {volume} {2}},\ \bibinfo
  {pages} {1--5} (\bibinfo {year} {2012}{\natexlab{b}})}\BibitemShut {NoStop}%
\bibitem [{\citenamefont {Sordi}\ \emph {et~al.}(2013)\citenamefont {Sordi},
  \citenamefont {S\'emon}, \citenamefont {Haule},\ and\ \citenamefont
  {Tremblay}}]{Sordi2013}%
  \BibitemOpen
  \bibfield  {author} {\bibinfo {author} {\bibfnamefont {G.}~\bibnamefont
  {Sordi}}, \bibinfo {author} {\bibfnamefont {P.}~\bibnamefont {S\'emon}},
  \bibinfo {author} {\bibfnamefont {K.}~\bibnamefont {Haule}}, \ and\ \bibinfo
  {author} {\bibfnamefont {A.-M.~S.}\ \bibnamefont {Tremblay}},\ }\bibfield
  {title} {\enquote {\bibinfo {title} {$c$-axis resistivity, pseudogap,
  superconductivity, and widom line in doped mott insulators},}\ }\href
  {\doibase 10.1103/PhysRevB.87.041101} {\bibfield  {journal} {\bibinfo
  {journal} {Phys. Rev. B}\ }\textbf {\bibinfo {volume} {87}},\ \bibinfo
  {pages} {041101(R)} (\bibinfo {year} {2013})}\BibitemShut {NoStop}%
\bibitem [{\citenamefont {Gunnarsson}\ \emph {et~al.}(2015)\citenamefont
  {Gunnarsson}, \citenamefont {Sch\"afer}, \citenamefont {LeBlanc},
  \citenamefont {Gull}, \citenamefont {Merino}, \citenamefont {Sangiovanni},
  \citenamefont {Rohringer},\ and\ \citenamefont {Toschi}}]{Gunnarsson2015}%
  \BibitemOpen
  \bibfield  {author} {\bibinfo {author} {\bibfnamefont {O.}~\bibnamefont
  {Gunnarsson}}, \bibinfo {author} {\bibfnamefont {T.}~\bibnamefont
  {Sch\"afer}}, \bibinfo {author} {\bibfnamefont {J.~P.~F.}\ \bibnamefont
  {LeBlanc}}, \bibinfo {author} {\bibfnamefont {E.}~\bibnamefont {Gull}},
  \bibinfo {author} {\bibfnamefont {J.}~\bibnamefont {Merino}}, \bibinfo
  {author} {\bibfnamefont {G.}~\bibnamefont {Sangiovanni}}, \bibinfo {author}
  {\bibfnamefont {G.}~\bibnamefont {Rohringer}}, \ and\ \bibinfo {author}
  {\bibfnamefont {A.}~\bibnamefont {Toschi}},\ }\bibfield  {title} {\enquote
  {\bibinfo {title} {{Fluctuation Diagnostics of the Electron Self-Energy:
  Origin of the Pseudogap Physics}},}\ }\href {\doibase
  10.1103/PhysRevLett.114.236402} {\bibfield  {journal} {\bibinfo  {journal}
  {Phys. Rev. Lett.}\ }\textbf {\bibinfo {volume} {114}},\ \bibinfo {pages}
  {236402} (\bibinfo {year} {2015})}\BibitemShut {NoStop}%
\bibitem [{\citenamefont {Wu}\ \emph {et~al.}(2018)\citenamefont {Wu},
  \citenamefont {Scheurer}, \citenamefont {Chatterjee}, \citenamefont
  {Sachdev}, \citenamefont {Georges},\ and\ \citenamefont {Ferrero}}]{Wu2018}%
  \BibitemOpen
  \bibfield  {author} {\bibinfo {author} {\bibfnamefont {Wei}\ \bibnamefont
  {Wu}}, \bibinfo {author} {\bibfnamefont {Mathias~S.}\ \bibnamefont
  {Scheurer}}, \bibinfo {author} {\bibfnamefont {Shubhayu}\ \bibnamefont
  {Chatterjee}}, \bibinfo {author} {\bibfnamefont {Subir}\ \bibnamefont
  {Sachdev}}, \bibinfo {author} {\bibfnamefont {Antoine}\ \bibnamefont
  {Georges}}, \ and\ \bibinfo {author} {\bibfnamefont {Michel}\ \bibnamefont
  {Ferrero}},\ }\bibfield  {title} {\enquote {\bibinfo {title} {{Pseudogap and
  Fermi-Surface Topology in the Two-Dimensional Hubbard Model}},}\ }\href
  {\doibase 10.1103/PhysRevX.8.021048} {\bibfield  {journal} {\bibinfo
  {journal} {Phys. Rev. X}\ }\textbf {\bibinfo {volume} {8}},\ \bibinfo {pages}
  {021048} (\bibinfo {year} {2018})}\BibitemShut {NoStop}%
\bibitem [{\citenamefont {Scheurer}\ \emph {et~al.}(2018)\citenamefont
  {Scheurer}, \citenamefont {Chatterjee}, \citenamefont {Wu}, \citenamefont
  {Ferrero}, \citenamefont {Georges},\ and\ \citenamefont
  {Sachdev}}]{Scheurer2018}%
  \BibitemOpen
  \bibfield  {author} {\bibinfo {author} {\bibfnamefont {Mathias~S.}\
  \bibnamefont {Scheurer}}, \bibinfo {author} {\bibfnamefont {Shubhayu}\
  \bibnamefont {Chatterjee}}, \bibinfo {author} {\bibfnamefont {Wei}\
  \bibnamefont {Wu}}, \bibinfo {author} {\bibfnamefont {Michel}\ \bibnamefont
  {Ferrero}}, \bibinfo {author} {\bibfnamefont {Antoine}\ \bibnamefont
  {Georges}}, \ and\ \bibinfo {author} {\bibfnamefont {Subir}\ \bibnamefont
  {Sachdev}},\ }\bibfield  {title} {\enquote {\bibinfo {title} {Topological
  order in the pseudogap metal},}\ }\href {\doibase 10.1073/pnas.1720580115}
  {\bibfield  {journal} {\bibinfo  {journal} {Proceedings of the National
  Academy of Sciences}\ }\textbf {\bibinfo {volume} {115}},\ \bibinfo {pages}
  {E3665--E3672} (\bibinfo {year} {2018})},\ \Eprint
  {http://arxiv.org/abs/https://www.pnas.org/content/115/16/E3665.full.pdf}
  {https://www.pnas.org/content/115/16/E3665.full.pdf} \BibitemShut {NoStop}%
\bibitem [{\citenamefont {Reymbaut}\ \emph {et~al.}(2019)\citenamefont
  {Reymbaut}, \citenamefont {Bergeron}, \citenamefont {Garioud}, \citenamefont
  {Th\'enault}, \citenamefont {Charlebois}, \citenamefont {S\'emon},\ and\
  \citenamefont {Tremblay}}]{Reymbaut2019}%
  \BibitemOpen
  \bibfield  {author} {\bibinfo {author} {\bibfnamefont {A.}~\bibnamefont
  {Reymbaut}}, \bibinfo {author} {\bibfnamefont {S.}~\bibnamefont {Bergeron}},
  \bibinfo {author} {\bibfnamefont {R.}~\bibnamefont {Garioud}}, \bibinfo
  {author} {\bibfnamefont {M.}~\bibnamefont {Th\'enault}}, \bibinfo {author}
  {\bibfnamefont {M.}~\bibnamefont {Charlebois}}, \bibinfo {author}
  {\bibfnamefont {P.}~\bibnamefont {S\'emon}}, \ and\ \bibinfo {author}
  {\bibfnamefont {A.-M.~S.}\ \bibnamefont {Tremblay}},\ }\bibfield  {title}
  {\enquote {\bibinfo {title} {{Pseudogap, van Hove singularity, maximum in
  entropy, and specific heat for hole-doped Mott insulators}},}\ }\href
  {\doibase 10.1103/PhysRevResearch.1.023015} {\bibfield  {journal} {\bibinfo
  {journal} {Phys. Rev. Research}\ }\textbf {\bibinfo {volume} {1}},\ \bibinfo
  {pages} {023015} (\bibinfo {year} {2019})}\BibitemShut {NoStop}%
\bibitem [{\citenamefont {Fratino}\ \emph {et~al.}(2017)\citenamefont
  {Fratino}, \citenamefont {S\'emon}, \citenamefont {Charlebois}, \citenamefont
  {Sordi},\ and\ \citenamefont {Tremblay}}]{Fratino2017}%
  \BibitemOpen
  \bibfield  {author} {\bibinfo {author} {\bibfnamefont {L.}~\bibnamefont
  {Fratino}}, \bibinfo {author} {\bibfnamefont {P.}~\bibnamefont {S\'emon}},
  \bibinfo {author} {\bibfnamefont {M.}~\bibnamefont {Charlebois}}, \bibinfo
  {author} {\bibfnamefont {G.}~\bibnamefont {Sordi}}, \ and\ \bibinfo {author}
  {\bibfnamefont {A.-M.~S.}\ \bibnamefont {Tremblay}},\ }\bibfield  {title}
  {\enquote {\bibinfo {title} {{Signatures of the Mott transition in the
  antiferromagnetic state of the two-dimensional Hubbard model}},}\ }\href
  {\doibase 10.1103/PhysRevB.95.235109} {\bibfield  {journal} {\bibinfo
  {journal} {Phys. Rev. B}\ }\textbf {\bibinfo {volume} {95}},\ \bibinfo
  {pages} {235109} (\bibinfo {year} {2017})}\BibitemShut {NoStop}%
\bibitem [{\citenamefont {{Vilk}}(1995)}]{Vilk1995}%
  \BibitemOpen
  \bibfield  {author} {\bibinfo {author} {\bibfnamefont {Y.}~\bibnamefont
  {{Vilk}}},\ }\bibfield  {title} {\enquote {\bibinfo {title} {{{Destruction of
  the Fermi liquid by spin fluctuations in two dimensions}}},}\ }\href
  {\doibase 10.1016/0022-3697(95)00168-9} {\bibfield  {journal} {\bibinfo
  {journal} {Journal of Physics and Chemistry of Solids}\ }\textbf {\bibinfo
  {volume} {56}},\ \bibinfo {pages} {1769--1771} (\bibinfo {year}
  {1995})}\BibitemShut {NoStop}%
\bibitem [{\citenamefont {Vilk}\ and\ \citenamefont
  {Tremblay}(1996)}]{Vilk1996}%
  \BibitemOpen
  \bibfield  {author} {\bibinfo {author} {\bibfnamefont {Y.~M.}\ \bibnamefont
  {Vilk}}\ and\ \bibinfo {author} {\bibfnamefont {A.-M.~S.}\ \bibnamefont
  {Tremblay}},\ }\bibfield  {title} {\enquote {\bibinfo {title} {{Destruction
  of Fermi-liquid quasiparticles in two dimensions by critical
  fluctuations}},}\ }\href {http://stacks.iop.org/0295-5075/33/i=2/a=159}
  {\bibfield  {journal} {\bibinfo  {journal} {EPL (Europhysics Letters)}\
  }\textbf {\bibinfo {volume} {33}},\ \bibinfo {pages} {159} (\bibinfo {year}
  {1996})}\BibitemShut {NoStop}%
\bibitem [{\citenamefont {Georges}\ and\ \citenamefont
  {Krauth}(1993)}]{Georges1993}%
  \BibitemOpen
  \bibfield  {author} {\bibinfo {author} {\bibfnamefont {Antoine}\ \bibnamefont
  {Georges}}\ and\ \bibinfo {author} {\bibfnamefont {Werner}\ \bibnamefont
  {Krauth}},\ }\bibfield  {title} {\enquote {\bibinfo {title} {{Physical
  properties of the half-filled Hubbard model in infinite dimensions}},}\
  }\href {\doibase 10.1103/PhysRevB.48.7167} {\bibfield  {journal} {\bibinfo
  {journal} {Phys. Rev. B}\ }\textbf {\bibinfo {volume} {48}},\ \bibinfo
  {pages} {7167--7182} (\bibinfo {year} {1993})}\BibitemShut {NoStop}%
\bibitem [{\citenamefont {Werner}\ \emph {et~al.}(2005)\citenamefont {Werner},
  \citenamefont {Parcollet}, \citenamefont {Georges},\ and\ \citenamefont
  {Hassan}}]{Werner2005}%
  \BibitemOpen
  \bibfield  {author} {\bibinfo {author} {\bibfnamefont {F.}~\bibnamefont
  {Werner}}, \bibinfo {author} {\bibfnamefont {O.}~\bibnamefont {Parcollet}},
  \bibinfo {author} {\bibfnamefont {A.}~\bibnamefont {Georges}}, \ and\
  \bibinfo {author} {\bibfnamefont {S.~R.}\ \bibnamefont {Hassan}},\ }\bibfield
   {title} {\enquote {\bibinfo {title} {{Interaction-Induced Adiabatic Cooling
  and Antiferromagnetism of Cold Fermions in Optical Lattices}},}\ }\href
  {\doibase 10.1103/PhysRevLett.95.056401} {\bibfield  {journal} {\bibinfo
  {journal} {Phys. Rev. Lett.}\ }\textbf {\bibinfo {volume} {95}},\ \bibinfo
  {pages} {056401} (\bibinfo {year} {2005})}\BibitemShut {NoStop}%
\bibitem [{\citenamefont {Dar\'e}\ \emph {et~al.}(2007)\citenamefont {Dar\'e},
  \citenamefont {Raymond}, \citenamefont {Albinet},\ and\ \citenamefont
  {Tremblay}}]{Dare2007}%
  \BibitemOpen
  \bibfield  {author} {\bibinfo {author} {\bibfnamefont {A.-M.}\ \bibnamefont
  {Dar\'e}}, \bibinfo {author} {\bibfnamefont {L.}~\bibnamefont {Raymond}},
  \bibinfo {author} {\bibfnamefont {G.}~\bibnamefont {Albinet}}, \ and\
  \bibinfo {author} {\bibfnamefont {A.-M.~S.}\ \bibnamefont {Tremblay}},\
  }\bibfield  {title} {\enquote {\bibinfo {title} {{Interaction-induced
  adiabatic cooling for antiferromagnetism in optical lattices}},}\ }\href
  {\doibase 10.1103/PhysRevB.76.064402} {\bibfield  {journal} {\bibinfo
  {journal} {Phys. Rev. B}\ }\textbf {\bibinfo {volume} {76}},\ \bibinfo
  {pages} {064402} (\bibinfo {year} {2007})}\BibitemShut {NoStop}%
\bibitem [{\citenamefont {{Taie}}\ \emph {et~al.}(2012)\citenamefont {{Taie}},
  \citenamefont {{Yamazaki}}, \citenamefont {{Sugawa}},\ and\ \citenamefont
  {{Takahashi}}}]{Taie2012}%
  \BibitemOpen
  \bibfield  {author} {\bibinfo {author} {\bibfnamefont {Shintaro}\
  \bibnamefont {{Taie}}}, \bibinfo {author} {\bibfnamefont {Rekishu}\
  \bibnamefont {{Yamazaki}}}, \bibinfo {author} {\bibfnamefont {Seiji}\
  \bibnamefont {{Sugawa}}}, \ and\ \bibinfo {author} {\bibfnamefont {Yoshiro}\
  \bibnamefont {{Takahashi}}},\ }\bibfield  {title} {\enquote {\bibinfo {title}
  {{{An SU(6) Mott insulator of an atomic Fermi gas realized by large-spin
  Pomeranchuk cooling}}},}\ }\href {\doibase 10.1038/nphys2430} {\bibfield
  {journal} {\bibinfo  {journal} {Nature Physics}\ }\textbf {\bibinfo {volume}
  {8}},\ \bibinfo {pages} {825--830} (\bibinfo {year} {2012})},\ \Eprint
  {http://arxiv.org/abs/1208.4883} {arXiv:1208.4883 [cond-mat.quant-gas]}
  \BibitemShut {NoStop}%
\bibitem [{\citenamefont {Galitskii}\ and\ \citenamefont
  {Migdal}(1958)}]{Galitskii1958}%
  \BibitemOpen
  \bibfield  {author} {\bibinfo {author} {\bibfnamefont {V.~M.}\ \bibnamefont
  {Galitskii}}\ and\ \bibinfo {author} {\bibfnamefont {A.~B.}\ \bibnamefont
  {Migdal}},\ }\bibfield  {title} {\enquote {\bibinfo {title} {{Application of
  quantum fiel theoretical methods to the many-body problem}},}\ }\href
  {http://www.jetp.ac.ru/cgi-bin/e/index/e/7/1/p96?a=list} {\bibfield
  {journal} {\bibinfo  {journal} {Sov. Phys. JETP}\ }\textbf {\bibinfo {volume}
  {34}},\ \bibinfo {pages} {139} (\bibinfo {year} {1958})}\BibitemShut
  {NoStop}%
\bibitem [{\citenamefont {van Loon}\ \emph {et~al.}(2016)\citenamefont {van
  Loon}, \citenamefont {Krien}, \citenamefont {Hafermann}, \citenamefont
  {Stepanov}, \citenamefont {Lichtenstein},\ and\ \citenamefont
  {Katsnelson}}]{vanLoon2016}%
  \BibitemOpen
  \bibfield  {author} {\bibinfo {author} {\bibfnamefont {Erik G. C.~P.}\
  \bibnamefont {van Loon}}, \bibinfo {author} {\bibfnamefont {Friedrich}\
  \bibnamefont {Krien}}, \bibinfo {author} {\bibfnamefont {Hartmut}\
  \bibnamefont {Hafermann}}, \bibinfo {author} {\bibfnamefont {Evgeny~A.}\
  \bibnamefont {Stepanov}}, \bibinfo {author} {\bibfnamefont {Alexander~I.}\
  \bibnamefont {Lichtenstein}}, \ and\ \bibinfo {author} {\bibfnamefont
  {Mikhail~I.}\ \bibnamefont {Katsnelson}},\ }\bibfield  {title} {\enquote
  {\bibinfo {title} {Double occupancy in dynamical mean-field theory and the
  dual boson approach},}\ }\href {\doibase 10.1103/PhysRevB.93.155162}
  {\bibfield  {journal} {\bibinfo  {journal} {Phys. Rev. B}\ }\textbf {\bibinfo
  {volume} {93}},\ \bibinfo {pages} {155162} (\bibinfo {year}
  {2016})}\BibitemShut {NoStop}%
\bibitem [{\citenamefont {Chakravarty}\ \emph {et~al.}(1988)\citenamefont
  {Chakravarty}, \citenamefont {Halperin},\ and\ \citenamefont
  {Nelson}}]{Chakravarty1988}%
  \BibitemOpen
  \bibfield  {author} {\bibinfo {author} {\bibfnamefont {Sudip}\ \bibnamefont
  {Chakravarty}}, \bibinfo {author} {\bibfnamefont {Bertrand~I.}\ \bibnamefont
  {Halperin}}, \ and\ \bibinfo {author} {\bibfnamefont {David~R.}\ \bibnamefont
  {Nelson}},\ }\bibfield  {title} {\enquote {\bibinfo {title} {Low-temperature
  behavior of two-dimensional quantum antiferromagnets},}\ }\href {\doibase
  10.1103/PhysRevLett.60.1057} {\bibfield  {journal} {\bibinfo  {journal}
  {Phys. Rev. Lett.}\ }\textbf {\bibinfo {volume} {60}},\ \bibinfo {pages}
  {1057--1060} (\bibinfo {year} {1988})}\BibitemShut {NoStop}%
\bibitem [{\citenamefont {Chakravarty}\ \emph {et~al.}(1989)\citenamefont
  {Chakravarty}, \citenamefont {Halperin},\ and\ \citenamefont
  {Nelson}}]{Chakravarty1989}%
  \BibitemOpen
  \bibfield  {author} {\bibinfo {author} {\bibfnamefont {Sudip}\ \bibnamefont
  {Chakravarty}}, \bibinfo {author} {\bibfnamefont {Bertrand~I.}\ \bibnamefont
  {Halperin}}, \ and\ \bibinfo {author} {\bibfnamefont {David~R.}\ \bibnamefont
  {Nelson}},\ }\bibfield  {title} {\enquote {\bibinfo {title} {{Two-dimensional
  quantum Heisenberg antiferromagnet at low temperatures}},}\ }\href {\doibase
  10.1103/PhysRevB.39.2344} {\bibfield  {journal} {\bibinfo  {journal} {Phys.
  Rev. B}\ }\textbf {\bibinfo {volume} {39}},\ \bibinfo {pages} {2344--2371}
  (\bibinfo {year} {1989})}\BibitemShut {NoStop}%
\bibitem [{\citenamefont {Borejsza}\ and\ \citenamefont
  {Dupuis}(2003)}]{Borejsza2003}%
  \BibitemOpen
  \bibfield  {author} {\bibinfo {author} {\bibfnamefont {K.}~\bibnamefont
  {Borejsza}}\ and\ \bibinfo {author} {\bibfnamefont {N.}~\bibnamefont
  {Dupuis}},\ }\bibfield  {title} {\enquote {\bibinfo {title}
  {{Antiferromagnetism and single-particle properties in the two-dimensional
  half-filled Hubbard model: Slater vs. Mott-Heisenberg}},}\ }\href {\doibase
  10.1209/epl/i2003-00584-7} {\bibfield  {journal} {\bibinfo  {journal}
  {Europhys. Lett.}\ }\textbf {\bibinfo {volume} {63}},\ \bibinfo {pages}
  {722--728} (\bibinfo {year} {2003})}\BibitemShut {NoStop}%
\bibitem [{\citenamefont {Borejsza}\ and\ \citenamefont
  {Dupuis}(2004)}]{Borejsza2004}%
  \BibitemOpen
  \bibfield  {author} {\bibinfo {author} {\bibfnamefont {K.}~\bibnamefont
  {Borejsza}}\ and\ \bibinfo {author} {\bibfnamefont {N.}~\bibnamefont
  {Dupuis}},\ }\bibfield  {title} {\enquote {\bibinfo {title}
  {{Antiferromagnetism and single-particle properties in the two-dimensional
  half-filled Hubbard model: A nonlinear sigma model approach}},}\ }\href
  {\doibase 10.1103/PhysRevB.69.085119} {\bibfield  {journal} {\bibinfo
  {journal} {Phys. Rev. B}\ }\textbf {\bibinfo {volume} {69}},\ \bibinfo
  {pages} {085119} (\bibinfo {year} {2004})}\BibitemShut {NoStop}%
\bibitem [{\citenamefont {Maier}\ \emph
  {et~al.}(2005{\natexlab{b}})\citenamefont {Maier}, \citenamefont {Jarrell},
  \citenamefont {Schulthess}, \citenamefont {Kent},\ and\ \citenamefont
  {White}}]{Maier2005a}%
  \BibitemOpen
  \bibfield  {author} {\bibinfo {author} {\bibfnamefont {T.~A.}\ \bibnamefont
  {Maier}}, \bibinfo {author} {\bibfnamefont {M.}~\bibnamefont {Jarrell}},
  \bibinfo {author} {\bibfnamefont {T.~C.}\ \bibnamefont {Schulthess}},
  \bibinfo {author} {\bibfnamefont {P.~R.~C.}\ \bibnamefont {Kent}}, \ and\
  \bibinfo {author} {\bibfnamefont {J.~B.}\ \bibnamefont {White}},\ }\bibfield
  {title} {\enquote {\bibinfo {title} {{Systematic Study of $d$-Wave
  Superconductivity in the 2D Repulsive Hubbard Model}},}\ }\href {\doibase
  10.1103/PhysRevLett.95.237001} {\bibfield  {journal} {\bibinfo  {journal}
  {Phys. Rev. Lett.}\ }\textbf {\bibinfo {volume} {95}},\ \bibinfo {pages}
  {237001} (\bibinfo {year} {2005}{\natexlab{b}})}\BibitemShut {NoStop}%
\bibitem [{\citenamefont {Toschi}\ \emph {et~al.}(2007)\citenamefont {Toschi},
  \citenamefont {Katanin},\ and\ \citenamefont {Held}}]{Toschi2007}%
  \BibitemOpen
  \bibfield  {author} {\bibinfo {author} {\bibfnamefont {A.}~\bibnamefont
  {Toschi}}, \bibinfo {author} {\bibfnamefont {A.~A.}\ \bibnamefont {Katanin}},
  \ and\ \bibinfo {author} {\bibfnamefont {K.}~\bibnamefont {Held}},\
  }\bibfield  {title} {\enquote {\bibinfo {title} {Dynamical vertex
  approximation; a step beyond dynamical mean-field theory},}\ }\href {\doibase
  10.1103/PhysRevB.75.045118} {\bibfield  {journal} {\bibinfo  {journal} {Phys
  Rev. B}\ }\textbf {\bibinfo {volume} {75}},\ \bibinfo {pages} {045118}
  (\bibinfo {year} {2007})}\BibitemShut {NoStop}%
\bibitem [{\citenamefont {Wentzell}\ \emph {et~al.}(2020)\citenamefont
  {Wentzell}, \citenamefont {Li}, \citenamefont {Tagliavini}, \citenamefont
  {Taranto}, \citenamefont {Rohringer}, \citenamefont {Held}, \citenamefont
  {Toschi},\ and\ \citenamefont {Andergassen}}]{Wentzell2020}%
  \BibitemOpen
  \bibfield  {author} {\bibinfo {author} {\bibfnamefont {Nils}\ \bibnamefont
  {Wentzell}}, \bibinfo {author} {\bibfnamefont {Gang}\ \bibnamefont {Li}},
  \bibinfo {author} {\bibfnamefont {Agnese}\ \bibnamefont {Tagliavini}},
  \bibinfo {author} {\bibfnamefont {Ciro}\ \bibnamefont {Taranto}}, \bibinfo
  {author} {\bibfnamefont {Georg}\ \bibnamefont {Rohringer}}, \bibinfo {author}
  {\bibfnamefont {Karsten}\ \bibnamefont {Held}}, \bibinfo {author}
  {\bibfnamefont {Alessandro}\ \bibnamefont {Toschi}}, \ and\ \bibinfo {author}
  {\bibfnamefont {Sabine}\ \bibnamefont {Andergassen}},\ }\bibfield  {title}
  {\enquote {\bibinfo {title} {{High-frequency asymptotics of the vertex
  function: Diagrammatic parametrization and algorithmic implementation}},}\
  }\href {\doibase 10.1103/PhysRevB.102.085106} {\bibfield  {journal} {\bibinfo
   {journal} {Phys. Rev. B}\ }\textbf {\bibinfo {volume} {102}},\ \bibinfo
  {pages} {085106} (\bibinfo {year} {2020})}\BibitemShut {NoStop}%
\bibitem [{\citenamefont {Kaufmann}\ \emph {et~al.}(2017)\citenamefont
  {Kaufmann}, \citenamefont {Gunacker},\ and\ \citenamefont
  {Held}}]{Kaufmann2017}%
  \BibitemOpen
  \bibfield  {author} {\bibinfo {author} {\bibfnamefont {Josef}\ \bibnamefont
  {Kaufmann}}, \bibinfo {author} {\bibfnamefont {Patrik}\ \bibnamefont
  {Gunacker}}, \ and\ \bibinfo {author} {\bibfnamefont {Karsten}\ \bibnamefont
  {Held}},\ }\bibfield  {title} {\enquote {\bibinfo {title} {{Continuous-time
  quantum Monte Carlo calculation of multiorbital vertex asymptotics}},}\
  }\href {\doibase 10.1103/PhysRevB.96.035114} {\bibfield  {journal} {\bibinfo
  {journal} {Phys. Rev. B}\ }\textbf {\bibinfo {volume} {96}},\ \bibinfo
  {pages} {035114} (\bibinfo {year} {2017})}\BibitemShut {NoStop}%
\bibitem [{\citenamefont {Katanin}(2020)}]{Katanin2020}%
  \BibitemOpen
  \bibfield  {author} {\bibinfo {author} {\bibfnamefont {A.}~\bibnamefont
  {Katanin}},\ }\bibfield  {title} {\enquote {\bibinfo {title} {{Improved
  treatment of fermion-boson vertices and Bethe-Salpeter equations in nonlocal
  extensions of dynamical mean field theory}},}\ }\href {\doibase
  10.1103/PhysRevB.101.035110} {\bibfield  {journal} {\bibinfo  {journal}
  {Phys. Rev. B}\ }\textbf {\bibinfo {volume} {101}},\ \bibinfo {pages}
  {035110} (\bibinfo {year} {2020})}\BibitemShut {NoStop}%
\bibitem [{\citenamefont {Ornstein}\ and\ \citenamefont
  {Zernike}(1916)}]{Zernike1916}%
  \BibitemOpen
  \bibfield  {author} {\bibinfo {author} {\bibfnamefont {L.~S.}\ \bibnamefont
  {Ornstein}}\ and\ \bibinfo {author} {\bibfnamefont {F.}~\bibnamefont
  {Zernike}},\ }\href@noop {} {\bibfield  {journal} {\bibinfo  {journal} {Proc.
  Roy. Acad. Amsterdam}\ }\textbf {\bibinfo {volume} {17}},\ \bibinfo {pages}
  {793} (\bibinfo {year} {1916})}\BibitemShut {NoStop}%
\bibitem [{\citenamefont {Moukouri}\ \emph {et~al.}(2000)\citenamefont
  {Moukouri}, \citenamefont {Allen}, \citenamefont {Lemay}, \citenamefont
  {Kyung}, \citenamefont {Poulin}, \citenamefont {Vilk},\ and\ \citenamefont
  {Tremblay}}]{Moukouri2000}%
  \BibitemOpen
  \bibfield  {author} {\bibinfo {author} {\bibfnamefont {S.}~\bibnamefont
  {Moukouri}}, \bibinfo {author} {\bibfnamefont {S.}~\bibnamefont {Allen}},
  \bibinfo {author} {\bibfnamefont {F.}~\bibnamefont {Lemay}}, \bibinfo
  {author} {\bibfnamefont {B.}~\bibnamefont {Kyung}}, \bibinfo {author}
  {\bibfnamefont {D.}~\bibnamefont {Poulin}}, \bibinfo {author} {\bibfnamefont
  {Y.~M.}\ \bibnamefont {Vilk}}, \ and\ \bibinfo {author} {\bibfnamefont
  {A.-M.~S.}\ \bibnamefont {Tremblay}},\ }\bibfield  {title} {\enquote
  {\bibinfo {title} {{Many-body theory versus simulations for the pseudogap in
  the Hubbard model}},}\ }\href {\doibase 10.1103/PhysRevB.61.7887} {\bibfield
  {journal} {\bibinfo  {journal} {Phys. Rev. B}\ }\textbf {\bibinfo {volume}
  {61}},\ \bibinfo {pages} {7887--7892} (\bibinfo {year} {2000})}\BibitemShut
  {NoStop}%
\bibitem [{\citenamefont {Sedrakyan}\ and\ \citenamefont
  {Chubukov}(2010)}]{chubukov2010}%
  \BibitemOpen
  \bibfield  {author} {\bibinfo {author} {\bibfnamefont {Tigran~A.}\
  \bibnamefont {Sedrakyan}}\ and\ \bibinfo {author} {\bibfnamefont {Andrey~V.}\
  \bibnamefont {Chubukov}},\ }\bibfield  {title} {\enquote {\bibinfo {title}
  {Pseudogap in underdoped cuprates and spin-density-wave fluctuations},}\
  }\href {\doibase 10.1103/PhysRevB.81.174536} {\bibfield  {journal} {\bibinfo
  {journal} {Phys. Rev. B}\ }\textbf {\bibinfo {volume} {81}},\ \bibinfo
  {pages} {174536} (\bibinfo {year} {2010})}\BibitemShut {NoStop}%
\bibitem [{\citenamefont {Vilk}(1997)}]{Vilk1997b}%
  \BibitemOpen
  \bibfield  {author} {\bibinfo {author} {\bibfnamefont {Y.~M.}\ \bibnamefont
  {Vilk}},\ }\bibfield  {title} {\enquote {\bibinfo {title} {Shadow features
  and shadow bands in the paramagnetic state of cuprate superconductors},}\
  }\href {\doibase 10.1103/PhysRevB.55.3870} {\bibfield  {journal} {\bibinfo
  {journal} {Phys. Rev. B}\ }\textbf {\bibinfo {volume} {55}},\ \bibinfo
  {pages} {3870--3875} (\bibinfo {year} {1997})}\BibitemShut {NoStop}%
\bibitem [{\citenamefont {Lemay}(2000)}]{Lemay2000}%
  \BibitemOpen
  \bibfield  {author} {\bibinfo {author} {\bibfnamefont {Fran{\c c}ois}\
  \bibnamefont {Lemay}},\ }\emph {\bibinfo {title} {Des propri{\'e}t{\'e}s de
  l'{\'e}tat normal du mod{\`e}le de Hubbard bidimensionnel}},\ \href
  {https://savoirs.usherbrooke.ca/bitstream/handle/11143/4993/NQ67111.pdf?sequence=1&isAllowed=y}
  {Ph.D. thesis},\ \bibinfo  {school} {Universit{\'e} de Sherbrooke} (\bibinfo
  {year} {2000})\BibitemShut {NoStop}%
\bibitem [{\citenamefont {Zlatic}(1997)}]{Zlatic1997}%
  \BibitemOpen
  \bibfield  {author} {\bibinfo {author} {\bibfnamefont {V}~\bibnamefont
  {Zlatic}},\ }\bibfield  {title} {\enquote {\bibinfo {title} {{Temperature
  dependence of the spectral properties for 2-D Hubbard model with anisotropic
  hopping}},}\ }\href {\doibase 10.1016/S0921-4526(96)00797-1} {\bibfield
  {journal} {\bibinfo  {journal} {PHYSICA B}\ }\textbf {\bibinfo {volume}
  {230}},\ \bibinfo {pages} {1034--1036} (\bibinfo {year} {1997})}\BibitemShut
  {NoStop}%
\bibitem [{\citenamefont {Katanin}\ \emph {et~al.}(2005)\citenamefont
  {Katanin}, \citenamefont {Kampf},\ and\ \citenamefont
  {Irkhin}}]{Katanin2005b}%
  \BibitemOpen
  \bibfield  {author} {\bibinfo {author} {\bibfnamefont {A.~A.}\ \bibnamefont
  {Katanin}}, \bibinfo {author} {\bibfnamefont {A.~P.}\ \bibnamefont {Kampf}},
  \ and\ \bibinfo {author} {\bibfnamefont {V.~Yu.}\ \bibnamefont {Irkhin}},\
  }\bibfield  {title} {\enquote {\bibinfo {title} {{Anomalous self-energy and
  Fermi surface quasisplitting in the vicinity of a ferromagnetic
  instability}},}\ }\href {\doibase 10.1103/PhysRevB.71.085105} {\bibfield
  {journal} {\bibinfo  {journal} {Phys. Rev. B}\ }\textbf {\bibinfo {volume}
  {71}},\ \bibinfo {pages} {085105} (\bibinfo {year} {2005})}\BibitemShut
  {NoStop}%
\bibitem [{\citenamefont {{M{\"u}ller-Hartmann}}(1989)}]{MuellerHartmann1989}%
  \BibitemOpen
  \bibfield  {author} {\bibinfo {author} {\bibfnamefont {E.}~\bibnamefont
  {{M{\"u}ller-Hartmann}}},\ }\bibfield  {title} {\enquote {\bibinfo {title}
  {{{The Hubbard model at high dimensions: some exact results and weak coupling
  theory}}},}\ }\href {\doibase 10.1007/BF01312686} {\bibfield  {journal}
  {\bibinfo  {journal} {Zeitschrift fur Physik B Condensed Matter}\ }\textbf
  {\bibinfo {volume} {76}},\ \bibinfo {pages} {211--217} (\bibinfo {year}
  {1989})}\BibitemShut {NoStop}%
\bibitem [{\citenamefont {Ku}\ and\ \citenamefont {Eguiluz}(2002)}]{Ku2002}%
  \BibitemOpen
  \bibfield  {author} {\bibinfo {author} {\bibfnamefont {Wei}\ \bibnamefont
  {Ku}}\ and\ \bibinfo {author} {\bibfnamefont {Adolfo~G.}\ \bibnamefont
  {Eguiluz}},\ }\bibfield  {title} {\enquote {\bibinfo {title} {Band-gap
  problem in semiconductors revisited: Effects of core states and many-body
  self-consistency},}\ }\href {\doibase 10.1103/PhysRevLett.89.126401}
  {\bibfield  {journal} {\bibinfo  {journal} {Phys. Rev. Lett.}\ }\textbf
  {\bibinfo {volume} {89}},\ \bibinfo {pages} {126401} (\bibinfo {year}
  {2002})}\BibitemShut {NoStop}%
\bibitem [{\citenamefont {Delaney}\ \emph {et~al.}(2004)\citenamefont
  {Delaney}, \citenamefont {Garc\'{\i}a-Gonz\'alez}, \citenamefont {Rubio},
  \citenamefont {Rinke},\ and\ \citenamefont {Godby}}]{Delaney2004}%
  \BibitemOpen
  \bibfield  {author} {\bibinfo {author} {\bibfnamefont {Kris}\ \bibnamefont
  {Delaney}}, \bibinfo {author} {\bibfnamefont {P.}~\bibnamefont
  {Garc\'{\i}a-Gonz\'alez}}, \bibinfo {author} {\bibfnamefont {Angel}\
  \bibnamefont {Rubio}}, \bibinfo {author} {\bibfnamefont {Patrick}\
  \bibnamefont {Rinke}}, \ and\ \bibinfo {author} {\bibfnamefont {R.~W.}\
  \bibnamefont {Godby}},\ }\bibfield  {title} {\enquote {\bibinfo {title}
  {Comment on ``band-gap problem in semiconductors revisited: Effects of core
  states and many-body self-consistency''},}\ }\href {\doibase
  10.1103/PhysRevLett.93.249701} {\bibfield  {journal} {\bibinfo  {journal}
  {Phys. Rev. Lett.}\ }\textbf {\bibinfo {volume} {93}},\ \bibinfo {pages}
  {249701} (\bibinfo {year} {2004})}\BibitemShut {NoStop}%
\bibitem [{\citenamefont {Ku}\ and\ \citenamefont {Eguiluz}(2004)}]{Ku2004}%
  \BibitemOpen
  \bibfield  {author} {\bibinfo {author} {\bibfnamefont {Wei}\ \bibnamefont
  {Ku}}\ and\ \bibinfo {author} {\bibfnamefont {A.~G.}\ \bibnamefont
  {Eguiluz}},\ }\bibfield  {title} {\enquote {\bibinfo {title} {{Ku and Eguiluz
  Reply}},}\ }\href {\doibase 10.1103/PhysRevLett.93.249702} {\bibfield
  {journal} {\bibinfo  {journal} {Phys. Rev. Lett.}\ }\textbf {\bibinfo
  {volume} {93}},\ \bibinfo {pages} {249702} (\bibinfo {year}
  {2004})}\BibitemShut {NoStop}%
\bibitem [{\citenamefont {Prokof'ev}\ and\ \citenamefont
  {Svistunov}(2007)}]{Prokofev2007}%
  \BibitemOpen
  \bibfield  {author} {\bibinfo {author} {\bibfnamefont {Nikolay}\ \bibnamefont
  {Prokof'ev}}\ and\ \bibinfo {author} {\bibfnamefont {Boris}\ \bibnamefont
  {Svistunov}},\ }\bibfield  {title} {\enquote {\bibinfo {title} {{Bold
  Diagrammatic Monte Carlo Technique: When the Sign Problem Is Welcome}},}\
  }\href {\doibase 10.1103/PhysRevLett.99.250201} {\bibfield  {journal}
  {\bibinfo  {journal} {Phys. Rev. Lett.}\ }\textbf {\bibinfo {volume} {99}},\
  \bibinfo {pages} {250201} (\bibinfo {year} {2007})}\BibitemShut {NoStop}%
\bibitem [{\citenamefont {Deng}\ \emph {et~al.}(2015)\citenamefont {Deng},
  \citenamefont {Kozik}, \citenamefont {Prokof'ev},\ and\ \citenamefont
  {Svistunov}}]{Deng2015}%
  \BibitemOpen
  \bibfield  {author} {\bibinfo {author} {\bibfnamefont {Youjin}\ \bibnamefont
  {Deng}}, \bibinfo {author} {\bibfnamefont {Evgeny}\ \bibnamefont {Kozik}},
  \bibinfo {author} {\bibfnamefont {Nikolay~V.}\ \bibnamefont {Prokof'ev}}, \
  and\ \bibinfo {author} {\bibfnamefont {Boris~V.}\ \bibnamefont {Svistunov}},\
  }\bibfield  {title} {\enquote {\bibinfo {title} {{Emergent {BCS} regime of
  the two-dimensional fermionic Hubbard model: Ground-state phase diagram}},}\
  }\href {\doibase 10.1209/0295-5075/110/57001} {\bibfield  {journal} {\bibinfo
   {journal} {{EPL} (Europhysics Letters)}\ }\textbf {\bibinfo {volume}
  {110}},\ \bibinfo {pages} {57001} (\bibinfo {year} {2015})}\BibitemShut
  {NoStop}%
\bibitem [{\citenamefont {Kozik}\ \emph {et~al.}(2015)\citenamefont {Kozik},
  \citenamefont {Ferrero},\ and\ \citenamefont {Georges}}]{Kozik2015}%
  \BibitemOpen
  \bibfield  {author} {\bibinfo {author} {\bibfnamefont {Evgeny}\ \bibnamefont
  {Kozik}}, \bibinfo {author} {\bibfnamefont {Michel}\ \bibnamefont {Ferrero}},
  \ and\ \bibinfo {author} {\bibfnamefont {Antoine}\ \bibnamefont {Georges}},\
  }\bibfield  {title} {\enquote {\bibinfo {title} {{Nonexistence of the
  Luttinger-Ward Functional and Misleading Convergence of Skeleton Diagrammatic
  Series for Hubbard-Like Models}},}\ }\href {\doibase
  10.1103/PhysRevLett.114.156402} {\bibfield  {journal} {\bibinfo  {journal}
  {Phys. Rev. Lett.}\ }\textbf {\bibinfo {volume} {114}},\ \bibinfo {pages}
  {156402} (\bibinfo {year} {2015})}\BibitemShut {NoStop}%
\bibitem [{\citenamefont {Stan}\ \emph {et~al.}(2015)\citenamefont {Stan},
  \citenamefont {Romaniello}, \citenamefont {Rigamonti}, \citenamefont
  {Reining},\ and\ \citenamefont {Berger}}]{Stan2015}%
  \BibitemOpen
  \bibfield  {author} {\bibinfo {author} {\bibfnamefont {A.}~\bibnamefont
  {Stan}}, \bibinfo {author} {\bibfnamefont {P.}~\bibnamefont {Romaniello}},
  \bibinfo {author} {\bibfnamefont {S.}~\bibnamefont {Rigamonti}}, \bibinfo
  {author} {\bibfnamefont {L.}~\bibnamefont {Reining}}, \ and\ \bibinfo
  {author} {\bibfnamefont {J.~A.}\ \bibnamefont {Berger}},\ }\bibfield  {title}
  {\enquote {\bibinfo {title} {Unphysical and physical solutions in many-body
  theories: from weak to strong correlation},}\ }\href
  {http://stacks.iop.org/1367-2630/17/i=9/a=093045} {\bibfield  {journal}
  {\bibinfo  {journal} {New J. Phys.}\ }\textbf {\bibinfo {volume} {17}},\
  \bibinfo {pages} {093045} (\bibinfo {year} {2015})}\BibitemShut {NoStop}%
\bibitem [{\citenamefont {Vu\v{c}i\v{c}evi\'c}\ \emph
  {et~al.}(2018)\citenamefont {Vu\v{c}i\v{c}evi\'c}, \citenamefont {Wentzell},
  \citenamefont {Ferrero},\ and\ \citenamefont {Parcollet}}]{Vucicevic2018}%
  \BibitemOpen
  \bibfield  {author} {\bibinfo {author} {\bibfnamefont {J.}~\bibnamefont
  {Vu\v{c}i\v{c}evi\'c}}, \bibinfo {author} {\bibfnamefont {N.}~\bibnamefont
  {Wentzell}}, \bibinfo {author} {\bibfnamefont {M.}~\bibnamefont {Ferrero}}, \
  and\ \bibinfo {author} {\bibfnamefont {O.}~\bibnamefont {Parcollet}},\
  }\bibfield  {title} {\enquote {\bibinfo {title} {{Practical consequences of
  the Luttinger-Ward functional multivaluedness for cluster DMFT methods}},}\
  }\href {\doibase 10.1103/PhysRevB.97.125141} {\bibfield  {journal} {\bibinfo
  {journal} {Phys. Rev. B}\ }\textbf {\bibinfo {volume} {97}},\ \bibinfo
  {pages} {125141} (\bibinfo {year} {2018})}\BibitemShut {NoStop}%
\bibitem [{\citenamefont {Rossi}\ and\ \citenamefont
  {Werner}(2015)}]{Rossi2015}%
  \BibitemOpen
  \bibfield  {author} {\bibinfo {author} {\bibfnamefont {Riccardo}\
  \bibnamefont {Rossi}}\ and\ \bibinfo {author} {\bibfnamefont {F{\'{e}}lix}\
  \bibnamefont {Werner}},\ }\bibfield  {title} {\enquote {\bibinfo {title}
  {{Skeleton series and multivaluedness of the self-energy functional in zero
  space-time dimensions}},}\ }\href {\doibase 10.1088/1751-8113/48/48/485202}
  {\bibfield  {journal} {\bibinfo  {journal} {Journal of Physics A:
  Mathematical and Theoretical}\ }\textbf {\bibinfo {volume} {48}},\ \bibinfo
  {pages} {485202} (\bibinfo {year} {2015})}\BibitemShut {NoStop}%
\bibitem [{\citenamefont {Kim}\ and\ \citenamefont
  {Sacksteder}(2020)}]{Kim2020b}%
  \BibitemOpen
  \bibfield  {author} {\bibinfo {author} {\bibfnamefont {Aaram~J.}\
  \bibnamefont {Kim}}\ and\ \bibinfo {author} {\bibfnamefont {Vincent}\
  \bibnamefont {Sacksteder}},\ }\bibfield  {title} {\enquote {\bibinfo {title}
  {{Multivaluedness of the Luttinger-Ward functional in the fermionic and
  bosonic system with replicas}},}\ }\href {\doibase
  10.1103/PhysRevB.101.115146} {\bibfield  {journal} {\bibinfo  {journal}
  {Phys. Rev. B}\ }\textbf {\bibinfo {volume} {101}},\ \bibinfo {pages}
  {115146} (\bibinfo {year} {2020})}\BibitemShut {NoStop}%
\bibitem [{\citenamefont {Gunnarsson}\ \emph {et~al.}(2017)\citenamefont
  {Gunnarsson}, \citenamefont {Rohringer}, \citenamefont {Sch\"afer},
  \citenamefont {Sangiovanni},\ and\ \citenamefont {Toschi}}]{Gunnarsson2017}%
  \BibitemOpen
  \bibfield  {author} {\bibinfo {author} {\bibfnamefont {O.}~\bibnamefont
  {Gunnarsson}}, \bibinfo {author} {\bibfnamefont {G.}~\bibnamefont
  {Rohringer}}, \bibinfo {author} {\bibfnamefont {T.}~\bibnamefont
  {Sch\"afer}}, \bibinfo {author} {\bibfnamefont {G.}~\bibnamefont
  {Sangiovanni}}, \ and\ \bibinfo {author} {\bibfnamefont {A.}~\bibnamefont
  {Toschi}},\ }\bibfield  {title} {\enquote {\bibinfo {title} {{Breakdown of
  Traditional Many-Body Theories for Correlated Electrons}},}\ }\href {\doibase
  10.1103/PhysRevLett.119.056402} {\bibfield  {journal} {\bibinfo  {journal}
  {Phys. Rev. Lett.}\ }\textbf {\bibinfo {volume} {119}},\ \bibinfo {pages}
  {056402} (\bibinfo {year} {2017})}\BibitemShut {NoStop}%
\bibitem [{\citenamefont {Sch\"afer}\ \emph {et~al.}(2013)\citenamefont
  {Sch\"afer}, \citenamefont {Rohringer}, \citenamefont {Gunnarsson},
  \citenamefont {Ciuchi}, \citenamefont {Sangiovanni},\ and\ \citenamefont
  {Toschi}}]{Schaefer2013}%
  \BibitemOpen
  \bibfield  {author} {\bibinfo {author} {\bibfnamefont {T.}~\bibnamefont
  {Sch\"afer}}, \bibinfo {author} {\bibfnamefont {G.}~\bibnamefont
  {Rohringer}}, \bibinfo {author} {\bibfnamefont {O.}~\bibnamefont
  {Gunnarsson}}, \bibinfo {author} {\bibfnamefont {S.}~\bibnamefont {Ciuchi}},
  \bibinfo {author} {\bibfnamefont {G.}~\bibnamefont {Sangiovanni}}, \ and\
  \bibinfo {author} {\bibfnamefont {A.}~\bibnamefont {Toschi}},\ }\bibfield
  {title} {\enquote {\bibinfo {title} {{Divergent Precursors of the
  Mott-Hubbard Transition at the Two-Particle Level}},}\ }\href {\doibase
  10.1103/PhysRevLett.110.246405} {\bibfield  {journal} {\bibinfo  {journal}
  {Phys. Rev. Lett.}\ }\textbf {\bibinfo {volume} {110}},\ \bibinfo {pages}
  {246405} (\bibinfo {year} {2013})}\BibitemShut {NoStop}%
\bibitem [{\citenamefont {Jani\v{s}}\ and\ \citenamefont
  {Pokorn\'y}(2014)}]{Janis2014}%
  \BibitemOpen
  \bibfield  {author} {\bibinfo {author} {\bibfnamefont {V.}~\bibnamefont
  {Jani\v{s}}}\ and\ \bibinfo {author} {\bibfnamefont {V.}~\bibnamefont
  {Pokorn\'y}},\ }\bibfield  {title} {\enquote {\bibinfo {title} {Critical
  metal-insulator transition and divergence in a two-particle irreducible
  vertex in disordered and interacting electron systems},}\ }\href {\doibase
  10.1103/PhysRevB.90.045143} {\bibfield  {journal} {\bibinfo  {journal} {Phys.
  Rev. B}\ }\textbf {\bibinfo {volume} {90}},\ \bibinfo {pages} {045143}
  (\bibinfo {year} {2014})}\BibitemShut {NoStop}%
\bibitem [{\citenamefont {Sch\"afer}\ \emph
  {et~al.}(2016{\natexlab{b}})\citenamefont {Sch\"afer}, \citenamefont
  {Ciuchi}, \citenamefont {Wallerberger}, \citenamefont {Thunstr{\"o}m},
  \citenamefont {Gunnarsson}, \citenamefont {Sangiovanni}, \citenamefont
  {Rohringer},\ and\ \citenamefont {Toschi}}]{Schaefer2016c}%
  \BibitemOpen
  \bibfield  {author} {\bibinfo {author} {\bibfnamefont {T.}~\bibnamefont
  {Sch\"afer}}, \bibinfo {author} {\bibfnamefont {S.}~\bibnamefont {Ciuchi}},
  \bibinfo {author} {\bibfnamefont {M.}~\bibnamefont {Wallerberger}}, \bibinfo
  {author} {\bibfnamefont {P.}~\bibnamefont {Thunstr{\"o}m}}, \bibinfo {author}
  {\bibfnamefont {O.}~\bibnamefont {Gunnarsson}}, \bibinfo {author}
  {\bibfnamefont {G.}~\bibnamefont {Sangiovanni}}, \bibinfo {author}
  {\bibfnamefont {G.}~\bibnamefont {Rohringer}}, \ and\ \bibinfo {author}
  {\bibfnamefont {A.}~\bibnamefont {Toschi}},\ }\bibfield  {title} {\enquote
  {\bibinfo {title} {{Non-perturbative landscape of the Mott-Hubbard
  transition: Multiple divergence lines around the critical endpoint}},}\
  }\href {\doibase 10.1103/PhysRevB.94.235108} {\bibfield  {journal} {\bibinfo
  {journal} {Phys. Rev. B}\ }\textbf {\bibinfo {volume} {94}},\ \bibinfo
  {pages} {235108} (\bibinfo {year} {2016}{\natexlab{b}})}\BibitemShut
  {NoStop}%
\bibitem [{\citenamefont {Chalupa}\ \emph {et~al.}(2018)\citenamefont
  {Chalupa}, \citenamefont {Gunacker}, \citenamefont {Sch\"afer}, \citenamefont
  {Held},\ and\ \citenamefont {Toschi}}]{Chalupa2017}%
  \BibitemOpen
  \bibfield  {author} {\bibinfo {author} {\bibfnamefont {P.}~\bibnamefont
  {Chalupa}}, \bibinfo {author} {\bibfnamefont {P.}~\bibnamefont {Gunacker}},
  \bibinfo {author} {\bibfnamefont {T.}~\bibnamefont {Sch\"afer}}, \bibinfo
  {author} {\bibfnamefont {K.}~\bibnamefont {Held}}, \ and\ \bibinfo {author}
  {\bibfnamefont {A.}~\bibnamefont {Toschi}},\ }\bibfield  {title} {\enquote
  {\bibinfo {title} {{Divergences of the irreducible vertex functions in
  correlated metallic systems: Insights from the Anderson impurity model}},}\
  }\href {\doibase 10.1103/PhysRevB.97.245136} {\bibfield  {journal} {\bibinfo
  {journal} {Phys. Rev. B}\ }\textbf {\bibinfo {volume} {97}},\ \bibinfo
  {pages} {245136} (\bibinfo {year} {2018})}\BibitemShut {NoStop}%
\bibitem [{\citenamefont {Thunstr\"om}\ \emph {et~al.}(2018)\citenamefont
  {Thunstr\"om}, \citenamefont {Gunnarsson}, \citenamefont {Ciuchi},\ and\
  \citenamefont {Rohringer}}]{Thunstrom2018}%
  \BibitemOpen
  \bibfield  {author} {\bibinfo {author} {\bibfnamefont {P.}~\bibnamefont
  {Thunstr\"om}}, \bibinfo {author} {\bibfnamefont {O.}~\bibnamefont
  {Gunnarsson}}, \bibinfo {author} {\bibfnamefont {Sergio}\ \bibnamefont
  {Ciuchi}}, \ and\ \bibinfo {author} {\bibfnamefont {G.}~\bibnamefont
  {Rohringer}},\ }\bibfield  {title} {\enquote {\bibinfo {title} {{Analytical
  investigation of singularities in two-particle irreducible vertex functions
  of the Hubbard atom}},}\ }\href {\doibase 10.1103/PhysRevB.98.235107}
  {\bibfield  {journal} {\bibinfo  {journal} {Phys. Rev. B}\ }\textbf {\bibinfo
  {volume} {98}},\ \bibinfo {pages} {235107} (\bibinfo {year}
  {2018})}\BibitemShut {NoStop}%
\bibitem [{\citenamefont {Springer}\ \emph {et~al.}(2020)\citenamefont
  {Springer}, \citenamefont {Chalupa}, \citenamefont {Ciuchi}, \citenamefont
  {Sangiovanni},\ and\ \citenamefont {Toschi}}]{Springer2019}%
  \BibitemOpen
  \bibfield  {author} {\bibinfo {author} {\bibfnamefont {D.}~\bibnamefont
  {Springer}}, \bibinfo {author} {\bibfnamefont {P.}~\bibnamefont {Chalupa}},
  \bibinfo {author} {\bibfnamefont {S.}~\bibnamefont {Ciuchi}}, \bibinfo
  {author} {\bibfnamefont {G.}~\bibnamefont {Sangiovanni}}, \ and\ \bibinfo
  {author} {\bibfnamefont {A.}~\bibnamefont {Toschi}},\ }\bibfield  {title}
  {\enquote {\bibinfo {title} {Interplay between local response and vertex
  divergences in many-fermion systems with on-site attraction},}\ }\href
  {\doibase 10.1103/PhysRevB.101.155148} {\bibfield  {journal} {\bibinfo
  {journal} {Phys. Rev. B}\ }\textbf {\bibinfo {volume} {101}},\ \bibinfo
  {pages} {155148} (\bibinfo {year} {2020})}\BibitemShut {NoStop}%
\bibitem [{\citenamefont {Reitner}\ \emph {et~al.}(2020)\citenamefont
  {Reitner}, \citenamefont {Chalupa}, \citenamefont {Del~Re}, \citenamefont
  {Springer}, \citenamefont {Ciuchi}, \citenamefont {Sangiovanni},\ and\
  \citenamefont {Toschi}}]{Reitner2020}%
  \BibitemOpen
  \bibfield  {author} {\bibinfo {author} {\bibfnamefont {M.}~\bibnamefont
  {Reitner}}, \bibinfo {author} {\bibfnamefont {P.}~\bibnamefont {Chalupa}},
  \bibinfo {author} {\bibfnamefont {L.}~\bibnamefont {Del~Re}}, \bibinfo
  {author} {\bibfnamefont {D.}~\bibnamefont {Springer}}, \bibinfo {author}
  {\bibfnamefont {S.}~\bibnamefont {Ciuchi}}, \bibinfo {author} {\bibfnamefont
  {G.}~\bibnamefont {Sangiovanni}}, \ and\ \bibinfo {author} {\bibfnamefont
  {A.}~\bibnamefont {Toschi}},\ }\bibfield  {title} {\enquote {\bibinfo {title}
  {{Attractive Effect of a Strong Electronic Repulsion: The Physics of Vertex
  Divergences}},}\ }\href {\doibase 10.1103/physrevlett.125.196403} {\bibfield
  {journal} {\bibinfo  {journal} {Physical Review Letters}\ }\textbf {\bibinfo
  {volume} {125}} (\bibinfo {year} {2020}),\
  10.1103/physrevlett.125.196403}\BibitemShut {NoStop}%
\bibitem [{\citenamefont {Chalupa}\ \emph {et~al.}(2021)\citenamefont
  {Chalupa}, \citenamefont {Sch\"afer}, \citenamefont {Reitner}, \citenamefont
  {Springer}, \citenamefont {Andergassen},\ and\ \citenamefont
  {Toschi}}]{Chalupa2020}%
  \BibitemOpen
  \bibfield  {author} {\bibinfo {author} {\bibfnamefont {P.}~\bibnamefont
  {Chalupa}}, \bibinfo {author} {\bibfnamefont {T.}~\bibnamefont {Sch\"afer}},
  \bibinfo {author} {\bibfnamefont {M.}~\bibnamefont {Reitner}}, \bibinfo
  {author} {\bibfnamefont {D.}~\bibnamefont {Springer}}, \bibinfo {author}
  {\bibfnamefont {S.}~\bibnamefont {Andergassen}}, \ and\ \bibinfo {author}
  {\bibfnamefont {A.}~\bibnamefont {Toschi}},\ }\bibfield  {title} {\enquote
  {\bibinfo {title} {{Fingerprints of the Local Moment Formation and its Kondo
  Screening in the Generalized Susceptibilities of Many-Electron Problems}},}\
  }\href {\doibase 10.1103/PhysRevLett.126.056403} {\bibfield  {journal}
  {\bibinfo  {journal} {Phys. Rev. Lett.}\ }\textbf {\bibinfo {volume} {126}},\
  \bibinfo {pages} {056403} (\bibinfo {year} {2021})}\BibitemShut {NoStop}%
\bibitem [{\citenamefont {Rohe}\ and\ \citenamefont
  {Honerkamp}(2020)}]{Rohe2020}%
  \BibitemOpen
  \bibfield  {author} {\bibinfo {author} {\bibfnamefont {D.}~\bibnamefont
  {Rohe}}\ and\ \bibinfo {author} {\bibfnamefont {C.}~\bibnamefont
  {Honerkamp}},\ }\bibfield  {title} {\enquote {\bibinfo {title}
  {{Quasi-particle functional Renormalisation Group calculations in the
  two-dimensional half-filled Hubbard model at finite temperatures}},}\ }\href
  {\doibase 10.21468/SciPostPhys.9.6.084} {\bibfield  {journal} {\bibinfo
  {journal} {SciPost Phys.}\ }\textbf {\bibinfo {volume} {9}},\ \bibinfo
  {pages} {84} (\bibinfo {year} {2020})}\BibitemShut {NoStop}%
\bibitem [{\citenamefont {{Afchain}}\ \emph {et~al.}(2005)\citenamefont
  {{Afchain}}, \citenamefont {{Magnen}},\ and\ \citenamefont
  {{Rivasseau}}}]{Afchain2005}%
  \BibitemOpen
  \bibfield  {author} {\bibinfo {author} {\bibfnamefont {St{\'e}phane}\
  \bibnamefont {{Afchain}}}, \bibinfo {author} {\bibfnamefont {Jacques}\
  \bibnamefont {{Magnen}}}, \ and\ \bibinfo {author} {\bibfnamefont {Vincent}\
  \bibnamefont {{Rivasseau}}},\ }\bibfield  {title} {\enquote {\bibinfo {title}
  {{The Hubbard Model at Half-Filling, Part III: the Lower Bound on the
  Self-Energy}},}\ }\href {\doibase 10.1007/s00023-005-0214-z} {\bibfield
  {journal} {\bibinfo  {journal} {Annales Henri Poincar{\'e}}\ }\textbf
  {\bibinfo {volume} {6}},\ \bibinfo {pages} {449--483} (\bibinfo {year}
  {2005})},\ \Eprint {http://arxiv.org/abs/cond-mat/0412401}
  {arXiv:cond-mat/0412401 [cond-mat.str-el]} \BibitemShut {NoStop}%
\bibitem [{\citenamefont {{Abanov}}\ \emph {et~al.}(2003)\citenamefont
  {{Abanov}}, \citenamefont {{Chubukov}},\ and\ \citenamefont
  {{Schmalian}}}]{Abanov2003}%
  \BibitemOpen
  \bibfield  {author} {\bibinfo {author} {\bibfnamefont {Ar.}\ \bibnamefont
  {{Abanov}}}, \bibinfo {author} {\bibfnamefont {Andrey~V.}\ \bibnamefont
  {{Chubukov}}}, \ and\ \bibinfo {author} {\bibfnamefont {J.}~\bibnamefont
  {{Schmalian}}},\ }\bibfield  {title} {\enquote {\bibinfo {title}
  {{{Quantum-critical theory of the spin-fermion model and its application to
  cuprates: normal state analysis}}},}\ }\href {\doibase
  10.1080/0001873021000057123} {\bibfield  {journal} {\bibinfo  {journal}
  {Advances in Physics}\ }\textbf {\bibinfo {volume} {52}},\ \bibinfo {pages}
  {119--218} (\bibinfo {year} {2003})}\BibitemShut {NoStop}%
\bibitem [{\citenamefont {Chubukov}\ and\ \citenamefont
  {Maslov}(2012)}]{Chubukov_2012}%
  \BibitemOpen
  \bibfield  {author} {\bibinfo {author} {\bibfnamefont {Andrey~V.}\
  \bibnamefont {Chubukov}}\ and\ \bibinfo {author} {\bibfnamefont {Dmitrii~L.}\
  \bibnamefont {Maslov}},\ }\bibfield  {title} {\enquote {\bibinfo {title}
  {First-matsubara-frequency rule in a fermi liquid. i. fermionic
  self-energy},}\ }\href {\doibase 10.1103/PhysRevB.86.155136} {\bibfield
  {journal} {\bibinfo  {journal} {Phys. Rev. B}\ }\textbf {\bibinfo {volume}
  {86}},\ \bibinfo {pages} {155136} (\bibinfo {year} {2012})}\BibitemShut
  {NoStop}%
\bibitem [{\citenamefont {Virosztek}\ and\ \citenamefont
  {Ruvalds}(1990)}]{Virosztek1990}%
  \BibitemOpen
  \bibfield  {author} {\bibinfo {author} {\bibfnamefont {A.}~\bibnamefont
  {Virosztek}}\ and\ \bibinfo {author} {\bibfnamefont {J.}~\bibnamefont
  {Ruvalds}},\ }\bibfield  {title} {\enquote {\bibinfo {title}
  {{Nested-Fermi-liquid theory}},}\ }\href {\doibase 10.1103/PhysRevB.42.4064}
  {\bibfield  {journal} {\bibinfo  {journal} {Phys. Rev. B}\ }\textbf {\bibinfo
  {volume} {42}},\ \bibinfo {pages} {4064--4072} (\bibinfo {year}
  {1990})}\BibitemShut {NoStop}%
\bibitem [{\citenamefont {Zheleznyak}\ \emph {et~al.}(1997)\citenamefont
  {Zheleznyak}, \citenamefont {Yakovenko},\ and\ \citenamefont
  {Dzyaloshinskii}}]{Zheleznyak1997}%
  \BibitemOpen
  \bibfield  {author} {\bibinfo {author} {\bibfnamefont {Anatoley~T.}\
  \bibnamefont {Zheleznyak}}, \bibinfo {author} {\bibfnamefont {Victor~M.}\
  \bibnamefont {Yakovenko}}, \ and\ \bibinfo {author} {\bibfnamefont {Igor~E.}\
  \bibnamefont {Dzyaloshinskii}},\ }\bibfield  {title} {\enquote {\bibinfo
  {title} {{Parquet solution for a flat Fermi surface}},}\ }\href {\doibase
  10.1103/PhysRevB.55.3200} {\bibfield  {journal} {\bibinfo  {journal} {Phys.
  Rev. B}\ }\textbf {\bibinfo {volume} {55}},\ \bibinfo {pages} {3200--3215}
  (\bibinfo {year} {1997})}\BibitemShut {NoStop}%
\bibitem [{\citenamefont {{F.Vistulo de Abreu}}\ and\ \citenamefont
  {{Dou\c{c}ot, B.}}(1997)}]{Vistulo1997}%
  \BibitemOpen
  \bibfield  {author} {\bibinfo {author} {\bibnamefont {{F.Vistulo de Abreu}}}\
  and\ \bibinfo {author} {\bibnamefont {{Dou\c{c}ot, B.}}},\ }\bibfield
  {title} {\enquote {\bibinfo {title} {Nesting effects on fermionic systems},}\
  }\href {\doibase 10.1209/epl/i1997-00279-7} {\bibfield  {journal} {\bibinfo
  {journal} {Europhys. Lett.}\ }\textbf {\bibinfo {volume} {38}},\ \bibinfo
  {pages} {533--538} (\bibinfo {year} {1997})}\BibitemShut {NoStop}%
\bibitem [{\citenamefont {Zanchi}\ and\ \citenamefont
  {Schulz}(2000)}]{Zanchi2000}%
  \BibitemOpen
  \bibfield  {author} {\bibinfo {author} {\bibfnamefont {D.}~\bibnamefont
  {Zanchi}}\ and\ \bibinfo {author} {\bibfnamefont {H.~J.}\ \bibnamefont
  {Schulz}},\ }\bibfield  {title} {\enquote {\bibinfo {title} {{Weakly
  correlated electrons on a square lattice: Renormalization-group theory}},}\
  }\href {\doibase 10.1103/PhysRevB.61.13609} {\bibfield  {journal} {\bibinfo
  {journal} {Phys. Rev. B}\ }\textbf {\bibinfo {volume} {61}},\ \bibinfo
  {pages} {13609--13632} (\bibinfo {year} {2000})}\BibitemShut {NoStop}%
\bibitem [{\citenamefont {Bertrand}\ \emph {et~al.}(2019)\citenamefont
  {Bertrand}, \citenamefont {Florens}, \citenamefont {Parcollet},\ and\
  \citenamefont {Waintal}}]{Bertrand2019}%
  \BibitemOpen
  \bibfield  {author} {\bibinfo {author} {\bibfnamefont {Corentin}\
  \bibnamefont {Bertrand}}, \bibinfo {author} {\bibfnamefont {Serge}\
  \bibnamefont {Florens}}, \bibinfo {author} {\bibfnamefont {Olivier}\
  \bibnamefont {Parcollet}}, \ and\ \bibinfo {author} {\bibfnamefont {Xavier}\
  \bibnamefont {Waintal}},\ }\bibfield  {title} {\enquote {\bibinfo {title}
  {{Reconstructing Nonequilibrium Regimes of Quantum Many-Body Systems from the
  Analytical Structure of Perturbative Expansions}},}\ }\href {\doibase
  10.1103/PhysRevX.9.041008} {\bibfield  {journal} {\bibinfo  {journal} {Phys.
  Rev. X}\ }\textbf {\bibinfo {volume} {9}},\ \bibinfo {pages} {041008}
  (\bibinfo {year} {2019})}\BibitemShut {NoStop}%
\bibitem [{\citenamefont {Vu\ifmmode \check{c}\else \v{c}\fi{}i\ifmmode
  \check{c}\else \v{c}\fi{}evi\ifmmode~\acute{c}\else \'{c}\fi{}}\ and\
  \citenamefont {Ferrero}(2020)}]{Vucicevic2020}%
  \BibitemOpen
  \bibfield  {author} {\bibinfo {author} {\bibfnamefont {J.}~\bibnamefont
  {Vu\ifmmode \check{c}\else \v{c}\fi{}i\ifmmode \check{c}\else
  \v{c}\fi{}evi\ifmmode~\acute{c}\else \'{c}\fi{}}}\ and\ \bibinfo {author}
  {\bibfnamefont {M.}~\bibnamefont {Ferrero}},\ }\bibfield  {title} {\enquote
  {\bibinfo {title} {{Real-frequency diagrammatic Monte Carlo at finite
  temperature}},}\ }\href {\doibase 10.1103/PhysRevB.101.075113} {\bibfield
  {journal} {\bibinfo  {journal} {Phys. Rev. B}\ }\textbf {\bibinfo {volume}
  {101}},\ \bibinfo {pages} {075113} (\bibinfo {year} {2020})}\BibitemShut
  {NoStop}%
\bibitem [{\citenamefont {Taheridehkordi}\ \emph
  {et~al.}(2020{\natexlab{a}})\citenamefont {Taheridehkordi}, \citenamefont
  {Curnoe},\ and\ \citenamefont {LeBlanc}}]{Taheridehkordi2020}%
  \BibitemOpen
  \bibfield  {author} {\bibinfo {author} {\bibfnamefont {Amir}\ \bibnamefont
  {Taheridehkordi}}, \bibinfo {author} {\bibfnamefont {S.~H.}\ \bibnamefont
  {Curnoe}}, \ and\ \bibinfo {author} {\bibfnamefont {J.~P.~F.}\ \bibnamefont
  {LeBlanc}},\ }\bibfield  {title} {\enquote {\bibinfo {title} {{Algorithmic
  approach to diagrammatic expansions for real-frequency evaluation of
  susceptibility functions}},}\ }\href {\doibase 10.1103/physrevb.102.045115}
  {\bibfield  {journal} {\bibinfo  {journal} {Physical Review B}\ }\textbf
  {\bibinfo {volume} {102}} (\bibinfo {year} {2020}{\natexlab{a}}),\
  10.1103/physrevb.102.045115}\BibitemShut {NoStop}%
\bibitem [{\citenamefont {Taheridehkordi}\ \emph
  {et~al.}(2020{\natexlab{b}})\citenamefont {Taheridehkordi}, \citenamefont
  {Curnoe},\ and\ \citenamefont {LeBlanc}}]{Taheridehkordi2020b}%
  \BibitemOpen
  \bibfield  {author} {\bibinfo {author} {\bibfnamefont {Amir}\ \bibnamefont
  {Taheridehkordi}}, \bibinfo {author} {\bibfnamefont {S.~H.}\ \bibnamefont
  {Curnoe}}, \ and\ \bibinfo {author} {\bibfnamefont {J.~P.~F.}\ \bibnamefont
  {LeBlanc}},\ }\bibfield  {title} {\enquote {\bibinfo {title} {{Optimal
  grouping of arbitrary diagrammatic expansions via analytic pole
  structure}},}\ }\href {\doibase 10.1103/PhysRevB.101.125109} {\bibfield
  {journal} {\bibinfo  {journal} {Phys. Rev. B}\ }\textbf {\bibinfo {volume}
  {101}},\ \bibinfo {pages} {125109} (\bibinfo {year}
  {2020}{\natexlab{b}})}\BibitemShut {NoStop}%
\bibitem [{\citenamefont {Kitatani}\ \emph {et~al.}(2019)\citenamefont
  {Kitatani}, \citenamefont {Sch\"afer}, \citenamefont {Aoki},\ and\
  \citenamefont {Held}}]{Kitatani2018}%
  \BibitemOpen
  \bibfield  {author} {\bibinfo {author} {\bibfnamefont {Motoharu}\
  \bibnamefont {Kitatani}}, \bibinfo {author} {\bibfnamefont {Thomas}\
  \bibnamefont {Sch\"afer}}, \bibinfo {author} {\bibfnamefont {Hideo}\
  \bibnamefont {Aoki}}, \ and\ \bibinfo {author} {\bibfnamefont {Karsten}\
  \bibnamefont {Held}},\ }\bibfield  {title} {\enquote {\bibinfo {title} {Why
  the critical temperature of high-${T}_{c}$ cuprate superconductors is so low:
  The importance of the dynamical vertex structure},}\ }\href {\doibase
  10.1103/PhysRevB.99.041115} {\bibfield  {journal} {\bibinfo  {journal} {Phys.
  Rev. B}\ }\textbf {\bibinfo {volume} {99}},\ \bibinfo {pages} {041115(R)}
  (\bibinfo {year} {2019})}\BibitemShut {NoStop}%
\bibitem [{\citenamefont {Tagliavini}\ \emph {et~al.}(2019)\citenamefont
  {Tagliavini}, \citenamefont {Hille}, \citenamefont {Kugler}, \citenamefont
  {Andergassen}, \citenamefont {Toschi},\ and\ \citenamefont
  {Honerkamp}}]{Tagliavini2019}%
  \BibitemOpen
  \bibfield  {author} {\bibinfo {author} {\bibfnamefont {Agnese}\ \bibnamefont
  {Tagliavini}}, \bibinfo {author} {\bibfnamefont {Cornelia}\ \bibnamefont
  {Hille}}, \bibinfo {author} {\bibfnamefont {Fabian}\ \bibnamefont {Kugler}},
  \bibinfo {author} {\bibfnamefont {Sabine}\ \bibnamefont {Andergassen}},
  \bibinfo {author} {\bibfnamefont {Alessandro}\ \bibnamefont {Toschi}}, \ and\
  \bibinfo {author} {\bibfnamefont {Carsten}\ \bibnamefont {Honerkamp}},\
  }\bibfield  {title} {\enquote {\bibinfo {title} {{Multiloop functional
  renormalization group for the two-dimensional Hubbard model: Loop convergence
  of the response functions}},}\ }\href {\doibase 10.21468/SciPostPhys.6.1.009}
  {\bibfield  {journal} {\bibinfo  {journal} {SciPost Physics}\ }\textbf
  {\bibinfo {volume} {6}},\ \bibinfo {pages} {009} (\bibinfo {year} {2019})},\
  \Eprint {http://arxiv.org/abs/1807.02697} {arXiv:1807.02697} \BibitemShut
  {NoStop}%
\bibitem [{\citenamefont {Kugler}\ and\ \citenamefont {von
  Delft}(2018{\natexlab{c}})}]{Kugler2018a}%
  \BibitemOpen
  \bibfield  {author} {\bibinfo {author} {\bibfnamefont {Fabian~B.}\
  \bibnamefont {Kugler}}\ and\ \bibinfo {author} {\bibfnamefont {Jan}\
  \bibnamefont {von Delft}},\ }\bibfield  {title} {\enquote {\bibinfo {title}
  {Multiloop functional renormalization group for general models},}\ }\href
  {\doibase 10.1103/PhysRevB.97.035162} {\bibfield  {journal} {\bibinfo
  {journal} {Phys. Rev. B}\ }\textbf {\bibinfo {volume} {97}},\ \bibinfo
  {pages} {035162} (\bibinfo {year} {2018}{\natexlab{c}})}\BibitemShut
  {NoStop}%
\bibitem [{\citenamefont {Taranto}\ \emph {et~al.}(2014)\citenamefont
  {Taranto}, \citenamefont {Andergassen}, \citenamefont {Bauer}, \citenamefont
  {Held}, \citenamefont {Katanin}, \citenamefont {Metzner}, \citenamefont
  {Rohringer},\ and\ \citenamefont {Toschi}}]{Taranto2014}%
  \BibitemOpen
  \bibfield  {author} {\bibinfo {author} {\bibfnamefont {C.}~\bibnamefont
  {Taranto}}, \bibinfo {author} {\bibfnamefont {S.}~\bibnamefont
  {Andergassen}}, \bibinfo {author} {\bibfnamefont {J.}~\bibnamefont {Bauer}},
  \bibinfo {author} {\bibfnamefont {K.}~\bibnamefont {Held}}, \bibinfo {author}
  {\bibfnamefont {A.}~\bibnamefont {Katanin}}, \bibinfo {author} {\bibfnamefont
  {W.}~\bibnamefont {Metzner}}, \bibinfo {author} {\bibfnamefont
  {G.}~\bibnamefont {Rohringer}}, \ and\ \bibinfo {author} {\bibfnamefont
  {A.}~\bibnamefont {Toschi}},\ }\bibfield  {title} {\enquote {\bibinfo {title}
  {{From Infinite to Two Dimensions through the Functional Renormalization
  Group}},}\ }\href {\doibase 10.1103/PhysRevLett.112.196402} {\bibfield
  {journal} {\bibinfo  {journal} {Phys. Rev. Lett.}\ }\textbf {\bibinfo
  {volume} {112}},\ \bibinfo {pages} {196402} (\bibinfo {year}
  {2014})}\BibitemShut {NoStop}%
\bibitem [{\citenamefont {Vilardi}\ \emph {et~al.}(2019)\citenamefont
  {Vilardi}, \citenamefont {Taranto},\ and\ \citenamefont
  {Metzner}}]{Vilardi2019}%
  \BibitemOpen
  \bibfield  {author} {\bibinfo {author} {\bibfnamefont {Demetrio}\
  \bibnamefont {Vilardi}}, \bibinfo {author} {\bibfnamefont {Ciro}\
  \bibnamefont {Taranto}}, \ and\ \bibinfo {author} {\bibfnamefont {Walter}\
  \bibnamefont {Metzner}},\ }\bibfield  {title} {\enquote {\bibinfo {title}
  {{Antiferromagnetic and d-wave pairing correlations in the strongly
  interacting two-dimensional Hubbard model from the functional renormalization
  group}},}\ }\href {\doibase 10.1103/PhysRevB.99.104501} {\bibfield  {journal}
  {\bibinfo  {journal} {Physical Review B}\ }\textbf {\bibinfo {volume} {99}},\
  \bibinfo {pages} {104501} (\bibinfo {year} {2019})},\ \Eprint
  {http://arxiv.org/abs/1810.02290} {arXiv:1810.02290} \BibitemShut {NoStop}%
\bibitem [{\citenamefont {Kitatani}\ \emph {et~al.}(2020)\citenamefont
  {Kitatani}, \citenamefont {Si}, \citenamefont {Janson}, \citenamefont
  {Arita}, \citenamefont {Zhong},\ and\ \citenamefont {Held}}]{Kitatani2020}%
  \BibitemOpen
  \bibfield  {author} {\bibinfo {author} {\bibfnamefont {Motoharu}\
  \bibnamefont {Kitatani}}, \bibinfo {author} {\bibfnamefont {Liang}\
  \bibnamefont {Si}}, \bibinfo {author} {\bibfnamefont {Oleg}\ \bibnamefont
  {Janson}}, \bibinfo {author} {\bibfnamefont {Ryotaro}\ \bibnamefont {Arita}},
  \bibinfo {author} {\bibfnamefont {Zhicheng}\ \bibnamefont {Zhong}}, \ and\
  \bibinfo {author} {\bibfnamefont {Karsten}\ \bibnamefont {Held}},\ }\bibfield
   {title} {\enquote {\bibinfo {title} {{Nickelate superconductors—a
  renaissance of the one-band Hubbard model}},}\ }\href {\doibase
  10.1038/s41535-020-00260-y} {\bibfield  {journal} {\bibinfo  {journal} {npj
  Quantum Materials}\ }\textbf {\bibinfo {volume} {5}} (\bibinfo {year}
  {2020}),\ 10.1038/s41535-020-00260-y}\BibitemShut {NoStop}%
\bibitem [{\citenamefont {Galler}\ \emph {et~al.}(2017)\citenamefont {Galler},
  \citenamefont {Thunstr\"om}, \citenamefont {Gunacker}, \citenamefont
  {Tomczak},\ and\ \citenamefont {Held}}]{Galler2016}%
  \BibitemOpen
  \bibfield  {author} {\bibinfo {author} {\bibfnamefont {Anna}\ \bibnamefont
  {Galler}}, \bibinfo {author} {\bibfnamefont {Patrik}\ \bibnamefont
  {Thunstr\"om}}, \bibinfo {author} {\bibfnamefont {Patrik}\ \bibnamefont
  {Gunacker}}, \bibinfo {author} {\bibfnamefont {Jan~M.}\ \bibnamefont
  {Tomczak}}, \ and\ \bibinfo {author} {\bibfnamefont {K.}~\bibnamefont
  {Held}},\ }\bibfield  {title} {\enquote {\bibinfo {title} {{Ab initio
  dynamical vertex approximation}},}\ }\href {\doibase
  10.1103/PhysRevB.95.115107} {\bibfield  {journal} {\bibinfo  {journal} {Phys.
  Rev. B}\ }\textbf {\bibinfo {volume} {95}},\ \bibinfo {pages} {115107}
  (\bibinfo {year} {2017})}\BibitemShut {NoStop}%
\bibitem [{\citenamefont {Strand}\ \emph
  {et~al.}(2019{\natexlab{a}})\citenamefont {Strand}, \citenamefont {Zingl},
  \citenamefont {Wentzell}, \citenamefont {Parcollet},\ and\ \citenamefont
  {Georges}}]{Strand2019}%
  \BibitemOpen
  \bibfield  {author} {\bibinfo {author} {\bibfnamefont {Hugo U.~R.}\
  \bibnamefont {Strand}}, \bibinfo {author} {\bibfnamefont {Manuel}\
  \bibnamefont {Zingl}}, \bibinfo {author} {\bibfnamefont {Nils}\ \bibnamefont
  {Wentzell}}, \bibinfo {author} {\bibfnamefont {Olivier}\ \bibnamefont
  {Parcollet}}, \ and\ \bibinfo {author} {\bibfnamefont {Antoine}\ \bibnamefont
  {Georges}},\ }\bibfield  {title} {\enquote {\bibinfo {title} {{Magnetic
  response of ${\mathrm{Sr}}_{2}{\mathrm{RuO}}_{4}$: Quasi-local spin
  fluctuations due to Hund's coupling}},}\ }\href {\doibase
  10.1103/PhysRevB.100.125120} {\bibfield  {journal} {\bibinfo  {journal}
  {Phys. Rev. B}\ }\textbf {\bibinfo {volume} {100}},\ \bibinfo {pages}
  {125120} (\bibinfo {year} {2019}{\natexlab{a}})}\BibitemShut {NoStop}%
\bibitem [{\citenamefont {Tamai}\ \emph {et~al.}(2019)\citenamefont {Tamai},
  \citenamefont {Zingl}, \citenamefont {Rozbicki}, \citenamefont {Cappelli},
  \citenamefont {Ricc\`o}, \citenamefont {de~la Torre}, \citenamefont
  {McKeown~Walker}, \citenamefont {Bruno}, \citenamefont {King}, \citenamefont
  {Meevasana}, \citenamefont {Shi}, \citenamefont
  {Radovi\ifmmode~\acute{c}\else \'{c}\fi{}}, \citenamefont {Plumb},
  \citenamefont {Gibbs}, \citenamefont {Mackenzie}, \citenamefont {Berthod},
  \citenamefont {Strand}, \citenamefont {Kim}, \citenamefont {Georges},\ and\
  \citenamefont {Baumberger}}]{Tamai2019}%
  \BibitemOpen
  \bibfield  {author} {\bibinfo {author} {\bibfnamefont {A.}~\bibnamefont
  {Tamai}}, \bibinfo {author} {\bibfnamefont {M.}~\bibnamefont {Zingl}},
  \bibinfo {author} {\bibfnamefont {E.}~\bibnamefont {Rozbicki}}, \bibinfo
  {author} {\bibfnamefont {E.}~\bibnamefont {Cappelli}}, \bibinfo {author}
  {\bibfnamefont {S.}~\bibnamefont {Ricc\`o}}, \bibinfo {author} {\bibfnamefont
  {A.}~\bibnamefont {de~la Torre}}, \bibinfo {author} {\bibfnamefont
  {S.}~\bibnamefont {McKeown~Walker}}, \bibinfo {author} {\bibfnamefont
  {F.~Y.}\ \bibnamefont {Bruno}}, \bibinfo {author} {\bibfnamefont {P.~D.~C.}\
  \bibnamefont {King}}, \bibinfo {author} {\bibfnamefont {W.}~\bibnamefont
  {Meevasana}}, \bibinfo {author} {\bibfnamefont {M.}~\bibnamefont {Shi}},
  \bibinfo {author} {\bibfnamefont {M.}~\bibnamefont
  {Radovi\ifmmode~\acute{c}\else \'{c}\fi{}}}, \bibinfo {author} {\bibfnamefont
  {N.~C.}\ \bibnamefont {Plumb}}, \bibinfo {author} {\bibfnamefont {A.~S.}\
  \bibnamefont {Gibbs}}, \bibinfo {author} {\bibfnamefont {A.~P.}\ \bibnamefont
  {Mackenzie}}, \bibinfo {author} {\bibfnamefont {C.}~\bibnamefont {Berthod}},
  \bibinfo {author} {\bibfnamefont {H.~U.~R.}\ \bibnamefont {Strand}}, \bibinfo
  {author} {\bibfnamefont {M.}~\bibnamefont {Kim}}, \bibinfo {author}
  {\bibfnamefont {A.}~\bibnamefont {Georges}}, \ and\ \bibinfo {author}
  {\bibfnamefont {F.}~\bibnamefont {Baumberger}},\ }\bibfield  {title}
  {\enquote {\bibinfo {title} {{High-Resolution Photoemission on
  ${\mathrm{Sr}}_{2}{\mathrm{RuO}}_{4}$ Reveals Correlation-Enhanced Effective
  Spin-Orbit Coupling and Dominantly Local Self-Energies}},}\ }\href {\doibase
  10.1103/PhysRevX.9.021048} {\bibfield  {journal} {\bibinfo  {journal} {Phys.
  Rev. X}\ }\textbf {\bibinfo {volume} {9}},\ \bibinfo {pages} {021048}
  (\bibinfo {year} {2019})}\BibitemShut {NoStop}%
\bibitem [{\citenamefont {Acharya}\ \emph {et~al.}(2019)\citenamefont
  {Acharya}, \citenamefont {Pashov}, \citenamefont {Weber}, \citenamefont
  {Park}, \citenamefont {Sponza},\ and\ \citenamefont
  {Schilfgaarde}}]{Acharya2019}%
  \BibitemOpen
  \bibfield  {author} {\bibinfo {author} {\bibfnamefont {Swagata}\ \bibnamefont
  {Acharya}}, \bibinfo {author} {\bibfnamefont {Dimitar}\ \bibnamefont
  {Pashov}}, \bibinfo {author} {\bibfnamefont {Cédric}\ \bibnamefont {Weber}},
  \bibinfo {author} {\bibfnamefont {Hyowon}\ \bibnamefont {Park}}, \bibinfo
  {author} {\bibfnamefont {Lorenzo}\ \bibnamefont {Sponza}}, \ and\ \bibinfo
  {author} {\bibfnamefont {Mark~Van}\ \bibnamefont {Schilfgaarde}},\ }\bibfield
   {title} {\enquote {\bibinfo {title} {{Evening out the spin and charge parity
  to increase ${T}_{c}$ in Sr$_{\text{2}}$RuO$_{\text{4}}$}},}\ }\href
  {\doibase 10.1038/s42005-019-0254-1} {\bibfield  {journal} {\bibinfo
  {journal} {Communications Physics}\ }\textbf {\bibinfo {volume} {2}}
  (\bibinfo {year} {2019}),\ 10.1038/s42005-019-0254-1}\BibitemShut {NoStop}%
\bibitem [{\citenamefont {Zantout}\ \emph {et~al.}(2019)\citenamefont
  {Zantout}, \citenamefont {Backes},\ and\ \citenamefont
  {Valent\'{\i}}}]{Zantout2019}%
  \BibitemOpen
  \bibfield  {author} {\bibinfo {author} {\bibfnamefont {Karim}\ \bibnamefont
  {Zantout}}, \bibinfo {author} {\bibfnamefont {Steffen}\ \bibnamefont
  {Backes}}, \ and\ \bibinfo {author} {\bibfnamefont {Roser}\ \bibnamefont
  {Valent\'{\i}}},\ }\bibfield  {title} {\enquote {\bibinfo {title} {{Effect of
  Nonlocal Correlations on the Electronic Structure of LiFeAs}},}\ }\href
  {\doibase 10.1103/PhysRevLett.123.256401} {\bibfield  {journal} {\bibinfo
  {journal} {Phys. Rev. Lett.}\ }\textbf {\bibinfo {volume} {123}},\ \bibinfo
  {pages} {256401} (\bibinfo {year} {2019})}\BibitemShut {NoStop}%
\bibitem [{\citenamefont {Gunnarsson}\ \emph {et~al.}(2016)\citenamefont
  {Gunnarsson}, \citenamefont {Sch\"afer}, \citenamefont {LeBlanc},
  \citenamefont {Merino}, \citenamefont {Sangiovanni}, \citenamefont
  {Rohringer},\ and\ \citenamefont {Toschi}}]{Gunnarsson2016}%
  \BibitemOpen
  \bibfield  {author} {\bibinfo {author} {\bibfnamefont {O.}~\bibnamefont
  {Gunnarsson}}, \bibinfo {author} {\bibfnamefont {T.}~\bibnamefont
  {Sch\"afer}}, \bibinfo {author} {\bibfnamefont {J.~P.~F.}\ \bibnamefont
  {LeBlanc}}, \bibinfo {author} {\bibfnamefont {J.}~\bibnamefont {Merino}},
  \bibinfo {author} {\bibfnamefont {G.}~\bibnamefont {Sangiovanni}}, \bibinfo
  {author} {\bibfnamefont {G.}~\bibnamefont {Rohringer}}, \ and\ \bibinfo
  {author} {\bibfnamefont {A.}~\bibnamefont {Toschi}},\ }\bibfield  {title}
  {\enquote {\bibinfo {title} {{Parquet decomposition calculations of the
  electronic self-energy}},}\ }\href {\doibase 10.1103/PhysRevB.93.245102}
  {\bibfield  {journal} {\bibinfo  {journal} {Phys. Rev. B}\ }\textbf {\bibinfo
  {volume} {93}},\ \bibinfo {pages} {245102} (\bibinfo {year}
  {2016})}\BibitemShut {NoStop}%
\bibitem [{\citenamefont {Wu}\ \emph {et~al.}(2017)\citenamefont {Wu},
  \citenamefont {Ferrero}, \citenamefont {Georges},\ and\ \citenamefont
  {Kozik}}]{Wu2016}%
  \BibitemOpen
  \bibfield  {author} {\bibinfo {author} {\bibfnamefont {Wei}\ \bibnamefont
  {Wu}}, \bibinfo {author} {\bibfnamefont {Michel}\ \bibnamefont {Ferrero}},
  \bibinfo {author} {\bibfnamefont {Antoine}\ \bibnamefont {Georges}}, \ and\
  \bibinfo {author} {\bibfnamefont {Evgeny}\ \bibnamefont {Kozik}},\ }\bibfield
   {title} {\enquote {\bibinfo {title} {{Controlling Feynman diagrammatic
  expansions: Physical nature of the pseudogap in the two-dimensional Hubbard
  model}},}\ }\href {\doibase 10.1103/PhysRevB.96.041105} {\bibfield  {journal}
  {\bibinfo  {journal} {Phys. Rev. B}\ }\textbf {\bibinfo {volume} {96}},\
  \bibinfo {pages} {041105(R)} (\bibinfo {year} {2017})}\BibitemShut {NoStop}%
\bibitem [{\citenamefont {Rohringer}(2018)}]{Rohringer2018b}%
  \BibitemOpen
  \bibfield  {author} {\bibinfo {author} {\bibfnamefont {Georg}\ \bibnamefont
  {Rohringer}},\ }\bibfield  {title} {\enquote {\bibinfo {title} {{Spectra of
  correlated many-electron systems: From a one- to a two-particle
  description}},}\ }\href {\doibase
  https://doi.org/10.1016/j.elspec.2018.11.003} {\bibfield  {journal} {\bibinfo
   {journal} {Journal of Electron Spectroscopy and Related Phenomena}\ }
  (\bibinfo {year} {2018}),\
  https://doi.org/10.1016/j.elspec.2018.11.003}\BibitemShut {NoStop}%
\bibitem [{\citenamefont {Arzhang}\ \emph {et~al.}(2020)\citenamefont
  {Arzhang}, \citenamefont {Antipov},\ and\ \citenamefont
  {LeBlanc}}]{Arzhang2020}%
  \BibitemOpen
  \bibfield  {author} {\bibinfo {author} {\bibfnamefont {Behnam}\ \bibnamefont
  {Arzhang}}, \bibinfo {author} {\bibfnamefont {A.~E.}\ \bibnamefont
  {Antipov}}, \ and\ \bibinfo {author} {\bibfnamefont {J.~P.~F.}\ \bibnamefont
  {LeBlanc}},\ }\bibfield  {title} {\enquote {\bibinfo {title} {{Fluctuation
  diagnostics of the finite-temperature quasi-antiferromagnetic regime of the
  two-dimensional Hubbard model}},}\ }\href {\doibase
  10.1103/physrevb.101.014430} {\bibfield  {journal} {\bibinfo  {journal}
  {Physical Review B}\ }\textbf {\bibinfo {volume} {101}},\ \bibinfo {pages}
  {014430} (\bibinfo {year} {2020})}\BibitemShut {NoStop}%
\bibitem [{\citenamefont {Gunnarsson}\ \emph {et~al.}(2018)\citenamefont
  {Gunnarsson}, \citenamefont {Merino}, \citenamefont {Sch\"afer},
  \citenamefont {Sangiovanni}, \citenamefont {Rohringer},\ and\ \citenamefont
  {Toschi}}]{Gunnarsson2018}%
  \BibitemOpen
  \bibfield  {author} {\bibinfo {author} {\bibfnamefont {O.}~\bibnamefont
  {Gunnarsson}}, \bibinfo {author} {\bibfnamefont {J.}~\bibnamefont {Merino}},
  \bibinfo {author} {\bibfnamefont {T.}~\bibnamefont {Sch\"afer}}, \bibinfo
  {author} {\bibfnamefont {G.}~\bibnamefont {Sangiovanni}}, \bibinfo {author}
  {\bibfnamefont {G.}~\bibnamefont {Rohringer}}, \ and\ \bibinfo {author}
  {\bibfnamefont {A.}~\bibnamefont {Toschi}},\ }\bibfield  {title} {\enquote
  {\bibinfo {title} {{Complementary views on electron spectra: From fluctuation
  diagnostics to real-space correlations}},}\ }\href {\doibase
  10.1103/PhysRevB.97.125134} {\bibfield  {journal} {\bibinfo  {journal} {Phys.
  Rev. B}\ }\textbf {\bibinfo {volume} {97}},\ \bibinfo {pages} {125134}
  (\bibinfo {year} {2018})}\BibitemShut {NoStop}%
\bibitem [{\citenamefont {Sch{\"a}fer}\ and\ \citenamefont
  {Toschi}(2021)}]{Schaefer2020}%
  \BibitemOpen
  \bibfield  {author} {\bibinfo {author} {\bibfnamefont {Thomas}\ \bibnamefont
  {Sch{\"a}fer}}\ and\ \bibinfo {author} {\bibfnamefont {Alessandro}\
  \bibnamefont {Toschi}},\ }\bibfield  {title} {\enquote {\bibinfo {title}
  {{How to read between the lines of electronic spectra: the diagnostics of
  fluctuations in strongly correlated electron systems}},}\ }\href
  {http://iopscience.iop.org/article/10.1088/1361-648X/abeb44} {\bibfield
  {journal} {\bibinfo  {journal} {Journal of Physics: Condensed Matter}\ }
  (\bibinfo {year} {2021})}\BibitemShut {NoStop}%
\bibitem [{\citenamefont {{J\"{u}lich Supercomputing
  Centre}}(2018)}]{Jureca2018}%
  \BibitemOpen
  \bibfield  {author} {\bibinfo {author} {\bibnamefont {{J\"{u}lich
  Supercomputing Centre}}},\ }\bibfield  {title} {\enquote {\bibinfo {title}
  {{JURECA: Modular supercomputer at J\"{u}lich Supercomputing Centre}},}\
  }\href {\doibase 10.17815/jlsrf-4-121-1} {\bibfield  {journal} {\bibinfo
  {journal} {Journal of large-scale research facilities}\ }\textbf {\bibinfo
  {volume} {4}} (\bibinfo {year} {2018}),\ 10.17815/jlsrf-4-121-1}\BibitemShut
  {NoStop}%
\bibitem [{\citenamefont {Huang}(1987)}]{Huang1987}%
  \BibitemOpen
  \bibfield  {author} {\bibinfo {author} {\bibfnamefont {Kerson}\ \bibnamefont
  {Huang}},\ }\href@noop {} {\emph {\bibinfo {title} {Statistical Mechanics,
  2nd Edition}}}\ (\bibinfo  {publisher} {John Wiley \& Sons},\ \bibinfo {year}
  {1987})\BibitemShut {NoStop}%
\bibitem [{\citenamefont {Sachdev}(1999)}]{Sachdev1999}%
  \BibitemOpen
  \bibfield  {author} {\bibinfo {author} {\bibfnamefont {S.}~\bibnamefont
  {Sachdev}},\ }\href@noop {} {\emph {\bibinfo {title} {Quantum Phase
  Transitions}}}\ (\bibinfo  {publisher} {Cambridge University Press},\
  \bibinfo {year} {1999})\BibitemShut {NoStop}%
\bibitem [{Note1()}]{Note1}%
  \BibitemOpen
  \bibinfo {note} {For obtaining the data on the continuous imaginary frequency
  axis, in a first step, $\protect \text {Im }\Sigma (\protect \mathbf
  {k},\omega )$ has been calculated on the real axis. After retrieving
  $\protect \text {Re }\Sigma (\protect \mathbf {k},\omega )$ via the
  Kramers-Kronig relations, the Cauchy formula has been used for the forward
  continuation onto the imaginary axis. The calculation on the real axis
  profits from analytical treatments of the delta function which reduces the
  quadrature dimensions.}\BibitemShut {Stop}%
\bibitem [{\citenamefont {Abrikosov}\ \emph {et~al.}(1975)\citenamefont
  {Abrikosov}, \citenamefont {Gorkov},\ and\ \citenamefont
  {Dzyaloshinski}}]{Abrikosov1975}%
  \BibitemOpen
  \bibfield  {author} {\bibinfo {author} {\bibfnamefont {A.~A.}\ \bibnamefont
  {Abrikosov}}, \bibinfo {author} {\bibfnamefont {L.~P.}\ \bibnamefont
  {Gorkov}}, \ and\ \bibinfo {author} {\bibfnamefont {I.~E.}\ \bibnamefont
  {Dzyaloshinski}},\ }\href@noop {} {\emph {\bibinfo {title} {Methods of
  Quantum Field Theory in Statistical Physics}}}\ (\bibinfo  {publisher}
  {Dover},\ \bibinfo {address} {New York},\ \bibinfo {year} {1975})\BibitemShut
  {NoStop}%
\bibitem [{\citenamefont {Rossi}\ \emph {et~al.}(2016)\citenamefont {Rossi},
  \citenamefont {Werner}, \citenamefont {Prokof'ev},\ and\ \citenamefont
  {Svistunov}}]{Rossi2016}%
  \BibitemOpen
  \bibfield  {author} {\bibinfo {author} {\bibfnamefont {Riccardo}\
  \bibnamefont {Rossi}}, \bibinfo {author} {\bibfnamefont {F\'elix}\
  \bibnamefont {Werner}}, \bibinfo {author} {\bibfnamefont {Nikolay}\
  \bibnamefont {Prokof'ev}}, \ and\ \bibinfo {author} {\bibfnamefont {Boris}\
  \bibnamefont {Svistunov}},\ }\bibfield  {title} {\enquote {\bibinfo {title}
  {{Shifted-action expansion and applicability of dressed diagrammatic
  schemes}},}\ }\href {\doibase 10.1103/PhysRevB.93.161102} {\bibfield
  {journal} {\bibinfo  {journal} {Phys. Rev. B}\ }\textbf {\bibinfo {volume}
  {93}},\ \bibinfo {pages} {161102(R)} (\bibinfo {year} {2016})}\BibitemShut
  {NoStop}%
\bibitem [{\citenamefont {Chen}\ and\ \citenamefont
  {Haule}(2019)}]{chen2019combined}%
  \BibitemOpen
  \bibfield  {author} {\bibinfo {author} {\bibfnamefont {Kun}\ \bibnamefont
  {Chen}}\ and\ \bibinfo {author} {\bibfnamefont {Kristjan}\ \bibnamefont
  {Haule}},\ }\bibfield  {title} {\enquote {\bibinfo {title} {A combined
  variational and diagrammatic quantum monte carlo approach to the
  many-electron problem},}\ }\href@noop {} {\bibfield  {journal} {\bibinfo
  {journal} {Nature communications}\ }\textbf {\bibinfo {volume} {10}},\
  \bibinfo {pages} {1--7} (\bibinfo {year} {2019})}\BibitemShut {NoStop}%
\bibitem [{\citenamefont {Rossi}\ \emph {et~al.}(2020)\citenamefont {Rossi},
  \citenamefont {Šimkovic},\ and\ \citenamefont {Ferrero}}]{Rossi2020}%
  \BibitemOpen
  \bibfield  {author} {\bibinfo {author} {\bibfnamefont {Riccardo}\
  \bibnamefont {Rossi}}, \bibinfo {author} {\bibfnamefont {Fedor}\ \bibnamefont
  {Šimkovic}}, \ and\ \bibinfo {author} {\bibfnamefont {Michel}\ \bibnamefont
  {Ferrero}},\ }\bibfield  {title} {\enquote {\bibinfo {title} {Renormalized
  perturbation theory at large expansion orders},}\ }\href {\doibase
  10.1209/0295-5075/132/11001} {\bibfield  {journal} {\bibinfo  {journal} {EPL
  (Europhysics Letters)}\ }\textbf {\bibinfo {volume} {132}},\ \bibinfo {pages}
  {11001} (\bibinfo {year} {2020})}\BibitemShut {NoStop}%
\bibitem [{\citenamefont {Šimkovic}\ \emph {et~al.}(2020)\citenamefont
  {Šimkovic}, \citenamefont {Rossi},\ and\ \citenamefont
  {Ferrero}}]{Simkovic2020b}%
  \BibitemOpen
  \bibfield  {author} {\bibinfo {author} {\bibfnamefont {Fedor}\ \bibnamefont
  {Šimkovic}}, \bibinfo {author} {\bibfnamefont {Riccardo}\ \bibnamefont
  {Rossi}}, \ and\ \bibinfo {author} {\bibfnamefont {Michel}\ \bibnamefont
  {Ferrero}},\ }\bibfield  {title} {\enquote {\bibinfo {title} {Efficient
  one-loop-renormalized vertex expansions with connected determinant
  diagrammatic monte carlo},}\ }\href {\doibase 10.1103/physrevb.102.195122}
  {\bibfield  {journal} {\bibinfo  {journal} {Physical Review B}\ }\textbf
  {\bibinfo {volume} {102}} (\bibinfo {year} {2020}),\
  10.1103/physrevb.102.195122}\BibitemShut {NoStop}%
\bibitem [{\citenamefont {{Van Houcke}}\ \emph {et~al.}(2010)\citenamefont
  {{Van Houcke}}, \citenamefont {Kozik}, \citenamefont {Prokof'ev},\ and\
  \citenamefont {Svistunov}}]{VanHoucke2010}%
  \BibitemOpen
  \bibfield  {author} {\bibinfo {author} {\bibfnamefont {Kris}\ \bibnamefont
  {{Van Houcke}}}, \bibinfo {author} {\bibfnamefont {Evgeny}\ \bibnamefont
  {Kozik}}, \bibinfo {author} {\bibfnamefont {Nikolay}\ \bibnamefont
  {Prokof'ev}}, \ and\ \bibinfo {author} {\bibfnamefont {Boris}\ \bibnamefont
  {Svistunov}},\ }\bibfield  {title} {\enquote {\bibinfo {title} {{Diagrammatic
  Monte Carlo}},}\ }\href {\doibase 10.1016/j.phpro.2010.09.034} {\bibfield
  {journal} {\bibinfo  {journal} {Physics Procedia}\ }\textbf {\bibinfo
  {volume} {6}},\ \bibinfo {pages} {95} (\bibinfo {year} {2010})}\BibitemShut
  {NoStop}%
\bibitem [{\citenamefont {Kozik}\ \emph {et~al.}(2010)\citenamefont {Kozik},
  \citenamefont {Houcke}, \citenamefont {Gull}, \citenamefont {Pollet},
  \citenamefont {Prokof'ev}, \citenamefont {Svistunov},\ and\ \citenamefont
  {Troyer}}]{Kozik2010}%
  \BibitemOpen
  \bibfield  {author} {\bibinfo {author} {\bibfnamefont {E.}~\bibnamefont
  {Kozik}}, \bibinfo {author} {\bibfnamefont {K.~Van}\ \bibnamefont {Houcke}},
  \bibinfo {author} {\bibfnamefont {E.}~\bibnamefont {Gull}}, \bibinfo {author}
  {\bibfnamefont {L.}~\bibnamefont {Pollet}}, \bibinfo {author} {\bibfnamefont
  {N.}~\bibnamefont {Prokof'ev}}, \bibinfo {author} {\bibfnamefont
  {B.}~\bibnamefont {Svistunov}}, \ and\ \bibinfo {author} {\bibfnamefont
  {M.}~\bibnamefont {Troyer}},\ }\bibfield  {title} {\enquote {\bibinfo {title}
  {{Diagrammatic Monte Carlo for correlated fermions}},}\ }\href
  {http://stacks.iop.org/0295-5075/90/i=1/a=10004} {\bibfield  {journal}
  {\bibinfo  {journal} {EPL (Europhysics Letters)}\ }\textbf {\bibinfo {volume}
  {90}},\ \bibinfo {pages} {10004} (\bibinfo {year} {2010})}\BibitemShut
  {NoStop}%
\bibitem [{\citenamefont {Rubtsov}\ \emph {et~al.}(2005)\citenamefont
  {Rubtsov}, \citenamefont {Savkin},\ and\ \citenamefont
  {Lichtenstein}}]{Rubtsov2005}%
  \BibitemOpen
  \bibfield  {author} {\bibinfo {author} {\bibfnamefont {A.~N.}\ \bibnamefont
  {Rubtsov}}, \bibinfo {author} {\bibfnamefont {V.~V.}\ \bibnamefont {Savkin}},
  \ and\ \bibinfo {author} {\bibfnamefont {A.~I.}\ \bibnamefont
  {Lichtenstein}},\ }\bibfield  {title} {\enquote {\bibinfo {title}
  {{Continuous-time quantum Monte Carlo method for fermions}},}\ }\href
  {\doibase 10.1103/PhysRevB.72.035122} {\bibfield  {journal} {\bibinfo
  {journal} {Phys. Rev. B}\ }\textbf {\bibinfo {volume} {72}},\ \bibinfo
  {pages} {035122} (\bibinfo {year} {2005})}\BibitemShut {NoStop}%
\bibitem [{\citenamefont {Rossi}(2018)}]{Rossi2018}%
  \BibitemOpen
  \bibfield  {author} {\bibinfo {author} {\bibfnamefont {Riccardo}\
  \bibnamefont {Rossi}},\ }\href@noop {} {\enquote {\bibinfo {title} {{Direct
  sampling of the self-energy with Connected Determinant Monte Carlo}},}\ }
  (\bibinfo {year} {2018}),\ \Eprint {http://arxiv.org/abs/1802.04743}
  {arXiv:1802.04743 [cond-mat.str-el]} \BibitemShut {NoStop}%
\bibitem [{\citenamefont {Van~Houcke}\ \emph {et~al.}(2012)\citenamefont
  {Van~Houcke}, \citenamefont {Werner}, \citenamefont {Kozik}, \citenamefont
  {Prokof'ev}, \citenamefont {Svistunov}, \citenamefont {Ku}, \citenamefont
  {Sommer}, \citenamefont {Cheuk}, \citenamefont {Schirotzek},\ and\
  \citenamefont {Zwierlein}}]{VanHoucke2012}%
  \BibitemOpen
  \bibfield  {author} {\bibinfo {author} {\bibfnamefont {Kris}\ \bibnamefont
  {Van~Houcke}}, \bibinfo {author} {\bibfnamefont {F}~\bibnamefont {Werner}},
  \bibinfo {author} {\bibfnamefont {E}~\bibnamefont {Kozik}}, \bibinfo {author}
  {\bibfnamefont {N}~\bibnamefont {Prokof'ev}}, \bibinfo {author}
  {\bibfnamefont {B}~\bibnamefont {Svistunov}}, \bibinfo {author}
  {\bibfnamefont {MJH}\ \bibnamefont {Ku}}, \bibinfo {author} {\bibfnamefont
  {AT}~\bibnamefont {Sommer}}, \bibinfo {author} {\bibfnamefont
  {LW}~\bibnamefont {Cheuk}}, \bibinfo {author} {\bibfnamefont {A}~\bibnamefont
  {Schirotzek}}, \ and\ \bibinfo {author} {\bibfnamefont {MW}~\bibnamefont
  {Zwierlein}},\ }\bibfield  {title} {\enquote {\bibinfo {title} {{Feynman
  diagrams versus Fermi-gas Feynman emulator}},}\ }\href
  {http://dx.doi.org/10.1038/NPHYS2273} {\bibfield  {journal} {\bibinfo
  {journal} {Nature Physics}\ }\textbf {\bibinfo {volume} {8}},\ \bibinfo
  {pages} {366--370} (\bibinfo {year} {2012})}\BibitemShut {NoStop}%
\bibitem [{\citenamefont {Prokof'ev}\ and\ \citenamefont
  {Svistunov}(2008)}]{Prokofev2008}%
  \BibitemOpen
  \bibfield  {author} {\bibinfo {author} {\bibfnamefont {N.~V.}\ \bibnamefont
  {Prokof'ev}}\ and\ \bibinfo {author} {\bibfnamefont {B.~V.}\ \bibnamefont
  {Svistunov}},\ }\bibfield  {title} {\enquote {\bibinfo {title} {{Bold
  diagrammatic Monte Carlo: A generic sign-problem tolerant technique for
  polaron models and possibly interacting many-body problems}},}\ }\href
  {\doibase 10.1103/PhysRevB.77.125101} {\bibfield  {journal} {\bibinfo
  {journal} {Phys. Rev. B}\ }\textbf {\bibinfo {volume} {77}},\ \bibinfo
  {pages} {125101} (\bibinfo {year} {2008})}\BibitemShut {NoStop}%
\bibitem [{\citenamefont {Šimkovic IV.}\ \emph {et~al.}(2019)\citenamefont
  {Šimkovic IV.}, \citenamefont {Deng},\ and\ \citenamefont
  {Kozik}}]{Simkovic2019b}%
  \BibitemOpen
  \bibfield  {author} {\bibinfo {author} {\bibfnamefont {Fedor}\ \bibnamefont
  {Šimkovic IV.}}, \bibinfo {author} {\bibfnamefont {Youjin}\ \bibnamefont
  {Deng}}, \ and\ \bibinfo {author} {\bibfnamefont {Evgeny}\ \bibnamefont
  {Kozik}},\ }\href@noop {} {\enquote {\bibinfo {title} {{Superfluid
  ground-state phase diagram of the $2d$ Hubbard Model in the emergent BCS
  regime}},}\ } (\bibinfo {year} {2019}),\ \Eprint
  {http://arxiv.org/abs/1912.13054} {arXiv:1912.13054 [cond-mat.str-el]}
  \BibitemShut {NoStop}%
\bibitem [{\citenamefont {Bj\"{o}rklund}\ \emph {et~al.}(2007)\citenamefont
  {Bj\"{o}rklund}, \citenamefont {Husfeldt}, \citenamefont {Kaski},\ and\
  \citenamefont {Koivisto}}]{Koivisto2007}%
  \BibitemOpen
  \bibfield  {author} {\bibinfo {author} {\bibfnamefont {Andreas}\ \bibnamefont
  {Bj\"{o}rklund}}, \bibinfo {author} {\bibfnamefont {Thore}\ \bibnamefont
  {Husfeldt}}, \bibinfo {author} {\bibfnamefont {Petteri}\ \bibnamefont
  {Kaski}}, \ and\ \bibinfo {author} {\bibfnamefont {Mikko}\ \bibnamefont
  {Koivisto}},\ }\bibfield  {title} {\enquote {\bibinfo {title} {{Fourier Meets
  M\"{o}bius: Fast Subset Convolution}},}\ }in\ \href {\doibase
  10.1145/1250790.1250801} {\emph {\bibinfo {booktitle} {Proceedings of the
  Thirty-Ninth Annual ACM Symposium on Theory of Computing}}},\ \bibinfo
  {series and number} {STOC ’07}\ (\bibinfo  {publisher} {Association for
  Computing Machinery},\ \bibinfo {address} {New York, NY, USA},\ \bibinfo
  {year} {2007})\ p.\ \bibinfo {pages} {67–74}\BibitemShut {NoStop}%
\bibitem [{\citenamefont {Rossi}\ \emph {et~al.}(2017)\citenamefont {Rossi},
  \citenamefont {Prokof'ev}, \citenamefont {Svistunov}, \citenamefont
  {Houcke},\ and\ \citenamefont {Werner}}]{Rossi2017b}%
  \BibitemOpen
  \bibfield  {author} {\bibinfo {author} {\bibfnamefont {R.}~\bibnamefont
  {Rossi}}, \bibinfo {author} {\bibfnamefont {N.}~\bibnamefont {Prokof'ev}},
  \bibinfo {author} {\bibfnamefont {B.}~\bibnamefont {Svistunov}}, \bibinfo
  {author} {\bibfnamefont {K.~Van}\ \bibnamefont {Houcke}}, \ and\ \bibinfo
  {author} {\bibfnamefont {F.}~\bibnamefont {Werner}},\ }\bibfield  {title}
  {\enquote {\bibinfo {title} {{Polynomial complexity despite the fermionic
  sign}},}\ }\href {\doibase 10.1209/0295-5075/118/10004} {\bibfield  {journal}
  {\bibinfo  {journal} {{EPL} (Europhysics Letters)}\ }\textbf {\bibinfo
  {volume} {118}},\ \bibinfo {pages} {10004} (\bibinfo {year}
  {2017})}\BibitemShut {NoStop}%
\bibitem [{\citenamefont {Brezinski}(1996)}]{Brezinski1996}%
  \BibitemOpen
  \bibfield  {author} {\bibinfo {author} {\bibfnamefont {C.}~\bibnamefont
  {Brezinski}},\ }\bibfield  {title} {\enquote {\bibinfo {title}
  {{Extrapolation algorithms and Padé approximations: a historical survey}},}\
  }\href {\doibase https://doi.org/10.1016/0168-9274(95)00110-7} {\bibfield
  {journal} {\bibinfo  {journal} {Applied Numerical Mathematics}\ }\textbf
  {\bibinfo {volume} {20}},\ \bibinfo {pages} {299 -- 318} (\bibinfo {year}
  {1996})}\BibitemShut {NoStop}%
\bibitem [{\citenamefont {Gonnet}\ \emph {et~al.}(2013)\citenamefont {Gonnet},
  \citenamefont {Güttel},\ and\ \citenamefont {Trefethen}}]{Gonnet2013}%
  \BibitemOpen
  \bibfield  {author} {\bibinfo {author} {\bibfnamefont {Pedro}\ \bibnamefont
  {Gonnet}}, \bibinfo {author} {\bibfnamefont {Stefan}\ \bibnamefont
  {Güttel}}, \ and\ \bibinfo {author} {\bibfnamefont {Lloyd~N.}\ \bibnamefont
  {Trefethen}},\ }\bibfield  {title} {\enquote {\bibinfo {title} {{Robust
  Pad{\'e} Approximation via SVD}},}\ }\href {\doibase 10.1137/110853236}
  {\bibfield  {journal} {\bibinfo  {journal} {SIAM Review}\ }\textbf {\bibinfo
  {volume} {55}},\ \bibinfo {pages} {101--117} (\bibinfo {year} {2013})},\
  \Eprint {http://arxiv.org/abs/https://doi.org/10.1137/110853236}
  {https://doi.org/10.1137/110853236} \BibitemShut {NoStop}%
\bibitem [{\citenamefont {Lenihan}\ \emph {et~al.}(2021)\citenamefont
  {Lenihan}, \citenamefont {Kim}, \citenamefont {\ifmmode
  \check{S}\else~\v{S}\fi{}imkovic IV.},\ and\ \citenamefont
  {Kozik}}]{Lenihan2021}%
  \BibitemOpen
  \bibfield  {author} {\bibinfo {author} {\bibfnamefont {Connor}\ \bibnamefont
  {Lenihan}}, \bibinfo {author} {\bibfnamefont {Aaram~J.}\ \bibnamefont {Kim}},
  \bibinfo {author} {\bibfnamefont {Fedor}\ \bibnamefont {\ifmmode
  \check{S}\else~\v{S}\fi{}imkovic IV.}}, \ and\ \bibinfo {author}
  {\bibfnamefont {Evgeny}\ \bibnamefont {Kozik}},\ }\bibfield  {title}
  {\enquote {\bibinfo {title} {{Entropy in the Non-Fermi-Liquid Regime of the
  Doped $2\mathrm{D}$ Hubbard Model}},}\ }\href {\doibase
  10.1103/PhysRevLett.126.105701} {\bibfield  {journal} {\bibinfo  {journal}
  {Phys. Rev. Lett.}\ }\textbf {\bibinfo {volume} {126}},\ \bibinfo {pages}
  {105701} (\bibinfo {year} {2021})}\BibitemShut {NoStop}%
\bibitem [{\citenamefont {Hirsch}(1983)}]{Hirsch1983}%
  \BibitemOpen
  \bibfield  {author} {\bibinfo {author} {\bibfnamefont {J.~E.}\ \bibnamefont
  {Hirsch}},\ }\bibfield  {title} {\enquote {\bibinfo {title} {{Discrete
  Hubbard-Stratonovich transformation for fermion lattice models}},}\ }\href
  {\doibase 10.1103/PhysRevB.28.4059} {\bibfield  {journal} {\bibinfo
  {journal} {Phys. Rev. B}\ }\textbf {\bibinfo {volume} {28}},\ \bibinfo
  {pages} {4059--4061} (\bibinfo {year} {1983})}\BibitemShut {NoStop}%
\bibitem [{\citenamefont {Assaad}\ and\ \citenamefont
  {Evertz}(2008)}]{AssaadEvertz2008}%
  \BibitemOpen
  \bibfield  {author} {\bibinfo {author} {\bibfnamefont {F.F.}\ \bibnamefont
  {Assaad}}\ and\ \bibinfo {author} {\bibfnamefont {H.G.}\ \bibnamefont
  {Evertz}},\ }\bibfield  {title} {\enquote {\bibinfo {title} {{World-line and
  Determinantal Quantum Monte Carlo Methods for Spins, Phonons and
  Electrons}},}\ }in\ \href {\doibase 10.1007/978-3-540-74686-7_10} {\emph
  {\bibinfo {booktitle} {Computational Many-Particle Physics}}},\ \bibinfo
  {series} {Lecture Notes in Physics}, Vol.\ \bibinfo {volume} {739},\ \bibinfo
  {editor} {edited by\ \bibinfo {editor} {\bibfnamefont {H.}~\bibnamefont
  {Fehske}}, \bibinfo {editor} {\bibfnamefont {R.}~\bibnamefont {Schneider}}, \
  and\ \bibinfo {editor} {\bibfnamefont {A.}~\bibnamefont {Wei{\ss}e}}}\
  (\bibinfo  {publisher} {Springer Berlin Heidelberg},\ \bibinfo {year}
  {2008})\ pp.\ \bibinfo {pages} {277--356}\BibitemShut {NoStop}%
\bibitem [{\citenamefont {Chang}\ \emph {et~al.}(2015)\citenamefont {Chang},
  \citenamefont {Gogolenko}, \citenamefont {Perez}, \citenamefont {Bai},\ and\
  \citenamefont {Scalettar}}]{Chang2015}%
  \BibitemOpen
  \bibfield  {author} {\bibinfo {author} {\bibfnamefont {Chia-Chen}\
  \bibnamefont {Chang}}, \bibinfo {author} {\bibfnamefont {Sergiy}\
  \bibnamefont {Gogolenko}}, \bibinfo {author} {\bibfnamefont {Jeffrey}\
  \bibnamefont {Perez}}, \bibinfo {author} {\bibfnamefont {Zhaojun}\
  \bibnamefont {Bai}}, \ and\ \bibinfo {author} {\bibfnamefont {Richard~T.}\
  \bibnamefont {Scalettar}},\ }\bibfield  {title} {\enquote {\bibinfo {title}
  {{Recent advances in determinant quantum Monte Carlo}},}\ }\href {\doibase
  10.1080/14786435.2013.845314} {\bibfield  {journal} {\bibinfo  {journal}
  {Philosophical Magazine}\ }\textbf {\bibinfo {volume} {95}},\ \bibinfo
  {pages} {1260--1281} (\bibinfo {year} {2015})}\BibitemShut {NoStop}%
\bibitem [{\citenamefont {He}\ \emph {et~al.}(2019{\natexlab{a}})\citenamefont
  {He}, \citenamefont {Qin}, \citenamefont {Shi}, \citenamefont {Lu},\ and\
  \citenamefont {Zhang}}]{Yuan19a}%
  \BibitemOpen
  \bibfield  {author} {\bibinfo {author} {\bibfnamefont {Yuan-Yao}\
  \bibnamefont {He}}, \bibinfo {author} {\bibfnamefont {Mingpu}\ \bibnamefont
  {Qin}}, \bibinfo {author} {\bibfnamefont {Hao}\ \bibnamefont {Shi}}, \bibinfo
  {author} {\bibfnamefont {Zhong-Yi}\ \bibnamefont {Lu}}, \ and\ \bibinfo
  {author} {\bibfnamefont {Shiwei}\ \bibnamefont {Zhang}},\ }\bibfield  {title}
  {\enquote {\bibinfo {title} {{Finite-temperature auxiliary-field quantum
  Monte Carlo: Self-consistent constraint and systematic approach to low
  temperatures}},}\ }\href {\doibase 10.1103/PhysRevB.99.045108} {\bibfield
  {journal} {\bibinfo  {journal} {Phys. Rev. B}\ }\textbf {\bibinfo {volume}
  {99}},\ \bibinfo {pages} {045108} (\bibinfo {year}
  {2019}{\natexlab{a}})}\BibitemShut {NoStop}%
\bibitem [{\citenamefont {He}\ \emph {et~al.}(2019{\natexlab{b}})\citenamefont
  {He}, \citenamefont {Shi},\ and\ \citenamefont {Zhang}}]{Yuan19b}%
  \BibitemOpen
  \bibfield  {author} {\bibinfo {author} {\bibfnamefont {Yuan-Yao}\
  \bibnamefont {He}}, \bibinfo {author} {\bibfnamefont {Hao}\ \bibnamefont
  {Shi}}, \ and\ \bibinfo {author} {\bibfnamefont {Shiwei}\ \bibnamefont
  {Zhang}},\ }\bibfield  {title} {\enquote {\bibinfo {title} {Reaching the
  continuum limit in finite-temperature ab initio field-theory computations in
  many-fermion systems},}\ }\href {\doibase 10.1103/PhysRevLett.123.136402}
  {\bibfield  {journal} {\bibinfo  {journal} {Phys. Rev. Lett.}\ }\textbf
  {\bibinfo {volume} {123}},\ \bibinfo {pages} {136402} (\bibinfo {year}
  {2019}{\natexlab{b}})}\BibitemShut {NoStop}%
\bibitem [{\citenamefont {Bohm}\ and\ \citenamefont {Pines}(1951)}]{Bohm1951}%
  \BibitemOpen
  \bibfield  {author} {\bibinfo {author} {\bibfnamefont {David}\ \bibnamefont
  {Bohm}}\ and\ \bibinfo {author} {\bibfnamefont {David}\ \bibnamefont
  {Pines}},\ }\bibfield  {title} {\enquote {\bibinfo {title} {{A Collective
  Description of Electron Interactions. I. Magnetic Interactions}},}\ }\href
  {\doibase 10.1103/PhysRev.82.625} {\bibfield  {journal} {\bibinfo  {journal}
  {Phys. Rev.}\ }\textbf {\bibinfo {volume} {82}},\ \bibinfo {pages} {625--634}
  (\bibinfo {year} {1951})}\BibitemShut {NoStop}%
\bibitem [{\citenamefont {Rubtsov}\ and\ \citenamefont
  {Lichtenstein}(2004)}]{Rubtsov2004}%
  \BibitemOpen
  \bibfield  {author} {\bibinfo {author} {\bibfnamefont {A.~N.}\ \bibnamefont
  {Rubtsov}}\ and\ \bibinfo {author} {\bibfnamefont {A.~I.}\ \bibnamefont
  {Lichtenstein}},\ }\bibfield  {title} {\enquote {\bibinfo {title}
  {{Continuous-time quantum Monte Carlo method for fermions: Beyond auxiliary
  field framework}},}\ }\href {\doibase 10.1134/1.1800216} {\bibfield
  {journal} {\bibinfo  {journal} {Journal of Experimental and Theoretical
  Physics Letters}\ }\textbf {\bibinfo {volume} {80}},\ \bibinfo {pages}
  {61–65} (\bibinfo {year} {2004})}\BibitemShut {NoStop}%
\bibitem [{\citenamefont {Gull}\ \emph
  {et~al.}(2011{\natexlab{a}})\citenamefont {Gull}, \citenamefont {Millis},
  \citenamefont {Lichtenstein}, \citenamefont {Rubtsov}, \citenamefont
  {Troyer},\ and\ \citenamefont {Werner}}]{Gull2011a}%
  \BibitemOpen
  \bibfield  {author} {\bibinfo {author} {\bibfnamefont {Emanuel}\ \bibnamefont
  {Gull}}, \bibinfo {author} {\bibfnamefont {Andrew~J.}\ \bibnamefont
  {Millis}}, \bibinfo {author} {\bibfnamefont {Alexander~I.}\ \bibnamefont
  {Lichtenstein}}, \bibinfo {author} {\bibfnamefont {Alexey~N.}\ \bibnamefont
  {Rubtsov}}, \bibinfo {author} {\bibfnamefont {Matthias}\ \bibnamefont
  {Troyer}}, \ and\ \bibinfo {author} {\bibfnamefont {Philipp}\ \bibnamefont
  {Werner}},\ }\bibfield  {title} {\enquote {\bibinfo {title} {{Continuous-time
  Monte Carlo methods for quantum impurity models}},}\ }\href {\doibase
  10.1103/RevModPhys.83.349} {\bibfield  {journal} {\bibinfo  {journal} {Rev.
  Mod. Phys.}\ }\textbf {\bibinfo {volume} {83}},\ \bibinfo {pages} {349}
  (\bibinfo {year} {2011}{\natexlab{a}})}\BibitemShut {NoStop}%
\bibitem [{\citenamefont {Parcollet}\ \emph {et~al.}(2015)\citenamefont
  {Parcollet}, \citenamefont {Ferrero}, \citenamefont {Ayral}, \citenamefont
  {Hafermann}, \citenamefont {Krivenko}, \citenamefont {Messio},\ and\
  \citenamefont {Seth}}]{TRIQS}%
  \BibitemOpen
  \bibfield  {author} {\bibinfo {author} {\bibfnamefont {Olivier}\ \bibnamefont
  {Parcollet}}, \bibinfo {author} {\bibfnamefont {Michel}\ \bibnamefont
  {Ferrero}}, \bibinfo {author} {\bibfnamefont {Thomas}\ \bibnamefont {Ayral}},
  \bibinfo {author} {\bibfnamefont {Hartmut}\ \bibnamefont {Hafermann}},
  \bibinfo {author} {\bibfnamefont {Igor}\ \bibnamefont {Krivenko}}, \bibinfo
  {author} {\bibfnamefont {Laura}\ \bibnamefont {Messio}}, \ and\ \bibinfo
  {author} {\bibfnamefont {Priyanka}\ \bibnamefont {Seth}},\ }\bibfield
  {title} {\enquote {\bibinfo {title} {{TRIQS: A toolbox for research on
  interacting quantum systems}},}\ }\href {\doibase
  http://dx.doi.org/10.1016/j.cpc.2015.04.023} {\bibfield  {journal} {\bibinfo
  {journal} {Computer Physics Communications}\ }\textbf {\bibinfo {volume}
  {196}},\ \bibinfo {pages} {398 -- 415} (\bibinfo {year} {2015})}\BibitemShut
  {NoStop}%
\bibitem [{\citenamefont {Rohringer}\ \emph
  {et~al.}(2018{\natexlab{b}})\citenamefont {Rohringer}, \citenamefont
  {Katanin}, \citenamefont {Sch{\"a}fer}, \citenamefont {Hausoel},
  \citenamefont {Held},\ and\ \citenamefont {Toschi}}]{Rohringer2018}%
  \BibitemOpen
  \bibfield  {author} {\bibinfo {author} {\bibfnamefont {G.}~\bibnamefont
  {Rohringer}}, \bibinfo {author} {\bibfnamefont {A.}~\bibnamefont {Katanin}},
  \bibinfo {author} {\bibfnamefont {T.}~\bibnamefont {Sch{\"a}fer}}, \bibinfo
  {author} {\bibfnamefont {A.}~\bibnamefont {Hausoel}}, \bibinfo {author}
  {\bibfnamefont {K.}~\bibnamefont {Held}}, \ and\ \bibinfo {author}
  {\bibfnamefont {A.}~\bibnamefont {Toschi}},\ }\bibfield  {title} {\enquote
  {\bibinfo {title} {{LadderD$\Gamma$A code}},}\ }\href
  {https://github.com/ladderDGA/ladderDGA} {\bibfield  {journal} {\bibinfo
  {journal} {github.com/ladderDGA}\ } (\bibinfo {year}
  {2018}{\natexlab{b}})}\BibitemShut {NoStop}%
\bibitem [{\citenamefont {Strand}\ \emph
  {et~al.}(2019{\natexlab{b}})\citenamefont {Strand}, \citenamefont
  {Wentzell},\ and\ \citenamefont {Parcollet}}]{tprf}%
  \BibitemOpen
  \bibfield  {author} {\bibinfo {author} {\bibfnamefont {H.~U.~R.}\
  \bibnamefont {Strand}}, \bibinfo {author} {\bibfnamefont {N.}~\bibnamefont
  {Wentzell}}, \ and\ \bibinfo {author} {\bibfnamefont {O.}~\bibnamefont
  {Parcollet}},\ }\href {https://triqs.github.io/tprf/2.1.x/index.html}
  {\enquote {\bibinfo {title} {{tprf - two-particle response function tools
  based on the TRIQS library}},}\ } (\bibinfo {year}
  {2019}{\natexlab{b}})\BibitemShut {NoStop}%
\bibitem [{\citenamefont {LeBlanc}\ and\ \citenamefont
  {Gull}(2013)}]{LeBlanc2013}%
  \BibitemOpen
  \bibfield  {author} {\bibinfo {author} {\bibfnamefont {J.~P.~F.}\
  \bibnamefont {LeBlanc}}\ and\ \bibinfo {author} {\bibfnamefont {Emanuel}\
  \bibnamefont {Gull}},\ }\bibfield  {title} {\enquote {\bibinfo {title}
  {{Equation of state of the fermionic two-dimensional Hubbard model}},}\
  }\href {\doibase 10.1103/PhysRevB.88.155108} {\bibfield  {journal} {\bibinfo
  {journal} {Phys. Rev. B}\ }\textbf {\bibinfo {volume} {88}},\ \bibinfo
  {pages} {155108} (\bibinfo {year} {2013})}\BibitemShut {NoStop}%
\bibitem [{\citenamefont {Gull}\ \emph
  {et~al.}(2008{\natexlab{b}})\citenamefont {Gull}, \citenamefont {Werner},
  \citenamefont {Parcollet},\ and\ \citenamefont {Troyer}}]{Gull2008a}%
  \BibitemOpen
  \bibfield  {author} {\bibinfo {author} {\bibfnamefont {E.}~\bibnamefont
  {Gull}}, \bibinfo {author} {\bibfnamefont {P.}~\bibnamefont {Werner}},
  \bibinfo {author} {\bibfnamefont {O.}~\bibnamefont {Parcollet}}, \ and\
  \bibinfo {author} {\bibfnamefont {M.}~\bibnamefont {Troyer}},\ }\bibfield
  {title} {\enquote {\bibinfo {title} {{Continuous-time auxiliary-field Monte
  Carlo for quantum impurity models}},}\ }\href
  {http://stacks.iop.org/0295-5075/82/i=5/a=57003} {\bibfield  {journal}
  {\bibinfo  {journal} {EPL (Europhysics Letters)}\ }\textbf {\bibinfo {volume}
  {82}},\ \bibinfo {pages} {57003} (\bibinfo {year}
  {2008}{\natexlab{b}})}\BibitemShut {NoStop}%
\bibitem [{\citenamefont {Gull}\ \emph
  {et~al.}(2011{\natexlab{b}})\citenamefont {Gull}, \citenamefont {Staar},
  \citenamefont {Fuchs}, \citenamefont {Nukala}, \citenamefont {Summers},
  \citenamefont {Pruschke}, \citenamefont {Schulthess},\ and\ \citenamefont
  {Maier}}]{Gull2011}%
  \BibitemOpen
  \bibfield  {author} {\bibinfo {author} {\bibfnamefont {Emanuel}\ \bibnamefont
  {Gull}}, \bibinfo {author} {\bibfnamefont {Peter}\ \bibnamefont {Staar}},
  \bibinfo {author} {\bibfnamefont {Sebastian}\ \bibnamefont {Fuchs}}, \bibinfo
  {author} {\bibfnamefont {Phani}\ \bibnamefont {Nukala}}, \bibinfo {author}
  {\bibfnamefont {Michael~S.}\ \bibnamefont {Summers}}, \bibinfo {author}
  {\bibfnamefont {Thomas}\ \bibnamefont {Pruschke}}, \bibinfo {author}
  {\bibfnamefont {Thomas~C.}\ \bibnamefont {Schulthess}}, \ and\ \bibinfo
  {author} {\bibfnamefont {Thomas}\ \bibnamefont {Maier}},\ }\bibfield  {title}
  {\enquote {\bibinfo {title} {{Submatrix updates for the continuous-time
  auxiliary-field algorithm}},}\ }\href {\doibase 10.1103/PhysRevB.83.075122}
  {\bibfield  {journal} {\bibinfo  {journal} {Phys. Rev. B}\ }\textbf {\bibinfo
  {volume} {83}},\ \bibinfo {pages} {075122} (\bibinfo {year}
  {2011}{\natexlab{b}})}\BibitemShut {NoStop}%
\bibitem [{\citenamefont {Gaenko}\ \emph {et~al.}(2017)\citenamefont {Gaenko},
  \citenamefont {Antipov}, \citenamefont {Carcassi}, \citenamefont {Chen},
  \citenamefont {Chen}, \citenamefont {Dong}, \citenamefont {Gamper},
  \citenamefont {Gukelberger}, \citenamefont {Igarashi}, \citenamefont
  {Iskakov}, \citenamefont {Könz}, \citenamefont {LeBlanc}, \citenamefont
  {Levy}, \citenamefont {Ma}, \citenamefont {Paki}, \citenamefont {Shinaoka},
  \citenamefont {Todo}, \citenamefont {Troyer},\ and\ \citenamefont
  {Gull}}]{ALPSCore}%
  \BibitemOpen
  \bibfield  {author} {\bibinfo {author} {\bibfnamefont {A.}~\bibnamefont
  {Gaenko}}, \bibinfo {author} {\bibfnamefont {A.E.}\ \bibnamefont {Antipov}},
  \bibinfo {author} {\bibfnamefont {G.}~\bibnamefont {Carcassi}}, \bibinfo
  {author} {\bibfnamefont {T.}~\bibnamefont {Chen}}, \bibinfo {author}
  {\bibfnamefont {X.}~\bibnamefont {Chen}}, \bibinfo {author} {\bibfnamefont
  {Q.}~\bibnamefont {Dong}}, \bibinfo {author} {\bibfnamefont {L.}~\bibnamefont
  {Gamper}}, \bibinfo {author} {\bibfnamefont {J.}~\bibnamefont {Gukelberger}},
  \bibinfo {author} {\bibfnamefont {R.}~\bibnamefont {Igarashi}}, \bibinfo
  {author} {\bibfnamefont {S.}~\bibnamefont {Iskakov}}, \bibinfo {author}
  {\bibfnamefont {M.}~\bibnamefont {Könz}}, \bibinfo {author} {\bibfnamefont
  {J.P.F.}\ \bibnamefont {LeBlanc}}, \bibinfo {author} {\bibfnamefont
  {R.}~\bibnamefont {Levy}}, \bibinfo {author} {\bibfnamefont {P.N.}\
  \bibnamefont {Ma}}, \bibinfo {author} {\bibfnamefont {J.E.}\ \bibnamefont
  {Paki}}, \bibinfo {author} {\bibfnamefont {H.}~\bibnamefont {Shinaoka}},
  \bibinfo {author} {\bibfnamefont {S.}~\bibnamefont {Todo}}, \bibinfo {author}
  {\bibfnamefont {M.}~\bibnamefont {Troyer}}, \ and\ \bibinfo {author}
  {\bibfnamefont {E.}~\bibnamefont {Gull}},\ }\bibfield  {title} {\enquote
  {\bibinfo {title} {{Updated core libraries of the ALPS project}},}\ }\href
  {\doibase https://doi.org/10.1016/j.cpc.2016.12.009} {\bibfield  {journal}
  {\bibinfo  {journal} {Computer Physics Communications}\ }\textbf {\bibinfo
  {volume} {213}},\ \bibinfo {pages} {235 -- 251} (\bibinfo {year}
  {2017})}\BibitemShut {NoStop}%
\bibitem [{\citenamefont {Wallerberger}\ \emph {et~al.}(2018)\citenamefont
  {Wallerberger}, \citenamefont {Iskakov}, \citenamefont {Gaenko},
  \citenamefont {Kleinhenz}, \citenamefont {Krivenko}, \citenamefont {Levy},
  \citenamefont {Li}, \citenamefont {Shinaoka}, \citenamefont {Todo},
  \citenamefont {Chen}, \citenamefont {Chen}, \citenamefont {LeBlanc},
  \citenamefont {Paki}, \citenamefont {Terletska}, \citenamefont {Troyer},\
  and\ \citenamefont {Gull}}]{ALPSCore_updated}%
  \BibitemOpen
  \bibfield  {author} {\bibinfo {author} {\bibfnamefont {Markus}\ \bibnamefont
  {Wallerberger}}, \bibinfo {author} {\bibfnamefont {Sergei}\ \bibnamefont
  {Iskakov}}, \bibinfo {author} {\bibfnamefont {Alexander}\ \bibnamefont
  {Gaenko}}, \bibinfo {author} {\bibfnamefont {Joseph}\ \bibnamefont
  {Kleinhenz}}, \bibinfo {author} {\bibfnamefont {Igor}\ \bibnamefont
  {Krivenko}}, \bibinfo {author} {\bibfnamefont {Ryan}\ \bibnamefont {Levy}},
  \bibinfo {author} {\bibfnamefont {Jia}\ \bibnamefont {Li}}, \bibinfo {author}
  {\bibfnamefont {Hiroshi}\ \bibnamefont {Shinaoka}}, \bibinfo {author}
  {\bibfnamefont {Synge}\ \bibnamefont {Todo}}, \bibinfo {author}
  {\bibfnamefont {Tianran}\ \bibnamefont {Chen}}, \bibinfo {author}
  {\bibfnamefont {Xi}~\bibnamefont {Chen}}, \bibinfo {author} {\bibfnamefont
  {James P.~F.}\ \bibnamefont {LeBlanc}}, \bibinfo {author} {\bibfnamefont
  {Joseph~E.}\ \bibnamefont {Paki}}, \bibinfo {author} {\bibfnamefont {Hanna}\
  \bibnamefont {Terletska}}, \bibinfo {author} {\bibfnamefont {Matthias}\
  \bibnamefont {Troyer}}, \ and\ \bibinfo {author} {\bibfnamefont {Emanuel}\
  \bibnamefont {Gull}},\ }\href@noop {} {\enquote {\bibinfo {title} {Updated
  core libraries of the alps project},}\ } (\bibinfo {year} {2018}),\ \Eprint
  {http://arxiv.org/abs/1811.08331} {arXiv:1811.08331 [physics.comp-ph]}
  \BibitemShut {NoStop}%
\bibitem [{\citenamefont {Sakai}\ \emph {et~al.}(2012)\citenamefont {Sakai},
  \citenamefont {Sangiovanni}, \citenamefont {Civelli}, \citenamefont {Motome},
  \citenamefont {Held},\ and\ \citenamefont {Imada}}]{Sakai2012}%
  \BibitemOpen
  \bibfield  {author} {\bibinfo {author} {\bibfnamefont {Shiro}\ \bibnamefont
  {Sakai}}, \bibinfo {author} {\bibfnamefont {Giorgio}\ \bibnamefont
  {Sangiovanni}}, \bibinfo {author} {\bibfnamefont {Marcello}\ \bibnamefont
  {Civelli}}, \bibinfo {author} {\bibfnamefont {Yukitoshi}\ \bibnamefont
  {Motome}}, \bibinfo {author} {\bibfnamefont {Karsten}\ \bibnamefont {Held}},
  \ and\ \bibinfo {author} {\bibfnamefont {Masatoshi}\ \bibnamefont {Imada}},\
  }\bibfield  {title} {\enquote {\bibinfo {title} {Cluster-size dependence in
  cellular dynamical mean-field theory},}\ }\href {\doibase
  10.1103/PhysRevB.85.035102} {\bibfield  {journal} {\bibinfo  {journal} {Phys.
  Rev. B}\ }\textbf {\bibinfo {volume} {85}},\ \bibinfo {pages} {035102}
  (\bibinfo {year} {2012})}\BibitemShut {NoStop}%
\bibitem [{\citenamefont {Stanescu}\ and\ \citenamefont
  {Kotliar}(2006)}]{Stanescu2006}%
  \BibitemOpen
  \bibfield  {author} {\bibinfo {author} {\bibfnamefont {Tudor~D.}\
  \bibnamefont {Stanescu}}\ and\ \bibinfo {author} {\bibfnamefont {Gabriel}\
  \bibnamefont {Kotliar}},\ }\bibfield  {title} {\enquote {\bibinfo {title}
  {{Fermi arcs and hidden zeros of the Green function in the pseudogap
  state}},}\ }\href {\doibase 10.1103/PhysRevB.74.125110} {\bibfield  {journal}
  {\bibinfo  {journal} {Phys. Rev. B}\ }\textbf {\bibinfo {volume} {74}},\
  \bibinfo {pages} {125110} (\bibinfo {year} {2006})}\BibitemShut {NoStop}%
\bibitem [{\citenamefont {Sakai}\ \emph {et~al.}(2009)\citenamefont {Sakai},
  \citenamefont {Motome},\ and\ \citenamefont {Imada}}]{Sakai2009}%
  \BibitemOpen
  \bibfield  {author} {\bibinfo {author} {\bibfnamefont {Shiro}\ \bibnamefont
  {Sakai}}, \bibinfo {author} {\bibfnamefont {Yukitoshi}\ \bibnamefont
  {Motome}}, \ and\ \bibinfo {author} {\bibfnamefont {Masatoshi}\ \bibnamefont
  {Imada}},\ }\bibfield  {title} {\enquote {\bibinfo {title} {{Evolution of
  Electronic Structure of Doped Mott Insulators: Reconstruction of Poles and
  Zeros of Green's Function}},}\ }\href {\doibase
  10.1103/PhysRevLett.102.056404} {\bibfield  {journal} {\bibinfo  {journal}
  {Phys. Rev. Lett.}\ }\textbf {\bibinfo {volume} {102}},\ \bibinfo {pages}
  {056404} (\bibinfo {year} {2009})}\BibitemShut {NoStop}%
\bibitem [{\citenamefont {Staar}\ \emph {et~al.}(2013)\citenamefont {Staar},
  \citenamefont {Maier}, \citenamefont {Summers}, \citenamefont {Fourestey},
  \citenamefont {Solca},\ and\ \citenamefont {Schulthess}}]{Staar2013a}%
  \BibitemOpen
  \bibfield  {author} {\bibinfo {author} {\bibfnamefont {Peter}\ \bibnamefont
  {Staar}}, \bibinfo {author} {\bibfnamefont {Thomas~A.}\ \bibnamefont
  {Maier}}, \bibinfo {author} {\bibfnamefont {Michael~S.}\ \bibnamefont
  {Summers}}, \bibinfo {author} {\bibfnamefont {Gilles}\ \bibnamefont
  {Fourestey}}, \bibinfo {author} {\bibfnamefont {Raffaele}\ \bibnamefont
  {Solca}}, \ and\ \bibinfo {author} {\bibfnamefont {Thomas~C.}\ \bibnamefont
  {Schulthess}},\ }\bibfield  {title} {\enquote {\bibinfo {title} {{Taking a
  Quantum Leap in Time to Solution for Simulations of high-Tc
  Superconductors}},}\ }in\ \href {\doibase 10.1145/2503210.2503282} {\emph
  {\bibinfo {booktitle} {Proceedings of the International Conference on High
  Performance Computing, Networking, Storage and Analysis}}},\ \bibinfo {series
  and number} {SC '13}\ (\bibinfo  {publisher} {ACM},\ \bibinfo {address} {New
  York, NY, USA},\ \bibinfo {year} {2013})\ pp.\ \bibinfo {pages}
  {1:1--1:11}\BibitemShut {NoStop}%
\bibitem [{\citenamefont {Sakai}\ \emph {et~al.}(2016)\citenamefont {Sakai},
  \citenamefont {Civelli},\ and\ \citenamefont {Imada}}]{Sakai2016}%
  \BibitemOpen
  \bibfield  {author} {\bibinfo {author} {\bibfnamefont {Shiro}\ \bibnamefont
  {Sakai}}, \bibinfo {author} {\bibfnamefont {Marcello}\ \bibnamefont
  {Civelli}}, \ and\ \bibinfo {author} {\bibfnamefont {Masatoshi}\ \bibnamefont
  {Imada}},\ }\bibfield  {title} {\enquote {\bibinfo {title} {{Hidden Fermionic
  Excitation Boosting High-Temperature Superconductivity in Cuprates}},}\
  }\href {\doibase 10.1103/PhysRevLett.116.057003} {\bibfield  {journal}
  {\bibinfo  {journal} {Phys. Rev. Lett.}\ }\textbf {\bibinfo {volume} {116}},\
  \bibinfo {pages} {057003} (\bibinfo {year} {2016})}\BibitemShut {NoStop}%
\bibitem [{\citenamefont {Verret}\ \emph {et~al.}(2019)\citenamefont {Verret},
  \citenamefont {Roy}, \citenamefont {Foley}, \citenamefont {Charlebois},
  \citenamefont {S\'en\'echal},\ and\ \citenamefont {Tremblay}}]{Verret2019}%
  \BibitemOpen
  \bibfield  {author} {\bibinfo {author} {\bibfnamefont {S.}~\bibnamefont
  {Verret}}, \bibinfo {author} {\bibfnamefont {J.}~\bibnamefont {Roy}},
  \bibinfo {author} {\bibfnamefont {A.}~\bibnamefont {Foley}}, \bibinfo
  {author} {\bibfnamefont {M.}~\bibnamefont {Charlebois}}, \bibinfo {author}
  {\bibfnamefont {D.}~\bibnamefont {S\'en\'echal}}, \ and\ \bibinfo {author}
  {\bibfnamefont {A.-M.~S.}\ \bibnamefont {Tremblay}},\ }\bibfield  {title}
  {\enquote {\bibinfo {title} {{Intrinsic cluster-shaped density waves in
  cellular dynamical mean-field theory}},}\ }\href {\doibase
  10.1103/PhysRevB.100.224520} {\bibfield  {journal} {\bibinfo  {journal}
  {Phys. Rev. B}\ }\textbf {\bibinfo {volume} {100}},\ \bibinfo {pages}
  {224520} (\bibinfo {year} {2019})}\BibitemShut {NoStop}%
\bibitem [{\citenamefont {Valli}\ \emph {et~al.}(2010)\citenamefont {Valli},
  \citenamefont {Sangiovanni}, \citenamefont {Gunnarsson}, \citenamefont
  {Toschi},\ and\ \citenamefont {Held}}]{Valli2010}%
  \BibitemOpen
  \bibfield  {author} {\bibinfo {author} {\bibfnamefont {A.}~\bibnamefont
  {Valli}}, \bibinfo {author} {\bibfnamefont {G.}~\bibnamefont {Sangiovanni}},
  \bibinfo {author} {\bibfnamefont {O.}~\bibnamefont {Gunnarsson}}, \bibinfo
  {author} {\bibfnamefont {A.}~\bibnamefont {Toschi}}, \ and\ \bibinfo {author}
  {\bibfnamefont {K.}~\bibnamefont {Held}},\ }\bibfield  {title} {\enquote
  {\bibinfo {title} {{Dynamical Vertex Approximation for Nanoscopic
  Systems}},}\ }\href {\doibase 10.1103/PhysRevLett.104.246402} {\bibfield
  {journal} {\bibinfo  {journal} {Phys. Rev. Lett.}\ }\textbf {\bibinfo
  {volume} {104}},\ \bibinfo {pages} {246402} (\bibinfo {year}
  {2010})}\BibitemShut {NoStop}%
\bibitem [{\citenamefont {Valli}\ \emph {et~al.}(2015)\citenamefont {Valli},
  \citenamefont {Sch\"afer}, \citenamefont {Thunstr\"om}, \citenamefont
  {Rohringer}, \citenamefont {Andergassen}, \citenamefont {Sangiovanni},
  \citenamefont {Held},\ and\ \citenamefont {Toschi}}]{Valli2015}%
  \BibitemOpen
  \bibfield  {author} {\bibinfo {author} {\bibfnamefont {A.}~\bibnamefont
  {Valli}}, \bibinfo {author} {\bibfnamefont {T.}~\bibnamefont {Sch\"afer}},
  \bibinfo {author} {\bibfnamefont {P.}~\bibnamefont {Thunstr\"om}}, \bibinfo
  {author} {\bibfnamefont {G.}~\bibnamefont {Rohringer}}, \bibinfo {author}
  {\bibfnamefont {S.}~\bibnamefont {Andergassen}}, \bibinfo {author}
  {\bibfnamefont {G.}~\bibnamefont {Sangiovanni}}, \bibinfo {author}
  {\bibfnamefont {K.}~\bibnamefont {Held}}, \ and\ \bibinfo {author}
  {\bibfnamefont {A.}~\bibnamefont {Toschi}},\ }\bibfield  {title} {\enquote
  {\bibinfo {title} {{Dynamical vertex approximation in its parquet
  implementation: Application to Hubbard nanorings}},}\ }\href {\doibase
  10.1103/PhysRevB.91.115115} {\bibfield  {journal} {\bibinfo  {journal} {Phys.
  Rev. B}\ }\textbf {\bibinfo {volume} {91}},\ \bibinfo {pages} {115115}
  (\bibinfo {year} {2015})}\BibitemShut {NoStop}%
\bibitem [{\citenamefont {Kauch}\ \emph {et~al.}(2020)\citenamefont {Kauch},
  \citenamefont {Pudleiner}, \citenamefont {Astleithner}, \citenamefont
  {Thunstr\"om}, \citenamefont {Ribic},\ and\ \citenamefont
  {Held}}]{Kauch2020}%
  \BibitemOpen
  \bibfield  {author} {\bibinfo {author} {\bibfnamefont {A.}~\bibnamefont
  {Kauch}}, \bibinfo {author} {\bibfnamefont {P.}~\bibnamefont {Pudleiner}},
  \bibinfo {author} {\bibfnamefont {K.}~\bibnamefont {Astleithner}}, \bibinfo
  {author} {\bibfnamefont {P.}~\bibnamefont {Thunstr\"om}}, \bibinfo {author}
  {\bibfnamefont {T.}~\bibnamefont {Ribic}}, \ and\ \bibinfo {author}
  {\bibfnamefont {K.}~\bibnamefont {Held}},\ }\bibfield  {title} {\enquote
  {\bibinfo {title} {{Generic Optical Excitations of Correlated Systems:
  $\ensuremath{\pi}$-tons}},}\ }\href {\doibase 10.1103/PhysRevLett.124.047401}
  {\bibfield  {journal} {\bibinfo  {journal} {Phys. Rev. Lett.}\ }\textbf
  {\bibinfo {volume} {124}},\ \bibinfo {pages} {047401} (\bibinfo {year}
  {2020})}\BibitemShut {NoStop}%
\bibitem [{\citenamefont {Kauch}\ \emph {et~al.}(2019)\citenamefont {Kauch},
  \citenamefont {Hörbinger}, \citenamefont {Li},\ and\ \citenamefont
  {Held}}]{Kauch2019}%
  \BibitemOpen
  \bibfield  {author} {\bibinfo {author} {\bibfnamefont {Anna}\ \bibnamefont
  {Kauch}}, \bibinfo {author} {\bibfnamefont {Felix}\ \bibnamefont
  {Hörbinger}}, \bibinfo {author} {\bibfnamefont {Gang}\ \bibnamefont {Li}}, \
  and\ \bibinfo {author} {\bibfnamefont {Karsten}\ \bibnamefont {Held}},\
  }\href@noop {} {\enquote {\bibinfo {title} {{Interplay between magnetic and
  superconducting fluctuations in the doped 2d Hubbard model}},}\ } (\bibinfo
  {year} {2019}),\ \Eprint {http://arxiv.org/abs/arXiv:1901.09743}
  {arXiv:1901.09743} \BibitemShut {NoStop}%
\bibitem [{\citenamefont {Sch\"afer}\ \emph
  {et~al.}(2015{\natexlab{b}})\citenamefont {Sch\"afer}, \citenamefont
  {Toschi},\ and\ \citenamefont {Tomczak}}]{Schaefer2015a}%
  \BibitemOpen
  \bibfield  {author} {\bibinfo {author} {\bibfnamefont {T.}~\bibnamefont
  {Sch\"afer}}, \bibinfo {author} {\bibfnamefont {A.}~\bibnamefont {Toschi}}, \
  and\ \bibinfo {author} {\bibfnamefont {Jan~M.}\ \bibnamefont {Tomczak}},\
  }\bibfield  {title} {\enquote {\bibinfo {title} {{Separability of dynamical
  and nonlocal correlations in three dimensions}},}\ }\href {\doibase
  10.1103/PhysRevB.91.121107} {\bibfield  {journal} {\bibinfo  {journal} {Phys.
  Rev. B}\ }\textbf {\bibinfo {volume} {91}},\ \bibinfo {pages} {121107(R)}
  (\bibinfo {year} {2015}{\natexlab{b}})}\BibitemShut {NoStop}%
\bibitem [{\citenamefont {Pudleiner}\ \emph {et~al.}(2016)\citenamefont
  {Pudleiner}, \citenamefont {Sch\"afer}, \citenamefont {Rost}, \citenamefont
  {Li}, \citenamefont {Held},\ and\ \citenamefont {Bl\"umer}}]{Pudleiner2016}%
  \BibitemOpen
  \bibfield  {author} {\bibinfo {author} {\bibfnamefont {P.}~\bibnamefont
  {Pudleiner}}, \bibinfo {author} {\bibfnamefont {T.}~\bibnamefont
  {Sch\"afer}}, \bibinfo {author} {\bibfnamefont {D.}~\bibnamefont {Rost}},
  \bibinfo {author} {\bibfnamefont {G.}~\bibnamefont {Li}}, \bibinfo {author}
  {\bibfnamefont {K.}~\bibnamefont {Held}}, \ and\ \bibinfo {author}
  {\bibfnamefont {N.}~\bibnamefont {Bl\"umer}},\ }\bibfield  {title} {\enquote
  {\bibinfo {title} {{Momentum structure of the self-energy and its
  parametrization for the two-dimensional Hubbard model}},}\ }\href {\doibase
  10.1103/PhysRevB.93.195134} {\bibfield  {journal} {\bibinfo  {journal} {Phys.
  Rev. B}\ }\textbf {\bibinfo {volume} {93}},\ \bibinfo {pages} {195134}
  (\bibinfo {year} {2016})}\BibitemShut {NoStop}%
\bibitem [{\citenamefont {Sch\"afer}\ \emph {et~al.}(2017)\citenamefont
  {Sch\"afer}, \citenamefont {Katanin}, \citenamefont {Held},\ and\
  \citenamefont {Toschi}}]{Schaefer2017}%
  \BibitemOpen
  \bibfield  {author} {\bibinfo {author} {\bibfnamefont {T.}~\bibnamefont
  {Sch\"afer}}, \bibinfo {author} {\bibfnamefont {A.~A.}\ \bibnamefont
  {Katanin}}, \bibinfo {author} {\bibfnamefont {K.}~\bibnamefont {Held}}, \
  and\ \bibinfo {author} {\bibfnamefont {A.}~\bibnamefont {Toschi}},\
  }\bibfield  {title} {\enquote {\bibinfo {title} {{Interplay of Correlations
  and Kohn Anomalies in Three Dimensions: Quantum Criticality with a Twist}},}\
  }\href {\doibase 10.1103/PhysRevLett.119.046402} {\bibfield  {journal}
  {\bibinfo  {journal} {Phys. Rev. Lett.}\ }\textbf {\bibinfo {volume} {119}},\
  \bibinfo {pages} {046402} (\bibinfo {year} {2017})}\BibitemShut {NoStop}%
\bibitem [{\citenamefont {Sch{\"a}fer}\ \emph {et~al.}(2019)\citenamefont
  {Sch{\"a}fer}, \citenamefont {Katanin}, \citenamefont {Kitatani},
  \citenamefont {Toschi},\ and\ \citenamefont {Held}}]{Schaefer2019}%
  \BibitemOpen
  \bibfield  {author} {\bibinfo {author} {\bibfnamefont {T.}~\bibnamefont
  {Sch{\"a}fer}}, \bibinfo {author} {\bibfnamefont {A.~A.}\ \bibnamefont
  {Katanin}}, \bibinfo {author} {\bibfnamefont {M.}~\bibnamefont {Kitatani}},
  \bibinfo {author} {\bibfnamefont {A.}~\bibnamefont {Toschi}}, \ and\ \bibinfo
  {author} {\bibfnamefont {K.}~\bibnamefont {Held}},\ }\bibfield  {title}
  {\enquote {\bibinfo {title} {{Quantum Criticality in the Two-Dimensional
  Periodic Anderson Model}},}\ }\href {\doibase 10.1103/PhysRevLett.122.227201}
  {\bibfield  {journal} {\bibinfo  {journal} {Phys. Rev. Lett.}\ }\textbf
  {\bibinfo {volume} {122}},\ \bibinfo {pages} {227201} (\bibinfo {year}
  {2019})}\BibitemShut {NoStop}%
\bibitem [{\citenamefont {Del~Re}\ \emph {et~al.}(2019)\citenamefont {Del~Re},
  \citenamefont {Capone},\ and\ \citenamefont {Toschi}}]{DelRe2019}%
  \BibitemOpen
  \bibfield  {author} {\bibinfo {author} {\bibfnamefont {Lorenzo}\ \bibnamefont
  {Del~Re}}, \bibinfo {author} {\bibfnamefont {Massimo}\ \bibnamefont
  {Capone}}, \ and\ \bibinfo {author} {\bibfnamefont {Alessandro}\ \bibnamefont
  {Toschi}},\ }\bibfield  {title} {\enquote {\bibinfo {title} {{Dynamical
  vertex approximation for the attractive Hubbard model}},}\ }\href {\doibase
  10.1103/PhysRevB.99.045137} {\bibfield  {journal} {\bibinfo  {journal} {Phys.
  Rev. B}\ }\textbf {\bibinfo {volume} {99}},\ \bibinfo {pages} {045137}
  (\bibinfo {year} {2019})}\BibitemShut {NoStop}%
\bibitem [{\citenamefont {Klebel-Knobloch}\ \emph {et~al.}(2021)\citenamefont
  {Klebel-Knobloch}, \citenamefont {Sch\"afer}, \citenamefont {Toschi},\ and\
  \citenamefont {Tomczak}}]{Klebel2021}%
  \BibitemOpen
  \bibfield  {author} {\bibinfo {author} {\bibfnamefont {B.}~\bibnamefont
  {Klebel-Knobloch}}, \bibinfo {author} {\bibfnamefont {T.}~\bibnamefont
  {Sch\"afer}}, \bibinfo {author} {\bibfnamefont {A.}~\bibnamefont {Toschi}}, \
  and\ \bibinfo {author} {\bibfnamefont {J.~M.}\ \bibnamefont {Tomczak}},\
  }\bibfield  {title} {\enquote {\bibinfo {title} {{Anisotropy of electronic
  correlations: On the applicability of local theories to layered
  materials}},}\ }\href {\doibase 10.1103/PhysRevB.103.045121} {\bibfield
  {journal} {\bibinfo  {journal} {Phys. Rev. B}\ }\textbf {\bibinfo {volume}
  {103}},\ \bibinfo {pages} {045121} (\bibinfo {year} {2021})}\BibitemShut
  {NoStop}%
\bibitem [{\citenamefont {Sch{\"a}fer}(2016)}]{SchaeferThesis}%
  \BibitemOpen
  \bibfield  {author} {\bibinfo {author} {\bibfnamefont {Thomas}\ \bibnamefont
  {Sch{\"a}fer}},\ }\emph {\bibinfo {title} {Classical and quantum phase
  transitions in strongly correlated electron systems}},\ \href
  {http://digital.obvsg.at/download/pdf/1492232} {Ph.D. thesis},\ \bibinfo
  {school} {TU Wien} (\bibinfo {year} {2016})\BibitemShut {NoStop}%
\bibitem [{\citenamefont {Rohringer}(2013)}]{Rohringer2013a}%
  \BibitemOpen
  \bibfield  {author} {\bibinfo {author} {\bibfnamefont {Georg}\ \bibnamefont
  {Rohringer}},\ }\emph {\bibinfo {title} {New routes towards a theoretical
  treatment of nonlocal electronic correlations}},\ \href
  {http://digital.obvsg.at/download/pdf/1631831} {Ph.D. thesis},\ \bibinfo
  {school} {Vienna University of Technology} (\bibinfo {year}
  {2013})\BibitemShut {NoStop}%
\bibitem [{\citenamefont {Ayral}\ and\ \citenamefont
  {Parcollet}(2015)}]{Ayral2015}%
  \BibitemOpen
  \bibfield  {author} {\bibinfo {author} {\bibfnamefont {Thomas}\ \bibnamefont
  {Ayral}}\ and\ \bibinfo {author} {\bibfnamefont {Olivier}\ \bibnamefont
  {Parcollet}},\ }\bibfield  {title} {\enquote {\bibinfo {title} {{Mott physics
  and spin fluctuations: a unified framework}},}\ }\href
  {http://link.aps.org/doi/10.1103/PhysRevB.92.115109} {\bibfield  {journal}
  {\bibinfo  {journal} {Phys Rev. B}\ }\textbf {\bibinfo {volume} {92}},\
  \bibinfo {pages} {115109} (\bibinfo {year} {2015})}\BibitemShut {NoStop}%
\bibitem [{\citenamefont {Ayral}\ and\ \citenamefont
  {Parcollet}(2016)}]{Ayral2016a}%
  \BibitemOpen
  \bibfield  {author} {\bibinfo {author} {\bibfnamefont {Thomas}\ \bibnamefont
  {Ayral}}\ and\ \bibinfo {author} {\bibfnamefont {Olivier}\ \bibnamefont
  {Parcollet}},\ }\bibfield  {title} {\enquote {\bibinfo {title} {{Mott physics
  and spin fluctuations: A functional viewpoint}},}\ }\href {\doibase
  10.1103/PhysRevB.93.235124} {\bibfield  {journal} {\bibinfo  {journal} {Phys.
  Rev. B}\ }\textbf {\bibinfo {volume} {93}},\ \bibinfo {pages} {235124}
  (\bibinfo {year} {2016})}\BibitemShut {NoStop}%
\bibitem [{\citenamefont {Ayral}\ \emph {et~al.}(2017)\citenamefont {Ayral},
  \citenamefont {Vu\ifmmode \check{c}\else \v{c}\fi{}i\ifmmode \check{c}\else
  \v{c}\fi{}evi\ifmmode~\acute{c}\else \'{c}\fi{}},\ and\ \citenamefont
  {Parcollet}}]{Ayral2017}%
  \BibitemOpen
  \bibfield  {author} {\bibinfo {author} {\bibfnamefont {Thomas}\ \bibnamefont
  {Ayral}}, \bibinfo {author} {\bibfnamefont {Jaksa}\ \bibnamefont {Vu\ifmmode
  \check{c}\else \v{c}\fi{}i\ifmmode \check{c}\else
  \v{c}\fi{}evi\ifmmode~\acute{c}\else \'{c}\fi{}}}, \ and\ \bibinfo {author}
  {\bibfnamefont {Olivier}\ \bibnamefont {Parcollet}},\ }\bibfield  {title}
  {\enquote {\bibinfo {title} {{Fierz Convergence Criterion: A Controlled
  Approach to Strongly Interacting Systems with Small Embedded Clusters}},}\
  }\href {\doibase 10.1103/PhysRevLett.119.166401} {\bibfield  {journal}
  {\bibinfo  {journal} {Phys. Rev. Lett.}\ }\textbf {\bibinfo {volume} {119}},\
  \bibinfo {pages} {166401} (\bibinfo {year} {2017})}\BibitemShut {NoStop}%
\bibitem [{\citenamefont {Vu\v{c}i\v{c}evi\'c}\ \emph
  {et~al.}(2017)\citenamefont {Vu\v{c}i\v{c}evi\'c}, \citenamefont {Ayral},\
  and\ \citenamefont {Parcollet}}]{Vucicevic2017}%
  \BibitemOpen
  \bibfield  {author} {\bibinfo {author} {\bibfnamefont {J.}~\bibnamefont
  {Vu\v{c}i\v{c}evi\'c}}, \bibinfo {author} {\bibfnamefont {T.}~\bibnamefont
  {Ayral}}, \ and\ \bibinfo {author} {\bibfnamefont {O.}~\bibnamefont
  {Parcollet}},\ }\bibfield  {title} {\enquote {\bibinfo {title} {{TRILEX and
  $GW$+EDMFT approach to $d$-wave superconductivity in the Hubbard model}},}\
  }\href {\doibase 10.1103/PhysRevB.96.104504} {\bibfield  {journal} {\bibinfo
  {journal} {Phys. Rev. B}\ }\textbf {\bibinfo {volume} {96}},\ \bibinfo
  {pages} {104504} (\bibinfo {year} {2017})}\BibitemShut {NoStop}%
\bibitem [{\citenamefont {Stepanov}\ \emph
  {et~al.}(2016{\natexlab{a}})\citenamefont {Stepanov}, \citenamefont {Huber},
  \citenamefont {van Loon}, \citenamefont {Lichtenstein},\ and\ \citenamefont
  {Katsnelson}}]{Stepanov2016a}%
  \BibitemOpen
  \bibfield  {author} {\bibinfo {author} {\bibfnamefont {E.~A.}\ \bibnamefont
  {Stepanov}}, \bibinfo {author} {\bibfnamefont {A.}~\bibnamefont {Huber}},
  \bibinfo {author} {\bibfnamefont {E.~G. C.~P.}\ \bibnamefont {van Loon}},
  \bibinfo {author} {\bibfnamefont {A.~I.}\ \bibnamefont {Lichtenstein}}, \
  and\ \bibinfo {author} {\bibfnamefont {M.~I.}\ \bibnamefont {Katsnelson}},\
  }\bibfield  {title} {\enquote {\bibinfo {title} {{From local to nonlocal
  correlations: The Dual Boson perspective}},}\ }\href {\doibase
  10.1103/PhysRevB.94.205110} {\bibfield  {journal} {\bibinfo  {journal} {Phys.
  Rev. B}\ }\textbf {\bibinfo {volume} {94}},\ \bibinfo {pages} {205110}
  (\bibinfo {year} {2016}{\natexlab{a}})}\BibitemShut {NoStop}%
\bibitem [{\citenamefont {Stepanov}\ \emph
  {et~al.}(2019{\natexlab{a}})\citenamefont {Stepanov}, \citenamefont
  {Harkov},\ and\ \citenamefont {Lichtenstein}}]{Stepanov2019}%
  \BibitemOpen
  \bibfield  {author} {\bibinfo {author} {\bibfnamefont {E.~A.}\ \bibnamefont
  {Stepanov}}, \bibinfo {author} {\bibfnamefont {V.}~\bibnamefont {Harkov}}, \
  and\ \bibinfo {author} {\bibfnamefont {A.~I.}\ \bibnamefont {Lichtenstein}},\
  }\bibfield  {title} {\enquote {\bibinfo {title} {{Consistent partial
  bosonization of the extended Hubbard model}},}\ }\href {\doibase
  10.1103/PhysRevB.100.205115} {\bibfield  {journal} {\bibinfo  {journal}
  {Phys. Rev. B}\ }\textbf {\bibinfo {volume} {100}},\ \bibinfo {pages}
  {205115} (\bibinfo {year} {2019}{\natexlab{a}})}\BibitemShut {NoStop}%
\bibitem [{\citenamefont {Krien}(2019)}]{Krien2019}%
  \BibitemOpen
  \bibfield  {author} {\bibinfo {author} {\bibfnamefont {Friedrich}\
  \bibnamefont {Krien}},\ }\bibfield  {title} {\enquote {\bibinfo {title}
  {Efficient evaluation of the polarization function in dynamical mean-field
  theory},}\ }\href {\doibase 10.1103/PhysRevB.99.235106} {\bibfield  {journal}
  {\bibinfo  {journal} {Phys. Rev. B}\ }\textbf {\bibinfo {volume} {99}},\
  \bibinfo {pages} {235106} (\bibinfo {year} {2019})}\BibitemShut {NoStop}%
\bibitem [{\citenamefont {Krien}\ \emph {et~al.}(2017)\citenamefont {Krien},
  \citenamefont {van Loon}, \citenamefont {Hafermann}, \citenamefont {Otsuki},
  \citenamefont {Katsnelson},\ and\ \citenamefont {Lichtenstein}}]{Krien2017}%
  \BibitemOpen
  \bibfield  {author} {\bibinfo {author} {\bibfnamefont {Friedrich}\
  \bibnamefont {Krien}}, \bibinfo {author} {\bibfnamefont {Erik G. C.~P.}\
  \bibnamefont {van Loon}}, \bibinfo {author} {\bibfnamefont {Hartmut}\
  \bibnamefont {Hafermann}}, \bibinfo {author} {\bibfnamefont {Junya}\
  \bibnamefont {Otsuki}}, \bibinfo {author} {\bibfnamefont {Mikhail~I.}\
  \bibnamefont {Katsnelson}}, \ and\ \bibinfo {author} {\bibfnamefont
  {Alexander~I.}\ \bibnamefont {Lichtenstein}},\ }\bibfield  {title} {\enquote
  {\bibinfo {title} {Conservation in two-particle self-consistent extensions of
  dynamical mean-field theory},}\ }\href {\doibase 10.1103/PhysRevB.96.075155}
  {\bibfield  {journal} {\bibinfo  {journal} {Phys. Rev. B}\ }\textbf {\bibinfo
  {volume} {96}},\ \bibinfo {pages} {075155} (\bibinfo {year}
  {2017})}\BibitemShut {NoStop}%
\bibitem [{\citenamefont {Rubtsov}\ \emph {et~al.}(2008)\citenamefont
  {Rubtsov}, \citenamefont {Katsnelson},\ and\ \citenamefont
  {Lichtenstein}}]{Rubtsov2008}%
  \BibitemOpen
  \bibfield  {author} {\bibinfo {author} {\bibfnamefont {A.~N.}\ \bibnamefont
  {Rubtsov}}, \bibinfo {author} {\bibfnamefont {M.~I.}\ \bibnamefont
  {Katsnelson}}, \ and\ \bibinfo {author} {\bibfnamefont {A.~I.}\ \bibnamefont
  {Lichtenstein}},\ }\bibfield  {title} {\enquote {\bibinfo {title} {{Dual
  fermion approach to nonlocal correlations in the Hubbard model}},}\ }\href
  {\doibase 10.1103/PhysRevB.77.033101} {\bibfield  {journal} {\bibinfo
  {journal} {Phys. Rev. B}\ }\textbf {\bibinfo {volume} {77}},\ \bibinfo
  {pages} {033101} (\bibinfo {year} {2008})}\BibitemShut {NoStop}%
\bibitem [{\citenamefont {Rubtsov}\ \emph {et~al.}(2009)\citenamefont
  {Rubtsov}, \citenamefont {Katsnelson}, \citenamefont {Lichtenstein},\ and\
  \citenamefont {Georges}}]{Rubtsov2009}%
  \BibitemOpen
  \bibfield  {author} {\bibinfo {author} {\bibfnamefont {A.~N.}\ \bibnamefont
  {Rubtsov}}, \bibinfo {author} {\bibfnamefont {M.~I.}\ \bibnamefont
  {Katsnelson}}, \bibinfo {author} {\bibfnamefont {A.~I.}\ \bibnamefont
  {Lichtenstein}}, \ and\ \bibinfo {author} {\bibfnamefont {A.}~\bibnamefont
  {Georges}},\ }\bibfield  {title} {\enquote {\bibinfo {title} {{Dual fermion
  approach to the two-dimensional Hubbard model: Antiferromagnetic fluctuations
  and Fermi arcs}},}\ }\href {\doibase 10.1103/PhysRevB.79.045133} {\bibfield
  {journal} {\bibinfo  {journal} {Phys. Rev. B}\ }\textbf {\bibinfo {volume}
  {79}},\ \bibinfo {pages} {045133} (\bibinfo {year} {2009})}\BibitemShut
  {NoStop}%
\bibitem [{\citenamefont {Hafermann}\ \emph {et~al.}(2009)\citenamefont
  {Hafermann}, \citenamefont {Li}, \citenamefont {Rubtsov}, \citenamefont
  {Katsnelson}, \citenamefont {Lichtenstein},\ and\ \citenamefont
  {Monien}}]{Hafermann2009}%
  \BibitemOpen
  \bibfield  {author} {\bibinfo {author} {\bibfnamefont {H.}~\bibnamefont
  {Hafermann}}, \bibinfo {author} {\bibfnamefont {G.}~\bibnamefont {Li}},
  \bibinfo {author} {\bibfnamefont {A.~N.}\ \bibnamefont {Rubtsov}}, \bibinfo
  {author} {\bibfnamefont {M.~I.}\ \bibnamefont {Katsnelson}}, \bibinfo
  {author} {\bibfnamefont {A.~I.}\ \bibnamefont {Lichtenstein}}, \ and\
  \bibinfo {author} {\bibfnamefont {H.}~\bibnamefont {Monien}},\ }\bibfield
  {title} {\enquote {\bibinfo {title} {{Efficient Perturbation Theory for
  Quantum Lattice Models}},}\ }\href {\doibase 10.1103/PhysRevLett.102.206401}
  {\bibfield  {journal} {\bibinfo  {journal} {Phys. Rev. Lett.}\ }\textbf
  {\bibinfo {volume} {102}},\ \bibinfo {pages} {206401} (\bibinfo {year}
  {2009})}\BibitemShut {NoStop}%
\bibitem [{\citenamefont {Antipov}\ \emph {et~al.}(2015)\citenamefont
  {Antipov}, \citenamefont {LeBlanc},\ and\ \citenamefont {Gull}}]{Antipov15}%
  \BibitemOpen
  \bibfield  {author} {\bibinfo {author} {\bibfnamefont {Andrey~E.}\
  \bibnamefont {Antipov}}, \bibinfo {author} {\bibfnamefont {James~P.F.}\
  \bibnamefont {LeBlanc}}, \ and\ \bibinfo {author} {\bibfnamefont {Emanuel}\
  \bibnamefont {Gull}},\ }\bibfield  {title} {\enquote {\bibinfo {title}
  {Opendf - an implementation of the dual fermion method for strongly
  correlated systems},}\ }\href {\doibase
  http://dx.doi.org/10.1016/j.phpro.2015.07.107} {\bibfield  {journal}
  {\bibinfo  {journal} {Physics Procedia}\ }\textbf {\bibinfo {volume} {68}},\
  \bibinfo {pages} {43 -- 51} (\bibinfo {year} {2015})},\ \bibinfo {note}
  {proceedings of the 28th Workshop on Computer Simulation Studies in Condensed
  Matter Physics (CSP2015)}\BibitemShut {NoStop}%
\bibitem [{\citenamefont {Sarker}(1988)}]{Sarker1988}%
  \BibitemOpen
  \bibfield  {author} {\bibinfo {author} {\bibfnamefont {S~K}\ \bibnamefont
  {Sarker}},\ }\bibfield  {title} {\enquote {\bibinfo {title} {{A new
  functional integral formalism for strongly correlated Fermi systems}},}\
  }\href {http://stacks.iop.org/0022-3719/21/i=18/a=002} {\bibfield  {journal}
  {\bibinfo  {journal} {J. Phys. C: Solid State Physics}\ }\textbf {\bibinfo
  {volume} {21}},\ \bibinfo {pages} {L667} (\bibinfo {year}
  {1988})}\BibitemShut {NoStop}%
\bibitem [{\citenamefont {Pairault}\ \emph {et~al.}(2000)\citenamefont
  {Pairault}, \citenamefont {S\'en\'echal},\ and\ \citenamefont
  {Tremblay}}]{Pairault2000}%
  \BibitemOpen
  \bibfield  {author} {\bibinfo {author} {\bibfnamefont {S.}~\bibnamefont
  {Pairault}}, \bibinfo {author} {\bibfnamefont {D.}~\bibnamefont
  {S\'en\'echal}}, \ and\ \bibinfo {author} {\bibfnamefont {A.-M.~S.}\
  \bibnamefont {Tremblay}},\ }\bibfield  {title} {\enquote {\bibinfo {title}
  {{Strong-coupling perturbation theory of the Hubbard model}},}\ }\href
  {http://dx.doi.org/10.1007/s100510070253} {\bibfield  {journal} {\bibinfo
  {journal} {Eur. Phys. J. B}\ }\textbf {\bibinfo {volume} {16}},\ \bibinfo
  {pages} {85} (\bibinfo {year} {2000})}\BibitemShut {NoStop}%
\bibitem [{\citenamefont {Bourbonnais}(1985)}]{Bourbonnais1985}%
  \BibitemOpen
  \bibfield  {author} {\bibinfo {author} {\bibfnamefont {Claude}\ \bibnamefont
  {Bourbonnais}},\ }\emph {\bibinfo {title} {{Fluctuations Quantiques dans les
  Syst{`e}mes a Basse Dimensionnalit{\'e}: Th{\'e}orie et Applications aux
  Conducteurs Organiques}}},\ \href
  {https://savoirs.usherbrooke.ca/handle/11143/15104} {Ph.D. thesis},\ \bibinfo
   {school} {Universit{\'e} de Sherbrooke} (\bibinfo {year} {1985})\BibitemShut
  {NoStop}%
\bibitem [{\citenamefont {Rohringer}\ \emph {et~al.}(2012)\citenamefont
  {Rohringer}, \citenamefont {Valli},\ and\ \citenamefont
  {Toschi}}]{Rohringer2012}%
  \BibitemOpen
  \bibfield  {author} {\bibinfo {author} {\bibfnamefont {G.}~\bibnamefont
  {Rohringer}}, \bibinfo {author} {\bibfnamefont {A.}~\bibnamefont {Valli}}, \
  and\ \bibinfo {author} {\bibfnamefont {A.}~\bibnamefont {Toschi}},\
  }\bibfield  {title} {\enquote {\bibinfo {title} {Local electronic correlation
  at the two-particle level},}\ }\href {\doibase 10.1103/PhysRevB.86.125114}
  {\bibfield  {journal} {\bibinfo  {journal} {Phys. Rev. B}\ }\textbf {\bibinfo
  {volume} {86}},\ \bibinfo {pages} {125114} (\bibinfo {year}
  {2012})}\BibitemShut {NoStop}%
\bibitem [{\citenamefont {Ribic}\ \emph {et~al.}(2017)\citenamefont {Ribic},
  \citenamefont {Gunacker}, \citenamefont {Iskakov}, \citenamefont
  {Wallerberger}, \citenamefont {Rohringer}, \citenamefont {Rubtsov},
  \citenamefont {Gull},\ and\ \citenamefont {Held}}]{Ribic2017b}%
  \BibitemOpen
  \bibfield  {author} {\bibinfo {author} {\bibfnamefont {T.}~\bibnamefont
  {Ribic}}, \bibinfo {author} {\bibfnamefont {P.}~\bibnamefont {Gunacker}},
  \bibinfo {author} {\bibfnamefont {S.}~\bibnamefont {Iskakov}}, \bibinfo
  {author} {\bibfnamefont {M.}~\bibnamefont {Wallerberger}}, \bibinfo {author}
  {\bibfnamefont {G.}~\bibnamefont {Rohringer}}, \bibinfo {author}
  {\bibfnamefont {A.~N.}\ \bibnamefont {Rubtsov}}, \bibinfo {author}
  {\bibfnamefont {E.}~\bibnamefont {Gull}}, \ and\ \bibinfo {author}
  {\bibfnamefont {K.}~\bibnamefont {Held}},\ }\bibfield  {title} {\enquote
  {\bibinfo {title} {{Role of three-particle vertex within dual fermion
  calculations}},}\ }\href {\doibase 10.1103/PhysRevB.96.235127} {\bibfield
  {journal} {\bibinfo  {journal} {Phys. Rev. B}\ }\textbf {\bibinfo {volume}
  {96}},\ \bibinfo {pages} {235127} (\bibinfo {year} {2017})}\BibitemShut
  {NoStop}%
\bibitem [{\citenamefont {Iskakov}\ \emph {et~al.}(2016)\citenamefont
  {Iskakov}, \citenamefont {Antipov},\ and\ \citenamefont
  {Gull}}]{Iskakov2016}%
  \BibitemOpen
  \bibfield  {author} {\bibinfo {author} {\bibfnamefont {Sergei}\ \bibnamefont
  {Iskakov}}, \bibinfo {author} {\bibfnamefont {Andrey~E.}\ \bibnamefont
  {Antipov}}, \ and\ \bibinfo {author} {\bibfnamefont {Emanuel}\ \bibnamefont
  {Gull}},\ }\bibfield  {title} {\enquote {\bibinfo {title} {{Diagrammatic
  Monte Carlo for dual fermions}},}\ }\href {\doibase
  10.1103/PhysRevB.94.035102} {\bibfield  {journal} {\bibinfo  {journal} {Phys.
  Rev. B}\ }\textbf {\bibinfo {volume} {94}},\ \bibinfo {pages} {035102}
  (\bibinfo {year} {2016})}\BibitemShut {NoStop}%
\bibitem [{\citenamefont {Gukelberger}\ \emph {et~al.}(2017)\citenamefont
  {Gukelberger}, \citenamefont {Kozik},\ and\ \citenamefont
  {Hafermann}}]{Gukelberger2017}%
  \BibitemOpen
  \bibfield  {author} {\bibinfo {author} {\bibfnamefont {Jan}\ \bibnamefont
  {Gukelberger}}, \bibinfo {author} {\bibfnamefont {Evgeny}\ \bibnamefont
  {Kozik}}, \ and\ \bibinfo {author} {\bibfnamefont {Hartmut}\ \bibnamefont
  {Hafermann}},\ }\bibfield  {title} {\enquote {\bibinfo {title} {{Diagrammatic
  Monte Carlo approach for diagrammatic extensions of dynamical mean-field
  theory: Convergence analysis of the dual fermion technique}},}\ }\href
  {\doibase 10.1103/PhysRevB.96.035152} {\bibfield  {journal} {\bibinfo
  {journal} {Phys. Rev. B}\ }\textbf {\bibinfo {volume} {96}},\ \bibinfo
  {pages} {035152} (\bibinfo {year} {2017})}\BibitemShut {NoStop}%
\bibitem [{\citenamefont {LeBlanc}\ \emph {et~al.}(2019)\citenamefont
  {LeBlanc}, \citenamefont {Li}, \citenamefont {Chen}, \citenamefont {Levy},
  \citenamefont {Antipov}, \citenamefont {Millis},\ and\ \citenamefont
  {Gull}}]{Leblanc2019}%
  \BibitemOpen
  \bibfield  {author} {\bibinfo {author} {\bibfnamefont {J.~P.~F.}\
  \bibnamefont {LeBlanc}}, \bibinfo {author} {\bibfnamefont {Shaozhi}\
  \bibnamefont {Li}}, \bibinfo {author} {\bibfnamefont {Xi}~\bibnamefont
  {Chen}}, \bibinfo {author} {\bibfnamefont {Ryan}\ \bibnamefont {Levy}},
  \bibinfo {author} {\bibfnamefont {A.~E.}\ \bibnamefont {Antipov}}, \bibinfo
  {author} {\bibfnamefont {Andrew~J.}\ \bibnamefont {Millis}}, \ and\ \bibinfo
  {author} {\bibfnamefont {Emanuel}\ \bibnamefont {Gull}},\ }\bibfield  {title}
  {\enquote {\bibinfo {title} {{Magnetic susceptibility and simulated neutron
  signal in the two-dimensional Hubbard model}},}\ }\href {\doibase
  10.1103/PhysRevB.100.075123} {\bibfield  {journal} {\bibinfo  {journal}
  {Phys. Rev. B}\ }\textbf {\bibinfo {volume} {100}},\ \bibinfo {pages}
  {075123} (\bibinfo {year} {2019})}\BibitemShut {NoStop}%
\bibitem [{\citenamefont {van Loon}\ \emph {et~al.}(2018)\citenamefont {van
  Loon}, \citenamefont {Hafermann},\ and\ \citenamefont
  {Katsnelson}}]{vanLoon2017}%
  \BibitemOpen
  \bibfield  {author} {\bibinfo {author} {\bibfnamefont {Erik G. C.~P.}\
  \bibnamefont {van Loon}}, \bibinfo {author} {\bibfnamefont {Hartmut}\
  \bibnamefont {Hafermann}}, \ and\ \bibinfo {author} {\bibfnamefont
  {Mikhail~I.}\ \bibnamefont {Katsnelson}},\ }\bibfield  {title} {\enquote
  {\bibinfo {title} {{Precursors of the insulating state in the square lattice
  Hubbard model}},}\ }\href {\doibase
  https://doi.org/10.1103/PhysRevB.97.085125} {\bibfield  {journal} {\bibinfo
  {journal} {Phys Rev. B}\ }\textbf {\bibinfo {volume} {97}},\ \bibinfo {pages}
  {085125} (\bibinfo {year} {2018})}\BibitemShut {NoStop}%
\bibitem [{\citenamefont {Astretsov}\ \emph {et~al.}(2020)\citenamefont
  {Astretsov}, \citenamefont {Rohringer},\ and\ \citenamefont
  {Rubtsov}}]{Astretsov2020}%
  \BibitemOpen
  \bibfield  {author} {\bibinfo {author} {\bibfnamefont {Grigory~V.}\
  \bibnamefont {Astretsov}}, \bibinfo {author} {\bibfnamefont {Georg}\
  \bibnamefont {Rohringer}}, \ and\ \bibinfo {author} {\bibfnamefont
  {Alexey~N.}\ \bibnamefont {Rubtsov}},\ }\bibfield  {title} {\enquote
  {\bibinfo {title} {{Dual parquet scheme for the two-dimensional Hubbard
  model: Modeling low-energy physics of high-${T}_{c}$ cuprates with high
  momentum resolution}},}\ }\href {\doibase 10.1103/PhysRevB.101.075109}
  {\bibfield  {journal} {\bibinfo  {journal} {Phys. Rev. B}\ }\textbf {\bibinfo
  {volume} {101}},\ \bibinfo {pages} {075109} (\bibinfo {year}
  {2020})}\BibitemShut {NoStop}%
\bibitem [{\citenamefont {Krien}\ \emph {et~al.}(2020)\citenamefont {Krien},
  \citenamefont {Valli}, \citenamefont {Chalupa}, \citenamefont {Capone},
  \citenamefont {Lichtenstein},\ and\ \citenamefont {Toschi}}]{Krien2020}%
  \BibitemOpen
  \bibfield  {author} {\bibinfo {author} {\bibfnamefont {Friedrich}\
  \bibnamefont {Krien}}, \bibinfo {author} {\bibfnamefont {Angelo}\
  \bibnamefont {Valli}}, \bibinfo {author} {\bibfnamefont {Patrick}\
  \bibnamefont {Chalupa}}, \bibinfo {author} {\bibfnamefont {Massimo}\
  \bibnamefont {Capone}}, \bibinfo {author} {\bibfnamefont {Alexander~I.}\
  \bibnamefont {Lichtenstein}}, \ and\ \bibinfo {author} {\bibfnamefont
  {Alessandro}\ \bibnamefont {Toschi}},\ }\bibfield  {title} {\enquote
  {\bibinfo {title} {Boson-exchange parquet solver for dual fermions},}\ }\href
  {\doibase 10.1103/physrevb.102.195131} {\bibfield  {journal} {\bibinfo
  {journal} {Physical Review B}\ }\textbf {\bibinfo {volume} {102}} (\bibinfo
  {year} {2020}),\ 10.1103/physrevb.102.195131}\BibitemShut {NoStop}%
\bibitem [{\citenamefont {Astleithner}\ \emph {et~al.}(2020)\citenamefont
  {Astleithner}, \citenamefont {Kauch}, \citenamefont {Ribic},\ and\
  \citenamefont {Held}}]{Astleithner2020}%
  \BibitemOpen
  \bibfield  {author} {\bibinfo {author} {\bibfnamefont {K.}~\bibnamefont
  {Astleithner}}, \bibinfo {author} {\bibfnamefont {A.}~\bibnamefont {Kauch}},
  \bibinfo {author} {\bibfnamefont {T.}~\bibnamefont {Ribic}}, \ and\ \bibinfo
  {author} {\bibfnamefont {K.}~\bibnamefont {Held}},\ }\bibfield  {title}
  {\enquote {\bibinfo {title} {{Parquet dual fermion approach for the
  Falicov-Kimball model}},}\ }\href {\doibase 10.1103/PhysRevB.101.165101}
  {\bibfield  {journal} {\bibinfo  {journal} {Phys. Rev. B}\ }\textbf {\bibinfo
  {volume} {101}},\ \bibinfo {pages} {165101} (\bibinfo {year}
  {2020})}\BibitemShut {NoStop}%
\bibitem [{\citenamefont {Rubtsov}\ \emph {et~al.}(2012)\citenamefont
  {Rubtsov}, \citenamefont {Katsnelson},\ and\ \citenamefont
  {Lichtenstein}}]{Rubtsov2012}%
  \BibitemOpen
  \bibfield  {author} {\bibinfo {author} {\bibfnamefont {A.N.}\ \bibnamefont
  {Rubtsov}}, \bibinfo {author} {\bibfnamefont {M.I.}\ \bibnamefont
  {Katsnelson}}, \ and\ \bibinfo {author} {\bibfnamefont {A.I.}\ \bibnamefont
  {Lichtenstein}},\ }\bibfield  {title} {\enquote {\bibinfo {title} {{Dual
  boson approach to collective excitations in correlated fermionic systems}},}\
  }\href {\doibase https://doi.org/10.1016/j.aop.2012.01.002} {\bibfield
  {journal} {\bibinfo  {journal} {Annals of Physics}\ }\textbf {\bibinfo
  {volume} {327}},\ \bibinfo {pages} {1320 -- 1335} (\bibinfo {year}
  {2012})}\BibitemShut {NoStop}%
\bibitem [{\citenamefont {van Loon}\ \emph
  {et~al.}(2014{\natexlab{a}})\citenamefont {van Loon}, \citenamefont
  {Lichtenstein}, \citenamefont {Katsnelson}, \citenamefont {Parcollet},\ and\
  \citenamefont {Hafermann}}]{vanLoon2014a}%
  \BibitemOpen
  \bibfield  {author} {\bibinfo {author} {\bibfnamefont {Erik G. C.~P.}\
  \bibnamefont {van Loon}}, \bibinfo {author} {\bibfnamefont {Alexander~I.}\
  \bibnamefont {Lichtenstein}}, \bibinfo {author} {\bibfnamefont {Mikhail~I.}\
  \bibnamefont {Katsnelson}}, \bibinfo {author} {\bibfnamefont {Olivier}\
  \bibnamefont {Parcollet}}, \ and\ \bibinfo {author} {\bibfnamefont {Hartmut}\
  \bibnamefont {Hafermann}},\ }\bibfield  {title} {\enquote {\bibinfo {title}
  {{Beyond extended dynamical mean-field theory: Dual boson approach to the
  two-dimensional extended Hubbard model}},}\ }\href {\doibase
  10.1103/PhysRevB.90.235135} {\bibfield  {journal} {\bibinfo  {journal} {Phys.
  Rev. B}\ }\textbf {\bibinfo {volume} {90}},\ \bibinfo {pages} {235135}
  (\bibinfo {year} {2014}{\natexlab{a}})}\BibitemShut {NoStop}%
\bibitem [{\citenamefont {Stepanov}\ \emph
  {et~al.}(2016{\natexlab{b}})\citenamefont {Stepanov}, \citenamefont {van
  Loon}, \citenamefont {Katanin}, \citenamefont {Lichtenstein}, \citenamefont
  {Katsnelson},\ and\ \citenamefont {Rubtsov}}]{Stepanov2016}%
  \BibitemOpen
  \bibfield  {author} {\bibinfo {author} {\bibfnamefont {E.~A.}\ \bibnamefont
  {Stepanov}}, \bibinfo {author} {\bibfnamefont {E.~G. C.~P.}\ \bibnamefont
  {van Loon}}, \bibinfo {author} {\bibfnamefont {A.~A.}\ \bibnamefont
  {Katanin}}, \bibinfo {author} {\bibfnamefont {A.~I.}\ \bibnamefont
  {Lichtenstein}}, \bibinfo {author} {\bibfnamefont {M.~I.}\ \bibnamefont
  {Katsnelson}}, \ and\ \bibinfo {author} {\bibfnamefont {A.~N.}\ \bibnamefont
  {Rubtsov}},\ }\bibfield  {title} {\enquote {\bibinfo {title} {Self-consistent
  dual boson approach to single-particle and collective excitations in
  correlated systems},}\ }\href {\doibase 10.1103/PhysRevB.93.045107}
  {\bibfield  {journal} {\bibinfo  {journal} {Phys. Rev. B}\ }\textbf {\bibinfo
  {volume} {93}},\ \bibinfo {pages} {045107} (\bibinfo {year}
  {2016}{\natexlab{b}})}\BibitemShut {NoStop}%
\bibitem [{\citenamefont {Sun}\ and\ \citenamefont {Kotliar}(2002)}]{Sun2002}%
  \BibitemOpen
  \bibfield  {author} {\bibinfo {author} {\bibfnamefont {Ping}\ \bibnamefont
  {Sun}}\ and\ \bibinfo {author} {\bibfnamefont {Gabriel}\ \bibnamefont
  {Kotliar}},\ }\bibfield  {title} {\enquote {\bibinfo {title} {Extended
  dynamical mean-field theory and $\mathrm{GW}$ method},}\ }\href {\doibase
  10.1103/PhysRevB.66.085120} {\bibfield  {journal} {\bibinfo  {journal} {Phys.
  Rev. B}\ }\textbf {\bibinfo {volume} {66}},\ \bibinfo {pages} {085120}
  (\bibinfo {year} {2002})}\BibitemShut {NoStop}%
\bibitem [{\citenamefont {Sun}\ and\ \citenamefont {Kotliar}(2004)}]{Sun2004}%
  \BibitemOpen
  \bibfield  {author} {\bibinfo {author} {\bibfnamefont {P.}~\bibnamefont
  {Sun}}\ and\ \bibinfo {author} {\bibfnamefont {G.}~\bibnamefont {Kotliar}},\
  }\bibfield  {title} {\enquote {\bibinfo {title} {{Many-Body Approximation
  Scheme beyond GW}},}\ }\href
  {http://dx.doi.org/10.1103/PhysRevLett.92.196402} {\bibfield  {journal}
  {\bibinfo  {journal} {Phys. Rev. Lett.}\ }\textbf {\bibinfo {volume} {92}},\
  \bibinfo {pages} {196402} (\bibinfo {year} {2004})}\BibitemShut {NoStop}%
\bibitem [{\citenamefont {Peters}\ \emph {et~al.}(2019)\citenamefont {Peters},
  \citenamefont {van Loon}, \citenamefont {Rubtsov}, \citenamefont
  {Lichtenstein}, \citenamefont {Katsnelson},\ and\ \citenamefont
  {Stepanov}}]{Peters2019}%
  \BibitemOpen
  \bibfield  {author} {\bibinfo {author} {\bibfnamefont {L.}~\bibnamefont
  {Peters}}, \bibinfo {author} {\bibfnamefont {E.~G. C.~P.}\ \bibnamefont {van
  Loon}}, \bibinfo {author} {\bibfnamefont {A.~N.}\ \bibnamefont {Rubtsov}},
  \bibinfo {author} {\bibfnamefont {A.~I.}\ \bibnamefont {Lichtenstein}},
  \bibinfo {author} {\bibfnamefont {M.~I.}\ \bibnamefont {Katsnelson}}, \ and\
  \bibinfo {author} {\bibfnamefont {E.~A.}\ \bibnamefont {Stepanov}},\
  }\bibfield  {title} {\enquote {\bibinfo {title} {Dual boson approach with
  instantaneous interaction},}\ }\href {\doibase 10.1103/PhysRevB.100.165128}
  {\bibfield  {journal} {\bibinfo  {journal} {Phys. Rev. B}\ }\textbf {\bibinfo
  {volume} {100}},\ \bibinfo {pages} {165128} (\bibinfo {year}
  {2019})}\BibitemShut {NoStop}%
\bibitem [{\citenamefont {van Loon}\ \emph
  {et~al.}(2014{\natexlab{b}})\citenamefont {van Loon}, \citenamefont
  {Hafermann}, \citenamefont {Lichtenstein}, \citenamefont {Rubtsov},\ and\
  \citenamefont {Katsnelson}}]{vanLoon2014}%
  \BibitemOpen
  \bibfield  {author} {\bibinfo {author} {\bibfnamefont {E.~G. C.~P.}\
  \bibnamefont {van Loon}}, \bibinfo {author} {\bibfnamefont {H.}~\bibnamefont
  {Hafermann}}, \bibinfo {author} {\bibfnamefont {A.~I.}\ \bibnamefont
  {Lichtenstein}}, \bibinfo {author} {\bibfnamefont {A.~N.}\ \bibnamefont
  {Rubtsov}}, \ and\ \bibinfo {author} {\bibfnamefont {M.~I.}\ \bibnamefont
  {Katsnelson}},\ }\bibfield  {title} {\enquote {\bibinfo {title} {{Plasmons in
  Strongly Correlated Systems: Spectral Weight Transfer and Renormalized
  Dispersion}},}\ }\href {\doibase 10.1103/PhysRevLett.113.246407} {\bibfield
  {journal} {\bibinfo  {journal} {Phys. Rev. Lett.}\ }\textbf {\bibinfo
  {volume} {113}},\ \bibinfo {pages} {246407} (\bibinfo {year}
  {2014}{\natexlab{b}})}\BibitemShut {NoStop}%
\bibitem [{\citenamefont {Stepanov}\ \emph
  {et~al.}(2019{\natexlab{b}})\citenamefont {Stepanov}, \citenamefont {Huber},
  \citenamefont {Lichtenstein},\ and\ \citenamefont
  {Katsnelson}}]{Stepanov2019a}%
  \BibitemOpen
  \bibfield  {author} {\bibinfo {author} {\bibfnamefont {E.~A.}\ \bibnamefont
  {Stepanov}}, \bibinfo {author} {\bibfnamefont {A.}~\bibnamefont {Huber}},
  \bibinfo {author} {\bibfnamefont {A.~I.}\ \bibnamefont {Lichtenstein}}, \
  and\ \bibinfo {author} {\bibfnamefont {M.~I.}\ \bibnamefont {Katsnelson}},\
  }\bibfield  {title} {\enquote {\bibinfo {title} {Effective ising model for
  correlated systems with charge ordering},}\ }\href {\doibase
  10.1103/PhysRevB.99.115124} {\bibfield  {journal} {\bibinfo  {journal} {Phys.
  Rev. B}\ }\textbf {\bibinfo {volume} {99}},\ \bibinfo {pages} {115124}
  (\bibinfo {year} {2019}{\natexlab{b}})}\BibitemShut {NoStop}%
\bibitem [{\citenamefont {Stepanov}\ \emph {et~al.}(2018)\citenamefont
  {Stepanov}, \citenamefont {Brener}, \citenamefont {Krien}, \citenamefont
  {Harland}, \citenamefont {Lichtenstein},\ and\ \citenamefont
  {Katsnelson}}]{Stepanov2018}%
  \BibitemOpen
  \bibfield  {author} {\bibinfo {author} {\bibfnamefont {E.~A.}\ \bibnamefont
  {Stepanov}}, \bibinfo {author} {\bibfnamefont {S.}~\bibnamefont {Brener}},
  \bibinfo {author} {\bibfnamefont {F.}~\bibnamefont {Krien}}, \bibinfo
  {author} {\bibfnamefont {M.}~\bibnamefont {Harland}}, \bibinfo {author}
  {\bibfnamefont {A.~I.}\ \bibnamefont {Lichtenstein}}, \ and\ \bibinfo
  {author} {\bibfnamefont {M.~I.}\ \bibnamefont {Katsnelson}},\ }\bibfield
  {title} {\enquote {\bibinfo {title} {{Effective Heisenberg Model and Exchange
  Interaction for Strongly Correlated Systems}},}\ }\href {\doibase
  10.1103/PhysRevLett.121.037204} {\bibfield  {journal} {\bibinfo  {journal}
  {Phys. Rev. Lett.}\ }\textbf {\bibinfo {volume} {121}},\ \bibinfo {pages}
  {037204} (\bibinfo {year} {2018})}\BibitemShut {NoStop}%
\bibitem [{\citenamefont {{Stepanov}}\ \emph {et~al.}(2018)\citenamefont
  {{Stepanov}}, \citenamefont {{Peters}}, \citenamefont {{Krivenko}},
  \citenamefont {{Lichtenstein}}, \citenamefont {{Katsnelson}},\ and\
  \citenamefont {{Rubtsov}}}]{Stepanov2018b}%
  \BibitemOpen
  \bibfield  {author} {\bibinfo {author} {\bibfnamefont {E.~A.}\ \bibnamefont
  {{Stepanov}}}, \bibinfo {author} {\bibfnamefont {L.}~\bibnamefont
  {{Peters}}}, \bibinfo {author} {\bibfnamefont {I.~S.}\ \bibnamefont
  {{Krivenko}}}, \bibinfo {author} {\bibfnamefont {A.~I.}\ \bibnamefont
  {{Lichtenstein}}}, \bibinfo {author} {\bibfnamefont {M.~I.}\ \bibnamefont
  {{Katsnelson}}}, \ and\ \bibinfo {author} {\bibfnamefont {A.~N.}\
  \bibnamefont {{Rubtsov}}},\ }\bibfield  {title} {\enquote {\bibinfo {title}
  {{Quantum spin fluctuations and evolution of electronic structure in
  cuprates}},}\ }\href {\doibase 10.1038/s41535-018-0128-x} {\bibfield
  {journal} {\bibinfo  {journal} {npj Quantum Materials}\ }\textbf {\bibinfo
  {volume} {3}},\ \bibinfo {eid} {54} (\bibinfo {year} {2018})},\ \Eprint
  {http://arxiv.org/abs/1806.05216} {arXiv:1806.05216 [cond-mat.str-el]}
  \BibitemShut {NoStop}%
\bibitem [{\citenamefont {Hafermann}\ \emph {et~al.}(2014)\citenamefont
  {Hafermann}, \citenamefont {van Loon}, \citenamefont {Katsnelson},
  \citenamefont {Lichtenstein},\ and\ \citenamefont
  {Parcollet}}]{Hafermann2014a}%
  \BibitemOpen
  \bibfield  {author} {\bibinfo {author} {\bibfnamefont {Hartmut}\ \bibnamefont
  {Hafermann}}, \bibinfo {author} {\bibfnamefont {Erik G. C.~P.}\ \bibnamefont
  {van Loon}}, \bibinfo {author} {\bibfnamefont {Mikhail~I.}\ \bibnamefont
  {Katsnelson}}, \bibinfo {author} {\bibfnamefont {Alexander~I.}\ \bibnamefont
  {Lichtenstein}}, \ and\ \bibinfo {author} {\bibfnamefont {Olivier}\
  \bibnamefont {Parcollet}},\ }\bibfield  {title} {\enquote {\bibinfo {title}
  {Collective charge excitations of strongly correlated electrons, vertex
  corrections, and gauge invariance},}\ }\href {\doibase
  10.1103/PhysRevB.90.235105} {\bibfield  {journal} {\bibinfo  {journal} {Phys.
  Rev. B}\ }\textbf {\bibinfo {volume} {90}},\ \bibinfo {pages} {235105}
  (\bibinfo {year} {2014})}\BibitemShut {NoStop}%
\bibitem [{\citenamefont {Hafermann}\ \emph {et~al.}(2013)\citenamefont
  {Hafermann}, \citenamefont {Werner},\ and\ \citenamefont
  {Gull}}]{Hafermann2013}%
  \BibitemOpen
  \bibfield  {author} {\bibinfo {author} {\bibfnamefont {Hartmut}\ \bibnamefont
  {Hafermann}}, \bibinfo {author} {\bibfnamefont {Philipp}\ \bibnamefont
  {Werner}}, \ and\ \bibinfo {author} {\bibfnamefont {Emanuel}\ \bibnamefont
  {Gull}},\ }\bibfield  {title} {\enquote {\bibinfo {title} {Efficient
  implementation of the continuous-time hybridization expansion quantum
  impurity solver},}\ }\href {\doibase 10.1016/j.cpc.2012.12.013} {\bibfield
  {journal} {\bibinfo  {journal} {Computer Physics Communications}\ }\textbf
  {\bibinfo {volume} {184}},\ \bibinfo {pages} {1280 -- 1286} (\bibinfo {year}
  {2013})}\BibitemShut {NoStop}%
\bibitem [{\citenamefont {Hafermann}(2014)}]{Hafermann2014}%
  \BibitemOpen
  \bibfield  {author} {\bibinfo {author} {\bibfnamefont {Hartmut}\ \bibnamefont
  {Hafermann}},\ }\bibfield  {title} {\enquote {\bibinfo {title} {{Self-energy
  and vertex functions from hybridization-expansion continuous-time quantum
  Monte Carlo for impurity models with retarded interaction}},}\ }\href
  {\doibase 10.1103/PhysRevB.89.235128} {\bibfield  {journal} {\bibinfo
  {journal} {Phys. Rev. B}\ }\textbf {\bibinfo {volume} {89}},\ \bibinfo
  {pages} {235128} (\bibinfo {year} {2014})}\BibitemShut {NoStop}%
\bibitem [{\citenamefont {Bauer}\ \emph {et~al.}(2011)\citenamefont {Bauer},
  \citenamefont {Carr}, \citenamefont {Evertz}, \citenamefont {Feiguin},
  \citenamefont {Freire}, \citenamefont {Fuchs}, \citenamefont {Gamper},
  \citenamefont {Gukelberger}, \citenamefont {Gull}, \citenamefont {Guertler},
  \citenamefont {Hehn}, \citenamefont {Igarashi}, \citenamefont {Isakov},
  \citenamefont {Koop}, \citenamefont {Ma}, \citenamefont {Mates},
  \citenamefont {Matsuo}, \citenamefont {Parcollet}, \citenamefont
  {Pawłowski}, \citenamefont {Picon}, \citenamefont {Pollet}, \citenamefont
  {Santos}, \citenamefont {Scarola}, \citenamefont {Schollwöck}, \citenamefont
  {Silva}, \citenamefont {Surer}, \citenamefont {Todo}, \citenamefont {Trebst},
  \citenamefont {Troyer}, \citenamefont {Wall}, \citenamefont {Werner},\ and\
  \citenamefont {Wessel}}]{ALPS2}%
  \BibitemOpen
  \bibfield  {author} {\bibinfo {author} {\bibfnamefont {B}~\bibnamefont
  {Bauer}}, \bibinfo {author} {\bibfnamefont {L~D}\ \bibnamefont {Carr}},
  \bibinfo {author} {\bibfnamefont {H~G}\ \bibnamefont {Evertz}}, \bibinfo
  {author} {\bibfnamefont {A}~\bibnamefont {Feiguin}}, \bibinfo {author}
  {\bibfnamefont {J}~\bibnamefont {Freire}}, \bibinfo {author} {\bibfnamefont
  {S}~\bibnamefont {Fuchs}}, \bibinfo {author} {\bibfnamefont {L}~\bibnamefont
  {Gamper}}, \bibinfo {author} {\bibfnamefont {J}~\bibnamefont {Gukelberger}},
  \bibinfo {author} {\bibfnamefont {E}~\bibnamefont {Gull}}, \bibinfo {author}
  {\bibfnamefont {S}~\bibnamefont {Guertler}}, \bibinfo {author} {\bibfnamefont
  {A}~\bibnamefont {Hehn}}, \bibinfo {author} {\bibfnamefont {R}~\bibnamefont
  {Igarashi}}, \bibinfo {author} {\bibfnamefont {S~V}\ \bibnamefont {Isakov}},
  \bibinfo {author} {\bibfnamefont {D}~\bibnamefont {Koop}}, \bibinfo {author}
  {\bibfnamefont {P~N}\ \bibnamefont {Ma}}, \bibinfo {author} {\bibfnamefont
  {P}~\bibnamefont {Mates}}, \bibinfo {author} {\bibfnamefont {H}~\bibnamefont
  {Matsuo}}, \bibinfo {author} {\bibfnamefont {O}~\bibnamefont {Parcollet}},
  \bibinfo {author} {\bibfnamefont {G}~\bibnamefont {Pawłowski}}, \bibinfo
  {author} {\bibfnamefont {J~D}\ \bibnamefont {Picon}}, \bibinfo {author}
  {\bibfnamefont {L}~\bibnamefont {Pollet}}, \bibinfo {author} {\bibfnamefont
  {E}~\bibnamefont {Santos}}, \bibinfo {author} {\bibfnamefont {V~W}\
  \bibnamefont {Scarola}}, \bibinfo {author} {\bibfnamefont {U}~\bibnamefont
  {Schollwöck}}, \bibinfo {author} {\bibfnamefont {C}~\bibnamefont {Silva}},
  \bibinfo {author} {\bibfnamefont {B}~\bibnamefont {Surer}}, \bibinfo {author}
  {\bibfnamefont {S}~\bibnamefont {Todo}}, \bibinfo {author} {\bibfnamefont
  {S}~\bibnamefont {Trebst}}, \bibinfo {author} {\bibfnamefont {M}~\bibnamefont
  {Troyer}}, \bibinfo {author} {\bibfnamefont {M~L}\ \bibnamefont {Wall}},
  \bibinfo {author} {\bibfnamefont {P}~\bibnamefont {Werner}}, \ and\ \bibinfo
  {author} {\bibfnamefont {S}~\bibnamefont {Wessel}},\ }\bibfield  {title}
  {\enquote {\bibinfo {title} {{The ALPS project release 2.0: open source
  software for strongly correlated systems}},}\ }\href
  {http://stacks.iop.org/1742-5468/2011/i=05/a=P05001} {\bibfield  {journal}
  {\bibinfo  {journal} {Journal of Statistical Mechanics: Theory and
  Experiment}\ }\textbf {\bibinfo {volume} {2011}},\ \bibinfo {pages} {P05001}
  (\bibinfo {year} {2011})}\BibitemShut {NoStop}%
\bibitem [{\citenamefont {Vilk}\ \emph {et~al.}(1994)\citenamefont {Vilk},
  \citenamefont {Chen},\ and\ \citenamefont {Tremblay}}]{Vilk1994}%
  \BibitemOpen
  \bibfield  {author} {\bibinfo {author} {\bibfnamefont {Y.~M.}\ \bibnamefont
  {Vilk}}, \bibinfo {author} {\bibfnamefont {Liang}\ \bibnamefont {Chen}}, \
  and\ \bibinfo {author} {\bibfnamefont {A.-M.~S.}\ \bibnamefont {Tremblay}},\
  }\bibfield  {title} {\enquote {\bibinfo {title} {{Theory of spin and charge
  fluctuations in the Hubbard model}},}\ }\href {\doibase
  10.1103/PhysRevB.49.13267} {\bibfield  {journal} {\bibinfo  {journal} {Phys.
  Rev. B}\ }\textbf {\bibinfo {volume} {49}},\ \bibinfo {pages} {13267--13270}
  (\bibinfo {year} {1994})}\BibitemShut {NoStop}%
\bibitem [{\citenamefont {Singwi}\ and\ \citenamefont
  {Tosi}(1981)}]{Singwi1981}%
  \BibitemOpen
  \bibfield  {author} {\bibinfo {author} {\bibfnamefont {K.~S.}\ \bibnamefont
  {Singwi}}\ and\ \bibinfo {author} {\bibfnamefont {M.P.}\ \bibnamefont
  {Tosi}},\ }\href@noop {} {\emph {\bibinfo {title} {Solid State Physics}}},\
  edited by\ \bibinfo {editor} {\bibfnamefont {H.}~\bibnamefont {Ehrenreich}},
  \bibinfo {editor} {\bibfnamefont {F.}~\bibnamefont {Seitz}}, \ and\ \bibinfo
  {editor} {\bibfnamefont {D.}~\bibnamefont {Turnbull}}\ (\bibinfo  {publisher}
  {Academic, New York},\ \bibinfo {year} {1981})\BibitemShut {NoStop}%
\bibitem [{\citenamefont {Ichimaru}(1982)}]{Ichimaru1982}%
  \BibitemOpen
  \bibfield  {author} {\bibinfo {author} {\bibfnamefont {Setsuo}\ \bibnamefont
  {Ichimaru}},\ }\bibfield  {title} {\enquote {\bibinfo {title} {Strongly
  coupled plasmas: high-density classical plasmas and degenerate electron
  liquids},}\ }\href {\doibase 10.1103/RevModPhys.54.1017} {\bibfield
  {journal} {\bibinfo  {journal} {Rev. Mod. Phys.}\ }\textbf {\bibinfo {volume}
  {54}},\ \bibinfo {pages} {1017--1059} (\bibinfo {year} {1982})}\BibitemShut
  {NoStop}%
\bibitem [{\citenamefont {Hedayati}\ and\ \citenamefont
  {Vignale}(1989)}]{Hedeyati1989}%
  \BibitemOpen
  \bibfield  {author} {\bibinfo {author} {\bibfnamefont {M.~R.}\ \bibnamefont
  {Hedayati}}\ and\ \bibinfo {author} {\bibfnamefont {G.}~\bibnamefont
  {Vignale}},\ }\bibfield  {title} {\enquote {\bibinfo {title} {{Ground-state
  energy of the one- and two-dimensional Hubbard model calculated by the method
  of Singwi, Tosi, Land, and Sj\"olander}},}\ }\href {\doibase
  10.1103/PhysRevB.40.9044} {\bibfield  {journal} {\bibinfo  {journal} {Phys.
  Rev. B}\ }\textbf {\bibinfo {volume} {40}},\ \bibinfo {pages} {9044--9051}
  (\bibinfo {year} {1989})}\BibitemShut {NoStop}%
\bibitem [{\citenamefont {Allen}\ \emph {et~al.}(2003)\citenamefont {Allen},
  \citenamefont {Tremblay},\ and\ \citenamefont {Vilk}}]{Allen2003}%
  \BibitemOpen
  \bibfield  {author} {\bibinfo {author} {\bibfnamefont {S.}~\bibnamefont
  {Allen}}, \bibinfo {author} {\bibfnamefont {A.-M.~S.}\ \bibnamefont
  {Tremblay}}, \ and\ \bibinfo {author} {\bibfnamefont {Y.~M.}\ \bibnamefont
  {Vilk}},\ }\bibfield  {title} {\enquote {\bibinfo {title} {Conserving
  approximations vs two-particle self-consistent approach},}\ }in\ \href@noop
  {} {\emph {\bibinfo {booktitle} {Theoretical Methods for Strongly Correlated
  Electrons}}},\ \bibinfo {editor} {edited by\ \bibinfo {editor} {\bibfnamefont
  {D.}~\bibnamefont {S{\'e}n{\'e}chal}}, \bibinfo {editor} {\bibfnamefont
  {C.}~\bibnamefont {Bourbonnais}}, \ and\ \bibinfo {editor} {\bibfnamefont
  {A.-M.~S.}\ \bibnamefont {Tremblay}}}\ (\bibinfo {year} {2003})\BibitemShut
  {NoStop}%
\bibitem [{\citenamefont {Kadanoff}\ and\ \citenamefont
  {Martin}(1961)}]{Kadanoff1962}%
  \BibitemOpen
  \bibfield  {author} {\bibinfo {author} {\bibfnamefont {Leo~P.}\ \bibnamefont
  {Kadanoff}}\ and\ \bibinfo {author} {\bibfnamefont {Paul~C.}\ \bibnamefont
  {Martin}},\ }\bibfield  {title} {\enquote {\bibinfo {title} {{Theory of
  Many-Particle Systems. II. Superconductivity}},}\ }\href {\doibase
  10.1103/PhysRev.124.670} {\bibfield  {journal} {\bibinfo  {journal} {Phys.
  Rev.}\ }\textbf {\bibinfo {volume} {124}},\ \bibinfo {pages} {670--697}
  (\bibinfo {year} {1961})}\BibitemShut {NoStop}%
\bibitem [{\citenamefont {Chen}\ \emph {et~al.}(2005)\citenamefont {Chen},
  \citenamefont {Stajic}, \citenamefont {Tan},\ and\ \citenamefont
  {Levin}}]{Chen_Stajic_Tan_Levin_2005}%
  \BibitemOpen
  \bibfield  {author} {\bibinfo {author} {\bibfnamefont {Q}~\bibnamefont
  {Chen}}, \bibinfo {author} {\bibfnamefont {J}~\bibnamefont {Stajic}},
  \bibinfo {author} {\bibfnamefont {S}~\bibnamefont {Tan}}, \ and\ \bibinfo
  {author} {\bibfnamefont {K}~\bibnamefont {Levin}},\ }\bibfield  {title}
  {\enquote {\bibinfo {title} {{BCS–BEC crossover: From high temperature
  superconductors to ultracold superfluids}},}\ }\href {\doibase
  10.1016/j.physrep.2005.02.005} {\bibfield  {journal} {\bibinfo  {journal}
  {Physics Reports}\ }\textbf {\bibinfo {volume} {412}},\ \bibinfo {pages}
  {1–88} (\bibinfo {year} {2005})}\BibitemShut {NoStop}%
\bibitem [{\citenamefont {Boyack}\ \emph {et~al.}(2018)\citenamefont {Boyack},
  \citenamefont {Chen}, \citenamefont {Varlamov},\ and\ \citenamefont
  {Levin}}]{Varlamov_Levin_2018}%
  \BibitemOpen
  \bibfield  {author} {\bibinfo {author} {\bibfnamefont {Rufus}\ \bibnamefont
  {Boyack}}, \bibinfo {author} {\bibfnamefont {Qijin}\ \bibnamefont {Chen}},
  \bibinfo {author} {\bibfnamefont {A.~A.}\ \bibnamefont {Varlamov}}, \ and\
  \bibinfo {author} {\bibfnamefont {K.}~\bibnamefont {Levin}},\ }\bibfield
  {title} {\enquote {\bibinfo {title} {Cuprate diamagnetism in the presence of
  a pseudogap: Beyond the standard fluctuation formalism},}\ }\href {\doibase
  10.1103/PhysRevB.97.064503} {\bibfield  {journal} {\bibinfo  {journal} {Phys.
  Rev. B}\ }\textbf {\bibinfo {volume} {97}},\ \bibinfo {pages} {064503}
  (\bibinfo {year} {2018})}\BibitemShut {NoStop}%
\bibitem [{\citenamefont {Salmhofer}(1999)}]{Salmhofer1999}%
  \BibitemOpen
  \bibfield  {author} {\bibinfo {author} {\bibfnamefont {M.}~\bibnamefont
  {Salmhofer}},\ }\href@noop {} {\emph {\bibinfo {title} {{Renormalization - An
  Introduction}}}},\ edited by\ \bibinfo {editor} {\bibfnamefont
  {R.}~\bibnamefont {Balian}}, \bibinfo {editor} {\bibfnamefont
  {W.}~\bibnamefont {Beiglb\"ock}}, \bibinfo {editor} {\bibfnamefont
  {H.}~\bibnamefont {Grosse}}, \bibinfo {editor} {\bibfnamefont {E.~H.}\
  \bibnamefont {Lieb}}, \bibinfo {editor} {\bibfnamefont {N.}~\bibnamefont
  {Reshetikhin}}, \bibinfo {editor} {\bibfnamefont {H.}~\bibnamefont {Spohn}},
  \ and\ \bibinfo {editor} {\bibfnamefont {W.}~\bibnamefont {Thirring}}\
  (\bibinfo  {publisher} {Springer-Verlag Berlin Heidelberg},\ \bibinfo {year}
  {1999})\BibitemShut {NoStop}%
\bibitem [{\citenamefont {Berges}\ \emph {et~al.}(2002)\citenamefont {Berges},
  \citenamefont {Tetradis},\ and\ \citenamefont {Wetterich}}]{Berges2002}%
  \BibitemOpen
  \bibfield  {author} {\bibinfo {author} {\bibfnamefont {Jürgen}\ \bibnamefont
  {Berges}}, \bibinfo {author} {\bibfnamefont {Nikolaos}\ \bibnamefont
  {Tetradis}}, \ and\ \bibinfo {author} {\bibfnamefont {Christof}\ \bibnamefont
  {Wetterich}},\ }\bibfield  {title} {\enquote {\bibinfo {title}
  {Non-perturbative renormalization flow in quantum field theory and
  statistical physics},}\ }\href {\doibase 10.1016/s0370-1573(01)00098-9}
  {\bibfield  {journal} {\bibinfo  {journal} {Physics Reports}\ }\textbf
  {\bibinfo {volume} {363}},\ \bibinfo {pages} {223–386} (\bibinfo {year}
  {2002})}\BibitemShut {NoStop}%
\bibitem [{\citenamefont {Kopietz}\ \emph {et~al.}(2010)\citenamefont
  {Kopietz}, \citenamefont {Bartosch},\ and\ \citenamefont
  {Sch{\"u}tz}}]{Kopietz2010}%
  \BibitemOpen
  \bibfield  {author} {\bibinfo {author} {\bibfnamefont {Peter}\ \bibnamefont
  {Kopietz}}, \bibinfo {author} {\bibfnamefont {Lorenz}\ \bibnamefont
  {Bartosch}}, \ and\ \bibinfo {author} {\bibfnamefont {Florian}\ \bibnamefont
  {Sch{\"u}tz}},\ }\href {\doibase 10.1007/978-3-642-05094-7} {\emph {\bibinfo
  {title} {{Introduction to the Functional Renormalization Group, Lecture Notes
  in Physics 798}}}}\ (\bibinfo  {publisher} {Springer-Verlag Berlin
  Heidelberg},\ \bibinfo {year} {2010})\BibitemShut {NoStop}%
\bibitem [{\citenamefont {Deisz}\ \emph {et~al.}(1996)\citenamefont {Deisz},
  \citenamefont {Hess},\ and\ \citenamefont {Serene}}]{Deisz1996}%
  \BibitemOpen
  \bibfield  {author} {\bibinfo {author} {\bibfnamefont {J.~J.}\ \bibnamefont
  {Deisz}}, \bibinfo {author} {\bibfnamefont {D.~W.}\ \bibnamefont {Hess}}, \
  and\ \bibinfo {author} {\bibfnamefont {J.~W.}\ \bibnamefont {Serene}},\
  }\bibfield  {title} {\enquote {\bibinfo {title} {{Incipient
  Antiferromagnetism and Low-Energy Excitations in the Half-Filled
  Two-Dimensional Hubbard Model}},}\ }\href {\doibase
  10.1103/physrevlett.76.1312} {\bibfield  {journal} {\bibinfo  {journal}
  {Physical Review Letters}\ }\textbf {\bibinfo {volume} {76}},\ \bibinfo
  {pages} {1312} (\bibinfo {year} {1996})}\BibitemShut {NoStop}%
\bibitem [{\citenamefont {Katanin}\ and\ \citenamefont
  {Kampf}(2004)}]{Katanin2004b}%
  \BibitemOpen
  \bibfield  {author} {\bibinfo {author} {\bibfnamefont {A.~A.}\ \bibnamefont
  {Katanin}}\ and\ \bibinfo {author} {\bibfnamefont {A.~P.}\ \bibnamefont
  {Kampf}},\ }\bibfield  {title} {\enquote {\bibinfo {title} {{Quasiparticle
  Anisotropy and Pseudogap Formation from the Weak-Coupling Renormalization
  Group Point of View}},}\ }\href {\doibase 10.1103/PhysRevLett.93.106406}
  {\bibfield  {journal} {\bibinfo  {journal} {Phys. Rev. Lett.}\ }\textbf
  {\bibinfo {volume} {93}},\ \bibinfo {pages} {106406} (\bibinfo {year}
  {2004})}\BibitemShut {NoStop}%
\bibitem [{\citenamefont {Rohe}\ and\ \citenamefont
  {Metzner}(2005)}]{Rohe2005}%
  \BibitemOpen
  \bibfield  {author} {\bibinfo {author} {\bibfnamefont {Daniel}\ \bibnamefont
  {Rohe}}\ and\ \bibinfo {author} {\bibfnamefont {Walter}\ \bibnamefont
  {Metzner}},\ }\bibfield  {title} {\enquote {\bibinfo {title} {{Pseudogap at
  hot spots in the two-dimensional Hubbard model at weak coupling}},}\ }\href
  {\doibase 10.1103/PhysRevB.71.115116} {\bibfield  {journal} {\bibinfo
  {journal} {Phys. Rev. B}\ }\textbf {\bibinfo {volume} {71}},\ \bibinfo
  {pages} {115116} (\bibinfo {year} {2005})}\BibitemShut {NoStop}%
\bibitem [{\citenamefont {Kampf}\ and\ \citenamefont
  {Katanin}(2006)}]{Kampf2006}%
  \BibitemOpen
  \bibfield  {author} {\bibinfo {author} {\bibfnamefont {A.}~\bibnamefont
  {Kampf}}\ and\ \bibinfo {author} {\bibfnamefont {A.A.}\ \bibnamefont
  {Katanin}},\ }\bibfield  {title} {\enquote {\bibinfo {title} {{Quasiparticle
  anisotropy and pseudogap formation: A weak-coupling renormalization-group
  analysis}},}\ }\href {\doibase https://doi.org/10.1016/j.jpcs.2005.10.044}
  {\bibfield  {journal} {\bibinfo  {journal} {Journal of Physics and Chemistry
  of Solids}\ }\textbf {\bibinfo {volume} {67}},\ \bibinfo {pages} {146 -- 149}
  (\bibinfo {year} {2006})},\ \bibinfo {note} {spectroscopies in Novel
  Superconductors 2004}\BibitemShut {NoStop}%
\bibitem [{Note2()}]{Note2}%
  \BibitemOpen
  \bibinfo {note} {Strongly correlated parameter regimes are beyond the 1-loop
  flow but might become accessible by exploiting the DMFT as a starting point
  for the fRG flow \cite {Taranto2014,Vilardi2019}.}\BibitemShut {Stop}%
\bibitem [{\citenamefont {Husemann}\ and\ \citenamefont
  {Salmhofer}(2009)}]{Husemann2009}%
  \BibitemOpen
  \bibfield  {author} {\bibinfo {author} {\bibfnamefont {C.}~\bibnamefont
  {Husemann}}\ and\ \bibinfo {author} {\bibfnamefont {M.}~\bibnamefont
  {Salmhofer}},\ }\bibfield  {title} {\enquote {\bibinfo {title} {{Efficient
  parametrization of the vertex function, $\ensuremath{\Omega}$ scheme, and the
  $t,{t}^{\ensuremath{'}}$ Hubbard model at van Hove filling}},}\ }\href
  {\doibase 10.1103/PhysRevB.79.195125} {\bibfield  {journal} {\bibinfo
  {journal} {Phys. Rev. B}\ }\textbf {\bibinfo {volume} {79}},\ \bibinfo
  {pages} {195125} (\bibinfo {year} {2009})}\BibitemShut {NoStop}%
\bibitem [{\citenamefont {Wang}\ \emph {et~al.}(2012)\citenamefont {Wang},
  \citenamefont {Han}, \citenamefont {de' Medici}, \citenamefont {Park},
  \citenamefont {Marianetti},\ and\ \citenamefont {Millis}}]{Wang2012}%
  \BibitemOpen
  \bibfield  {author} {\bibinfo {author} {\bibfnamefont {Xin}\ \bibnamefont
  {Wang}}, \bibinfo {author} {\bibfnamefont {M.~J.}\ \bibnamefont {Han}},
  \bibinfo {author} {\bibfnamefont {Luca}\ \bibnamefont {de' Medici}}, \bibinfo
  {author} {\bibfnamefont {Hyowon}\ \bibnamefont {Park}}, \bibinfo {author}
  {\bibfnamefont {C.~A.}\ \bibnamefont {Marianetti}}, \ and\ \bibinfo {author}
  {\bibfnamefont {Andrew~J.}\ \bibnamefont {Millis}},\ }\bibfield  {title}
  {\enquote {\bibinfo {title} {Covalency, double-counting, and the
  metal-insulator phase diagram in transition metal oxides},}\ }\href {\doibase
  10.1103/PhysRevB.86.195136} {\bibfield  {journal} {\bibinfo  {journal} {Phys.
  Rev. B}\ }\textbf {\bibinfo {volume} {86}},\ \bibinfo {pages} {195136}
  (\bibinfo {year} {2012})}\BibitemShut {NoStop}%
\bibitem [{\citenamefont {Lichtenstein}\ \emph {et~al.}(2017)\citenamefont
  {Lichtenstein}, \citenamefont {{S{\'{a}}nchez de la Pe{\~{n}}a}},
  \citenamefont {Rohe}, \citenamefont {{Di Napoli}}, \citenamefont
  {Honerkamp},\ and\ \citenamefont {Maier}}]{Lichtenstein2017}%
  \BibitemOpen
  \bibfield  {author} {\bibinfo {author} {\bibfnamefont {J}~\bibnamefont
  {Lichtenstein}}, \bibinfo {author} {\bibfnamefont {D}~\bibnamefont
  {{S{\'{a}}nchez de la Pe{\~{n}}a}}}, \bibinfo {author} {\bibfnamefont
  {D}~\bibnamefont {Rohe}}, \bibinfo {author} {\bibfnamefont {E}~\bibnamefont
  {{Di Napoli}}}, \bibinfo {author} {\bibfnamefont {C}~\bibnamefont
  {Honerkamp}}, \ and\ \bibinfo {author} {\bibfnamefont {S.~A.}\ \bibnamefont
  {Maier}},\ }\bibfield  {title} {\enquote {\bibinfo {title} {{High-performance
  functional Renormalization Group calculations for interacting fermions}},}\
  }\href {\doibase 10.1016/j.cpc.2016.12.013} {\bibfield  {journal} {\bibinfo
  {journal} {Computer Physics Communications}\ }\textbf {\bibinfo {volume}
  {213}},\ \bibinfo {pages} {100--110} (\bibinfo {year} {2017})},\ \Eprint
  {http://arxiv.org/abs/1604.06296} {arXiv:1604.06296 [cond-mat.str-el]}
  \BibitemShut {NoStop}%
\bibitem [{\citenamefont {Vilardi}\ \emph {et~al.}(2017)\citenamefont
  {Vilardi}, \citenamefont {Taranto},\ and\ \citenamefont
  {Metzner}}]{Vilardi2017}%
  \BibitemOpen
  \bibfield  {author} {\bibinfo {author} {\bibfnamefont {Demetrio}\
  \bibnamefont {Vilardi}}, \bibinfo {author} {\bibfnamefont {Ciro}\
  \bibnamefont {Taranto}}, \ and\ \bibinfo {author} {\bibfnamefont {Walter}\
  \bibnamefont {Metzner}},\ }\bibfield  {title} {\enquote {\bibinfo {title}
  {{Nonseparable frequency dependence of the two-particle vertex in interacting
  fermion systems}},}\ }\href {\doibase 10.1103/PhysRevB.96.235110} {\bibfield
  {journal} {\bibinfo  {journal} {Physical Review B}\ }\textbf {\bibinfo
  {volume} {96}},\ \bibinfo {pages} {235110} (\bibinfo {year} {2017})},\
  \Eprint {http://arxiv.org/abs/1708.03539} {arXiv:1708.03539} \BibitemShut
  {NoStop}%
\bibitem [{\citenamefont {Katanin}(2004)}]{Katanin2004}%
  \BibitemOpen
  \bibfield  {author} {\bibinfo {author} {\bibfnamefont {A.~A.}\ \bibnamefont
  {Katanin}},\ }\bibfield  {title} {\enquote {\bibinfo {title} {{Fulfillment of
  Ward identities in the functional renormalization group approach}},}\ }\href
  {\doibase 10.1103/PhysRevB.70.115109} {\bibfield  {journal} {\bibinfo
  {journal} {Phys. Rev. B}\ }\textbf {\bibinfo {volume} {70}},\ \bibinfo
  {pages} {115109} (\bibinfo {year} {2004})}\BibitemShut {NoStop}%
\bibitem [{\citenamefont {{De Dominicis}}\ and\ \citenamefont
  {Martin}(1964{\natexlab{b}})}]{DeDominicis1964}%
  \BibitemOpen
  \bibfield  {author} {\bibinfo {author} {\bibfnamefont {Cyrano}\ \bibnamefont
  {{De Dominicis}}}\ and\ \bibinfo {author} {\bibfnamefont {Paul~C.}\
  \bibnamefont {Martin}},\ }\bibfield  {title} {\enquote {\bibinfo {title}
  {{Stationary Entropy Principle and Renormalization in Normal and Superfluid
  Systems. {I}. {Algebraic} Formulation}},}\ }\href {\doibase
  10.1063/1.1704062} {\bibfield  {journal} {\bibinfo  {journal} {J. Math.
  Phys.}\ }\textbf {\bibinfo {volume} {5}},\ \bibinfo {pages} {14--30}
  (\bibinfo {year} {1964}{\natexlab{b}})}\BibitemShut {NoStop}%
\bibitem [{\citenamefont {Vasil'ev}\ \emph {et~al.}(1974)\citenamefont
  {Vasil'ev}, \citenamefont {Kazanskii},\ and\ \citenamefont
  {Pis'mak}}]{Vasiliev74}%
  \BibitemOpen
  \bibfield  {author} {\bibinfo {author} {\bibfnamefont {A.~N.}\ \bibnamefont
  {Vasil'ev}}, \bibinfo {author} {\bibfnamefont {A.~K.}\ \bibnamefont
  {Kazanskii}}, \ and\ \bibinfo {author} {\bibfnamefont {Yu.~M.}\ \bibnamefont
  {Pis'mak}},\ }\bibfield  {title} {\enquote {\bibinfo {title} {{Diagrammatic
  analysis of the fourth Legendre transform}},}\ }\href {\doibase
  10.1007/BF01037327} {\bibfield  {journal} {\bibinfo  {journal} {Theoretical
  and Mathematical Physics}\ }\textbf {\bibinfo {volume} {20}},\ \bibinfo
  {pages} {754--762} (\bibinfo {year} {1974})}\BibitemShut {NoStop}%
\bibitem [{\citenamefont {Yang}\ \emph {et~al.}(2009)\citenamefont {Yang},
  \citenamefont {Fotso}, \citenamefont {Liu}, \citenamefont {Maier},
  \citenamefont {Tomko}, \citenamefont {D'Azevedo}, \citenamefont {Scalettar},
  \citenamefont {Pruschke},\ and\ \citenamefont {Jarrell}}]{Yang09}%
  \BibitemOpen
  \bibfield  {author} {\bibinfo {author} {\bibfnamefont {S.~X.}\ \bibnamefont
  {Yang}}, \bibinfo {author} {\bibfnamefont {H.}~\bibnamefont {Fotso}},
  \bibinfo {author} {\bibfnamefont {J.}~\bibnamefont {Liu}}, \bibinfo {author}
  {\bibfnamefont {T.~A.}\ \bibnamefont {Maier}}, \bibinfo {author}
  {\bibfnamefont {K.}~\bibnamefont {Tomko}}, \bibinfo {author} {\bibfnamefont
  {E.~F.}\ \bibnamefont {D'Azevedo}}, \bibinfo {author} {\bibfnamefont {R.~T.}\
  \bibnamefont {Scalettar}}, \bibinfo {author} {\bibfnamefont {T.}~\bibnamefont
  {Pruschke}}, \ and\ \bibinfo {author} {\bibfnamefont {M.}~\bibnamefont
  {Jarrell}},\ }\bibfield  {title} {\enquote {\bibinfo {title} {{Parquet
  approximation for the $4\ifmmode\times\else\texttimes\fi{}4$ Hubbard
  cluster}},}\ }\href {\doibase 10.1103/PhysRevE.80.046706} {\bibfield
  {journal} {\bibinfo  {journal} {Phys. Rev. E}\ }\textbf {\bibinfo {volume}
  {80}},\ \bibinfo {pages} {046706} (\bibinfo {year} {2009})}\BibitemShut
  {NoStop}%
\bibitem [{\citenamefont {Tam}\ \emph {et~al.}(2013)\citenamefont {Tam},
  \citenamefont {Fotso}, \citenamefont {Yang}, \citenamefont {Lee},
  \citenamefont {Moreno}, \citenamefont {Ramanujam},\ and\ \citenamefont
  {Jarrell}}]{Tam2013}%
  \BibitemOpen
  \bibfield  {author} {\bibinfo {author} {\bibfnamefont {Ka-Ming}\ \bibnamefont
  {Tam}}, \bibinfo {author} {\bibfnamefont {H.}~\bibnamefont {Fotso}}, \bibinfo
  {author} {\bibfnamefont {S.-X.}\ \bibnamefont {Yang}}, \bibinfo {author}
  {\bibfnamefont {Tae-Woo}\ \bibnamefont {Lee}}, \bibinfo {author}
  {\bibfnamefont {J.}~\bibnamefont {Moreno}}, \bibinfo {author} {\bibfnamefont
  {J.}~\bibnamefont {Ramanujam}}, \ and\ \bibinfo {author} {\bibfnamefont
  {M.}~\bibnamefont {Jarrell}},\ }\bibfield  {title} {\enquote {\bibinfo
  {title} {{Solving the parquet equations for the Hubbard model beyond weak
  coupling}},}\ }\href {\doibase 10.1103/PhysRevE.87.013311} {\bibfield
  {journal} {\bibinfo  {journal} {Phys. Rev. E}\ }\textbf {\bibinfo {volume}
  {87}},\ \bibinfo {pages} {013311} (\bibinfo {year} {2013})}\BibitemShut
  {NoStop}%
\bibitem [{\citenamefont {Li}\ \emph {et~al.}(2019)\citenamefont {Li},
  \citenamefont {Kauch}, \citenamefont {Pudleiner},\ and\ \citenamefont
  {Held}}]{Li19}%
  \BibitemOpen
  \bibfield  {author} {\bibinfo {author} {\bibfnamefont {Gang}\ \bibnamefont
  {Li}}, \bibinfo {author} {\bibfnamefont {Anna}\ \bibnamefont {Kauch}},
  \bibinfo {author} {\bibfnamefont {Petra}\ \bibnamefont {Pudleiner}}, \ and\
  \bibinfo {author} {\bibfnamefont {Karsten}\ \bibnamefont {Held}},\ }\bibfield
   {title} {\enquote {\bibinfo {title} {The victory project v1.0: An efficient
  parquet equations solver},}\ }\href {\doibase
  https://doi.org/10.1016/j.cpc.2019.03.008} {\bibfield  {journal} {\bibinfo
  {journal} {Computer Physics Communications}\ }\textbf {\bibinfo {volume}
  {241}},\ \bibinfo {pages} {146 -- 154} (\bibinfo {year} {2019})}\BibitemShut
  {NoStop}%
\bibitem [{\citenamefont {Pudleiner}\ \emph
  {et~al.}(2019{\natexlab{a}})\citenamefont {Pudleiner}, \citenamefont
  {Thunstr\"om}, \citenamefont {Valli}, \citenamefont {Kauch}, \citenamefont
  {Li},\ and\ \citenamefont {Held}}]{Pudleiner19a}%
  \BibitemOpen
  \bibfield  {author} {\bibinfo {author} {\bibfnamefont {P.}~\bibnamefont
  {Pudleiner}}, \bibinfo {author} {\bibfnamefont {P.}~\bibnamefont
  {Thunstr\"om}}, \bibinfo {author} {\bibfnamefont {A.}~\bibnamefont {Valli}},
  \bibinfo {author} {\bibfnamefont {A.}~\bibnamefont {Kauch}}, \bibinfo
  {author} {\bibfnamefont {G.}~\bibnamefont {Li}}, \ and\ \bibinfo {author}
  {\bibfnamefont {K.}~\bibnamefont {Held}},\ }\bibfield  {title} {\enquote
  {\bibinfo {title} {{Parquet approximation for molecules: Spectrum and optical
  conductivity of the Pariser-Parr-Pople model}},}\ }\href {\doibase
  10.1103/PhysRevB.99.125111} {\bibfield  {journal} {\bibinfo  {journal} {Phys.
  Rev. B}\ }\textbf {\bibinfo {volume} {99}},\ \bibinfo {pages} {125111}
  (\bibinfo {year} {2019}{\natexlab{a}})}\BibitemShut {NoStop}%
\bibitem [{\citenamefont {Pudleiner}\ \emph
  {et~al.}(2019{\natexlab{b}})\citenamefont {Pudleiner}, \citenamefont {Kauch},
  \citenamefont {Held},\ and\ \citenamefont {Li}}]{Pudleiner19b}%
  \BibitemOpen
  \bibfield  {author} {\bibinfo {author} {\bibfnamefont {Petra}\ \bibnamefont
  {Pudleiner}}, \bibinfo {author} {\bibfnamefont {Anna}\ \bibnamefont {Kauch}},
  \bibinfo {author} {\bibfnamefont {Karsten}\ \bibnamefont {Held}}, \ and\
  \bibinfo {author} {\bibfnamefont {Gang}\ \bibnamefont {Li}},\ }\bibfield
  {title} {\enquote {\bibinfo {title} {{Competition between antiferromagnetic
  and charge density wave fluctuations in the extended Hubbard model}},}\
  }\href {\doibase 10.1103/PhysRevB.100.075108} {\bibfield  {journal} {\bibinfo
   {journal} {Phys. Rev. B}\ }\textbf {\bibinfo {volume} {100}},\ \bibinfo
  {pages} {075108} (\bibinfo {year} {2019}{\natexlab{b}})}\BibitemShut
  {NoStop}%
\bibitem [{\citenamefont {Li}\ \emph {et~al.}(2016)\citenamefont {Li},
  \citenamefont {Wentzell}, \citenamefont {Pudleiner}, \citenamefont
  {Thunstr\"om},\ and\ \citenamefont {Held}}]{Li16}%
  \BibitemOpen
  \bibfield  {author} {\bibinfo {author} {\bibfnamefont {Gang}\ \bibnamefont
  {Li}}, \bibinfo {author} {\bibfnamefont {Nils}\ \bibnamefont {Wentzell}},
  \bibinfo {author} {\bibfnamefont {Petra}\ \bibnamefont {Pudleiner}}, \bibinfo
  {author} {\bibfnamefont {Patrik}\ \bibnamefont {Thunstr\"om}}, \ and\
  \bibinfo {author} {\bibfnamefont {Karsten}\ \bibnamefont {Held}},\ }\bibfield
   {title} {\enquote {\bibinfo {title} {{Efficient implementation of the
  parquet equations: Role of the reducible vertex function and its kernel
  approximation}},}\ }\href {\doibase 10.1103/PhysRevB.93.165103} {\bibfield
  {journal} {\bibinfo  {journal} {Phys. Rev. B}\ }\textbf {\bibinfo {volume}
  {93}},\ \bibinfo {pages} {165103} (\bibinfo {year} {2016})}\BibitemShut
  {NoStop}%
\bibitem [{\citenamefont {Kaufmann}\ \emph {et~al.}(2021)\citenamefont
  {Kaufmann}, \citenamefont {Eckhardt}, \citenamefont {Pickem}, \citenamefont
  {Kitatani}, \citenamefont {Kauch},\ and\ \citenamefont
  {Held}}]{Kaufmann2020}%
  \BibitemOpen
  \bibfield  {author} {\bibinfo {author} {\bibfnamefont {Josef}\ \bibnamefont
  {Kaufmann}}, \bibinfo {author} {\bibfnamefont {Christian}\ \bibnamefont
  {Eckhardt}}, \bibinfo {author} {\bibfnamefont {Matthias}\ \bibnamefont
  {Pickem}}, \bibinfo {author} {\bibfnamefont {Motoharu}\ \bibnamefont
  {Kitatani}}, \bibinfo {author} {\bibfnamefont {Anna}\ \bibnamefont {Kauch}},
  \ and\ \bibinfo {author} {\bibfnamefont {Karsten}\ \bibnamefont {Held}},\
  }\bibfield  {title} {\enquote {\bibinfo {title} {Self-consistent ladder
  dynamical vertex approximation},}\ }\href {\doibase
  10.1103/PhysRevB.103.035120} {\bibfield  {journal} {\bibinfo  {journal}
  {Phys. Rev. B}\ }\textbf {\bibinfo {volume} {103}},\ \bibinfo {pages}
  {035120} (\bibinfo {year} {2021})}\BibitemShut {NoStop}%
\bibitem [{\citenamefont {Krien}\ \emph {et~al.}(2021)\citenamefont {Krien},
  \citenamefont {Kauch},\ and\ \citenamefont {Held}}]{Krien2020b}%
  \BibitemOpen
  \bibfield  {author} {\bibinfo {author} {\bibfnamefont {Friedrich}\
  \bibnamefont {Krien}}, \bibinfo {author} {\bibfnamefont {Anna}\ \bibnamefont
  {Kauch}}, \ and\ \bibinfo {author} {\bibfnamefont {Karsten}\ \bibnamefont
  {Held}},\ }\bibfield  {title} {\enquote {\bibinfo {title} {{Tiling with
  triangles: parquet and $GW\ensuremath{\gamma}$ methods unified}},}\ }\href
  {\doibase 10.1103/PhysRevResearch.3.013149} {\bibfield  {journal} {\bibinfo
  {journal} {Phys. Rev. Research}\ }\textbf {\bibinfo {volume} {3}},\ \bibinfo
  {pages} {013149} (\bibinfo {year} {2021})}\BibitemShut {NoStop}%
\end{thebibliography}%
\end{document}